\documentclass[dvips,a4paper,11pt]{book}
\usepackage{epsfig}
\usepackage[utf8]{inputenc}
\usepackage[T1]{fontenc}
\usepackage{lmodern}
\usepackage{lscape}
\usepackage{esvect}
\usepackage{amsmath}
\usepackage{amssymb}
\usepackage{multirow}
\usepackage[frenchb]{babel}
\usepackage{natbib}
\usepackage{color}
\usepackage{wasysym}
\usepackage{fancyvrb}
\usepackage{array}

\newcommand{\aap}{Astronomy and Astrophysics}
\newcommand{\aj}{The Astronomical Journal}
\newcommand{\apj}{The Astrophysical Journal}
\newcommand{\apjl}{The Astrophysical Journal Letters}
\newcommand{\asr}{Advances in Space Research}
\newcommand{\cmda}{Celestial Mechanics and Dynamical Astronomy}
\newcommand{\celmec}{Celestial Mechanics}
\newcommand{\mnras}{The Monthly Notices of the Royal Astronomical Society}
\newcommand{\mnrasl}{The Monthly Notices of the Royal Astronomical Society Letters}
\newcommand{\pepi}{Physics of the Earth and Planetary Interiors}
\newcommand{\planss}{Planetary and Space Science}
\newcommand{\grl}{Geophysical Research Letters}
\newcommand{\jgr}{Journal of Geophysical Research}
\newcommand{\ssr}{Space Science Review}
\newcommand{\epsl}{Earth and Planetary Science Letters}
\newcommand{\dexp}[1]{\times 10^{#1}}

\def\sgn{\mathop{\rm sgn}\nolimits} 

\begin{document}

\thispagestyle{empty}

\includegraphics[scale=.5]{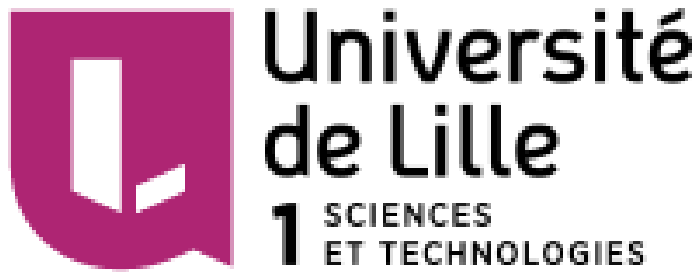} \hspace{8cm} N$^o$ d'ordre : 41602

{\large

\vspace*{1cm}

\begin{center}

{\bf TH\`ESE D'HABILITATION \`A DIRIGER DES RECHERCHES DE \\ L'UNIVERSIT\'E DE LILLE 1}

\vspace*{0.5cm}

Sp\'ecialit\'e \\ [2ex]
{\bf Astronomie}\ \\

\vspace*{1.5cm}

Pr\'esent\'ee par \ \\

\vspace*{0.5cm}

{\Large {\bf Beno\^it NOYELLES}}

\vspace*{2cm}

\end{center}

\begin{center}
{\Large {\bf Contribution \`a l'\'etude de la rotation r\'esonnante  \\ 
dans le Syst\`eme Solaire \\ }}
\end{center}

\vspace*{1.5cm} 
\flushleft{soutenue \`a l'Observatoire de Lille le 8 d\'ecembre 2014}\\[2ex]
\flushleft{devant le jury compos\'e de :  }\\[1ex]
\flushleft{\begin{tabular}{r@{\ }lll}
  & M. Hugues {\sc Leroux} & Pr\'esident & Universit\'e de Lille 1 \\  
  & M. Carl D. {\sc Murray} & Rapporteur & Queen Mary University of London \\
  & M. Jean {\sc Souchay} & Rapporteur  & Observatoire de Paris\\
  & M. Tim {\sc Van Hoolst} & Rapporteur & KU Leuven \\
  & Mme Anne {\sc Lema\^itre} & Examinateur & Universit\'e de Namur \\
  & M. Alain  {\sc Vienne} & Garant & Universit\'e de Lille 1 \\
\end{tabular}}

}



\tableofcontents

\listoffigures

\listoftables

\newpage

\thispagestyle{empty}

\begin{center}
\textbf{\Huge Remerciements}
\end{center}

\vspace{3cm}

\par Je tiens tout d'abord \`a remercier les membres du jury qui ont accepté d'examiner cette Habilitation \`a Diriger des Recherches. 
Il est de tradition de commencer par les rapporteurs, donc je remercie Carl Murray, Tim Van Hoolst et Jean Souchay d'avoir
accepté de rédiger un rapport sur ce document. Je remercie Anne Lema\^itre d'avoir accepté la charge d'examinateur(rice), 
Alain Vienne de s'\^etre porté garant auprès de l'Universit\'e de Lille 1 pour que je puisse m'inscrire en HDR, et Hugues
Leroux pour avoir pr\'esid\'e ce jury. Merci aussi \`a toutes les personnes qui se sont d\'eplac\'ees pour assister \`a 
cette soutenance. Et je n'oublie pas Damya Souami et Jonathan Besserer, qui se sont propos\'es pour relire le manuscrit\ldots
m\^eme si je ne les ai finalement pas sollicit\'es.

\par Cette recherche a pu se faire, tout au long de ces ann\'ees, gr\^ace aux sponsors. Donc merci \`a l'Universit\'e
de Namur, BELSPO et le FNRS de m'avoir financ\'e, ainsi qu'aux personnes qui m'ont permis d'obtenir ces financements. D'un point de 
vue scientifique, ces travaux sur la rotation ont \'et\'e faits \`a l'Universit\'e de Namur, dans la continuit\'e d'une tradition
locale initi\'ee par le regrett\'e Jacques Henrard, que j'ai eu le grand honneur de conna\^itre. Je remercie par cette occasion
mes collaborateurs locaux Anne Lema\^itre, Sandrine D'Hoedt, Julien Dufey, Nicolas Delsate, Timoteo Carletti, Christoph Lhotka
et Julien Frouard. Ces travaux ont \'et\'e rendus possibles notamment gr\^ace \`a l'\'enorme travail de Fr\'ed\'eric Wautelet sur
les serveurs de calcul. Je remercie \'egalement mes collaborateurs ext\'erieurs \"Ozg\"ur Karatekin, Nicolas Rambaux, Valeri Makarov, et Radwan Tajeddine. 
Enfin, une mention sp\'eciale pour les personnes qui m'ont permis d'effectuer plusieurs s\'ejours scientifiques \`a l'\'etranger, 
 Michael Efroimsky, Tadashi Yokoyama et Francis Nimmo.

\par L'Habilitation \`a Diriger des Recherches sanctionne toute une exp\'erience de chercheur, qui ne se limite pas \`a ce manuscrit.
J'ai aussi travaill\'e sur la dynamique orbitale des satellites naturels, d\`es ma th\`ese de doctorat, avant de le faire au sein du groupe
transversal Encelade. Merci en particulier \`a Val\'ery Lainey, S\'ebastien Charnoz et K\'evin Bailli\'e pour cette collaboration sur les anneaux de Saturne
(qui n'est pas encore termin\'ee). Merci \'egalement \`a \'Emilie Verheylewegen pour notre collaboration sur les satellites d'Uranus.

\par Cette soutenance a eu lieu dans un lieu particulier, l'Observatoire de Lille, 80 ans jour pour jour apr\`es son inauguration.
J'en profite pour rendre hommage aux b\'en\'evoles de l'Association Jonckh\`eere pour leur temps consacr\'e \`a l'entretien du 
patrimoine scientifique de ce b\^atiment, ainsi que pour toutes leurs observations \`a la lunette, utiles \`a la science (notamment
les ph\'enom\`enes mutuels). Un hommage aussi \`a Madame Ballenghien, la gardienne des lieux.

\par Et puis un grand merci \`a mes parents, qui m'ont toujours soutenu.

\newpage

\thispagestyle{empty}

\chapter{Introduction}

\par L'objet de cette th\`ese d'Habilitation \`a Diriger des Recherches est de pr\'esenter mes travaux sur la rotation des corps r\'esonnants, qui sont soit des satellites naturels des
plan\`etes g\'eantes, en rotation synchrone, soit Mercure, qui effectue 3 rotations sur elle-m\^eme en exactement 2 r\'evolutions autour du Soleil. Cette th\'ematique se situe \`a un 
carrefour entre la m\'ecanique c\'eleste, la plan\'etologie, et la g\'eophysique, car l'amplitude de la r\'eponse d'un corps en rotation \`a une excitation ext\'erieure, dans la plupart
des cas gravitationnelle, d\'epend de la structure interne de ce corps. Cette d\'ependance est d'autant plus importante dans le cas de corps en rotation r\'esonnante, car cette
configuration est un \'equilibre dynamique, c'est-\`a-dire un minimum d'\'energie. Ce minimum d'\'energie est atteint gr\^ace \`a des ph\'enom\`enes dissipatifs, essentiellement
des forces de mar\'ee et des ph\'enom\`enes de friction interne fluide-solide, agissant depuis la formation de l'objet. Ainsi, les d\'eviations par rapport \`a la rotation strictement 
r\'esonnante sont la signature d'excitation par une source externe, tandis que des corps non r\'esonnants, c'est-\`a-dire dont la rotation n'a pas \'et\'e enti\`erement amortie, 
ont un mouvement essentiellement d\^u aux ph\'enom\`enes s'\'etant produits pendant leur formation (accr\'etion, impacts,\ldots)

\par Mod\'eliser la rotation de tels objets n\'ecessite donc d'avoir des informations sur leur structure interne, ce qui n'\'etait pendant longtemps possible que pour la Lune. 
Ceci a justifi\'e un certain nombre d'\'etudes sur le sujet \citep{k1967,wsek1973,m1980,e1981,m1982a,m1982b,wbyrd2001,rw2011}, m\^elant aussi bien \'etudes th\'eoriques 
qu'observations de la rotation, en particulier gr\^ace \`a des dispositifs de type Laser-Lune, consistant \`a mesurer tr\`es pr\'ecis\'ement la distance entre un \'emetteur
laser et un panneau r\'eflecteur situ\'e sur la surface de la Lune, \`a l'aide du temps d'aller-retour des photons.

\par Ceci dit, la rotation de la Lune est tr\`es diff\'erente de celle des satellites naturels des plan\`etes g\'eantes, essentiellement car la Lune orbite autour d'une plan\`ete de 
taille relativement modeste (notre Terre) et subit fortement la perturbation gravitationnelle du Soleil. Une cons\'equence est que l'orientation de son moment cin\'etique, ou 
\emph{obliquit\'e}, correspond \`a l'\emph{\'Etat de Cassini 2}, alors que l'obliquit\'e des satellites des plan\`etes g\'eantes est suppos\'ee \^etre \`a l'\'Etat de Cassini 1.

\par Jusqu'\`a la mission spatiale Galileo qui a visit\'e le syst\`eme de Jupiter de 1995 \`a 2003, les satellites des plan\`etes g\'eantes n'\'etaient connus que comme des masses 
sph\'eriques avec certaines propri\'et\'es de surface; les sondes Galileo, puis Cassini en orbite autour de Saturne depuis 2004, ont permis de conna\^itre les champ de gravit\'e
\`a l'ordre 2 des principaux satellites, voire parfois de mesurer un \'etat de rotation. En parall\`ele, des \'etudes th\'eoriques et exp\'erimentales se sont d\'evelopp\'ees,
en particulier pour mod\'eliser l'int\'erieur d'un satellite naturel, pour \'etudier le comportement d'une couche fluide en son sein, ou pour comprendre la r\'eponse rotationnelle
d'un corps qui a un oc\'ean global.

\par Ce manuscrit a pour but de pr\'esenter ma contribution \`a certaines de ces \'etudes. Il s'agit du point de vue d'un m\'ecanicien c\'eleste qui s'est lentement mais s\^urement
rapproch\'e des sciences plan\'etaires. Je pr\'esenterai d'abord le cas le plus simple, c'est-\`a-dire celui de la rotation d'un corps triaxial rigide (Chapitre \ref{chap:rigide}), 
ainsi que la fa\c{c}on dont j'ai appliqu\'e ce mod\`ele \`a certains satellites de Jupiter et de Saturne (Chapitre \ref{chap:apprigide}). Je pr\'esenterai ensuite des complications 
de ce mod\`ele afin de se diriger vers une structure interne de plus en plus r\'ealiste, d'abord en consid\'erant un noyau enti\`erement fluide (Chapitre \ref{chap:poincarehough}), 
puis en tenant compte d'un oc\'ean global compris entre 2 couches rigides, \`a savoir un noyau et une cro\^ute (Chapitre \ref{chap:oceanglobal}).

\par Je montrerai ensuite comment j'ai appliqu\'e ces m\'ethodes \`a Mercure, en r\'esonance spin-orbite 3:2, dans le cadre de la pr\'eparation de la mission de l'ESA BepiColombo. 
J'explique en particulier comment on peut utiliser l'obliquit\'e (Chapitre \ref{chap:mercobliq}) et les librations en longitude (Chapitre \ref{chap:merclibra}) pour d\'eterminer
la taille du noyau fluide de la plan\`ete. Mon expertise sur Mercure m'a \'egalement amen\'e \`a collaborer avec une \'equipe de l'US Naval Observatory (Washington, DC, USA)
sp\'ecialis\'ee dans les effets de mar\'ee, pour comprendre la fa\c{c}on dont la plan\`ete a \'et\'e captur\'ee dans son actuelle r\'esonance spin-orbite (Chapitre \ref{chap:tides}).

\par Ayant \'et\'e form\'e dans un laboratoire sp\'ecialis\'e dans les \'eph\'em\'erides, je me suis toujours attach\'e \`a utiliser la mod\'elisation la plus r\'ealiste possible
du mouvement orbital du corps que j'\'etudiais, ceci notamment afin d'inclure toute r\'esonance \'eventuelle avec une fr\'equence de for\c{c}age. Un mod\`ele orbital complet est 
complexe, ceci a n\'ecessit\'e la mise au point d'algorithmes num\'eriques sans lesquels certaines des \'etudes que je pr\'esente ici n'auraient pas \'et\'e men\'ees \`a bien.
J'expose ici un algorithme, NAFFO pour Numerical Algorithm For Forced Oscillations, qui se base sur le fait que les mouvements de rotation sont quasip\'eriodiques pour
d\'eterminer it\'erativement et avec une grande pr\'ecision l'\'equilibre dynamique d'un corps en r\'esonance spin-orbite pour lequel on tient compte du mouvement
orbital complet (Chapitre \ref{chap:naffo}).

\part{La rotation des satellites naturels}


  \chapter{Un mod\`ele Hamiltonien de rotation rigide\label{chap:rigide}}

  \par Ce chapitre donne une formulation Hamiltonienne de la rotation rigide d'un corps triaxial. Il est fortement inspir\'e des travaux de Jacques Henrard \citep{hs2004,h2005,h2005b}, \`a 
  la suite desquels mes travaux se situent.
  
  \section{Exposition du probl\`eme}
  
  \par On consid\`ere ici que l'objet d'\'etude $S$ est un ellipso\"ide triaxial, rigide, en orbite autour de sa plan\`ete parente $P$. La matrice d'inertie $I$ du satellite $S$
  est d\'efinie ainsi:
  
  \begin{equation}
    \label{eq:matinertie}
    I=\left(
    \begin{tabular}{ccc}
     $A$ & $0$ & $0$ \\
     $0$ & $B$ & $0$ \\
     $0$ & $0$ & $C$ 
    \end{tabular}\right)
  \end{equation}
dans un rep\`ere normal $\left(\vec{f_1},\vec{f_2},\vec{f_3}\right)$ li\'e au satellite, avec $0 \leq A \leq B \leq C$. Les axes $\left(\vec{f_1},\vec{f_2},\vec{f_3}\right)$ sont les 
\emph{axes principaux d'inertie} du satellite. On a, par d\'efinition des moments principaux d'inertie $A$, $B$ et $C$:

\begin{eqnarray}
  A & = & \iiint\rho(x,y,z)\left(y^2+z^2\right)\,\textrm{d}x\,\textrm{d}y\,\textrm{d}z, \label{eq:momA} \\
  B & = & \iiint\rho(x,y,z)\left(x^2+z^2\right)\,\textrm{d}x\,\textrm{d}y\,\textrm{d}z, \label{eq:momB} \\
  C & = & \iiint\rho(x,y,z)\left(x^2+y^2\right)\,\textrm{d}x\,\textrm{d}y\,\textrm{d}z, \label{eq:momC}
\end{eqnarray}
o\`u $\rho(x,y,z)$ est la masse volumique locale, chaque \'el\'ement de masse du satellite \'etant rep\'er\'e par ses coordonn\'ees cart\'esiennes $(x,y,z)$ dans le rep\`ere
$\left(\vec{f_1},\vec{f_2},\vec{f_3}\right)$. Pour un corps homog\`ene, la quantit\'e $\rho$ est constante, on a $M_S=4\pi/3\rho abc$ et

\begin{eqnarray}
  A & = & \frac{M_S}{5}\left(b^2+c^2\right), \label{eq:momA2} \\
  B & = & \frac{M_S}{5}\left(a^2+c^2\right), \label{eq:momB2} \\
  C & = & \frac{M_S}{5}\left(a^2+b^2\right), \label{eq:momC2}
\end{eqnarray}
o\`u $M_S$ est la masse du satellite, et $0 \leq c \leq b \leq a$ sont les 3 rayons. $c$ est le rayon polaire, il correspond \`a l'axe $\vv{f_3}$. $a$ et $b$ sont les rayons equatoriaux
li\'es aux axes $\vv{f_1}$ et $\vv{f_2}$. Pour un corps en rotation synchrone, ils sont dirig\'es respectivement vers la plan\`ete et le long de l'orbite.

\par
\begin{figure}[ht]
  \centering
  \includegraphics[width=0.85\textwidth]{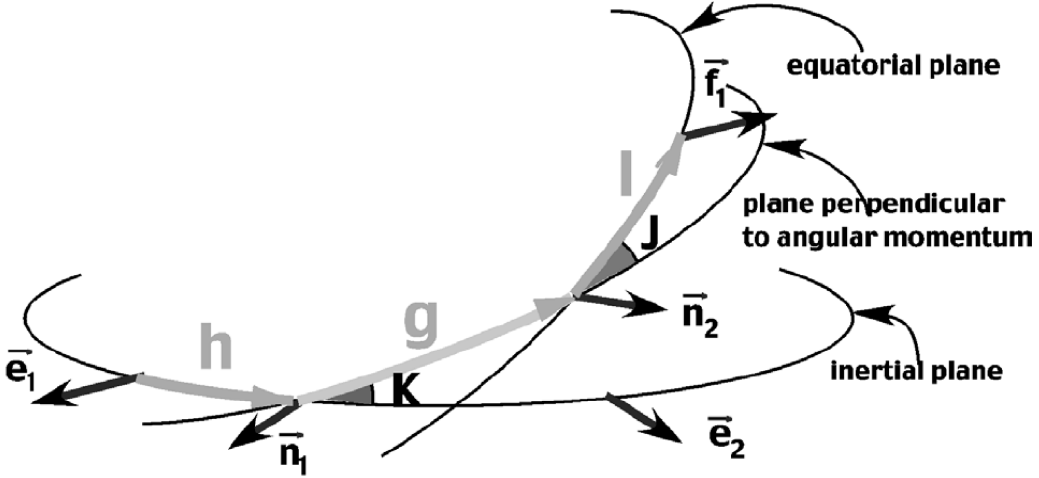}
  \caption{Les angles d'Euler du probl\`eme (reproduit de \citep{h2005}).\label{fig:anglesEuler}}
\end{figure}

 Dans notre \'etude nous distinguerons l'orientation du moment cin\'etique $\vec{G}$ et celle des axes principaux d'inertie, ou axes de figure, $(\vv{f_1},\vv{f_2},\vv{f_3})$.
Nous avons besoin de 2 autres rep\`eres de r\'ef\'erence: un rep\`ere inertiel $(\vv{e_1},\vv{e_2},\vv{e_3})$, et un rep\`ere li\'e au moment cin\'etique $\vec{G}$
$(\vv{n_1},\vv{n_2},\vv{n_3})$, avec $\vec{G}=G\vv{n_3}$. Dans la pratique, le rep\`ere inertiel sera le plan \'equatorial du corps parent \`a la date J2000. Le passage d'un 
rep\`ere \`a un autre se fait gr\^ace \`a 2 jeux d'angles d'Euler : $(h,K,g)$ du rep\`ere inertiel $(\vv{e_1},\vv{e_2},\vv{e_3})$ vers le rep\`ere du moment cin\'etique 
$(\vv{n_1},\vv{n_2},\vv{n_3})$, et $(-,J,l)$ de $(\vv{n_1},\vv{n_2},\vv{n_3})$ vers le rep\`ere li\'e au satellite $(\vv{f_1},\vv{f_2},\vv{f_3})$ (Fig.\ref{fig:anglesEuler}).
    
\par $h$ repr\'esente l'orientation de la ligne des noeuds du rep\`ere du moment cin\'etique par rapport au rep\`ere inertiel. Nous appellerons $K$ l'\emph{obliquit\'e inertielle},
cet angle repr\'esente l'orientation du moment cin\'etique du satellite par rapport \`a la normale au plan de r\'ef\'erence, alors que classiquement l'obliquit\'e est d\'efinie par
rapport \`a la normale \`a l'orbite du satellite. $g$ est un angle de spin tournant avec le satellite. L'angle $J$ repr\'esente l'amplitude du \emph{mouvement polaire}, 
c'est-\`a-dire la d\'eviation de l'orientation du moment cin\'etique par rapport \`a l'axe de figure, et $l$ est l'orientation de ce mouvement polaire.
    
\par \`A ces angles sont associ\'ees les variables canoniques d'Andoyer \citep{a1926}:

\begin{equation}
\label{eq:andoyer}
\begin{tabular}{lll}
 $l$ & \hspace{4cm} & $L=G\cos J$ \\
 $g$ & \hspace{4cm} & $G$ \\
 $h$ & \hspace{4cm} & $H=G\cos K$,
\end{tabular}
\end{equation}
il d\'ecoule naturellement des d\'efinitions de $J$ et $K$ que $L$ et $H$ sont les projections du moment cin\'etique $\vec{G}$ sur $\vv{f_3}$ et $\vv{e_3}$, respectivement. Si
l'obliquit\'e $K$ est en g\'en\'eral significative, il n'en est pas de m\^eme pour l'amplitude du mouvement polaire $J$. Quand $J$ est nul, alors les angles $l$ et $g$ ne sont 
pas clairement d\'efinis, alors que leur somme l'est toujours. Nous contournons ce probl\`eme en introduisant les \'el\'ements d'Andoyer modifi\'es:
  
\begin{equation}
\label{eq:andoyermodif}
\begin{tabular}{lll}
 $p=l+g+h$ & \hspace{0.8cm} & $P=\frac{G}{nC}$ \\
 $r=-h$ & \hspace{0.8cm} & $R=\frac{G-H}{nC}=2P\sin^2\frac{K}{2}$ \\
 $\xi=-\sqrt{\frac{2(G-L)}{nC}}\sin l=-\sqrt{\frac{2G}{nC}}\sin\frac{J}{2}\sin l$ & \hspace{0.8cm} & $\eta=\sqrt{\frac{2(G-L)}{nC}}\cos l=\sqrt{\frac{2G}{nC}}\sin\frac{J}{2}\cos l$.
\end{tabular}
\end{equation}
Ici, les variables cart\'esiennes $\xi$ et $\eta$ sont nulles lorsque $J$ est nul, l'ind\'etermination est donc lev\'ee. $nC$ est un facteur de normalisation, c'est la norme moyenne 
du moment cin\'etique $\vec{G}$ si le satellite est en rotation synchrone.

  \section{Les \'equations du mouvement}  

\par Comme dit pr\'ec\'edemment, nous consid\'erons que la seule force exerc\'ee sur le satellite est la perturbation gravitationnelle de sa plan\`ete parente. Nos variables 
(Eq.\ref{eq:andoyermodif}) \'etant canoniques, nous pouvons utiliser une formulation Hamiltonienne. Pour cela, nous devons exprimer l'\'energie cin\'etique $\mathcal{T}$ et l'\'energie
potentielle $\mathcal{V}$ due \`a la perturbation gravitationnelle de la plan\`ete.

\subsection{L'\'energie cin\'etique $\mathcal{T}$}

\par On a 

\begin{equation}
  \label{eq:kinetikgen}
  \mathcal{T} = \frac{1}{2}\vec{\omega}\cdot\vec{G},
\end{equation}
o\`u $\vec{\omega}$ est le vecteur rotation instantan\'ee. Il d\'ecoule de la d\'efinition des moments d'inertie et des angles d'Euler:

\begin{eqnarray}
  \vec{G} & = & G\sin J\sin l\vv{f_1}+G\sin J\cos l\vv{f_2}+G\cos J\vv{f_3}, \label{eq:defG} \\
  \vec{\omega} & = & \frac{G}{A}\sin J\sin l\vv{f_1}+\frac{G}{B}\sin J\cos l\vv{f_2}+\frac{G}{C}\cos J\vv{f_3}. \label{eq:defomega}
\end{eqnarray}

Dans ce cas, on obtient 

\begin{equation}
\label{eq:T}
\mathcal{T}=\frac{G^2\sin^2J}{2}\left(\frac{\sin^2 l}{A}+\frac{\cos^2 l}{B}\right)+\frac{G^2\cos^2 J}{2C}
\end{equation}
ce qui donne, dans les variables d'Andoyer :
\begin{equation}
\label{eq:Tando}
\mathcal{T}=\frac{G^2-L^2}{2}\left(\frac{\sin^2 l}{A}+\frac{\cos^2 l}{B}\right)+\frac{L^2}{2C}
\end{equation}
et, dans les variables d'Andoyer modifi\'ees\footnote{cette formule est donn\'ee avec une erreur typographique dans \citep{h2005b}} :

\begin{equation}
\label{eq:Tandomod}
\mathcal{T}=nC\left(\frac{nP^2}{2}+\frac{n}{8}\left(4P-\xi^2-\eta^2\right)\left(\frac{\gamma_1+\gamma_2}{1-\gamma_1-\gamma_2}\xi^2+\frac{\gamma_1-\gamma_2}{1-\gamma_1+\gamma_2}\eta^2\right)\right)
\end{equation}
avec 

\begin{eqnarray}
  \gamma_1 & = & \frac{2C-A-B}{2C}=J_2\frac{M_SR_S^2}{C}, \label{eq:gamma1} \\
  \gamma_2 & = & \frac{B-A}{2C}=2C_{22}\frac{M_SR_S^2}{C}. \label{eq:gamma2}
\end{eqnarray}
$J_2=-C_{20}$ et $C_{22}$ sont les classiques harmoniques de degr\'e 2 du champ de gravit\'e du satellite, ou \emph{coefficients de Stokes}, et $R_S$ son rayon moyen.

\subsection{L'\'energie potentielle $\mathcal{V}$}

\par Le potentiel gravitationnel de l'interaction entre le satellite et sa plan\`ete, vue comme une masse sph\'erique\footnote{\citet{h2005c} \'etudie l'influence 
de l'aplatissement de la plan\`ete, introduit comme un terme correctif. Cela a une influence n\'egligeable.}, s'\'ecrit

\begin{equation}
  \label{eq:potentialV}
  \mathcal{V} = -\mathcal{G}M_P\iiint_W\frac{\rho}{r'}dW,
\end{equation}
o\`u $\mathcal{G}$ est la constante gravitationnelle, $W$ le volume du satellite, et $d'$ la distance entre le centre de masse de la plan\`ete et l'\'el\'ement de masse
du satellite. Ce potentiel (\ref{eq:potentialV}) peut \^etre d\'evelopp\'e en harmoniques sph\'eriques sous la forme 

\begin{equation}
  \label{eq:expanspherik}
  \mathcal{V} = -\frac{\mathcal{G}M_PM_S}{r}\sum_{n=0}^{\infty}\sum_{m=0}^n \left(\frac{R_S}{r}\right)^nP_n^m\left(\sin\phi\right)\left(C_{nm}\cos\left(m\lambda\right)+S_{nm}\sin\left(m\lambda\right)\right),
\end{equation}
o\`u $r$ est la distance satellite-plan\`ete. Les fonctions $P_n^m$ sont les fonctions de Legendre associ\'ees, d\'efinies par rapport aux polyn\^omes de Legendre 
classiques $P_n$ par:

\begin{equation}
  \label{eq:legendreassociees}
  P_n^m(u) = \left(1-u^2\right)^{m/2}\frac{d^mP_n}{du^m}(u).
\end{equation}
Les angles $\phi$ et $\lambda$ sont respectivement la latitude et la longitude, et les coefficients $C_{nm}$ et $S_{nm}$ sont les coefficients de Stokes zonaux ($m=0$),
tesseraux ($m<n$) et sectoriaux ($m=n$). Avec les hypoth\`eses que nous avons faites sur le champ de gravit\'e du satellite, c'est-\`a-dire un ellipso\"ide triaxial
rigide homog\`ene, et un choix optimal du rep\`ere de r\'ef\'erence dans lequel on travaille, ici celui des axes principaux d'inertie $(\vv{f_1},\vv{f_2},\vv{f_3})$, 
le potentiel (\ref{eq:expanspherik}) prend une forme simplifi\'ee

\begin{equation}
  \label{eq:expanspherik2}
  \mathcal{V} = -\frac{\mathcal{G}M_PM_S}{r}\left(1+\left(\frac{R_S}{r}\right)^2\left(C_{20}P_2\left(\sin\phi\right)+C_{22}P_2^2\left(\sin\phi\right)\cos2\lambda\right)\right),
\end{equation}
avec

\begin{eqnarray}
  P_2(u) & = & \frac{3}{2}u^2-\frac{1}{2}, \label{eq:legendre20} \\
  P_2^2(u) & = & 3-3u^2. \label{eq:legendre22}
\end{eqnarray}
Dans le potentiel (\ref{eq:expanspherik2}), le premier terme n'a aucun effet sur la rotation car il correspond \`a un satellite ponctuel, on a donc maintenant

\begin{equation}
  \label{eq:expanspherik3}
  \mathcal{V} = -\mathcal{G}M_PM_S\frac{R_S^2}{r^3}\left(-\frac{3}{2}C_{20}\cos^2\phi+3C_{22}\cos^2\phi\cos2\lambda\right),
\end{equation}
o\`u $P_2(\sin\phi)=3/2\sin^2\phi-1/2$ a \'et\'e remplac\'e par $3/2\sin^2\phi-3/2$ par simple ajout de constante, ce qui permet d'\'ecrire $P_2(\sin\phi)=-3/2\cos^2\phi$.
Soit $\vec{r}$ le vecteur satellite-plan\`ete, posons 

\begin{eqnarray}
  \vec{r} & = & x_S\vv{f_1}+y_S\vv{f_2}+z_S\vv{f_3}, \label{eq:r1} \\
          & = & r\left(\hat{x}_S\vv{f_1}+\hat{y}_S\vv{f_2}+\hat{z}_S\vv{f_3}\right), \label{eq:r2}
\end{eqnarray}
avec

\begin{eqnarray}
  \hat{x}_S & = & \cos\phi\cos\lambda, \label{eq:xchapeau} \\
  \hat{y}_S & = & \cos\phi\sin\lambda, \label{eq:ychapeau} \\
  \hat{z}_S & = & \sin\phi.            \label{eq:zchapeau}
\end{eqnarray}
On a alors

\begin{eqnarray}
  \cos^2\phi & = & \hat{x}_S^2+\hat{y}_S^2, \label{eq:cos2phi} \\
  \cos^2\phi\cos2\lambda & = & \hat{x}_S^2-\hat{y}_S^2, \label{eq:cos2phi2lambda}
\end{eqnarray}
et

\begin{equation}
  \label{eq:expanspherik4}
  \mathcal{V} = -\frac{3}{2}\mathcal{G}M_PM_S\frac{R_S^2}{r^3}\left(J_2\left(\hat{x}_S^2+\hat{y}_S^2\right)+2C_{22}\left(\hat{x}_S^2-\hat{y}_S^2\right)\right),
\end{equation}
ou encore

\begin{equation}
  \label{eq:pothamil}
  \mathcal{V} = -\frac{3}{2}C\frac{\mathcal{G}M_P}{r^3}\left(\gamma_1\left(\hat{x}_S^2+\hat{y}_S^2\right)+\gamma_2\left(\hat{x}_S^2-\hat{y}_S^2\right)\right).
\end{equation}

\subsection{Le Hamiltonien du probl\`eme}

\par Le Hamiltonien du probl\`eme $\mathcal{H} = (\mathcal{T}+\mathcal{V})/(nC)$ s'\'ecrit alors

\begin{eqnarray}
  \mathcal{H}(p,r,\xi,P,R,\eta,t) & = & \frac{nP^2}{2}+\frac{n}{8}\left(4P-\xi^2-\eta^2\right)\left(\frac{\gamma_1+\gamma_2}{1-\gamma_1-\gamma_2}\xi^2+\frac{\gamma_1-\gamma_2}{1-\gamma_1+\gamma_2}\eta^2\right) \nonumber \\
                                  & - & \frac{3}{2}\frac{\mathcal{G}M_P}{nr(t)^3}\left(\gamma_1\left(\hat{x}_S(t)^2+\hat{y}_S(t)^2\right)+\gamma_2\left(\hat{x}_S(t)^2-\hat{y}_S(t)^2\right)\right). \label{eq:hamil}
\end{eqnarray}

\par Ici le choix a \'et\'e fait de l'\'ecrire sous forme non autonome, c'est-\`a-dire d\'ependant explicitement du temps. Ceci est implicitement une cons\'equence du fait que le 
mouvement orbital du satellite autour de la plan\`ete, plut\^ot vu ici comme le mouvement de la plan\`ete autour du satellite, est consid\'er\'e comme un for\c{c}age externe, qui 
agit sur la rotation mais sans lui-m\^eme \^etre affect\'e. Cette approximation est tout-\`a-fait l\'egitime dans le cas des satellites de plan\`etes g\'eantes o\`u l'\'energie de 
rotation est de plusieurs ordres de grandeur inf\'erieure \`a l'\'energie orbitale, il suffit pour le voir de comparer les masses de la plan\`ete $M_P$ et du satellite $M_S$. Cette
approximation perd de sa l\'egitimit\'e dans le cas des ast\'ero\"ides binaires o\`u un mod\`ele complet d'interactions entre la rotation et l'orbite doit \^etre consid\'er\'e
\citep{m1995,fs2008,bl2009}.

\par En g\'en\'eral, le mouvement orbital du satellite est donn\'e dans un rep\`ere inertiel $(\vv{e_1},\vv{e_2},\vv{e_3})$ par des \'eph\'em\'erides comme L1 pour les satellites 
Galil\'eens de Jupiter \citep{ldv2006} ou TASS pour les principaux satellites de Saturne \citep{vd1995}, il est donc n\'ecessaire d'effectuer des rotations pour se mettre dans le 
rep\`ere li\'e au satellite $(\vv{f_1},\vv{f_2},\vv{f_3})$. En appelant $(\hat{x}_I,\hat{y}_I,\hat{z}_I)$ les coordonn\'ees du vecteur unitaire pointant vers la plan\`ete dans le
rep\`ere inertiel, il d\'ecoule de la d\'efinition des angles d'Euler (Fig.\ref{fig:anglesEuler}):

\begin{equation}
  \label{eq:lesrotations}
  \left(\begin{tabular}{c}
         $\hat{x}_S$ \\
         $\hat{y}_S$ \\
         $\hat{z}_S$
        \end{tabular}\right)
        = R_3(-l)R_1(-J)R_3(-g)R_1(-K)R_3(-h)
  \left(\begin{tabular}{c}
         $\hat{x}_I$ \\
         $\hat{y}_I$ \\
         $\hat{z}_I$
        \end{tabular}\right),
\end{equation}
o\`u les matrices de rotation sont d\'efinies par\footnote{Il est important de les d\'efinir, car il existe 2 conventions de signe pour les angles\ldots}:

\begin{equation}
  \label{eq:R3}
  R_3(\phi) = \left(\begin{tabular}{ccc}
                     $\cos\phi$ & $-\sin\phi$ & $0$ \\
                     $\sin\phi$ &  $\cos\phi$ & $0$ \\
                     $0$        &  $0$        & $1$
                    \end{tabular}\right),
\end{equation}
et

\begin{equation}
  \label{eq:R1}
  R_1(\phi) = \left(\begin{tabular}{ccc}
                     $1$ & $0$        & $0$ \\
                     $0$ & $\cos\phi$ & $-\sin\phi$ \\
                     $0$ & $\sin\phi$ &  $\cos\phi$ 
                    \end{tabular}\right).
\end{equation}

\par Ce Hamiltonien donne directement les \'equations du probl\`eme:

\begin{equation}
  \label{eq:equhamil}
  \begin{tabular}{lll}
   \(\displaystyle\frac{dp}{dt}=\frac{\partial\mathcal{H}}{\partial P}\)      & \hspace{2cm} & \(\displaystyle\frac{dP}{dt}=-\frac{\partial\mathcal{H}}{\partial p}\) \\
                                                                              & \hspace{2cm} & \\
   \(\displaystyle\frac{dr}{dt}=\frac{\partial\mathcal{H}}{\partial R}\)      & \hspace{2cm} & \(\displaystyle\frac{dR}{dt}=-\frac{\partial\mathcal{H}}{\partial r}\) \\
                                                                              & \hspace{2cm} & \\
   \(\displaystyle\frac{d\xi}{dt}=\frac{\partial\mathcal{H}}{\partial \eta}\) & \hspace{2cm} & \(\displaystyle\frac{d\eta}{dt}=-\frac{\partial\mathcal{H}}{\partial \xi}\).
  \end{tabular}
\end{equation}

\par Dans toutes les \'etudes de rotation rigide que je montre par la suite, les simulations num\'eriques ont consist\'e \`a propager les \'equations (\ref{eq:equhamil}).
Nulle part ces \'equations n'imposent que le satellite soit en rotation synchrone. La normalisation par $nC$ le sugg\`ere, mais n'est en aucun cas une contrainte. Ces \'equations
sont donc valables pour tout corps rigide dont la rotation n'influe pas significativement sur l'orbite, ce sont les conditions initiales qui vont d\'eterminer si la rotation est 
r\'esonnante ou non, et de quelle r\'esonance il s'agit.

\section{Caract\'erisation de la rotation r\'esonnante}
  
\par Le but de cette section est de caract\'eriser l'\'etat d'\'equilibre dynamique dans lequel les satellites naturels sont suppos\'es \^etre. Il s'agit d'une \'etude
analytique visant \`a comprendre le fonctionnement global du syst\`eme. Pour cela, nous faisons quelques approximations, mais dans les applications qui vont suivre, nous 
utiliserons notamment un code num\'erique propageant les \'equations compl\`etes (\ref{eq:equhamil}) et utilisant le mouvement orbital r\'eel du satellite. L'\'etat 
d'\'equilibre implique une rotation en moyenne synchrone, une obliquit\'e correspondant \`a l'\'Etat de Cassini 1, et un mouvement polaire tr\`es petit. Le choix qui 
a \'et\'e fait ici est de consid\'erer la dynamique d'un point de vue global, c'est-\`a-dire en consid\'erant les 3 degr\'es de libert\'e du syst\`eme simultan\'ement, 
sans n\'egliger leurs couplages. Une telle approche peut faire intervenir un grand nombre de termes, c'est la raison pour laquelle nous n\'egligeons ici tout effet non 
indispensable \`a l'\'etablissement de cet \'equilibre, en particulier nous consid\'erons que l'orbite du satellite est circulaire. 

\par Cette approximation a l'inconv\'enient d'emp\^echer la mod\'elisation des librations forc\'ees en longitude, souvent mises en avant en plan\'etologie. Ces librations 
sont tr\`es bien mod\'elis\'ees par une approche plus classique consistant \`a consid\'erer un probl\`eme plan (inclinaison orbitale nulle, $K=J=0$) et une excentricit\'e constante. 
Sous ces hypoth\`eses, le Hamiltonien $\mathcal{H}$ devient \citep{gp1966,c1990,md1999}:

\begin{equation}
  \label{eq:hamilplan}
  \mathcal{H}(p,-,-,P,-,-,t) = \frac{nP^2}{2}-\frac{3n}{4}\frac{B-A}{C}\sum_i H\left(\frac{i}{2},e\right)\cos(2p-i\lambda),
\end{equation}
o\`u $\lambda=nt$ est la longitude moyenne du satellite, et les fonctions $H\left(\frac{i}{2},e\right)=G_{20(i-2)}(e)$ sont des fonctions classiques en excentricit\'e d\'efinies par \citep{k1966}. 
On a, sous forme d\'evelopp\'ee en excentricit\'e :

\begin{eqnarray}
  \sum_i H\left(\frac{i}{2},e\right)\cos(2p-i\lambda) & = & \frac{15625}{129024}e^7\cos(2p+5\lambda) \nonumber \\
                                                      & + & \frac{4}{45}e^6\cos(2p+4\lambda) \nonumber \\
                                                      & + & \left(\frac{81}{1280}e^5+\frac{81}{2048}e^7\right)\cos(2p+3\lambda) \nonumber \\
                                                      & + & \left(\frac{1}{24}e^4+\frac{7}{240}e^6\right)\cos(2p+2\lambda) \nonumber \\
                                                      & + & \left(\frac{1}{48}e^3+\frac{11}{768}e^5+\frac{313}{30720}e^7\right)\cos(2p+\lambda) \nonumber \\
                                                      & + & \left(-\frac{1}{2}e+\frac{1}{16}e^3-\frac{5}{384}e^5-\frac{143}{18432}e^7\right)\cos(2p-\lambda) \nonumber \\
                                                      & + & \left(1-\frac{5}{2}e^2+\frac{13}{16}e^4-\frac{35}{288}e^6\right)\cos(2p-2\lambda) \nonumber \\
                                                      & + & \left(\frac{7}{2}e-\frac{123}{16}e^3+\frac{489}{128}e^5-\frac{1763}{2048}e^7\right)\cos(2p-3\lambda) \label{eq:celletti} \\
                                                      & + & \left(\frac{17}{2}e^2-\frac{115}{6}e^4+\frac{601}{48}e^6\right)\cos(2p-4\lambda) \nonumber \\
                                                      & + & \left(\frac{845}{48}e^3-\frac{32525}{768}e^5+\frac{208225}{6144}e^7\right)\cos(2p-5\lambda) \nonumber \\
                                                      & + & \left(\frac{533}{16}e^4-\frac{13827}{160}e^6\right)\cos(2p-6\lambda) \nonumber \\
                                                      & + & \left(\frac{228347}{3840}e^5-\frac{3071075}{18432}e^7\right)\cos(2p-7\lambda) \nonumber \\
                                                      & + & \frac{73369}{720}e^6\cos(2p-8\lambda) \nonumber \\
                                                      & + & \frac{12144273}{71680}e^7\cos(2p-9\lambda)+\mathcal{O}(e^8). \nonumber
\end{eqnarray}

On voit ainsi qu'\`a excentricit\'e nulle (orbite circulaire), seule la r\'esonance synchrone d'argument $2p-2\lambda$ subsiste. Le seul autre cas de r\'esonance connu
dans le Syst\`eme Solaire est la r\'esonance 3:2 de Mercure, d'argument $2p-3\lambda$. On voit que le pr\'efacteur est de degr\'e 1 en excentricit\'e, Mercure a d'ailleurs
une excentricit\'e significative, actuellement $0.206$, alors que l'excentricit\'e des satellites est g\'en\'eralement d'un ordre de grandeur plus faible.

  \subsection{La rotation d'\'equilibre des satellites des plan\`etes g\'eantes}
  
  \subsubsection{La position d'\'equilibre}
  
  \par Nous repartons ici du Hamiltonien (\ref{eq:hamil}) \`a partir duquel nous allons \'etudier l'\'etat de rotation attendu des satellites naturels, 
  c'est-\`a-dire l'\'Etat de Cassini 1 avec rotation synchrone. L'orbite du satellite autour de sa plan\`ete parente est circulaire, avec une inclinaison
  constante et une pr\'ecession uniforme de son n{\oe}ud ascendant. En notant $a$, $I$, $\lambda$ et $\ascnode$ les \'el\'ements orbitaux du satellite, 
  respectivement le demi-grand axe, l'inclinaison, la longitude moyenne et le n{\oe}ud ascendant, les coordonn\'ees du vecteur unitaire pointant vers la 
  plan\`ete s'\'ecrivent, dans le rep\`ere inertiel $(\vv{e_1},\vv{e_2},\vv{e_3})$:
  
  \begin{eqnarray}
  	  \hat{x}_I & = & -\cos\ascnode\cos(\lambda-\ascnode)+\cos I\sin\ascnode\sin(\lambda-\ascnode), \label{eq:xicircul} \\
  	  \hat{y}_I & = & -\sin\ascnode\cos(\lambda-\ascnode)-\cos I\cos\ascnode\sin(\lambda-\ascnode), \label{eq:yicircul} \\
  	  \hat{z}_I & = & -\sin I\sin(\lambda-\ascnode). \label{eq:zicircul}
  \end{eqnarray}

\par On cherche ici \`a caract\'eriser un comportement moyen du syst\`eme, ce qui n\'ecessite de moyenner le Hamiltonien du probl\`eme. Pour cela, on commence 
par exprimer les angles en libration. \`A la r\'esonance synchrone, l'angle $p-\lambda$ libre autour de $0$ ou $\pi$, ce qui veut dire que l'axe le plus 
long du satellite pointe en moyenne vers la plan\`ete (en fait vers l'antifoyer de son orbite, la distinction n'a pas de sens ici puisque l'orbite est 
	circulaire). De plus, il d\'ecoule de la d\'efinition de l'\'Etat de Cassini 1, et plus pr\'ecis\'ement de la troisi\`eme loi de Cassini que, si le 
plan du rep\`ere inertiel $(\vv{e_1},\vv{e_2})$ est choisi de fa\c{c}on \`a optimiser la repr\'esentation de la rotation (cf. Sect.\ref{sec:callisto}), alors 
l'argument $\ascnode-h=r+\ascnode$ est lui aussi en libration.

\par Afin de tenir compte de ces librations, nous faisons le changement de variables suivant:

\begin{equation}
	\label{eq:changreson}
\begin{tabular}{lll}
$\sigma=p-\lambda$ & \hspace{0.8cm} & $P$ \\
$\rho=r+\ascnode$ & \hspace{0.8cm} & $R$ \\
 $\xi$ & \hspace{0.8cm} & $\eta$.
\end{tabular}
\end{equation}
Ce changement de variables d\'epend lin\'eairement du temps, nous devons donc ajouter au Hamiltonien $\mathcal{H}$ la quantit\'e $-nP+\dot{\ascnode}R$, o\`u
$\dot{\ascnode}$ est la vitesse de pr\'ecession constante du n{\oe}ud $\ascnode$. 

\par Tous calculs faits, on obtient, apr\`es moyennisation :

\begin{equation}
	\label{eq:x2py2}
	\hat{x}_S^2+\hat{y}_S^2=a_1\sin^2K+a_2\sin K\cos K\cos\rho+a_3\cos2\rho(1-\cos2K)+\mathcal{O}\left(\xi^2,\xi\eta,\eta^2\right)
\end{equation}
et

\begin{eqnarray}
	\hat{x}_S^2-\hat{y}_S^2 & = & b_1(1+\cos K)^2\cos2\sigma+b_2\sin K(1+\cos K)\cos(2\sigma+\rho)+b_3\sin^2K\cos(2\sigma+2\rho) \nonumber \\
                                & + & b_4\sin K(1-\cos K)\cos(2\sigma+3\rho)+b_5(1-\cos K)^2\cos(2\sigma+4\rho) \label{eq:x2my2} \\
                                & + & \mathcal{O}\left(\xi^2,\xi\eta,\eta^2\right) \nonumber
\end{eqnarray}
avec

\begin{eqnarray}
	a_1 & = & \frac{\sin^2I}{2}-\frac{1+\cos^2I}{4}, \label{eq:a1} \\
	a_2 & = & \frac{\sin2I}{2}, \label{eq:a2} \\
	a_3 & = & \frac{\sin^2I}{8}, \label{eq:a3} \\
	b_1 & = & \frac{1+2\cos I+\cos^2I}{16}, \label{eq:b1} \\
	b_2 & = & \frac{2\sin I+\sin2I}{8}, \label{eq:b2} \\
	b_3 & = & \frac{3}{8}\sin^2I, \label{eq:b3} \\
	b_4 & = & \frac{2\sin I-\sin2I}{8}, \label{eq:b4} \\
	b_5 & = & \frac{1-2\cos I+\cos^2I}{16}. \label{eq:b5}
\end{eqnarray}
  
Il en d\'ecoule le Hamiltonien moyenn\'e $\mathcal{H}_1$:

\begin{eqnarray}
\mathcal{H}_1 & = & \frac{nP^2}{2}-nP+\dot{\ascnode}R-\frac{3n}{2}\gamma_1[a_1\sin^2K+a_2\sin K\cos K\cos\rho +a_3\cos2\rho(1-\cos2K)] \nonumber \\
            & - & \frac{3n}{2}\gamma_2[b_1(1+\cos K)^2\cos2\sigma+b_2\sin K(1+\cos K)\cos(2\sigma+\rho) \label{eq:hammoy} \\
            &   & +b_3\sin^2K\cos(2\sigma+2\rho)+b_4\sin K(1-\cos K)\cos(2\sigma+3\rho) \nonumber \\
            &   & +b_5(1-\cos K)^2\cos(2\sigma+4\rho)]+\mathcal{O}(\xi^2,\xi\eta,\eta^2), \nonumber
\end{eqnarray}
la notation $K$ est conserv\'ee uniquement par commodit\'e d'\'ecriture. Lorsqu'on est \`a l'\'equilibre strict, on a, par hypoth\`ese\footnote{ce qui 
signifie qu'on ne cherche pas tous les \'equilibres, mais juste \`a caract\'eriser l'\'Etat de Cassini 1}, $\sigma=\rho=0$, et par d\'efinition de l'\'equilibre:

\begin{eqnarray}
  \frac{\partial\mathcal{H}_1}{\partial P} & = & 0, \label{eq:dHdP} \\
  \frac{\partial\mathcal{H}_1}{\partial R} & = & 0, \label{eq:dHdR}
\end{eqnarray}
ce qui donne

\begin{eqnarray}
n\left(P-1-\frac{3}{2}\Delta\frac{\cos K-1}{P\sin K}\right) & = & 0, \label{eq:E10} \\
\dot{\ascnode}-\frac{3}{2}\frac{n\Delta}{P\sin K}           & = & 0, \label{eq:E20}
\end{eqnarray}
avec

\begin{eqnarray}
\Delta & = & \gamma_1\left(a_1\sin2K+a_2\cos2K+2a_3\sin2K\right) \nonumber \\
       & + & \gamma_2\left(-2b_1\sin K(1+\cos K)+b_2(\cos K+\cos2K)+b_3\sin2K\right. \label{eq:gd} \\
       &   & \left.+b_4(\cos K-\cos2K)+2b_5\sin K(1-\cos K)\right). \nonumber
\end{eqnarray}

\par Dans le cadre d'une application, on peut r\'esoudre num\'eriquement le syst\`eme compos\'e des \'equations (\ref{eq:E10}) et (\ref{eq:E20}). Dans un cadre plus 
g\'en\'eral, en supposant les quantit\'es $K$ et $I$ petites, l'\'equation (\ref{eq:E10}) donne imm\'ediatement $P^*=1$, l'\'etoile d\'esignant la valeur \`a
l'\'equilibre.

%


Soit $\epsilon$ l'obliquit\'e mesur\'ee par rapport \`a la normale \`a l'orbite. La troisi\`eme loi de Cassini \citep{c1693,c1966}, initialement \'ecrite pour la Lune,
affirme que l'axe de rotation, la normale \`a l'orbite, et la normale au plan de r\'ef\'erence sont coplanaires, ce qui permet d'\'ecrire $\epsilon = K^*-I$. En 
substituant $P^*$ dans l'\'equation (\ref{eq:E10}) \`a partir de son expression d\'ecoulant de (Eq.\ref{eq:E20}), on a :

\begin{eqnarray}
  0 & = & -1+\frac{\dot{\ascnode}}{n}\left(1-\cos(I+\epsilon)\right)-\frac{3}{8}\frac{n}{\dot{\ascnode}}\frac{2\gamma_1\sin2\epsilon+\gamma_2\left(\sin2\epsilon+2\sin\epsilon\right)}{\sin\left(I+\epsilon\right)} \nonumber \\
    &   & -\frac{3}{8}\gamma_1\frac{n}{\dot{\ascnode}}\frac{\sin\left(I+2\epsilon\right)+\sin\left(3I+2\epsilon\right)}{\sin\left(I+\epsilon\right)}. \label{eq:cassinigeneral}
\end{eqnarray}

\subsubsection{Les petites oscillations autour de l'\'equilibre}

\par Une mani\`ere de caract\'eriser le comportement d'un syst\`eme \`a l'\'equilibre est d'\'evaluer les fr\'equences des petites oscillations (ou librations) autour.
On s'attend a priori \`a ce que ces petites oscillations aient une amplitude suffisamment faible pour \^etre n\'egligeable et ind\'etectable. N\'eanmoins, conna\^itre
leur fr\'equence de libration peut permettre de d\'etecter si elles subsistent, et aussi d'estimer l'amplitude de r\'eponse aux sollicitations forc\'ees. 

\par On part du Hamiltonien moyenn\'e $\mathcal{H}_1$ (Eq.\ref{eq:hammoy}) et on introduit de nouvelles variables afin de se centrer sur la position d'\'equilibre :

\begin{equation}
\label{eq:centrage}
\begin{array}{lll}
\xi_1=\sigma, & \hspace{2cm} & \eta_1=P-P^*, \\
\xi_2=\rho,   & \hspace{2cm} & \eta_2=R-R^*, \\
\xi_3=\xi,    & \hspace{2cm} & \eta_3=\eta.
\end{array}
\end{equation}

\par Exprim\'e dans ces variables, le Hamiltonien $\mathcal{H}$ ne contient plus de termes lin\'eaires\footnote{Le lecteur avis\'e aura remarqu\'e que c'\'etait 
d\'ej\`a le cas pour le troisi\`eme degr\'e de libert\'e, li\'e au mouvement polaire. Nous n'excluons pas que son amplitude $J$ ait une valeur d'\'equilibre non nulle, mais 
cette quantit\'e est en g\'en\'eral n\'eglig\'ee. De plus, en tenir compte rend les calculs bien plus complexes, donc nous laissons ce probl\`eme de c\^ot\'e dans le cadre 
de cette \'etude analytique. N\'eanmoins, cet effet est bien pr\'esent dans les simulations num\'eriques que nous pr\'esentons au Chap.\ref{chap:apprigide}.}. 
Soit $\mathcal{N}$ la somme des termes quadratiques, on peut \'ecrire :

\begin{equation}
\label{eq:N}
\mathcal{N}(\xi_1,\xi_2,\xi_3,\eta_1,\eta_2,\eta_3)=n\left(\gamma_{11}\xi_1^2+2\gamma_{12}\xi_1\xi_2+\gamma_{22}\xi_2^2+\gamma_{33}\xi_3^2+\mu_{11}\eta_1^2+2\mu_{12}\eta_1\eta_2+\mu_{2}\eta_2^2+\mu_{33}\eta_3^2\right)
\end{equation}
avec

\begin{eqnarray}
\gamma_{11} & = & 3\gamma_2(b_1(1+\cos K^*)^2+b_2\sin K^*(1+\cos K^*) \nonumber \\
            &   & +b_3\sin^2K^*+b_4\sin K^*(1-\cos K^*)+b_5(1-\cos K^*)^2), \label{eq:g11} \\
\gamma_{12} & = & \frac{3}{2}\gamma_2(b_2\sin K^*(1+\cos K^*)+b_3\sin^2K^* \nonumber \\
            &   & +3b_4\sin K^*(1-\cos K^*)+4b_5(1-\cos K^*)^2), \label{eq:g12} \\
\gamma_{22} & = & \gamma_1\Big(\frac{3}{8}a_2\sin 2K^*+6a_3\sin^2K^*\Big)+\gamma_2\Big(\frac{3}{4}b_2\sin K^*(1+\cos K^*)+3b_3\sin^2K^* \nonumber \\
            &   & +\frac{27}{4}b_4\sin K^*(1-\cos K^*)+12b_5(1-\cos K^*)^2\Big) \label{eq:g22} \\
\gamma_{33} & = & \frac{P^*}{2}\frac{\gamma_1+\gamma_2}{1-\gamma_1-\gamma_2}+\frac{3}{2}(\gamma_1+\gamma_2)\left(\frac{\cos(K^*-I)}{4}+\frac{7}{16}\cos(2(K^*-I))+\frac{5}{16}\right), \label{eq:g33}
\end{eqnarray}
et

\begin{eqnarray}
\mu_{11} & = & \frac{1}{2}-\frac{3\gamma_1}{2P^{*2}}\Big((a_1+2a_3)(1-\cos K^*)(3\cos K^*-1) \nonumber \\
         &   & +\frac{a_2}{2}\frac{\sin K^*}{(1+\cos K^*)^2}(6\cos^3K^*+4\cos^2K^*-5\cos K^*-2)\Big) \nonumber \\
         & - & \frac{3\gamma_2}{2P^{*2}}\Big(b_1(3\cos K^*+1)(\cos K^*-1)+\frac{3}{2}b_2\frac{\sin K^*\cos 2K^*}{1+\cos K^*} \label{eq:mu11} \\
         &   & +b_3(1-\cos K^*)(3\cos K^*-1)+\frac{b_4}{2}\frac{1-\cos K^*}{1+\cos K^*}\sin K^* (1+8\cos K^*+6\cos^2K^*)+3b_5\Big), \nonumber \\
\mu_{12} & = & -\frac{3\gamma_1}{2P^{*2}}\Big((a_1+2a_3)(1-2\cos K^*)+\frac{a_2}{2}\frac{1+4\cos K^*-2\cos^2 K^*-4\cos^3 K^*}{\sin K^* (1+\cos K^*)}\Big) \nonumber \\
         & - &  \frac{3\gamma_2}{2P^{*2}}\Big(2b_1\cos K^*-b_2\frac{4\cos^2K^*-\cos K^*-2}{2\sin K^*}-b_3(2\cos K^*-1) \nonumber \\
         &   & +\frac{b_4}{2}\frac{\cos K^*}{\sin K^*}\frac{\cos K^* -1}{\cos K^* +1}(4\cos K^*+5)-2b_5\frac{\sin^2K^*}{1+\cos K^*}\Big), \label{eq:mu12} \\
\mu_{22} & = & \frac{3\gamma_1}{2P^{*2}}\Big(a_1+\frac{a_2}{2}\frac{3\cos K^*-2\cos^3K^*}{\sin^3K^*}+2a_3\Big) \nonumber \\
         & - & \frac{3\gamma_2}{2P^{*2}}\Big(b_1+\frac{b_2}{2}\frac{2\cos^3K^*-3\cos K^*-1}{\sin^3K^*}-b_3-\frac{b_4}{2}\frac{1-3\cos K^*+2\cos^3K^*}{\sin^3K^*}+b_5\Big), \label{eq:mu22} \\
\mu_{33} & = & \frac{P^*}{2}\frac{\gamma_1-\gamma_2}{1-\gamma_1+\gamma_2}. \label{eq:mu33}
\end{eqnarray}

\par On peut remarquer la pr\'esence de termes crois\'es $\xi_1\xi_2$ et $\eta_1\eta_2$ qu'il serait int\'eressant d'\'eliminer pour d\'ecoupler les 2 degr\'es de libert\'e associ\'es.
Pour cela nous introduisons la \emph{untangling transformation}\footnote{En fran\c{c}ais on dirait qu'on diagonalise le syst\`eme. Je remercie Philippe Robutel pour 
m'avoir inspir\'e cette traduction.}\citep{hl2005} :

\begin{equation}
\label{eq:untangling}
\begin{array}{lll}
\xi_1=x_1-\beta x_2                 & \hspace{2cm} & \eta_1=(1-\alpha\beta)y_1-\alpha y_2 \\
\xi_2=\alpha x_1+(1-\alpha\beta)x_2 & \hspace{2cm} & \eta_2=\beta y_1+y_2 \\
\xi_3=x_3                           & \hspace{2cm} & \eta_3=y_3,
\end{array}
\end{equation}
en choisissant $\alpha$ et $\beta$ de fa\c{c}on \`a ce que $\mathcal{N}$ s'\'ecrive comme une somme de carr\'es, c'est-\`a-dire :

\begin{equation}
\label{eq:nuntang}
\mathcal{N}_1=n\left(\zeta_1x_1^2+\zeta_2x_2^2+\zeta_3x_3^2+\psi_1y_1^2+\psi_2y_2^2+\psi_3y_3^2\right)
\end{equation}
avec

\begin{eqnarray}
\zeta_1 & = & \gamma_{11}+2\gamma_{12}\alpha+\gamma_{22}\alpha^2, \label{eq:zeta1} \\
\zeta_2 & = & \gamma_{11}\beta^2-2\beta(1-\alpha\beta)\gamma_{12}+\gamma_{22}(1-\alpha\beta)^2, \label{eq:zeta2} \\
\psi_1  & = & \mu_{11}(1-\alpha\beta)^2+2\beta(1-\alpha\beta)\mu_{12}+\beta^2\mu_{22}, \label{equ:psi1} \\
\psi_2  & = & \alpha^2\mu_{11}-2\alpha\mu_{12}+\mu_{22}, \label{eq:psi2} \\
\zeta_3 & = & \gamma_{33}, \label{equ:zeta3} \\
\psi_3  & = & \mu_{33}. \label{equ:psi3}
\end{eqnarray}

\par On trouve $\alpha$ et $\beta$ en r\'esolvant les \'equations :

\begin{eqnarray}
\gamma_{12}-\gamma_{11}\beta+\gamma_{22}\alpha-2\gamma_{12}\alpha\beta-\gamma_{22}\alpha^2\beta & = & 0, \label{eq:alfbet1} \\
\mu_{12}-\mu_{11}\alpha+\mu_{22}\beta+\mu_{11}\alpha^2\beta-2\mu_{12}\alpha\beta                & = & 0, \label{equ:alfbet2}
\end{eqnarray}
qui s'\'ecrivent \'egalement

\begin{eqnarray}
\beta & = & \frac{\gamma_{12}+\gamma_{22}\alpha}{\gamma_{11}+2\gamma_{12}\alpha+\gamma_{22}\alpha^2}=\frac{\mu_{11}\alpha-\mu_{12}}{\mu_{22}+\mu_{11}\alpha^2-2\mu_{12}\alpha}, \label{eq:beta} \\
0     & = & (\gamma_{12}\mu_{11}+\gamma_{22}\mu_{12})\alpha^2+(\mu_{11}\gamma_{11}-\gamma_{22}\mu_{22})\alpha-\gamma_{12}\mu_{22}-\gamma_{11}\mu_{12}. \label{eq:alpha}
\end{eqnarray}

Le syst\`eme d'\'equations (\ref{eq:beta}-\ref{eq:alpha}) a 2 solutions. Le choix du couple $(\alpha,\beta)$ d\'eterminera si les variables $(x_1,y_1)$ repr\'esenterons 
plut\^ot le mouvement en longitude, ou plut\^ot en obliquit\'e. En pratique, le couplage entre ces 2 degr\'es de libert\'e est faible et n\'eglig\'e par de nombreux auteurs,
il existe une solution o\`u $\alpha$ est proche de 1, ce choix permet de consid\'erer que $(x_1,y_1)$ mod\'elise le mouvement en longitude.

\par Ceci fait, on peut introduire, apr\`es changement d'\'echelle, les variables polaires canoniques suivantes : 

\begin{equation}
\label{eq:chgpolaires}
\begin{array}{lll}
x_1=\sqrt{2UU^*}\sin u & \hspace{2cm} & y_1=\sqrt{\frac{2U}{U^*}}\cos u \\
x_2=\sqrt{2VV^*}\sin v & \hspace{2cm} & y_2=\sqrt{\frac{2V}{V^*}}\cos v \\
x_3=\sqrt{2WW^*}\sin w & \hspace{2cm} & y_3=\sqrt{\frac{2W}{W^*}}\cos w 
\end{array}
\end{equation}
avec :

\begin{eqnarray}
U^* & = & \sqrt{\frac{\psi_1}{\zeta_1}}, \label{eq:Ustar} \\
V^* & = & \sqrt{\frac{\psi_2}{\zeta_2}}, \label{eq:Vstar} \\
W^* & = & \sqrt{\frac{\psi_3}{\zeta_3}}, \label{eq:Wstar}
\end{eqnarray}
ce qui permet d'\'ecrire le Hamiltonien quadratique $\mathcal{N}$ ainsi :

\begin{equation}
\label{equ:Nafk}
\mathcal{N}_2(-,-,-,U,V,W)=\omega_uU+\omega_vV+\omega_wW
\end{equation}
avec :

\begin{eqnarray}
\omega_u & = & 2n\sqrt{\psi_1\zeta_1}, \label{eq:omegu} \\
\omega_v & = & 2n\sqrt{\psi_2\zeta_2}, \label{eq:omegv} \\
\omega_w & = & 2n\sqrt{\psi_3\zeta_3}. \label{eq:omegw}
\end{eqnarray}

\par Les angles $u$, $v$ et $w$ sont les 3 angles des librations libres autour de l'\'equilibre, et $U$, $V$ et $W$ sont les amplitudes associ\'ees. Dans ce Hamiltonien 
$\mathcal{N}_2$, les termes d'ordre sup\'erieur ont \'et\'e n\'eglig\'es, ce qui signifie que les amplitudes sont suppos\'ees petites. On a aussi

\begin{eqnarray}
  \frac{du}{dt} & = & \frac{\partial\mathcal{N}_2}{\partial U}=\omega_u, \label{eq:dudt} \\
  \frac{dv}{dt} & = & \frac{\partial\mathcal{N}_2}{\partial V}=\omega_v, \label{eq:dvdt} \\
  \frac{dw}{dt} & = & \frac{\partial\mathcal{N}_2}{\partial W}=\omega_w, \label{eq:dwdt}
\end{eqnarray}
les quantit\'es $\omega_{u,v,w}$ sont donc les fr\'equences\footnote{Les puristes reconna\^itront l\`a un abus de langage, les $\omega_{u,v,w}$ \'etant en fait des pulsations. 
La fr\'equence est l'inverse de la p\'eriode, et diff\`ere de la pulsation d'un facteur $2\pi$.} des petites librations, plus exactement ce sont les limites des fr\'equences 
lorsque le syst\`eme est \`a l'\'equilibre dynamique.

\par En n\'egligeant les couplages ($\alpha\approx1$, $\beta\approx0$), l'inclinaison $I$ et l'obliquit\'e $K$, on a 

\begin{eqnarray}
  \psi_1  & \approx & \mu_{11}\approx\frac{1}{2}, \nonumber \\
  \zeta_1 & \approx & \gamma_{11}\approx12\gamma_2b_1\frac{3}{2}\frac{B-A}{C}, \nonumber \\
\end{eqnarray}
ce qui donne, pour la fr\'equence des petites librations en longitude

\begin{equation}
  \label{eq:omegau}
  \omega_u \approx n\sqrt{3\frac{B-A}{C}}=2n\sqrt{3\frac{C_{22}}{C/(M_SR_S^2)}}.
\end{equation}

Cette expression est \`a rapprocher de celle donn\'ee dans la litt\'erature pour les probl\`emes plans \citep{md1999}:

\begin{eqnarray}
  \omega_u & =       & n\sqrt{3\frac{B-A}{C}H(1,e)} \label{eq:omegaumd} \\
           & \approx & n\sqrt{3\frac{B-A}{C}\left(1-\frac{5e^2}{2}+\frac{13e^4}{16}-\frac{35e^6}{288}\right)}. \label{eq:omegaumd2}
\end{eqnarray}

Pour le mouvement polaire, on a

\begin{eqnarray}
  \psi_3  & = & \mu_{33}=\frac{C-B}{2B}, \nonumber \\
  \zeta_3 & = & \gamma_{33} \approx \frac{1}{2}\frac{C-A}{A}+\frac{3}{2}\frac{C-A}{C}\approx2\frac{C-A}{A}, \nonumber \\
\end{eqnarray}
ce qui donne

\begin{equation}
  \label{eq:omegaw}
  \omega_w \approx 2n\sqrt{\frac{C-B}{A}\frac{C-A}{B}},
\end{equation}
cette formule \'etant notamment pr\'esente dans \citep{rw2011}. L'obtention d'une formule analytique simple pour la fr\'equence d'oscillation de l'obliquit\'e
$\omega_v$ est plus complexe car elle n\'ecessiterait, par notre m\'ethode, de comparer les ordres de grandeur du coefficient de couplage $\beta$ avec l'inclinaison
$I$ et l'obliquit\'e $K$. \citet[Eq.C6]{n2010} donne :

\begin{equation}
  \label{eq:noyelles2010}
  \omega_v = n\left(\frac{3}{2}\frac{C-A}{C}+\frac{\dot{\ascnode}}{n}\cos I\right).
\end{equation}

  \subsection{Les librations forc\'ees}
  
  \par Nous consid\'erons ici que le satellite a une orbite k\'epl\'erienne autour de sa plan\`ete parente. En cons\'equence, le probl\`eme est plan, et l'excentricit\'e
  orbitale est une constante non nulle, en g\'en\'eral comprise entre $0.001$ et $0.05$.
  
  \par Les librations forc\'ees en longitude sont un mouvement oscillatoire de l'axe de plus faible inertie $\vv{f_1}$ dans le plan orbital du satellite, dues \`a un for\c{c}age
  gravitationnel. Implicitement on d\'esigne les oscillations \`a la fr\'equence orbitale, elles sont dues \`a l'excentricit\'e de l'orbite, donc aux variations de la distance
  plan\`ete-satellite, ainsi qu'\`a la rigidit\'e du satellite,  ou dans un cas plus g\'en\'eral de sa cro\^ute. 
  
  \par
\begin{figure}[ht]
  \centering
  \includegraphics[width=0.85\textwidth]{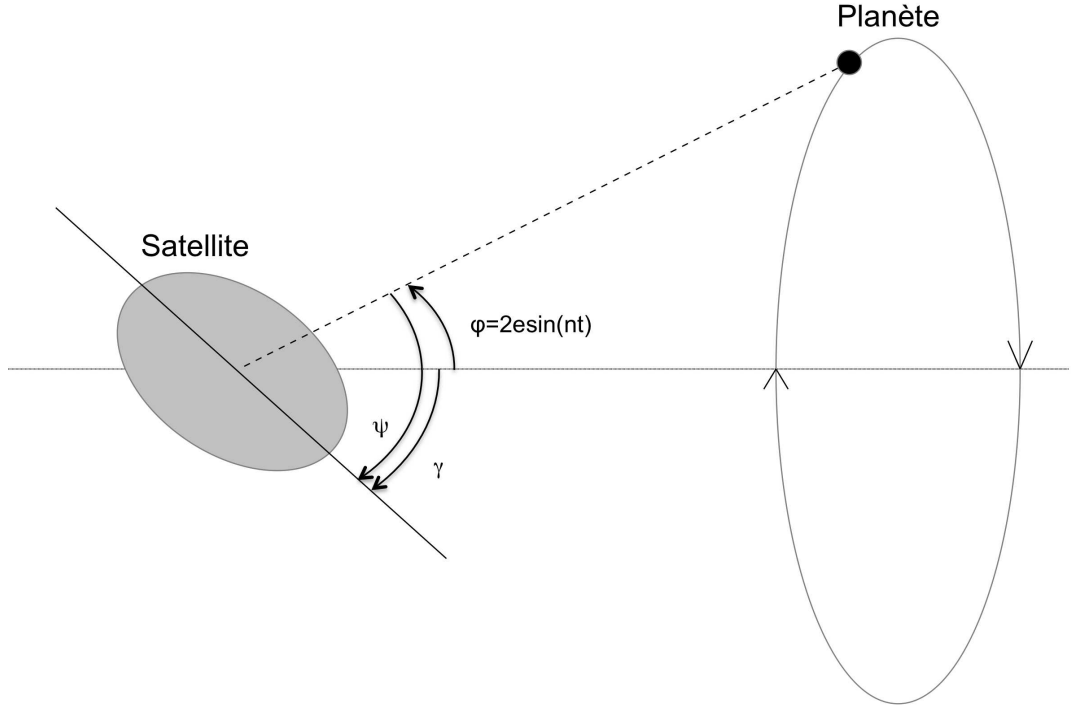}
  \caption[Les 3 diff\'erents types de libration en longitude]{Les 3 diff\'erents types de libration en longitude, exprim\'es dans un rep\`ere li\'e au satellite (figure inspir\'ee de \citep{ttb2009}). De cette 
  figure d\'ecoule imm\'ediatement la relation $\phi=\gamma-\psi$.\label{fig:librations}}
  \end{figure}

 On distingue g\'en\'eralement 3 quantit\'es (cf. Fig.\ref{fig:librations}):
  
  \begin{itemize}
  
    \item[les \emph{librations de mar\'ee} $\psi$]: Il s'agit de l'angle entre la direction de l'axe de plus faible inertie $\vv{f_1}$ et la direction satellite-plan\`ete.
    
    \item[les \emph{librations optiques} $\phi$]: Au degr\'e 1 en excentricit\'e, le satellite pointe naturellement vers l'antifoyer de son orbite autour de sa plan\`ete parente. Ceci est
    une cons\'equence de son excentricit\'e, qui induit des variations de sa vitesse orbitale instantan\'ee. Il en r\'esulte des librations entre la direction instantan\'ee 
    plan\`ete-satellite et la stricte rotation synchrone, exprim\'ees ainsi :
    
    \begin{equation}
      \label{eq:libroptik}
      \phi(t)=2e\sin nt.
    \end{equation}

    Ces librations ne sont pas affect\'ees par l'int\'erieur du satellite.
    
    \item[les \emph{librations physiques} $\gamma$]: Ces librations sont d\'efinies par l'angle entre $\vv{f_1}$ et la stricte rotation synchrone. Elles sont domin\'ees par
    l'influence de l'int\'erieur, d'o\`u leur nom.
  
  \end{itemize}

  \par Afin d'obtenir une expression analytique des librations physiques $\gamma$, partons du Hamiltonien plan (Eq.\ref{eq:hamilplan}) limit\'e au degr\'e 2 en excentricit\'e:
  
  \begin{eqnarray}
   \mathcal{H}_2(p,P,t) & = & \frac{nP^2}{2}-\frac{3n}{4}\frac{B-A}{C}\times \label{eq:hamilplan2} \\
                      &   & \left(-\frac{e}{2}\cos(2p-\lambda)+\left(1-\frac{5e^2}{2}\right)\cos(2p-2\lambda)+\frac{7e}{2}\cos(2p-3\lambda)+\frac{17e^2}{2}\cos(2p-4\lambda)\right) \nonumber
  \end{eqnarray}
avec $\lambda = nt$. En exprimant l'argument r\'esonnant $\sigma=p-\lambda$, on obtient, \`a partir des \'equations de Hamilton :

\begin{equation}
  \label{eq:d2sigma}
  \ddot{\sigma}+\frac{3}{4}n^2\frac{B-A}{C}\left(-e\sin(2\sigma+\lambda)+2\left(1-\frac{5e^2}{2}\right)\sin2\sigma+7e\sin(2\sigma-\lambda)+17e^2\sin(2\sigma-2\lambda)\right)=0.
\end{equation}
\`A la r\'esonance, $\sigma$ est proche de $0$, on peut donc d\'evelopper l'\'equation (\ref{eq:d2sigma}) au degr\'e 1 en $\sigma$, ce qui donne

\begin{equation}
  \label{eq:d2sigma2}
  \ddot{\sigma}+\frac{3}{4}n^2\frac{B-A}{C}\left(-8e\sin\lambda+4\left(1-\frac{5e^2}{2}\right)\sigma-17e^2\sin2\lambda\right)=0,
\end{equation}
donc on d\'eduit imm\'ediatement

\begin{equation}
  \label{eq:sigmaresolu}
  \sigma(t) = \mathcal{A}\cos(\omega_0't+\alpha)-\frac{2e\omega_0^2}{n^2-\omega_0^{'2}}\sin\lambda-\frac{17}{4}\frac{e^2\omega_0^2}{4n^2-\omega_0^{'2}}\sin2\lambda,
\end{equation}
avec 

\begin{eqnarray}
  \omega_0^2    & = & 3n^2\frac{B-A}{C}, \label{eq:omega02} \\
  \omega_0^{'2} & = & \omega_0^2\left(1-\frac{5e^2}{2}\right), \label{eq:omega0p2}
\end{eqnarray}
et o\`u $\mathcal{A}$ et $\alpha$ sont des constantes d'int\'egration relatives aux librations libres. On s'attend \`a ce qu'elles soient suffisamment amorties pour que 
$\mathcal{A}$ soit n\'egligeable. Ceci donne l'expression des librations physiques

\begin{equation}
  \label{eq:librphys}
  \gamma(t) = -\frac{2e\omega_0^2}{n^2-\omega_0^{'2}}\sin\lambda = -\frac{2e\omega_0^2}{n^2-\omega_0^{'2}}\sin nt.
\end{equation}

\par Ce calcul permet d'appr\'ehender une m\'ethode de calcul des librations, pr\'esent\'ee de fa\c{c}on plus g\'en\'erale par exemple dans \citep{cb2003}. Comme sugg\'er\'e par 
l'Eq.(\ref{eq:sigmaresolu}), l'amplitude des librations \`a la fr\'equence $kn$ \`a une amplitude proportionnelle \`a $e^k$, ce qui la rend en pratique ind\'etectable pour $k\geq2$ 
et \`a la limite des possibilit\'es des observations pour $k=1$. Quand $\gamma$ est mesur\'e, alors on peut obtenir $\omega_0$ et donc le param\`etre $(B-A)/C$, une sorte d'ellipticit\'e
\'equatoriale. On observe aussi un diviseur qui devient petit si la fr\'equence de for\c{c}age est proche de la fr\'equence des oscillations naturelles $\omega_0'$. Dans ce cas
on a un ph\'enom\`ene de r\'esonance qui fait radicalement cro\^itre l'amplitude de r\'eponse du satellite, et qui par exemple a permis la d\'etection des librations 
d'\'Epim\'eth\'ee \citep{ttb2009}. Ceci est en fait vrai pour tout for\c{c}age externe.

  \subsection{Les \'Etats de Cassini\label{sec:etatscassini}}
  
  \par Les \'Etats de Cassini sont les orientations possibles du moment cin\'etique du satellite \`a l'\'equilibre. \citet{c1966} a montr\'e qu'ils sont au nombre de 2 ou 4,
  selon la vitesse de pr\'ecession du n{\oe}ud orbital $\dot{\ascnode}$ par rapport au moyen mouvement $n$. L'\'Etat de Cassini attendu pour la plupart des satellites naturels est 
  l'\'Etat $1$, consid\'er\'e comme le plus stable, mais cet \'etat n'existe pas pour la Lune, qui est dans l'\'Etat de Cassini 2. Je pr\'esente ici un calcul de ces \'etats 
  que j'ai pr\'esent\'e dans \citep[Annexe B]{n2010}, en m'inspirant de \citep{wh2004}. Ce sujet est \'egalement trait\'e dans d'autres r\'ef\'erences, par exemple
  \citep{p1969,b1972,hm1987,dlr2006}.
  
  \par Ici on utilise des \'equations restreintes pour d\'eterminer uniquement l'orientation du moment cin\'etique. On consid\`ere que l'orbite a une inclinaison constante
  et pr\'ecesse uniform\'ement par rapport au plan de r\'ef\'erence $(\vv{e_1},\vv{e_2})$\footnote{On peut ainsi consid\'erer ce plan comme un Plan de Laplace. Cette notion 
  sera l'objet de la Sec.\ref{sec:callisto}}. Soit $\vec{n}$ la normale \`a l'orbite et $\vec{s}$ le vecteur unitaire colin\'eaire au moment cin\'etique ($\vec{s}=\vec{G}/G$).
  
  On a 
  
  \begin{equation}
    \label{eq:wardhamilton}
    \frac{d\vec{s}}{dt} = \alpha\left(\vec{s}\cdot\vec{n}\right)\left(\vec{s}\times\vec{n}\right)+\dot{\ascnode}\left(\vec{s}\times\vv{e_3}\right)
  \end{equation}

avec 

\begin{equation}
  \label{eq:alphawh}
  \alpha = \frac{3}{2}n\frac{J_2+2C_{22}}{C/(M_SR_S^2)} = \frac{3}{2}n\frac{C-A}{C}.
\end{equation}
On a \`a l'\'equilibre $d\vec{s}/dt=\vec{0}$ dans le rep\`ere de r\'ef\'erence pr\'ecessant d\'efini par les vecteurs $\vec{n}$, $\vv{e_3}$ et leur produit vectoriel. La projection
de l'\'equation (\ref{eq:wardhamilton}) sur la direction normale au plan ($\vec{s}$, $\vv{e_3}$) donne:

\begin{equation}
  \label{eq:wardhamiltonproj}
  \frac{\alpha}{2\dot{\ascnode}}\sin2\epsilon+\sin\left(I+\epsilon\right).
\end{equation}
On peut montrer \citep{wh2004} que cette \'equation a 4 z\'eros si $|\alpha/\dot{\ascnode}|>\left(\sin^{2/3}I+\cos^{2/3}I\right)^{3/2}\approx1$, donc que le satellite a 
4 \'Etats de Cassini, contre seulement 2 sinon.

\begin{figure}[ht]
  \centering
  \begin{tabular}{cc}
  \includegraphics[width=0.47\textwidth]{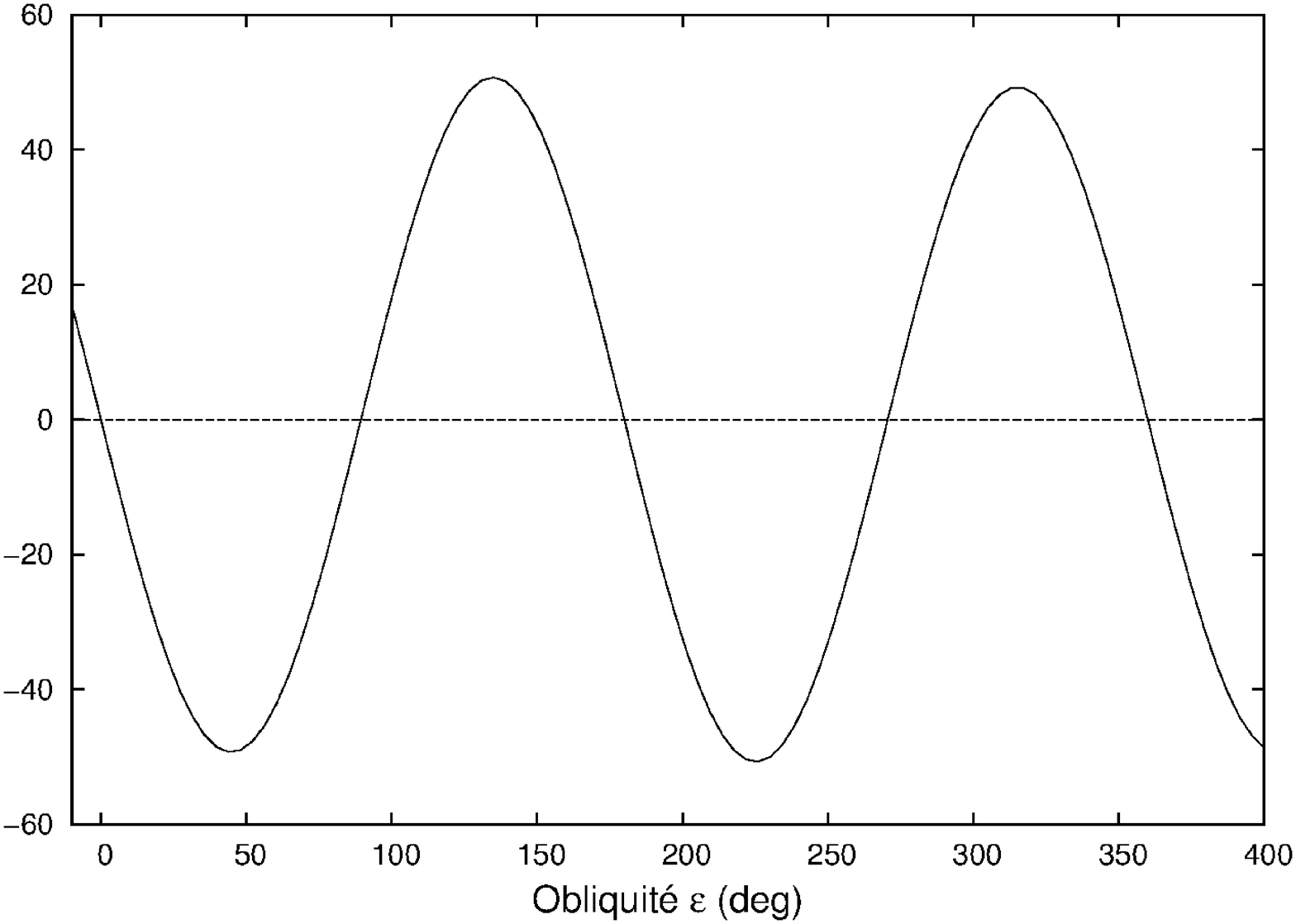} & \includegraphics[width=0.47\textwidth]{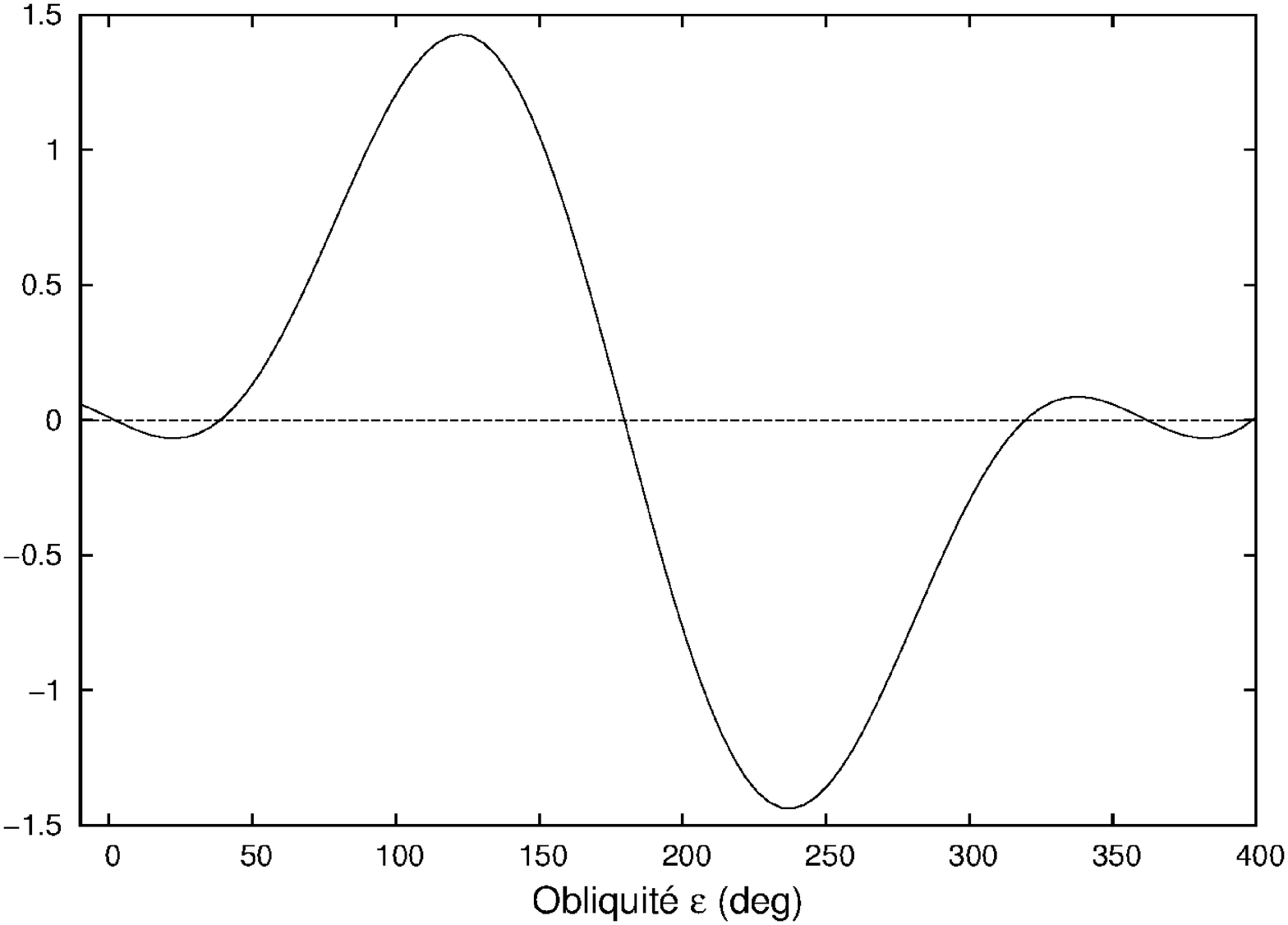}
  \end{tabular}
  \caption[2 ou 4 \'Etats de Cassini]{Positions des \'Etats de Cassini donn\'es par l'Eq.(\ref{eq:wardhamiltonproj}) en ordonn\'ee avec $\alpha/\dot{\ascnode}=-100$ (gauche)
  et $\alpha/\dot{\ascnode}=-1.3$ (droite) pour une inclinaison orbitale $I=0.01$ rad. Les \'Etats de Cassini sont respectivement, de la gauche vers la droite : 1, 4, 3, et 2.
  Lorsque le rapport $|\alpha/\dot{\ascnode}|$ est important alors les 4 \'Etats de Cassini sont proches de multiples de $\pi/2$, alors que les \'Etats 1, 4 et 2 tendent \`a se
  confondre lorsque ce rapport devient petit.\label{fig:2o4cassini}}
\end{figure}

\par Pour les satellites naturels des plan\`etes g\'eantes, il est classique d'avoir $|\alpha/\dot{\ascnode}>>1|$, dans ce cas on peut obtenir la position des 4 \'Etats de 
Cassini en d\'eveloppant l'Eq.(\ref{eq:wardhamiltonproj}) autour des multiples de $\pi/2$ et on obtient :

\begin{eqnarray}
  \epsilon & \approx & -\frac{\sin I}{\alpha/\dot{\ascnode}+\cos I} \hspace{2.4cm} \textrm{(\'Etat de Cassini 1)}, \label{eq:cs1} \\
  \epsilon & \approx & -\frac{\pi}{2}+\frac{\cos I}{\sin I-\alpha/\dot{\ascnode}} \hspace{1.6cm} \textrm{(\'Etat de Cassini 2)}, \label{eq:cs2} \\
  \epsilon & \approx & \pi+\frac{\sin I}{\alpha/\dot{\ascnode}-\cos I} \hspace{2cm} \textrm{(\'Etat de Cassini 3)}, \label{eq:cs3} \\
  \epsilon & \approx & \frac{\pi}{2}+\frac{\cos I}{\sin I+\alpha/\dot{\ascnode}} \hspace{2cm} \textrm{(\'Etat de Cassini 4)}. \label{eq:cs4}
\end{eqnarray}

     \section{Conclusion}
     
\par Ce chapitre a pr\'esent\'e des \'equations Hamiltoniennes d\'ecrivant la rotation rigide d'un corps subissant une perturbation gravitationnelle. Le Hamiltonien (\ref{eq:hamil})
admet pour seules autres hypoth\`eses que le perturbateur peut \^etre consid\'er\'e comme sph\'erique et que son mouvement n'est pas influenc\'e par la rotation de
notre corps. L'hypoth\`ese de la rotation synchrone avec \'equilibre de Cassini n'a \'et\'e faite que par la suite. Le Hamiltonien (\ref{eq:hamil}) peut donc d\'ecrire la rotation
de nombreux corps, m\^eme non r\'esonnants comme Hyp\'erion (Sec.\ref{sec:hyperion}), ou pi\'eg\'es dans une autre r\'esonance comme Mercure (Part.\ref{part:mercure}).

\par L'\'etude analytique de la rotation synchrone que j'ai pr\'esent\'ee ensuite permet notamment de mettre en lumi\`ere des liens entre l'int\'erieur et des grandeurs 
observables, comme les librations en longitude et l'obliquit\'e. \`A l'instant o\`u j'\'ecris ces lignes, les librations ont \'et\'es d\'etect\'ees de fa\c{c}on incontest\'ee
pour la Lune \citep{k1967}, le satellite de Mars Phobos \citep{bu1972}, et le satellite de Saturne \'Epim\'eth\'ee \citep{ttb2009}. On peut ajouter Mercure \`a cette liste 
\citep{mpjsh2007} en s'\'etendant \`a la r\'esonance 3:2. Des librations ont \'et\'e annonc\'ees pour Encelade \citep{ghrhtn2011} mais elles sont controvers\'ees car d'amplitude 
comparable aux incertitudes. Je fais partie d'une \'equipe qui s'appr\^ete \`a annoncer la mesure de librations de Mimas \citep{trlcrrn2014}. L'obliquit\'e dans l'\'Etat de Cassini
a \'et\'e mesur\'ee au moins pour la Lune, Mercure \citep{mpjsh2007} et Titan \citep{sklhloacgiphjw2008,mi2012}.

  \chapter{Applications de la rotation rigide\label{chap:apprigide}}

    \section{La m\'ethode g\'en\'erale}
    
    \par J'expose ici la mani\`ere g\'en\'erale dont j'ai r\'esolu des probl\`emes de rotation rigide des satellites naturels. Dans le chapitre pr\'ec\'edent j'ai expos\'e des
    aspects analytiques du probl\`eme, mais mes r\'esolutions ont \'et\'e essentiellement num\'eriques afin de n\'egliger le moins d'effets possibles. Je me suis notamment attach\'e 
    \`a avoir une mod\'elisation la plus compl\`ete possible du couple gravitationnel de la plan\`ete sur le satellite. Ceci a n\'ecessit\'e de conna\^itre pr\'ecis\'ement le 
    mouvement orbital du satellite autour de la plan\`ete, pour cela j'ai utilis\'e des \'eph\'em\'erides orbitales. J'ai essentiellement utilis\'e les \'eph\'em\'erides L1.2
    \citep{ldv2006} pour les satellites Galil\'eens de Jupiter, et TASS1.7 \citep{vd1995,dv1997} pour les principaux satellites de Saturne. J'ai coupl\'e ces \'eph\'em\'erides
    aux \'equations diff\'erentielles de rotation pour obtenir une repr\'esentation temporelle des observables de la rotation.
    
    \subsection{Les \'eph\'em\'erides orbitales des satellites naturels}
    
    \par Il existe essentiellement 2 centres de ressources dans le monde \'elaborant des \'eph\'em\'erides orbitales : le Jet Propulsion Laboratory (JPL) \`a Pasadena, CA, 
    qui d\'epend de la NASA et du Californian Institute of Technology (CalTech), et qui a notamment pour mission de livrer des \'eph\'em\'erides suffisamment pr\'ecises pour piloter 
    les sondes spatiales, et l'Institut de M\'ecanique C\'eleste et de Calcul des \'Eph\'em\'erides (IMCCE), o\`u j'ai fait ma th\`ese de doctorat, et qui est en charge des 
    \'eph\'em\'erides nationales. Les 2 entit\'es ont des approches assez diff\'erentes; notamment le JPL tend \`a fournir les \'eph\'em\'erides sous une forme num\'erique avec
    une validit\'e limit\'ee dans le temps (quelques si\`ecles), alors que l'IMCCE les d\'elivre aussi sous forme de s\'eries sinuso\"idales, ajust\'ees sur les observations,
    et extrapolables sans divergence. Ceci pr\'esente les avantages de s\'eparer les diff\'erents effets dans les \'eph\'em\'erides (l'influence de l'aplatissement de la plan\`ete, 
    d'un autre satellite,\ldots), ainsi que de pouvoir \^etre utilis\'ees sur une longue dur\'ee, permettant, dans la rotation, de s\'eparer les oscillations dues au for\c{c}age des
    oscillations libres, influenc\'ees par un choix non optimal des conditions initiales.
    
    \par Les \'eph\'em\'erides L1.2 et TASS1.7 donnent les solutions orbitales sous la forme suivante:
    
    \begin{eqnarray}
      a(t)       & = & a_0+\sum_i a_i\cos\left(\omega_i^{(1)}t+\phi_i^{(1)}\right), \label{eq:smaephem} \\
      \lambda(t) & = & nt+\lambda_0+\sum_i \lambda_i\sin\left(\omega_i^{(2)}t+\phi_i^{(2)}\right), \label{eq:lambdaephem} \\
      z(t)       & = & \sum_i z_i\exp\left(\imath\left(\omega_i^{(3)}t+\phi_i^{(3)}\right)\right), \label{eq:zephem} \\
      \zeta(t)   & = & \sum_i \zeta_i\exp\left(\imath\left(\omega_i^{(4)}t+\phi_i^{(4)}\right)\right), \label{eq:zetaephem}
    \end{eqnarray}
    avec 
    
    \begin{eqnarray}
      z & = & e\exp\imath\varpi \nonumber \\
        & = & k+\imath h,     \label{eq:defz} \\
      \zeta & = & \gamma\exp\imath\ascnode \nonumber \\
            & = & \sin\frac{I}{2}\exp\imath\ascnode \nonumber \\
            & = & q+\imath p,
    \end{eqnarray}
2 exemples sont donn\'es dans les Tables \ref{tab:lambmimas} et \ref{tab:zetacallisto}.

\begin{table}[ht]
\centering
\caption[Longitude moyenne de Mimas]{Longitude moyenne $\lambda(t)$ de Mimas donn\'ee par TASS1.7, avec $\lambda_0=0.1822485$ et $n=2435.14429644$ radians par an, l'origine des dates \'etant J1980. $\omega_1$,
$\phi_1$, $\phi_3$, $\Phi_1$, $\lambda_{01}$ et $\rho_1$ sont des \'el\'ements propres utilis\'es pour d\'ecrire la dynamique orbitale des principaux satellites
de Saturne. En particulier, $\omega_1$ repr\'esente la libration de la r\'esonance Mimas-T\'ethys.\label{tab:lambmimas}}
\begin{tabular}{rrrrrcr}
\hline
N  & Amplitude    & Phase     & Fr\'equence     & P\'eriode & Identification               & Amplitude   \\
   & (rad)        & (deg)     & (rad /an)       & (ans)     &                              & (km)        \\
\hline
 1 &  $0.7574073$ &  $39.325$ &    $0.08904538$ &   $70.56$ &  $\omega_1$                  & $140528.59$ \\
 2 &  $0.0124330$ & $117.974$ &    $0.26713613$ &   $23.52$ & $3\omega_1$                  &   $2306.81$ \\
 3 &  $0.0022664$ & $126.606$ &   $10.19765304$ &    $0.62$ & $\phi_1$                     &    $420.51$ \\
 4 &  $0.0010599$ & $267.281$ &   $10.10860767$ &    $0.62$ & $\phi_1-\omega_1$            &    $196.65$ \\
 5 &  $0.0010228$ & $165.931$ &   $10.28669842$ &    $0.61$ & $\phi_1+\omega_1$            &    $189.77$ \\
 6 &  $0.0007266$ &  $78.649$ &    $0.17809075$ &   $35.28$ & $2\omega_1$                  &    $134.81$ \\
 7 &  $0.0005061$ & $259.757$ &    $0.05765338$ &  $108.98$ & $\phi_3+2\Phi_1+\omega_1$    &     $93.90$ \\
 8 &  $0.0003590$ & $196.624$ &    $0.44522688$ &   $14.11$ & $5\omega_1$                  &     $66.61$ \\
 9 &  $0.0002628$ &   $3.120$ &    $0.06492496$ &   $96.78$ & $\phi_1+4\Phi_1+\omega_1$    &     $48.76$ \\
10 &  $0.0002459$ & $178.892$ &    $0.12043737$ &   $52.17$ & $-\phi_3-2\Phi_1+\omega_1$   &     $45.62$ \\
11 &  $0.0002237$ &  $47.956$ &   $10.01956229$ &    $0.63$ & $\phi_1-2\omega_1$           &     $41.51$ \\
12 &  $0.0002097$ & $205.255$ &   $10.37574380$ &    $0.61$ & $\phi_1+2\omega_1$           &     $38.91$ \\
13 &  $0.0001970$ & $255.529$ &    $0.11316579$ &   $55.52$ & $-\phi_1-4\Phi_1+\omega_1$   &     $36.55$ \\
14 &  $0.0001456$ &  $13.921$ & $2428.76308172$ & $0.94$ j  & $\lambda_{01}+\rho_1-\phi_1$ &     $27.01$ \\
15 &  $0.0001276$ & $312.080$ &    $5.02184135$ &    $1.25$ & $-2\Phi_1-\omega_1$          &     $23.67$ \\
16 &  $0.0001164$ & $210.730$ &    $5.19993211$ &    $1.21$ & $-2\Phi_1+\omega_1$          &     $21.60$ \\
\hline
\end{tabular}
\end{table}

\begin{table}[ht]
\centering
\caption[Inclinaison de Callisto]{$\zeta(t)$ de Callisto donn\'e par L1.2, l'origine des dates \'etant J2000. Les modes propres $\Phi_1$, $\Phi_2$, $\Phi_3$ et $\Phi_4$ 
sont dus \`a l'influence de l'aplatissement de Jupiter sur la pr\'ecession des n{\oe}uds de Io, Europe, Ganym\`ede et Callisto (satellites J-1 \`a J-4), $\Phi_0$
est un terme constant d\^u \`a l'influence s\'eculaire du Soleil, et $\lambda_{\sun}$ est le for\c{c}age annuel, d\^u au mouvement de Jupiter autour du Soleil.\label{tab:zetacallisto}}
\begin{tabular}{rrrrrcr}
\hline
N  & Amplitude    & Phase     & Fr\'equence     & P\'eriode & Identification                   & Inclinaison  \\
   &              & (deg)     & (rad /an)       & (ans)     &                                  & (arcmin)     \\
\hline
 1 &  $0.0038423$ & $138.277$ &   $0$          &   $\infty$ &  $\Phi_0$                        & $26.418$     \\
 2 &  $0.0022454$ & $342.180$ &  $-0.01116251$ &   $562.88$ &  $\Phi_4$                        & $15.438$     \\
 3 &  $0.0002604$ &  $11.189$ &  $-0.04562451$ &   $137.72$ &  $\Phi_3$                        &  $1.790$     \\
 4 &  $0.0000332$ & $318.588$ &   $1.05936266$ &     $5.93$ &  $2\lambda_{\sun}$               &  $0.228$     \\
 5 &  $0.0000050$ &  $16.323$ &  $-0.52963339$ &    $11.86$ &  $-\lambda_{\sun}$               &  $0.034$     \\
 6 &  $0.0000049$ & $240.028$ &  $-0.20790140$ &    $30.22$ &  $\Phi_2$                        &  $0.034$     \\
 7 &  $0.0000044$ & $101.323$ &   $0.52966163$ &    $11.86$ &  $\lambda_{\sun}$                &  $0.030$     \\
 8 &  $0.0000038$ & $261.084$ &   $1.58900716$ &     $3.95$ &  $3\lambda_{\sun}$               &  $0.026$     \\
 9 &  $0.0000031$ & $295.132$ &   $1.07065919$ &     $5.87$ &  $2\lambda_{\sun}-\Phi_4+\Phi_0$ &  $0.021$     \\
\hline
\end{tabular}
\end{table}

    \par Ces solutions orbitales permettent d'exprimer les quantit\'es $\hat{x}_I$, $\hat{y}_I$ et $\hat{z}_I$, coordonn\'ees du vecteur unitaire pointant vers la plan\`ete
    dans le rep\`ere inertiel.

    \subsection{Une r\'esolution num\'erique du probl\`eme}

    \par Apr\`es une \'evaluation num\'erique des quantit\'es donn\'ees par l'\'etude analytique que j'ai pr\'esent\'ee dans le Chapitre pr\'ec\'edent, en particulier 
    l'obliquit\'e d'\'equilibre, l'amplitude des librations forc\'ees, et les p\'eriodes des oscillations libres $\omega_u$, $\omega_v$ et $\omega_w$, j'int\`egre num\'eriquement 
    les \'equations de la dynamique de rotation (\ref{eq:equhamil}) \`a l'aide de l'int\'egrateur d'Adams-Bashforth-Moulton d'ordre 10 \citep{hnw1993}. Cet int\'egrateur est 
    en fait compos\'e d'un pr\'edicteur d'ordre 10, dit d'Adams-Bashforth, et d'un correcteur d'ordre 11, dit d'Adams-Moulton. Ces int\'egrateurs reposent sur les sch\'emas d'Adams,
    qui consistent \`a calculer la valeur $x_{n+1}$ de la fonction $x$ au temps $t_{n+1}$ \`a partir de sa valeur $x_n$ au temps $t_n$ et d'\'evaluations de sa d\'eriv\'ee $f$, donn\'ee
    par l'\'equation diff\'erentielle ordinaire utilis\'ee. On pose 
    
    \begin{eqnarray}
      x_{n+1}-x_n & = & \int_{t_n}^{t_{n+1}}f(\tau,x(\tau))d\tau \nonumber \\
                  & = & h\sum_{i=0}^k\beta_if_{n+i-k+1}, \label{eq:adams}
    \end{eqnarray}
avec $\sum\beta_i=1$, et o\`u les $f_{n+i-k+1}$ sont des \'evaluations ant\'erieures de la fonction de force $f$. \`A l'ordre 10, le pr\'edicteur d'Adams-Bashforth est le sch\`ema
implicite suivant :

\begin{equation}
  \label{eq:adamsbashforth}
  \begin{split}
  x_{n+1}=x_n+\frac{h}{7257600}\big(30277247f_n-104995189f_{n-1}+265932680f_{n-2} \\
  -454661776f_{n-3}+538363838f_{n-4}-444772162f_{n-5}+252618224f_{n-6} \\
  -94307320f_{n-7}+20884811f_{n-8}-2082753f_{n-9}\big)+\mathcal{O}(h^{10}),
  \end{split}
\end{equation}
et le correcteur d'Adams-Moulton est le sch\`ema explicite suivant :

\begin{equation}
  \label{eq:adamsmoulton}
  \begin{split}
  x_{n+1}=x_n+\frac{h}{479001600}\big(134211265f_{n+1}+656185652f_n-890175549f_{n-1} \\
  +1446205080f_{n-2}-1823311566f_{n-3}+1710774528f_{n-4}-1170597042f_{n-5} \\
  +567450984f_{n-6}-184776195f_{n-7}+36284876f_{n-8}-3250433f_{n-9}\big)+\mathcal{O}(h^{11}).
  \end{split}
\end{equation}  
    
    \par Par \emph{sch\`ema implicite} il faut comprendre que l'\'evaluation de $x_{n+1}$ par le pr\'edicteur d'Adams-Bashforth n'utilise pas la valeur de la d\'eriv\'ee en $t_{n+1}$, on 
    est donc dans une logique d'extrapolation, tandis que le correcteur d'Adams-Moulton utilise cette valeur, elle-même \'evalu\'ee \`a l'aide de la valeur pr\'edite de $x_{n+1}$, ce 
    \emph{sch\`ema explicite} r\'ealise donc une interpolation polynomiale de la d\'eriv\'ee $f$. Si le correcteur est inconditionnellement stable, le pr\'edicteur d'Adams-Bashforth 
    (Eq.\ref{eq:adamsbashforth}) est, lui, conditionnellement stable, ce qui signifie qu'un choix d'un pas de temps trop grand peut le faire diverger. En pratique, un pas \'egal \`a 
    $1/80$ de la p\'eriode de rotation du corps consid\'er\'e est un bon choix.
    
    \par Le choix des conditions initiales doit se faire de fa\c{c}on \`a ce qu'au temps initial de la simulation, le syst\`eme soit dans l'\'etat qu'on veut simuler, en g\'en\'eral la 
    rotation synchrone et l'\'Etat de Cassini 1 pour les satellites naturels. En g\'en\'eral on obtient cet \'etat avec une bonne approximation en posant, au temps initial, $P=1$, 
    $p=\lambda$, $R=1-\cos I$ et $r=-\ascnode$. Dans ce cas la trajectoire oscille autour de l'\'equilibre avec une relativement faible amplitude d'oscillations libres. Cet \'etat 
    peut \^etre optimis\'e par it\'erations apr\`es identification, par analyse en fr\'equences, de ces oscillations. Un algorithme adapt\'e, NAFFO, mis au point avec Nicolas Delsate
    et Timoteo Carletti \citep{ndc20xx}, fera l'objet du Chapitre \ref{chap:naffo}.
    
    \par Une fois cette trajectoire optimis\'ee connue, les solutions du probl\`eme sont exprim\'ees sous forme d'une d\'ecomposition quasi-p\'eriodique.
  
    \section{Titan}
    
    \par Ma premi\`ere \'etude de la rotation rigide a concern\'e Titan \citep{nlv2008,n2008}. Cette \'etude a en fait eu lieu en 2006-2007, \`a l'\'epoque nous ne 
    disposions que de peu de donn\'ees sur ce satellite. Cassini avait effectu\'e un survol d\'edi\'e \`a la gravitation, T11, le 27 f\'evrier 2006, qui avait donn\'e 
    les valeurs suivantes :
    
    \begin{eqnarray}
      J_2    & = & (3.15\pm0.32)\times10^{-5}, \label{eq:J2tortora} \\
      C_{22} & = & (1.1235\pm0.0061)\times10^{-5}. \label{eq:C22tortora}
    \end{eqnarray}

\par Des premi\`eres observations de la rotation par Cassini avaient \'et\'e effectu\'ees mais non encore publi\'ees, elles sugg\'eraient une rotation l\'eg\`erement super-synchrone
ainsi qu'une obliquit\'e significative. Ces r\'esultats ont \'et\'e publi\'es presque simultan\'ement \`a mon \'etude, je n'ai donc pas pu en tenir compte. Depuis, la rotation 
super-synchrone a \'et\'e d\'enonc\'ee mais l'obliquit\'e a \'et\'e confirm\'ee. Quant au champ de gravit\'e, 2 m\'ethodes de r\'eduction des observations ont permis d'obtenir 2
solutions significativement diff\'erentes. Toutes ces donn\'ees seront discut\'ees dans le Chapitre \ref{chap:oceanglobal}, notamment personne n'a r\'eussi, \`a ce jour, \`a expliquer
l'obliquit\'e de Titan sans prendre en compte un oc\'ean global.

\par L'\'etude que j'ai conduite \`a l'\'epoque doit donc \^etre consid\'er\'ee pour les m\'ethodes pr\'esent\'ees, mais pas pour les r\'esultats sur la rotation de Titan.
    
    \subsection{Rotation rigide}
    
    \par Cette \'etude consistait essentiellement en des essais analytiques et num\'eriques pour diff\'erentes valeurs du moment d'inertie polaire $C$, les coefficients 
    $J_2$ et $C_{22}$ \'etant consid\'er\'es comme connus (Eq.\ref{eq:J2tortora} \& \ref{eq:C22tortora}). 
    
    \begin{table}[ht]
     \centering
     \caption{Fr\'equences propres des petites oscillations autour de l'\'equilibre, pour un Titan rigide.\label{tab:titrigfrek}}
     \begin{tabular}{l|rrrr}
     \hline
      Modes propres & \multicolumn{2}{c}{$C=0.31M_SR_S^2$}       & \multicolumn{2}{c}{$C=0.35M_SR_S^2$} \\
                    & Fr\'equence  & P\'eriode    & Fr\'equence  & P\'eriode \\ 
                    & (rad/an)     & (ans)        & (rad/an)     & (ans) \\
     \hline
     $u$            & $2.999838$              &   $2.094508$   & $2.822839$              &   $2.225839$ \\
     $v$            & $3.754149\times10^{-2}$ & $167.366424$   & $3.324655\times10^{-2}$ & $188.987571$ \\
     $w$            & $2.049150\times10^{-2}$ & $306.623991$   & $1.814709\times10^{-2}$ & $346.236493$ \\
     \hline     
     \end{tabular}
    \end{table}

    \begin{table}[ht]
    \centering
    \caption[Modes propres pr\'esents dans le mouvement orbital de Titan autour de Saturne]{Modes propres pr\'esents dans le mouvement orbital de Titan autour de Saturne. Ces modes se retrouvent dans les variables de rotation et constituent le mouvement forc\'e
    de l'\'equilibre dynamique. Les $\lambda_i$ repr\'esentent le mouvement orbital des corps consid\'er\'es autour de Saturne, $\phi_i$ le mouvement de leur p\'ericentre, li\'e \`a
    l'excentricit\'e $e_i$, et $\Phi_i$ celui de leur n{\oe}ud ascendant, li\'e \`a l'inclinaison $I_i$.\label{tab:modorbtitan}}
    \begin{tabular}{lrrl}
     \hline
     Modes propres & Fr\'equence  & P\'eriode  & Cause   \\
     orbitaux      & (rad/an)     &            &         \\
     \hline
     $\lambda_5$   & $508.009320$ &    $4.52$ j & Rh\'ea \\
     $\lambda_6$   & $143.924047$ &   $15.95$ j & Titan  \\
     $\lambda_8$   &  $28.928522$ &   $79.33$ j & Japet  \\
     $\phi_5$      &   $0.175549$ &   $35.79$ a & $e_5$  \\
     $\Phi_5$      &  $-0.175468$ &   $35.81$ a & $I_5$  \\
     $\phi_6$      &   $0.008934$ &  $703.30$ a & $e_6$  \\
     $\Phi_6$      &  $-0.008931$ &  $703.51$ a & $I_6$  \\
     $\phi_8$      &   $0.001975$ & $3181.86$ a & $e_8$  \\
     $\Phi_8$      &  $-0.001926$ & $3263.07$ a & $I_8$  \\
     $\lambda_9$   &   $0.213299$ &   $29.46$ a & Soleil \\
     \hline
    \end{tabular}

    \end{table}

    \par La Table \ref{tab:titrigfrek} donne des exemples des fr\'equences propres d'oscillation autour de l'\'equilibre. Il est int\'eressant de comparer ces fr\'equences entre elles,
    et avec les fr\'equences du mouvement orbital, \'egalement pr\'esentes dans la dynamique de rotation. Lorsque le syst\`eme est \`a l'\'equilibre strict, alors les amplitudes associ\'ees
    aux modes propres du mouvement de rotation (Tab.\ref{tab:titrigfrek}) sont nulles tandis que celles dues au mouvement orbital sont en premi\`ere approximation ind\'ependantes des 
    conditions initiales. On constate plusieurs \'echelles de temps dans la dynamique, notamment les perturbations sur la longitude moyenne ont une p\'eriode bien plus courte que celles
    sur les n{\oe}uds et p\'ericentres, et aussi que les oscillations libres (Tab.\ref{tab:titrigfrek}). Ceci a pour cons\'equence que les librations forc\'ees repr\'esentent un mouvement 
    rapide, alors que les variations de l'obliquit\'e sont bien plus lentes. Le mouvement polaire a en g\'en\'eral une amplitude n\'egligeable, la Tab.\ref{tab:polairetitan} donne sa 
    d\'ecomposition quasi-p\'eriodique.
    
    \begin{table}[ht]
      \centering
      \caption[D\'ecomposition quasi-p\'eriodique de $\eta+\imath\xi$]{D\'ecomposition quasi-p\'eriodique de la variable complexe $\eta+\imath\xi$, pour $C=0.31M_SR_S^2$. $V$ et $W$ sont li\'es aux oscillations libres, leurs amplitudes 
      ne sont donc dues qu'au choix des conditions initiales. Seuls les termes $3$ et $4$ restent lorsque Titan est r\'eellement dans l'\'Etat de Cassini 1.\label{tab:polairetitan}}
      \begin{tabular}{r|rrrr}
       \hline
       N   & Amplitude         & P\'eriode  & Identification  & Cause         \\
           & $(\times10^{-4})$ & (ans)      &                 &               \\
       \hline
       $1$ & $9.123917$        &  $306.336$ & $w$             & $\sqrt{W}$    \\
       $2$ & $6.016886$        & $-306.336$ & $-w$            & $\sqrt{W}$    \\
       $3$ & $5.730335$        &  $351.703$ & $\phi_6-\Phi_6$ & $e_6I_6$      \\
       $4$ & $3.832129$        & $-351.703$ & $\Phi_6-\phi_6$ & $e_6I_6$      \\
       $5$ & $0.636430$        &  $135.274$ & $v-\Phi_6$      & $\sqrt{V}I_6$ \\
       $6$ & $0.383955$        & $-135.274$ & $\Phi_6-v$      & $\sqrt{V}I_6$ \\
       \hline
       \end{tabular}
    \end{table}

    \par On constate que la p\'eriode du for\c{c}age, $351.703$ ans, est proche de celle des oscillations libres, $306.336$ ans. En faisant varier le moment d'inertie polaire $C$, 
    cette p\'eriode propre peut devenir tr\`es proche de celle de for\c{c}age et peut induire un ph\'enom\`ene r\'esonnant ou quasi-r\'esonnant, rendant significatif le mouvement 
    polaire.

    \subsection{Un mouvement polaire r\'esonnant}
    
    \par
\begin{figure}[ht]
    \centering
    \begin{tabular}{cc}
     \includegraphics[width=0.45\textwidth]{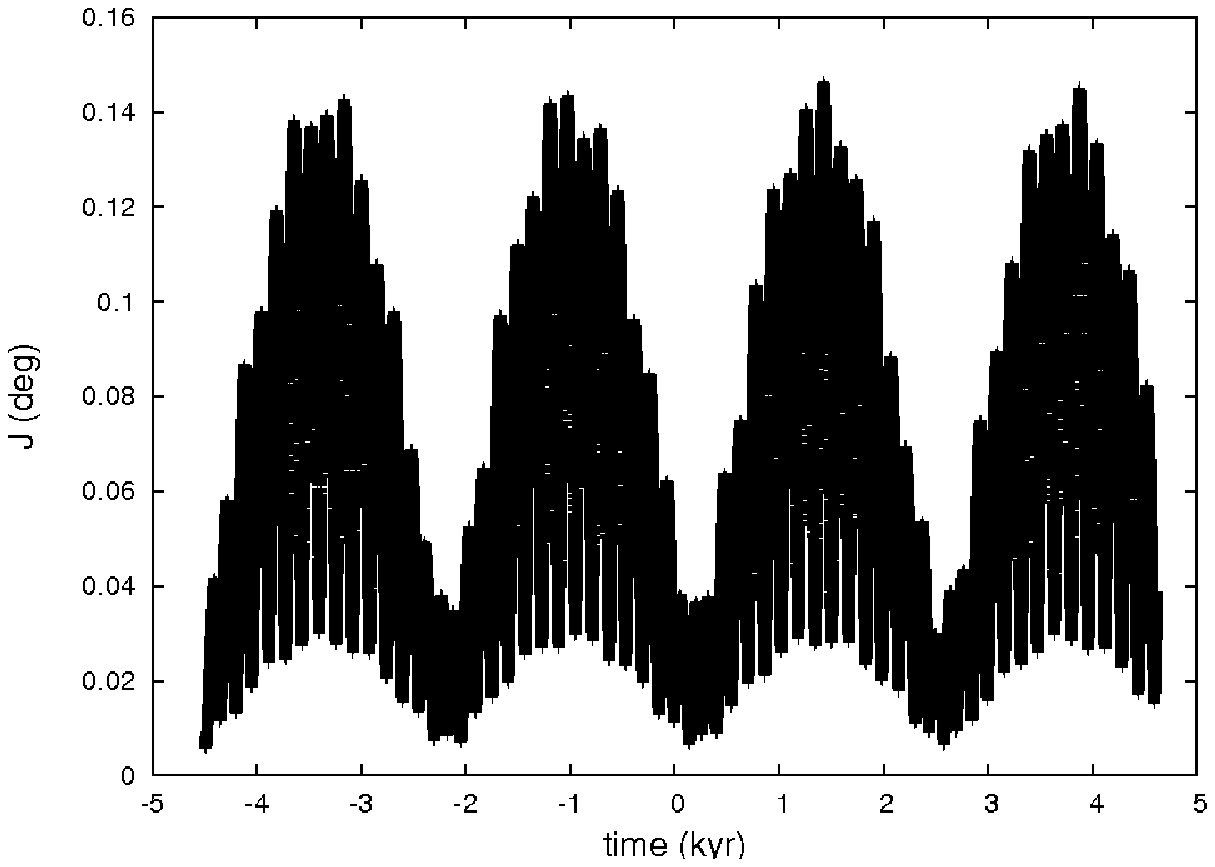} & \includegraphics[width=0.45\textwidth]{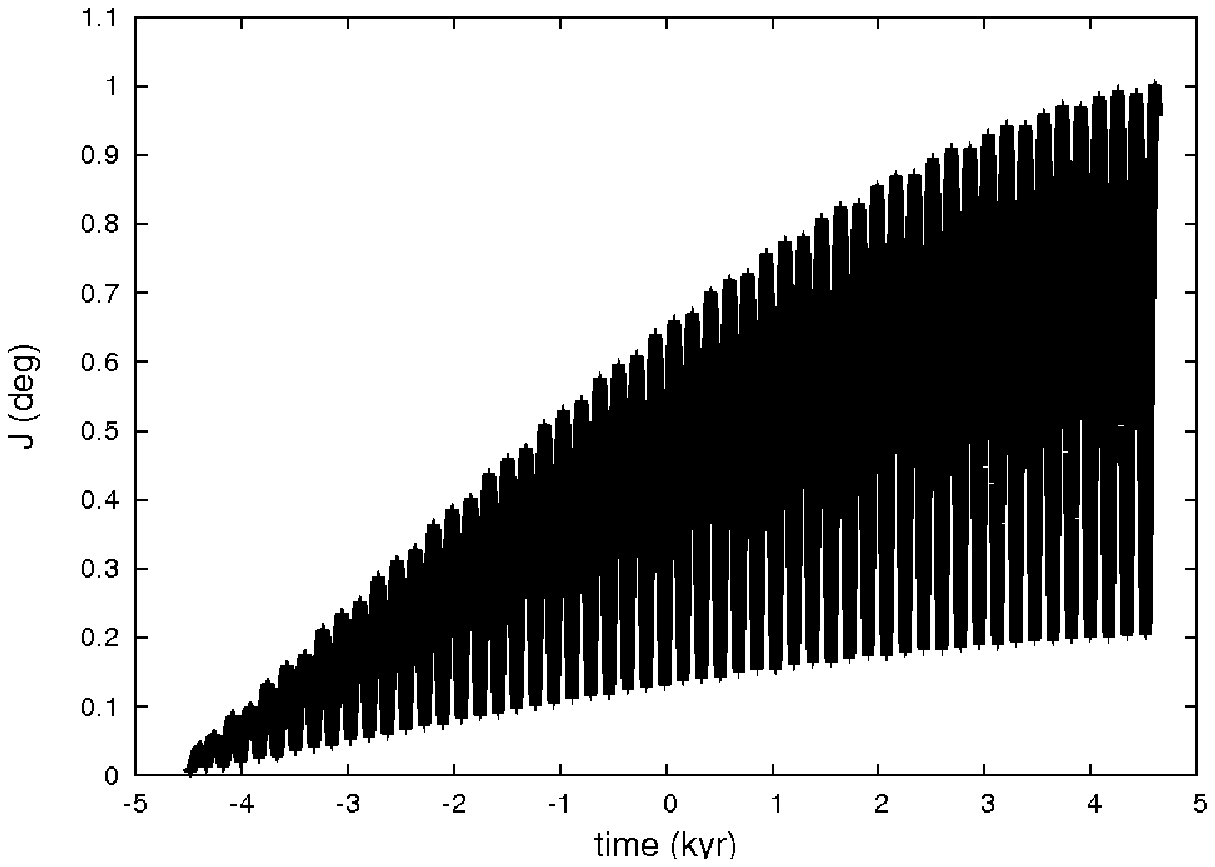} \\
     $C=0.31M_SR_S^2$ & $C=0.35M_SR_S^2$
    \end{tabular}
    \caption{Comportement quasi-r\'esonnant du mouvement polaire de Titan.\label{fig:Jtitres}}
    \end{figure}

 La Figure \ref{fig:Jtitres} montre le comportement que peut avoir l'amplitude du mouvement polaire $J$ au cours du temps, en fonction du moment d'inertie polaire $C$. On voit bien 
    notamment que cette amplitude peut d\'epasser le degr\'e (figure de droite).
    
    \par Pour repr\'esenter cette dynamique, il faut \'ecrire l'excentricit\'e et l'inclinaison de Titan ainsi : 
    
    \begin{eqnarray}
      z(t)     & = & e_1\cos\phi_6, \label{eq:ztitrig}\\
      \zeta(t) & = & \Gamma_0\cos\Phi_0+\Gamma_1\cos\Phi_6, \label{eq:zetatitrig}
    \end{eqnarray}
o\`u $\Phi_0$ est un terme constant. Lorsque le syst\`eme est strictement dans cette r\'esonance, d'argument $w+\Phi_6-\phi_6$, le mouvement polaire a une amplitude forc\'ee 
non nulle, que j'ai cherch\'ee \`a exprimer. Pour cela, je suis parti du Hamiltonien quadratique avec perturbation
    
    \begin{eqnarray}
      \mathcal{K}(u,v,w,U,V,W,t) & = & \mathcal{N}(-,-,-,U,V,W)+\mathcal{P}(u,v,w,U,V,W,t) \nonumber \\
                                 & = & \omega_uU+\omega_vV+\omega_wW+\mathcal{P}(u,v,w,U,V,W,t), \label{eq:Hquadrapert}
    \end{eqnarray}
    et j'ai effectu\'e un changement de variable pour exprimer l'argument r\'esonnant $\theta$ :

    \begin{equation}
    \label{eq:chgwobble}
    \begin{array}{lll}
    u & \hspace{2cm} & U \\
    v & \hspace{2cm} & V \\
    \theta=w+\Phi_6-\phi_6 & \hspace{2cm} & \Theta=W.
    \end{array}
    \end{equation}
    $\phi_6$ et $\Phi_6$ sont ici des fonctions lin\'eaires du temps, ce changement de variable implique donc d'ajouter la quantit\'e $(\dot{\Phi}_6-\dot{\phi}_6)\Theta$ au nouveau
    Hamiltonien $\mathcal{T}$, qui devient :
    
    \begin{equation}
      \label{eq:Hamiltau2}
      \mathcal{T} = \omega_uU+\omega_vV+(\omega_w+\dot{\Phi}_6-\dot{\phi}_6)\Theta+\mathcal{T}_2.
    \end{equation}
    En consid\'erant que $\theta$ est le seul angle lent dans le mod\`ele, on obtient, apr\`es moyennisation :
    
    \begin{equation}
      \label{eq:Hamiltau3}
      \mathcal{T} = \psi\Theta+\mu\Theta^2+\epsilon\sqrt{2\Theta}\cos\theta,
    \end{equation}
avec

\begin{equation}
  \label{eq:psi}
  \psi=\omega_{\theta}+\dot{\Phi}_6-\dot{\phi}_6,
\end{equation}


\begin{eqnarray}
    \mu & = & \frac{3n(1+\delta_s)}{128P^{*4}W^{*2}}\bigg(\gamma_1\Big(120R^{*2}\Gamma_0^2-12R^{*2}W^{*2}+48\Gamma_0^2P^{*2}W^{*2} \nonumber \\
& + & 96\Gamma_0^2R^{*2}W^{*2}-228R^*P^*\Gamma_0^2+48W^{*4}R^*P^*-21W^{*4}R^{*2} \label{eq:mu} \\
& + & 96W^{*4}\Gamma_0^2P^{*2}+168W^{*4}\Gamma_0^2R^{*2}-348W^{*4}\Gamma_0^2R^*P^*+24R^*P^* \nonumber \\
& - & 192R^*\Gamma_0^2P^*W^{*2}-15R^{*2}-24W^{*4}P^{*2}+48\Gamma_0^2P^{*2}+24R^*P^*W^{*2}-8P^{*2}W^{*2} \nonumber \\
& + & \left(2\frac{R^*}{P^*}-\left(\frac{R^*}{P^*}\right)^2\right)\big(-96\Gamma_0W^{*4}P^{*2}+84R^*\Gamma_0W^{*4}P^* \nonumber \\
& - & 48\Gamma_0P^{*2}+48\Gamma_0P^*R^*W^{*2}+60R^*P^*\Gamma_0-48\Gamma_0P^{*2}W^{*2}\big)\Big) \nonumber \\
& + & \gamma_2\Big(-120R^{*2}\Gamma_0^2+2R^{*2}W^{*2}-16\Gamma_0^2P^{*2}W^{*2}-16\Gamma_0^2R^{*2}W^{*2} \nonumber \\
& + & 228R^*P^*\Gamma_0^2+48R^*P^*W^{*4}-21R^{*2}W^{*4}+96P^{*2}W^{*4}\Gamma_0^2+168R^{*2}W^{*4}\Gamma_0^2-348R^*P^*W^{*4}\Gamma_0^2 \nonumber \\
& - & 24R^*P^*+40R^*P^*\Gamma_0^2W^{*2}+15R^{*2}-24W^{*4}P^{*2}-48\Gamma_0^2P^{*2}-8R^*P^*W^{*2}+8P^{*2}W^{*2} \nonumber \\
& + & \left(2\frac{R^*}{P^*}-\left(\frac{R^*}{P^*}\right)^2\right)\big(-96\Gamma_0W^{*4}P^{*2}+84R^*\Gamma_0W^{*4}P^* \nonumber \\
& + & 48\Gamma_0P^{*2}-8R^*\Gamma_0P^*W^{*2}-60R^*P^*\Gamma_0+16\Gamma_0P^{*2}W^{*2}\big)\Big)\bigg), \nonumber 
\end{eqnarray}

et


\begin{eqnarray}
  \epsilon & = & \frac{3ne_1\Gamma_1}{8\sqrt{W^*}P^{*\frac{5}{2}}}(1+\delta_s)\left(\gamma_1\left(8P^{*2}-46R^*P^*W^*+4R^{*2}+16P^{*2}W^*-14R^*P^*+20R^{*2}W^*\right.\right. \nonumber \\
 & + & \left.\left(P^*\sqrt{R^*(2P^*-R^*)}\right)\left(21\Gamma_0+69\Gamma_0W^*-60W^*\Gamma_0\frac{R^*}{P^*}-12\Gamma_0\frac{R^*}{P^*}\right)\right) \nonumber \\
 & + & \gamma_2\left(-8P^{*2}-46R^*P^*W^*-4R^{*2}+16P^{*2}W^*+14R^*P^*+20R^{*2}W^*\right. \label{eq:epsilon} \\
 & + & \left.\left.\left(P^*\sqrt{R^*(2P^*-R^*)}\right)\left(-21\Gamma_0+69\Gamma_0W^*-60W^*\Gamma_0\frac{R^*}{P^*}+12\Gamma_0\frac{R^*}{P^*}\right)\right)\right), \nonumber
\end{eqnarray}

o\`u $e_1=0.0289265$, $\Gamma_0=5.6024\times10^{-3}$ et $\Gamma_1=2.7899\times10^{-3}$. $\delta_s$ est un terme correctif qui tient compte de l'aplatissement de Saturne, 
$R^*$ et $P^*\approx1$ sont les valeurs de $R$ et $P$ \`a l'\'equilibre, et $W^*$ la constante donn\'ee par l'Eq.(\ref{eq:Wstar}).

\par Le Hamiltonien (\ref{eq:Hamiltau3}) est celui du Second Mod\`ele Fondamental de la R\'esonance \citep{hl1983}. Il a 2 \'equilibres si la quantit\'e
$\delta=-1-sign(\psi\mu)\left|\frac{4}{27}\frac{\psi^3}{\mu\epsilon^2}\right|^{1/3}$ est positive, ces \'equilibres sont les racines de l'\'equation cubique
$x^3-3(\delta+1)x-2=0$ avec $x=\left|\frac{2\mu}{\epsilon}\right|^{1/3}\sqrt{2\Theta}$. Une seule de ces racines est positive, il s'agit de l'\'equilibre stable de la r\'esonance.

\par En utilisant les relations

\begin{eqnarray}
  \xi^2+\eta^2 & = & 4\sin^2\frac{J}{2}, \label{eq:wob1} \\
  \xi^2+\eta^2 & = & \frac{2W_0}{W^*}\left(\cos^2\left(\omega_wt+\omega_0\right)+W^{*2}\sin^2\left(\omega_wt+\omega_0\right)\right), \label{eq:wob2}
\end{eqnarray}
o\`u $W_0$ est la valeur de $W=\Theta$ \`a l'\'equilibre de la r\'esonance, on a le mouvement polaire moyen $<J>$ :

\begin{equation}
  \label{eq:meanwob}
  <J> = \frac{1}{\pi}\int_0^{2\pi}\arcsin\left(\sqrt{\frac{W_0}{2W^*}}\sqrt{\cos^2\left(\tau\right)+W^{*2}\sin^2\left(\tau\right)}\right)\,\textrm{d}\tau.
\end{equation}

\begin{table}[ht]
  \centering
  \caption[Valeurs de l'\'equilibre r\'esonnant]{Valeurs de l'\'equilibre r\'esonnant. $W_0$ correspond \`a la valeur d'\'equilibre de $\Theta=W$, et l'amplitude moyenne 
  du mouvement polaire $<J>$ est donn\'ee par l'Eq.\ref{eq:meanwob}.\label{tab:resowob}}
  \begin{tabular}{lccc}
  \hline
    $\frac{C}{M_SR_S^2}$ & $\delta$ & $W_0$ & $<J>$ \\
\hline
$0.34$ & $5349.4$ & (pas de solution r\'eelle) & \\
$0.35$ & $1909.6$ & $0.342$ & $80.368^{\circ}$ \\
$0.355$ & $189.69$ & $0.108$ & $40.702^{\circ}$ \\
$0.3555$ & $17.705$ & $0.034$ & $22.337^{\circ}$ \\
$0.355551$ & $0.1616$ & $0.010$ & $12.034^{\circ}$ \\
$0.35555146967191$ & $1.6\times10^{-10}$ & $0.009$ & $11.413^{\circ}$ \\
$0.35555146967192$ & $-3.3\times10^{-9}$ & (pas de r\'esonance) & \\
\hline
  \end{tabular}
\end{table}

\par Les valeurs de l'\'equilibre r\'esonnant sont donn\'ees dans la Tab.\ref{tab:resowob}. Ces chiffres sont \`a prendre avec pr\'ecaution pour au moins 3 raisons :

\begin{itemize}
 \item ils supposent que Titan est rigide,
 \item ils indiquent que la r\'esonance existe, mais pas que le syst\`eme s'y trouve effectivement,
 \item ils ne rendent pas compte du comportement de Titan lorsqu'il se trouve proche de la r\'esonance. Dans ce cas, plus probable que le cas strictement r\'esonnant, 
 on peut avoir un comportement o\`u le mouvement polaire est amplifi\'e (cf.Fig.\ref{fig:Jtitres}).
\end{itemize}

\subsection{L'influence du mouvement polaire sur la dissipation de mar\'ee}

\par Les effets de mar\'ee entre un satellite et sa plan\`ete parente cr\'eent un bourrelet de mar\'ee sur chacun de ces corps, dont l'alignement avec la direction instantan\'ee
plan\`ete-satellite pr\'esente un temps de retard. Ce d\'esaxage induit un couple de rappel qui se traduira par une dissipation. Le calcul de la dissipation moyenne doit tenir 
compte de la rotation synchrone du satellite, qui pr\'esente des variations oscillantes dues \`a l'excentricit\'e de son orbite, elle-m\^eme responsable de variations de la vitesse
du satellite. Dans ce cas, la formule classique de dissipation d'\'energie $E$ dans le satellite est \citep{pc1978,pcr1979,p1999} :

\begin{equation}
  \label{eq:pealecassen}
  \frac{dE}{dt}=\frac{21}{2}\frac{k_s}{Q_s}f\frac{\mathcal{G}M_PnR_S^5e^2}{a^6},
\end{equation}
o\`u $k_s$ est le nombre de Love du champ de gravit\'e, $Q_s$ la fonction de dissipation du satellite, et $f>1$ un facteur qui permet de tenir compte d'un int\'erieur en partie
visqueux. Dans cette formule, il est suppos\'e que les librations du satellite sont les librations optiques d'amplitude $2e$, et que le mouvement est strictement plan. \citet{w2004}
propose une extension de cette formule qui tient compte de librations forc\'ees, de librations dues \`a une r\'esonance secondaire, et de l'obliquit\'e, \`a partir d'un raisonnement
bas\'e sur l'orientation spatiale du bulbe de mar\'ee :

\begin{equation}
  \label{eq:dissipwisdom}
  \frac{dE}{dt} = \frac{k_s}{Q_s}f\frac{\mathcal{G}M_PnR_S^5}{a^6}\left(\frac{9}{2}e^2+\frac{3}{2}(2e+F)^2+\frac{1}{2}S^2+\frac{3}{2}\sin^2\epsilon\right),
\end{equation}
o\`u $F$ est l'amplitude de la libration forc\'ee, $S$ est un terme suppl\'ementaire de libration, correspondant \`a une \'eventuelle r\'esonance secondaire dans la libration en
longitude, et $\epsilon$ est l'obliquit\'e du satellite. Avec $\epsilon=F=S=0$, on retrouve la formule de Peale \& Cassen (Eq.\ref{eq:pealecassen}).

\par J'\'etends ici ce raisonnement pour tenir compte du mouvement polaire. Soit $U_T$ le potentiel d\^u aux mar\'ees. La dissipation d'\'energie au sein du satellite
s'obtient par int\'egration sur son volume $V$ :

\begin{equation}
  \label{eq:dEdttitrig}
  \frac{dE}{dt} = -\iiint_V \rho\vec{v}\cdot\vv{\textrm{grad}}(U_T)\,\textrm{d}V,
\end{equation}
o\`u $\rho$ est la densit\'e d'un \'el\'ement de volume $dV$ de vitesse $\vec{v}$. Si le satellite est incompressible et de densit\'e constante, alors on a 

\begin{eqnarray}
  \frac{dE}{dt} & = & -\rho\iiint_V \textrm{div}\left(U_T\vec{v}\right)\,\textrm{d}V, \label{eq:dEdttitrig2} \\
                & = & -\rho\iint_{\mathcal{S}}U_T\vec{v}\cdot\vec{n}\,\textrm{d}\mathcal{S}, \label{eq:greenostro}
\end{eqnarray}
d'apr\`es le th\'eor\`eme de Green-Ostrogradski, ou th\'eor\`eme de flux-divergence, $\mathcal{S}$ \'etant la surface du satellite. La hauteur de mar\'ee \`a un 
point donn\'e de la surface peut s'exprimer ainsi :

\begin{equation}
  \label{eq:deltartid}
  \Delta r = -\frac{h_s}{g}U_T^*,
\end{equation}
o\`u $U_T^*$ est le potentiel de mar\'ee avec un retard temporel d\^u au temps de r\'eponse du mat\'eriau \`a l'excitation de mar\'ee. $h_s$ est le nombre de Love
li\'e au d\'eplacement, et $g$ l'acc\'el\'eration locale de la gravit\'e. Ceci donne :

\begin{equation}
  \label{eq:dEdtdeltar}
  \frac{dE}{dt} = -\frac{\rho h_s}{g}\iint_{\mathcal{S}}U_T\frac{dU_T^*}{dt}d\mathcal{S},
\end{equation}
l'\'el\'ement de surface $d\mathcal{S}$ correspondant \`a la sph\`ere de rayon unitaire.

\par Le potentiel de mar\'ee \`a un point donn\'e du satellite s'exprime par :

\begin{eqnarray}
  U_T & =       & -\frac{\mathcal{G}M_P}{r}\sum_{l=2}^{\infty}\left(\frac{R}{r}\right)^lP_l(\cos\alpha) \label{eq:potentialtides} \\
      & \approx & -\frac{\mathcal{G}M_PR^2}{r^3}P_2(\cos\alpha), \label{eq:potentialtides2}
\end{eqnarray}
o\`u $\alpha$ est l'angle entre la direction plan\`ete-satellite et satellite-point o\`u le potentiel est \'evalu\'e. $P_l$ sont les polyn\^omes de Legendre, on a
notamment $P_2(x)=(3x^2-1)/2$. $r$ est la distance satellite-plan\`ete, et $R$ la distance entre le centre du satellite et le point d'\'evaluation. On a 

\begin{equation}
  \label{eq:cosalpha}
  \cos\alpha = \frac{\vec{o}\cdot\vec{s}}{rR},
\end{equation}
o\`u $\vec{o}$ et $\vec{s}$ sont, respectivement, la position de la plan\`ete et celle de l'\'el\'ement de surface consid\'er\'e par rapport au centre du satellite. Les coordonn\'ees
cart\'esiennes d'un \'el\'ement du satellite dans le rep\`ere de figure $(\vv{f_1},\vv{f_2},\vv{f_3})$ sont :

\begin{equation}
  \label{eq:szerotitrig}
  \vv{s_0} = R\left(\sin\theta\cos\lambda, \sin\theta\sin\lambda, \cos\theta\right),
\end{equation}
o\`u $\lambda$ est la longitude plan\'etocentrique, et $\theta$ la colatitude. On a (cf. Eq.\ref{eq:lesrotations})

\begin{equation}
  \label{eq:sfroms0}
  \vec{s} = R_3(-l)R_1(-J)R_3(-g)R_1(-K)R_3(-h)\vv{s_0}.
\end{equation}
Mon but n'\'etant d'avoir que la contribution du mouvement polaire $J$, j'ai pos\'e $K=0$, $h=0$, $l=-wt$ et $g=(n+w)t$. Avec nos conventions, nous avons

\begin{equation}
  \label{eq:vectoro}
  \vec{o}=r\left(\cos f,-\sin f,0\right),
\end{equation}
o\`u $f$ est l'anomalie vraie. Au degr\'e 1 en excentricit\'e, on a

\begin{eqnarray}
  r^{-1} & = & a^{-1}\left(1+e\cos nt\right), \label{eq:ratitrig} \\
  \cos f & = & \cos nt+e\left(\cos 2nt - 1\right), \label{eq;cosftitrig} \\
  \sin f & = & \sin nt+e\sin 2nt, \label{eq:sinftitrig}
\end{eqnarray}
ce qui donne finalement d$\mathcal{S}$ :

\begin{equation}
  \label{eq:dStitrig}
  d\mathcal{S} = \sin\theta\,\textrm{d}\theta\,\textrm{d}\lambda.
\end{equation}
J'obtiens ainsi

\begin{equation}
  \label{eq:dEdtmoititrig}
  \frac{dE}{dt} = -\frac{3h_s\Delta}{5}f\frac{\mathcal{G}M_PnR_s^5}{a^6}\left(\frac{21}{2}e^2+\frac{3}{2}J_0^2\left(\frac{n+w}{n}\right)^2\right),
\end{equation}
o\`u $\Delta$ est le retard temporel du bulbe de mar\'ee. Avec $k_s = 3h_s/5$ et $\Delta=-1/Q_s$, on obtient :

\begin{equation}
  \label{eq:dEdtmoititrig2}
  \frac{dE}{dt} = \frac{k_s}{Q_s}f\frac{\mathcal{G}M_PnR_s^5}{a^6}\left(\frac{21}{2}e^2+\frac{3}{2}J_0^2\left(\frac{n+w}{n}\right)^2\right),
\end{equation}
ce qui permet de compl\'eter la formule de \citet{w2004} :

\begin{equation}
  \label{eq:dissipnoyelles}
  \frac{dE}{dt} = \frac{k_s}{Q_s}f\frac{\mathcal{G}M_PnR_S^5}{a^6}\left(\frac{9}{2}e^2+\frac{3}{2}(2e+F)^2+\frac{1}{2}S^2+\frac{3}{2}\sin^2\epsilon+\frac{3}{2}J_0^2\left(\frac{n+w}{n}\right)^2\right).
\end{equation}
Il est en g\'en\'eral consid\'er\'e que la principale source de dissipation est l'excentricit\'e. En r\'ealit\'e, l'excentricit\'e est en souvent le seul  param\`etre connu avec pr\'ecision,
notamment l'obliquit\'e des satellites naturels n'est en g\'en\'eral pas connue. On peut remarquer la pr\'esence du terme $n+w$ dans la contribution du mouvement polaire, ce qui signifie 
que dans un mod\`ele de dissipation physiquement r\'ealiste o\`u $Q_s$ d\'epend de la fr\'equence d'excitation de mar\'ee, alors une fonction de dissipation diff\'erente doit \^etre 
consid\'er\'ee pour la contribuion du mouvement polaire.

\par Ce ph\'enom\`ene de composition de vitesses mis en valeur par cette \'etude des mar\'ees sugg\`ere que si Titan a un mouvement polaire significatif mais qu'il est a priori n\'eglig\'e 
au moment d'observer sa rotation, alors il peut laisser l'impression que la rotation de Titan est l\'eg\`erement super-synchrone\ldots et c'est justement ce que l'\'equipe RADAR de Cassini
a annonc\'e \citep{sklhloacgiphjw2008}. Apr\`es une r\'evision \`a la baisse de cette super-synchronicit\'e \citep{sklhloacgiphjw2010}, les derni\`eres mesures sugg\`erent que la rotation
de Titan est en fait synchrone \citep{mi2012}.

    \section{Callisto \label{sec:callisto}}

    \par J'ai eu la curiosit\'e de faire tourner mon code de rotation rigide pour le satellite de Jupiter Callisto. J'ai pour cela utilis\'e les \'eph\'em\'erides L1.2 \citep{ldv2006}, 
    ainsi que le champ de gravit\'e sugg\'er\'e par les observations de la sonde Galileo \citep[Tab.\ref{tab:gravcallisto}]{ajmmst2001}. Les modes propres de la dynamique sont donn\'es
    dans la Tab.\ref{tab:propmodes}, et les solutions de la rotation sont trac\'ees dans la Fig.\ref{fig:numsimcallisto}.
    
    \begin{table}[ht]
    \centering
    \caption[Champ de gravit\'e de Callisto obtenu par Galileo]{Coefficients du champ de gravit\'e de Callisto obtenus \`a partir des survols de Galileo \citep{ajmmst2001}.
    La pr\'ecision des mesures n'a pas permis de mesurer ind\'ependamment $J_2$ et $C_{22}$, ils ont donc \'et\'e consid\'er\'es li\'es par la relation d\'ecoulant de 
    l'\'equilibre hydrostatique $J_2=10C_{22}/3$. De ces r\'esultats ainsi que de la forme de Callisto observ\'ee, un moment d'inertie polaire $C$ a \'et\'e d\'eduit.\label{tab:gravcallisto}}
     \begin{tabular}{lll}
      \hline
      Param\`etre & Valeur & R\'ef\'erence \\
      \hline
      $\mathcal{G}M_P$ & $1.26712767\times10^8$ km$^3$.s$^{-2}$ & Pioneer \& Voyager \citep{cs1985} \\
      $\mathcal{G}M_S$ & $7179.292$ km$^3$.s$^{-2}$             & Galileo \citep{ajmmst2001} \\
      $J_2$            & $3.27\times10^{-5}$                    & Galileo \citep{ajmmst2001} \\
      $C_{22}$         & $1.02\times10^{-5}$                    & Galileo \citep{ajmmst2001} \\
      $C$              & $0.3549M_SR_S^2$                       & \citet{ajmmst2001} \\
      \hline
     \end{tabular}
    \end{table}

    \begin{table}[ht]
      \centering
      \caption[Les modes propres de la rotation de Callisto]{Les modes propres de la rotation de Callisto. Ils viennent tous de son mouvement orbital,
      sauf les 3 derniers qui sont les oscillations libres du mouvement de rotation (Eq.\ref{eq:omegu}-\ref{eq:omegw}). La r\'esonance laplacienne est la r\'esonance
      de moyen mouvement Io-Europe-Ganym\`ede, les modes propres $\nu$ et $\rho$ ne sont pas lin\'eairement ind\'ependants des autres, ils correspondent \`a des 
      quasi-r\'esonances, et ont donc des effets dynamiques significatifs. $\nu$ est une cons\'equence de la r\'esonance laplacienne, tandis que $\rho$ est li\'e 
      \`a l'in\'egalit\'e de De Haerdtl 3:7 entre Ganym\`ede et Callisto \citep{d1892,l1973,n2005,nv2007}. $e_i$ et $I_i$ d\'esignent respectivement
      l'excentricit\'e et l'inclinaison du satellite J-i.\label{tab:propmodes}}
      \begin{tabular}{lrrl}
      \hline
      Mode   & Fr\'equence     & P\'eriode & Cause \\
      propre & (rad.an$^{-1}$) &           & \\
      \hline
      $\lambda_1$      & $1297.20447253$ &   $1.769$ j & J-1 Io \\
      $\lambda_2$      &  $646.24512024$ &   $3.551$ j & J-2 Europe \\
      $\lambda_3$      &  $320.76544409$ &   $7.154$ j & J-3 Ganym\`ede \\
      $\lambda_4$      &  $137.51159676$ &  $16.689$ j & J-4 Callisto \\
      $\lambda_{\sun}$ &    $0.52967961$ &  $11.863$ a & Soleil \\
      $\Psi$           &     $1.1142494$ &   $5.639$ a & R\'esonance laplacienne \\
      $\nu$            &     $4.7142321$ &   $1.333$ a & $\lambda_1-2\lambda_2$ \\
      $\rho$           &    $-0.2848446$ &  $22.058$ a & $3\lambda_3-7\lambda_4$ \\
      $\phi_1$         &     $0.9731185$ &   $6.457$ a & $e_1$ \\
      $\phi_2$         &     $0.2476298$ &  $25.373$ a & $e_2$ \\
      $\phi_3$         &     $0.0464869$ & $135.160$ a & $e_3$ \\
      $\phi_4$         &     $0.0117118$ & $536.484$ a & $e_4$ \\
      $\Phi_1$         &    $-0.8455888$ &   $7.431$ a & $I_1$ \\
      $\Phi_2$         &    $-0.2079026$ &  $30.222$ a & $I_2$ \\
      $\Phi_3$         &    $-0.0456245$ & $137.715$ a & $I_3$ \\
      $\Phi_4$         &    $-0.0111625$ & $562.883$ a & $I_4$ \\
      $\Phi_0$         &     $0.0000000$ &    $\infty$ & $I_{\sun}$ \\
      \hline
      $u$              & $2.5535931$     &   $2.461$ a & $\sqrt{U}$ \\
      $v$              & $0.0308604$     & $203.600$ a & $\sqrt{V}$ \\
      $w$              & $0.0198050$     & $317.253$ a & $\sqrt{W}$ \\
      \hline
      \end{tabular}
    \end{table}

         \subsection{Un comportement surprenant}
         
    \begin{figure}[ht]
     \centering
      \begin{tabular}{cc}
      \includegraphics[width=0.46\textwidth]{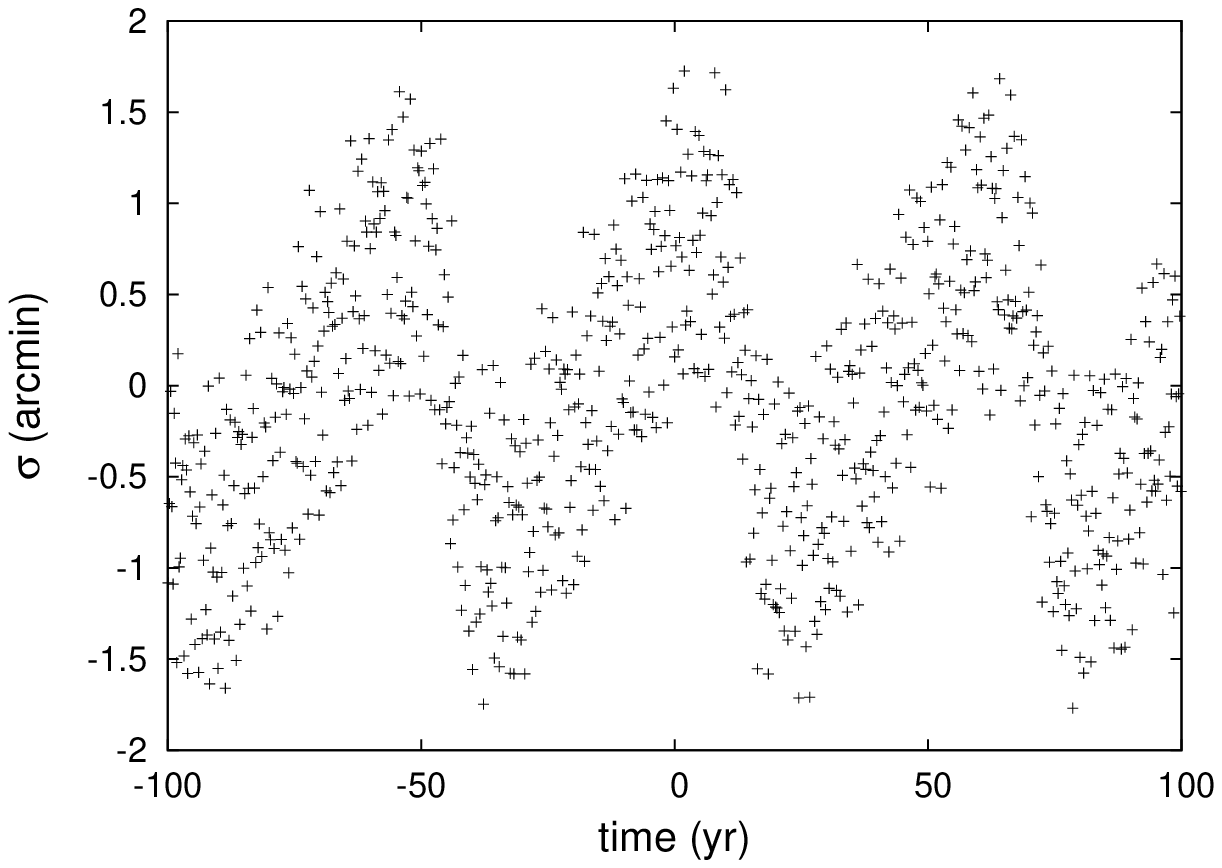} & \includegraphics[width=0.46\textwidth]{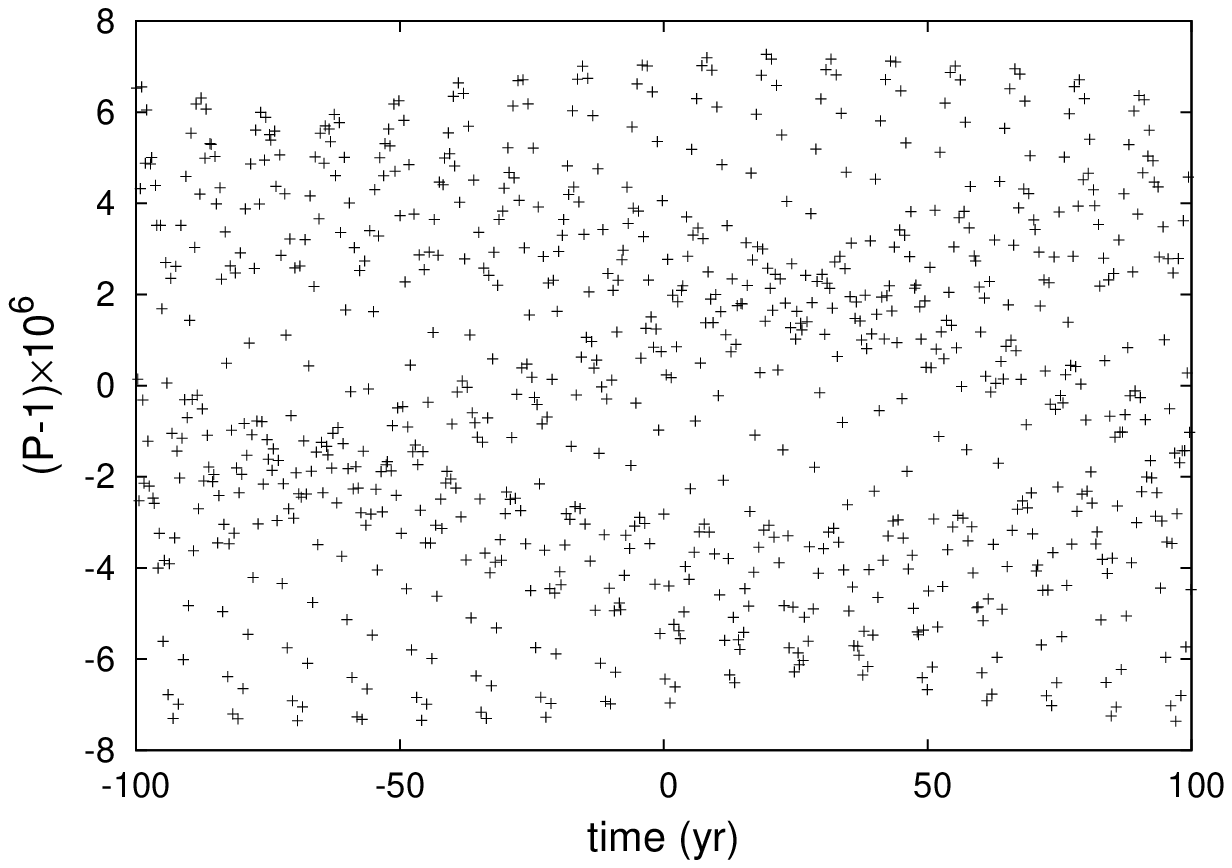} \\
      \includegraphics[width=0.46\textwidth]{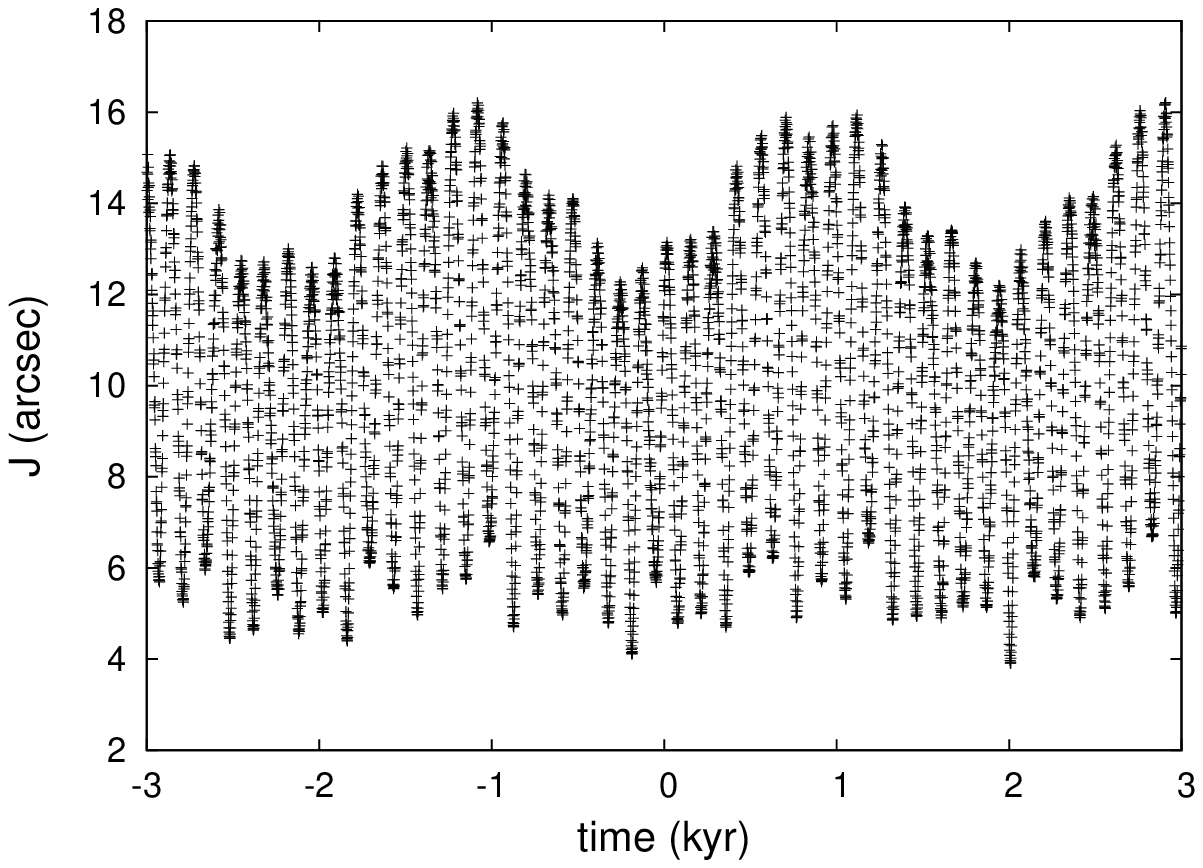} & \includegraphics[width=0.46\textwidth]{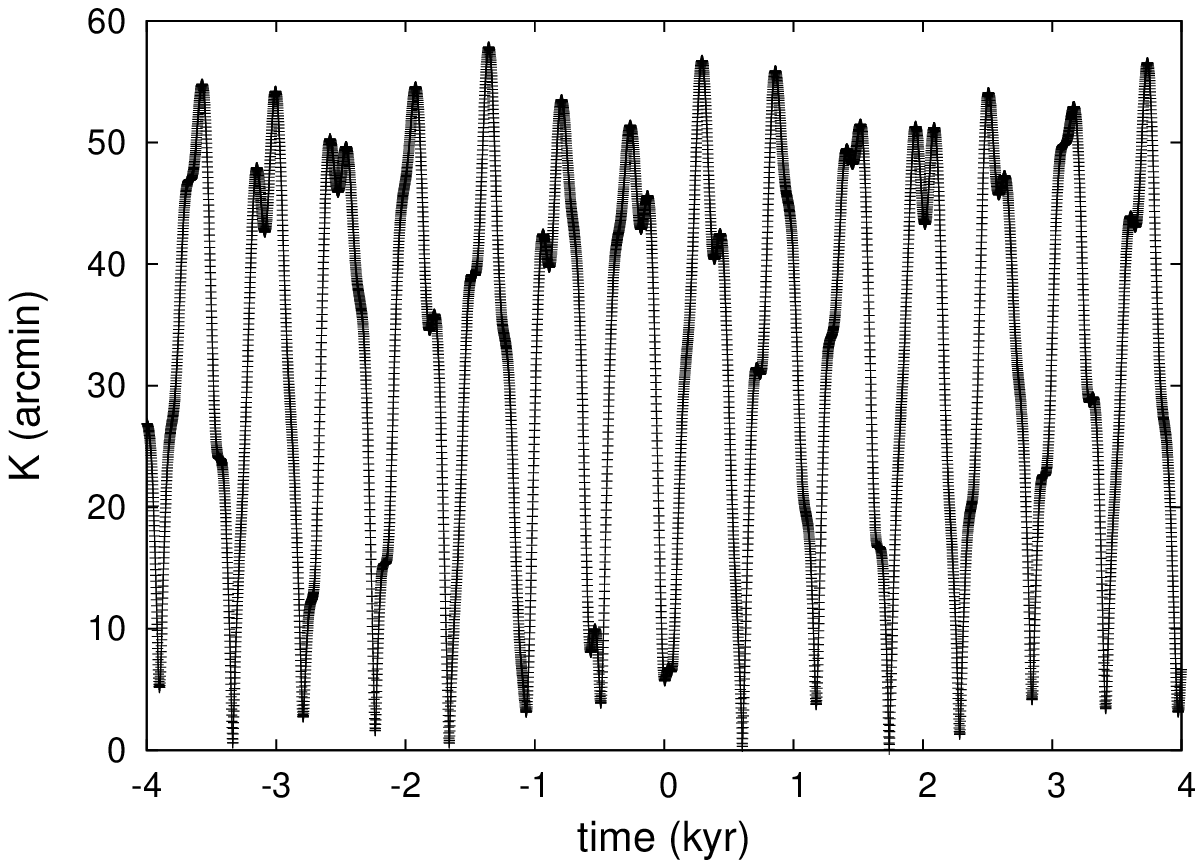} \\
      \includegraphics[width=0.46\textwidth]{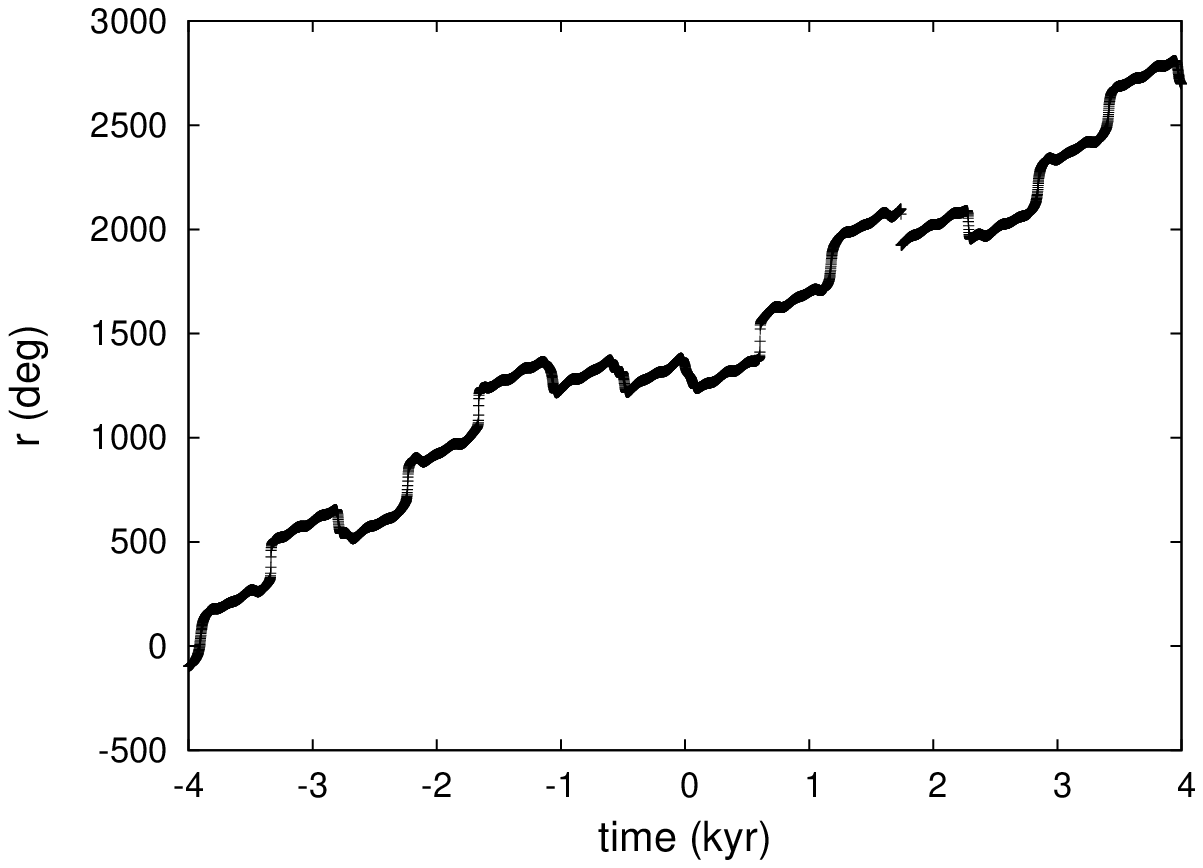} & \includegraphics[width=0.46\textwidth]{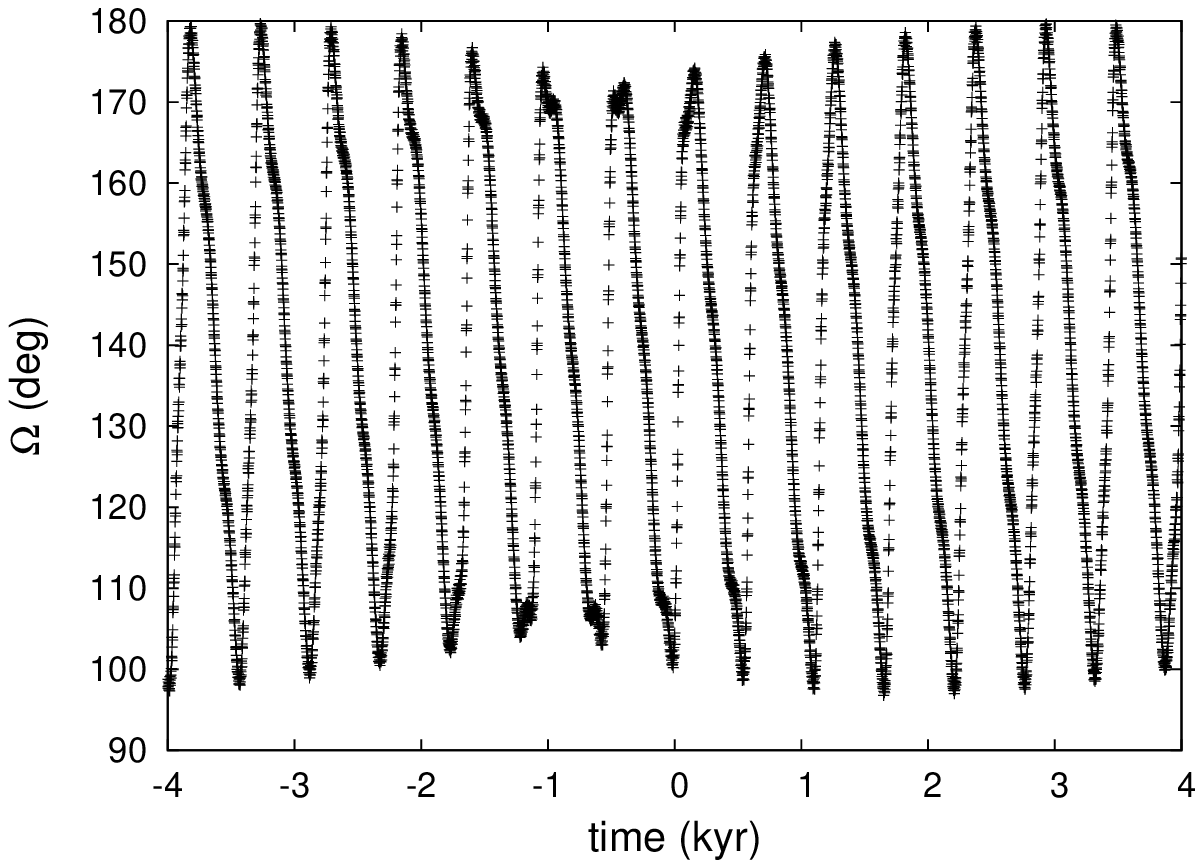}
      \end{tabular}
      \caption[Simulations num\'eriques de la rotation de Callisto.]{Simulations num\'eriques de la rotation de Callisto. L'origine des temps
      est J1950 = JD 2433282.5.\label{fig:numsimcallisto}}
\end{figure}

\clearpage

\par Un r\'esultat interpelant est le comportement du n{\oe}ud de la rotation $r$, a priori suppos\'e se comporter comme le n{\oe}ud orbital ascendant $\Omega$, selon la Troisi\`eme Loi 
de Cassini. $\Omega$ oscille avec une amplitude d'environ $40^{\circ}$ autour de la valeur moyenne $\approx140^{\circ}$ (en fait $138.277^{\circ}$), alors que le n{\oe}ud $r=-h$ 
pr\'ecesse, mais de fa\c{c}on non uniforme, comme si son comportement \'etait chaotique.

\par En fait, la Troisi\`eme Loi de Cassini \citep{c1693,c1966} dit que l'axe de rotation\footnote{Il y a ici ambiguit\'e avec le moment cin\'etique, 
le mouvement polaire \'etant n\'eglig\'e.} d'un corps dans un \'Etat de Cassini, la normale \`a son orbite, et la normale au plan de r\'ef\'erence sont coplanaires. Le 
probl\`eme est la d\'efinition du plan de r\'ef\'erence. On adopte en g\'en\'eral un plan dit de Laplace, qu'on ne d\'efinit pas toujours. La Fig.\ref{fig:numsimcallisto} 
utilise le plan de r\'ef\'erence des \'eph\'em\'erides L1.2, c'est-\`a-dire le plan \'equatorial de Jupiter \`a la date J2000. Le comportement en apparence chaotique du n{\oe}ud $r$
est en fait d\^u au choix de ce rep\`ere de r\'ef\'erence, mal adapt\'e \`a la repr\'esentation du mouvement de rotation.
    
    
      \subsection{La probl\'ematique du Plan de Laplace\label{sec:laplcall}}
      
      \par Le Plan de Laplace est, par d\'efinition, le plan qui, choisi comme r\'ef\'erence, va minimiser les variations de l'inclinaison orbitale d'un corps consid\'er\'e. 
      L'id\'ee qui rend populaire son utilisation est que l'obliquit\'e qui en r\'esulte a directement un sens physique, c'est-\`a-dire que sa valeur est domin\'ee par la
      r\'eponse de l'int\'erieur du corps consid\'er\'e, alors que pour un plan de r\'ef\'erence arbitraire, l'obliquit\'e serait domin\'ee par un effet g\'eom\'etrique, 
      signature de l'inclinaison de ce plan par rapport \`a l'orbite moyenne. Dans le contexte de la d\'etermination de l'obliquit\'e de Mercure, \citet{ym2006} proposent
      3 m\'ethodes non rigoureusement \'equivalentes pour le d\'eterminer, une bas\'ee sur une th\'eorie s\'eculaire, une sur un ajustement num\'erique sur les \'eph\'em\'erides,
      et une sur une d\'etermination analytique du pole de Laplace, \`a partir de valeurs d'inclinaison et de n{\oe}ud orbital ainsi que de leurs d\'eriv\'ees. J'ai ici
      utilis\'e la th\'eorie s\'eculaire de \citep{ym2006} pour d\'eterminer un Plan de Laplace pour Callisto.
      
      \par L'\'equation de Laplace-Lagrange r\'egissant la pr\'ecession du n{\oe}ud orbital ascendant $\Omega$ sous l'effet de l'aplatissement de Jupiter $\omega_{J4}$ et de 
      l'action gravitationnelle des autres satellites Galil\'eens $\omega_{i4}$ et du Soleil $\omega_{{\sun}4}$ s'\'ecrit :
      
      \begin{equation}
        \label{eq:dOmega}
        \frac{d\Omega}{dt} = \omega_{J4}+\omega_{14}+\omega_{24}+\omega_{34}+\omega_{{\sun}4}
      \end{equation}
avec

\begin{eqnarray}
  \omega_{J4}      & = & \frac{3}{2}J_2n_4\left(\frac{R_P}{a_4}\right)^2, \label{eq:omegaJ4call} \\
  \omega_{i4}      & = & -\frac{1}{4}\frac{\mathcal{G}M_i}{\mathcal{G}M_P}\frac{a_i}{a_4}b_{3/2}^{(1)}\left(\frac{a_i}{a_4}\right), \label{eq:omegai4} \\
  \omega_{{\sun}4} & = & -\frac{1}{4}\frac{\mathcal{G}M_{\sun}}{\mathcal{G}M_P}\left(\frac{a_{\sun}}{a_4}\right)^2b_{3/2}^{(1)}\left(\frac{a_{\sun}}{a_4}\right), \label{eq:omegas4}
\end{eqnarray}
o\`u ici $J_2=1.4736\times10^{-2}$ \citep{cs1985} est celui de Jupiter, $n_4$ est la fr\'equence orbitale de Callisto, $M_i$ est la masse du satellite $i$, $M_{\sun}$ celle
du Soleil, $a_i$ et $a_{\sun}$ les demi-grands axes des satellites et du Soleil en orbite autour de Jupiter\footnote{Dire que le Soleil tourne autour de Jupiter est 
simplement une repr\'esentation non conventionnelle du mouvement de Jupiter autour du Soleil.}, et $b_{3/2}^{(1)}$ un coefficient de Laplace classique d\'efini par \citep{bc1960} :

\begin{equation}
  \label{eq:coeflaplace}
  b_{s}^{(j)}(\alpha) = \frac{1}{\pi}\int_0^{2\pi}\frac{\cos(j\psi)}{\left(1-2\alpha\cos\psi+\alpha^2\right)^s}\,\textrm{d}\alpha.
\end{equation}

\par Appelons $p_k=\sin i_k\sin\Omega_k$ et $q_k=\sin i_k\cos\Omega_k$ pour les satellites 1 \`a 5 (le cinqui\`eme \'etant le Soleil), et $p_0$ et $q_0$ ces m\^emes quantit\'es
pour l'orientation du Plan de Laplace. On a \citep{ym2006} :

\begin{eqnarray}
  p_0 & = & \frac{p_1\omega_{14}+p_2\omega_{24}+p_3\omega_{34}+p_4\omega_{J4}+p_5\omega_{{\sun}4}}{\omega_{14}+\omega_{24}+\omega_{34}+\omega_{J4}+\omega_{{\sun}4}}, \label{eq:p0marie} \\
  q_0 & = & \frac{q_1\omega_{14}+q_2\omega_{24}+q_3\omega_{34}+q_4\omega_{J4}+q_5\omega_{{\sun}4}}{\omega_{14}+\omega_{24}+\omega_{34}+\omega_{J4}+\omega_{{\sun}4}}, \label{eq:q0marie}
\end{eqnarray}
et la d\'ecomposition quasi-p\'eriodique du r\'esultat est donn\'ee dans la Table \ref{tab:laplcall}. Il s'agit du Plan de Laplace instantan\'e, dont l'orientation d\'epend du temps.
Une fa\c{c}on directe d'obtenir un plan de r\'ef\'erence \`a partir de ce r\'esultat est d'en prendre la partie constante.

       \begin{table}[ht]
       \centering
        \caption[Le Plan de Laplace de Callisto.]{D\'ecomposition quasi-p\'eriodique de l'orientation du Plan de Laplace instantan\'e de Callisto $q_0+\imath p_0$. On obtient l'orientation du Plan de 
        Laplace s\'eculaire en conservant la premi\`ere ligne, correspondant \`a la partie constante. L'origine des dates est J2000.\label{tab:laplcall}}
        \begin{tabular}{r|rrrrrr}
        \hline
        N & Amplitude & Amplitude & Fr\'equence & Phase & P\'eriode & Identification \\
          & (deg) & & (rad/an) & (deg) & (an) & \\
        \hline
        1 &   $0.234718$ & $14.083$ arcmin & $0$          &  $138.277$ &   $\infty$ & $\Phi_0$ \\
        2 &   $0.128245$ &  $7.695$ arcmin & $-0.0111625$ &  $-17.821$ & $562.8834$ & $\Phi_4$ \\
        3 &   $0.083042$ &  $4.983$ arcmin & $-0.0456245$ & $-168.810$ & $137.7152$ & $\Phi_3$ \\
        4 &   $0.010548$ & $37.972$ arcsec & $-0.2079027$ &   $60.029$ &  $30.2218$ & $\Phi_2$ \\
        5 &   $0.002558$ &  $9.208$ arcsec &  $1.0593612$ &  $-41.412$ &   $5.9311$ & $2\lambda_{\sun}$ \\
        6 &   $0.000817$ &  $2.940$ arcsec & $-0.8455888$ &  $160.220$ &  $-7.4305$ & $\Phi_1$ \\
        7 &   $0.000392$ &  $1.413$ arcsec & $-0.5296331$ &   $16.320$ & $-11.8633$ & $-\lambda_{\sun}$ \\
        8 &   $0.000336$ &  $1.208$ arcsec &  $0.5296621$ &  $101.254$ &  $11.8626$ & $\lambda_{\sun}$ \\
        9 &   $0.000291$ &  $1.047$ arcsec &  $1.5890069$ &  $-98.914$ &   $3.9542$ & $3\lambda_{\sun}$ \\
        \hline
       \end{tabular}
       \end{table}
       
       \par La Table \ref{tab:seclaplace} donne le Plan de Laplace s\'eculaire des 4 satellites Galil\'eens. On remarque que son inclinaison augmente \`a mesure qu'on s'\'eloigne
       de la plan\`ete. En fait, l'orientation du Plan de Laplace r\'esulte essentiellement d'un \'equilibre entre la contribution de l'aplatissement de la plan\`ete et la
       perturbation gravitationnelle du Soleil. Si le satellite est proche de sa plan\`ete alors le Plan de Laplace sera proche du plan \'equatorial de celle-ci, alors que s'il est loin 
       son inclinaison sera proche de celle du Soleil par rapport \`a l'\'equateur de la plan\`ete.

       \begin{table}[ht]
        \centering
        \caption[Le Plan de Laplace des 4 satellites Galil\'eens]{Le Plan de Laplace s\'eculaire des 4 satellites Galil\'eens de Jupiter, obtenu à l'aide des formules 
        (\ref{eq:p0marie}) \& (\ref{eq:q0marie}).\label{tab:seclaplace}}
        \begin{tabular}{lcc}
        \hline
          Satellite & Amplitude & Phase \\
                    & (deg)     & (deg) \\
        \hline
          J-1 Io         & $3.41\times10^{-3}$ & $138.333$ \\
          J-2 Europe     & $2.93\times10^{-2}$ & $138.281$ \\
          J-3 Ganym\`ede & $0.121$             & $138.277$ \\
          J-4 Callisto   & $0.235$             & $138.277$ \\
          \hline
        \end{tabular}
       \end{table}

       \par Calculer l'orientation du Plan de Laplace est en pratique relativement difficile, aussi j'ai cherch\'e une mani\`ere plus directe de d\'eterminer un plan 
       de r\'ef\'erence acceptable.
      
      \subsection{Ma proposition de plan de r\'ef\'erence acceptable\label{sec:propplanref}}
      
      \par L'Union Astronomique Internationale donne l'orientation du P\^ole Nord de rotation des corps en fonction du temps sous la forme d'une s\'erie
      hybride compos\'ee d'un polyn\^ome et de termes sinuso\"idaux (cf. e.g. les recommandations de 2009 \citep{aabcccfhhknossttw2011}). Ces recommandations
      sont donn\'ees dans l'International Celestial Reference Frame (ICRF) \citep{maefgjsac1998}, lui-m\^eme proche du rep\`ere des axes principaux d'inertie de la Terre \`a J2000.
      Ce rep\`ere a l'avantage d'avoir une d\'efinition tr\`es robuste, condition indispensable pour un rep\`ere de r\'ef\'erence, mais son manque de lien avec la dynamique du 
      corps consid\'er\'e fait que le r\'esultat est domin\'e par un effet g\'eom\'etrique. La physique li\'ee \`a la r\'eponse de l'int\'erieur dans l'obliquit\'e n'est donc pas 
      directement accessible.
      
      \par La Fig.\ref{fig:numsimcallisto} montre une \'evolution irr\'eguli\`ere du n{\oe}ud de la rotation, rendant impossible une repr\'esentation sous forme de s\'erie 
      compos\'ee d'un polyn\^ome et de termes oscillants. Pour comprendre ce qui se passe, il faut exprimer les variables d'inclinaison orbitale en fonction du temps, i.e.
      
      \begin{equation}
        \label{eq:zetacallisto}
        \zeta(t) \approx 0.0038423\exp\imath\Phi_0+0.0022454\exp\imath\Phi_4,
      \end{equation}
      d'apr\`es \citep{ldv2006} et la Tab.\ref{tab:zetacallisto}. On constate que les amplitudes associ\'ees au terme constant et au terme oscillant sont du m\^eme ordre de grandeur,
      alors que dans la logique d'une repr\'esentation bas\'ee sur la Plan de Laplace, le terme oscillant aurait une amplitude largement dominante.
      
      \par J'ai calcul\'e la variable de rotation $R=P(1-\cos K)$ \`a l'aide d'une th\'eorie des perturbations \citep{d1969}. Pour cela, j'ai consid\'er\'e que Callisto
      \'etait sur une orbite circulaire, que son inclinaison \'etait r\'egie par la formule (\ref{eq:zetacallisto}) o\`u $\Phi_0$ est constant et $\Phi_4$ d\'epend 
      lin\'eairement du temps, et que le mouvement polaire \'etait nul. De fa\c{c}on similaire au Chapitre \ref{chap:rigide}, le Hamiltonien $\mathcal{H}(p,r,P,R,t)$ devient 
      $\mathcal{K}(u,v,U,V,t)$ :
      
      \begin{equation}
        \label{eq:hamilkcallisto}
        \mathcal{K}(u,v,U,V,t)=\omega_uU+\omega_vV+\mathcal{P}(u,v,U,V,t),
      \end{equation}
o\`u $\omega_u$ et $\omega_v$ sont les fr\'equences des petites oscillations autour de l'\'equilibre (cf.Tab.\ref{tab:propmodes}). On a besoin, pour exprimer $R$, du
g\'en\'erateur du premier ordre $\mathcal{W}_1$, solution de moyenne nulle de l'\'equation homologique :

\begin{equation}
  \label{eq:homologikcallisto}
  \omega_u\frac{\partial\mathcal{W}_1}{\partial u}+\omega_v\frac{\partial\mathcal{W}_1}{\partial v}+n_4\frac{\partial\mathcal{W}_1}{\partial \lambda_4}+\dot{\Phi}_4\frac{\partial\mathcal{W}_1}{\partial \Phi_4}=\mathcal{P}.
\end{equation}
On peut exprimer $R$ comme fonction des variables canoniques en inversant les expressions de $u$, $v$, $U$, $V$ (Eq.\ref{eq:chgpolaires}), i.e. $R=f(u,v,U,W)$, et on
obtient finalement $R$ d\'ependant du temps par :

\begin{eqnarray}
  R & = & R^*+\left(f(u,v,U,V)\right)|_{U=0,V=0} \nonumber \\
    & = & R^*+\frac{\partial f}{\partial u}\frac{\partial\mathcal{W}_1}{\partial U}+\frac{\partial f}{\partial v}\frac{\partial\mathcal{W}_1}{\partial V}-\frac{\partial f}{\partial U}\frac{\partial\mathcal{W}_1}{\partial u}-\frac{\partial f}{\partial V}\frac{\partial\mathcal{W}_1}{\partial v}, \label{eq:poissoncallisto}
\end{eqnarray}
o\`u $R^*=P^*(1-\cos K^*)$ est donn\'e par le syst\`eme d'\'equations (\ref{eq:E10})-(\ref{eq:E20}), et $(;)$ est le crochet de Poisson. Tous calculs faits, on obtient num\'eriquement

\begin{eqnarray}
 R(t) & = & 2.95297\times10^{-5} \nonumber \\
      & + & 5.40736\times10^{-5}\cos(\Phi_4-\Phi_0) \nonumber \\
      & - & 3.87221\times10^{-9}\cos(\Phi_4-2\lambda_4+\Phi_0) \nonumber \\
      & + & 2.56663\times10^{-9}\cos(2\Phi_4-2\Phi_0), \label{eq:Rnum}
\end{eqnarray}
ce qui est incompatible avec la d\'efinition de $R=P(1-\cos K)$ qui sugg\`ere une quantit\'e toujours positive. En \'ecrivant 

\begin{eqnarray}
  \zeta(t) & = & \Gamma_0\exp\imath\Phi_0+\Gamma_1\exp\imath\Phi_4, \label{eq:zetacallistoencore} \\
  R(t)     & = & A_0+A_1\cos\left(\Phi_4-\Phi_0\right), \label{eq:Rcallistoencore}
\end{eqnarray}
on obtient, apr\`es calculs :

\begin{eqnarray}
  A_0 & = & 2\Gamma_0^2,\label{eq:A0callisto} \\ 
  A_1 & = & 6\Gamma_0\Gamma_1\frac{n_4}{\big|\omega_v^2-\dot{\Phi}_4^2\big|}(\gamma_1+\gamma_2)\left(\frac{3}{4}n_4(2\gamma_1+\gamma_2)\left(1+\frac{1}{2}\left(\frac{\Gamma_1}{\Gamma_0}\right)^2\right)-\dot{\Phi}_4\right), \label{A1callisto}
\end{eqnarray}
et pour avoir une variable $R$ toujours positive alors on doit avoir $A_1<A_0$, ce qui n'est pas le cas ici. \`A l'instar des librations forc\'ees en longitude (Eq.\ref{eq:sigmaresolu}),
on a une diff\'erence de fr\'equences au d\'enominateur, ce qui veut dire que si la fr\'equence des oscillations libres $\omega_v$ est proche de la fr\'equence de pr\'ecession
du n{\oe}ud orbital ascendant $\dot{\Phi}_4$, alors un effet de quasi-r\'esonance va amplifier fortement l'amplitude de r\'eponse $A_1$. On peut directement exprimer une condition 
en fonction des quantit\'es orbitales $\Gamma_0$ et $\Gamma_1$, i.e. $A_1<A_0$ revient \`a $\Gamma_1<\alpha\Gamma_0$ avec

\begin{equation}
\label{eq:alphacallisto}
\alpha=\frac{\big|\omega_v^2-\dot{\Phi}_4^2\big|}{3n(\gamma_1+\gamma_2)\left(\frac{9}{8}n_4(2\gamma_1+\gamma_2)-\dot{\Phi}_4\right)}.
\end{equation}
Ici, $\alpha$ est en fait une borne sup\'erieure o\`u la quantit\'e $1+(\Gamma_1/\Gamma_0)^2/2$ a \'et\'e major\'ee par $3/2$. En effet, ce calcul n'a de sens que 
lorsque $\Gamma_1<\Gamma_0$. Pour r\'esumer, on a

\begin{itemize}
  \item si $\Gamma_1>\Gamma_0$, alors le n{\oe}ud orbital et le n{\oe}ud de la rotation pr\'ecessent tous les 2 \`a la même fr\'equence,
  \item si $\Gamma_1<\alpha\Gamma_0$, alors les 2 n{\oe}uds oscillent autour d'une valeur constante,
  \item si $1>\Gamma_1/\Gamma_0>\alpha$, alors on est dans une sorte de zone interdite o\`u le n{\oe}ud orbital oscille alors que le 
  n{\oe}ud de rotation peut pr\'ecesser. Cet effet purement g\'eom\'etrique peut r\'esulter en un comportement en apparence chaotique, d\'ej\`a identifi\'e
  par \citet{c1966} \emph{the relative motion is in some cases a pure regular precessional motion but, in many cases, it may be very different from regular
  precessional motion}.
\end{itemize}

\par En une phrase, tout ceci signifie que pour repr\'esenter de fa\c{c}on quasi-p\'eriodique l'orientation du moment cin\'etique d'un corps dans 
l'\'Etat de Cassini 1, le plan de r\'ef\'erence doit \^etre soit suffisamment proche du plan orbital moyen, soit suffisamment \'eloign\'e (c'est en
g\'en\'eral le cas de l'ICRF), mais une zone interm\'ediaire est \`a \'eviter. Je sugg\`ere de prendre comme plan de r\'ef\'erence le plan d\'efini par
le mode propre constant $\Phi_0$, qui est une sorte de plan orbital moyen. Ceci permet d'annuler la nouvelle quantit\'e $\Gamma_0$. \`A partir du rep\`ere des 
\'eph\'em\'erides, c'est-\`a-dire le plan \'equatorial de Jupiter \`a J2000, il est obtenu par une rotation d'axe $z$ (une $R_3$) d'angle $138.277^{\circ}$, puis une 
rotation d'axe $x$ (une $R_1$) d'angle $2\arcsin\Gamma_0\approx0.44^{\circ}$. Le choix de ce rep\`ere peut sembler naturel, et il est m\^eme parfois abusivement 
consid\'er\'e comme le Plan de Laplace dans la litt\'erature. Mon travail montre que ce n'est pas le Plan de Laplace, mais que son utilisation est tout-\`a-fait l\'egitime.

\begin{figure}[ht]
\centering
\begin{tabular}{cc}
\includegraphics[width=0.45\textwidth]{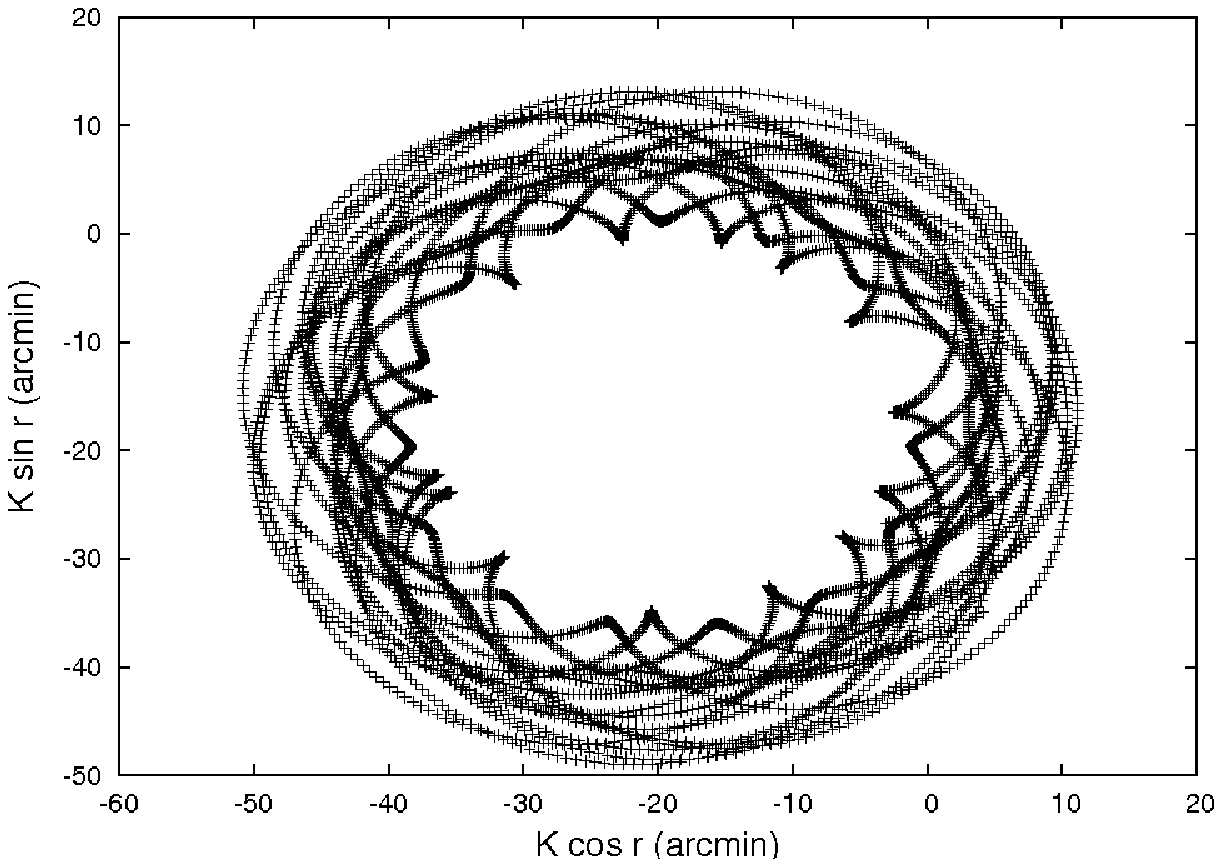} & \includegraphics[width=0.45\textwidth]{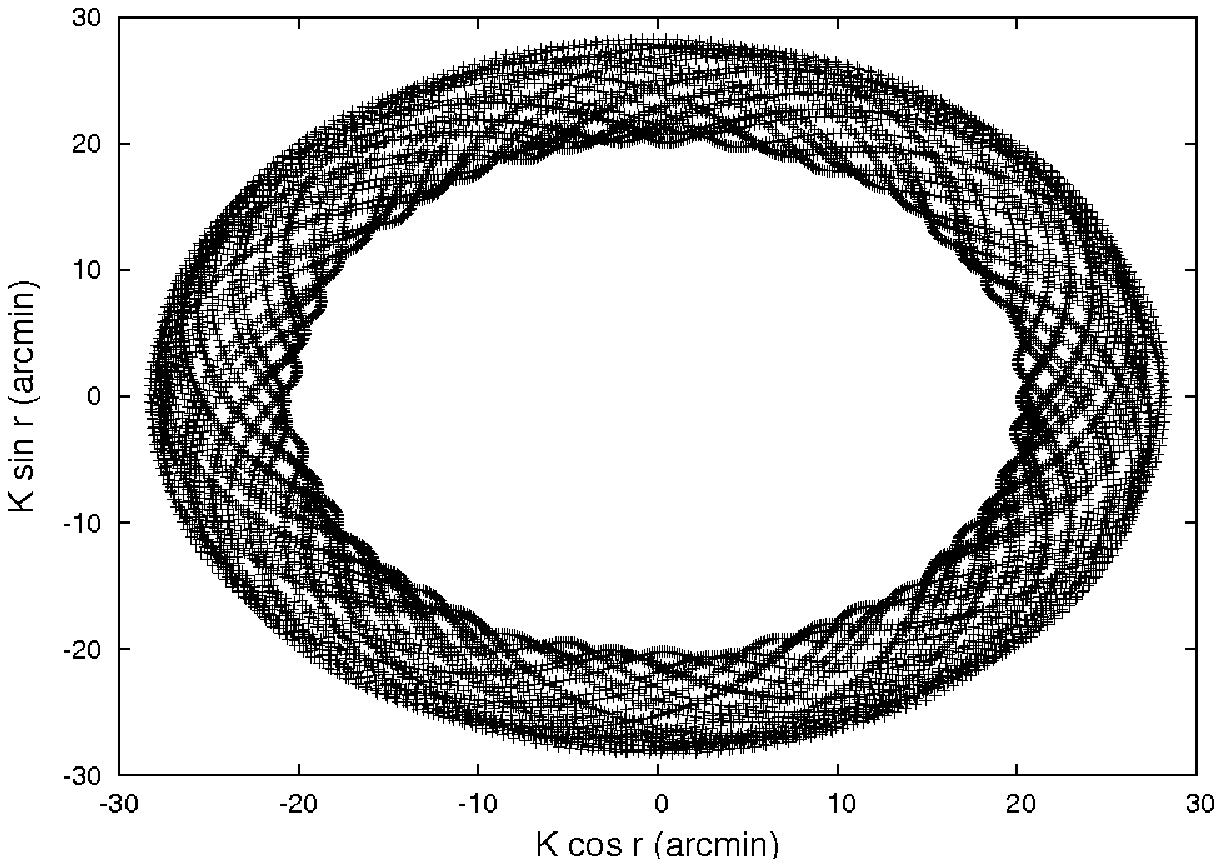}
\end{tabular}
\caption[Mouvement du moment cin\'etique de Callisto $\vec{G}$]{Mouvement du moment cin\'etique de Callisto $\vec{G}$, par rapport au plan \'equatorial jovien \`a J2000 (\`a gauche) et par rapport au plan d\'efini par $\Phi_0$ (\`a droite). Ceci 
permet d'illustrer aussi le fait que le mouvement en apparence irr\'egulier obtenu pour le n{\oe}ud \'etait d\^u au fait que le centre du rep\`ere \'etait dans l'\'epaisseur de la 
trajectoire.\label{fig:Krcallisto}}
\end{figure}

\par La Fig.\ref{fig:Krcallisto} illustre ce changement de rep\`ere, qui consiste en pratique \`a centrer le rep\`ere de r\'ef\'erence par rapport au mouvement de pr\'ecession 
du moment cin\'etique $\vec{G}$.

      \subsection{Les quantit\'es de rotation de Callisto}
      
      \par
\begin{figure}[ht]
     \centering
     \begin{tabular}{cc}
     \includegraphics[width=0.46\textwidth]{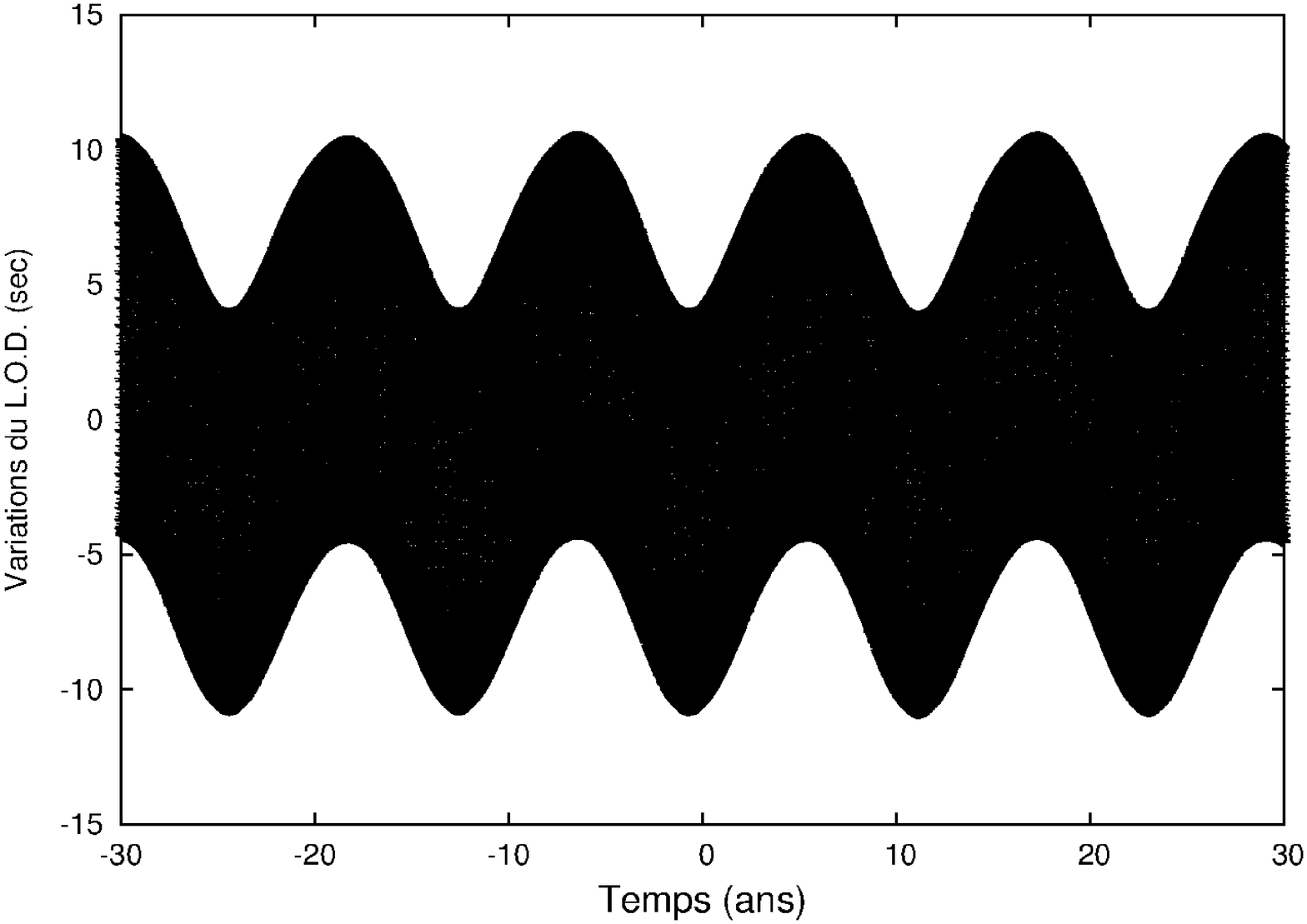} & \includegraphics[width=0.46\textwidth]{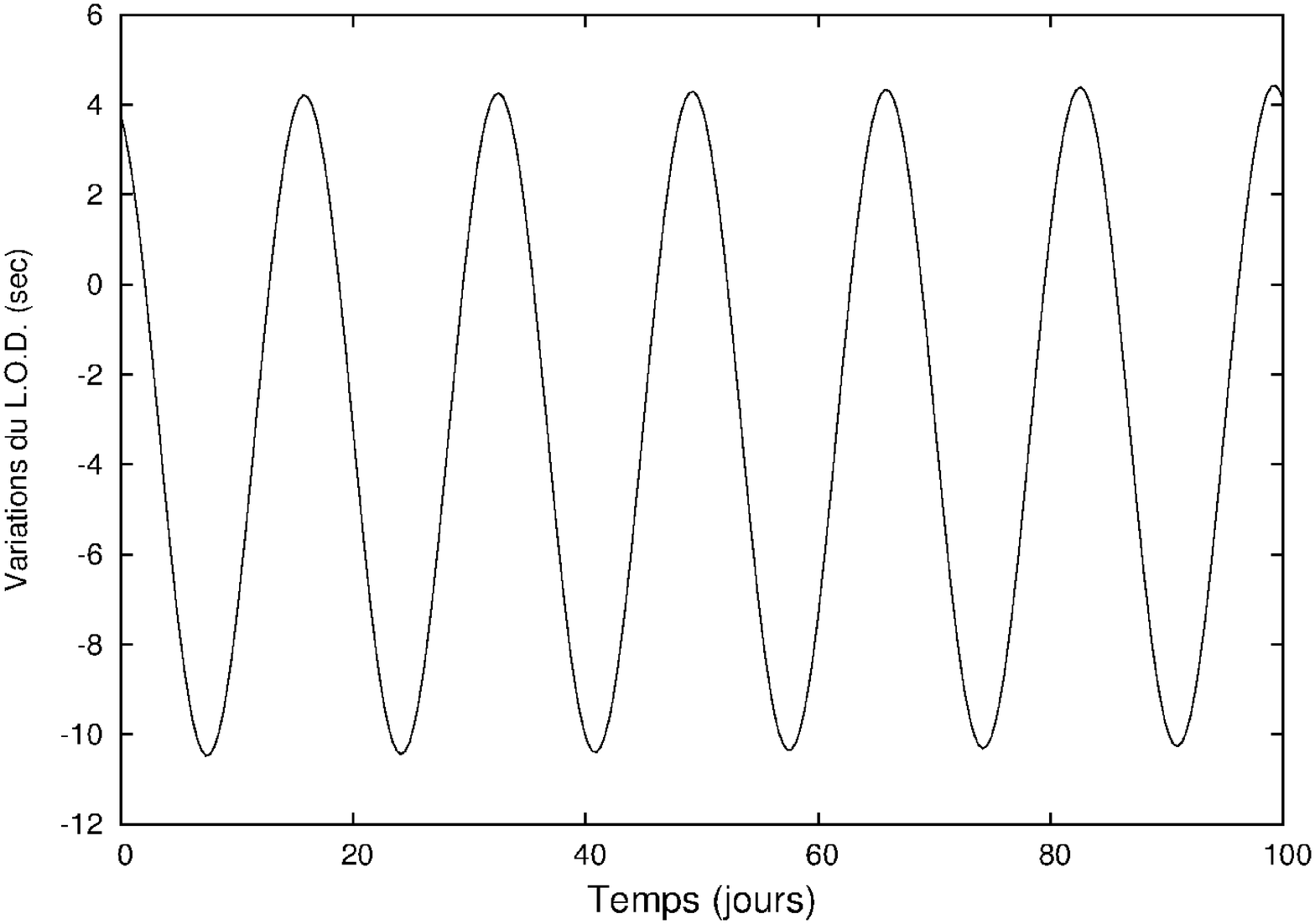}
     \end{tabular}
     \caption[Variations de la dur\'ee du jour de Callisto]{Variations de la dur\'ee du jour (L.O.D.) de Callisto. La valeur moyenne est $16.6860174175$ jours, et la 
     date de r\'ef\'erence J2000.\label{fig:lodcallisto}}
    \end{figure}

 Le lecteur trouvera dans \citep{n2009} la d\'ecomposition quasi-p\'eriodique des diff\'erentes variables de rotation, notamment l'obliquit\'e, le mouvement
      polaire, et les librations en longitude. J'obtiens un mouvement polaire d'une amplitude inf\'erieure \`a 200 m\`etres, une obliquit\'e moyenne de $9.65$ arcmin avec
      des oscillations de $5.24$ arcmin en 182 ans, correspondant \`a la combinaison de modes propres $\Phi_3-\Phi_4$, et des librations en longitude \`a la fr\'equence $n_4$
      d'amplitude $\approx1$ arcsec, soit $\approx12$ m\`etres. J'ai ici consid\'er\'e Callisto comme un corps rigide, \citep{rvk2011} obtiennent un r\'esultat environ 10
      fois plus grand en supposant l'existence d'un oc\'ean global qui d\'ecouple le mouvement de la cro\^ute, de relativement faible inertie, du reste de l'int\'erieur.
      Mais la principale composante des librations vient du for\c{c}age annuel, de p\'eriode $11.86$ ans, et d'amplitude $\approx2$ arcmin, soit $\approx140$ m\`etres.
      La Fig.\ref{fig:lodcallisto} illustre les variations de la dur\'ee du jour sur Callisto, et illustre l'importance de ce for\c{c}age annuel.
    
    \section{Les coorbitaux Janus et \'Epim\'eth\'ee}

    \par Cette \'etude a \'et\'e motiv\'ee par l'observation par Cassini de la rotation de Janus et \'Epim\'eth\'ee \citep{ttb2009}.
    
    \subsection{Une dynamique orbitale int\'eressante}
    
    \par
\begin{figure}[ht]
      \centering
      \begin{tabular}{cc}
      \includegraphics[width=0.46\textwidth]{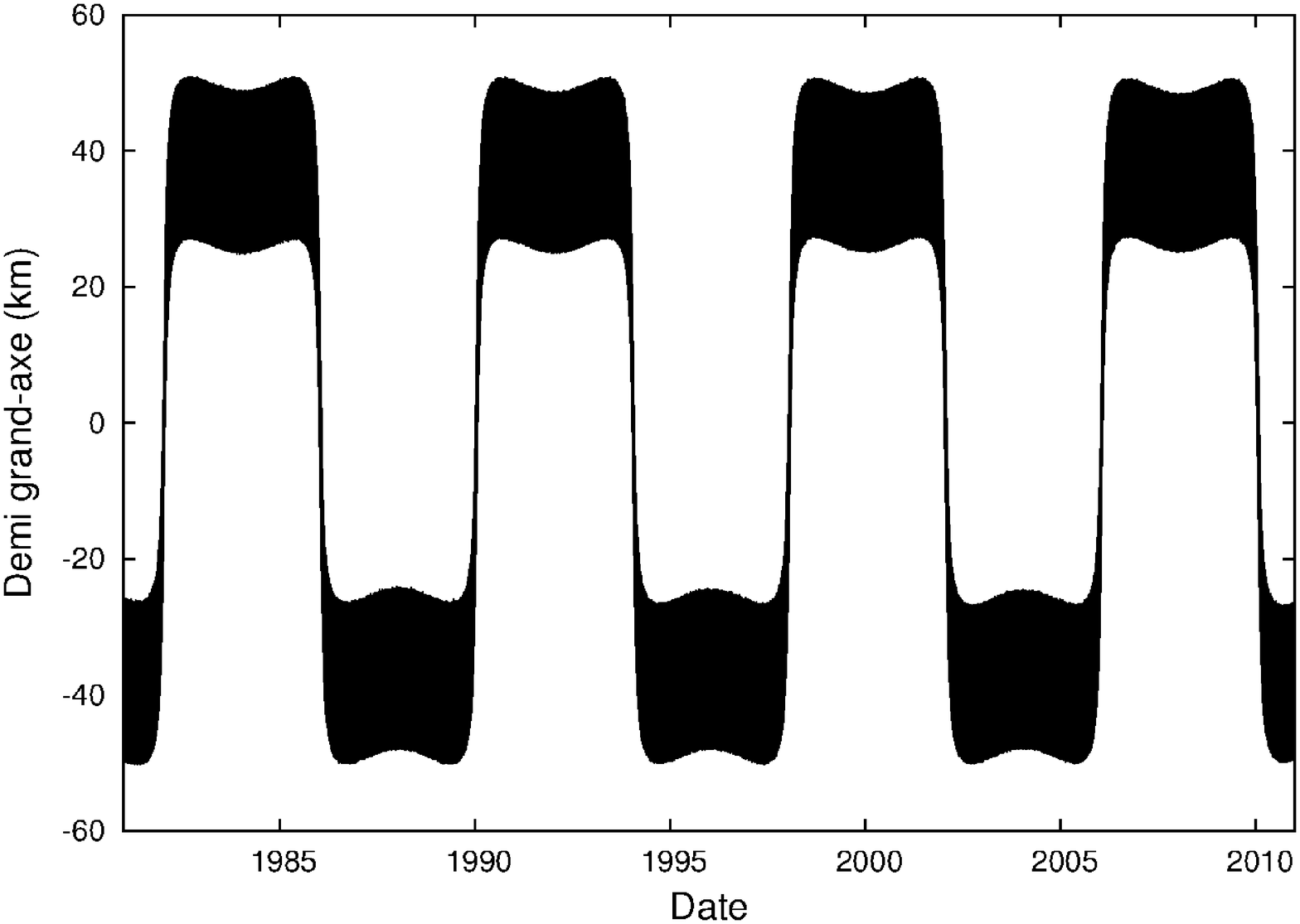} & \includegraphics[width=0.46\textwidth]{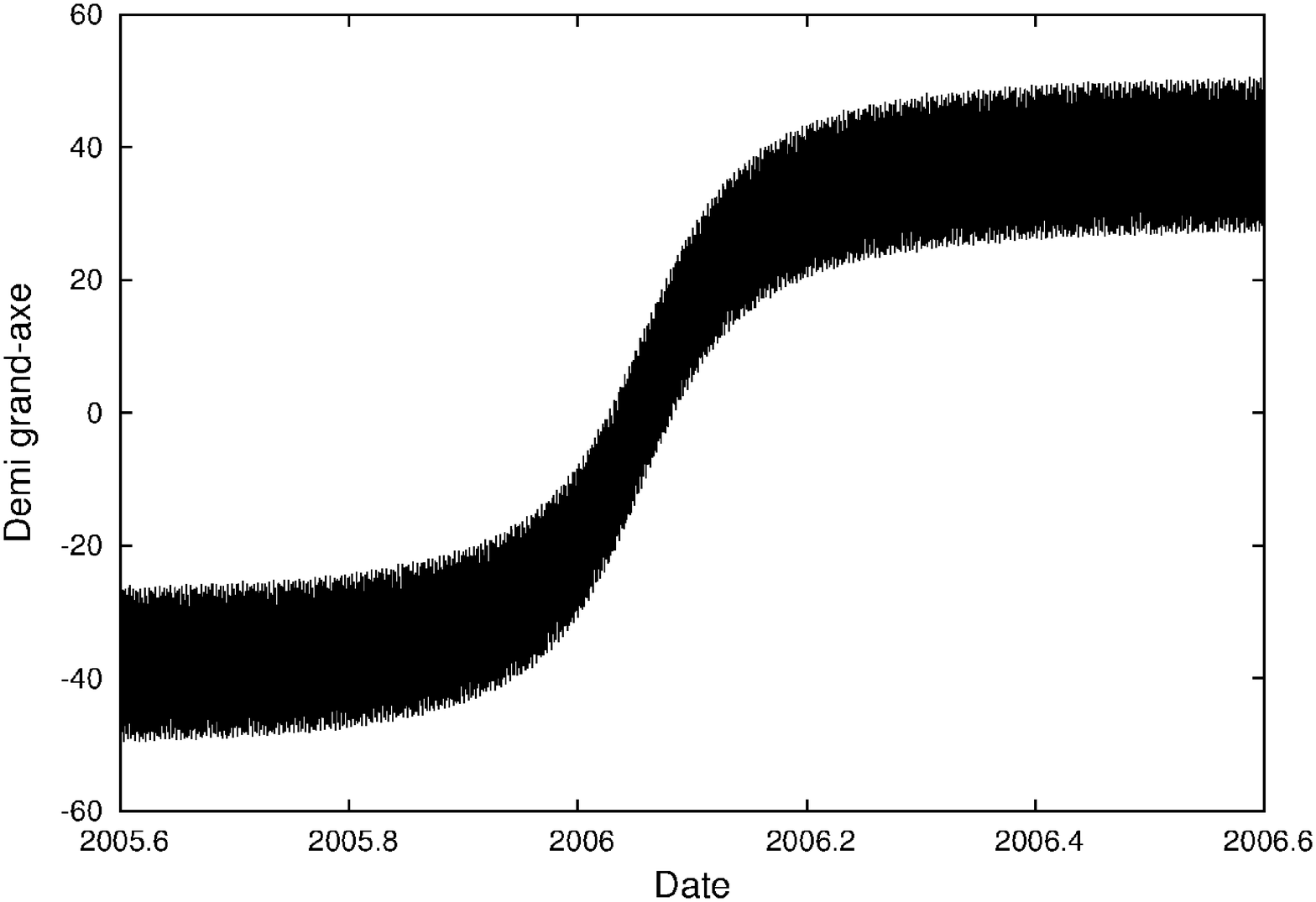} \\
      \'Epim\'eth\'ee & \'Epim\'eth\'ee \\
      \includegraphics[width=0.46\textwidth]{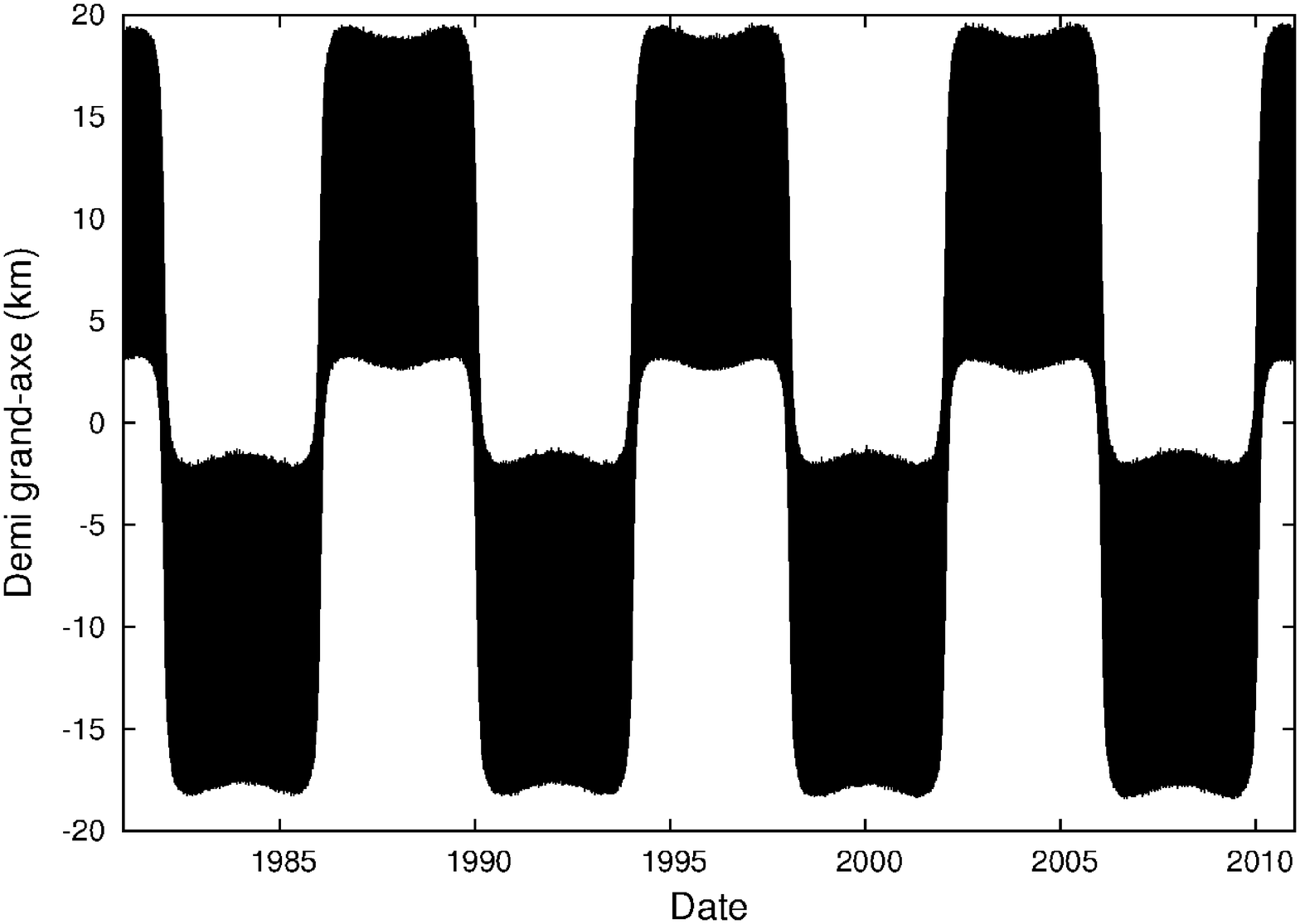} & \includegraphics[width=0.46\textwidth]{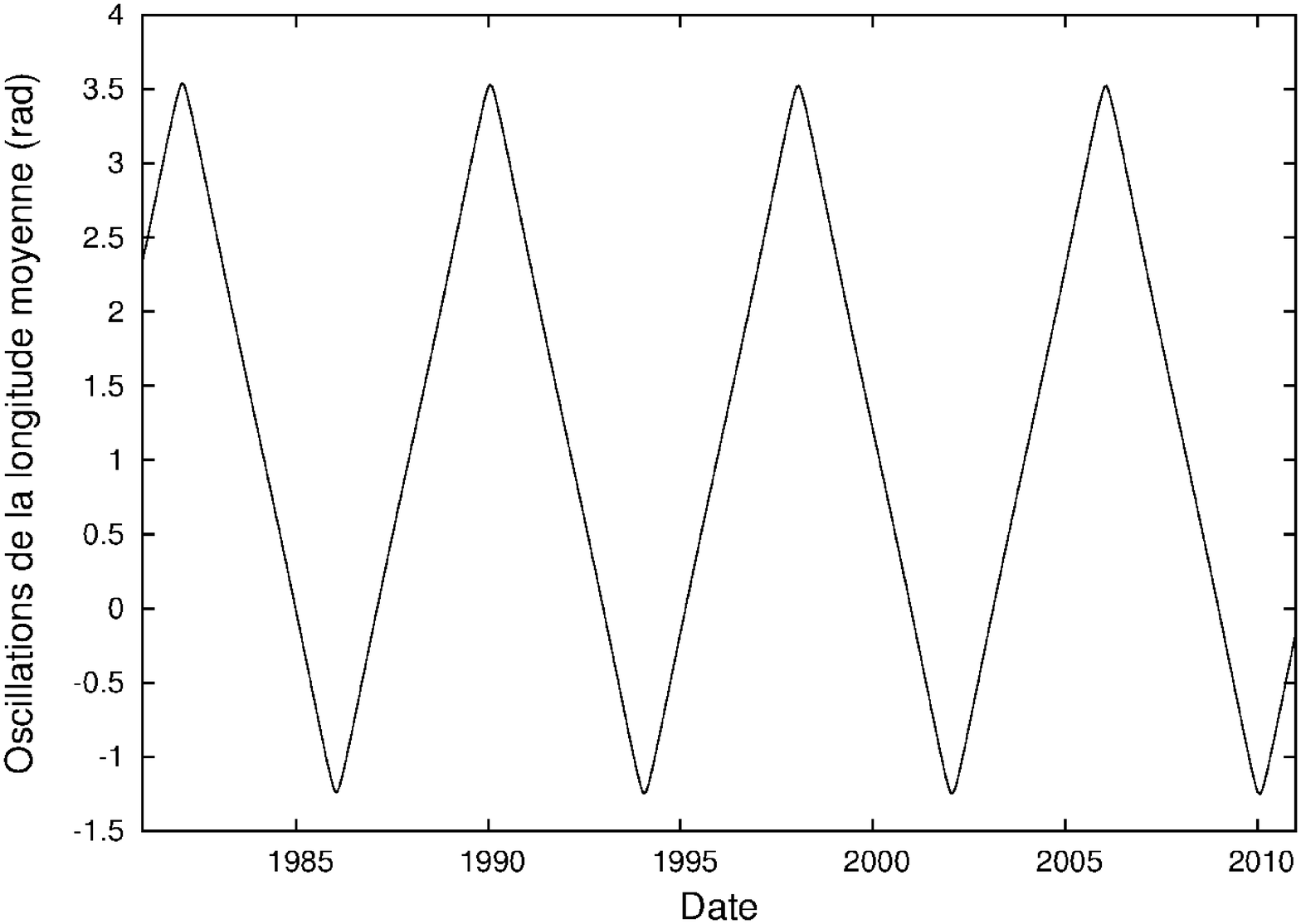} \\
      Janus & \'Epim\'eth\'ee
      \end{tabular}
      \caption[\'Echanges d'orbites entre Janus et \'Epim\'eth\'ee]{Les \'echanges d'orbites entre Janus et \'Epim\'eth\'ee, trac\'es \`a l'aide des \'eph\'em\'erides du JPL.
      Les variations de la longitude moyenne d'\'Epim\'eth\'ee ont \'et\'e obtenues apr\`es la soustraction d'une pente de $3304.01449932$ rad/an.\label{fig:swapjanepim}}
    \end{figure}

 Les satellites de Saturne $S-10$ Janus et $S-11$ \'Epim\'eth\'ee sont des petits corps ($\approx160$ km de diam\`etre pour Janus et $\approx120$ km pour \'Epim\'eth\'ee) 
    orbitant pr\`es de Saturne en environ 17 heures, et en r\'esonance de moyen mouvement $1:1$. Ceci signifie qu'ils cohabitent \`a la m\^eme distance moyenne de Saturne, 
    le demi-grand axe moyen \'etant $152043$ km. R\'eguli\`erement, \`a l'occasion de proches rencontres, ils \'echangent leurs orbites, c'est-\`a-dire que le satellite le plus 
    proche de Saturne devient le plus \'eloign\'e, et inversement. Ces proches rencontres ont lieu tous les 4 ans, ce ph\'enom\`ene a donc une p\'eriode de 8 ans. Dans un rep\`ere tournant, les orbites ont la forme 
    de fer \`a cheval, d'o\`u le nom \emph{horseshoe orbits}. Les \'el\'ements orbitaux sont trac\'es Fig.\ref{fig:swapjanepim}, on y voit notamment que l'amplitude des \'echanges
    est plus importante pour \'Epim\'eth\'ee, en fait le rapport des amplitudes est aussi le rapport des masses entre les 2 satellites ($\approx3.6$). On voit \'egalement que 
    l'\'echange d'orbites n'est pas un processus instantan\'e mais dure environ 6 mois.
    
    \clearpage
    
    \par En utilisant des valeurs tabul\'ees des \'el\'ements orbitaux de Janus et \'Epim\'eth\'ee obtenues \`a partir du portail JPL HORIZONS, l'analyse en 
    fr\'equence m'a donn\'e les 7 modes propres recens\'es Tab.\ref{tab:propmodjanepim}. La r\'esonance de moyen mouvement entre Janus et \'Epim\'eth\'ee a pour 
    cons\'equence que les 2 longitudes moyennes ne sont pas des modes propres ind\'ependants (ils ont la m\^eme fr\'equence), par contre les librations de l'argument r\'esonnant, $\phi$, apparaissent,
    et leur p\'eriode est celle des \'echanges d'orbite, soit $\approx8$ ans. Les 4 modes propres suivants sont ceux li\'es aux p\'ericentres $\varpi$ de Janus (J) et
    \'Epim\'eth\'ee (E), et aux n{\oe}uds ascendants $\ascnode$. Le dernier mode propre, $\omega$, a \'et\'e obtenu num\'eriquement dans l'inclinaison de Janus, il est possible
    qu'il soit un artefact d\^u \`a une somme d'autres effets, comme les perturbations gravitationnelles des autres satellites. HORIZONS donne les \'eph\'em\'erides orbitales
    de Janus et \'Epim\'eth\'ee sur la p\'eriode [1950-2050], avec le noyau SAT299, la d\'etection d'une oscillation de p\'eriode 136 ans sur un intervalle de 100 ans
    doit \^etre consid\'er\'ee avec pr\'ecaution.

    \begin{table}[ht]
     \centering
     \caption[Modes propres du mouvement orbital de Janus et \'Epim\'eth\'ee]{Modes propres du mouvement orbital de Janus et \'Epim\'eth\'ee.\label{tab:propmodjanepim}}
     \begin{tabular}{lrrr}
     \hline
       & Fr\'equence (rad/an) & Phase \`a J2000 & P\'eriode \\
     \hline
     $\lambda$ &      $3304.0143278$ & $-114.564^{\circ}$ &     $0.69459$ j \\
     $\phi$ &            $0.7847244$ &  $177.674^{\circ}$ &     $8.00687$ a \\
     $\varpi_J$ &       $13.0908741$ &  $129.064^{\circ}$ &   $175.30788$ j \\
     $\varpi_E$ &       $13.0928523$ & $-121.751^{\circ}$ &   $175.28140$ j \\
     $\ascnode_J$ &    $-13.0386776$ &  $114.152^{\circ}$ &  $-176.00968$ j \\
     $\ascnode_E$ &    $-13.0400438$ &  $152.811^{\circ}$ &  $-175.99124$ j \\
     $\omega$ &          $0.0461439$ & $-120.692^{\circ}$ &   $136.16498$ a \\
     \hline
     \end{tabular}
    \end{table}

    \par La Tab.\ref{tab:smajanepim} donne la d\'ecomposition quasi-p\'eriodique des demi-grands axes de Janus et \'Epim\'eth\'ee. On y observe l'importance
    du terme $\phi$, li\'e aux \'echanges d'orbite. On peut remarquer \'egalement l'opposition de phase, ou le changement de signe de l'amplitude, pour les termes 
    en $\phi$ dans Janus et \'Epim\'eth\'ee. Les \'el\'ements orbitaux moyens des satelites sont donn\'es dans la Tab.\ref{tab:moyensjanepim}, qui indique que les
    excentricit\'es et inclinaisons sont limit\'ees.
    
    \begin{table}[ht]
     \centering
     \caption[Demi grand-axe de Janus et \'Epim\'eth\'ee]{D\'ecomposition quasi-p\'eriodique des demi-grands axes de Janus et \'Epim\'eth\'ee.
     Les s\'eries sont en cosinus. $\varpi$ d\'esigne $\varpi_J$ pour Janus, et $\varpi_E$ pour \'Epim\'eth\'ee.\label{tab:smajanepim}}
     \begin{tabular}{rrrrrrr}
     \hline
     $\lambda$ & $\phi$ & $\varpi$ & Janus (km) & \'Epim\'eth\'ee (km) & P\'eriode \\
     \hline
     -   & -     & -    & $152043.049$ & $152043.602$ & $\infty$ \\
     -   & $1$   & -    &     $13.182$ &    $-47.500$ &   $8.00687$ a \\
     $1$ & -     & $-1$ &      $7.397$ &     $-3.114$ &   $0.69735$ j \\
     -   & $3$   & -    &     $-4.538$ &     $16.353$ &   $2.66892$ a \\
     -   & $5$   & -    &      $2.403$ &     $-8.660$ &   $1.60135$ a \\
     $1$ & $-1$  & $-1$ &      $2.098$ &     $-6.251$ &   $0.69752$ j \\
     $1$ & $1$   & $-1$ &      $2.097$ &     $-6.234$ &   $0.69719$ j \\
     -   & $7$   & -    &     $-1.547$ &      $5.577$ &   $1.14382$ a \\
     -   & $9$   & -    &      $1.087$ &     $-3.918$ & $324.94105$ j \\
     -   & $11$  & -    &     $-0.802$ &      $2.889$ & $265.86068$ j \\
     -   & $13$  & -    &      $0.611$ &     $-2.201$ & $224.95882$ j \\
     -   & $15$  & -    &     $-0.476$ &      $1.715$ & $194.96413$ j \\
     -   & $17$  & -    &      $0.377$ &     $-1.359$ & $172.02726$ j \\
     $1$ &  $2$  & $-1$ &      $0.365$ &     $-4.731$ &   $0.69702$ j \\
     $1$ & $-2$  & $-1$ &      $0.359$ &     $-4.726$ &   $0.69769$ j \\
     -   & $19$  & -    &     $-0.302$ &      $1.090$ & $153.91905$ j \\
     $1$ & $-3$  & $-1$ &      $0.203$ &     $-1.293$ &   $0.69785$ j \\
     $1$ &  $3$  & $-1$ &      $0.200$ &     $-1.319$ &   $0.69685$ j \\
     \hline
     \end{tabular}
    \end{table}

    \begin{table}[ht]
     \centering
     \caption{\'El\'ements orbitaux moyens de Janus et \'Epim\'eth\'ee.\label{tab:moyensjanepim}}
     \begin{tabular}{lrr}
     \hline
     \'El\'ement & Janus & \'Epim\'eth\'ee \\
     \hline
     Demi-grand axe & $152043.049$ km    & $152043.602$ km \\
     Excentricit\'e & $9.8\times10^{-3}$ & $1.61\times10^{-2}$ \\
     Inclinaison    & $0.165^{\circ}$    & $0.353^{\circ}$ \\
     \hline
     \end{tabular}
    \end{table}

    \par La disponibilit\'e d'\'eph\'em\'erides pour Janus et \'Epim\'eth\'ee uniquement de la part du JPL m'a amen\'e \`a t\'el\'echarger des valeurs tabul\'ees des coordonn\'ees
    cart\'esiennes, puis \`a les interpoler par des splines cubiques, \`a l'aide de routines pr\'esentes dans la GNU Scientific Library \citep{gdtgjabr2009}. L'erreur associ\'ee
    est d'ordre 4 \citep{s1969}, ce qui signifie qu'une r\'eduction du pas d'\'echantillonnage des donn\'ees par un facteur $\alpha$ r\'eduira l'erreur d'un facteur $\alpha^4$. J'ai
    utilis\'e un pas d'\'echantillonnage d'une heure pour obtenir une erreur inf\'erieure \`a 9 km. J'ai \'echantillonn\'e les coordonn\'ees cart\'esiennes pour obtenir ce r\'esultat;
    un premier essai sur les \'el\'ements orbitaux pr\'esentait un probl\`eme de convergence lors des \'echanges d'orbite. L'interpolation des orbites des satellites m'a permis 
    d'\'ecrire des routines C++ donnant leur position \`a n'importe quelle date, m\^eme si elle ne correspondait pas exactement \`a une donn\'ee tabul\'ee. J'ai ainsi pu les utiliser dans mon
    code num\'erique de rotation rigide\footnote{La m\'ethode d'utilisation de ces \'eph\'em\'erides officiellement conseill\'ee est d'utiliser la librairie 
    SPICE \verb!http://naif.jpl.nasa.gov/naif/toolkit.html! \citep{a1996}. Ayant jug\'e le temps d'apprentissage prohibitif pour cet outil aux nombreuses fonctionnalit\'es, j'ai 
    utilis\'e ma propre m\'ethode, centr\'ee sur l'usage que je souhaitais faire des \'eph\'em\'erides. Je signale \'egalement la disponibilit\'e de la librairie CALCEPH, 
    r\'ealis\'ee par Mickael Gastineau, \verb!http://www.imcce.fr/inpop/calceph/!, destin\'ee \`a lire les noyaux SPICE. Son usage est peut-\^etre plus direct.}.
    
    \subsection{La rotation observ\'ee \citep{ttb2009}}
    
    \par
\begin{figure}[ht]
      \centering
      \includegraphics[width=\textwidth]{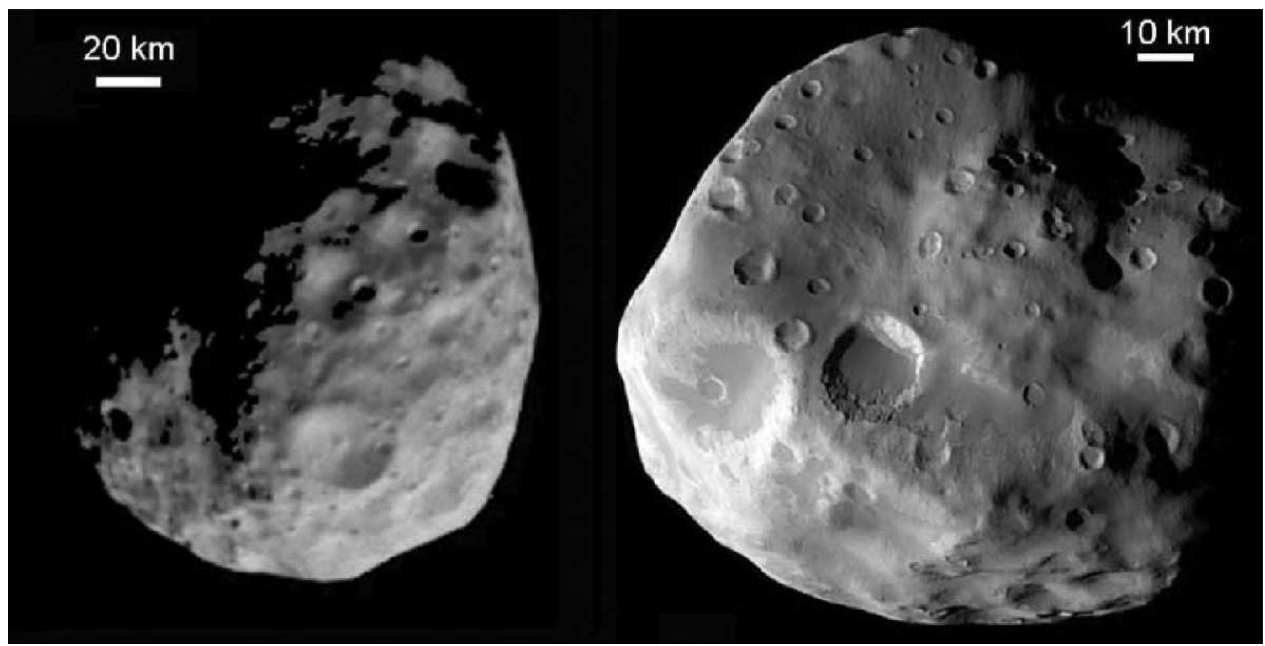}
      \caption[Janus et \'Epim\'eth\'ee vus par Cassini]{Images Cassini N1537923147 pour Janus (\`a gauche) et N1575363491 pour \'Epim\'eth\'ee (\`a droite). Figure reproduite de \citep{ttb2009}.\label{fig:janepimtisca}}
    \end{figure}

 \citet{ttb2009} ont publi\'e des observations de la rotation de Janus et \'Epim\'eth\'ee \`a partir d'images Cassini ISS (Imaging Science Subsystem) acquises entre 2005 et 
    2008 (Fig.\ref{fig:janepimtisca}), d'une r\'esolution inf\'erieure au kilom\`etre. Le principe consiste \`a rep\'erer des points remarquables \`a la surface de ces satellites, par 
    exemple des crat\`eres, pr\'esents sur au moins 2 images, et \`a ajuster un mouvement de rotation. Pour Janus, 66 points de contr\^ole ont \'et\'e rep\'er\'es \`a partir de 
    104 images, et 49 points de contr\^ole \`a partir de 50 images pour \'Epim\'eth\'ee. Cette \'etude ajuste \'egalement une forme triaxiale sur Janus et \'Epim\'eth\'ee 
    (Tab.\ref{tab:formjanepim}), et la compare avec la forme d\'eduite du mouvement de rotation (Tab.\ref{tab:rotajanepim}).
    
    \begin{table}[ht]
      \centering
      \caption[Formes triaxiales de Janus et \'Epim\'eth\'ee observ\'ees par \citet{ttb2009}]{Formes triaxiales de Janus et \'Epim\'eth\'ee observ\'ees par \citet{ttb2009}. 
      $a$, $b$ et $c$ sont les trois rayons de l'ellipso\"ide, $\bar{R}$ le rayon moyen, et les moments d'inertie sont ici calcul\'es \`a partir de la forme, en supposant 
      une densit\'e constante.\label{tab:formjanepim}}
      \begin{tabular}{l|r|r}
         & Janus & \'Epim\'eth\'ee \\
         \hline
              $a$ (km) &   $101.5\pm1.9$ & $64.9\pm2.0$ \\
              $b$ (km) &    $92.5\pm1.2$ & $57.0\pm3.7$ \\
              $c$ (km) &    $76.3\pm1.2$ & $53.1\pm0.7$ \\
        $\bar{R}$ (km) &    $89.5\pm1.4$ & $58.1\pm1.8$ \\
            $A/(MR^2)$ &         $0.360$ &      $0.328$ \\
            $B/(MR^2)$ &         $0.407$ &      $0.469$ \\
            $C/(MR^2)$ &         $0.470$ &      $0.476$ \\
             $(B-A)/C$ & $0.100\pm0.012$ & $0.296\substack{+0.019 \\ -0.027}$ \\
             \hline
      \end{tabular}
    \end{table}

    \begin{table}[ht]
      \centering
      \caption[Libration physique diurne de Janus et \'Epim\'eth\'ee]{Libration physique diurne, c'est-\`a-dire \`a la fr\'equence orbitale, de Janus et \'Epim\'eth\'ee, et comparaison avec ce qu'indique la forme observ\'ee. 
      On constate qu'une libration non nulle est mesur\'ee pour \'Epim\'eth\'ee, et que les observations de la rotation sont plut\^ot en accord avec celles de la 
      forme.\label{tab:rotajanepim}}
      \begin{tabular}{l|r|r}
                                 & Janus & \'Epim\'eth\'ee \\
        \hline
        Libration physique       & & \\
        d\'eduite de la forme    & $-0.33^{\circ}\pm0.06^{\circ}$ & $-8.9^{\circ}\substack{-10.4^{\circ} \\ +4.2^{\circ}}$ \\
        mesur\'ee                & $-0.3^{\circ}\pm0.9^{\circ}$   & $-5.9^{\circ}\pm1.9^{\circ}$ \\
        \hline
        $(B-A)/C$                & & \\
        d\'eduit de la forme     & $0.100\pm0.012$ & $0.296\substack{+0.019 \\ -0.027}$ \\
        d\'eduit de la libration & $0.09\substack{+0.11 \\ -0.09}$ & $0.280\substack{+0.008 \\ -0.011}$ \\
        \hline
      \end{tabular}
    \end{table}

    \par Le premier r\'esultat est la confirmation que ces corps sont bien en rotation synchrone. De plus, pour la troisi\`eme fois apr\`es la Lune \citep{k1967} et le 
    satellite de Mars Phobos \citep{b1972}, des librations diurnes ont \'et\'e mesur\'ees pour un satellite naturel : \'Epim\'eth\'ee. Les barres d'erreur sont bien plus 
    petites que le signal, ce qui permet de d\'eterminer le param\`etre d'int\'erieur $(B-A)/C$ \`a l'aide de la formule (\ref{eq:librphys}). La forte \'elongation de la 
    forme d'\'Epim\'eth\'ee favorise ces librations. Un signal a \'egalement \'et\'e d\'etect\'e pour Janus, qui semble en accord avec la forme observ\'ee, mais les barres 
    d'erreur sont plus importantes que le signal lui-m\^eme, donc ce dernier r\'esultat doit \^etre accueilli avec prudence.
    
    \par Cette \'etude annonce \'egalement la mesure d'un d\'ecalage de l'orientation moyenne de l'axe le plus long de Janus par rapport \`a la direction Janus-Saturne, de
    $5.2^{\circ}\pm1^{\circ}$ en longitude et $2.3^{\circ}\pm1^{\circ}$ en latitude. Cette mesure est tr\`es surprenante car elle semble tr\`es difficile \`a expliquer,
    m\^eme en supposant des anomalies de masse dans Janus \citep{rrc2011}. Des mesures plus r\'ecentes, r\'ealis\'ees par la m\^eme \'equipe et \`a ce jour non publi\'ees,
    sugg\`erent que ce d\'ecalage est en fait bien plus faible qu'initialement annonc\'e.
    
    \par Cette observation de la forme et de la rotation de Janus et \'Epim\'eth\'ee m'a incit\'e \`a simuler leur rotation en 3 dimensions \`a l'aide d'\'eph\'em\'erides r\'ealistes,
    notamment dans la gestion des \'echanges d'orbites. Le but est d'essayer d'am\'eliorer l'explication des observations, de pr\'edire des quantit\'es non encore mesur\'ees comme
    les obliquit\'es, ainsi que de comprendre la fa\c{c}on dont les \'echanges d'orbite affectent la rotation.
    
    \subsection{Simulation de la rotation}
    
    \par
\begin{figure}[ht]
      \centering
      \begin{tabular}{cc}
      \includegraphics[width=.47\textwidth]{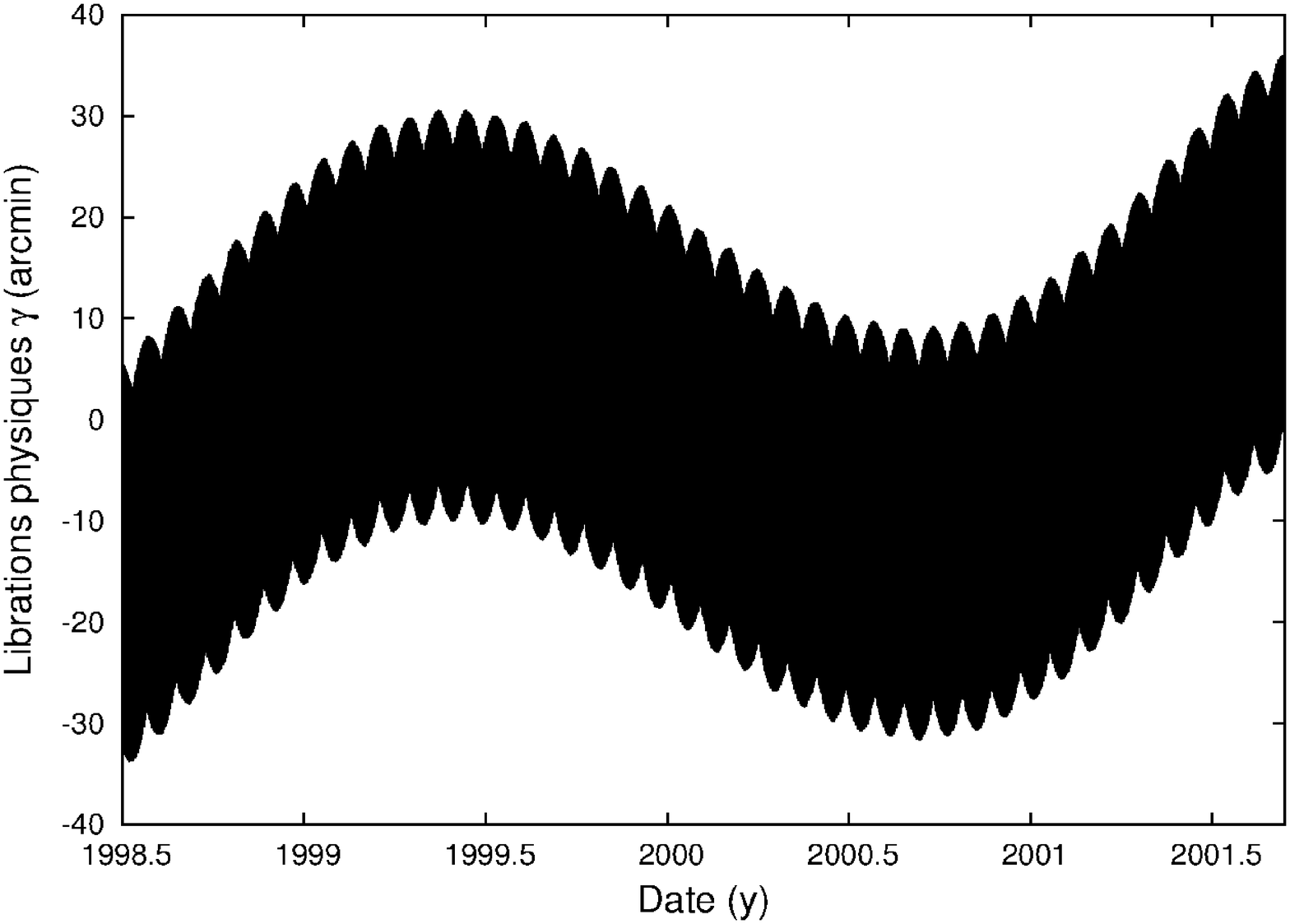} & \includegraphics[width=.47\textwidth]{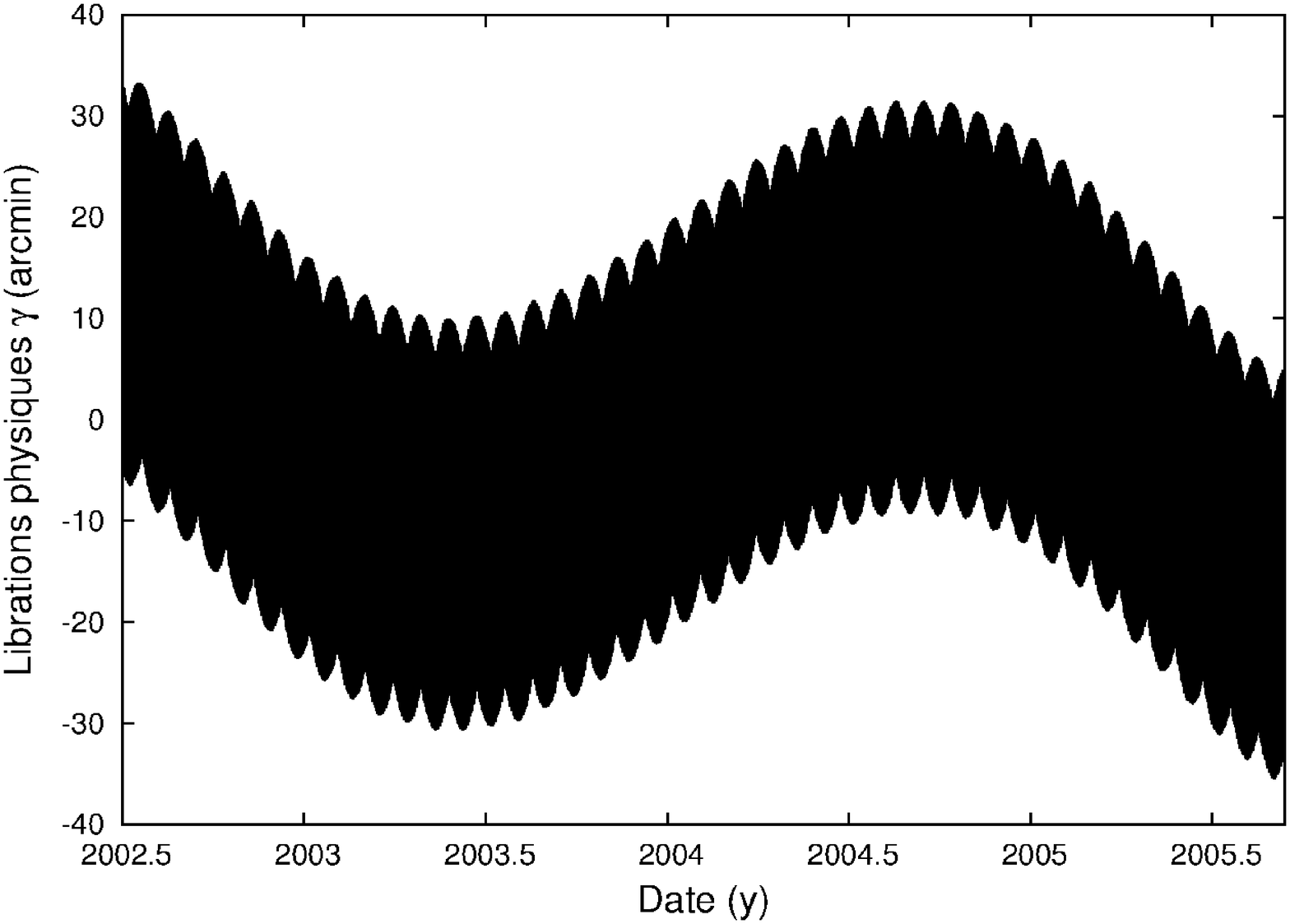} \\
      \includegraphics[width=.47\textwidth]{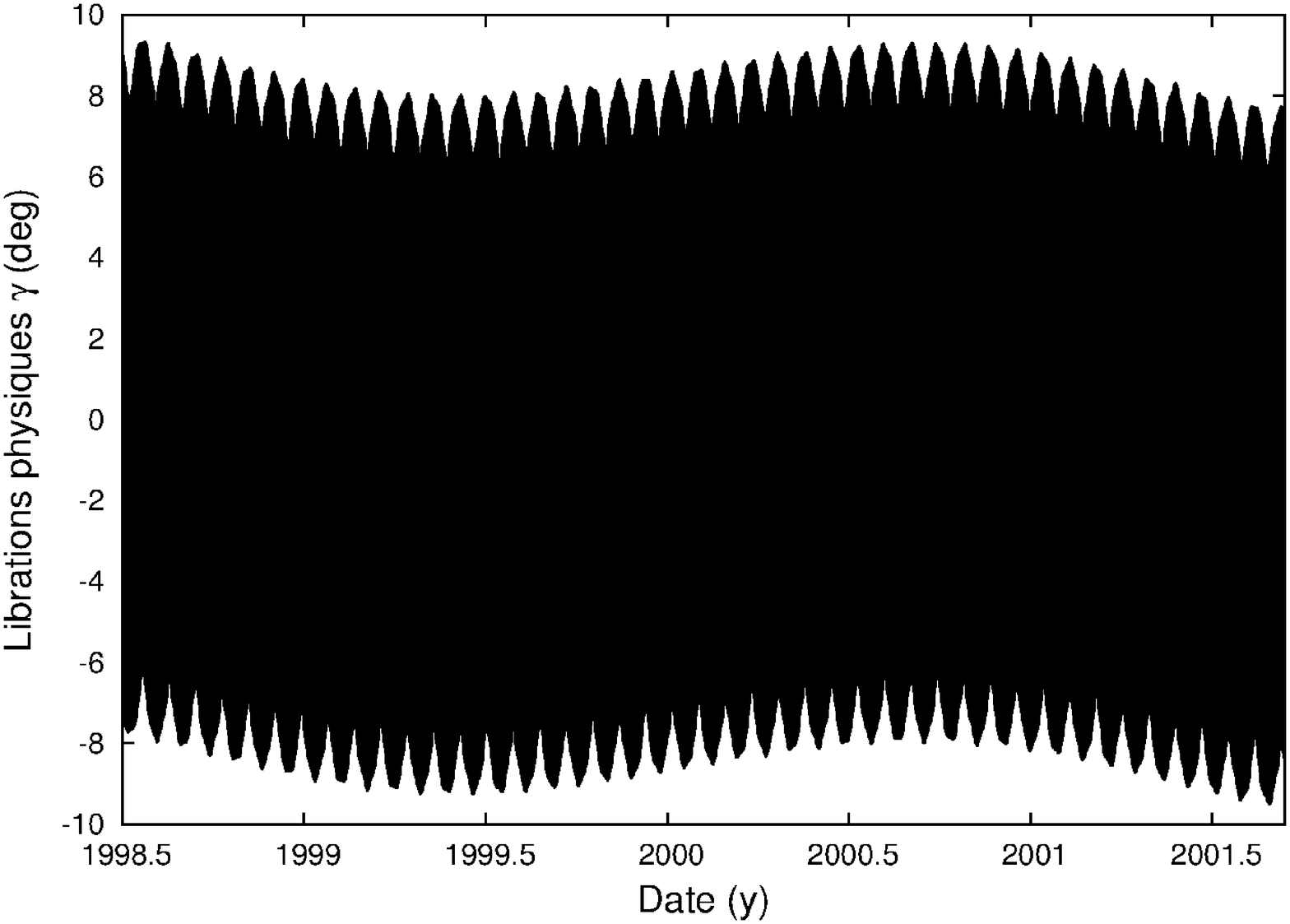} & \includegraphics[width=.47\textwidth]{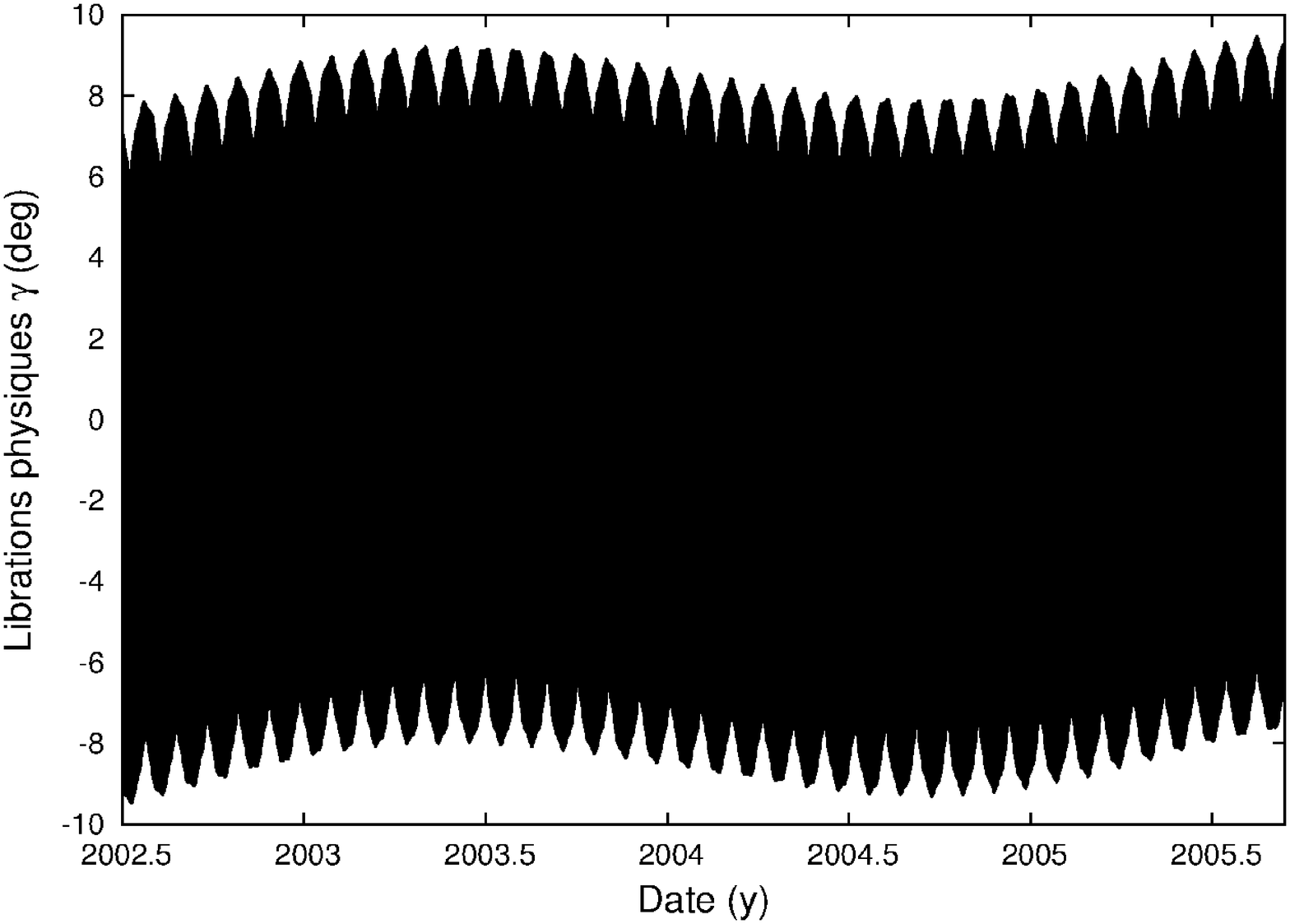}
      \end{tabular}
      \caption[Librations physiques de Janus et \'Epim\'eth\'ee]{Librations physiques de Janus et \'Epim\'eth\'ee, de part et d'autre d'un \'echange d'orbites.
      Ces figures ont \'et\'e obtenues apr\`es la soustraction d'une pente constante, respectivement de $3304.356301$ et $3303.673150$ radians par an pour Janus,
      et $3302.7839016$ et $3305.24602$ radians par an pour \'Epim\'eth\'ee. \label{fig:longijanepis}}
    \end{figure}

 J'ai utilis\'e mon code num\'erique de rotation rigide en 3 dimensions ainsi que les \'eph\'em\'erides obtenues \`a partir de JPL HORIZONS pour simuler la rotation de
    Janus et \'Epim\'eth\'ee \`a l'\'Etat de Cassini 1. Mes param\`etres d'int\'erieur sont ceux d\'eduits de la forme (Tab.\ref{tab:formjanepim}). La figure \ref{fig:longijanepis} 
    montre les librations physiques, de part et d'autre d'un \'echange d'orbites.
    
  \par Ces librations pr\'esentent au moins 3 \'echelles de p\'eriode. L'\'epaisseur de la courbe repr\'esente les librations diurnes, de p\'eriode la p\'eriode orbitale, 
  soit $\approx16.7$ heures. L'amplitude est de l'ordre de $0.3^{\circ}$ pour Janus et $8^{\circ}$ pour \'Epim\'eth\'ee. On remarque \'egalement des pics en haut et en bas 
  de ces courbes, d'amplitude quelques minutes d'arc pour Janus et de l'ordre du degr\'e pour \'Epim\'eth\'ee. Ces pics peuvent perturber la mesure de la libration diurne,
  et peut-\^etre expliquer en partie les barres d'erreur sur les observations. Enfin, on observe une modulation \`a plus longue p\'eriode, signature de la p\'eriodicit\'e
  \`a 8 ans des \'echanges d'orbite.
  
  \par
\begin{figure}[ht]
      \centering
      \includegraphics[width=.7\textwidth]{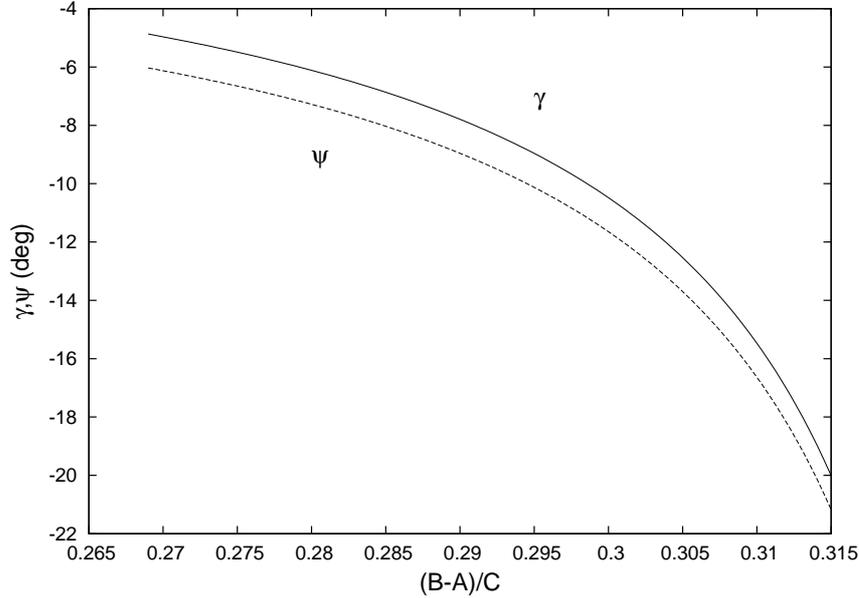}
      \caption[Libration diurne d'\'Epim\'eth\'ee]{Amplitude de la libration diurne d'\'Epim\'eth\'ee en fonction de son ellipticit\'e \'equatoriale $(B-A)/C$. On peut 
      remarquer un comportement asymptotique \`a proximit\'e de la valeur $1/3$, ceci correspond \`a une r\'esonance entre la fr\'equence orbitale $n$ et la fr\'equence 
      des librations libres $\omega_u$.\label{fig:errtwojanepim}}
    \end{figure}

 La mesure \'elev\'ee de la libration diurne pour \'Epim\'eth\'ee est int\'eressante. Il s'agit en fait d'un cas de quasi-r\'esonance, i.e. la p\'eriode des librations libres 
  est proche de la p\'eriode de for\c{c}age. C'est une cons\'equence de l'ellipticit\'e \'equatoriale du satellite $(B-A)/C$, cf. la figure \ref{fig:errtwojanepim}, obtenue \`a 
  l'aide de l'amplitude exprim\'ee dans la formule (\ref{eq:librphys}).

    \par Ces simulations indiquent une forte influence des \'echanges d'orbite sur les librations en longitude. Je l'ai montr\'e num\'eriquement, et \citet{rrc2011} l'ont 
    confirm\'e analytiquement. Cette derni\`ere \'etude propose une approche originale du calcul des librations par une approche perturbative, en consid\'erant le rapport
    des fr\'equences $\dot{\phi}/n$ comme un petit param\`etre ($\approx 2.38\times10^{-4}$) sur lequel le d\'eveloppement peut s'appuyer.
    
    Mes simulations num\'eriques ont \'egalement permis de confirmer que le mouvement polaire doit \^etre tr\`es petit, en fait de l'ordre du m\`etre. Les obliquit\'es 
    attendues sont de l'ordre de 6 secondes d'arc pour Janus et 11 secondes d'arc pour \'Epim\'eth\'ee.

    \section{Mimas}
    
    \par Ce travail sur Mimas \citep{nkr2011} s'est inscrit dans le cadre du groupe de travail Encelade \footnote{\verb!http://www.imcce.fr/~lainey/Encelade.htm!}, anim\'e par V.~Lainey 
    et regroupant diverses comp\'etences europ\'eennes en plan\'etologie (structure interne, m\'ecanique c\'eleste, astrom\'etrie, mar\'ees,\ldots). Ce groupe s'est cr\'e\'e \`a 
    la suite de la d\'ecouverte de geysers sur Encelade, et a pour but de mieux comprendre le syst\`eme de Saturne \`a la lumi\`ere des d\'ecouvertes de Cassini, des nouvelles
    observations, et des nouveaux d\'eveloppements th\'eoriques. On y trouve des membres de diff\'erents centres de recherches europ\'eens, comme l'Observatoire de Paris-Meudon, 
    le CEA, l'Universit\'e de Namur, l'Observatoire Royal de Belgique, ou encore la Queen Mary University of London.
    
    \par Dans ce cadre, nous avons choisi de nous int\'eresser \`a la rotation de Mimas. Avant de disposer de suffisamment de donn\'ees pour observer cette rotation (ce qu'a 
    finalement fait R.~Tajeddine \`a l'issue de sa th\`ese de doctorat), nous l'avons simul\'ee th\'eoriquement.

    \subsection{Les mod\`eles d'int\'erieur}
    
    \par
\begin{figure}[ht]
     \centering
     \begin{tabular}{lcr}
      \includegraphics[width=0.4\textwidth]{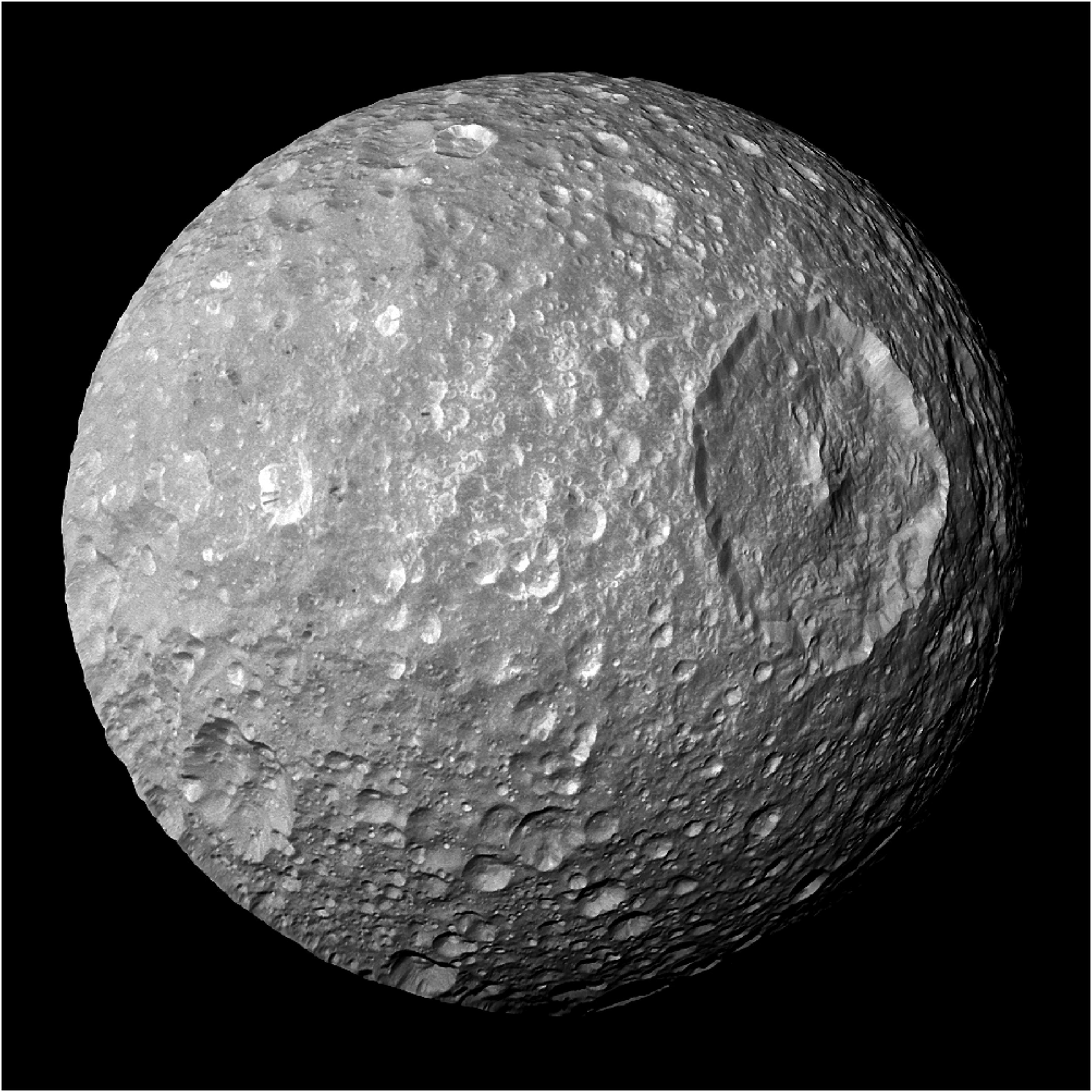} & \hspace{1cm} & \includegraphics[width=0.43\textwidth]{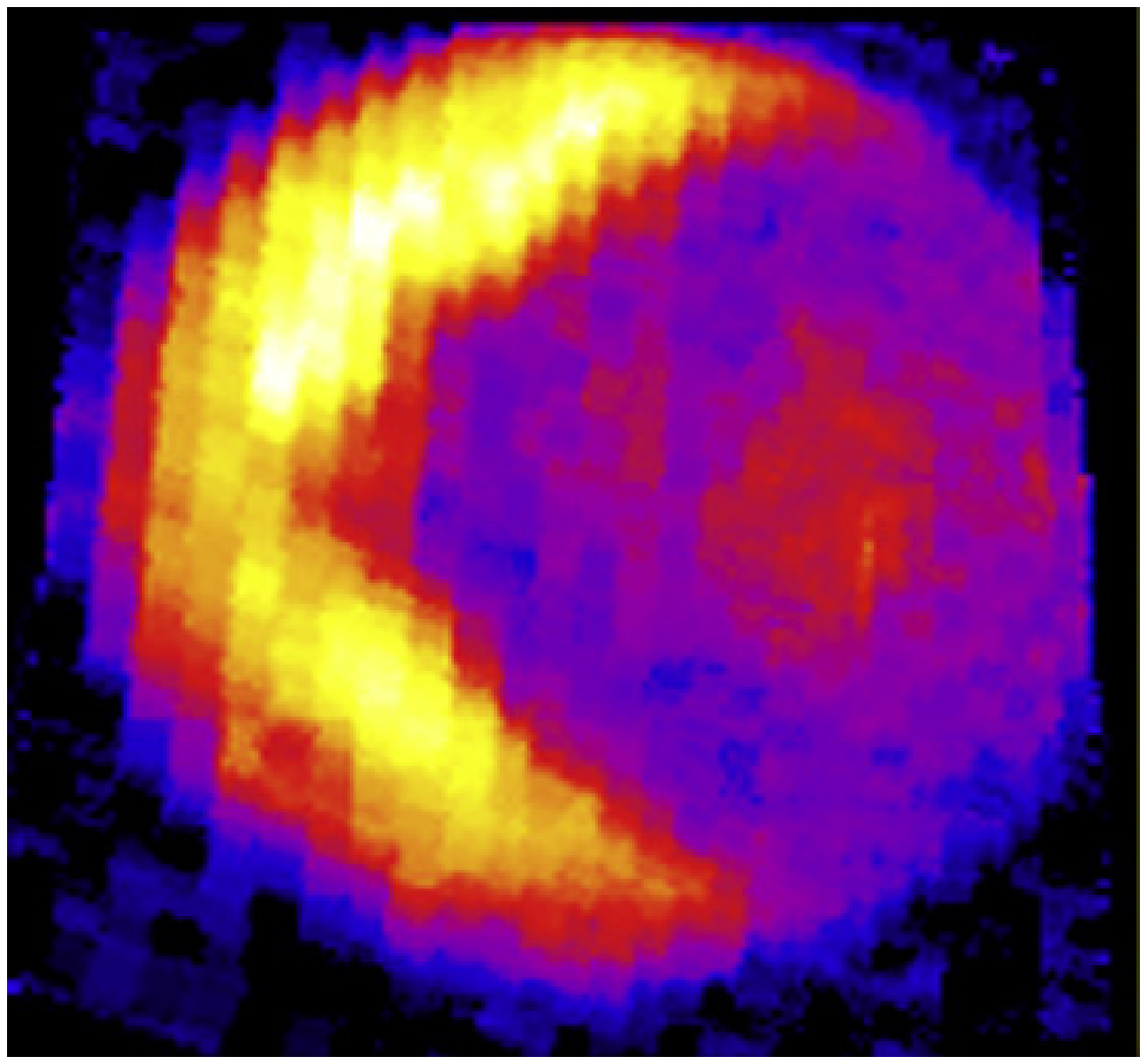}
     \end{tabular} 
     \caption[Mimas vu par Cassini]{Mimas vu par Cassini. \`A gauche dans le visible, on distingue bien le crat\`ere Herschel. \`A droite dans l'infra-rouge moyen 
     (instrument CIRS, ici entre 15.4 et 16.7 $\mu m$ de longueur d'onde), o\`u le croissant jaune (ou PacMan\textregistered) repr\'esente l'anomalie de temp\'erature.\label{fig:mimas2}}
    \end{figure}

 Le champ de gravit\'e de Mimas n'a \'et\'e mesur\'e qu'\`a l'ordre 0, c'est-\`a-dire que sa masse est connue\footnote{Elle est actuellement d\'etermin\'ee par 
    analyse des librations des r\'esonances de moyen mouvement avec T\'ethys et M\'ethone \citep{jspo2006}. Les survols de Cassini sont trop lointains pour d\'eterminer la masse de Mimas.} 
    mais nous ne savons rien de sa r\'epartition dans le satellite. Cette r\'epartition est tr\`es probablement inhomog\`ene, ne serait-ce que par l'anomalie de masse li\'ee au 
    crat\`ere Herschel. Sa forme est incompatible avec l'\'equilibre hydrostatique si on consid\`ere sa densit\'e constante. La d\'etection d'anomalies de temp\'erature sur Mimas
    par l'instrument CIRS (Cassini Composite Infrared Spectrometer) (Fig.\ref{fig:mimas2} \`a droite, reprise de \citet{hssjphvs2011}) a \'et\'e envisag\'ee comme la d\'etection d'anomalies de structure
    interne. Mais l'explication la plus probable est un bombardement \'electronique externe. Cette anomalie de temp\'erature a \'egalement \'et\'e d\'etect\'ee pour T\'ethys 
    \citep{hshvs2012}, et \`a chaque fois elle est tr\`es bien corr\'el\'ee avec le mouvement orbital du satellite. L'anomalie de temp\'erature correspond \`a l'h\'emisph\`ere
    qui m\`ene le mouvement.

    \par Suivant de pr\'ec\'edents auteurs, notamment \citet{e1990}, nous avons consid\'er\'e Mimas comme un corps glac\'e, relativement poreux \`a l'ext\'erieur et moins en son 
    c{\oe}ur, sous l'effet de la compaction. Ce ph\'enom\`ene de compaction dans les satellites de glace a r\'ecemment \'et\'e confirm\'e par des exp\'eriences de laboratoire
    r\'ealis\'ees au Japon \citep{ya2008,ya2009}. Nous avons mod\'elis\'e cette compaction en consid\'erant Mimas comme un corps \`a 2 couches rigides, concentriques triaxiales et 
    coaxiales, la couche externe \'etant moins dense que la couche interne. Il s'agit donc de mod\`eles \`a 8 param\`etres en entr\'ee, \`a savoir une densit\'e (ou masse volumique) $\rho$ 
    et 3 rayons $a\times b\times c$ pour chaque couche $c$ (noyau/\emph{core}) et $s$ (cro\^ute/\emph{shell}), et 3 param\`etres en sortie, dont nous avons besoin pour mod\'eliser la 
    rotation d'un corps rigide : les coefficients de Stokes $J_2$ et $C_{22}$, et le moment d'inertie polaire normalis\'e $C/(mR^2)$.
    
    \par Nous avons impos\'e \`a ces mod\`eles d'int\'erieur d'avoir la m\^eme masse totale que Mimas, et afin d'avoir une autre contrainte liant les param\`etres d'int\'erieur nous avons 
    soit consid\'er\'e que Mimas \'etait \`a l'\'equilibre hydrostatique, violant ainsi l'observation de la forme, soit impos\'e comme rayons externes les rayons observ\'es, et 
    consid\'er\'e l'interface noyau-cro\^ute comme homoth\'etique de la surface. Aucune de ces hypoth\`eses n'est pleinement satisfaisante, mais le manque de donn\'ees incite \`a 
    simplifier le mod\`ele. Nos param\`etres sont regroup\'es Tab.\ref{tab:paramimas}.

     \begin{table}[ht]
      \centering
      \caption[Param\`etres physiques et dynamiques de Mimas]{Param\`etres physiques et dynamiques de Mimas. Le moyen mouvement de Mimas $n$ a \'et\'e mis \`a jour 
      depuis TASS 1.6, mais nous avons utilis\'e cette valeur car elle est coh\'erente avec l'utilisation des \'eph\'em\'erides TASS 1.6 et c'est la seule d\'etermination 
      dans la litt\'erature qui est moyenn\'ee convenablement, c'est-\`a-dire apr\`es identification de toutes les fr\'equences qui r\'egissent le syst\`eme orbital des 
      satellites de Saturne.\label{tab:paramimas}}
      \begin{tabular}{lrr}
        \hline
        Param\`etres               & Valeurs                          & Sources \\
        \hline
        Moyen mouvement $n$        & $2435.14429644$ rad/an           & TASS 1.6 \citep{vd1995}   \\
        Masse $m$                  & $3.7495\times10^{19}$ kg         & \citet{jspo2006} \\
        Masse volumique $\rho$     & $1150.03$ $kg/m^3$               & \citet{tbhsvptmdgrjj2007} \\
        Rayon moyen $R$            & $198.2$ km                       & \citet{tbhsvptmdgrjj2007} \\
        Rayons $a\times b\times c$ & $207.8\times196.7\times190.6$ km & \citet{t2010} \\
        \hline
      \end{tabular}
    \end{table}

    \par Tous les mod\`eles bas\'es sur la forme sont \'equivalents pour la mod\'elisation de la dynamique de la rotation, car cette dynamique est la signature de la r\'epartition 
    relative de la masse au sein du corps consid\'er\'e. C'est pourquoi nous n'avons qu'un seul mod\`ele de Mimas y correspondant, le 23 (Tab.\ref{tab:mimaslescas}).
    
    \par Si Mimas est \`a l'\'equilibre hydrostatique, alors sa forme repr\'esente un \'equilibre entre sa propre gravit\'e, sa rotation, et les mar\'ees exerc\'ees par Saturne.
    Un param\`etre important est 
    
    \begin{equation}
      \label{eq:qr}
      q_r = \frac{\Omega^2R^3}{\mathcal{G}m},
    \end{equation}
o\`u $\Omega$ est la vitesse de rotation du corps ($\Omega=n$ pour les corps en synchronisation spin-orbite). $q_r$ permet de quantifier l'effet de la d\'eformation de rotation.
Un autre param\`etre important est le nombre de Love fluide $k_f$, qui caract\'erise la r\'eaction du satellite \`a un potentiel perturbateur apr\`es relaxation des 
contraintes visqueuses. La forme du satellite est donc stabilis\'ee. On obtient $k_f$ avec

\begin{eqnarray}
  I   & = & \frac{2}{3}\iiint\rho\left(x^2+y^2+z^2\right) \,\textrm{d}x\,\textrm{d}y\,\textrm{d}z, \label{eq:Imimas} \\
  MOI & = & \frac{I}{mR^2}, \label{eq:MOImimas} \\
  MOI & = & \frac{2}{3}\left[1-\frac{2}{5}\sqrt{\frac{4-k_f}{1+k_f}}\right], \label{eq:MOI2mimas}
\end{eqnarray}
ce qui permet d'obtenir \citep{rbga1997}:

\begin{eqnarray}
  C_{22} & = & \frac{k_f}{4}q_r+\mathcal{O}(q_r^2), \label{eq:C22mimas} \\
  J_2    & = & \frac{5k_f}{6}q_r+\mathcal{O}(q_r^2). \label{eq:J2mimas}
\end{eqnarray}
De plus, il vient de la d\'efinition des moments d'inertie (Eq.\ref{eq:momA}-\ref{eq:momC}) et de $J_2=(2C-A-B)/(2C)$ :

\begin{equation}
  \label{eq:CfromImimas}
  C=I+\frac{2}{3}J_2mR^2.
\end{equation}
Il est \`a noter la relation classique pour un satellite naturel \`a l'\'equilibre hydrostatique : $J_2=10/3\,C_{22}$. En pratique, cette relation a \'et\'e v\'erifi\'ee pour Io
\citep{ajlms2001}, d\'enonc\'ee pour Rh\'ea \citep{mitr2008}, et peut \^etre compatible avec le champ de gravit\'e de Titan \citep{irjrstaa2010}. Mais dans le cas de Titan, la combinaison
du champ de gravit\'e et de la forme n'est pas compatible avec l'\'equilibre hydrostatique \citep{zshlkl2009}. Cette hypoth\`ese doit donc \^etre consid\'er\'ee au mieux comme une 
approximation; n\'eanmoins son utilisation me para\^it l\'egitime dans le sens o\`u le manque de donn\'ees par rapport au nombre de param\`etres n\'ecessaires \`a l'\'elaboration
d'un mod\`ele d'int\'erieur n\'ecessite l'introduction de contraintes suppl\'ementaires.

Ici les calculs ont \'et\'e faits \`a faible excentricit\'e et faible $q_r$, donc rotation lente. L'influence de l'excentricit\'e est donn\'ee dans \citep{mn2009}, et 
d'un $q_r$ significatif dans \citep{t2014}.

Nous avons ainsi suffisamment de param\`etres pour simuler la rotation d'un tel corps. Nous avons ainsi cr\'e\'e 22 mod\`eles de Mimas hydrostatiques (Tab.\ref{tab:mimaslescas}),
pour diff\'erentes densit\'es du noyau et de la cro\^ute.

    \begin{table}[ht]
      \centering
       \caption[Nos mod\`eles d'int\'erieur pour Mimas]{Nos mod\`eles d'int\'erieur pour Mimas. Les 22 premi\`eres lignes correspondent \`a un Mimas \`a l'\'equilibre 
       hydrostatique, tandis que le dernier est bas\'e sur la forme observ\'ee. Les masses volumiques sont en $kg.m^{-3}$.\label{tab:mimaslescas}}
       \begin{tabular}{l|lllllll}
       \hline\hline
       N & $\rho_c$ & $\rho_s$ & $k_f$ & $MOI$ & $J_2$ $(10^{-2})$ & $C_{22}$ $(10^{-3})$ & $C/(mR^2)$ \\
       \hline
        1 & $1200$ &  $800$ & $1.40473$ & $0.389636$ & $2.17051$ & $6.51152$ & $0.404106$ \\
        2 & $1500$ &  $800$ & $1.11293$ & $0.354953$ & $1.71963$ & $5.15889$ & $0.366418$ \\
        3 & $2000$ &  $800$ & $0.94032$ & $0.331801$ & $1.45293$ & $4.35878$ & $0.341487$ \\
        4 & $2500$ &  $800$ & $0.86349$ & $0.320705$ & $1.33422$ & $4.00267$ & $0.329600$ \\
        5 & $3000$ &  $800$ & $0.81885$ & $0.314001$ & $1.26524$ & $3.79571$ & $0.322436$ \\
        6 & $3500$ &  $800$ & $0.78921$ & $0.309439$ & $1.21944$ & $3.65831$ & $0.317569$ \\
        7 & $4000$ &  $800$ & $0.76788$ & $0.306100$ & $1.18649$ & $3.55946$ & $0.314010$ \\
       \hline
        8 & $1200$ & $1000$ & $1.41613$ & $0.390899$ & $2.18812$ & $6.56437$ & $0.405486$ \\
        9 & $1500$ & $1000$ & $1.24455$ & $0.371206$ & $1.92301$ & $5.76902$ & $0.384026$ \\
       10 & $2000$ & $1000$ & $1.17336$ & $0.362551$ & $1.81300$ & $5.43901$ & $0.374638$ \\
       11 & $2500$ & $1000$ & $1.14536$ & $0.359061$ & $1.76975$ & $5.30925$ & $0.370860$ \\
       12 & $3000$ & $1000$ & $1.12980$ & $0.357099$ & $1.74570$ & $5.23711$ & $0.368737$ \\
       13 & $3500$ & $1000$ & $1.11969$ & $0.355816$ & $1.73009$ & $5.19026$ & $0.367350$ \\
       14 & $4000$ & $1000$ & $1.11251$ & $0.354901$ & $1.71899$ & $5.15698$ & $0.366361$ \\
       \hline
       15 & $1200$ & $1100$ & $1.44040$ & $0.393565$ & $2.22563$ & $6.67688$ & $0.408403$ \\
       16 & $1500$ & $1100$ & $1.38066$ & $0.386951$ & $2.13331$ & $6.39993$ & $0.401173$ \\
       17 & $2000$ & $1100$ & $1.36451$ & $0.385134$ & $2.10836$ & $6.32508$ & $0.399189$ \\
       18 & $2500$ & $1100$ & $1.35879$ & $0.384487$ & $2.09953$ & $6.29858$ & $0.398484$ \\
       19 & $3000$ & $1100$ & $1.35572$ & $0.384139$ & $2.09478$ & $6.28435$ & $0.398105$ \\
       20 & $3500$ & $1100$ & $1.35376$ & $0.383917$ & $2.09176$ & $6.27527$ & $0.397862$ \\
       21 & $4000$ & $1100$ & $1.35239$ & $0.383761$ & $2.08963$ & $6.26890$ & $0.397692$ \\
       \hline
       22 & $1150.03$ & $1150.03$ & $1.5$ & $0.400000$ & $2.30951$ & $6.92854$ & $0.415397$ \\
       \hline
       23 & $1200$ &  $800$ & $1.40473$ & $0.389636$ & $2.28639$ & $5.57013$ & $0.406273$ \\
       \hline
       \end{tabular}
     \end{table}

    \clearpage

    \subsection{La rotation simul\'ee}
    
    \par Comme pr\'ec\'edemment j'ai fait des simulations num\'eriques de la rotation de Mimas, en confirmant analytiquement les r\'esultats. J'ai utilis\'e pour cela les 
    \'eph\'em\'erides TASS 1.6 \citep{vd1995}. Les modes propres orbitaux sont pr\'esent\'es Tab.\ref{tab:propmodmimas}.
    
    \begin{table}[ht]
      \centering
      \caption[Les fr\'equences propres du mouvement orbital de Mimas]{Les fr\'equences propres du mouvement orbital de Mimas d'apr\`es TASS1.6 \citep{vd1995}).\label{tab:propmodmimas}}
      \begin{tabular}{l|rrr}
      \hline
        & Fr\'equence (rad/an) & P\'eriode (j) & P\'eriode (a) \\
      \hline
      $\lambda_{01}$ & $2435.14429644$ & $0.942421949$ & $2.580211\times10^{-3}$ \\
      $\omega_1$  &    $0.08904538$ & $25772.62777$ & $70.561609$ \\
      $\phi_1$    &   $10.19765304$ & $225.0452555$ & $0.616140$ \\
      $\rho_1$   &    $3.81643833$ & $601.3285779$ & $1.646348$ \\
      $\Phi_1$    &   $-2.55544336$ & $898.0568575$ & $2.458746$ \\
      \hline
      \end{tabular}
    \end{table}

    \par Le premier, $\lambda_{01}$, correspond au mouvement k\'epl\'erien de Mimas autour de Saturne. $\omega_1$ est l'argument de la libration de la r\'esonance de moyen 
    mouvement avec T\'ethys, tandis que les 3 autres modes propres apparaissent dans le n{\oe}ud et le p\'ericentre. Comme le montre la d\'ecomposition quasi-p\'eriodique 
    de l'\'evolution de la longitude moyenne $\lambda$ (Tab.\ref{tab:lambmimas}), la libration de la r\'esonance prend une grande importance dans le mouvement en longitude, 
    on s'attend donc \`a le retrouver dans la rotation. C'est bien ce qui se passe dans les librations physiques (Tab.\ref{tab:mimasphysic}) o\`u ce terme a une amplitude
    de $\approx44^{\circ}$.
    
    \begin{table}[ht]
      \centering
      \caption[Librations physiques de Mimas]{Librations physiques de Mimas, pour le mod\`ele 23, bas\'e sur la forme. Les s\'eries sont en cosinus.\label{tab:mimasphysic}}
      \begin{tabular}{rrrrrrrr}
      \hline
        &  & & & Fr\'equence & P\'eriode & Amplitude & Phase \\
      $\lambda_{01}$ & $\omega_1$ & $\phi_1$ & $\rho_1$ & (rad/an) & (j) & (arcmin) & at J2000 \\
      \hline
       -  & $1$ &  -   &  -  &    $0.08904538$ & $25772.62777$ & $43.61^{\circ}$ & $51.354^{\circ}$ \\
       -  & $3$ &  -   &  -  &    $0.26713614$ &  $8590.87592$ &  $43.261$ arcmin & $-25.913^{\circ}$ \\
      $1$ &  -  & $-1$ & $1$ & $2428.763080$   &    $0.944898$ &  $26.075$ arcmin & $101.355^{\circ}$ \\
       -  &  -  &  $1$ &  -  &   $10.19765304$ &   $225.04526$ &   $7.828$ arcmin & $-157.744^{\circ}$ \\
       -  & $1$ & $-1$ &  -  &   $10.10860766$ &   $227.02728$ &   $3.657$ arcmin & $-119.032^{\circ}$ \\
       -  & $1$ &  $1$ &  -  &   $10.28669842$ &   $223.09718$ &   $3.532$ arcmin & $-16.309^{\circ}$ \\
      \hline
      \end{tabular}
    \end{table}

    \par
\begin{figure}[ht]
     \centering
     \begin{tabular}{lcr}
      \includegraphics[width=0.45\textwidth]{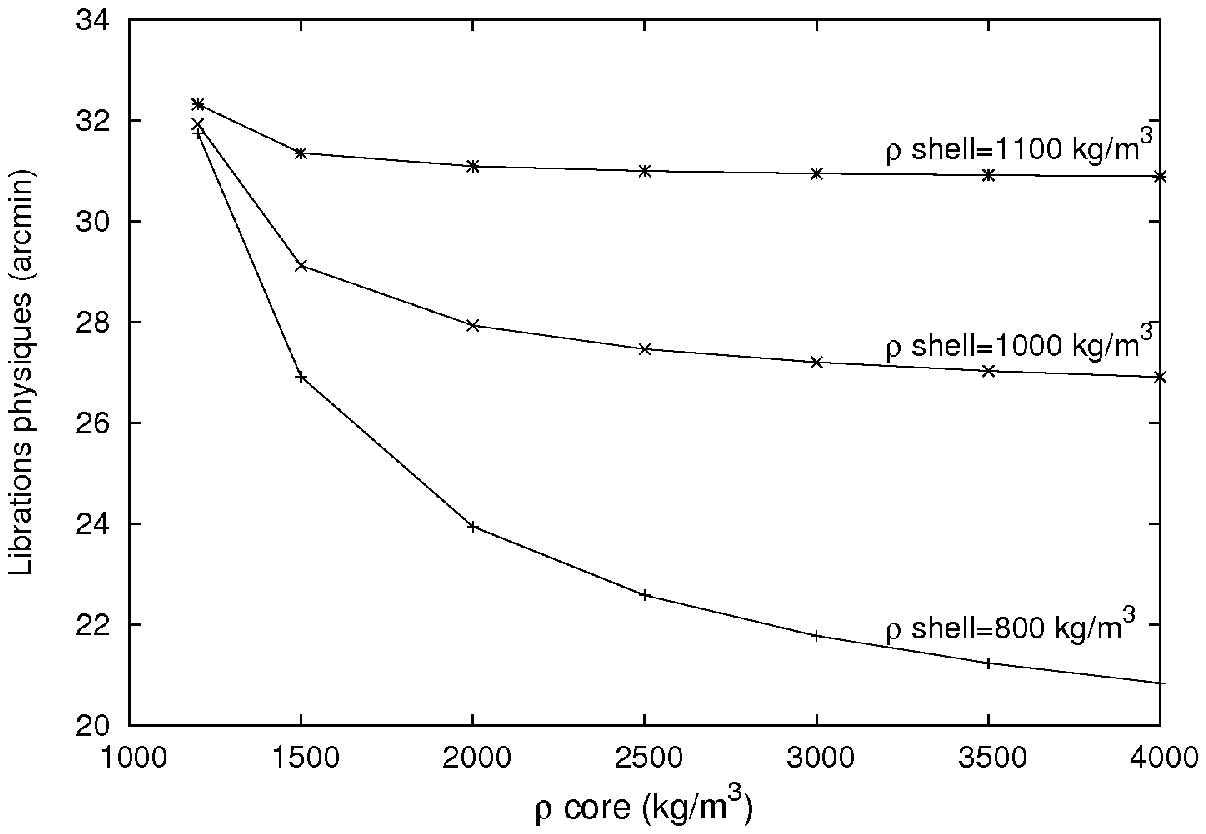} & \hspace{.1cm} & \includegraphics[width=0.45\textwidth]{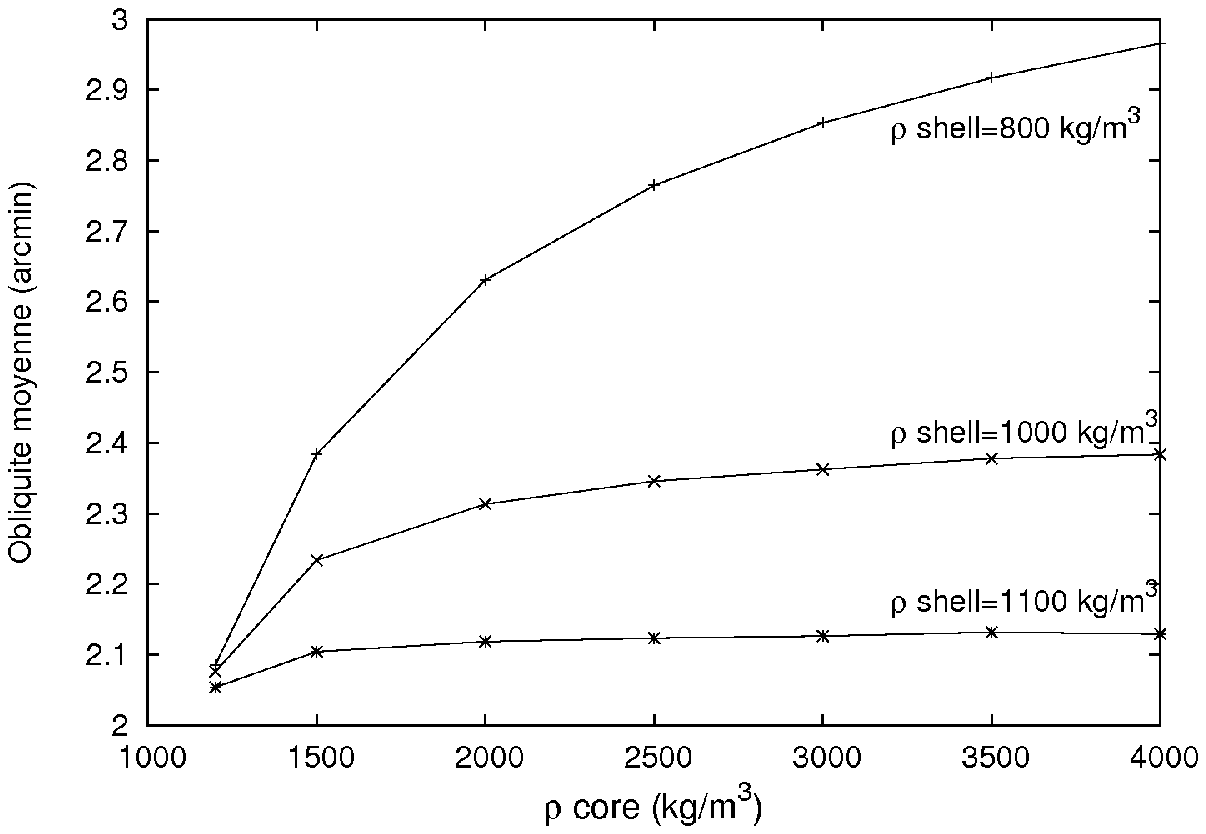}
     \end{tabular} 
     \caption[Rotation de Mimas simul\'ee]{La rotation de Mimas, simul\'ee num\'eriquement.\label{fig:rotamimas}}
    \end{figure}

 En fait, le terme qui a un sens en terme de structure interne est celui li\'e \`a la libration diurne, par la formule (\ref{eq:librphys}), de p\'eriode $0.944898$ jour.
    Un autre terme int\'eressant est l'obliquit\'e moyenne, par la formule (\ref{eq:cs1}). La figure \ref{fig:rotamimas} et la table \ref{tab:qrotamimas} donnent ces quantit\'es,
    toutes d\'etermin\'ees num\'eriquement puis v\'erifi\'ees analytiquement.

    \begin{table}[ht]
      \centering
      \caption[Les quantit\'es de rotation pour nos mod\`eles de Mimas]{Les quantit\'es de rotation pour nos mod\`eles de Mimas.\label{tab:qrotamimas}}
      \begin{tabular}{l|rrrrrrr}
      \hline
         &       &       &       & Librations & Librations  & Obliquit\'e & Librations \\
       N & $T_u$ & $T_v$ & $T_w$ & de mar\'ee & en latitude & moyenne & physiques \\
         & (j) & (j) & (j) & (arcmin)   & (arcmin)   & (arcmin)  & (arcmin) \\
      \hline
       1 & $2.143878$ &  $7.885550$ & $11.621674$ & $163.398$ & $2.016$ & $2.086$ & $31.744$ \\
       2 & $2.294081$ &  $8.997072$ & $13.222674$ & $158.577$ & $2.314$ & $2.384$ & $26.914$ \\
       3 & $2.407777$ &  $9.908107$ & $12.763086$ & $155.609$ & $2.559$ & $2.631$ & $23.944$ \\
       4 & $2.468518$ & $10.416095$ & $13.618627$ & $154.248$ & $2.693$ & $2.765$ & $22.582$ \\
       5 & $2.507237$ & $10.742236$ & $14.181571$ & $153.442$ & $2.780$ & $2.853$ & $21.776$ \\
       6 & $2.534519$ & $10.975456$ & $14.591169$ & $152.900$ & $2.843$ & $2.917$ & $21.234$ \\
       7 & $2.555063$ & $11.152890$ & $14.906138$ & $152.508$ & $2.891$ & $2.966$ & $20.841$ \\
      \hline 
       8 & $2.138844$ &  $7.849766$ & $11.569966$ & $163.583$ & $2.006$ & $2.076$ & $31.922$ \\
       9 & $2.220696$ &  $8.443608$ & $12.426948$ & $160.777$ & $2.166$ & $2.234$ & $29.115$ \\
      10 & $2.257477$ &  $8.730512$ & $12.839826$ & $159.593$ & $2.243$ & $2.313$ & $27.930$ \\
      11 & $2.274983$ &  $8.851150$ & $13.013162$ & $159.124$ & $2.275$ & $2.346$ & $27.461$ \\
      12 & $2.284054$ &  $8.920304$ & $13.112486$ & $158.862$ & $2.293$ & $2.363$ & $27.199$ \\
      13 & $2.290031$ &  $8.966015$ & $13.178106$ & $158.692$ & $2.306$ & $2.378$ & $27.028$ \\
      14 & $2.294326$ &  $8.998974$ & $13.225425$ & $158.571$ & $2.315$ & $2.384$ & $26.907$ \\
      \hline
      15 & $2.128297$ &  $7.775062$ & $11.461869$ & $163.975$ & $1.986$ & $2.054$ & $32.314$ \\
      16 & $2.154684$ &  $7.962727$ & $11.733240$ & $163.008$ & $2.037$ & $2.104$ & $31.347$ \\
      17 & $2.162067$ &  $8.015705$ & $11.809777$ & $162.745$ & $2.051$ & $2.119$ & $31.084$ \\
      18 & $2.164709$ &  $8.034707$ & $11.837202$ & $162.652$ & $2.056$ & $2.124$ & $30.991$ \\
      19 & $2.166135$ &  $8.044989$ & $11.852091$ & $162.602$ & $2.059$ & $2.127$ & $30.941$ \\
      20 & $2.167044$ &  $8.051527$ & $11.861492$ & $162.570$ & $2.060$ & $2.132$ & $30.909$ \\
      21 & $2.167685$ &  $8.056158$ & $11.868221$ & $162.548$ & $2.062$ & $2.130$ & $30.886$ \\
      \hline
      22 & $2.106951$ &  $7.625231$ & $11.244996$ & $164.792$ & $1.946$ & $2.014$ & $33.132$ \\
      \hline
      23 & $2.323129$ &  $8.075914$ & $10.113182$ & $157.734$ & $2.071$ & $2.135$ & $26.075$ \\
      \hline
      \end{tabular}
    \end{table}

    \par  Les simulations indiquent que si Mimas est \`a l'\'equilibre hydrostatique, alors on peut s'attendre \`a une obliquit\'e entre 2 et 3 minutes d'arc 
    (entre $0.033$ et $0.05$ degr\'es), et des librations diurnes entre 21 et 32 minutes d'arc, soit entre $0.34$ et $0.54$ degr\'es. L'amplitude des autres librations 
    n'est pratiquement pas affect\'ee. Le cas homog\`ene repr\'esente un extremum, c'est-\`a-dire que c'est l'obliquit\'e la moins \'el\'ev\'ee et la libration la
    plus grande. Nous avons \'egalement estim\'e que les mar\'ees n'affectaient pas significativement la rotation d'\'equilibre (l'\'Etat de Cassini 1).
    
    \par Ces r\'esultats sont des simulations. Ils demandent \`a \^etre confront\'es aux observations. C'est ce que nous avons fait dans une \'etude men\'ee par Radwan Tajeddine.
    
    \clearpage
    
    \subsection{La rotation observ\'ee \citep{trlcrrn2014}}
    
    \par Je suis volontairement bref dans cette section car ma contribution \`a cette \'etude a \'et\'e relativement limit\'ee. Nous avons utilis\'e des images Cassini 
    ISS\footnote{Image Science Subsystem} acquises par la cam\'era NAC\footnote{Narrow Angle Camera} pour rep\'erer un r\'eseau de 260 points de contr\^ole. Il s'agit de 
    motifs remarquables sur la surface de Mimas, par exemple des crat\`eres, qui apparaissent sur au moins 2 images. \`A partir de l'observation de ces points de contr\^ole,
    nous avons ajust\'e un mouvement en longitude, qui confirme la rotation synchrone et mesure des librations physiques (Tab.\ref{tab:rotamimaobserv}).

    \begin{table}[ht]
      \centering
      \caption[Librations de Mimas observ\'ees]{Librations physiques de Mimas observ\'ees.\label{tab:rotamimaobserv}}
      \begin{tabular}{r|rr|rr}
      \hline
      P\'eriode & \multicolumn{2}{c}{Amplitude (arcmin)} & \multicolumn{2}{c}{Phase \`a J2000 ($^{\circ}$)} \\
      (j)       &  Th\'eorie & Observation               & Th\'eorie & Observation \\
      \hline
      25772.62  & $43.61^{\circ}$ & $43.86^{\circ}\pm0.05^{\circ}$ &   $51.35$ & $52.9\pm0.9$ \\
       8590.87  & $43.26$         & $44.5\pm1.1$             &  $-25.91$ & $-18\pm3.2$ \\
          0.945 & $26-33$         & $50.3\pm1.0$             &  $101.35$ & $107.7\pm0.8$ \\
        225.04  & $7.82$          &  $7.5\pm0.8$             & $-157.74$ & -- \\
        227.02  & $3.65$          &  $2.9\pm0.9$             & $-119.03$ & -- \\
        223.09  & $3.53$          &  $3.3\pm0.8$             &  $-16.31$ & -- \\
        \hline
      \end{tabular}
    \end{table}

    \par
\begin{figure}[ht]
    \centering
    \includegraphics[width=.6\textwidth]{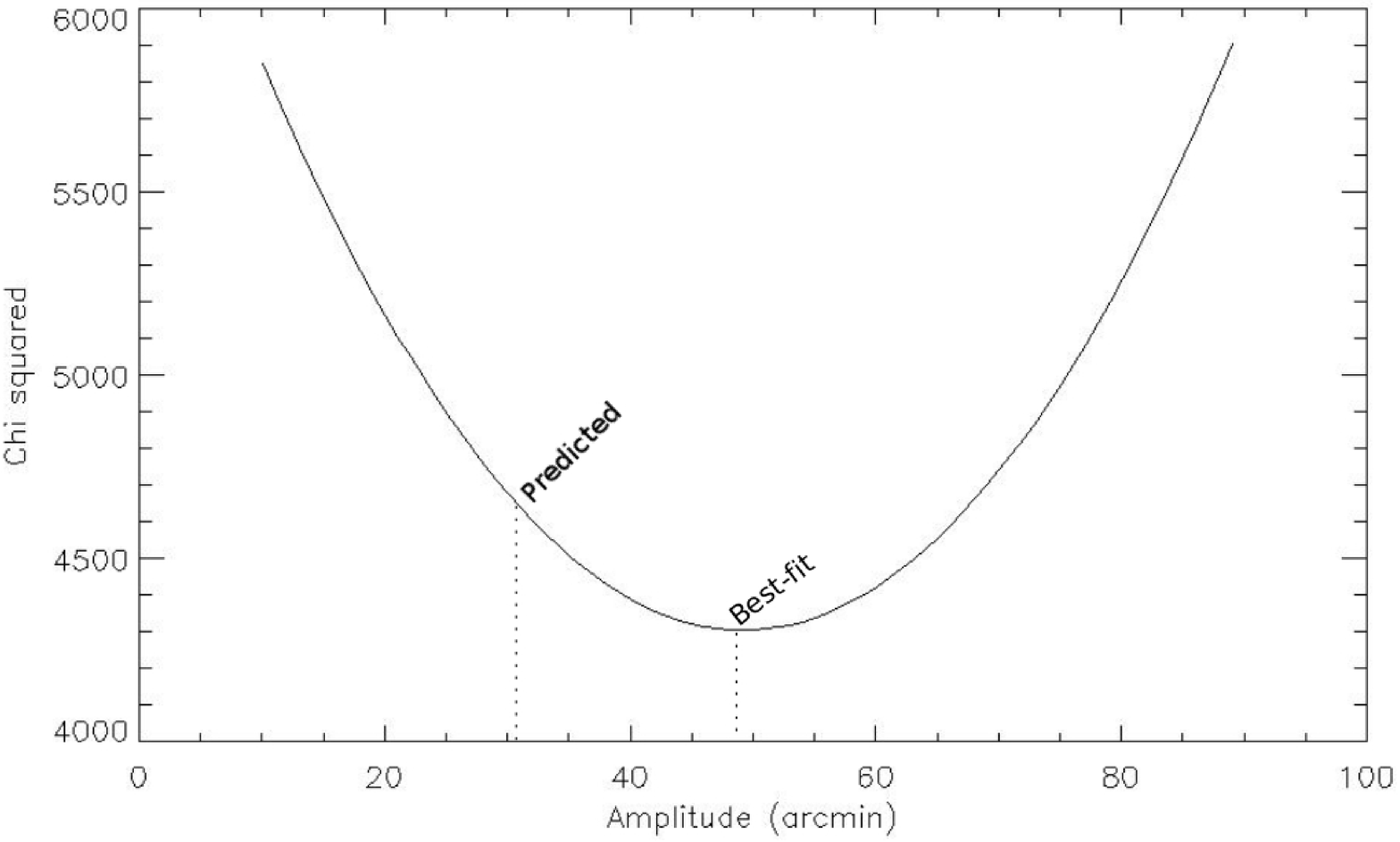}
    \caption[Minimisation de l'erreur en ajustant l'amplitude des librations]{Minimisation de l'erreur en ajustant l'amplitude des librations diurnes de Mimas. 
    Figure extraite de \citep{trlcrrn2014}.\label{fig:radwan}}
    \end{figure}

 On constate que les amplitudes des librations non diurnes sont plut\^ot confirm\'ees, ce qui donne confiance en les mesures. Par contre l'amplitude des librations 
    diurnes est plus \'elev\'ee que pr\'evu, ce qui sugg\`ere que l'int\'erieur de Mimas ne soit pas ce qu'on croyait jusqu'\`a pr\'esent. La Fig.\ref{fig:radwan}
    donne l'\'evaluation de l'erreur en fonction de cette amplitude.
    
    \par On consid\`ere en g\'en\'eral que d'importantes librations diurnes sont la signature d'un oc\'ean global. Cei voudrait dire que Mimas n'est en fait pas un corps gel\'e.
    Une autre possibilit\'e serait que Mimas soit bien un corps solide, mais avec un noyau d'\'el\'ements lourds tr\`es allong\'e, comme s'il s'\'etait form\'e bien plus pr\`es de 
    Saturne, au bord ext\'erieur des anneaux, avant d'\^etre enrob\'e de silicates. Une telle structure interne confirmerait le mod\`ele de formation des satellites \`a partir des 
    anneaux de \citet{cccldktmls2011}.
    
    \section{Un satellite non r\'esonnant : Hyp\'erion \label{sec:hyperion}}
    
    \par Je vais clore ce chapitre avec le cas d'un satellite rigide dont la rotation n'est pas amortie. En cons\'equence, il tourne sur 3 axes, sa rotation ne correspond 
    pas \`a une r\'esonance. J'ai toujours consid\'er\'e Hyp\'erion comme un corps int\'eressant car il s'agit, dans la limite de la connaissance actuelle, du plus grand objet de forme 
    irr\'eguli\`ere dans le Syst\`eme Solaire, ceci indiquerait qu'il serait issu d'une collision relativement r\'ecente \citep{fmnpz1983}. D'un point de vue de la dynamique,
    son orbite est relativement allong\'ee avec une excentricit\'e de l'ordre de $0.1$. Il est de plus en r\'esonance de moyen-mouvement 4:3 avec Titan. Son mouvement orbital
    est assez complexe, notamment la d\'ecomposition quasi-p\'eriodique de ses \'el\'ements orbitaux converge assez lentement \citep{dv1997}, ce qui a n\'ecessit\'e un traitement
    \`a part dans les \'eph\'em\'erides TASS.
    
    \par Suite aux observations de la rotation par Cassini \citep{htn2011}, j'ai voulu moi-m\^eme \'etudier le chaos pr\'esent dans ce mouvement de rotation. Je ne suis pour l'instant 
    pas all\'e au-del\`a de quelques simulations num\'eriques, c'est pourquoi ce travail n'a encore \'et\'e soumis nulle part.
    
    \subsection{Une premi\`ere d\'etection de chaos dans le Syst\`eme Solaire?}
    
    \par
\begin{figure}[ht]
      \centering
      \begin{tabular}{ccc}
      \includegraphics[width=0.42\textwidth]{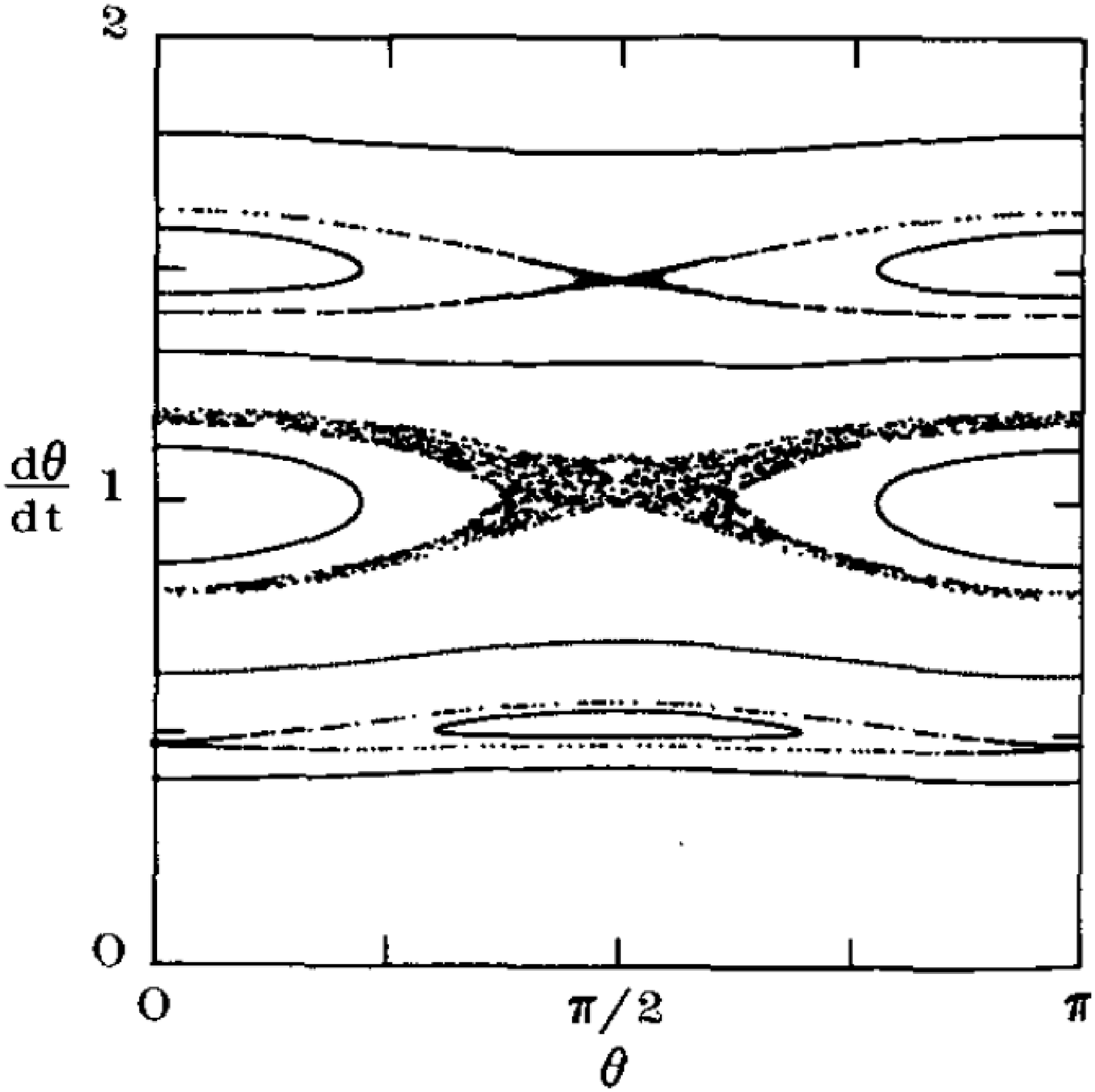} & \hspace{1cm} & \includegraphics[width=0.42\textwidth]{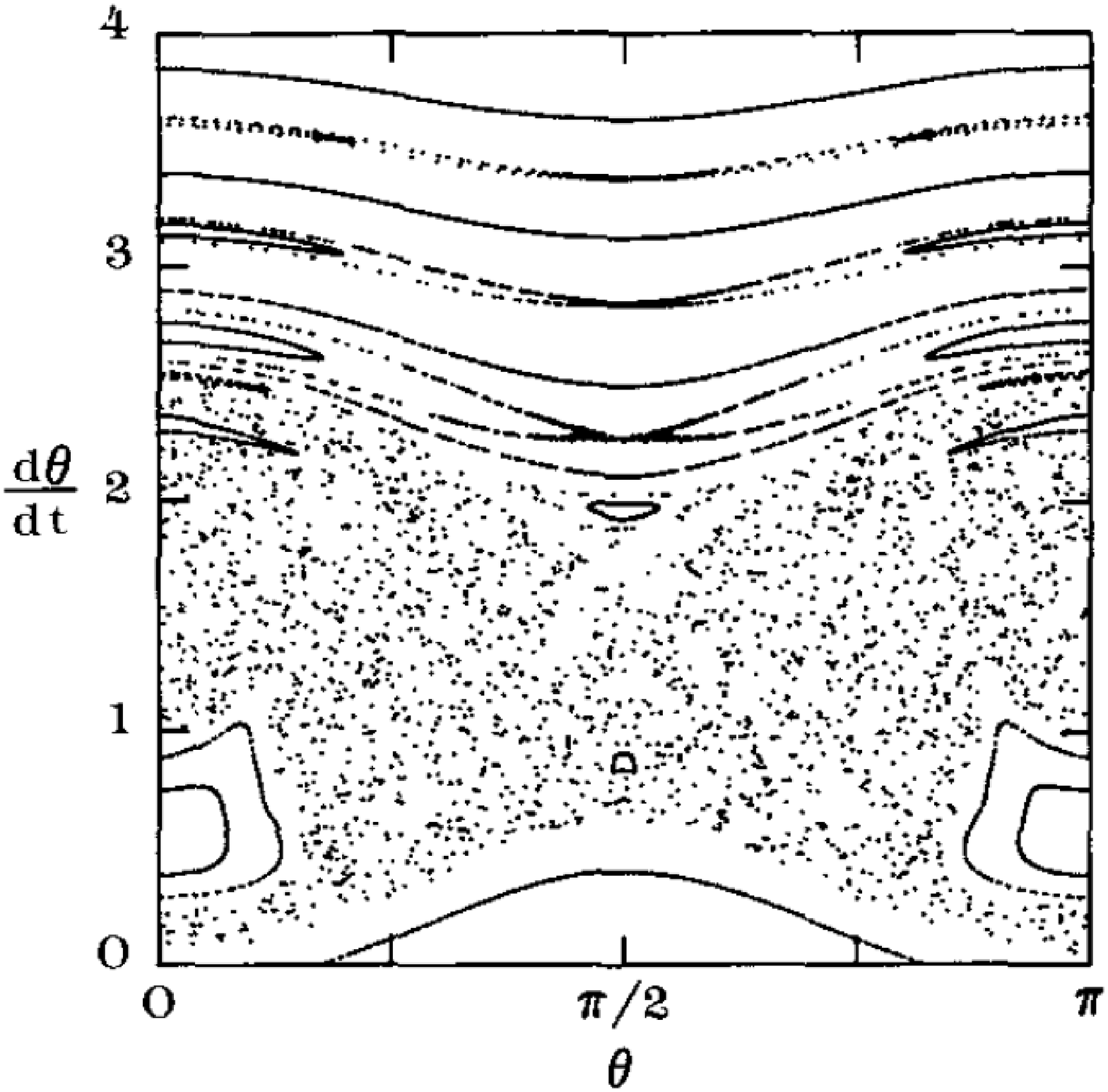}
      \end{tabular}
      \caption[Portraits de phase pour la rotation d'Hyp\'erion]{Portraits de phase de la r\'esonance spin-orbite pour un satellite classique (\`a gauche) et pour Hyp\'erion (\`a droite),
      avec $e=0.1$ et $(B-A)/C=0.264$, en supposant que le mouvement polaire et l'obliquit\'e sont amortis, reproduits de \citep{wpm1984}. On remarque une large zone de chaos couvrant 
      notamment les r\'esonances $1:2$, $1:1$, $3:2$ et $2:1$.\label{fig:mapwisdom}}
    \end{figure}

 Suite aux observations de la forme irr\'eguli\`ere d'Hyp\'erion \citep{ssbbimsbbbbchjmtccpdidhmosvss1982} par Voyager 2, et en connaissance de son excentricit\'e, 
    \citet{wpm1984} ont \'etudi\'e la stabilit\'e des r\'esonances spin-orbite dans son cas, cf. les portraits de phase Fig.\ref{fig:mapwisdom}, obtenus \`a l'aide d'un Hamiltonien
    similaire \`a l'Eq.(\ref{eq:hamilplan2}).
    
    \par Dans le cas le plus commun, c'est-\`a-dire un satellite de forme relativement sph\'erique et d'orbite relativement circulaire, les r\'esonances sont s\'epar\'ees, et ce sont 
    des zones de stabilit\'e. Par contre, dans le cas d'une excentricit\'e et d'un aplatissement significatifs comme pour Hyp\'erion, les r\'esonances sont suffisamment larges pour 
    s'intersecter, ce qui est source de chaos \citep{c1979}. En cons\'equence les r\'esonances $1:2$, $1:1$, $3:2$ et $2:1$ dans le sens o\`u on les entend g\'en\'eralement ne sont pas 
    stables, d'autres zones de stabilit\'e apparaissent mais elles sont plus petites et entour\'ees de chaos. Leur pr\'esence ne signifie donc pas que si la rotation d'Hyp\'erion 
    avait \'et\'e suffisamment amortie, il s'y trouverait bloqu\'e. Les auteurs ne s'attardent pas sur ces zones, mais on peut voir proche de la r\'esonance synchrone une petite zone
    de stabilit\'e autour de $\pi/2$, c'est-\`a-dire que la r\'esonance synchrone o\`u l'axe le plus long est tangent \`a l'orbite est stable dans ce cas. Cette configuration a \'et\'e
    nomm\'ee quelques ann\'ees plus tard \emph{Effet Amalth\'ee} \citep{ms1998}, du nom du satellite de Jupiter pour lequel la th\'eorie n'interdit pas son existence.
    
    \par En 1984, on savait d\'ej\`a que Japet, satellite de Saturne plus \'eloign\'e qu'Hyp\'erion, avait une rotation suffisamment amortie pour \^etre synchrone \citep{w1950}. Il 
    \'etait donc naturel de penser que la rotation d'Hyp\'erion \'etait amortie. \citet{wpm1984} en ont d\'eduit qu'Hyp\'erion se trouvait n\'ecessairement dans la zone chaotique, et 
    qu'il tournait m\^eme probablement sur 3 axes. Ceci est parfois consid\'er\'e comme la premi\`ere d\'etection de chaos dans le Syst\`eme Solaire.
    
    \par S'il s'est av\'er\'e par la suite que la rotation d'Hyp\'erion \'etait bien sur 3 axes et chaotique, elle n'est en fait pas suffisamment amortie pour atteindre la zone de 
    chaos identifi\'ee par \citet{wpm1984}, sa fr\'equence de rotation \'etant plus de 4 fois son moyen mouvement moyen. \citet{wpm1984} ne peuvent donc \^etre cr\'edit\'es de la 
    pr\'ediction de chaos dans la rotation d'Hyp\'erion.

    \subsection{La v\'erit\'e des observations}
    
    \par Nous disposons d'observations de la rotation d'Hyp\'erion depuis Voyager. Entre Voyager et Cassini, des tentatives de mesures depuis la Terre ont \'et\'e faites, 
    \`a partir des variations de la courbe de lumi\`ere. En voici les r\'esultats :
    
    \begin{itemize}
    
     \item \citet{tvwdd1984} ont annonc\'e une p\'eriode de rotation de 13.1 jours \`a partir de donn\'ees Voyager. La p\'eriode orbitale \'etant de 21.25 jours, on a un 
     rapport de $1.62$, et on serait dans la zone chaotique de \citet{wpm1984}.
    
     \item \citet{tv1985} confirment cette mesure avec plus de donn\'ees.
    
     \item \citet{k1989}, en utilisant aussi des donn\'ees terrestres, incite \`a la prudence. Il dit ne pas d\'etecter clairement de p\'eriode, ce qui semble coh\'erent pour
     un corps tournant sur 3 axes. En cherchant \`a ajuster une sinuso\"ide sur ses observations, il trouve 2 optimaux, \`a $6.6$ et $13.8$ jours, mais pr\'ef\`ere rester prudent 
     quant \`a leur interpr\'etation. \`A partir de cette \'etude, tous les auteurs consid\'ereront qu'Hyp\'erion tourne sur 3 axes. Dans ce cas, parler de p\'eriode est un abus de
     langage. Il faut la comprendre comme une valeur d\'eduite de la norme du moment cin\'etique.
     
     \item \citet{bnt1995}, \`a l'aide de donn\'ees Voyager, trouvent une p\'eriode de $5\substack{+0.29 \\ -0.2}$ jours, ce qui sugg\`ere un rapport d'environ $4.25$ 
     entre la fr\'equence de rotation et l'orbitale. La chute de la p\'eriode de rotation par rapport aux \'etudes pr\'ec\'edentes refl\`ete l'un des dangers de l'observation de
     ph\'enom\`enes ayant une p\'eriodicit\'e : l'aliasing. Le th\'eor\`eme de Shannon nous dit que si l'intervalle de temps entre 2 observations est sup\'erieur \`a la moiti\'e
     de la p\'eriode du ph\'enom\`ene, alors l'analyse des donn\'ees d\'etectera bien une p\'eriode, mais qui sera plus longue que la p\'eriode r\'eelle (cf.Fig.\ref{fig:aliasing}). 
     Il faut garder \`a l'esprit qu'une p\'eriode si courte pour Hyp\'erion n'\'etait pas attendue.
     
     \begin{figure}[ht]
       \centering
       \begin{tabular}{cc}
        \includegraphics[width=.47\textwidth]{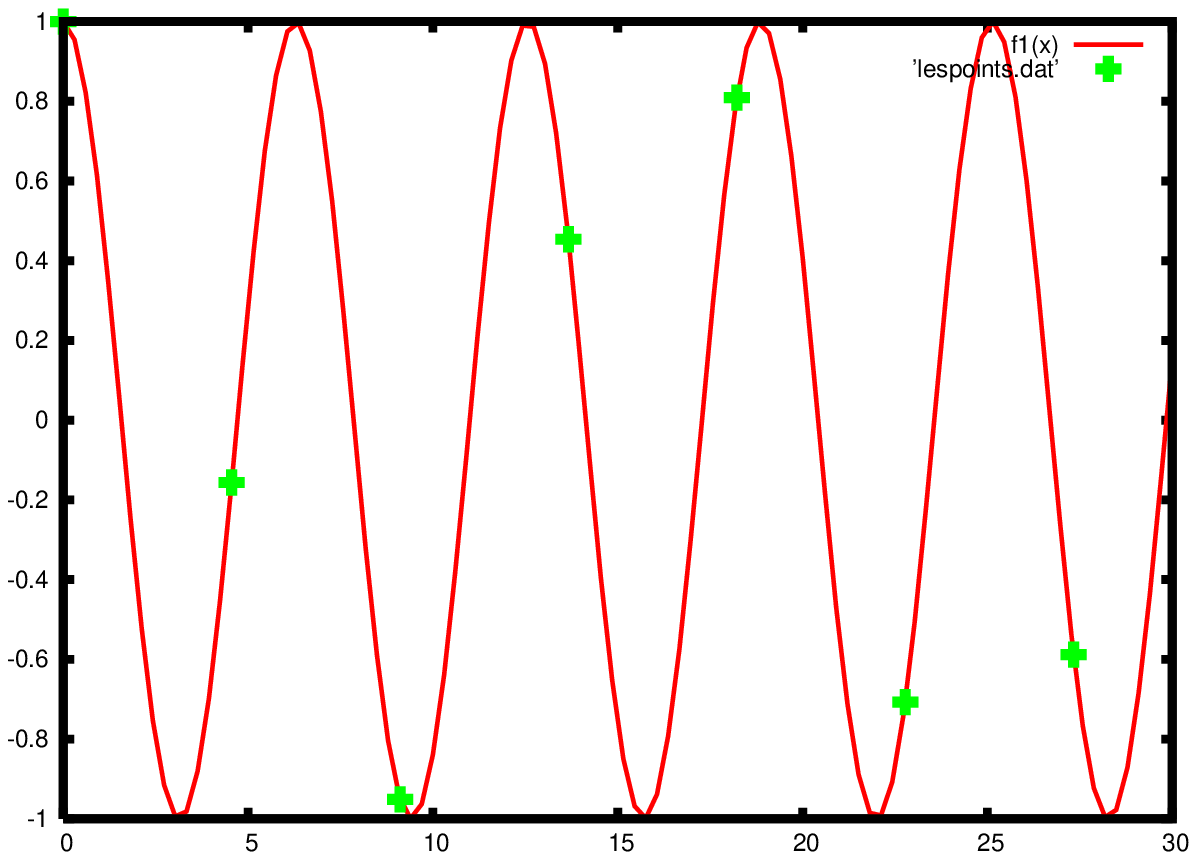} & \includegraphics[width=.47\textwidth]{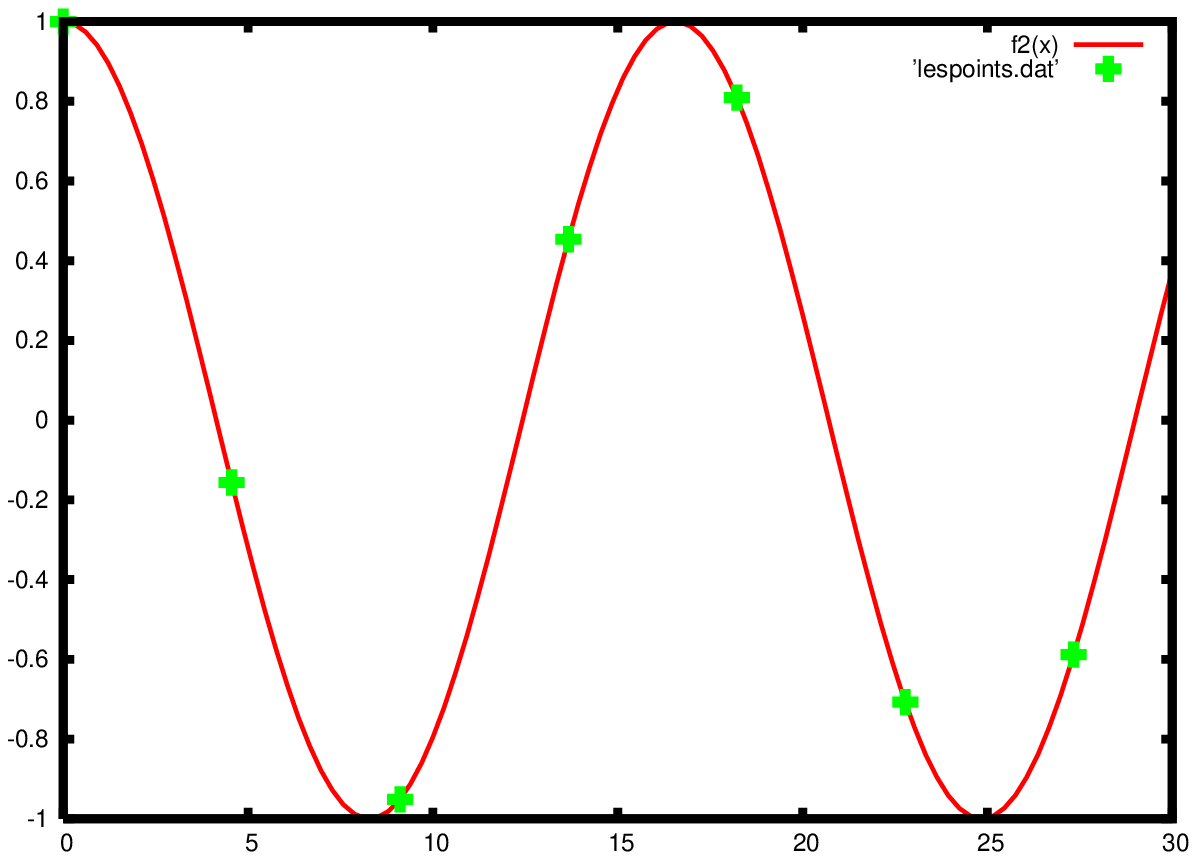}
       \end{tabular}
       \caption[Exemple d'aliasing]{Illustration du probl\`eme d'aliasing. L'analyse en fr\'equence d'un \'echantillon de mesures issu d'un probl\`eme p\'eriodique donne
       une p\'eriode trop longue si le pas d'\'echantilonnage est trop grand. Cet exemple est issu d'un s\'eminaire sur l'analyse en fr\'equences que j'ai donn\'e en octobre 2007
       \`a Namur, lors d'une formation doctorale.\label{fig:aliasing}} 
     \end{figure}

     \item \citet{htn2011}, \`a l'aide de donn\'ees Cassini, trouvent une p\'eriode de rotation entre $4.8$ et $5$ jours, et essaient de simuler l'\'evolution temporelle de
     la rotation d'Hyp\'erion. Ils trouvent que ce mouvement est chaotique, avec un temps de Lyapounov de l'ordre de 100 jours. 
 
    \end{itemize}

    \par Ces observations confirment que la dynamique de la rotation d'Hyp\'erion est un probl\`eme complexe. Ce serait bien un mouvement chaotique, et Hyp\'erion tournerait bien
    sur 3 axes.

    \subsection{Quelques simulations\ldots}
    
    \par Ayant un code de rotation rigide, j'ai tent\'e moi aussi de m'attaquer au probl\`eme de la rotation d'Hyp\'erion. Je disposais pour cela des param\`etres d'int\'erieur 
    et des observations (Tab.\ref{tab:dataharbison}) de \citet{htn2011}.
    
    \begin{table}[ht]
      \centering
      \caption[Moments principaux d'inertie d'Hyp\'erion]{Moments principaux d'inertie d'Hyp\'erion, bas\'es sur la forme et sur la rotation. Ici Hyp\'erion est 
      consid\'er\'e comme un ellipso\"ide triaxial, ce qui est probablement une approximation simpliste de sa forme.\label{tab:momhyperion}}
      \begin{tabular}{r|rr}
       \hline
        & Forme & Rotation \\
        & \citep{taabdghijmnprrsttv2007} & \citep{htn2011} \\
        \hline
      A/C & $0.58\pm0.03$ & $\approx0.61$ \\
      B/C & $0.87\pm0.03$ & $\approx0.80$ \\
      C   & $0.542MR^2$ & -- \\
      \hline
      \end{tabular}
    \end{table}

    \begin{table}[ht]
      \centering
      \caption[Orientations du vecteur rotation d'Hyp\'erion observ\'ees par Voyager et Cassini]{Orientations du vecteur rotation d'Hyp\'erion observ\'ees par Voyager et Cassini, 
      \`a 5 dates \citep{htn2011}. Les coordonn\'ees $(\omega_A,\omega_B,\omega_C)$ se rapportent au rep\`ere des axes principaux d'inertie d'Hyp\'erion, tandis que 
      $(\omega_x,\omega_y,\omega_z)$ sont dans un rep\`ere li\'e au mouvement orbital d'Hyp\'erion.\label{tab:dataharbison}}
      \begin{tabular}{rrrrrrrrr}
       \hline
       N & Date & $\omega_A/|\omega|$ & $\omega_B/|\omega|$ & $\omega_C/|\omega|$ & $\omega_x$ & $\omega_y$ & $\omega_z$ & $|\omega|(^{\circ}/j)$ \\
       \hline
       1 & 1981-08-23 & 0.986 & 0.160 & -0.049 & -2.457 & -2.501 & 2.409 & $72\pm3$ \\
       2 & 2005-06-10 & 0.890 & 0.067 &  0.451 &  3.399 & -1.511 & 2.411 & $75\pm1$ \\
       3 & 2005-08-16 & 0.907 & 0.162 &  0.389 &  3.026 &  1.909 & 2.303 & $72\pm1$ \\
       4 & 2005-09-25 & 0.902 & 0.133 &  0.411 &  1.151 &  2.018 & 3.565 & $72\pm1$ \\
       5 & 2007-02-16 & 0.749 & 0.080 &  0.659 & -3.797 &  1.905 & 0.250 & $72\pm1$ \\
       \hline
      \end{tabular}
  \end{table}

  \par Lorsque j'ai entrepris cette \'etude num\'erique, j'avais pour objectif d'essayer de trouver une ou des trajectoires qui permettaient d'expliquer les observations. 
  J'utilisais pour cela un mouvement orbital complet, donn\'e par TASS1.7 \citep{dv1997}. Mes conditions initiales utilisent les quantit\'es ($p$,$h$,$l$,$P$,$K$,$J$) d\'efinies au
  Chap.\ref{chap:rigide}. Pour celles li\'ees \`a l'angle de spin $p$ et \`a la norme du moment cin\'etique normalis\'e $P$, j'ai voulu les laisser libres. Notamment, $p$ peut
  prendre n'importe quelle valeur, et $P$ reste dans les barres d'erreur de la fr\'equence de rotation. L'obliquit\'e $K$, le mouvement polaire $J$, et les angles de 
  pr\'ecession associ\'es $h$ et $l$ sont facilement d\'eductibles des donn\'ees de la Tab.\ref{tab:dataharbison}, et sont rassembl\'ees dans la Tab.\ref{tab:KJhyperion}.
  
  \begin{table}[ht]
    \centering
    \caption{Angles d\'efinissant l'orientation du moment cin\'etique d'Hyp\'erion.\label{tab:KJhyperion}}
    \begin{tabular}{r|rr|rr|rr}
      \hline
        & & & \multicolumn{2}{c}{$A/C=0.58$, $B/C=0.87$} & \multicolumn{2}{c}{$A/C=0.61$, $B/C=0.80$} \\
      N & $K$ & $h$ & $J$ & $l$ & $J$ & $l$ \\
      \hline
      1 & $55.218^{\circ}$ &   $49.788^{\circ}$ & $94.759^{\circ}$ & $76.320^{\circ}$ & $94.556^{\circ}$ & $77.986^{\circ}$ \\
      2 & $57.922^{\circ}$ &    $9.829^{\circ}$ & $49.036^{\circ}$ & $83.557^{\circ}$ & $50.419^{\circ}$ & $84.361^{\circ}$ \\
      3 & $58.271^{\circ}$ &  $-42.896^{\circ}$ & $54.463^{\circ}$ & $75.002^{\circ}$ & $55.606^{\circ}$ & $76.817^{\circ}$ \\
      4 & $33.899^{\circ}$ &  $-67.356^{\circ}$ & $52.509^{\circ}$ & $77.528^{\circ}$ & $53.744^{\circ}$ & $79.055^{\circ}$ \\
      5 & $86.251^{\circ}$ & $-135.557^{\circ}$ & $33.728^{\circ}$ & $80.898^{\circ}$ & $34.995^{\circ}$ & $82.026^{\circ}$ \\
      \hline
    \end{tabular}
  \end{table}

  \par \`A partir de 5 jeux de conditions initiales et 2 jeux de param\`etres int\'erieurs, j'ai r\'ealis\'e 10 groupes de 72000 simulations num\'eriques, chacune diff\'erant des autres
  par le couple de conditions initiales ($p$,$P$). Chacune des simulations \'etait r\'ealis\'ee sur 100 ans, et affichait en sortie les variations relatives des diff\'erentes variables 
  de rotation. J'ai utilis\'e pour cela le cluster de calculs ISCF \emph{Interuniversity Scientific Computing Facility} qui me permettait d'utiliser jusqu'\`a 64 processeurs simultan\'ement.
  
  \par Les r\'esultats ne montrent malheureusement pas grand-chose d'int\'eressant, on peut parler de r\'esultat n\'egatif.
  
  \begin{figure}[ht]
   \centering
   \begin{tabular}{cc}
    \includegraphics[width=0.47\textwidth]{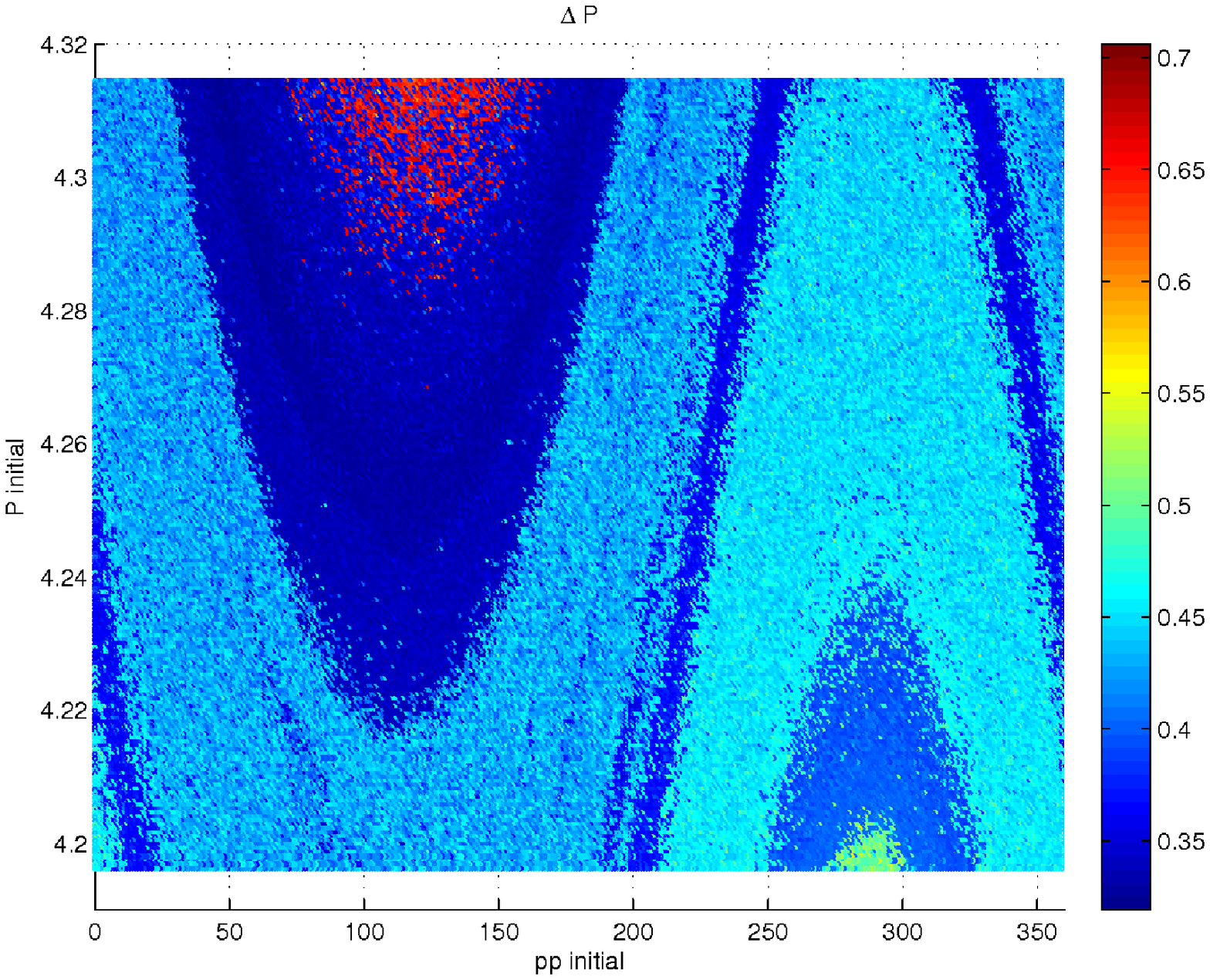} & \includegraphics[width=0.47\textwidth]{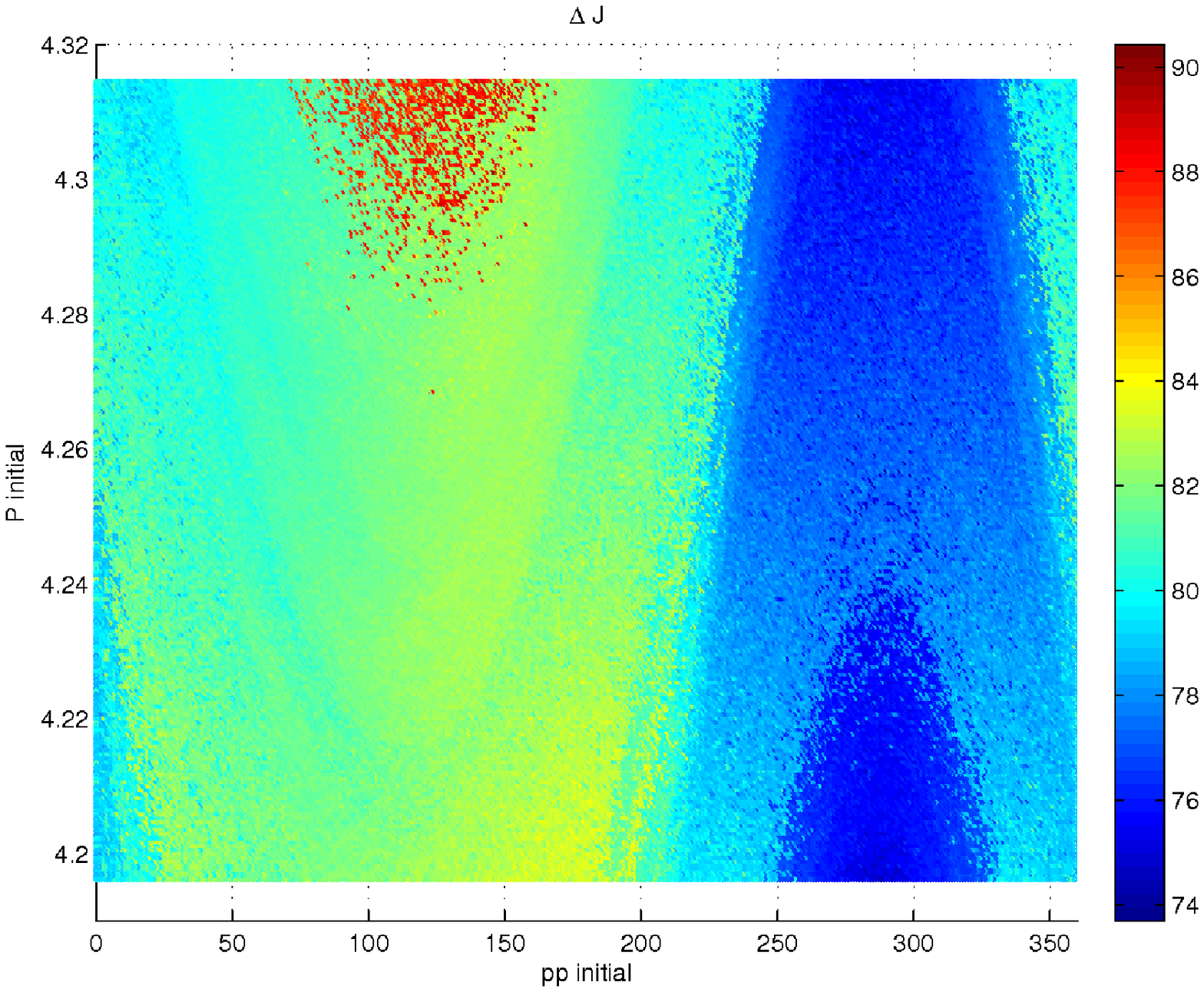}
   \end{tabular}
   \caption[Variations des quantit\'es de rotation pour l'observation 3 et le mod\`ele bas\'e sur la forme]{Variations de la norme du moment cin\'etique $P$ et de l'amplitude 
   du mouvement polaire $J$ d'Hyp\'erion pour l'observation 3 et le mod\`ele bas\'e sur la forme. On observe une structure, mais qui ne stabilise pas vraiment la 
   dynamique.\label{fig:toto_l5}}
  \end{figure}

  \begin{figure}[ht]
   \centering
   \begin{tabular}{cc}
    \includegraphics[width=0.47\textwidth]{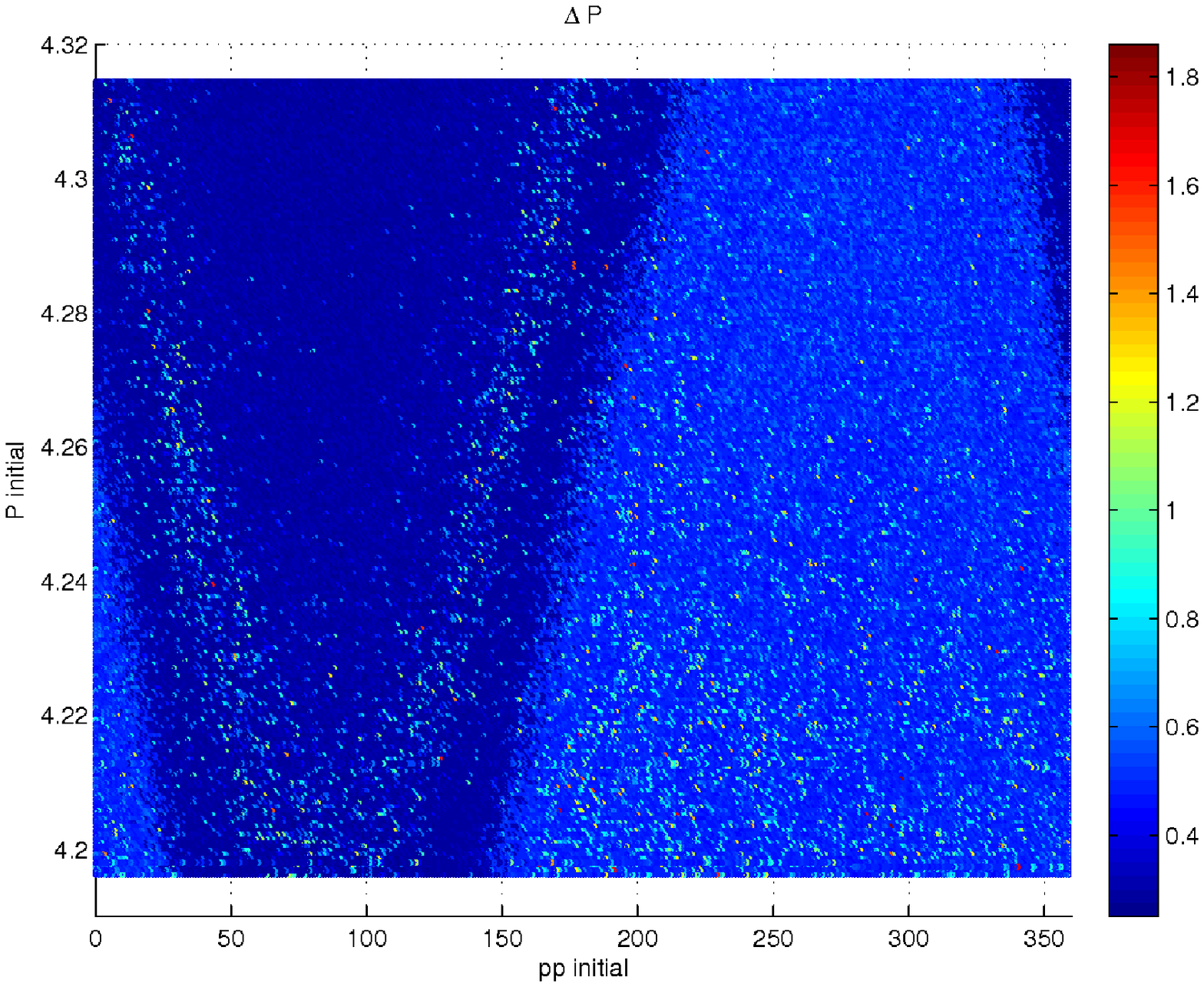} & \includegraphics[width=0.47\textwidth]{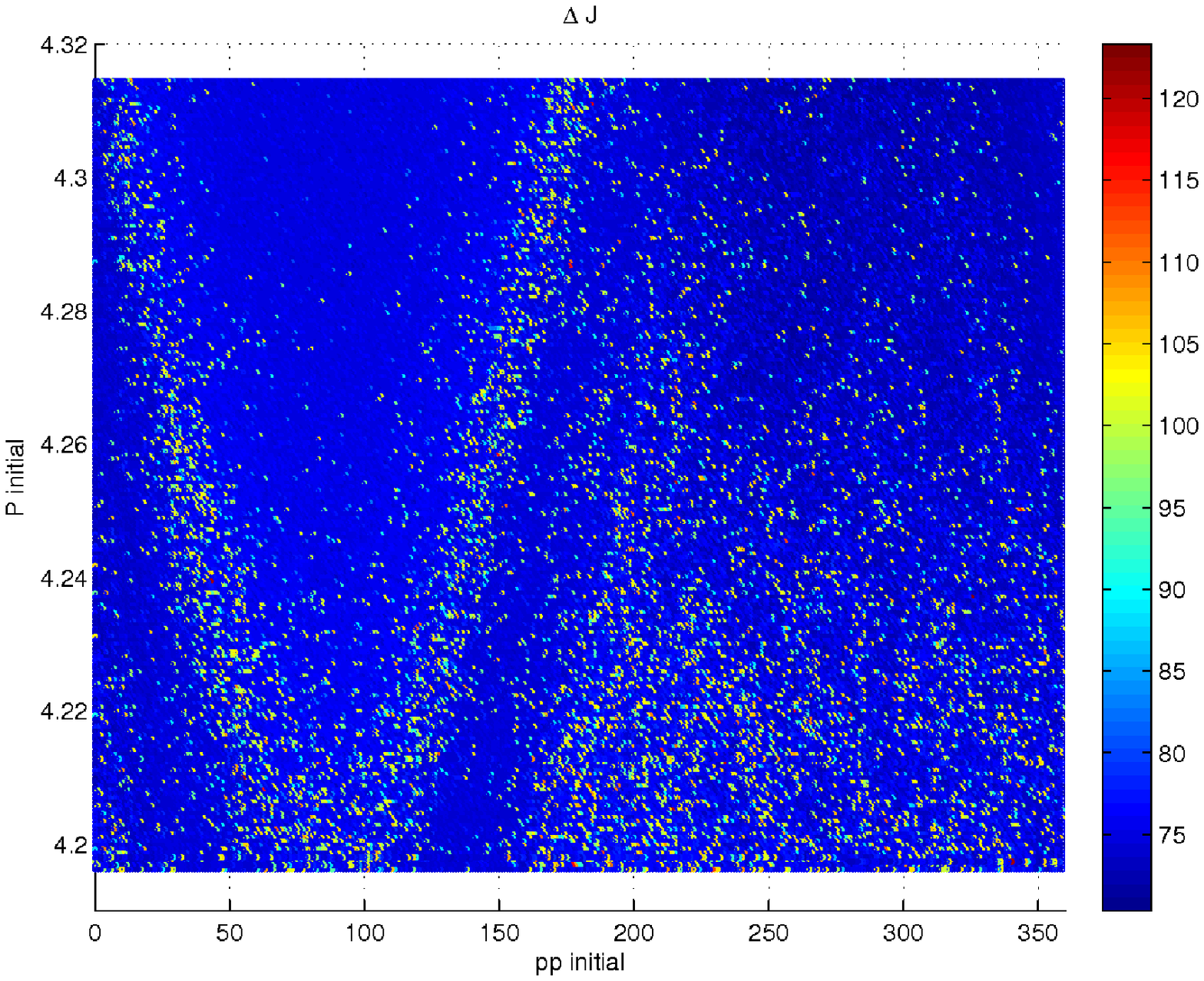}
   \end{tabular}
   \caption[Variations des quantit\'es de rotation pour l'observation 3 et le mod\`ele bas\'e sur la rotation]{Variations de la norme du moment cin\'etique $P$ et de l'amplitude 
   du mouvement polaire $J$ d'Hyp\'erion pour l'observation 3 et le mod\`ele bas\'e sur la rotation. On n'observe rien de bien int\'eressant\ldots\label{fig:toto_l6}}
  \end{figure}

  \par Je ne montre ici que les cartes issues de 2 jeux de conditions initiales / param\`etres (Fig.\ref{fig:toto_l5} \& \ref{fig:toto_l6}), il n'est pas utile d'afficher les 10. 
  Certaines semblent montrer des structures, mais aucune ne montre une possibilit\'e de stabiliser les quantit\'es de rotation. On a notamment un mouvement polaire qui varie de 
  plusieurs dizaines de degr\'es. Je n'ai pas non plus trouv\'e de trajectoire qui ait l'air relativement proche des 5 conditions initiales.
  
  \par Il semble difficile de trouver une solution suffisamment pr\'ecise pour la rotation d'Hyp\'erion, du moins en le consid\'erant comme un corps triaxial. Ceci semble 
  confirmer la chaoticit\'e de ce mouvement. 
  
  \section{Conclusion}
  
  \par Dans ce chapitre, j'ai pr\'esent\'e quelques applications de la rotation rigide. De tels mod\`eles semblent parfaitement adapt\'es aux petits satellites comme Janus 
  ou \'Epim\'eth\'ee. N\'eanmoins, \`a partir d'une certaine taille, le corps se diff\'erencie et peut cr\'eer un noyau fluide ou un oc\'ean global. Dans un tel cas, un mod\`ele
  rigide ne suffit plus. En effet, une couche fluide global d\'ecouple la rotation de la cro\^ute ou du manteau du reste de l'int\'erieur. C'est pourquoi je pr\'esente, dans la 
  suite, les effets d'un noyau fluide (Chap.\ref{chap:poincarehough}), puis d'un oc\'ean global (Chap.\ref{chap:oceanglobal}).

  \chapter{La prise en compte d'un noyau fluide\label{chap:poincarehough}}
  
  \section{Introduction}
  
  \par \`A partir d'une certaine taille, typiquement de l'ordre du millier de kilom\`etres de rayon, les corps c\'elestes ont une structure diff\'erenci\'ee, c'est-\`a-dire 
  que les \'el\'ements les plus lourds comme le fer migrent vers le centre, alors qu'au voisinage de la surface la cro\^ute est un m\'elange de glace et de roches. Sous l'effet 
  de la pression augmentant avec la profondeur, une partie de ce m\'elange glac\'e peut devenir fluide. La pr\'esence d'une couche fluide globale alt\`ere la r\'eponse rotationnelle
  du corps \`a un for\c{c}age gravitationnel, en d\'ecouplant (ou d\'esolidarisant) la cro\^ute solide, ou le manteau, du reste de l'int\'erieur. Dans ce chapitre nous allons 
  mod\'eliser la rotation des corps compos\'es d'une couche externe rigide et d'une couche interne, que nous appellerons noyau, fluide. En pratique cela revient \`a n\'egliger 
  l'influence du noyau solide, \'egalement appel\'e graine, s'il existe.
  
  \par Je vais utiliser le mod\`ele dit de Poincar\'e-Hough \citep{p1910,h1895}, qui consid\`ere un comportement laminaire du fluide, que je pr\'esente (Sec.\ref{sec:poincarehough})
  sous une forme Hamiltonienne inspir\'ee de \citep{tw2001} et revisit\'ee par \citep{h2008}. J'\'etudie ensuite le comportement d'un satellite naturel (Sec.\ref{sec:comportement}), 
  c'est-\`a-dire un corps en rotation synchrone sur une orbite faiblement excentrique, ayant un tel int\'erieur \citep{n2012}, avant de l'appliquer \`a Io \citep{n2013} 
  (Sec.\ref{sec:Io}). Je discute ensuite des \'eventuels effets non lin\'eaires du fluide (Sec.\ref{sec:fluideturbulent}), encore tr\`es mal connus. Un tel mod\`ele d'int\'erieur
  est actuellement utilis\'e, sous une forme non Hamiltonienne, pour inverser les observations Laser-Lune, avec juste l'ajout des forces de mar\'ee \citep{wbyrd2001,rw2011}.
  
  \section{Le mod\`ele de Poincar\'e-Hough\label{sec:poincarehough}}
  
  \par Ce mod\`ele a \'et\'e \'elabor\'e ind\'ependamment par \citet{h1895} et \citet{p1910}. Il consid\`ere le satellite comme un corps triaxial contenant une cavit\'e elle aussi
  triaxiale, remplie d'un fluide non visqueux de champ de vitesse et de densit\'e uniformes. La surface du satellite et l'interface noyau-manteau sont concentriques et coaxiales.
  Le fluide exerce un couple de pression \`a l'interface noyau-manteau.
  
  \subsection{G\'eom\'etrie du probl\`eme}
  
  \par La formulation Hamiltonienne a ceci de frustrant que, si elle est tr\`es bien adapt\'ee pour un traitement math\'ematique pr\'ecis et efficace des probl\`emes, elle n\'ecessite
  un jeu de variables canoniques dont l'interpr\'etation physique n'est pas toujours directe, mais n\'ecessite une transformation math\'ematique bijective pour passer des 
  calculs \`a l'interpr\'etation. Dans notre cas, nous souhaitons mod\'eliser l'orientation du manteau, c'est-\`a-dire de la surface observable, ainsi que du champ de vitesse
  dans le fluide, nous avons donc besoin pour cela de 2 jeux d'angles d'Euler. Mais au lieu de les utiliser pour le manteau et le noyau, l'un de ces jeux se rapportera au
  satellite dans son ensemble, et l'autre \`a un pseudo-noyau, objet virtuel ressemblant math\'ematiquement au noyau, ceci afin d'avoir des variables canoniques. Cette 
  repr\'esentation est celle de \citep{tw2001}, \`a l'origine de la formulation Hamiltonienne du mod\`ele de Poincar\'e-Hough.
  
  \par Il ne s'agit pas d'un mod\`ele \`a 6 degr\'es de libert\'e mais \`a 4. S'il y a bien 3 degr\'es de libert\'e r\'egissant l'orientation du manteau rigide comme dans le cas
  de la rotation rigide (Chap.\ref{chap:rigide}), les contraintes physiques que nous imposons (fluide non visqueux, interface noyau-manteau rigide donc ind\'eformable,\ldots) r\'eduisent
  le mouvement du fluide \`a un seul degr\'e de libert\'e ind\'ependant des autres : l'orientation de son moment cin\'etique. Nous d\'efinissons donc 4 rep\`eres de r\'ef\'erence :
  
    \begin{figure}[ht]
   \centering
   \begin{tabular}{cc}
   \includegraphics[width=0.47\textwidth]{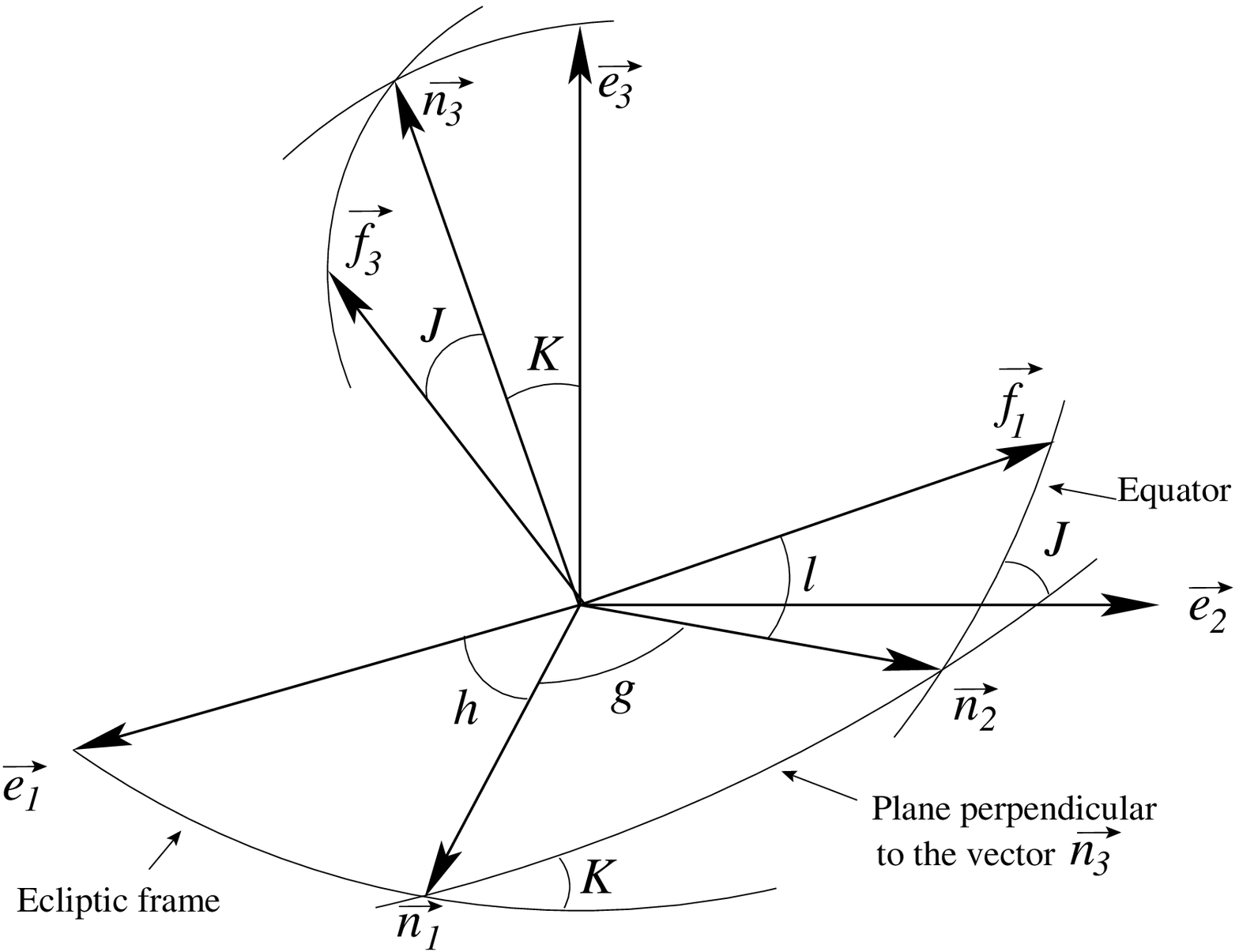} & \includegraphics[width=0.47\textwidth]{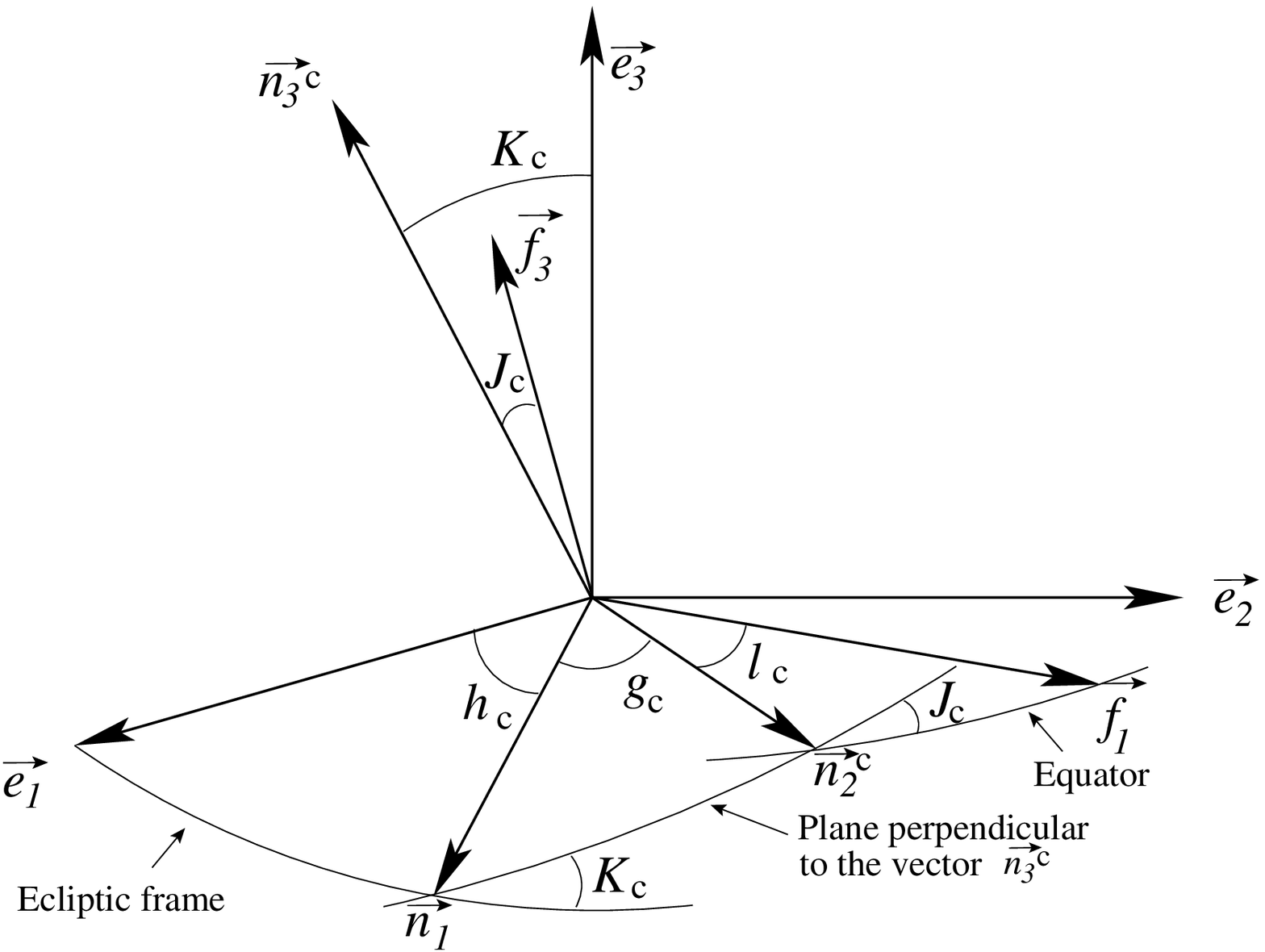}
   \end{tabular}
   \caption[Param\'etrisation de l'orientation du satellite et de son pseudo-noyau]{Param\'etrisation de l'orientation du satellite (gauche) et de son pseudo-noyau (droite). Je remercie Julien Dufey pour avoir r\'ealis\'e cette figure, \`a l'origine
   pour \citep{ndl2010}.\label{fig:2Eulerpoincare}}
  \end{figure}
  
\begin{figure}[ht]
   \centering
   \includegraphics[width=0.55\textwidth]{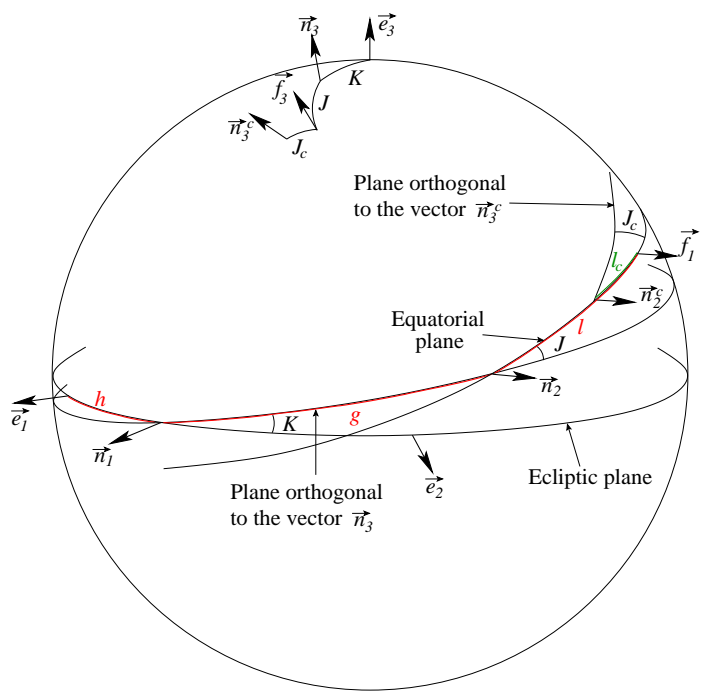}
   \caption[Les 2 param\'etrisations sur la m\^eme figure]{Les 2 param\'etrisations (cf.Fig.\ref{fig:2Eulerpoincare}) sur la m\^eme figure. Encore merci \`a Julien D.\label{fig:2Eulerpoincarebig}}
  \end{figure}

  \begin{itemize}
  
    \item le rep\`ere inertiel ($\vv{e_1}$,$\vv{e_2}$,$\vv{e_3}$), id\'ealement le m\^eme que les \'eph\'em\'erides orbitales. Il est en g\'en\'eral li\'e au plan \'equatorial de
    la plan\`ete \`a la date J2000,
    
    \item le rep\`ere li\'e au moment cin\'etique du satellite ($\vv{n_1}$,$\vv{n_2}$,$\vv{n_3}$). Il est li\'e au rep\`ere inertiel par les angles d'Euler $h$ et $K$, 
    d\'efinis comme dans le cas rigide, c'est-\`a-dire que $h$ est un n{\oe}ud et $K$ une obliquit\'e par rapport \`a la normale au plan inertiel 
    (cf. Fig.\ref{fig:2Eulerpoincare} \& \ref{fig:2Eulerpoincarebig}),
    
    \item le rep\`ere li\'e au moment cin\'etique du pseudo-noyau ($\vv{n_1^c}$,$\vv{n_2^c}$,$\vv{n_3^c}$). Il est li\'e au rep\`ere inertiel par les angles d'Euler $h_c$ et $K_c$, 
    d\'efinis comme pr\'ec\'edemment,
    
    \item le rep\`ere li\'e aux axes principaux d'inertie du satellite, qui sont aussi ceux de la cavit\'e et du manteau par d\'efinition du mod\`ele, 
    ($\vv{f_1}$,$\vv{f_2}$,$\vv{f_3}$). On y passe par le jeu d'angles d'Euler ($g$, $J$, $l$) o\`u $g$ est li\'e au spin du satellite, $J$ est l'amplitude du mouvement polaire, et 
    $l$ l'angle de pr\'ecession associ\'e. On peut aussi d\'efinir analogiquement, pour le pseudo-noyau, les angles $g_c$, $J_c$ et $l_c$. $J_c$ est d\'efini dans le sens oppos\'e 
    de $J$, encore une fois pour des raisons li\'ees \`a la formulation Hamiltonienne.
  
  \end{itemize}

  \par Dans le rep\`ere des axes principaux d'inertie ($\vv{f_1}$,$\vv{f_2}$,$\vv{f_3}$), la matrice d'inertie du satellite $I$ est d\'efinie ainsi :
  
  \begin{equation}
  \label{eq:inertiefluidcore}
  I=\left(\begin{array}{ccc}
    A & 0 & 0 \\
    0 & B & 0 \\
    0 & 0 & C
    \end{array}\right)
  \end{equation}
avec $0<A\leq B \leq C$ d\'efinis par les \'Eq.(\ref{eq:momA}) \`a (\ref{eq:momC}). De m\^eme celle du noyau $I_c$ est d\'efinie par

 \begin{equation}
  \label{eq:inertiecorefluid}
  I_c=\left(\begin{array}{ccc}
    A_c & 0 & 0 \\
    0 & B_c & 0 \\
    0 & 0 & C_c
    \end{array}\right),
  \end{equation}
 avec
 
 \begin{eqnarray}
   A_c & = & \displaystyle\iiint_{noyau}(x_2^2+x_3^2)\rho_c\,\textrm{d}x_1\,\textrm{d}x_2\,\textrm{d}x_3 \label{eq:Ac} \\
       & = & \frac{M_c}{5}(b_c^2+c_c^2), \label{eq:Ac2} \\
   B_c & = & \displaystyle\iiint_{noyau}(x_1^2+x_3^2)\rho_c\,\textrm{d}x_1\,\textrm{d}x_2\,\textrm{d}x_3 \label{eq:Bc} \\
       & = &  \frac{M_c}{5}(a_c^2+c_c^2), \label{eq:Bc2} \\
   C_c & = & \displaystyle\iiint_{noyau}(x_1^2+x_2^2)\rho_c\,\textrm{d}x_1\,\textrm{d}x_2\,\textrm{d}x_3 \label{eq:Cc} \\
       & = &  \frac{M_c}{5}(a_c^2+b_c^2), \label{eq:Cc2}
 \end{eqnarray}
o\`u $a_c$, $b_c$ et $c_c$ sont les 3 rayons de l'interface noyau-manteau, $\rho_c$ est la masse volumique du fluide, et $M_c=4\pi/3\times a_cb_cc_c$ sa masse. Il d\'ecoule directement de 
la d\'efinition des moments d'inertie, qui sont des sommes continues, et du fait que le noyau et le manteau sont coaxiaux, que les 3 moments d'inertie du manteau sont
$A_m=A-A_c$, $B_m=B-B_c$ et $C_m=C-C_c$.
  
  \subsection{L'\'energie cin\'etique}
  
  \par Soit $\vec{v}(x_1,x_2,x_3,t)=v_1\vv{f_1}+v_2\vv{f_2}+v_3\vv{f_3}$ le vecteur vitesse d'une particule de fluide dans le noyau, o\`u ($x_1$, $x_2$, $x_3$) sont
  les coordonn\'ees cart\'esiennes de la particule dans le rep\`ere des axes principaux d'inertie ($\vv{f_1}$,$\vv{f_2}$,$\vv{f_3}$). Une simple \'equation 
  cin\'ematique donne : 
  
  \begin{eqnarray}
    v_1(-,x_2,x_3,t) & = & \left(\omega_2(t)+\frac{a_c}{c_c}\nu_2(t)\Big)x_3-\Big(\omega_3(t)+\frac{a_c}{b_c}\nu_3(t)\right)x_2, \label{eq:v1core} \\
    v_2(x_1,-,x_3,t) & = & \left(\omega_3(t)+\frac{b_c}{a_c}\nu_3(t)\Big)x_1-\Big(\omega_1(t)+\frac{b_c}{c_c}\nu_1(t)\right)x_3, \label{eq:v2core} \\
    v_3(x_1,x_2,-,t) & = & \left(\omega_1(t)+\frac{c_c}{b_c}\nu_1(t)\Big)x_2-\Big(\omega_2(t)+\frac{c_c}{a_c}\nu_2(t)\right)x_1, \label{eq:v3core}
  \end{eqnarray}
o\`u $\vec{\omega}(t)=\omega_1(t)\vv{f_1}+\omega_2(t)\vv{f_2}+\omega_3(t)\vv{f_3}$ est le vecteur-rotation instantan\'e du manteau, et 
$\vec{\nu}(t)=\nu_1(t)\vv{f_1}(t)+\nu_2(t)\vv{f_2}(t)+\nu_3(t)\vv{f_3}(t)$ est le vecteur d\'efinissant le champ de vitesse du fluide par rapport au manteau. Nous 
avons fait l'hypoth\`ese que $\vec{\nu}$ \'etait ind\'ependant des coordonn\'ees spatiales (champ de vitesses uniforme).

\par Le moment cin\'etique du noyau $\vv{N_c'}$ s'\'ecrit

\begin{equation}
\label{eq:intNpc}
\vv{N'_c}=\iiint_{noyau}(\vec{x}\times\vec{v})\rho_c\,dx_1\,dx_2\,dx_3,
\end{equation}
ce qui donne directement

\begin{eqnarray}
\vv{N'_c} & = & \frac{M_c}{5}\left[\left(\frac{c_c}{b_c}\nu_1+\omega_1\right)b_c^2+\left(\frac{b_c}{c_c}\nu_1+\omega_1\right)c_c^2\right]\vec{f_1} \nonumber \\
           & + & \frac{M_c}{5}\left[\left(\frac{c_c}{a_c}\nu_2+\omega_2\right)a_c^2+\left(\frac{a_c}{c_c}\nu_2+\omega_2\right)c_c^2\right]\vec{f_2} \label{eq:Npc} \\
           & + & \frac{M_c}{5}\left[\left(\frac{b_c}{a_c}\nu_3+\omega_3\right)a_c^2+\left(\frac{a_c}{b_c}\nu_3+\omega_3\right)b_c^2\right]\vec{f_3}. \nonumber
\end{eqnarray}

\par En posant maintenant

\begin{eqnarray}
D_1 & = & \frac{2M_c}{5}b_cc_c=\sqrt{\left(A_c-B_c+C_c\right)\left(A_c+B_c-C_c\right)}, \\
D_2 & = & \frac{2M_c}{5}a_cc_c=\sqrt{\left(-A_c+B_c+C_c\right)\left(A_c+B_c-C_c\right)}, \\
D_3 & = & \frac{2M_c}{5}a_cb_c=\sqrt{\left(-A_c+B_c+C_c\right)\left(A_c-B_c+C_c\right)},
\end{eqnarray}
nous pouvons \'ecrire le moment cin\'etique du noyau sous une forme plus compacte :

\begin{equation}
  \label{eq:NpcD}
  \vv{N_c'} = (A_c\omega_1+D_1\nu_1)\vv{f_1}+(B_c\omega_2+D_2\nu_2)\vv{f_2}+(C_c\omega_3+D_3\nu_3)\vv{f_3}.
\end{equation}

\par Comme le moment cin\'etique du manteau s'\'ecrit 

\begin{equation}
  \label{eq:Nm}
  \vv{N_m}=A_m\omega_1\vv{f_1}+B_m\omega_2\vv{f_2}+C_m\omega_3\vv{f_3},
\end{equation}
le moment cin\'etique total du satellite est

\begin{equation}
  \vec{N} = \left(A\omega_1+D_1\nu_1\right)\vv{f_1}+\left(B\omega_2+D_2\nu_2\right)\vv{f_2}+\left(C\omega_3+D_3\nu_3\right)\vv{f_3}.
\end{equation}

\par L'\'energie cin\'etique du noyau est

\begin{equation}
  \label{eq:nrjcore}
  T_c = \iiint_{noyau}\rho_cv^2\,\textrm{d}x_1\,\textrm{d}x_2\,\textrm{d}x_3,
\end{equation}
c'est-\`a-dire

\begin{equation}
  \label{eq:nrjcore2}
  T_c = \frac{1}{2}\left(A_c\left(\omega_1^2+\nu_1^2\right)+B_c\left(\omega_2^2+\nu_2^2\right)+C_c\left(\omega_3^2+\nu_3^2\right)+2D_1\omega_1\nu_1+2D_2\omega_2\nu_2+2D_3\omega_3\nu_3\right),
\end{equation}
alors que l'\'energie cin\'etique du manteau $T_m$ est

\begin{equation}
  \label{eq:nrjmanteau}
  T_m = \frac{1}{2}\vv{N_m}\cdot\vec{\omega} = \frac{A_m\omega_1^2+B_m\omega_2^2+C_m\omega_3^2}{2}.
\end{equation}
Nous pouvons ainsi d\'eduire facilement l'\'energie cin\'etique totale du satellite $T=T_m+T_c$

\begin{equation}
  \label{eq:nrjtotale}
  T = \frac{1}{2}\left(A\omega_1^2+B\omega_2^2+C\omega_3^2+A_c\nu_1^2+B_c\nu_2^2+C_c\nu_3^2+2D_1\omega_1\nu_1+2D_2\omega_2\nu_2+2D_3\omega_3\nu_3\right).
\end{equation}

\par Afin d'appr\'ehender le formalisme Hamiltonien, nous pouvons \'ecrire les d\'eriv\'ees partielles de l'\'energie cin\'etique totale $T$:

\begin{eqnarray}
  \frac{\partial T}{\partial \omega_1} & = & A\omega_1+D_1\nu_1=N_1, \label{eq:dTdomeg1} \\
  \frac{\partial T}{\partial \omega_2} & = & B\omega_2+D_2\nu_2=N_2, \label{eq:dTdomeg2} \\
  \frac{\partial T}{\partial \omega_3} & = & C\omega_3+D_3\nu_3=N_3, \label{eq:dTdomeg3} \\
  \frac{\partial T}{\partial \nu_1}    & = & D_1\omega_1+A_c\nu_1=N_1^c, \label{eq:dTdnu1} \\
  \frac{\partial T}{\partial \nu_2}    & = & D_2\omega_2+B_c\nu_2=N_2^c, \label{eq:dTdnu2} \\
  \frac{\partial T}{\partial \nu_3}    & = & D_3\omega_3+C_c\nu_3=N_3^c, \label{eq:dTdnu3}
\end{eqnarray}
o\`u les $N_i$ sont les composantes du moment cin\'etique total $\vec{N}$. Les $N_i^c$ sont proches des composantes du moment cin\'etique du noyau, on a en fait :

\begin{eqnarray}
  N_1-N_1^{'c} & = & (A_c-D_1)(\omega_1-\nu_1) = \frac{M_c}{5}(c_c-b_c)^2(\omega_1-\nu_1), \label{eq:N1psc} \\
  N_2-N_2^{'c} & = & (B_c-D_2)(\omega_2-\nu_2) = \frac{M_c}{5}(c_c-a_c)^2(\omega_2-\nu_2), \label{eq:N2psc} \\
  N_3-N_3^{'c} & = & (C_c-D_3)(\omega_3-\nu_3) = \frac{M_c}{5}(a_c-b_c)^2(\omega_3-\nu_3), \label{eq:N3psc}
\end{eqnarray}
la diff\'erence est donc d'ordre 2 en l'asph\'ericit\'e de la cavit\'e. Appelons moment cin\'etique du pseudo-noyau la quantit\'e $\vv{N^c}=N_1^c\vv{e_1}+N_2^c\vv{e_2}+N_3^c\vv{e_3}$.

\par Avec ces notations, les \'equations de Poincar\'e-Hough s'\'ecrivent, en l'absence de for\c{c}age externe \citep{tw2001} : 

\begin{eqnarray}
  \frac{d\vec{N}}{dt}  & = & \vec{N}  \times \vv{\nabla}_{\vec{N}}\mathcal{T}, \label{eq:dNsdt} \\
  \frac{d\vv{N^c}}{dt} & = & \vv{N^c} \times \vv{\nabla}_{-\vv{N^c}}\mathcal{T}, \label{eq:dNcsdt}
\end{eqnarray}
avec

\begin{eqnarray}
  \vv{\nabla}_{\vec{N}}\mathcal{T} & = & \frac{\partial\mathcal{T}}{\partial N_1}\vv{e_1}+\frac{\partial\mathcal{T}}{\partial N_2}\vv{e_2}+\frac{\partial\mathcal{T}}{\partial N_3}\vv{e_3}, \label{eq:nablan} \\
  \vv{\nabla}_{-\vv{N^c}}\mathcal{T} & = & -\frac{\partial\mathcal{T}}{\partial N_1^c}\vv{e_1}-\frac{\partial\mathcal{T}}{\partial N_2^c}\vv{e_2}-\frac{\partial\mathcal{T}}{\partial N_3^c}\vv{e_3}. \label{eq:nablamnc}
\end{eqnarray}

\par Ici $\mathcal{T}$ d\'esigne l'\'energie cin\'etique du satellite exprim\'ee dans les composantes des moments cin\'etiques $\vec{N}$ et $\vv{N^c}$, c'est-\`a-dire

\begin{eqnarray}
\mathcal{T} & = & \frac{1}{2\alpha}\left(A_cN_1^2+A(N_1^c)^2-2D_1N_1N_1^c\right)+\frac{1}{2\beta}\left(B_cN_2^2+B(N_2^c)^2-2D_2N_2N_2^c\right) \nonumber \\
            & + & \frac{1}{2\gamma}\left(C_cN_3^2+C(N_3^c)^2-2D_3N_3N_3^c\right) \label{eq:Thamilton},
\end{eqnarray}
avec $\alpha=AA_c-D_1^2$, $\beta=BB_c-D_2^2$ et $\gamma=CC_c-D_3^2$.

  \subsection{Le Hamiltonien du syst\`eme}

  \par Il est \`a ce stade commode d'introduire les param\`etres d'asph\'ericit\'e du satellite et du noyau : 
  
  \begin{eqnarray}
    \epsilon_1 & = & \frac{2C-A-B}{2C} = J_2\frac{MR^2}{C}, \label{eq:epsiloncm1} \\
    \epsilon_2 & = & \frac{B-A}{2C} = 2C_{22}\frac{MR^2}{C}, \label{eq:epsiloncm2} \\
    \epsilon_3 & = & \frac{2C_c-A_c-B_c}{2C_c}, \label{eq:epsiloncm3} \\
    \epsilon_4 & = & \frac{B_c-A_c}{2C_c}. \label{eq:epsiloncm4}
  \end{eqnarray}

  La surface du satellite et l'interface noyau-manteau \'etant en g\'en\'eral proches de la sph\`ere, les $\epsilon_i$ sont petits.
  
  \subsubsection{Le Hamiltonien de libre rotation}
  
  \par Introduisons maintenant 2 jeux d'\'el\'ements d'Andoyer \citep{a1926} ($l$,$g$,$h$,$L$,$G$,$H$) pour le satellite, et ($l_c$,$g_c$,$h_c$,$L_c$,$G_c$,$H_c$)
  pour le pseudo-noyau. Les angles d'Euler $l$, $g$, $h$, $l_c$, $g_c$, $h_c$ ont d\'ej\`a \'et\'e d\'efinis. On a de plus, de la m\^eme fa\c{c}on que dans le chapitre \ref{chap:rigide} :
  $G=N$, $G^c=N^c$ (les normes des 2 moments cin\'etiques), $H=G\cos K$, $H_c=G_c\cos K_c$, $L=G\cos J$, $L_c=G_c\cos J_c$, o\`u $K$ et $K^c$ sont les obliquit\'es mesur\'ees
  par rapport au vecteur inertiel $\vv{e_3}$, et $J$ et $J_c$ sont les amplitudes des mouvements polaires. Les composantes des moments cin\'etiques s'\'ecrivent :
  
  \begin{center}
  $\begin{array}{lll}
  N_1=\sqrt{G^2-L^2}\sin l, & \hspace{2cm} & N_1^c=\sqrt{G_c^2-L_c^2}\sin l_c, \\
  N_2=\sqrt{G^2-L^2}\cos l, & \hspace{2cm} & N_2^c=\sqrt{G_c^2-L_c^2}\cos l_c, \\
  N_3=L, & \hspace{2cm} & N_3^c=L_c. \\
  \end{array}$
  \end{center}

  \par Le Hamiltonien de libre rotation $\mathcal{H}_0$ de notre satellite est son \'energie cin\'etique \'ecrite dans un jeu de variables canoniques. Nous avons, dans
  les variables d'Andoyer, et apr\`es lin\'earisation par rapport aux param\`etres de forme $\epsilon_i$ :
  
  \begin{eqnarray}
   \mathcal{H}_0 & = & \frac{1}{2C(1-\delta)}\left(G^2+\frac{G_c^2}{\delta}+2\sqrt{(G^2-L^2)(G_c^2-L_c^2)}\cos(l-l_c)+2LL_c\right) \nonumber \\
                 & + & \frac{\epsilon_1}{2C(1-\delta)^2}\left(G^2-L^2+G_c^2-L_c^2+2\sqrt{(G^2-L^2)(G_c^2-L_c^2)}\cos(l-l_c)\right) \nonumber \\
                 & - & \frac{\epsilon_2}{2C(1-\delta)^2}\left((G^2-L^2)\cos(2l)+(G_c^2-L_c^2)\cos(2l_c)+2\sqrt{(G^2-L^2)(G_c^2-L_c^2)}\cos(l+l_c)\right) \nonumber \\
                 & - & \frac{\epsilon_3}{2C(1-\delta)^2}\left(\delta (G^2-L^2)+(G_c^2-L_c^2)(2-\frac{1}{\delta})+2\delta\sqrt{(G^2-L^2)(G_c^2-L_c^2)}\cos(l-l_c)\right) \nonumber \\
                 & + & \frac{\epsilon_4}{2C(1-\delta)^2}\times \label{eq:HG} \\
                 &   & \left(\delta(G^2-L^2)\cos(2l)+(G_c^2-L_c^2)(2-\frac{1}{\delta})\cos(2l_c)+2\delta\sqrt{(G^2-L^2)(G_c^2-L_c^2)}\cos(l+l_c)\right), \nonumber
  \end{eqnarray}
avec $\delta=C_c/C$. Ce param\`etre peut \^etre vu comme l'inertie relative du noyau.

\par Nous introduisons maintenant le changement de variables canoniques suivant, de multiplicit\'e $1/(nC)$, proche de celui que nous utilisions pour la rotation rigide 
(Eq.\ref{eq:andoyermodif}) :

  \begin{equation}\label{eq:chvar}
  \begin{array}{lll}
   p=l+g+h, & \hspace{2cm} & P=\frac{G}{nC}, \\
   r=-h, & \hspace{2cm} & R=P(1-\cos K), \\
   \xi_1=-\sqrt{2P(1-\cos J)}\sin l, & \hspace{2cm} & \eta_1=\sqrt{2P(1-\cos J)}\cos l, \\
   p_c=-l_c+g_c+h_c, & \hspace{2cm} & P_c=\frac{G_c}{nC}, \\
   r_c=-h_c, & \hspace{2cm} & R_c=P_c(1-\cos K_c), \\
   \xi_2=\sqrt{2P_c(1+\cos J_c)}\sin l_c, & \hspace{2cm} & \eta_2=\sqrt{2P_c(1+\cos J_c)}\cos l_c. \\
  \end{array} \\
  \end{equation}
  
  \par Ce changement de variables a l'avantage de ne pas pr\'esenter de singularit\'es pour de petites obliquit\'es et de petits mouvements polaires. Comme le pr\'ec\'edent,
  c'est un jeu de variables \`a 6 degr\'es de libert\'e, alors que notre syst\`eme n'en contient que 4. Cette \'ecriture permet une uniformisation de traitement entre le noyau 
  et le manteau, mais les 2 degr\'es inutiles ($p_c$/$P_c$ et $r_c$/$R_c$) seront bien identifi\'es et supprim\'es dans le Hamiltonien final.
  
  \par Afin d'\^etre coh\'erent avec le signe moins dans les \'equations (\ref{eq:dNsdt})-(\ref{eq:dNcsdt}) et devant $l_c$, $J_c$ doit \^etre remplac\'e par $\pi-Jc$. Les 
  composantes des moments cin\'etiques s'\'ecrivent donc maintenant
  
  \begin{center}
  $\begin{array}{lll}
  N_1=-nC\sqrt{P^2-\Big(P-\frac{\xi_1^2+\eta_1^2}{2}\Big)^2}\frac{\xi_1}{\xi_1^2+\eta_1^2}, & \hspace{0.6cm} & N_1^c=nC\sqrt{P_c^2-\Big(\frac{\xi_2^2+\eta_2^2}{2}-P_c\Big)^2}\frac{\xi_2}{\xi_2^2+\eta_2^2}, \\
  N_2=nC\sqrt{P^2-\Big(P-\frac{\xi_1^2+\eta_1^2}{2}\Big)^2}\frac{\eta_1}{\xi_1^2+\eta_1^2}, & \hspace{0.6cm} & N_2^c=nC\sqrt{P_c^2-\Big(\frac{\xi_2^2+\eta_2^2}{2}-P_c\Big)^2}\frac{\eta_2}{\xi_2^2+\eta_2^2}, \\
  N_3=nC\Big(P-\frac{\xi_1^2+\eta_1^2}{2}\Big), & \hspace{0.6cm} & N_3^c=nC\Big(\frac{\xi_2^2+\eta_2^2}{2}-Pc\Big), \\
  \end{array}$ \\
  \end{center}
  et le Hamiltonien de libre rotation, devenu $\mathcal{H}_1$, est, apr\`es division par $nC$ :

  \begin{eqnarray}
    \mathcal{H}_1 & = & \frac{n}{2(1-\delta)}\Bigg(P^2+\frac{P_c^2}{\delta}+2\sqrt{\Big(P-\frac{\xi_1^2+\eta_1^2}{4}\Big)\Big(P_c-\frac{\xi_2^2+\eta_2^2}{4}\Big)}\big(\eta_1\eta_2-\xi_1\xi_2\big) \nonumber \\
                  & & +2\Big(P-\frac{\xi_1^2+\eta_1^2}{2}\Big)\Big(\frac{\xi_2^2+\eta_2^2}{2}-P_c\Big)\Bigg) \nonumber \\
                  & + &\frac{n\epsilon_1}{2(1-\delta)^2}\Bigg(P_c^2-\Big(\frac{\xi_2^2+\eta_2^2}{2}-P_c\Big)^2+P^2-\Big(P-\frac{\xi_1^2+\eta_1^2}{2}\Big)^2 \nonumber \\
                  & + & 2\sqrt{\Big(P-\frac{\xi_1^2+\eta_1^2}{4}\Big)\Big(P_c-\frac{\xi_2^2+\eta_2^2}{4}\Big)}\big(\eta_1\eta_2-\xi_1\xi_2\big)\Bigg) \nonumber \\
                  & + & \frac{n\epsilon_2}{2(1-\delta)^2}\Bigg(\frac{1}{4}\big(4P-\xi_1^2-\eta_1^2\big)\big(\xi_1^2-\eta_1^2\big)+\frac{1}{4}\big(4P_c-\xi_2^2-\eta_2^2\big)\big(\xi_2^2-\eta_2^2\big) \nonumber \\
                  & - & 2\sqrt{\Big(P-\frac{\xi_1^2+\eta_1^2}{4}\Big)\Big(P_c-\frac{\xi_2^2+\eta_2^2}{4}\Big)}\big(\eta_1\eta_2+\xi_1\xi_2\big)\Bigg) \nonumber \\
                  & - & \frac{n\epsilon_3}{2(1-\delta)^2}\Bigg(\delta \Big(P^2-\Big(P-\frac{\xi_1^2+\eta_1^2}{2}\Big)^2\Big)+\Big(P_c^2-(\frac{\xi_2^2+\eta_2^2}{2}-P_c\Big)^2\Big)\Big(2-\frac{1}{\delta}\Big) \nonumber \\
                  & + & 2\delta\sqrt{\Big(P-\frac{\xi_1^2+\eta_1^2}{4}\Big)\Big(P_c-\frac{\xi_2^2+\eta_2^2}{4}\Big)}\big(\eta_1\eta_2-\xi_1\xi_2\big)\Bigg) \nonumber \\
                  & + &\frac{n\epsilon_4}{2(1-\delta)^2}\Bigg(\frac{\delta}{4}\big(4P-\xi_1^2-\eta_1^2\big)\big(\eta_1^2-\xi_1^2\big)+\Big(2-\frac{1}{\delta}\Big) \frac{1}{4}\big(4P_c-\xi_2^2-\eta_2^2\big)\big(\eta_2^2-\xi_2^2\big)  \nonumber \\
                  & + & 2\delta\sqrt{\Big(P-\frac{\xi_1^2+\eta_1^2}{4}\Big)\Big(P_c-\frac{\xi_2^2+\eta_2^2}{4}\Big)}\big(\eta_1\eta_2+\xi_1\xi_2\big)\Bigg). \label{eq:HG3}
  \end{eqnarray}

  \par Afin d'obtenir une formule plus lisible, on peut d\'evelopper ce Hamiltonien \`a l'ordre 2 en ($\xi_1$,$\xi_2$,$\eta_1$,$\eta_2$) :
  
  \begin{eqnarray}
   \mathcal{H}_1 & \approx & \frac{n}{2(1-\delta)}\left(P^2+\frac{P_c^2}{\delta}+2\sqrt{PP_c}\left(\eta_1\eta_2-\xi_1\xi_2\right)+2\left(P\frac{\xi_2^2+\eta_2^2}{2}+P_c\frac{\xi_1^2+\eta_1^2}{2}-PP_c\right)\right) \nonumber \\
                 & +       & \frac{n\epsilon_1}{2(1-\delta)^2}\left(P\left(\xi_1^2+\eta_1^2\right)+P_c\left(\xi_2^2+\eta_2^2\right)+2\sqrt{PP_c}\left(\eta_1\eta_2-\xi_1\xi_2\right)\right) \nonumber \\
                 & +       & \frac{n\epsilon_2}{2(1-\delta)^2}\left(P\left(\xi_1^2-\eta_1^2\right)+P_c\left(\xi_2^2-\eta_2^2\right)-2\sqrt{PP_c}\left(\eta_1\eta_2+\xi_1\xi_2\right)\right) \label{eq:HG4} \\
                 & -       & \frac{n\epsilon_3}{2(1-\delta)^2}\left(\delta P\left(\xi_1^2+\eta_1^2\right)+\left(2-\frac{1}{\delta}\right)P_c \left(\xi_2^2+\eta_2^2\right)+2\delta\sqrt{PP_c}\left(\eta_1\eta_2-\xi_1\xi_2\right)\right) \nonumber \\
                 & +       & \frac{n\epsilon_4}{2(1-\delta)^2}\left(\delta P\left(\eta_1^2-\xi_1^2\right)+\left(2-\frac{1}{\delta}\right)P_c\left(\eta_2^2-\xi_2^2\right)+2\delta\sqrt{PP_c}\left(\eta_1\eta_2+\xi_1\xi_2\right)\right). \nonumber
  \end{eqnarray}
Il s'agit en fait d'un d\'eveloppement \`a l'ordre 3, le terme de degr\'e 3 \'etant nul.

\par Nous avons maintenant le Hamiltonien de libre rotation, il nous faut exprimer le potentiel gravitationnel de la plan\`ete parente.

  \subsubsection{Le potentiel gravitationnel du perturbateur}
  
  \par Le perturbateur est la plan\`ete parente du satellite (Jupiter pour Io), qui lui applique un couple gravitationnel. Sans surprise, il s'agit du m\^eme couple que dans 
  le cas rigide (Eq.\ref{eq:expanspherik4}). Le potentiel perturbateur s'\'ecrit donc, avec nos notations :
  
  \begin{equation}
    \label{eq:potperturbliquidcore}
    V(l,g,h,J,K,t) = -\frac{3}{2}C\frac{\mathcal{G}M_p}{r^3}\left(\epsilon_1\left(\hat{x}_S^2+\hat{y}_S^2\right)+\epsilon_2\left(\hat{x}_S^2-\hat{y}_S^2\right)\right),
  \end{equation}
o\`u, comme pr\'ec\'edemment, ($\hat{x}_S$, $\hat{y}_S$, $\hat{z}_S$) sont les coordonn\'ees du vecteur unitaire pointant vers le perturbateur, dans le rep\`ere li\'e aux
axes principaux d'inertie du satellite ($\vv{f_1}$, $\vv{f_2}$, $\vv{f_3}$). On en d\'eduit la forme Hamiltonienne, impliquant l'utilisation des variables canoniques et la 
division par $nC$ :

  \begin{equation}
    \label{eq:hamilpertliquidcore}
    \mathcal{H}_2(p,P,r,R,\xi_1,\eta_1,t) = -\frac{3}{2}\frac{\mathcal{G}M_p}{nr^3}\left(\epsilon_1\left(\hat{x}_S^2+\hat{y}_S^2\right)+\epsilon_2\left(\hat{x}_S^2-\hat{y}_S^2\right)\right).
  \end{equation}

  \par On en d\'eduit le Hamiltonien total du syst\`eme $\mathcal{H}$ \`a l'aide des formules (\ref{eq:HG3}) et (\ref{eq:hamilpertliquidcore}) :
  
  \begin{eqnarray}
    \mathcal{H}(-,P_c,-,-,\xi_2,\eta_2,p,P,r,R,\xi_1,\eta_1,t) & = & \mathcal{H}_1(-,P_c,-,-,\xi_2,\eta_2,-,P,-,-,\xi_1,\eta_1,-) \nonumber \\
                                                               & + & \mathcal{H}_2(-,-,-,-,-,-,p,P,r,R,\xi_1,\eta_1,t). \label{eq:htotalliquidcore}
  \end{eqnarray}

  \subsection{Seulement 4 degr\'es de libert\'e}
  
  \par Notre satellite est compos\'e de 2 couches, un manteau rigide et un noyau fluide, il y a donc a priori 6 degr\'es de libert\'e, \`a savoir 3 pour orienter 
  le manteau et 3 pour orienter le noyau. Par abus de langage, je passe ici sous silence la d\'ependance en temps, pr\'esente dans la position du perturbateur.
  
  \par Le jeu de variables li\'e \`a l'obliquit\'e du champ de vitesses dans le fluide, ($r_c$, $R_c$) n'est pas pr\'esent dans le Hamiltonien du probl\`eme, on peut donc 
  l'exclure comme degr\'e de libert\'e. La raison est que la contrainte que la surface du satellite et l'interface noyau-manteau soient coaxiaux donne une liaison entre 
  les obliquit\'es de chacune des 2 couches et leurs mouvements polaires. Ils repr\'esentent donc, \`a eux 4, 3 degr\'es de libert\'e ind\'ependants.
  
  \par La norme du moment cin\'etique normalis\'e du noyau, $P_c$, appara\^it, mais pas la variable associ\'ee $p_c$. $P_c$ est donc une constante. Nous poserons
  
  \begin{equation}
    \label{eq:Pcdelta}
    P_c=\delta=C_c/C,
  \end{equation}
ce qui signifie que la vitesse angulaire du fluide est en moyenne celle du manteau. Cela signifie physiquement que le fluide, visqueux, r\'epond aux sollicitations lentes 
mais pas aux rapides, comme les librations diurnales. Nous verrons plus tard que cette hypoth\`ese est \`a la base de l'exp\'erience de Peale \citep{p1976}, visant \`a d\'etecter la 
pr\'esence d'un noyau fluide pour Mercure. Les r\'ecentes observations de la rotation de Mercure \citep{mpjsh2007,mpshgjygpc2012} tendent \`a confirmer la validit\'e de cette 
exp\'erience, donc de l'hypoth\`ese de base.
  
  \subsection{Retrouver les quantit\'es physiques \citep{ndl2010}}
  
  \par Je donne ici un formulaire permettant de retrouver les quantit\'es de rotation ayant un sens physique \`a partir des variables canoniques. Il est extrait d'un article sur 
  Mercure, en r\'esonance spin-orbite 3:2. Ces formules n'utilisant pas la r\'esonance $3:2$, elles s'appliquent \'egalement aux corps en rotation synchrone.
  
  \par Les \'equations (\ref{eq:dTdomeg1}) \`a (\ref{eq:dTdnu3}) donnent :
  
  \begin{eqnarray}
    \omega_1 & = & \frac{D_1N_1^c+A_cN_1}{D_1^2-AA_c}, \label{eq:omega1} \\
    \omega_2 & = & \frac{D_2N_2^c+B_cN_2}{D_2^2-BB_c}, \label{eq:omega2} \\
    \omega_3 & = & \frac{D_3N_3^c+C_cN_3}{D_3^2-CC_c}, \label{eq:omega3} \\
    \nu_1    & = & \frac{D_1N_1+AN_1^c}{D_1^2-AA_c}, \label{eq:nu1} \\
    \nu_2    & = & \frac{D_2N_2+BN_2^c}{D_2^2-BB_c}, \label{eq:nu2} \\
    \nu_3    & = & \frac{D_3N_3+CN_3^c}{D_3^2-CC_c}, \label{eq:nu3}
  \end{eqnarray}
ce qui permet d'\'ecrire

\begin{eqnarray}
  \vv{N_m}  & = & A_m\frac{D_1N_1^c-A_cN_1}{D_1^2-AA_c}\vv{f_1}+B_m\frac{D_2N_2^c-B_cN_2}{D_2^2-BB_c}\vv{f_2}+C_m\frac{D_3N_3^c-C_cN_3}{D_3^2-CC_c}\vv{f_3}, \label{eq:Nmformulaire} \\
  \vv{N_c'} & = & \left(A_c\frac{D_1N_1^c-A_cN_1}{D_1^2-AA_c}+D_1\frac{D_1N_1+AN_1^c}{D_1^2-AA_c}\right)\vv{f_1}+
  \left(B_c\frac{D_2N_2^c-B_cN_2}{D_2^2-BB_c}+D_2\frac{D_2N_2+BN_2^c}{D_2^2-BB_c}\right)\vv{f_2} \nonumber \\
            & + & \left(C_c\frac{D_3N_3^c-C_cN_3}{D_3^2-CC_c}+D_3\frac{D_3N_3+CN_3^c}{D_3^2-CC_c}\right)\vv{f_3}. \label{eq:Npcformulaire}
\end{eqnarray}

\par Par d\'efinition de la norme du moment cin\'etique $G$, de l'amplitude du mouvement polaire $J$ et de l'angle de pr\'ecession $l$, on a

\begin{eqnarray}
  \vv{N_m}   & = & G_m\sin J_m\sin l_m\vv{f_1}+G_m\sin J_m\cos l_m\vv{f_2}+G_m\cos J_m\vv{f_3}, \label{eq:Nmdefinition} \\
  \vv{N_c'} & = & G_c'\sin J_c'\sin l_c'\vv{f_1}+G_c'\sin J_c'\cos l_c'\vv{f_2}+G_c'\cos J_c'\vv{f_3}, \label{eq:Npcdefinition}
\end{eqnarray}
ce qui donne les variables $G_m$, $J_m$, $l_m$, $G_c'$, $J_c'$ et $l_c'$ par identification.

\par La d\'etermination des angles d'Euler li\'es \`a l'obliquit\'e et au spin du manteau $h_m$, $K_m$ et $g_m$ est un peu plus technique, elle n\'ecessite de
consid\'erer les rotations entre les diff\'erents rep\`eres. En posant

\begin{eqnarray}
  \vv{N_m}  & = & T_1\vv{e_1}+T_2\vv{e_2}+T_3\vv{e_3}, \label{eq:NmT123} \\
  \vv{N_c'} & = & T_1'^c\vv{e_1}+T_2'^c\vv{e_2}+T_3'^c\vv{e_3}, \label{eq:NpcT123}
\end{eqnarray}
on a

\begin{equation}
\label{eq:passage}
\left(\begin{array}{c}
T_1 \\
T_2 \\
T_3
\end{array}\right)
=R_3(h_m)R_1(K_m)R_3(g_m)\left(\begin{array}{c}
0 \\
0 \\
G^m
\end{array}\right)
\end{equation}
ce qui donne :

\begin{eqnarray}
T_1 &=& G^m\sin(K_m)\sin(h_m), \label{eq:t1cm} \\
T_2 &=& -G^m\sin(K_m)\cos(h_m), \label{eq:t2cm} \\
T_3 &=& G^m\cos(K_m). \label{eq:t3cm}
\end{eqnarray}

\par On a de plus :

\begin{equation}
\label{eq:passage2}
\left(\begin{array}{c}
T_1'^c \\
T_2'^c \\
T_3'^c
\end{array}\right)
=R_3(h)R_1(K)R_3(g)R_1(J)R_3(l)\left(\begin{array}{c}
N_1'^c \\
N_2'^c \\
N_3'^c
\end{array}\right),
\end{equation}
ce qui permet d'obtenir, en se rappelant que $\vec{N}=\vv{N_m}+\vv{N_c'}$ :

\begin{eqnarray}
G^m\sin(K_m)\sin(h_m)&=&G\sin(K)\sin(h)-T_1'^c, \label{eq:km1} \\
G^m\sin(K_m)\cos(h_m)&=&G\sin(K)\cos(h)+T_2'^c, \label{eq:km2} \\
G^m\cos(K_m)&=&G\cos(K)-T_3'^c, \label{eq:km3}
\end{eqnarray}
et finalement

\begin{eqnarray}
K_m & = & \arccos\left(\frac{G\cos(K)-T_3'^c}{G^m}\right), \label{eq:Km} \\
h_m & = & \arctan\left(\frac{G\sin(K)\sin(h)-T_1'^c}{G\sin(K)\cos(h)+T_2'^c}\right). \label{eq:hm}
\end{eqnarray}

  \section{Comportement d'un satellite naturel\label{sec:comportement}}
  
  \par Dans cette section je pr\'esente le comportement attendu d'un satellite en rotation synchrone assujetti au mod\`ele de Poincar\'e-Hough. Il s'agit d'une \'etude 
  num\'erique que j'ai r\'ealis\'ee \citep{n2012} \`a la suite d'un travail analytique de Jacques Henrard \citep{h2008}. Il s'agissait, \`a ma connaissance, de la 
  premi\`ere application du mod\`ele de Poincar\'e-Hough en formulation hamiltonienne \`a un corps en r\'esonance spin-orbite.
  
  \subsection{\'Etude analytique \citep{h2008}}
  
  \par Jacques voulait simuler la rotation d'Io \`a l'aide d'une m\'ethode de perturbations et d'\'eph\'em\'erides sous une forme synth\'etique, c'est-\`a-dire 
  quasi-p\'eriodique, avec des amplitudes num\'eriques et des angles analytiques, d\'ependant explicitement des modes propres du syst\`eme orbital. Il utilisait 
  directement des coordonn\'ees cart\'esiennes, calcul\'ees par Val\'ery Lainey \`a partir des \'el\'ements orbitaux de sa th\'eorie L1.2 \citep{ldv2006}.
  
  \par Cette \'etude mettait en lumi\`ere le fait que le syst\`eme \'etait \`a 4 degr\'es de libert\'e, et montrait que les 2 degr\'es de libert\'e li\'es aux mouvements polaires,
  du satellite et du pseudo-noyau, \'etaient coupl\'es de la m\^eme fa\c{c}on que le sont la longitude et l'obliquit\'e. Par contre, 2 modifications \'etaient faites par rapport 
  \`a ce que j'ai pr\'esent\'e plus haut :
  
  \begin{itemize}
  
    \item Le champ de gravit\'e du fluide \'etait proportionnel \`a celui du satellite total, ce qui donnait $\epsilon_1=\epsilon_3$ et $\epsilon_2=\epsilon_4$. 
    La complexification de l'int\'erieur fait intervenir de plus en plus de param\`etres, il est donc n\'ecessaire de simplifier les choses. Une autre possibilit\'e 
    d'introduire une liaison entre les param\`etres d'int\'erieur est de faire l'hypoth\`ese de l'\'equilibre hydrostatique.
    
    \item Jacques consid\'erait que le couple du perturbateur ne s'appliquait que sur le manteau. C'est malheureusement une erreur\footnote{initialement relev\'ee 
    par Rose-Marie Baland, alors qu'elle commen\c{c}ait sa th\`ese de doctorat \`a l'Observatoire Royal de Belgique}, qui a pour cons\'equence que la fr\'equence des 
    librations en longitude, ainsi que l'amplitude des librations diurnes, ne semblent pas \^etre affect\'ees par le noyau.
    
  \end{itemize}

  \par J.~Henrard est d\'ec\'ed\'e en mars 2008 pendant le processus de \emph{reviewing} de son article, et ne l'a pas vu para\^itre.

  \subsection{\'Etude num\'erique \citep{n2012}}
  
  \par J'ai pour ma part r\'ealis\'e une \'etude num\'erique du comportement d'un tel satellite. J'ai consid\'er\'e un satellite fictif, largement inspir\'e d'Io 
  (Tab.~\ref{tab:pseudoIo}). Son orbite est une ellipse \`a excentricit\'e mod\'er\'ee, uniform\'ememt pr\'ecessante sur un plan lui-m\^eme pr\'ecessant uniform\'ement. 
  Son demi-grand axe $a$, son excentricit\'e $e$ et son inclinaison par rapport \`a l'\'equateur de la plan\`ete \`a J2000 $I$ sont constantes. On n'a donc pas, dans ce 
  cas fictif, les perturbations gravitationnelles des autres satellites de la plan\`ete parente.
  
  \begin{table}[ht]
   \centering
    \caption[Param\`etres physiques et dynamiques de mon pseudo-Io]{Param\`etres physiques et dynamiques de mon pseudo-Io. Le moyen mouvement moyen $n$
    ainsi que les vitesses de pr\'ecession du n{\oe}ud $\dot{\ascnode}$ et du p\'ericentre $\dot{\varpi}$ sont issus des \'eph\'em\'erides L1.2 \citep{ldv2006}. 
    Les phases, c'est-\`a-dire les valeurs initiales des angles, sont choisies arbitrairement. Ces quantit\'es \'etant pr\'ecessantes et non r\'esonnantes entre
    elles, les phases initiales ne sont pas critiques.\label{tab:pseudoIo}}
    \begin{tabular}{lr}
    \hline
    Param\`etres & Valeurs \\
    \hline
    $\mathcal{G}M_p$ (plan\`ete) & $1.26712765\times10^8$ km$^3$.s$^{-2}$ \\
    $\mathcal{G}M$ (satellite) & $5955.5$ km$^3$.s$^{-2}$ \\
    $R_p$ & $71492$ km \\
    $J_{2p}$ & $1.4736\times10^{-2}$ \\
    $J_2$ & $1.828\times10^{-3}$ \\
    $C_{22}$ & $5.537\times10^{-4}$ \\
    $C/(MR^2)$ & $0.376856$ \\
    $a$ & $422029.958$ km \\
    $e$ & $4.15\times10^{-3}$ \\
    $I$ & $2.16$ arcmin \\
    $n$ & $1297.2044725279755$ rad/an \\
    $\dot{\varpi}$ & $0.97311853791375$ rad/an \\
    $\dot{\ascnode}$ & $-0.8455888497945$ rad/an \\
    $\lambda_o(0)$ & 0 \\
    $\varpi_o(0)$ & $2$ rad \\
    $\ascnode_o(0)$ & $0.1$ rad \\
    $\epsilon_1=J_2\frac{MR^2}{C}$ & $4.85066\times10^{-3}$ \\
    $\epsilon_2=2C_{22}\frac{MR^2}{C}$ & $2.93852\times10^{-3}$ \\
    \hline
    \end{tabular}
   \end{table}

  \par L'id\'ee \'etait d'essayer de comprendre dans quelle mesure les param\`etres d'int\'erieur $\delta$, $\epsilon_3$ et $\epsilon_4$ influaient sur la rotation, 
  avec en arri\`ere-pens\'ee l'id\'ee d'\'etudier la faisabilit\'e de la d\'etermination de param\`etres d'int\'erieur \`a partir d'observations de la rotation. J'ai bien
  trouv\'e qu'un \'etat de type Cassini 1, c'est-\`a-dire ressemblant fortement \`a l'\'Etat de Cassini 1 tel qu'il a \'et\'e d\'efini pour les corps rigides, \'etait
  stable pour la plupart des mod\`eles d'Io acceptables, avec une rotation synchrone, une obliquit\'e mod\'er\'ee, et un mouvement polaire de faible amplitude, correspondant
  \`a une oscillation de l'axe de figure $\vv{f_3}$ autour de la direction du moment cin\'etique $\vv{N_m}$. 
  
  \par Techniquement, l'\'etude \'etait essentiellement num\'erique. Elle consistait, comme d'autres \'etudes pr\'ec\'edentes, en des simulations num\'eriques de la 
  rotation au voisinage de l'\'equilibre attendu, pour diff\'erents mod\`eles d'int\'erieur, en fait diff\'erents noyaux fluides dans notre cas.
  
  \par Une bonne fa\c{c}on d'appr\'ehender l'influence des param\`etres d'int\'erieur sur le comportement du syst\`eme est d'\'etudier l'\'evolution des fr\'equences 
  propres de la dynamique de rotation $\omega_u$, $\omega_v$, $\omega_w$ et $\omega_z$. Je pr\'esenterai ensuite un cas n\'ecessitant une r\'evision de la th\'eorie des 
  \'Etats de Cassini, o\`u l'axe de figure peut \^etre significativement inclin\'e par rapport au moment cin\'etique.

  \subsubsection{Influence des param\`etres d'int\'erieur}
  
  \par Je me place ici dans les conditions o\`u l'\'etat de type Cassini 1 existe. Ces conditions sont assez g\'en\'erales, car elles supposent que l'aplatissement
  polaire du noyau n'est pas d'un ordre de grandeur sup\'erieur \`a celui du satellite. Pour $\delta=C_c/C=0.5$, la limite se situe vers $\epsilon_3\approx\epsilon_1$.
  
  \par Je pr\'esente les d\'ependances des p\'eriodes propres du syst\`eme $T_{u,v,w,z}=2\pi/\omega_{u,v,w,z}$ en les param\`etres d'int\'erieur $\delta$ 
  (Tab.\ref{tab:infludeltae}), $\epsilon_3$ (Tab.\ref{tab:influeps3e}) et $\epsilon_4$ (Tab.\ref{tab:influeps4e}). $T_u$ est \emph{plus ou moins} la fr\'equence
  des oscillations libres en longitude, $T_v$ est \emph{plus ou moins} celle des oscillations de l'obliquit\'e, $T_w$ est \emph{plus ou moins} celle des 
  oscillations du mouvement polaire du satellite, et $T_z$ \emph{plus ou moins} celle des oscillations de l'orientation du champ de vitesses dans le fluide. Je tiens
  beaucoup \`a nuancer cette interpr\'etation physique, car en r\'ealit\'e tous ces degr\'es de libert\'e sont coupl\'es entre eux, m\^eme si le couplage peut \^etre
  n\'egligeable en pratique. La simulation d'une variable de rotation quelconque peut en th\'eorie faire intervenir les 4 fr\'equences $\omega_{u,v,w,z}$. La prise en 
  compte de ce couplage entre les diff\'erents degr\'es de libert\'e de la dynamique de rotation constitue l'une des originalit\'es de mes travaux par rapport \`a ce qui
  se fait ailleurs. Mon exp\'erience m'indique que les \'etudes qui regardent le mouvement en longitude ind\'ependamment des autres mouvements sont en g\'en\'eral tr\`es 
  pr\'ecises, et que celles qui ne regardent que l'obliquit\'e donnent de bons r\'esultats.
  
  \par Je montre \'egalement la valeur d'\'equilibre $R^*$ de la variable $R=P(1-\cos K)$. Cette valeur est donc li\'ee \`a l'obliquit\'e d'\'equilibre. Les quantit\'es
  obtenues ont \'et\'e identifi\'ees par analyse en fr\'equences dans les solutions num\'eriques.

  \begin{table}[ht]
   \centering
   \caption{Influence de la taille du noyau $\delta=C_c/C$, avec $\epsilon_3=\epsilon_1$ et $\epsilon_4=\epsilon_2$.\label{tab:infludeltae}}
   \begin{tabular}{r|rrrrr}
   $\delta$ & $T_u$ (j) & $T_v$ (j) & $T_w$ (j) & $T_z$ (j) & $R^*$ \\
   \hline
   $0.1$ & $12.650$ & $453.259$ & $208.551$ & $1.745$ & $2.304\times10^{-7}$ \\
   $0.2$ & $11.926$ & $480.369$ & $185.790$ & $1.741$ & $2.365\times10^{-7}$ \\
   $0.3$ & $11.156$ & $504.669$ & $163.028$ & $1.736$ & $2.450\times10^{-7}$ \\
   $0.4$ & $10.328$ & $526.965$ & $140.264$ & $1.729$ & $2.525\times10^{-7}$ \\
   $0.5$ &  $9.428$ & $547.734$ & $117.496$ & $1.720$ & $2.597\times10^{-7}$ \\
   $0.6$ &  $8.433$ & $567.278$ &  $94.723$ & $1.706$ & $2.680\times10^{-7}$ \\
   $0.7$ &  $7.303$ & $585.780$ &  $71.939$ & $1.685$ & $2.760\times10^{-7}$ \\
   $0.8$ &  $5.963$ & $603.294$ &  $49.130$ & $1.645$ & $2.842\times10^{-7}$ \\
   $0.9$ &  $4.216$ & $619.394$ &  $26.230$ & $1.540$ & $2.922\times10^{-7}$ \\
   \hline
   \end{tabular}
  \end{table}

  \par La taille du noyau influe essentiellement sur la p\'eriode $T_u$, ce qui r\'esulte en une cons\'equence pr\'epond\'erante sur l'amplitude des oscillations diurnes. 
  Plus le noyau est grand, plus la p\'eriode de ces oscillations est courte, et plus l'amplitude des oscillations diurnes sera importante. Pour m\'emoire, 
  $T_u=13.25$ jours \citep{h2005} lorsque Io est rigide ($\delta=0$). L'influence sur les autres degr\'es de libert\'e est plus mod\'er\'ee.

  \begin{table}[ht]
   \centering
   \caption{Influence de l'aplatissement polaire du noyau $\epsilon_3$, avec $\delta=0.5$ et $\epsilon_4=\epsilon_2$.
   \label{tab:influeps3e}}
   \begin{tabular}{r|rrrrr}
   $\epsilon_3/\epsilon_1$ & $T_u$ (j) & $T_v$ (j) & $T_w$ (j) & $T_z$ (j) & $R^*$ \\
   \hline
   $0.2$ & $9.428$ & $6414.819$ & $117.118$ & $1.728$ & $1.040\times10^{-6}$ \\
   $0.3$ & $9.428$ & $2491.673$ & $117.112$ & $1.727$ & $4.612\times10^{-7}$ \\
   $0.4$ & $9.428$ & $1572.784$ & $117.121$ & $1.726$ & $3.592\times10^{-7}$ \\
   $0.5$ & $9.428$ & $1163.452$ & $117.147$ & $1.725$ & $3.177\times10^{-7}$ \\
     $1$ & $9.428$ &  $547.734$ & $117.496$ & $1.720$ & $2.597\times10^{-7}$ \\
     $3$ & $9.428$ &  $254.876$ & $122.879$ & $1.695$ & $2.326\times10^{-7}$ \\
     $5$ & $9.428$ &  $210.742$ & $136.657$ & $1.668$ & $2.277\times10^{-7}$ \\ 
     $6$ & $9.428$ &  $200.875$ & $148.926$ & $1.655$ & $2.268\times10^{-7}$ \\
     $7$ & $9.428$ &  $194.454$ & $168.492$ & $1.641$ & $2.260\times10^{-7}$ \\
     $8$ & $9.428$ &  $189.284$ & $201.639$ & $1.628$ & $2.256\times10^{-7}$ \\
     $9$ & $9.428$ &  $185.621$ & $278.943$ & $1.616$ & $2.251\times10^{-7}$ \\
   \hline
   \end{tabular}
  \end{table}

  \par L'aplatissement polaire du noyau $\epsilon_3$ semble avoir surtout une influence sur l'obliquit\'e, comme le sugg\`erent les colonnes
  $T_v$ et $R^*$. En particulier, l'obliquit\'e semble plus importante lorsque le noyau a une section m\'eridienne proche du cercle. Il faut garder \`a l'esprit
  que j'ai impos\'e au champ de gravit\'e de mon pseudo-Io de conserver les valeurs mesur\'ees par Galileo, tout en faisant varier la forme de son noyau. Un Io
  r\'ealiste qui aurait un noyau sph\'erique aurait probablement un autre champ de gravit\'e, donc ce r\'esultat doit \^etre plus consid\'er\'e comme un 
  comportement math\'ematique du syst\`eme dynamique r\'egi par le Hamiltonien (\ref{eq:htotalliquidcore}) que comme un r\'esultat plan\'etologique. L'actuel
  manque de donn\'ees sur l'int\'erieur et la rotation des satellites naturels n'exclut pas de bonnes surprises pour le futur.

  \begin{table}[ht]
   \centering
   \caption{Influence de l'ellipticit\'e \'equatoriale du noyau $\epsilon_4$, avec $\delta=0.5$ et $\epsilon_3=\epsilon_1$.
   \label{tab:influeps4e}}
   \begin{tabular}{r|rrrrr}
   $\epsilon_4/\epsilon_2$ & $T_u$ (j) & $T_v$ (j) & $T_w$ (j) & $T_z$ (j) & $R^*$ \\
   \hline
     $0$ & $9.428$ &  $545.949$ & $117.771$ & $1.7199$ & $2.5996\times10^{-7}$ \\
   $0.1$ & $9.428$ &  $546.128$ & $117.718$ & $1.7199$ & $2.5998\times10^{-7}$ \\
   $0.5$ & $9.428$ &  $546.841$ & $117.563$ & $1.7199$ & $2.5954\times10^{-7}$ \\
     $1$ & $9.428$ &  $547.734$ & $117.496$ & $1.7198$ & $2.5967\times10^{-7}$ \\
     $3$ & $9.428$ &  $551.316$ & $118.652$ & $1.7195$ & $2.6070\times10^{-7}$ \\
     $5$ & $9.428$ &  $554.914$ & $122.283$ & $1.7193$ & $2.6069\times10^{-7}$ \\
    $10$ & $9.428$ &  $564.010$ & $149.248$ & $1.7186$ & $2.6248\times10^{-7}$ \\
   \hline
   \end{tabular}
  \end{table}

  \par L'ellipticit\'e \'equatoriale du noyau $\epsilon_4$ semble, elle, avoir une influence tr\`es limit\'ee. On peut constater num\'eriquement des variations,
  mais sans r\'eelle influence.
  
  \par La principale information \`a retenir est que les librations diurnes en longitude sont une signature de la taille du noyau, mais pas de sa forme. Ainsi, il est 
  pratiquement \'equivalent, dans des \'etudes se focalisant sur le mouvement en longitude, de consid\'erer le noyau comme sph\'erique. L'\'etude est dans ce cas plus
  facile car la r\'esultante du couple de pression \`a l'interface noyau-manteau est nulle, on peut donc consid\'erer le satellite comme un corps creux, en substituant
  dans les \'equations la rotation rigide le moment d'inertie polaire du manteau $C_m=C-C_c$ \`a celui du satellite $C$. Dans ce cas, on a 
  
  \begin{equation}
    \label{eq:Tupoincare}
    T_u \approx \frac{\pi}{n}\sqrt{\frac{C_m/(MR^2)}{3C_{22}}} \propto \sqrt{1-\delta}.
  \end{equation}

  \par On peut \'egalement remarquer que la p\'eriode li\'ee aux oscillations libres dans le fluide, $T_z$, est proche de la p\'eriode orbitale $1.769$ jour. 
  J'ai longtemps pens\'e que cela pouvait \^etre une source de ph\'enom\`enes r\'esonnants, mais n'y ai jamais \'et\'e confront\'e dans les simulations. Mon
  sentiment est que s'il peut y avoir une situation de quasi-r\'esonance entre $\omega_z$ et la fr\'equence, elle n'ira pas plus loin que de rendre le mouvement
  polaire dans le noyau significatif. Ce ph\'enom\`ene serait difficile \`a observer, une piste pour cela est de mesurer le champ magn\'etique.

  \subsubsection{Une bifurcation \ldots}
  
  \par
\begin{figure}[ht]
   \centering
   \begin{tabular}{cc}
    \includegraphics[width=.47\textwidth]{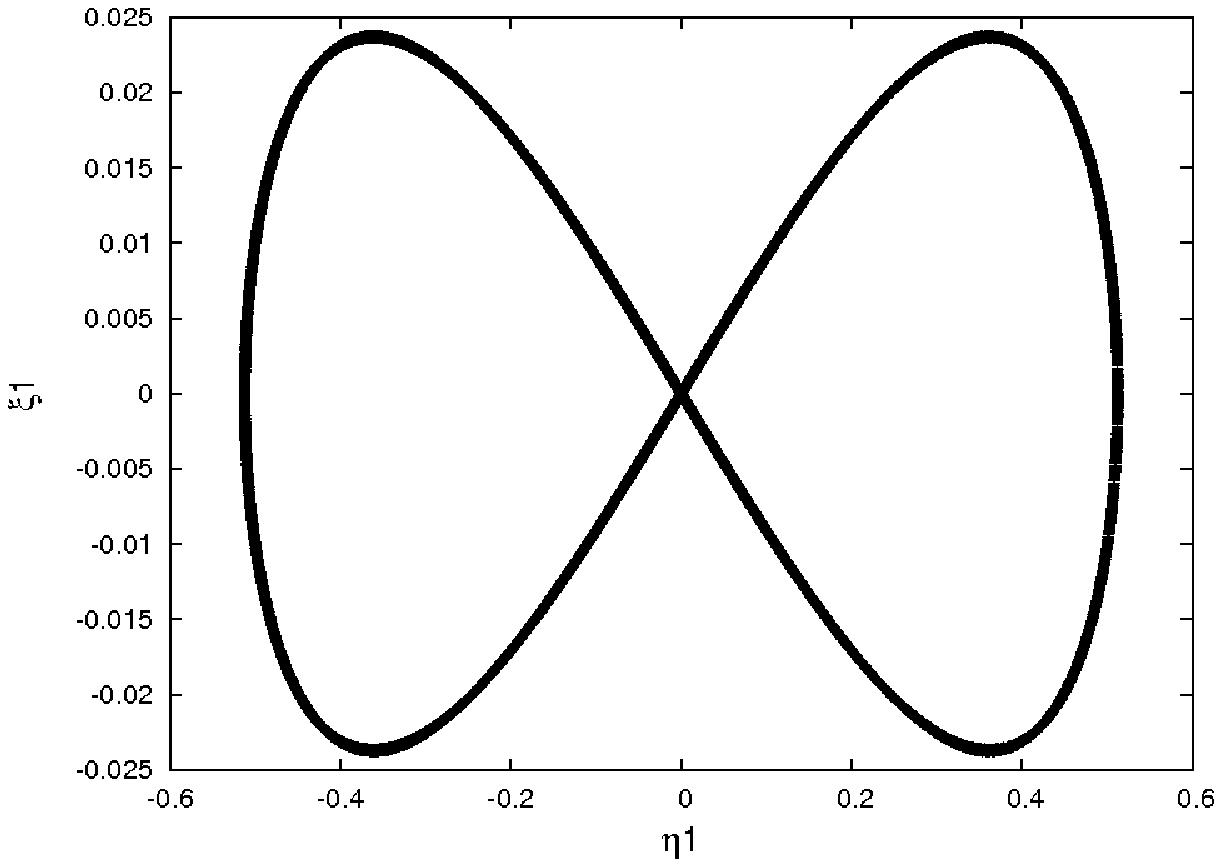} & \includegraphics[width=.47\textwidth]{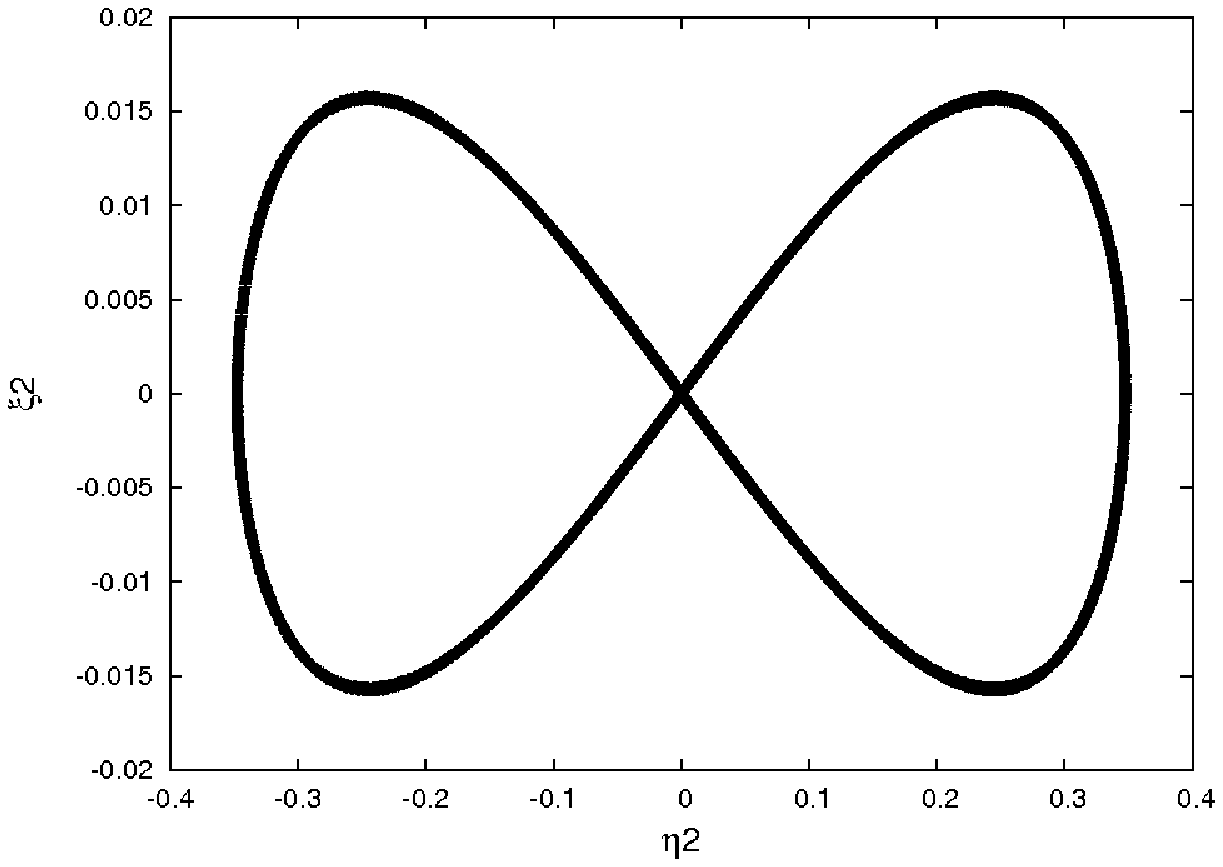}
   \end{tabular}
   \caption[Trajectoire passant proche de l'\'equilibre $J=J^c=0$ pour un noyau tr\`es aplati]{Trajectoire passant proche de l'\'equilibre $J=J^c=0$ pour 
   $\delta=0.5$, $\epsilon_3=10\epsilon_1$ et $\epsilon_4=0$. La trajectoire en papillon sugg\`ere que cet \'equilibre est instable, mais que 2 \'equilibres 
   stables sont pr\'esents dans les lobes.\label{fig:papillon}}
  \end{figure}

 La simulation num\'erique d'une trajectoire proche de l'\'equilibre dynamique doit donner, en 2 dimensions, une section de forme elliptique, traduisant des 
  petites librations autour de cet \'equilibre. Pour un noyau tr\`es allong\'e, j'obtiens la Fig.\ref{fig:papillon}.
  
  \par Je n'ai compris cette figure que gr\^ace \`a une discussion avec Nicolas Delsate. Elle sugg\`ere en fait que l'\'equilibre attendu au centre du rep\`ere est
  bien un \'equilibre, mais qu'il est instable. Par contre, 2 \'equilibres stables apparaissent, qui impliquent que pour un tel int\'erieur, l'axe de figure $\vv{f_3}$
  peut \^etre en moyenne significativement d\'esax\'e par rapport au moment cin\'etique. Afin de comprendre et pr\'edire ce ph\'enom\`ene, j'en ai r\'ealis\'e une 
  \'etude analytique, en partie \`a l'aide du logiciel Maple.
  
  \par Nous partons pour cela du Hamiltonien $\mathcal{H}$ (Eq.\ref{eq:htotalliquidcore}), dans lequel les coordonn\'ees du vecteur unitaire plan\`ete-satellite sont,
  dans le rep\`ere inertiel :
  
  \begin{eqnarray}
    x_i & = & -\cos\lambda_o, \\
    y_i & = & -\sin\lambda_o, \\
    z_i & = & 0.
  \end{eqnarray}
Ici je n\'eglige l'excentricit\'e $e$, l'inclinaison $I$, les pr\'ecessions $\dot{\varpi}$ et $\dot{\ascnode}$. Ceci rend invalide le calcul de l'obliquit\'e,
je n\'eglige donc \'egalement ce degr\'e de libert\'e.

\par Afin de faire appara\^itre la r\'esonance spin-orbite, j'introduis le changement de variables canonique suivant :

  \begin{equation}
   \label{eq:chrescm}
   \begin{array}{lll}
   \sigma=p-\lambda_o+\pi, & \hspace{2cm} & P, \\
   \xi_1, & \hspace{2cm} & \eta_1, \\
   \xi_2, & \hspace{2cm} & \eta_2, \\
   \end{array} \\
  \end{equation}
o\`u $\sigma$ est l'argument r\'esonnant. Ce changement de variable d\'ependant lin\'eairement du temps, il n\'ecessite d'ajouter $-nP$ au Hamiltonien du probl\`eme.

\par Apr\`es moyennisation des courtes p\'eriodes pour faire dispara\^itre $\lambda_o$, j'ai exprim\'e les \'equations de Hamilton :

  \begin{equation}
   \label{eq:hequations}
   \begin{array}{lll}
   \frac{d\sigma}{dt}=\frac{\partial \mathcal{H}}{\partial P}, & \hspace{2cm} & \frac{dP}{dt}=-\frac{\partial \mathcal{H}}{\partial \sigma}, \\
   & & \\
   \frac{d\xi_1}{dt}=\frac{\partial \mathcal{H}}{\partial \eta_1}, & \hspace{2cm} & \frac{d\eta_1}{dt}=-\frac{\partial \mathcal{H}}{\partial \xi_1}, \\
   & & \\
   \frac{d\xi_2}{dt}=\frac{\partial \mathcal{H}}{\partial \eta_2}, & \hspace{2cm} & \frac{d\eta_2}{dt}=-\frac{\partial \mathcal{H}}{\partial \xi_2}. \\
   \end{array} \\
  \end{equation}

  \par Je cherche \`a \'etudier la stabilit\'e de l'\'equilibre d\'efini par $\xi_1=\eta_1=\xi_2=\eta_2=0$. Je pose $\xi_1=\xi_2=0$ pour simplifier les \'equations et 
  j'obtiens
  
  \begin{eqnarray}
   \frac{1}{n}\frac{d\sigma}{dt} & =& -1+\frac{P-P_c}{1-\delta} +\frac{\eta_1^2}{2(1-\delta)^2} \left(\epsilon_1-\epsilon_2-\delta\epsilon_3+\delta\epsilon_4\right)+\frac{\eta_2^2}{2(1-\delta)} \label{eq:ndpdt} \\
   & +& \frac{\eta_1\eta_2\left(P_c-\eta_2^2/4\right)}{2\left(1-\delta\right)\sqrt{PP_c-P\eta_2^2/4-P_c\eta_1^2/4+\eta_1^2\eta_2^2/16}}\left(1+\frac{\epsilon_1-\epsilon_2-\delta\epsilon_3+\delta\epsilon_4}{1-\delta}\right),\nonumber
  \end{eqnarray}

  \begin{eqnarray}
   \frac{1}{n}\frac{d\xi_1}{dt} & = & \frac{\eta_1P}{(1-\delta)^2}\left(\epsilon_1-\epsilon_2-\delta\epsilon_3+\delta\epsilon_4\right)+\frac{\eta_1P_c}{1-\delta}+\frac{\eta_1^3}{2(1-\delta)^2}\left(-\epsilon_1+\epsilon_2+\delta\epsilon_3-\delta\epsilon_4\right)-   \frac{\eta_1\eta_2^2}{2(1-\delta)} \nonumber \\
   & + & \frac{\eta_1^2\eta_2\left(\eta_2^2/4-P_c\right)}{4(1-\delta)\sqrt{PP_c-P\eta_2^2/4-P_c\eta_1^2/4+\eta_1^2\eta_2^2/16}}\left(1+\frac{\epsilon_1-\epsilon_2-\delta\epsilon_3+\delta\epsilon_4}{1-\delta}\right) \label{eq:ndxi1dt} \\
   & + & \frac{\eta_2}{1-\delta}\sqrt{PP_c-P\eta_2^2/4-P_c\eta_1^2/4+\eta_1^2\eta_2^2/16}\left(1+\frac{\epsilon_1-\epsilon_2-\delta\epsilon_3+\delta\epsilon_4}{1-\delta}\right), \nonumber 
\end{eqnarray}
et 

  \begin{eqnarray}
   \frac{1}{n}\frac{d\xi_2}{dt} & = & \frac{\eta_2P_c}{(1-\delta)^2}\left(\epsilon_1-\epsilon_2+\left(\frac{1}{\delta}-2\right)\epsilon_3+\left(2-\frac{1}{\delta}\right)\epsilon_4\right)+\frac{\eta_2P}{1-\delta} \nonumber \\
   & + &\frac{\eta_2^3}{2(1-\delta)^2}\left(-\epsilon_1+\epsilon_2+\left(2-\frac{1}{\delta}\right)\epsilon_3+\left(\frac{1}{\delta}-2\right)\epsilon_4\right)-\frac{\eta_1^2\eta_2}{2(1-\delta)} \label{eq:ndxi2dt} \\
   & + & \frac{\eta_1\eta_2^2\left(\eta_1^2/4-P\right)}{4(1-\delta)\sqrt{PP_c-P\eta_2^2/4-P_c\eta_1^2/4+\eta_1^2\eta_2^2/16}}\left(1+\frac{\epsilon_1-\epsilon_2-\delta\epsilon_3+\delta\epsilon_4}{1-\delta}\right) \nonumber \\
   & + & \frac{\eta_1}{1-\delta}\sqrt{PP_c-P\eta_2^2/4-P_c\eta_1^2/4+\eta_1^2\eta_2^2/16}\left(1+\frac{\epsilon_1-\epsilon_2-\delta\epsilon_3+\delta\epsilon_4}{1-\delta}\right)\nonumber.
\end{eqnarray}

\par Pour $\epsilon_3=10\epsilon_1$, $\epsilon_4=0$ et $\delta=0.5$, les racines r\'eelles de ce syst\`eme sont

\begin{itemize}
  \item $P=1.046772470$, $\eta_1=1.446908787$, $\eta_2=-1.023119016$
  \item $P=1.002812138$, $\eta_1=0.2502391659$, $\eta_2=-0.1690724173$
  \item $P=1$, $\eta_1=\eta_2=0$
  \item $P=1.002812138$, $\eta_1=-0.2502391659$, $\eta_2=0.1690724173$
  \item $P=0.3489241565$, $\eta_1=0.8353731582$, $\eta_2=-0.5606980247$
\end{itemize}
alors qu'elles sont, pour $\epsilon_3=9\epsilon_1$, $\epsilon_4=0$ et $\delta=0.5$:

\begin{itemize}
  \item $P=1.041484268$, $\eta_1=-1.443249317$, $\eta_2=-1.020531379$
  \item $P=0.3471614224$, $\eta_1=-0.8332603709$, $\eta_2=-0.5892040580$
  \item $P=1$, $\eta_1=\eta_2=0$
  \item $P=0.9978852209$, $\eta_1=-2.010693353$, $\eta_2=-1.421011855$.
\end{itemize}

\par On remarque que dans les 2 cas, la position d'\'equilibre classiquement attendue, ($P=1$, $\eta_1=\eta_2=0$) est pr\'esente. Pour $\epsilon_3=10\epsilon_1$, 2
positions d'\'equilibre sym\'etriques apparaissent avec $P$ qui reste tr\`es proche de $1$. Ces positions sont visibles sur la Fig.\ref{fig:papillon}.

\par Pour \'etudier la stabilit\'e de l'\'equilibre classique, posons $\xi_1=\xi_2=0$, $P=1$ et $P_c=\delta$ dans le Hamiltonien moyenn\'e, ceci donne la quantit\'e $\mathcal{S}$ :

\begin{eqnarray}
  \mathcal{S}(\eta_1,\eta_2) & = & \alpha-1+\frac{1-\delta+\eta_1^2\delta+\eta_2^2-\eta_1^2\eta_2^2/2}{2(1-\delta)} \nonumber \\
                             & + & \epsilon_1\left(-\frac{3}{2}+\frac{\eta_1^2+\eta_2^2\delta-\left(\eta_1^4+\eta_2^4\right)/4}{2(1-\delta)^2}+\frac{\alpha}{1-\delta}\right) \nonumber \\
                             & + & \epsilon_2\left(-\frac{3}{2}-\frac{\eta_1^2+\eta_2^2\delta-\left(\eta_1^4+\eta_2^4\right)/4}{2(1-\delta)^2}-\frac{\alpha}{1-\delta}\right) \label{eq:surfacepoincare} \\
                             & + & (\epsilon_3-\epsilon_4)\left(\frac{-\delta\eta_1^2+\eta_2^2(1-2\delta)+\delta\eta_1^4/4+\eta_2^4/2(1-1/(2\delta))}{2(1-\delta)^2}-\frac{\delta\alpha}{1-\delta}\right) \nonumber
\end{eqnarray}
avec

\begin{equation}
  \label{eq:alphapoincare}
  \alpha=\frac{\eta_1\eta_2}{1-\delta}\sqrt{\delta-\frac{\eta_1^2\delta+\eta_2^2}{4}+\frac{\eta_1^2\eta_2^2}{16}}.
\end{equation}
$\mathcal{S}$ n'est pas un Hamiltonien car les variables $\xi_1$ et $\xi_2$ sont consid\'er\'ees comme des constantes alors que leurs moments associ\'es 
$\eta_1$ et $\eta_2$ peuvent varier. Pour conna\^itre la stabilit\'e du point ($\eta_1=0$,$\eta_2=0$), il faut \'etudier les d\'eriv\'ees secondes de $\mathcal{S}$
en ce point. Soit la matrice hessienne

 \begin{eqnarray}
  \mathcal{M} & = & \left(\begin{array}{cc}
  \frac{\partial^2\mathcal{S}}{\partial\eta_1^2} &  \frac{\partial^2\mathcal{S}}{\partial\eta_1\partial\eta_2} \\
  \frac{\partial^2\mathcal{S}}{\partial\eta_2\partial\eta_1} &  \frac{\partial^2\mathcal{S}}{\partial\eta_2^2}
  \end{array}\right) \label{eq:matrixx} \\
  &=&\frac{1}{(\delta-1)^2}\left(\begin{array}{cc}
  \epsilon_1-\epsilon_2+\delta(\epsilon_4-\epsilon_3+1-\delta) & \sqrt{\delta}(1-\delta+\epsilon_1-\epsilon_2+\delta(\epsilon_4-\epsilon_3)) \\
  \sqrt{\delta}(1-\delta+\epsilon_1-\epsilon_2+\delta(\epsilon_4-\epsilon_3)) & 1-\delta+\delta(\epsilon_1-\epsilon_2-2\epsilon_3+2\epsilon_4)
\end{array}\right),\nonumber
\end{eqnarray}
un minimum de $\mathcal{S}$, correspondant \`a un \'equilibre stable, est atteint lorsque les 2 valeurs propres de $\mathcal{M}$, $\lambda_1$ et 
$\lambda_2$, sont r\'eelles et positives. J'ai

\begin{eqnarray}
  \lambda_1 & = & \beta+\frac{\sqrt{\Delta}}{2} \\
  \lambda_2 & = & \beta-\frac{\sqrt{\Delta}}{2}
\end{eqnarray}
avec

\begin{equation}
  \label{eq:betapoincare}
  \beta=\frac{1-\delta^2+(\epsilon_1-\epsilon_2)(1+\delta)+(\epsilon_3-\epsilon_4)(1-3\delta)}{2}
\end{equation}
et

\begin{eqnarray}
  \Delta & = & \left(1-\delta^2\right)^2-2(\epsilon_1-\epsilon_2)(1-7\delta+7\delta^2-\delta^3)+2(\epsilon_3-\epsilon_4)(1-3\delta-\delta^2+3\delta^3) \nonumber \\
         & + & (\epsilon_1-\epsilon_2)^2(1+\delta)^2+2(\epsilon_1\epsilon_4+\epsilon_2\epsilon_3-\epsilon_1\epsilon_3-\epsilon_2\epsilon_4)(1-2\delta+5\delta^2) \nonumber \\
         & + & (\epsilon_3-\epsilon_4)^2(1-2\delta+\delta^2+4\delta^3). \label{eq:gdeltapoincare}
\end{eqnarray}

\par Des \'evaluations num\'eriques montrent que $\lambda_1$ est toujours positive, et que $\lambda_2$ l'est en g\'en\'eral, sauf si $\epsilon_3$ est suffisamment grand,
c'est-\`a-dire au del\`a de $\approx9.4\epsilon_1$ pour $\delta=0.5$. Le point d\'efini par ($\eta_1=0$, $\eta_2=0$) est un point-col, et l'\'equilibre associ\'e est instable.
$\lambda_2>0$ est donc une condition de stabilit\'e de la position d'\'equilibre pressentie.

\par Ceci fait penser au d\'ebut d'un ph\'enom\`ene de cascade de type Feigenbaum \citep{f1979}, o\`u en augmentant le param\`etre incrimin\'e, ici $\epsilon_3$, on 
aurait la cr\'eation successive d'\'equilibres stables et instables, de plus en plus nombreux. Il s'agit d'un ph\'enom\`eme relativement classique dans les syst\`emes
dynamiques, un exemple r\'ecent en m\'ecanique c\'eleste est la stabilit\'e de la r\'esonance coorbitale lorsque le petit corps pi\'eg\'e au voisinage d'un des 
points de Lagrange $L_4$ et $L_5$ a une masse sup\'erieure \`a la limite de Gascheau \citep{s2010}. Bien entendu, ceci ne peut \^etre affirm\'e sans \'etude pr\'ealable,
or je n'ai pas explor\'e cette partie de l'espace des param\`etres, un $\epsilon_3$ mod\'er\'e me paraissant plus r\'ealiste\ldots

\par La question de l'application plan\'etologique de cette bifurcation se pose r\'eellement. Si le corps est \`a l'\'equilibre hydrostatique, alors 
la forme de l'interface noyau-manteau d\'epend de la forme de la surface, de la distribution de masse au sein du satellite, et de sa fr\'equence orbitale, 
\'egale \`a sa vitesse de rotation. Dans ces conditions, il para\^it difficile d'obtenir une interface noyau-manteau aussi allong\'ee. Le mod\`ele de 
formation des satellites de Saturne de \citet{cccldktmls2011} autorise la formation du noyau d'\'el\'ements lourds au bord des anneaux, qui sera plus allong\'e 
car ayant une vitesse orbitale plus importante, avant enrobage de silicates. Cependant, ce noyau contiendra une graine, c'est-\`a-dire un noyau interne solide. 
Pour que notre mod\`ele soit acceptable, il faudrait que la graine soit suffisamment petite pour que son influence puisse \^etre n\'eglig\'ee.

\par Apr\`es cette \'etude th\'eorique du comportement d'un satellite assujetti au mod\`ele de Poincar\'e-Hough, je l'applique dans le cas r\'ealiste d'Io.

  \section{Application \`a Io\label{sec:Io}}
  
  \par Le passage d'une \'etude purement math\'ematique comme celle que je viens de montrer \`a une application n\'ecessite notamment de mod\'eliser le mouvement
  orbital et l'int\'erieur de fa\c{c}on r\'ealiste. Pour le mouvement orbital d'Io autour de Jupiter, j'ai utilis\'e les \'eph\'em\'erides L1.2 \citep{ldv2006}. Les
  modes propres orbitaux sont les m\^emes que pour Callisto, cf.Tab.\ref{tab:propmodes}. Il est \`a noter que Io est dans une r\'esonance de moyen mouvement $4:2:1$
  dite laplacienne, ou de Laplace, impliquant \'egalement Europe et Ganym\`ede. Cette r\'esonance, d'argument $\lambda_1-3\lambda_2+2\lambda_3$, a une libration d'amplitude
  de l'ordre du demi-degr\'e, et de p\'eriode $5.639$ ans. De plus, la proximit\'e de la r\'esonance de moyen mouvement $2:1$ avec Europe implique 2 oscillations
  significatives dans la longitude d'Io, d'amplitudes respectivement $81$ et $40$ km, et de p\'eriodes $463$ et $483$ jours.
  
  \subsection{6 mod\`eles d'int\'erieur}
  
  \par Il faut consid\'erer ici une mod\'elisation r\'ealiste de l'int\'erieur d'Io. Pour cela, j'ai pris comme contraintes les donn\'ees observ\'ees (Tab.\ref{tab:paramio}).
  
  \begin{table}[ht]
   \caption[Param\`etres de forme et champ de gravit\'e d'Io]{Param\`etres de forme et champ de gravit\'e d'Io. $I$ n'a pas \'et\'e mesur\'e ind\'ependamment des
   autres param\`etres, il est d\'eduit des valeurs mesur\'ees de $J_2$ et $C_{22}$ combin\'ees avec l'hypoth\`ese de l'\'equilibre hydrostatique.\label{tab:paramio}}
   \centering
   \begin{tabular}{lrr}
   \hline
   Param\`etre & Valeur & Source \\
   \hline
   Masse volumique moyenne $\bar{\rho}$ & $3527.8\pm2.9$ kg/m$^3$             & \citet{ajlms2001}\\
                                  $J_2$ & $(1.8459\pm0.0042)\times10^{-3}$    & \citet{ajlms2001}\\
                               $C_{22}$ & $(5.537\pm0.012)\times10^{-4}$      & \citet{ajlms2001}\\
                  $I/\left(MR^2\right)$ & $0.37685\pm0.00035$                 & \citet{ajlms2001}\\
                          Rayon moyen R & $1821.49$ km                        & \citet{aabcccfhhknossttw2011} \\
             Rayons $a\times b\times c$ & $1829.4\times1819.4\times1815.7$ km & \citet{aabcccfhhknossttw2011} \\
   \hline          
   \end{tabular}
  \end{table}

  \par D'apr\`es la litt\'erature, nous devons consid\'erer un noyau fluide d'\'el\'ements lourds, contenant du fer, et peut-\^etre aussi du soufre \citep{ass1996}. Ce noyau
  a une masse volumique $\rho_c$ comprise entre $8000$ kg/m$^3$ (fer pur) et $5150$ kg/m$^3$ (m\'elange eutectique
  \footnote{L'eutectique est la limite entre la phase liquide et la phase solide. Si le noyau contenait plus de soufre, alors il serait solide.} de fer et de soufre, \citet{u1975}). 
  Pour le manteau, \citet{ajlms2001} ont consid\'er\'e 2 possibilit\'es : soit une cro\^ute fine (<50 km) de faible densit\'e (<2600 kg/m$^3$), ou une cro\^ute tr\`es fine (n\'egligeable)
  surmontant une asth\'enosph\`ere plus \'epaisse (de 100 \`a 200 km) d'une masse volumique comprise entre 3000 et 3200 kg/m$^3$.
  
  \par Afin d'int\'egrer ces diff\'erentes possibilit\'es, j'ai consid\'er\'e 6 mod\`eles d'int\'erieur pour Io :
  
  \begin{itemize} 
  
    \item \emph{Mod\`ele 1 :} noyau de fer pur et manteau homog\`ene,
    
    \item \emph{Mod\`ele 2 :} noyau eutectique de fer et soufre, manteau homog\`ene,
    
    \item \emph{Mod\`ele 3 :} noyau de fer pur, cro\^ute de 30 km,
    
    \item \emph{Mod\`ele 4 :} noyau eutectique de fer et soufre, cro\^ute de 30 km,
    
    \item \emph{Mod\`ele 5 :} noyau de fer pur, asth\'enosph\`ere de 150 km,
    
    \item \emph{Mod\`ele 6 :} noyau eutectique de fer et soufre, asth\'enosph\`ere de 150 km.
    
  \end{itemize}
  \par Pour cela, nous avons besoin de 8 ou 12 param\`etres d'int\'erieur :
  
  \begin{itemize}
  
   \item $\rho_c$, $\rho_m$, $\rho_s$ : masses volumiques du noyau, du manteau, et \'eventuellement de la couche externe (cro\^ute ou asth\'enosph\`ere),
   
   \item $a_c\times b_c\times c_c$ : rayons de l'interface noyau-manteau,
   
   \item $a_m\times b_m\times c_m$ : rayons de l'interface manteau-couche externe, si elle existe.
   
  \end{itemize}

  \par Les contraintes que j'ai impos\'ees permettent de trouver de fa\c{c}on unique les param\`etres manquants, sans utiliser l'hypoth\`ese de l'\'equilibre hydrostatique. Un 
  autre choix possible aurait pu \^etre d'autoriser des variations des param\`etres d'int\'erieur, par exemple permettre \`a la masse volumique du noyau $\rho_c$ de prendre 
  n'importe quelle valeur entre les 2 extr\^emes que constituent le noyau de fer pur et le m\'elange eutectique fer-soufre, autrement dit permettre au noyau de contenir un peu 
  de soufre, sans atteindre la limite eutectique, d'imposer l'\'equilibre hydrostatique, et de conserver ensuite uniquement les mod\`eles d'Io qui satisfont les contraintes 
  observationnelles (Tab.\ref{tab:paramio}).
  
  \par Les contraintes observationnelles donnent directement
  
  \begin{eqnarray}
    A & = & 1.11149\times 10^{20}\,\textrm{kg.km}^2, \label{eq:AIo} \\
    B & = & 1.11805\times 10^{20}\,\textrm{kg.km}^2, \label{eq:BIo} \\
    C & = & 1.12024\times 10^{20}\,\textrm{kg.km}^2, \label{eq:CIo} \\
    \epsilon_1 & = & 4.88230\times10^{-3}, \label{eq:epsilon1Io} \\
    \epsilon_2 & = & 2.92901\times10^{-3}. \label{eq:epsilon2Io}
  \end{eqnarray}

  \par \`A partir des d\'efinitions de la masse et des moments d'inertie (Eq. \ref{eq:momA}-\ref{eq:momC} \& \ref{eq:Ac}-\ref{eq:Cc2}) on obtient
  
  \begin{eqnarray}
   R_c & = & \left(\frac{R^3\left(\bar{\rho}-\rho_s\right)+R_m^3\left(\rho_s-\rho_m\right)}{\rho_c-\rho_m}\right)^{1/3}, \label{eq:Rc3Io} \\
   \frac{I}{MR^2} & = & \frac{2}{5} \frac{\rho_cR_c^5+\rho_m\left(R_m^5-R_c^5\right)+\rho_s\left(R^5-R_m^5\right)}{\bar{\rho}R^2}, \label{eq:MOI3Io} \\
   A & = & \frac{4\pi}{15}\left(\rho_ca_cb_cc_c\left(b_c^2+c_c^2\right)+\rho_m\left(a_mb_mc_m\left(b_m^2+c_m^2\right)-a_cb_cc_c\left(b_c^2+c_c^2\right)\right)\right. \nonumber \\
     &   & \left.+\rho_s\left(abc\left(b^2+c^2\right)-a_mb_mc_m\left(b_m^2+c_m^2\right)\right)\right), \label{eq:A3Io} \\
   B & = & \frac{4\pi}{15}\left(\rho_ca_cb_cc_c\left(a_c^2+c_c^2\right)+\rho_m\left(a_mb_mc_m\left(a_m^2+c_m^2\right)-a_cb_cc_c\left(a_c^2+c_c^2\right)\right)\right. \nonumber \\
     &   & \left.+\rho_s\left(abc\left(a^2+c^2\right)-a_mb_mc_m\left(a_m^2+c_m^2\right)\right)\right), \label{eq:B3Io} \\
   C & = & \frac{4\pi}{15}\left(\rho_ca_cb_cc_c\left(a_c^2+b_c^2\right)+\rho_m\left(a_mb_mc_m\left(a_m^2+b_m^2\right)-a_cb_cc_c\left(a_c^2+b_c^2\right)\right)\right. \nonumber \\
     &   & \left.+\rho_s\left(abc\left(a^2+b^2\right)-a_mb_mc_m\left(a_m^2+b_m^2\right)\right)\right), \label{eq:C3Io}     
\end{eqnarray}
ce qui permet de retrouver de fa\c{c}on unique les param\`etres manquants, \`a savoir $R_c$ (rayon moyen du noyau), $\rho_m$, $a_c$, $b_c$ et $c_c$ (Tab.\ref{tab:param6intIo}).
Les param\`etres requis par le mod\`ele de rotation de Poincar\'e-Hough sont donn\'es dans la Tab.\ref{tab:param6phIo}.

\begin{table}[ht]
 \centering
 \caption[Param\`etres physiques de nos 6 mod\`eles d'Io]{Param\`etres physiques de nos 6 mod\`eles d'Io. 2 extrema sont consid\'er\'es pour le noyau : 
 fer pur (mod\`eles 1, 3 \& 5), et m\'elange eutectique de fer et soufre (mod\`eles 2, 4, \& 6). $d_s$ est l'\'epaisseur constante de la cro\^ute ou
 de l'asth\'enosph\`ere.\label{tab:param6intIo}}
 \begin{tabular}{r|rrrrrrrr}
 Mod\`eles & $\rho_c$   & $\rho_m$   & $\rho_s$  & $d_s$ & $a_c$     & $b_c$     & $c_c$     & $R_c$ \\
           & (kg/m$^3$) & (kg/m$^3$) & (kg/m$^3$ & (km)  & (km)      &  (km)     & (km)      &  (km) \\
 \hline
         1 & $8000$     & $3291.465$ & --        & $0$   & $690.411$ & $660.545$ & $663.908$ & $671.904$ \\
         3 & $8000$     & $3377.242$ & $2400$    & $30$  & $657.994$ & $623.714$ & $629.205$ & $635.355$ \\
         5 & $8000$     & $3409.044$ & $3100$    & $150$ & $650.283$ & $614.596$ & $620.705$ & $628.935$ \\
 \hline
         2 & $5150$     & $3243.815$ & --        & $0$   & $978.846$ & $958.372$ & $959.124$ & $965.144$ \\
         4 & $5150$     & $3337.988$ & $2400$    & $30$  & $936.963$ & $913.946$ & $915.956$ & $922.517$ \\
         6 & $5150$     & $3351.865$ & $3100$    & $150$ & $936.495$ & $913.320$ & $915.379$ & $921.963$ \\
 \hline
 \end{tabular}
\end{table}

  \begin{table}[ht]
   \centering
   \caption[Les param\`etres du mod\`ele de Poincar\'e-Hough]{Les param\`etres du mod\`ele de Poincar\'e-Hough. Pour chacun d'entre eux, on a 
   $\epsilon_1=4.88230\times10^{-3}$ et $\epsilon_2=2.92901\times10^{-3}$.\label{tab:param6phIo}}
   \begin{tabular}{r|ccccc}
Mod\`eles & $\epsilon_3$         & $\epsilon_4$         & $\delta$             & $\epsilon_3/\epsilon_1$ & $\epsilon_4/\epsilon_2$ \\
   \hline
        1 & $1.722\times10^{-2}$ & $2.210\times10^{-2}$ & $1.654\times10^{-2}$ &                 $3.527$ & $7.544$ \\
        3 & $1.836\times10^{-2}$ & $2.673\times10^{-2}$ & $1.270\times10^{-2}$ &                 $3.760$ & $9.125$ \\
        5 & $1.877\times10^{-2}$ & $2.819\times10^{-2}$ & $1.188\times10^{-2}$ &                 $3.844$ & $9.625$ \\
   \hline
        2 & $0.980\times10^{-2}$ & $1.057\times10^{-2}$ & $6.503\times10^{-2}$ &                 $2.007$ & $3.608$ \\
        4 & $1.029\times10^{-2}$ & $1.243\times10^{-2}$ & $5.175\times10^{-2}$ &                 $2.107$ & $4.245$ \\
        6 & $1.033\times10^{-2}$ & $1.253\times10^{-2}$ & $5.160\times10^{-2}$ &                 $2.115$ & $4.277$ \\
   \hline
   \end{tabular}
  \end{table}

  \par On peut remarquer la faible valeur du param\`etre $\delta=C_c/C$. Il faut garder \`a l'esprit que le moment d'inertie est sensible \`a la puissance 5 de la distance au centre. 
  Une masse importante proche du centre aura une influence n\'egligeable sur les moments d'inertie. Ceci permet d'anticiper que l'amplitude des librations en longitude sera 
  dans tous les cas relativement proche de l'amplitude donn\'ee par le mod\`ele rigide.

  \subsection{Les variables de rotation}
  
  \par Comme dans mes \'etudes pr\'ec\'edentes, j'ai r\'ealis\'e, pour chacun des 6 mod\`eles d'int\'erieur, des simulations num\'eriques en propageant des 
  conditions initiales correspondant \`a l'\'equilibre dynamique qui s'apparente \`a l'\'Etat de Cassini 1. Les p\'eriodes des oscillations propres sont donn\'ees
  Tab.\ref{tab:rotpropmodesIo}.
  
  \begin{table}[ht]
   \centering
   \caption{P\'eriodes des oscillations propres autour de la rotation d'\'equilibre.\label{tab:rotpropmodesIo}}
   \begin{tabular}{r|rrrr}
   \hline
   Mod\`eles & $T_u$ & $T_v$ & $T_w$ & $T_z$ \\
             &   (j) &   (j) &   (j) & (j) \\
   \hline
           1 & $13.2322$ & $166.3520$ & $225.0927$ & $1.7382$ \\
           3 & $13.2580$ & $164.5679$ & $225.7442$ & $1.7368$ \\
           5 & $13.2634$ & $164.2023$ & $225.8303$ & $1.7362$ \\
   \hline 
           2 & $12.9018$ & $127.7468$ & $213.9738$ & $1.7545$ \\
           4 & $12.9931$ & $121.0550$ & $216.9747$ & $1.7537$ \\
           6 & $12.9942$ & $120.8621$ & $217.0118$ & $1.7536$ \\
   \hline        
   \end{tabular}
  \end{table}

  \par On distingue clairement 2 groupes, selon la pr\'esence ou non de soufre dans le noyau. On peut donc imaginer au mieux 2 types de comportement, chacun associ\'e \`a un de ces groupes.
  On peut notamment remarquer que la p\'eriode des oscillations libres dans le fluide, $T_z$, est bien plus proche de la p\'eriode de rotation ($1.769$ jour) avec du soufre que sans. 
  On peut donc anticiper que, dans ce cas, le d\'esaxage du champ de vitesses dans le fluide est bien plus important.
  
  \par
\begin{figure}[ht]
   \centering
   \begin{tabular}{cc}
   \includegraphics[width=0.47\textwidth]{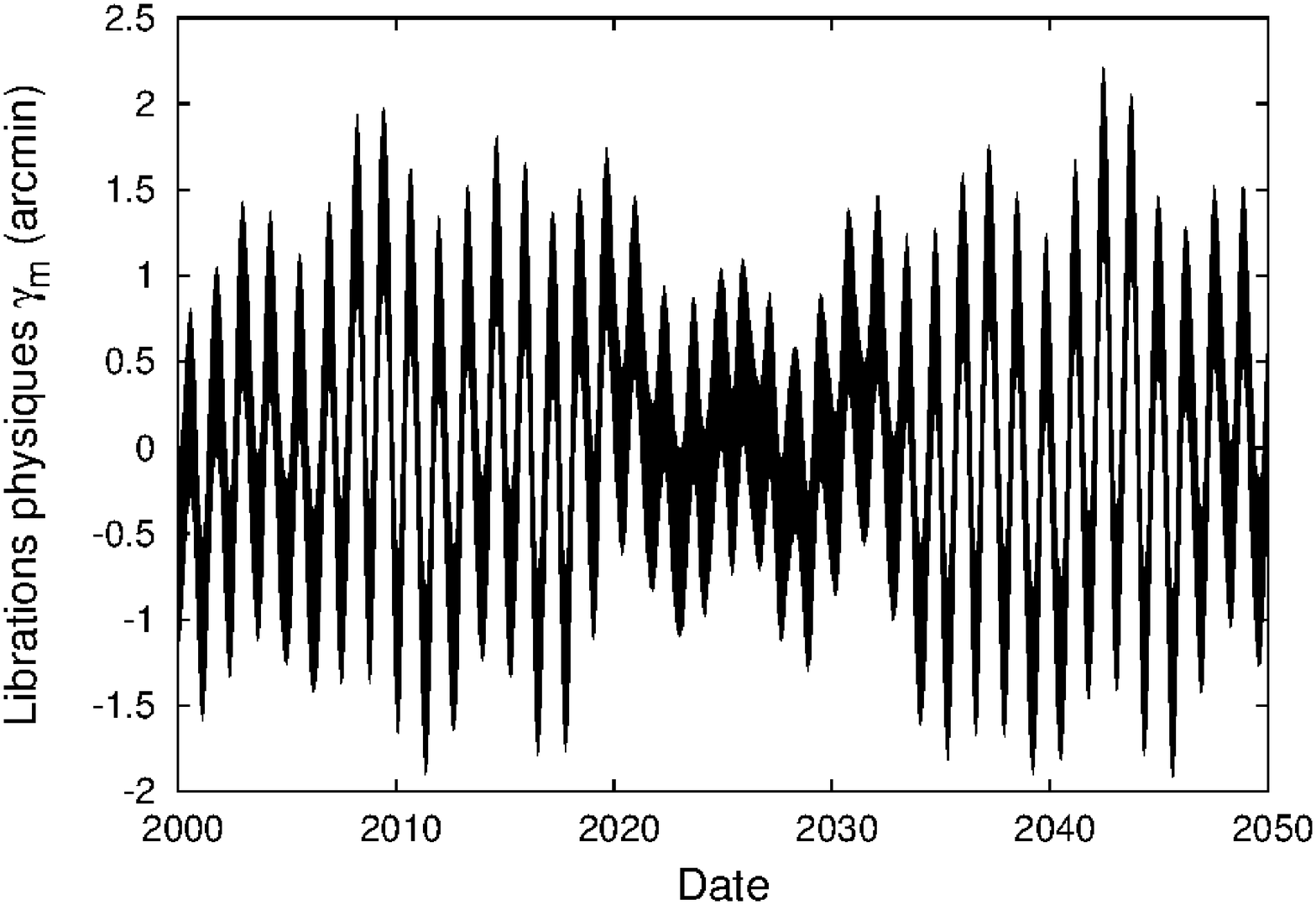} & \includegraphics[width=0.47\textwidth]{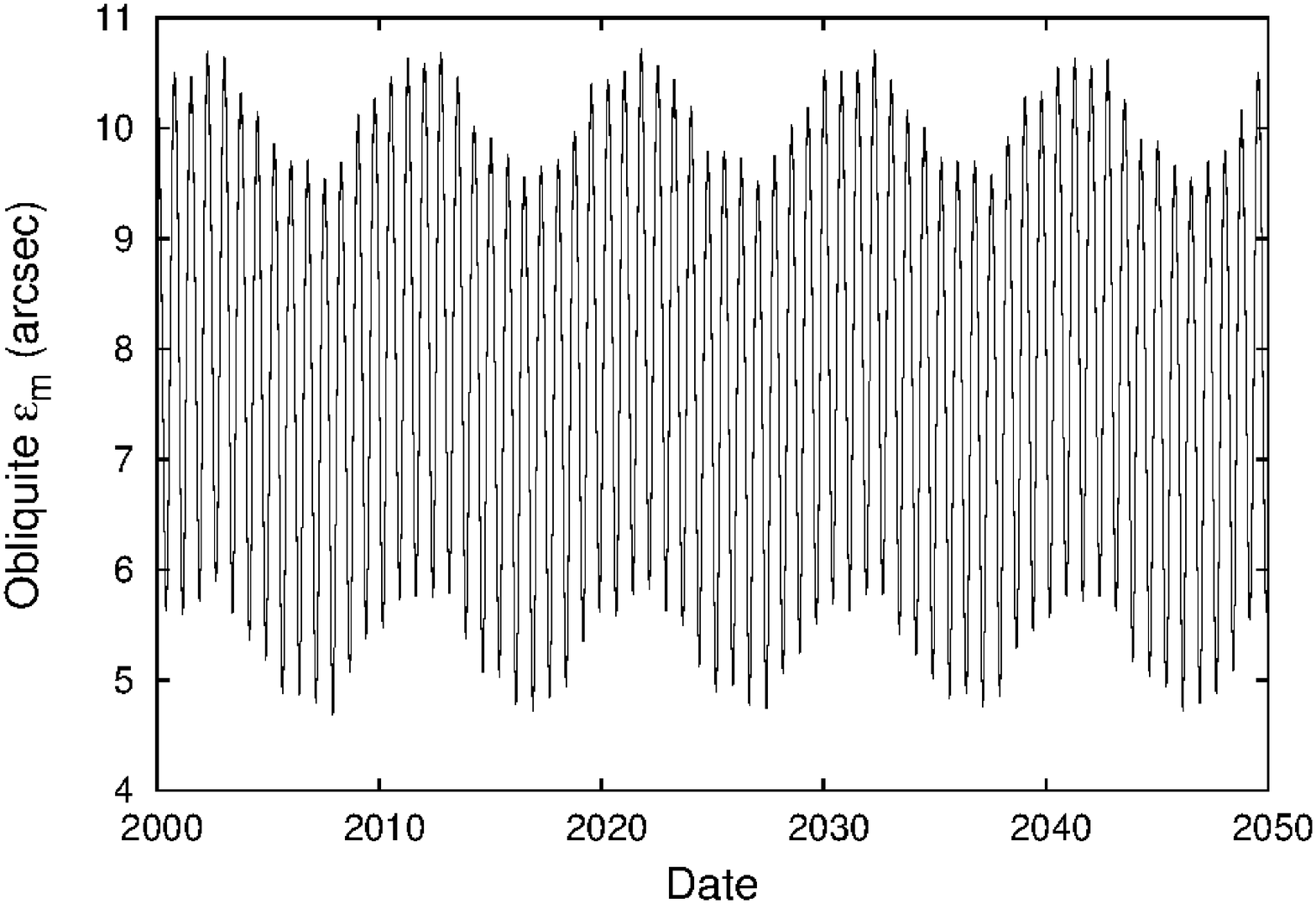} \\
   \includegraphics[width=0.47\textwidth]{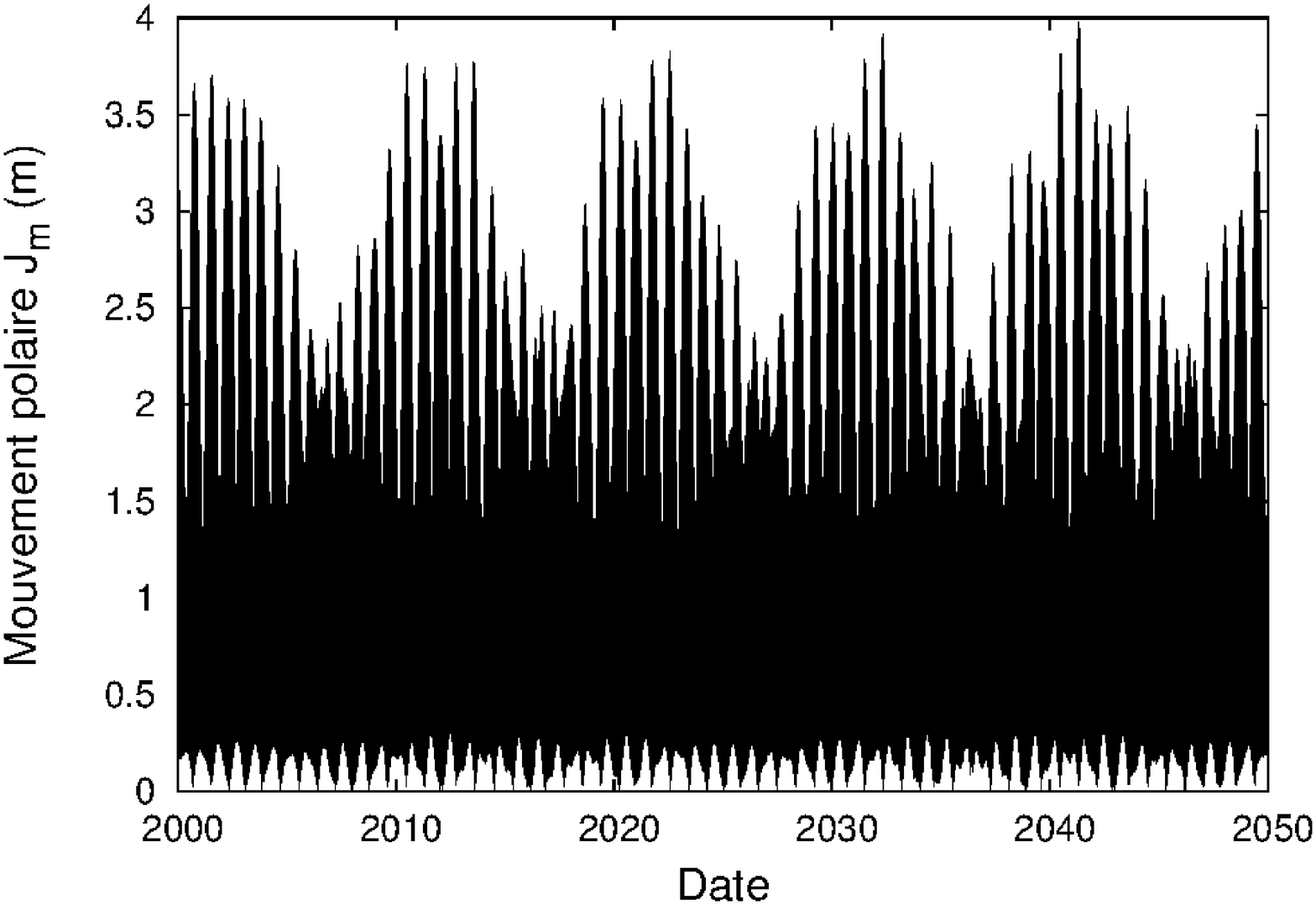} & \includegraphics[width=0.47\textwidth]{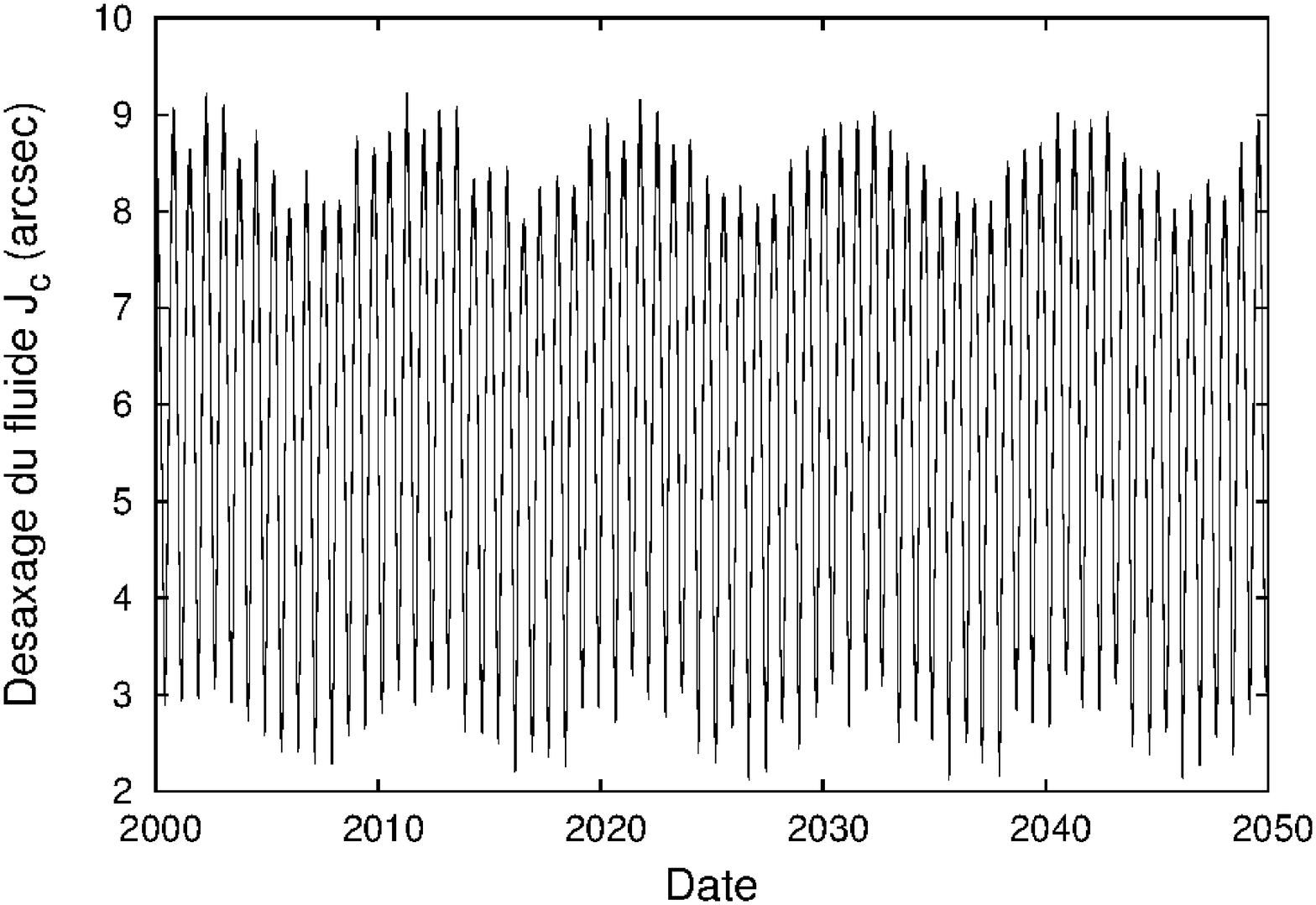}
   \end{tabular}
   \caption{La rotation du mod\`ele 1 (noyau compos\'e uniquement de fer).\label{fig:out1Io}}
  \end{figure}

\begin{figure}[ht]
   \centering
   \begin{tabular}{cc}
   \includegraphics[width=0.47\textwidth]{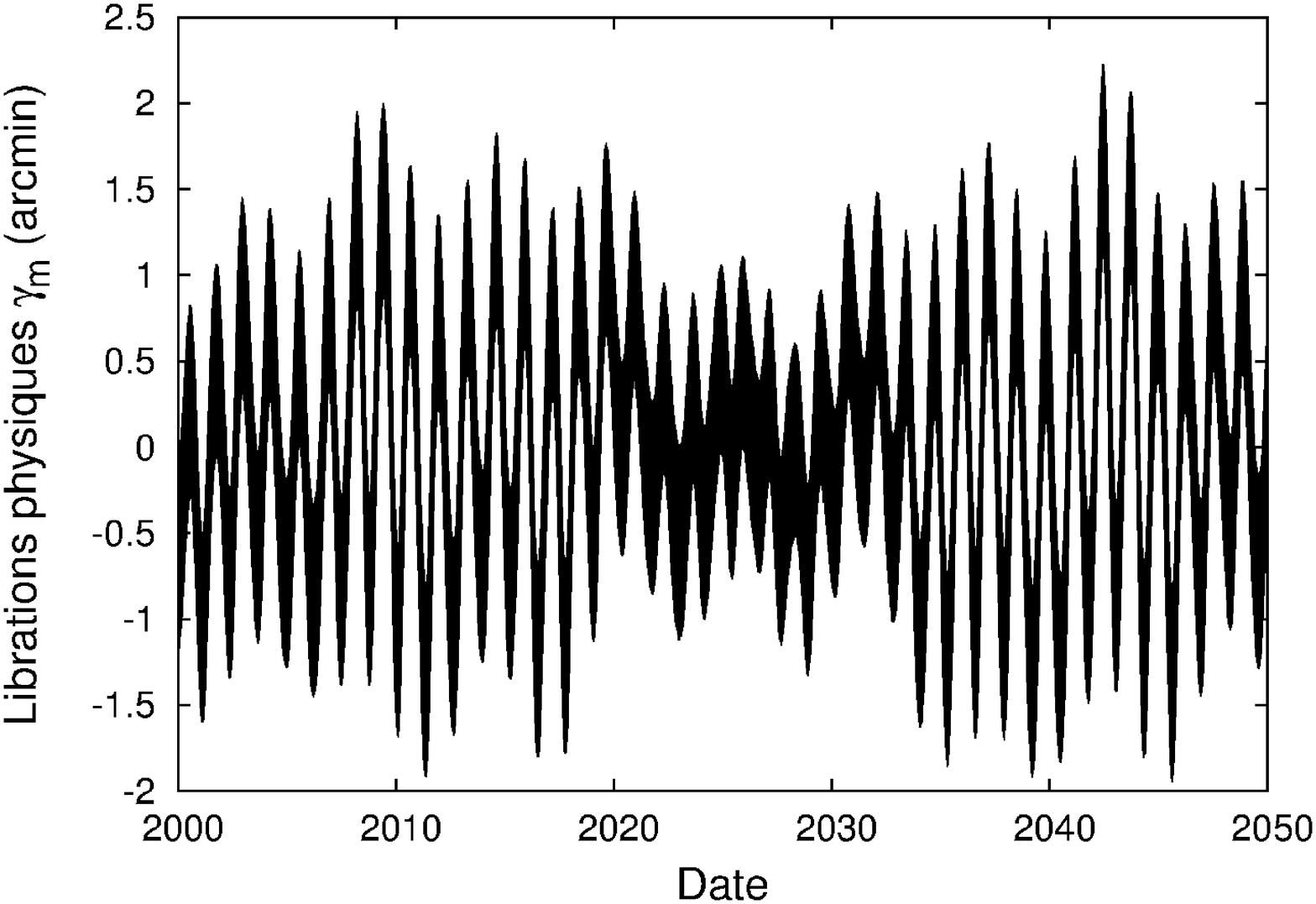} & \includegraphics[width=0.47\textwidth]{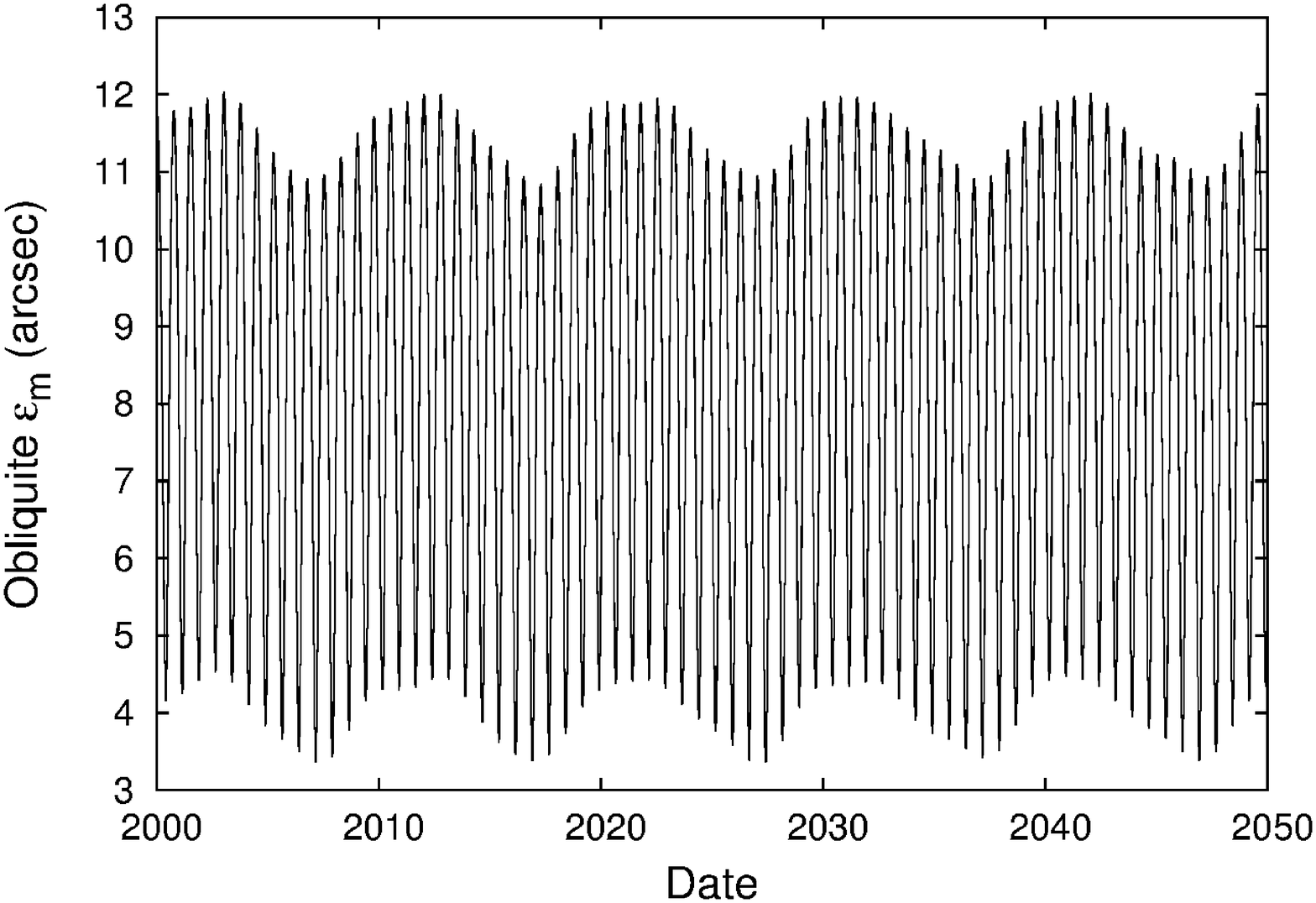} \\
   \includegraphics[width=0.47\textwidth]{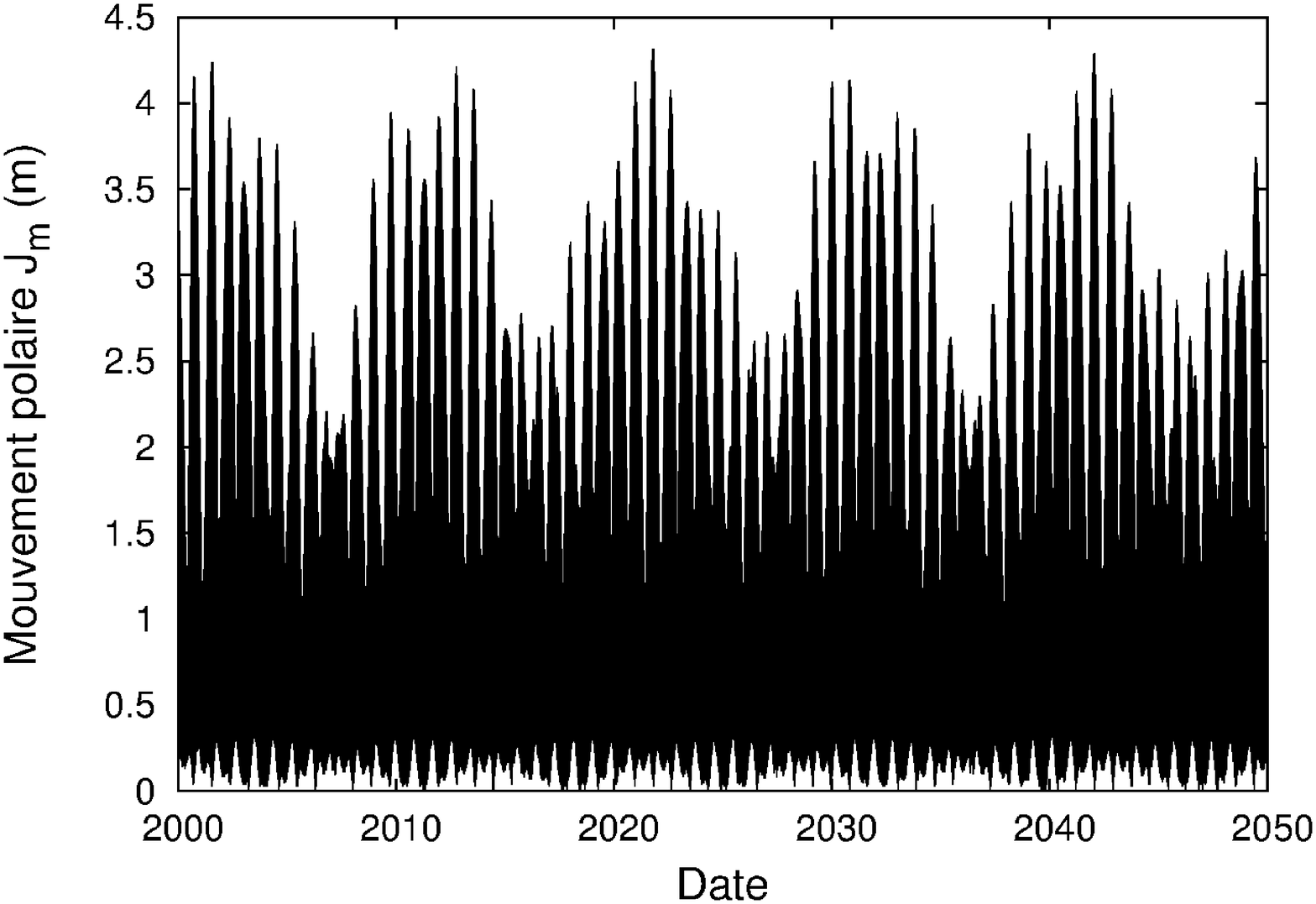} & \includegraphics[width=0.47\textwidth]{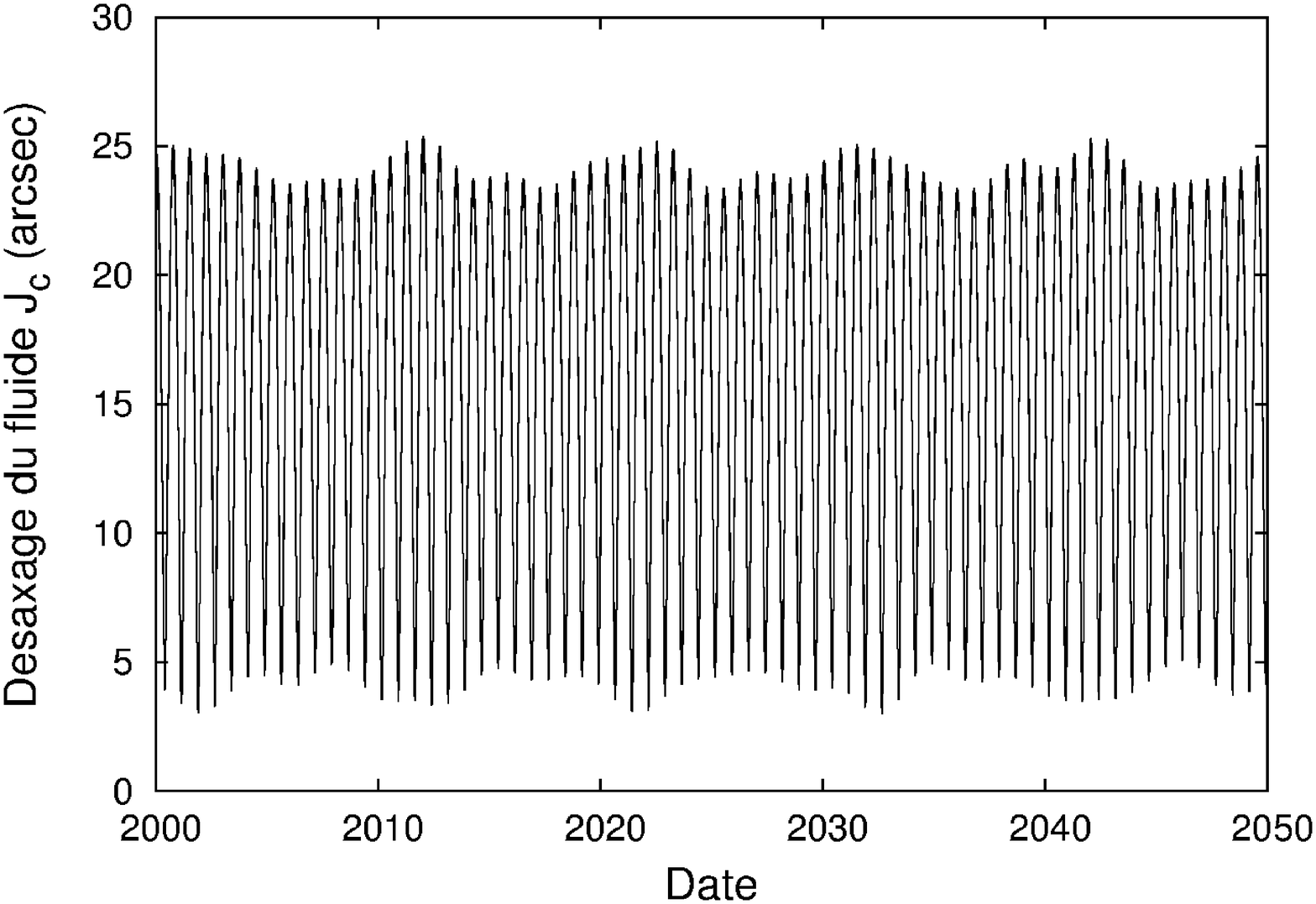}
   \end{tabular}
   \caption{La rotation du mod\`ele 2 (le noyau est un m\'elange eutectique de fer et de soufre).\label{fig:out2Io}}
  \end{figure}

 Pour repr\'esenter la rotation \`a l'aide de quantit\'es ayant un sens physique direct, j'ai consid\'er\'e les variables suivantes :
  
  \begin{itemize}
   \item librations physiques du manteau $\gamma_m$,
   \item obliquit\'e du manteau $\epsilon_m$ par rapport \`a la normale instantan\'ee \`a l'orbite,
   \item amplitude du mouvement polaire $J_m$,
   \item d\'esaxage du champ de vitesses dans le fluide $J_c$.
  \end{itemize}
L'\'evolution temporelle de ces variables est donn\'ee dans la Fig.\ref{fig:out1Io} pour le Mod\`ele 1 et dans la Fig.\ref{fig:out2Io} pour le Mod\`ele 2.
  
  \par On remarque effectivement que le d\'esaxage du fluide $J_c$ est bien plus important quand le noyau contient du soufre, mais les quantit\'es restent tr\`es faibles 
  (25 secondes d'arc contre 9), et donc leur implication plan\'etologique est probablement tr\`es limit\'ee. Le mouvement polaire du manteau est au maximum de $\approx4$
  m\`etres, ce qui est n\'egligeable, et l'obliquit\'e est au maximum de l'ordre de 10 secondes d'arc dans les 2 cas. Les librations physiques pr\'esentent des termes \`a
  longue p\'eriode, c'est pourquoi j'en propose une d\'ecomposition quasi-p\'eriodique, obtenue par analyse en fr\'equence (Tab.\ref{tab:gammam1Io}).
  

  \begin{table}[ht]
   \centering
   \caption[Librations physiques du Mod\`ele 1]{D\'ecomposition quasi-p\'eriodique des librations physiques du manteau $\gamma_m$. Les s\'eries sont en sinus.\label{tab:gammam1Io}}
   \begin{tabular}{r|rrrrr}
   \hline
      & Amplitude & Fr\'equence & Phase              & P\'eriode  & Identification \\
      & (arcsec)  &    (rad/an) & (J2000.0)          &    (j)     &                \\
  \hline
  $1$ & $-39.792$ &   $4.96186$ & $-172.463^{\circ}$ &  $462.514$ & $\lambda_1-2\lambda_2+\varpi_2$ \\
  $2$ & $-30.731$ &  $1301.919$ &   $60.879^{\circ}$ &  $1.76273$ & $2\lambda_1-2\lambda_2$ \\
  $3$ & $-20.139$ &   $4.76032$ &  $-66.209^{\circ}$ &  $482.097$ & $\lambda_1-2\lambda_2+\varpi_3$ \\
  $4$ & $-17.601$ &   $1.11413$ &  $151.694^{\circ}$ & $2059.835$ & $\psi$ \\
  \hline
   \end{tabular}
  \end{table}

  \par On peut voir que ces librations sont tr\`es fortement affect\'ees par les perturbations Europ\'eennes (indice $2$), ce qui illustre l'importance de 
  la r\'esonance laplacienne et de la quasi-r\'esonance $2:1$ associ\'ee. Le deuxi\`eme terme, de p\'eriode $1.76276$ jour, sera celui susceptible de donner
  des informations plan\'etologiques. D'une part, sa p\'eriode est la plus proche de la p\'eriode des oscillations propres $T_u$, donc son amplitude est la plus 
  sensible \`a l'int\'erieur, plus pr\'ecis\'ement \`a la taille du noyau (param\`etre $\delta=C_c/C$). D'autre part, sa courte p\'eriode en facilite la d\'etection.

  \begin{table}[ht]
   \centering
   \caption[Obliquit\'e du Mod\`ele 1]{D\'ecomposition quasi-p\'eriodique de l'obliquit\'e du manteau $\epsilon_m$ dans le Mod\`ele 1. Les s\'eries sont en cosinus.\label{tab:epsm1Io}}
   \begin{tabular}{r|rrrrr}
   \hline
       & Amplitude & Fr\'equence & Phase              & P\'eriode  & Identification \\
       & (arcsec)  &   (rad/an)  & (J2000.0)          &       (j)  &                \\
   \hline
   $1$ & $7.88143$ &        $0$ &        $0^{\circ}$  &    $\infty$ & cst \\
   $2$ & $2.35206$ & $8.374973$ &  $-15.890^{\circ}$  &   $274.023$ & $2\lambda_1-4\lambda_2+\Omega_1+\Omega_2$ \\
   $3$ & $0.50730$ & $0.637684$ &  $-73.351^{\circ}$  &  $3598.855$ & $\Omega_2-\Omega_1$ \\
   $4$ & $0.17966$ & $16.74995$ &  $148.219^{\circ}$  &   $137.011$ & $3\lambda_1-5\lambda_2-2\lambda_3+2\Omega_1+2\Omega_2$ \\
   $5$ & $0.07709$ & $7.737289$ & $-122.540^{\circ}$  &   $296.607$ & $\lambda_1-\lambda_2-2\lambda_3+2\Omega_1$ \\
   $6$ & $0.06353$ & $1.904935$ &   $35.863^{\circ}$  &  $1204.731$ & $2\lambda_{\sun}-\Omega_1-\Omega_{\sun}$ \\
   \hline
   \end{tabular}
  \end{table}

  \par La Tab. \ref{tab:epsm1Io} donne la d\'ecomposition quasi-p\'eriodique de l'obliquit\'e du manteau $\epsilon_m$. Il s'agit d'un petit signal, mais qui pr\'esente 
  des variations significatives de sa valeur moyenne. La Tab.\ref{tab:outputIo} rassemble les quantit\'es de rotation simul\'ees pour les 6 diff\'erents mod\`eles.
  
  \begin{table}[ht]
   \centering
   \caption{Les variables de rotation pour les diff\'erents mod\`eles.\label{tab:outputIo}}
    \begin{tabular}{r|rrrrrr}
    \hline
    Mod\`eles & $\gamma_m$ & $\gamma_m$ & $<J_m>$   & $<J_c>$    & $<\epsilon_m>$ & $\epsilon_m$ \\
              & ($462$ j)  & ($1.76$ j) &           &            &                & ($274$ j)    \\
              & (arcsec)   & (arcsec)   & (mas)     & (arcsec)   & (arcsec)       & (arcsec)     \\
   \hline
          $1$ & $39.82462$ & $30.73077$ & $136.592$ &  $5.89884$ & $7.88443$      & $2.35306$    \\
          $3$ & $39.82475$ & $30.60912$ & $135.689$ &  $5.52936$ & $7.87640$      & $2.33113$    \\
          $5$ & $39.82475$ & $30.58332$ & $138.148$ &  $5.41307$ & $7.87537$      & $2.32659$    \\
   \hline
          $2$ & $39.82303$ & $32.35491$ & $139.912$ & $15.82812$ & $8.14896$      & $3.60967$    \\
          $4$ & $39.82346$ & $31.89343$ & $140.550$ & $13.13468$ & $8.06166$      & $3.08910$    \\
          $6$ & $39.82345$ & $31.88809$ & $136.018$ & $12.78276$ & $8.04716$      & $3.07908$    \\
   \hline
   \end{tabular}
  \end{table}

  \par Ce dernier tableau confirme que le param\`etre d'int\'erieur le plus discriminant est bien la masse volumique du noyau $\rho_c$. Cependant, la 
  variation dans l'amplitude du terme \`a $1.76$ jour dans les librations est inf\'erieure \`a 2 secondes d'arc, soit $\approx18$ m\`etres. Les mesurer 
  avec une telle pr\'ecision n'est pas envisag\'e pour l'instant, leur d\'etection \'etant en soi un d\'efi.
  
  \par On peut, pour cette \'etude, parler de r\'esultat n\'egatif. C'est pourquoi il a fait l'objet d'une note. Je tiens \`a remercier Alessandro
  Morbidelli qui, agissant en sa qualit\'e d'\'editeur associ\'e d'\emph{Icarus}, en a accept\'e la publication. L'un des reviewers en avait en effet propos\'e le
  rejet, estimant qu'un r\'esultat n\'egatif avait plut\^ot sa place dans une revue plus sp\'ecialis\'ee.
  
  
  \section{Vers une mod\'elisation non lin\'eaire du fluide \ldots\label{sec:fluideturbulent}}
  
  \par Pour l'instant nous avons consid\'er\'e que le fluide avait le comportement le plus simple possible, c'est-\`a-dire que son champ de vitesse \'etait uniforme et
  que son moment cin\'etique \'etait de norme $P_cnC$ constante. Nous avons en fait n\'eglig\'e l'influence de comportements non lin\'eaires. Nous allons ici en discuter.
  
  \subsection{Pourquoi consid\'erer $P_c$ comme constant}
  
  \par C'est l'hypoth\`ese de $P_c=\delta nC$ constant qui permet d'obtenir une amplitude de librations diurnes qui d\'epend de la taille du noyau. Ceci semble confirm\'e
  par les observations des librations de la Lune \citep{k1967,wsek1973} et de Mercure \citep{mpjsh2007,mpshgjygpc2012}. Poser $P_c(t)=\delta P(t)$ revient \`a dire que le 
  noyau suit rigidement le manteau, on aurait ainsi le m\^eme comportement que si Io \'etait rigide.
  
  \par La condition de non-glissement indique qu'\`a l'interface noyau-manteau, le fluide doit suivre le mouvement du manteau. Il y a donc une fine couche turbulente, dite 
  couche d'Ekman, \`a la limite externe du fluide, qui sert de zone de transition entre le manteau et la zone o\`u la norme du moment cin\'etique est constante. L'\'epaisseur
  typique de la couche d'Ekman est $d=\sqrt{\nu/\Omega}$ \citep{g1968}, o\`u $\nu$ est la viscosit\'e cin\'ematique du fluide et $\Omega=n$ est sa fr\'equence de rotation.
  Pour Io, \citet{km1998} utilisent $\nu=10^{-6}$ m$^2$/s, ce qui est compatible avec m\'elange de fer et de soufre, ce qui donne $d=0.16$ m. Pour obtenir une couche d'Ekman
  d'un kilom\`etre, il faut $\nu=36$ m$^2$/s. En fait, la viscosit\'e ne devrait pas \^etre constante mais devrait augmenter \`a mesure qu'on se rapproche du centre du satellite
  \citep{rsuhwrs2002}. Pour l'anecdote, l'extremum de viscosit\'e qui ait \'et\'e mesur\'e est $\nu=(2.09\times10^5\pm4.6\times10^4)$ m$^2$/s pour le bitume, lors de 
  l'exp\'erience dite de la goutte de poix \citep{edp1984}. Cette exp\'erience est en cours depuis 1927 \`a l'Universit\'e du Queensland \`a Brisbane, QLD, Australie. Elle
  consiste \`a observer la formation et chute des gouttes de bitume. La neuvi\`eme goutte est tomb\'ee le 24 avril 2014. Cette exp\'erience a re\c{c}u le prix Ig Nobel de 
  physique en 2005.

  \subsection{Lien avec l'\'equation de Navier-Stokes}
  
  \par La dynamique d'un fluide est en fait r\'egie par l'\'equation de Navier-Stokes. Cette \'equation est une r\'ef\'erence d'\'equation aux d\'eriv\'ees partielles
  non lin\'eaires. Je la pr\'esente ici sous la forme donn\'ee par \citet{g1968} :
  
  \begin{equation}
   \label{eq:navierstokes}
   \frac{\partial}{\partial t}\vec{q}+\left(\vec{q}\cdot\vv{grad}\right)\vec{q}+2\vec{\omega}\times\vec{q}=
   -\frac{1}{\rho}\vv{grad}(p)-\nu\,\vv{rot}\left(\vv{rot}\left(\vec{q}\right)\right)-\vec{r}\times\frac{d\vec{\omega}}{dt},
  \end{equation}
o\`u elle s'applique \`a une particule de fluide, avec

  \begin{itemize}
  
   \item $\vec{q}$ : vitesse de la particule dans un rep\`ere tournant,
   
   \item $\vec{\omega}$ : vecteur rotation instantan\'ee du syst\`eme tournant, ses coordonn\'ees \'etant ($\omega_1$, $\omega_2$, $\omega_3$),
  
   \item $\rho$ : masse volumique du fluide,
   
   \item $\vec{r}$ : vecteur position de la particule dans le rep\`ere tournant,
   
   \item $p=P+\rho\mathcal{U}-\rho/2\left(\vec{\omega}\times\vec{r}\right)$ o\`u $P$ est la pression du fluide, et $\mathcal{U}$ un potentiel externe.
   $p$ est homog\`ene \`a une pression, on l'appelle \emph{pression r\'eduite},
   
   \item $\nu$ est la viscosit\'e cin\'ematique du fluide.
  
  \end{itemize}

  \par Nous avons, dans notre cas :
  
  \begin{equation}
   \label{eq:vecq}
   \vec{q}=\left(\begin{array}{c}
   (a/c)\nu_2x_3-(a/b)\nu_3x_2 \\
   (b/a)\nu_3x_1-(b/c)\nu_1x_3 \\
   (c/b)\nu_1x_2-(c/a)\nu_2x_1
   \end{array}\right).
  \end{equation}

  \par Dans un cas extr\^emement simplifi\'e o\`u on n\'eglige la viscosit\'e $\nu$, l'acc\'el\'eration convective $\left(\vec{q}\cdot\vv{grad}\right)\vec{q}$ et la pression
  r\'eduite $p$, la formule (\ref{eq:navierstokes}) s'\'ecrit
  
  \begin{equation}
    \label{eq:navierstokessimp}
    \frac{\partial}{\partial t}\vec{q}+2\vec{\omega}\times\vec{q}=\vec{0},
  \end{equation}
c'est-\`a-dire

\begin{eqnarray}
  \frac{d\nu_1}{dt}+2(\omega_2\nu_3-\omega_3\nu_2) & = & 0, \nonumber \\
  \frac{d\nu_2}{dt}+2(\omega_3\nu_1-\omega_1\nu_3) & = & 0, \label{eq:nv4} \\
  \frac{d\nu_3}{dt}+2(\omega_1\nu_2-\omega_2\nu_1) & = & 0. \nonumber
\end{eqnarray}

Pour comparaison, la formule (\ref{eq:dNsdt}) s'\'ecrit :

\begin{eqnarray}
  A\frac{d\omega_1}{dt}+D_1\frac{d\nu_1}{dt} & = & (B\omega_2+D_2\nu_2)\omega_3-(C\omega_3+D_3\nu_3)\omega_2, \nonumber \\
  B\frac{d\omega_2}{dt}+D_2\frac{d\nu_2}{dt} & = & (C\omega_3+D_3\nu_3)\omega_1-(A\omega_1+D_1\nu_1)\omega_3, \label{eq:nv5} \\
  C\frac{d\omega_3}{dt}+D_3\frac{d\nu_3}{dt} & = & (A\omega_1+D_1\nu_1)\omega_2-(B\omega_2+D_2\nu_2)\omega_1. \nonumber
\end{eqnarray}

Les syst\`emes d'\'equations (\ref{eq:nv4}) et (\ref{eq:nv5}) pr\'esentent des similarit\'es, la principale diff\'erence \'etant que les moments d'inertie apparaissent dans le
syst\`eme (\ref{eq:nv5}). Ces derni\`eres \'equations doivent \^etre consid\'er\'ees comme globales, c'est-\`a-dire qu'elles consid\`erent le noyau fluide dans son ensemble,
tandis que le syst\`eme (\ref{eq:nv4}) se rapporte \`a une particule de fluide.

  \subsection{L'instabilit\'e elliptique}
  
  \par L'instabilit\'e elliptique est un ph\'enom\`ene observ\'e exp\'erimentalement par \citet{p1986} et expliqu\'e pour la premi\`ere fois par \citet{b1986}, il s'agit 
  d'un ph\'enom\`ene r\'esonnant dans le fluide, provoqu\'e par un for\c{c}age p\'eriodique, qui produit des ondes inertielles dont l'amplitude peut cro\^itre 
  exponentiellement. Il en r\'esulte des turbulences dans le fluide.  Nous pouvons \'etudier la stabilit\'e du champ de vitesses dans le fluide \`a partir de la vitesse
  de croissance $\sigma$ pour une perturbation arbitraire $\vec{v}$ \citep{km1998} :
  
  \begin{equation}
   \label{eq:km1998}
   \sigma(t)=\frac{1}{2}\frac{d\ln<\vec{v}^2/2>}{dt}=-\frac{<\vec{v}\cdot\vec{\nabla} U\cdot\vec{v}>}{<\vec{v}^2>},
  \end{equation}
o\`u $\vec{U}$ est le champ de vitesses dans le fluide, a priori suppos\'e laminaire, et $<>$ est une moyennisation sur le volume. Le flux est stable pour $\sigma<0$, 
et instable sinon. \citet{clml2012} ont r\'ecemment propos\'e la formule suivante, dans le contexte d'un corps triaxial perturb\'e par un primaire, en n\'egligeant 
l'obliquit\'e et le mouvement polaire, et sous l'influence d'un champ magn\'etique :

  \begin{equation}
   \label{eq:cebron2012}
   \sigma=n\left(\frac{17}{64}\epsilon\beta-2.62(1-\eta)\frac{1+\eta^4}{1-\eta^5}\sqrt{E}-\frac{\Lambda}{16}\right),
  \end{equation}
avec

  \begin{itemize}
   \item $\epsilon$ : amplitude des librations physiques diurnes (cf. Tab.\ref{tab:outputIo}),
   \item $\beta=(a_c^2-b_c^2)/(a_c^2+b_c^2)$ est un param\`etre d'\'elongation \'equatoriale du noyau,
   \item $\eta$ : rapport entre les rayons interne et externe de la couche liquide. $\eta=0$ dans notre cas puisque nous n'avons pas de graine,
   \item $E=\nu/(nR_c^2)$ : nombre d'Ekman, une sorte de viscosit\'e adimensionn\'ee,
   \item $\Lambda=\sigma_eB_0^2/(n\rho_c)$, o\`u $\sigma_e$ est la conductivit\'e \'electrique du fluide et $B_0$ l'intensit\'e du champ magn\'etique : nombre d'Elsasser.
   Cette contribution du champ magn\'etique vient de la g\'en\'eralisation par \citet{clml2012} d'un r\'esultat de \citet{hll2009}.
  \end{itemize}

  \begin{table}[ht]
   \centering
   \caption[Instabilit\'e elliptique d'Io]{Vitesse de croissance $\sigma$, et le temps associ\'e.\label{tab:instabresult}}
   \begin{tabular}{r|rr}
   Mod\`eles & $\sigma$ (/an) & $1/\sigma$ (ans) \\
   \hline
      $1$ & $1.142\times10^{-3}$ &  $875$ \\
      $3$ & $1.561\times10^{-3}$ &  $640$ \\
      $5$ & $1.700\times10^{-3}$ &  $588$ \\
   \hline
      $2$ & $3.325\times10^{-4}$ &  $3008$ \\
      $4$ & $2.260\times10^{-4}$ &  $4425$ \\
      $6$ & $2.353\times10^{-4}$ &  $4250$ \\
   \end{tabular}
  \end{table}

  \par Les r\'esultats sont donn\'es dans la Tab.\ref{tab:instabresult}, et indiquent que $\sigma$ est toujours positif, autrement dit que le noyau d'Io doit \^etre sujet
  \`a l'instabilit\'e elliptique. N\'eanmoins, il est ``moins instable'' lorsque le noyau contient du soufre. Cette instabilit\'e pour Io a \'et\'e pour la premi\`ere fois
  sugg\'er\'ee par \citet{km1998}, et confirm\'ee par \citet{hll2009} en incluant l'effet du champ magn\'etique. Ces derniers auteurs obtiennent un temps caract\'eristique
  $1/\sigma$ d'environ $63$ ans, donc leur Io est bien plus instable que les miens. \citet{clml2012} trouvent en revanche un Io stable, c'est-\`a-dire $\sigma<0$. La diff\'erence
  entre leur \'etude et la mienne r\'eside principalement dans l'\'elongation \'equatoriale de l'interface noyau-manteau. Les miennes sont plus allong\'ees que la leur, ce qui 
  r\'esulte d'hypoth\`eses diff\'erentes sur la construction des mod\`eles d'int\'erieur.
  
  \par Ces d\'esaccords entre auteurs montrent que la question est toujours ouverte. D'ailleurs des travaux exp\'erimentaux \citep{nhwba2009} ainsi que th\'eoriques 
  \citep{br2013,v2014} sont toujours d'actualit\'e.
  
  \par Cette instabilit\'e du champ de vitesses dans le fluide visqueux est source de dissipation d'\'energie. En r\'egime continu, \citet{llllr2010} proposent 
  la formule suivante :
  
  \begin{equation}
   \label{eq:lebars2}
   \dot{E} = -\frac{8\pi}{3}\rho_cR_c^4\sqrt{\nu}\left|\Omega^s-\Omega^o\right|^{5/2}\left|\omega_{SO}\right|^{5/2},
  \end{equation}
o\`u $\Omega^s=nP$ est la fr\'equence de rotation du manteau, et $\omega_{SO}=1/P$ une fr\'equence normalis\'ee du mode dit de \emph{spin-over}, qui appara\^it en 
cas d'instabilit\'e. Avec $|\Omega^s-\Omega^o|=2e|\cos nt|$ et $e\approx 0.004$, j'obtiens une dissipation d'\'energie de $8.47\times10^8$ W pour le Mod\`ele 1, et
$2.32\times10^9$ W pour le Mod\`ele 2. \citet{llllr2010} l'estiment au maximum \`a $\approx4\times10^9$ W. Ces chiffres sont tr\`es faibles par rapport \`a ce que les
mar\'ees dissipent, soit $(9.33\pm1.84)\times10^{13}$ W \citep{lakv2009}. Un bilan \'energ\'etique est pr\'esent\'e dans la Tab.\ref{tab:energyIo}.

  \begin{table}[ht]
   \centering
   \caption[Bilan \'energ\'etique d'Io]{Bilan \'energ\'etique d'Io. Les mar\'ees sont la seule source d'\'energie significative et elles expliquent \`a elles 
   seules le flux de chaleur observ\'e\label{tab:energyIo}}
   \begin{tabular}{lrr}
   \hline
	Source de dissipation & $\dot{E}$ & R\'ef\'erence \\
   \hline
   Flux de chaleur \`a la surface & $8.33-10.83\times10^{13}$ W & \citet{rstmbt2004} \\
   Dissipation de mar\'ee & $(9.33\pm1.84)\times10^{13}$ W & \citet{lakv2009} \\
   Chauffage radiog\'enique & $3.08-5.14\times10^{11}$ W & \citet{hclmstv2010} \\
   Instabilit\'e elliptique & $8.47\times10^8-2.32\times10^9$ W & \citet{n2013} \\
   \hline
   \end{tabular}
  \end{table}
 
 \par Le probl\`eme des non lin\'earit\'es dans le fluide est pour l'instant tr\`es mal ma\^itris\'e. On ne sait pas si le flux dans le noyau d'Io est laminaire ou 
 turbulent. Par contre, il appara\^it clair qu'en r\'egime continu, ce ph\'enom\`ene ne peut pas contribuer significativement \`a la dissipation d'\'energie.
 
  \section{Conclusion}
  
  \par J'ai pr\'esent\'e ici le mod\`ele de Poincar\'e-Hough, qui permet de mod\'eliser la rotation d'un corps constitu\'e d'un manteau rigide et d'un noyau
  fluide, qui a un comportement laminaire. Il faut voir ce mod\`ele comme une approximation. S'il y a une graine, c'est-\`a-dire un noyau interne solide, elle
  doit \^etre suffisamment petite pour que son influence soit n\'egligeable. Il est pr\'ef\'erable dans ce cas d'avoir un manteau assez \'epais, donc d'inertie 
  importante. Dans ce cas, son \'elasticit\'e n'aura que tr\`es peu d'influence. Un tel mod\`ele semble valide pour Io et la Lune, et peut-\^etre aussi pour Mercure,
  en r\'esonance $3:2$.
  
  \par Ce mod\`ele permet de poser la question des effets des non-lin\'earit\'es dans le fluide. Il s'agit d'un probl\`eme hautement complexe, sur le plan
  math\'ematique il consiste ni plus ni moins \`a r\'esoudre l'\'equation de Navier-Stokes. Plusieurs pistes existent, comme les exp\'erimentations, les
  simulations num\'eriques par \'el\'ements finis, ou encore l'\'etude analytique d'\'equations simplifi\'ees au voisinage d'un ph\'enom\`ene pr\'ecis.
  
  \par La communaut\'e scientifique est de plus en plus pri\'ee de s'int\'eresser aux lieux potentiellement habitables, comme le sont les oc\'eans sous-surfaciques. C'est
  par exemple une priorit\'e de la prospective Cosmic Vision. Plusieurs d'entre eux ont \'et\'e d\'etect\'es ou sugg\'er\'es, pour Titan, Europe, Ganym\`ede, ou encore Callisto \ldots
  Le mod\`ele de Poincar\'e-Hough est devenu insuffisant pour ces corps, soumis notamment \`a un couplage gravitationnel entre la graine et la cro\^ute. Nous nous y int\'eressons
  dans le chapitre qui arrive.
  
  \chapter{La pr\'esence d'un oc\'ean global\label{chap:oceanglobal}}
  
  \section{Introduction}
  
  \par Plusieurs grands satellites naturels dans le Syst\`eme Solaire sont soup\c{c}onn\'es d'abriter un oc\'ean global sous-surfacique. La principale raison est leur taille, 
  qui implique une certaine diff\'erenciation. Ici, les \'el\'ements lourds migrent vers le centre, et la surface est essentiellement compos\'ee d'eau glac\'ee. Le gradient de
  pression dans le satellite implique qu'\`a une certaine profondeur, l'eau subsiste sous forme liquide. Le chauffage de mar\'ee peut contribuer \`a maintenir cet \'etat liquide, 
  comme il semblerait que ce soit le cas pour Europe \citep{rs1987,hsw2002} et pour Encelade \citep{rn2008,t2011,m2013}. Dans le cas de Titan, l'oc\'ean global sous-surfacique a \'et\'e
  pr\'edit par de nombreuses \'etudes th\'eoriques \citep{ls1987,gs1996,gsd2000,tglms2005,fgtv2007} et pourrait avoir \'et\'e d\'etect\'e indirectement par sa signature dans
  le champ \'electromagn\'etique de l'atmosph\`ere \citep{sghlsbbbbcfffhjjmrstt2007,bsh2010,brhkswbgs2012}, ainsi que par un nombre de Love $k_2$ de l'ordre de $0.6$ \citep{ijdslaarrt2012}.
  Les oc\'eans internes font des corps concern\'es des cibles prioritaires pour les futures programmes spatiaux, car ils sont susceptibles d'abriter des formes de
  vie bact\'eriologique \citep{lbckkgpreywgsckgr2009,gdcbetbcdfhjklpttv2013}.
  
  \par
\begin{figure}[ht]
   \centering
   \begin{tabular}{m{.47\textwidth} m{.47\textwidth}}
    \includegraphics[height = 0.3\textheight,width=0.47\textwidth]{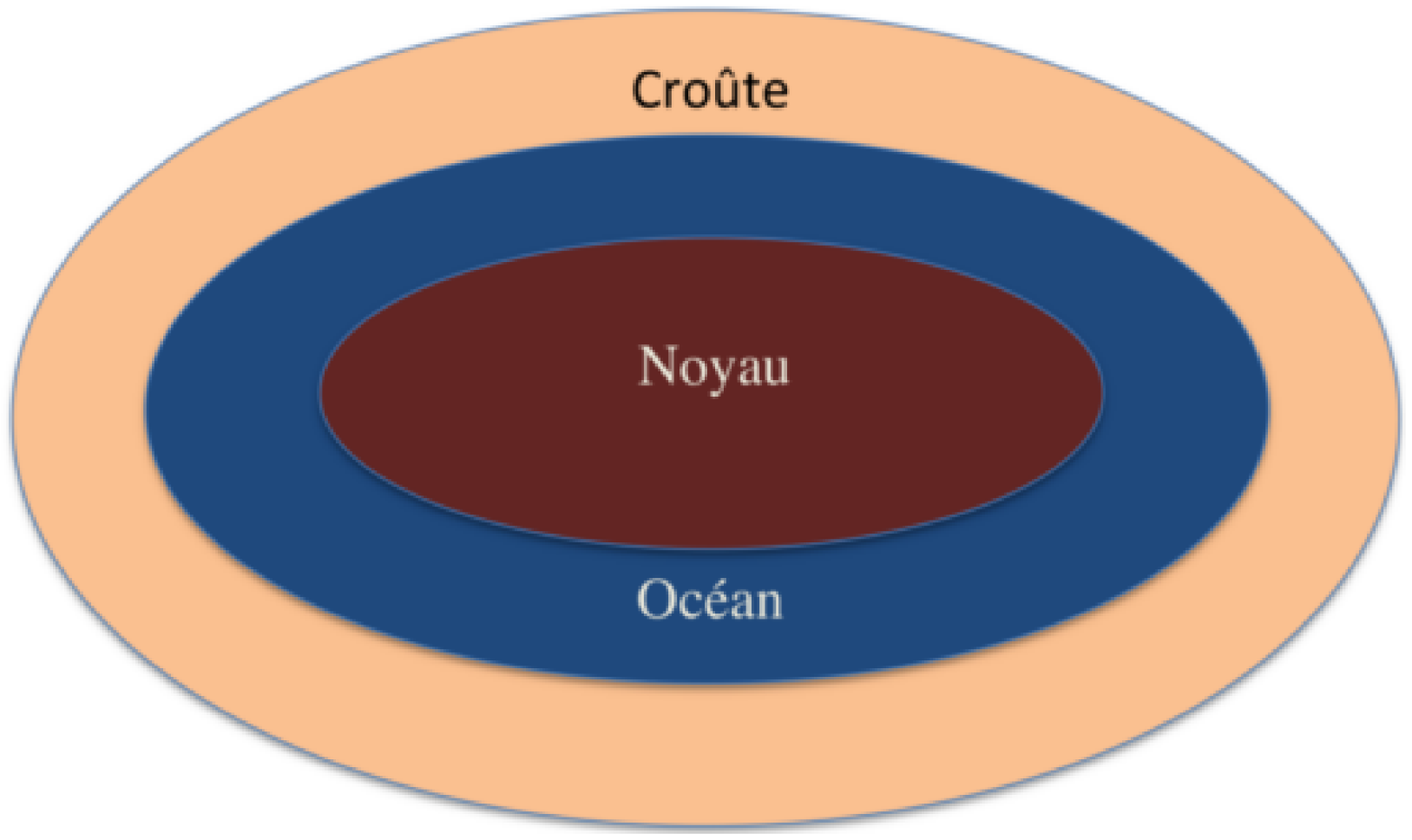} & \includegraphics[width=0.47\textwidth]{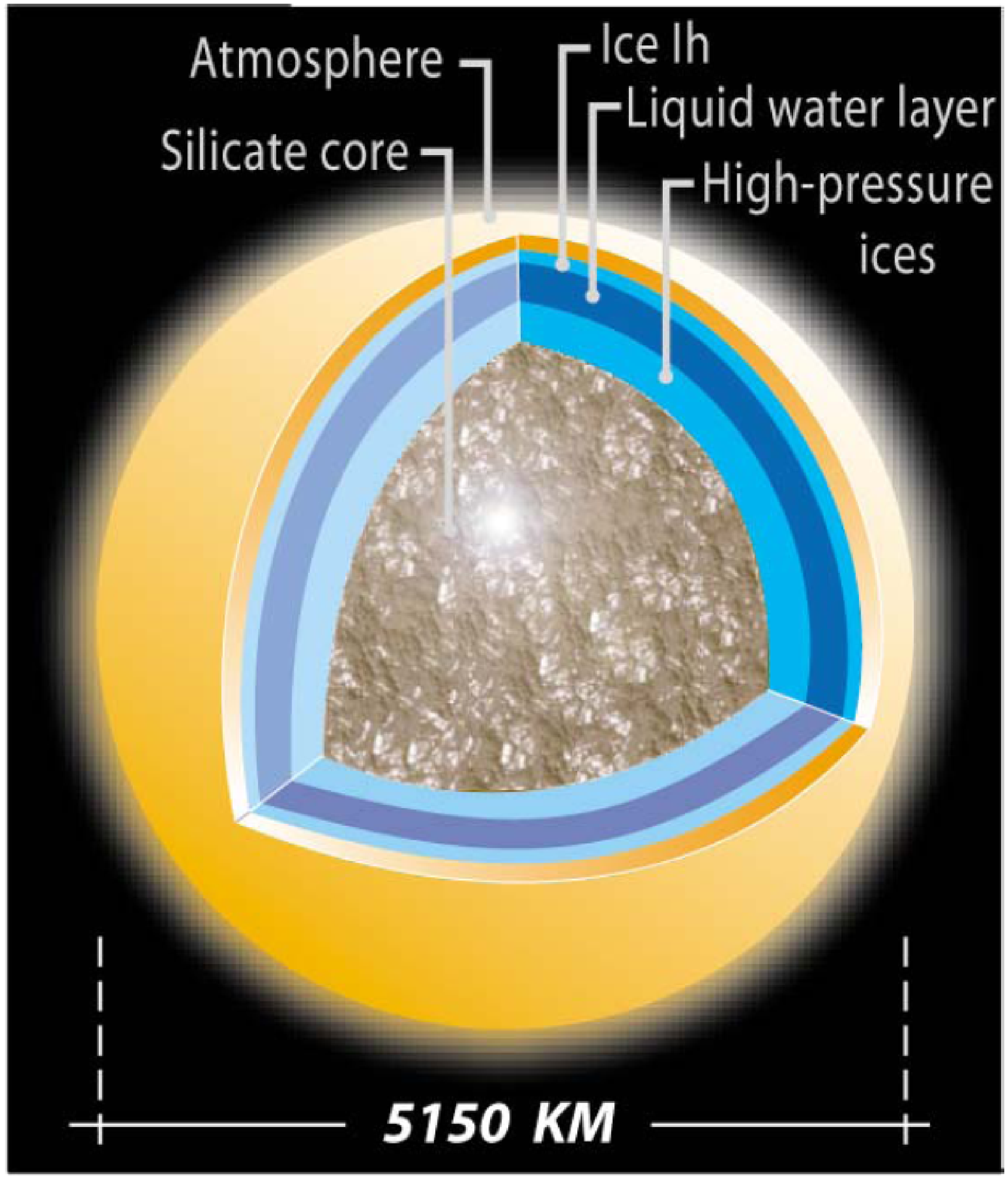}
   \end{tabular}
   \caption[Mod\`ele d'int\'erieur de Titan]{Mod\`eles d'int\'erieur de Titan. \`A gauche : notre mod\`ele. \`A droite : l'int\'erieur de Titan vu par 
   \citet{tglms2005}.\label{fig:interiortitan}}
  \end{figure}

 D'un point de vue purement dynamique, une couche fluide globale situ\'ee entre 2 couches solides tend \`a les d\'ecoupler l'une de l'autre. Il est donc n\'ecessaire
  de construire un mod\`ele dynamique tenant compte de cet oc\'ean. Je consid\`ere ici le satellite comme un ellipso\"ide triaxial, constitu\'e de 3 couches concentriques : un noyau
  rigide interne, un oc\'ean global fluide non visqueux, et une cro\^ute rigide (Fig.\ref{fig:interiortitan}). Le cas o\`u la cro\^ute est d\'eformable est abord\'e \`a la 
  Sect.\ref{sec:elastic}. Chacune de ces couches a une densit\'e constante.
  
  \section{Mise en \'equations du probl\`eme}
  
  \par Je pr\'esente ici des \'equations publi\'ees dans \citep{nn2014}. Elles sont inspir\'ees de \citep{sx1997} et de \citep{bvyk2011}. Le mod\`ele d'int\'erieur utilise 12
  param\`etres, c'est-\`a-dire 3 rayons et une masse volumique pour chaque couche. Ils sont :
  
  \begin{itemize}
  
   \item $\rho_c$, $\rho_o$, $\rho_s$ : masses volumiques du noyau (\emph{core}), de l'oc\'ean et de la cro\^ute (\emph{shell}),
   
   \item $a_c$, $b_c$, $c_c$ : rayons de l'interface noyau-oc\'ean
   
   \item $a_o$, $b_o$, $c_o$ : rayons de l'interface oc\'ean-cro\^ute,
   
   \item $a$, $b$, $c$ : rayons de la surface du satellite.
   
  \end{itemize}
  
  \par Contrairement \`a pr\'ec\'edemment, le formalisme Hamiltonien n'est pas utilis\'e. L'objectif \'etait d'\'ecrire un syst\`eme d'\'equations appropri\'e
  \`a la description du comportement du syst\`eme. Le passage en formalisme Hamiltonien sera pour une prochaine \'etape.
  
  \subsection{Param\'etrisation du probl\`eme}
  
  \par
\begin{figure}[ht]
  \centering
  \includegraphics[width = 0.6\textwidth]{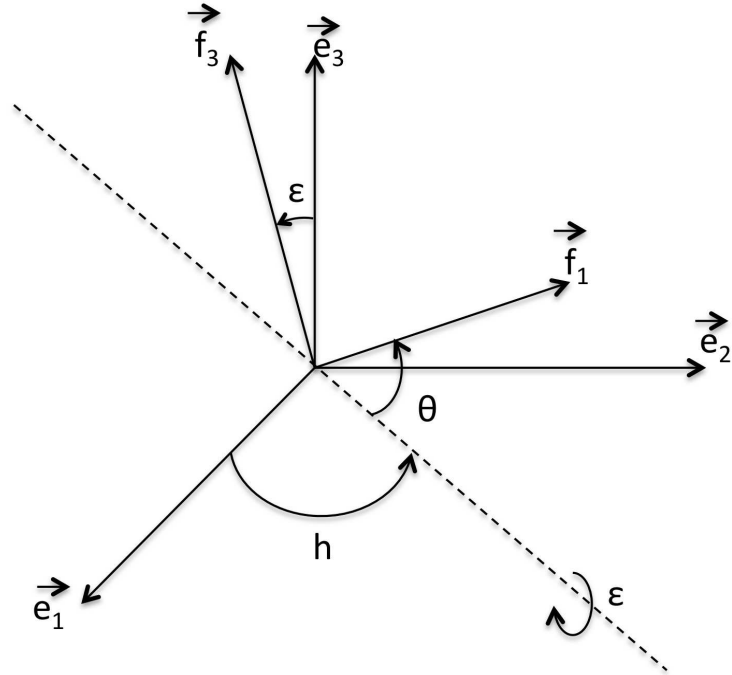}
  \caption[Angles d'Euler]{Les angles d'Euler utilis\'es dans notre probl\`eme.\label{fig:eulertitan}}
  \end{figure}

 \citet{bvyk2011} ont montr\'e que le bilan des forces sur l'oc\'ean \'etait nul, si on faisait l'approximation qu'il \'etait \`a l'\'equilibre hydrostatique. On doit
  donc mod\'eliser l'orientation des 2 couches rigides, \`a savoir le noyau et la cro\^ute. Pour \c{c}a, on a besoin de 2 jeux d'angles d'Euler, $(h^c,\epsilon^c,\theta^c)$
  et $(h^s,\epsilon^s,\theta^s)$ qui vont rep\'erer respectivement l'orientation des axes principaux d'inertie du noyau $(\vv{f_1^c},\vv{f_2^c},\vv{f_3^c})$ et de la cro\^ute
  $(\vv{f_1^s},\vv{f_2^s},\vv{f_3^s})$ par rapport au rep\`ere inertiel $(\vv{e_1},\vv{e_2},\vv{e_3})$ (Fig.\ref{fig:eulertitan}).

  \par Ces angles d'Euler pr\'esentent une singularit\'e virtuelle : lorsque l'obliquit\'e $\epsilon$ est nulle, on ne peut pas distinguer de fa\c{c}on unique les angles $h$ et $\theta$,
  alors que la somme $h+\theta$ est toujours d\'efinie. Pour contourner ce probl\`eme, nous introduisons des coordonn\'ees r\'egularis\'ees :
  
  \begin{eqnarray}
	\xi^{s,c}  & = & 2\sin\frac{\epsilon^{s,c}}{2}\sin h^{s,c}, \nonumber \\
	\eta^{s,c} & = & 2\sin\frac{\epsilon^{s,c}}{2}\cos h^{s,c}, \nonumber \\
	p^{s,c}    & = & h^{s,c}+\theta^{s,c}.	\nonumber
  \end{eqnarray}
  Lorsque l'obliquit\'e $\epsilon$ est nulle, alors $\eta$ et $\xi$ sont nuls aussi. Ils restent donc d\'efinis et ne pr\'esentent plus de singularit\'e.
  
  \par Pour repr\'esenter le moment cin\'etique $\vec{G}$ de chacune de ces couches, nous utiliserons comme variables les 3 composantes du vecteur rotation 
  de la couche consid\'er\'ee $\vec{\omega}$ dans le rep\`ere des axes principaux d'inertie de cette m\^eme couche. On a ainsi
  
  \begin{equation}
    \label{eq:Gs}
    \vv{G^s} = A^s\omega_1^s\vv{f_1^s}+B^s\omega_2^s\vv{f_2^s}+C^s\omega_3^s\vv{f_3^s}
  \end{equation}
pour la cro\^ute et 

  \begin{equation}
    \label{eq:Gc}
    \vv{G^c} = A^c\omega_1^c\vv{f_1^c}+B^c\omega_2^c\vv{f_2^c}+C^c\omega_3^c\vv{f_3^c}
  \end{equation}
pour le noyau. $A$, $B$ et $C$ sont les moments principaux d'inertie de la couche consid\'er\'ee, d\'efinis comme pr\'ec\'ed\'emment (Eq.\ref{eq:momA} \`a \ref{eq:momC}).

  \subsection{Les \'equations cin\'ematiques}
  
  \par Pour d\'eterminer les \'equations r\'egissant la variation des angles d'Euler en fonction des composantes des vecteurs rotation $\vec{\omega^s}$ et $\vec{\omega^c}$, nous
  partons de la g\'eom\'etrie du probl\`eme, en nous inspirant de \citep{fc1999}.
  
  \par Le vecteur rotation $\vec{\omega}=\omega_1\vec{f_1}+\omega_2\vec{f_2}+\omega_3\vec{f_3}$ repr\'esente 3 rotations successives du rep\`ere inertiel 
  $(\vv{e_1},\vv{e_2},\vv{e_3})$ au rep\`ere de figure $(\vv{f_1},\vv{f_2},\vv{f_3})$ (cf. Fig.\ref{fig:eulertitan}):
  
  \begin{enumerate}
  
   \item une rotation d'axe $z$ (ici $\vv{e_3}$) d'angle $h$,
   
   \item une rotation autour de l'axe $x$ provisoire,
   
   \item une rotation autour de l'axe z final, c'est-\`a-dire l'axe de figure $\vv{f_3}$, d'angle $\theta$.
   
  \end{enumerate}
Ceci s'\'ecrit math\'ematiquement

\begin{equation}
	\label{eq:kinetic1}
	\left(\begin{array}{c}
	\omega_1 \\
	\omega_2 \\
	\omega_3 
	\end{array}\right) = 
	\left(\begin{array}{c}
	0 \\
	0 \\
	\dot{\theta} 
	\end{array}\right)+R_3(-\theta)\left(\begin{array}{c}
	\dot{\epsilon} \\
	0 \\
	0
	\end{array}\right)+R_3(-\theta)R_1(-\epsilon)\left(\begin{array}{c}
	0 \\
	0 \\
	\dot{h} 
	\end{array}\right),
\end{equation}
les matrices de rotation $R_1$ et $R_3$ ayant \'et\'e pr\'ec\'edemment d\'efinies (Eq. \ref{eq:R1} \& \ref{eq:R3}). Ceci donne

\begin{eqnarray}
	\omega_1 & = & \dot{\epsilon}\cos\theta+\dot{h}\sin\epsilon\sin\theta, \label{eq:omega1t} \\
	\omega_2 & = & -\dot{\epsilon}\sin\theta+\dot{h}\sin\epsilon\cos\theta, \label{eq:omega2t} \\
	\omega_3 & = & \dot{\theta}+\dot{h}\cos\epsilon, \label{eq:omega3t}
\end{eqnarray}
et donc

\begin{eqnarray}
	\dot{h} & = & \frac{\omega_1\sin\theta+\omega_2\cos\theta}{\sin\epsilon}, \label{eq:doth} \\
	\dot{\epsilon} & = & \omega_1\cos\theta-\omega_2\sin\theta, \label{eq:dotepsilon} \\
	\dot{\theta} & = & \omega_3-\frac{\omega_1\sin\theta+\omega_2\cos\theta}{\tan\epsilon}. \label{eq:dottheta}
\end{eqnarray}

\par Les \'equations (\ref{eq:doth}) et (\ref{eq:dottheta}) illustrent la singularit\'e virtuelle induite par les angles d'Euler. Ces formules sont les m\^emes 
que dans \citep{bnt1995,htn2011} mais sont pr\'esent\'ees avec un signe oppos\'e dans d'autres \'etudes comme \citep{wbyrd2001}. En variables r\'eguli\`eres, 
ceci donne

\begin{eqnarray}
	\dot{\xi}  & = & \omega_1\left(\cos\theta\cos\frac{\epsilon}{2}\sin h+\frac{\sin\theta\cos h}{\cos\left(\epsilon/2\right)}\right)
                        +\omega_2\left(\frac{\cos\theta\cos h}{\cos\left(\epsilon/2\right)}-\sin\theta\cos\frac{\epsilon}{2}\sin h\right), \label{eq:dotxi} \\
	\dot{\eta} & = & \omega_1\left(\cos\theta\cos\frac{\epsilon}{2}\cos h-\frac{\sin\theta\sin h}{\cos\left(\epsilon/2\right)}\right)
                        -\omega_2\left(\sin\theta\cos\frac{\epsilon}{2}\cos h+\frac{\cos\theta\sin h}{\cos\left(\epsilon/2\right)}\right), \label{eq:doteta} \\
	\dot{p}    & = & \omega_3+\left(\omega_1\sin\theta+\omega_2\cos\theta\right)\tan\frac{\epsilon}{2}, \label{eq:dotp}
\end{eqnarray}
et la singularit\'e virtuelle a disparu.

  \subsection{Les forces intervenant}
  
  \par Les forces que nous consid\'erons sont :
  
  \begin{itemize}
  
   \item la force gravitationnelle de Saturne sur chacune des couches,
   
   \item les forces de pression aux 2 interfaces fluide-solide,
   
   \item le couplage gravitationnel entre le noyau et la cro\^ute, qui tend \`a aligner les axes de ces 2 couches rigides.
   
  \end{itemize}

  \par Le mod\`ele pourrait \^etre encore complexifi\'e. Notamment, comme pr\'ec\'edemment (Chap.\ref{chap:poincarehough}) nous ne tenons pas compte d'effets non lin\'eaires
  dus \`a la viscosit\'e du fluide. Notre \'etude doit donc \^etre consid\'er\'ee comme pr\'eliminaire \`a d'autres \'etudes plus compl\`etes sur le sujet. D\'etaillons
  maintenant ces diff\'erentes forces.
  
  \subsubsection{Le for\c{c}age gravitationnel de Saturne}
  
  \par Pour un corps rigide de moments principaux d'inertie $A$, $B$ et $C$, le couple gravitationnel d\^u \`a la perturbation de Saturne s'\'ecrit classiquement
  
  \begin{equation}
	\label{eq:pullrigid}
	\vv{\Gamma}_{\saturn} = \frac{3\mathcal{G}M_{\saturn}}{\|\vv{r_{\saturn}}\|^5}\left((C-B)y_{\saturn}z_{\saturn}\vec{f_1}+(A-C)x_{\saturn}z_{\saturn}\vec{f_2}+(B-A)x_{\saturn}y_{\saturn}\vec{f_3}\right),
  \end{equation}
  o\`u le vecteur $\vv{r_{\saturn}}=x_{\saturn}\vv{f_1}+y_{\saturn}\vv{f_2}+z_{\saturn}\vv{f_3}$ repr\'esente la position du perturbateur dans le rep\`ere de figure du corps dont
  on \'etudie la rotation. Dans notre cas, les couples de Saturne s'appliquant sur le noyau et la cro\^ute s'\'ecrivent
  
  \begin{eqnarray}
	\vv{\Gamma^c_{\saturn}} & = & \frac{3\mathcal{G}M_{\saturn}}{\|\vv{r^c_{\saturn}}\|^5}\left((C^c-B^c)y^c_{\saturn}z^c_{\saturn}\vv{f^c_1}+(A^c-C^c)x^c_{\saturn}z^c_{\saturn}\vv{f^c_2}+(B^c-A^c)x^c_{\saturn}y^c_{\saturn}\vv{f^c_3}\right), \label{eq:pullcore} \\
	\vv{\Gamma^s_{\saturn}} & = & \frac{3\mathcal{G}M_{\saturn}}{\|\vv{r^s_{\saturn}}\|^5}\left((C^s-B^s)y^s_{\saturn}z^s_{\saturn}\vv{f^s_1}+(A^s-C^s)x^s_{\saturn}z^s_{\saturn}\vv{f^s_2}+(B^s-A^s)x^s_{\saturn}y^s_{\saturn}\vv{f^s_3}\right). \label{eq:pullshell}
  \end{eqnarray}

  \subsubsection[Le couplage gravitationnel]{Le couplage gravitationnel entre les 2 couches rigides}

  \par Nous nous basons ici sur l'\'etude de \citet{sx1997} qui mod\'elisait, pour la Terre, le couplage entre la cro\^ute et le noyau, en n\'egligeant toute influence 
  du fluide remplissant l'espace entre ces 2 couches. La triaxialit\'e \'etait prise en compte dans cette \'etude.
  
  \par Le couple gravitationnel de la cro\^ute agissant sur le noyau s'\'ecrit
  
  \begin{equation}
	\label{eq:torquesx}
	\vv{\Gamma^c_{sh}} = \iiint_{noyau}\rho_c \vec{r}\times\vec{\nabla}\Phi \,\textrm{d}V,
  \end{equation}
  o\`u $\vec{r}$ pointe vers un \'el\'ement de masse du noyau, et $\Phi$ est le potentiel gravitationnel de la cro\^ute donn\'e par
  
  \begin{equation}
	\label{eq:phisx}
	\Phi = \alpha+\beta r^2P_2\left(\cos\psi\right)+\gamma r^2P_2^2\left(\cos\psi\right)\cos 2\phi+\mathcal{O}(f_{1,2}^2,\kappa_{1,2}^2),
\end{equation}
o\`u

\begin{itemize}

	\item $\alpha$ est une constante qui n'affectera pas le r\'esultat,
	
	\item $\beta=\frac{4\pi}{15}\mathcal{G}\rho_s\left(\kappa_2-\kappa_1-2(f_2-f_1)\right)$,
		
	\item $\gamma=\frac{2\pi}{15}\mathcal{G}\rho_s(\kappa_2-\kappa_1)$,
		
	\item $f_1=(a_o-c_o)/a_o$ est l'aplatissement polaire de l'interface cro\^ute-oc\'ean,
		
	\item $f_2=(a-c)/a$ est l'aplatissement polaire de la surface du satellite,
		
	\item $\kappa_1=(a_o-b_o)/a_o$ est l'ellipticit\'e \'equatoriale de l'interface cro\^ute-oc\'ean,
		
	\item $\kappa_2=(a-b)/a$ est l'ellipticit\'e \'equatoriale de la surface de Titan,
		
	\item $P_2(x)=(3x^2-1)/2$ est le polyn\^ome de Legendre d'ordre 2,
		
	\item $P_2^2(x)=3(1-x^2)$ est une fonction associ\'ee de Legendre,
		
	\item $\psi$ et $\phi$ sont respectivement la colatitude et la longitude compt\'ee positivement vers l'est de l'\'el\'ement de masse consid\'er\'e, dans le rep\`ere 
	de figure de la cro\^ute $\left(\vv{f_1^s},\vv{f_2^s},\vv{f_3^s}\right)$.
	
\end{itemize}

On a 

  \begin{eqnarray}
	\vec{r} & = & X\vv{f_1^s}+Y\vv{f_2^s}+Z\vv{f_3^s} \nonumber \\
	& = & x\vv{f_1^c}+y\vv{f_2^c}+z\vv{f_3^c} \nonumber
\end{eqnarray}	
avec

\begin{equation}
	\label{eq:l1n3}
	\left(\begin{array}{c}
	X \\
	Y \\
	Z \end{array}\right) = \left(\begin{array}{ccc}
	l_1 & l_2 & l_3 \\
	m_1 & m_2 & m_3 \\
	n_1 & n_2 & n_3 \end{array}\right)\left(\begin{array}{c}
	x \\
	y \\
	z \end{array}\right)
\end{equation}
et

\begin{equation}
	\label{eq:defl1n3}
	\left(\begin{array}{ccc}
	l_1 & l_2 & l_3 \\
	m_1 & m_2 & m_3 \\
	n_1 & n_2 & n_3 \end{array}\right) = R_3(-\theta^s)R_1(-\epsilon^s)R_3(h^c-h^s)R_1(\epsilon^c)R_3(\theta^c).
\end{equation}

\par Ici, $(X,Y,Z)$ et $(x,y,z)$ sont les coordonn\'ees cart\'esiennes de l'\'el\'ement de masse exprim\'e respectivement dans les rep\`eres de figure de la 
cro\^ute et du noyau. \`A partir de

\begin{eqnarray}
	r^2 & = & X^2+Y^2+Z^2, \nonumber \\
	\cos\psi & = & Z/\sqrt{X^2+Y^2+Z^2}, \nonumber \\
	\sin^2\psi\cos 2\phi & = & (X^2-Y^2)/(X^2+Y^2+Z^2), \nonumber \\
	P_2(\cos\psi) & = & (2Z^2-X^2-Y^2)/(2(X^2+Y^2+Z^2)), \nonumber \\
	P_2^2(\cos\psi)\cos 2\phi & = & 3(X^2-Y^2)/(X^2+Y^2+Z^2), \nonumber
\end{eqnarray}
on obtient

\begin{equation}
	\label{eq:potPhi}
	\Phi = \beta(2Z^2-X^2-Y^2)/2+3\gamma(X^2-Y^2).
\end{equation}

\par Apr\`es expression du potentiel $\Phi$ dans le rep\`ere du noyau, d\'erivation pour obtenir le gradient $\vec{\nabla}\Phi$ et le produit vectoriel 
$\vec{r}\times\vec{\nabla}\Phi$, et \'elimination des termes crois\'es par sommation, on obtient :

\begin{eqnarray}
	\label{eq:torquesx2}
	\vv{\Gamma^c_{sh}} & = & \beta \left[\begin{array}{c}
	(-2n_2n_3+m_2m_3+l_2l_3)\iiint_{noyau}\rho_c\left(z^2-y^2\right)\,\textrm{d}x\,\textrm{d}y\,\textrm{d}z \vv{f_1^c} \\
	(2n_1n_3-m_1m_3-l_1l_3)\iiint_{noyau}\rho_c\left(z^2-x^2\right)\,\textrm{d}x\,\textrm{d}y\,\textrm{d}z \vv{f_2^c} \\
	(-2n_1n_2+m_1m_2+l_1l_2)\iiint_{noyau}\rho_c\left(y^2-x^2\right)\,\textrm{d}x\,\textrm{d}y\,\textrm{d}z \vv{f_3^c} \end{array}\right] \nonumber \\
	& + & 6\gamma\left[\begin{array}{c}
	(m_2m_3-l_2l_3)\iiint_{noyau}\rho_c\left(z^2-y^2\right)\,\textrm{d}x\,\textrm{d}y\,\textrm{d}z \vv{f_1^c} \\
	(l_1l_3-m_1m_3)\iiint_{noyau}\rho_c\left(z^2-x^2\right)\,\textrm{d}x\,\textrm{d}y\,\textrm{d}z \vv{f_2^c} \\
	(m_1m_2-l_1l_2)\iiint_{noyau}\rho_c\left(y^2-x^2\right)\,\textrm{d}x\,\textrm{d}y\,\textrm{d}z \vv{f_3^c} \end{array}\right].
\end{eqnarray}

Le passage de $\left(\vv{f_1^s},\vv{f_2^s},\vv{f_3^s}\right)$ \`a $\left(\vv{f_1^c},\vv{f_2^c},\vv{f_3^c}\right)$ \'etant une transformation orthonormale, on a

\begin{eqnarray}
	l_1l_2+m_1m_2+n_1n_2 & = & 0, \nonumber \\
	l_1l_3+m_1m_3+n_1n_3 & = & 0, \nonumber \\
	l_2l_3+m_2m_3+n_2n_3 & = & 0. \nonumber
\end{eqnarray}
En utilisant aussi

\begin{eqnarray}
	\iiint_{noyau}\rho_c\left(z^2-y^2\right)\,\textrm{d}x\,\textrm{d}y\,\textrm{d}z & = & B^c-C^c, \nonumber \\
	\iiint_{noyau}\rho_c\left(z^2-x^2\right)\,\textrm{d}x\,\textrm{d}y\,\textrm{d}z & = & A^c-C^c, \nonumber \\
	\iiint_{noyau}\rho_c\left(y^2-x^2\right)\,\textrm{d}x\,\textrm{d}y\,\textrm{d}z & = & A^c-B^c, \nonumber
\end{eqnarray}
on obtient finalement

\begin{eqnarray}
	\label{eq:torquesx3}
	\vv{\Gamma^c_{sh}} & = & 3\beta \left[\begin{array}{c}
	n_2n_3(C^c-B^c) \vv{f_1^c} \\
	-n_1n_3(C^c-A^c) \vv{f_2^c} \\
	n_1n_2(B^c-A^c) \vv{f_3^c} \end{array}\right] + 6\gamma\left[\begin{array}{c}
	(l_2l_3-m_2m_3)(C^c-B^c) \vv{f_1^c} \\
	-(l_1l_3-m_1m_3)(C^c-A^c) \vv{f_2^c} \\
	(l_1l_2-m_1m_2)(B^c-A^c) \vv{f_3^c} \end{array}\right],
\end{eqnarray}	
on retrouve ainsi \citep[Eq.7]{sx1997}. En l'absence d'oc\'ean, le couple d\^u au noyau agissant sur la cro\^ute $\vv{\Gamma^s_{co}}$ doit \^etre l'oppos\'e de $\vv{\Gamma^c_{sh}}$,
on a donc

\begin{equation}
	\label{eq:torqueshell}
	\vv{\Gamma^s_{co}}  =  -3\beta \left[\begin{array}{c}
	n_2n_3(C^c-B^c) \vv{f_1^c} \\
	-n_1n_3(C^c-A^c) \vv{f_2^c} \\
	n_1n_2(B^c-A^c) \vv{f_3^c} \end{array}\right] - 6\gamma\left[\begin{array}{c}
	(l_2l_3-m_2m_3)(C^c-B^c) \vv{f_1^c} \\
	-(l_1l_3-m_1m_3)(C^c-A^c) \vv{f_2^c} \\
	(l_1l_2-m_1m_2)(B^c-A^c) \vv{f_3^c} \end{array}\right].
\end{equation}

  \subsubsection{L'influence de l'oc\'ean}
  
  \par Cette section est largement inspir\'ee de \citep{bvyk2011}. Ici nous n\'egligeons les effets non lin\'eaires dans le fluide. L'oc\'ean agit sur les interfaces 
  fluide-solide par le biais d'un couplage de pression. Lorsque l'oc\'ean est \`a l'\'equilibre hydrostatique\footnote{Dans le cas particulier de Titan, l'\'ecart \`a 
  l'\'equilibre hydrostatique est en g\'en\'eral consid\'er\'e comme d\^u essentiellement \`a la cro\^ute.}, alors la pression $P_{ext}$ appliqu\'ee sur l'oc\'ean par 
  Saturne v\'erifie $\vec{\nabla}P_{ext} = -\rho_o\vec{\nabla}W$ o\`u $W$ est le potentiel gravitationnel d\^u au perturbateur Saturne. Il en r\'esulte des termes 
  additionnels, li\'es \`a l'oc\'ean, dans les couples pr\'ec\'edemment donn\'es.
  
  \par Pour cela nous avons besoin de scinder l'oc\'ean en 2 parties : une partie haute \emph{top ocean} de moments d'inertie $A_t^o$, $B_t^o$ et $C_t^o$ dans le rep\`ere de 
  la cro\^ute $\left(\vv{f_1^s},\vv{f_2^s},\vv{f_3^s}\right)$, et une partie basse \emph{bottom ocean} de moments $A_b^o$, $B_b^o$ et $C_b^o$ dans le rep\`ere du noyau
  $\left(\vv{f_1^c},\vv{f_2^c},\vv{f_3^c}\right)$. La limite entre ces 2 oc\'eans est sph\'erique et de rayon arbitraire. La r\'esultante des couplages s'appliquant sur une 
  limite sph\'erique est nulle, ce qui simplifie les calculs.
  
  \par Le plus commode est de r\'e\'ecrire les couples externes $\vv{\Gamma^c_{\saturn}}$ et $\vv{\Gamma^s_{\saturn}}$ et internes $\vv{\Gamma^c_{sh}}$ et $\vv{\Gamma^s_{co}}$
  du satellite sans oc\'ean en ajoutant les termes dus \`a l'oc\'ean, les couples deviennent donc
  
  \begin{equation}
	\label{eq:pulloccore}
	\begin{split}
	\vv{\Gamma^{c,o}_{\saturn}}  = \frac{3\mathcal{G}M_{\saturn}}{\|\vv{r^c_{\saturn}}\|^5}((C^c-B^c+C^o_b-B^o_b)y^c_{\saturn}z^c_{\saturn}\vv{f^c_1}+(A^c-C^c+A^o_b-C^o_b)x^c_{\saturn}z^c_{\saturn}\vv{f^c_2} \\
	 +  (B^c-A^c+B^o_b-A^o_b)x^c_{\saturn}y^c_{\saturn}\vv{f^c_3}),
	\end{split}
  \end{equation}

  \begin{equation}
	\label{eq:pullocshell}
	\begin{split}
	\vv{\Gamma^{s,o}_{\saturn}}  =  \frac{3\mathcal{G}M_{\saturn}}{\|\vv{r^s_{\saturn}}\|^5}((C^s-B^s+C^o_t-B^o_t)y^s_{\saturn}z^s_{\saturn}\vv{f^s_1}+(A^s-C^s+A^o_t-C^o_t)x^s_{\saturn}z^s_{\saturn}\vv{f^s_2} \\
	 +  (B^s-A^s+B^o_t-A^o_t)x^s_{\saturn}y^s_{\saturn}\vv{f^s_3}), 
	\end{split}
  \end{equation}

  \begin{equation}
	\label{eq:torqueocinner}
	\vv{\Gamma^{c,o}_{sh}}  =  \left[\begin{array}{c}
	\left(3\beta^o n_2n_3+6\gamma^o(l_2l_3-m_2m_3)\right)(C^c-B^c+C^o_b-B^o_b) \vv{f_1^c} \\
	-\left(3\beta^on_1n_3+6\gamma^o(l_1l_3-m_1m_3)\right) (C^c-A^c+C^o_b-A^o_b) \vv{f_2^c} \\
	\left(3\beta^on_1n_2+6\gamma^o(l_1l_2-m_1m_2)\right)(B^c-A^c+B^o_b-A^o_b) \vv{f_3^c} \end{array}\right],
  \end{equation}	

  et

  \begin{equation}
	\label{eq:torqueocshell}
	\vv{\Gamma^{s,o}_{co}}  =  \left[\begin{array}{c}
	-\left(3\beta^o n_2n_3+6\gamma^o(l_2l_3-m_2m_3)\right)(C^c-B^c+C^o_b-B^o_b) \vv{f_1^c} \\
	\left(3\beta^on_1n_3+6\gamma^o(l_1l_3-m_1m_3)\right) (C^c-A^c+C^o_b-A^o_b) \vv{f_2^c} \\
	-\left(3\beta^on_1n_2+6\gamma^o(l_1l_2-m_1m_2)\right)(B^c-A^c+B^o_b-A^o_b) \vv{f_3^c} \end{array}\right],
  \end{equation}	
  avec
  
  \begin{eqnarray}
	\beta^o  & = & \frac{4\pi}{15}\mathcal{G}\left(\rho_s\left(\kappa_2-\kappa_1-2(f_2-f_1)\right)+\rho_o\left(\kappa_1-2f_1\right)\right), \label{eq:betao} \\
	\gamma^o & = & \frac{2\pi}{15}\mathcal{G}\left(\rho_s(\kappa_2-\kappa_1)+\rho_o\kappa_1\right). \label{eq:gammao}
  \end{eqnarray}
  
  \par Nous avons maintenant tous les \'el\'ements pour \'ecrire le syst\`eme d'\'equations complet.

  \subsection{Le syst\`eme d'\'equations}
  
  \par Nous obtenons
  
  \begin{eqnarray}
  	\frac{d\xi^c}{dt}      & = & \omega_1^c\left(\cos\theta^c\cos\frac{\epsilon^c}{2}\sin h^c+\frac{\sin\theta^c\cos h^c}{\cos\left(\epsilon^c/2\right)}\right) \nonumber \\
                               & + & \omega_2^c\left(\frac{\cos\theta^c\cos h^c}{\cos\left(\epsilon^c/2\right)}-\sin\theta^c\cos\frac{\epsilon^c}{2}\sin h^c\right), \label{eq:dotxic} \\
	\frac{d\eta^c}{dt}     & = & \omega_1^c\left(\cos\theta^c\cos\frac{\epsilon^c}{2}\cos h^c-\frac{\sin\theta^c\sin h^c}{\cos\left(\epsilon^c/2\right)}\right) \nonumber \\
                               & - & \omega_2^c\left(\sin\theta^c\cos\frac{\epsilon^c}{2}\cos h^c+\frac{\cos\theta^c\sin h^c}{\cos\left(\epsilon^c/2\right)}\right), \label{eq:dotetac} \\
	\frac{dp^c}{dt}        & = & \omega_3^c+\left(\omega_1^c\sin\theta^c+\omega_2^c\cos\theta^c\right)\tan\frac{\epsilon^c}{2}, \label{eq:dotpc} \\
	\frac{d\omega_1^c}{dt} & = & \left(3\beta^on_2n_3+6\gamma^o\left(l_2l_3-m_2m_3\right)\right)\frac{\Delta_1}{A^c} \nonumber \\
	                       & + & 3\frac{\mathcal{G}M_{\saturn}}{||\vv{r_{\saturn}^c}||^5}\frac{C^c-B^c+C_b^o-B_b^o}{A^c}y_{\saturn}^cz_{\saturn}^c-\frac{\left(\vec{\omega^c}\times\vec{G^c}\right)\cdot\vv{f_1^c}}{A^c}, \label{eq:dotomega1c} \\
	\frac{d\omega_2^c}{dt} & = & -\left(3\beta^on_1n_3+6\gamma^o\left(l_1l_3-m_1m_3\right)\right)\frac{\Delta_2}{B^c} \nonumber \\
	                       & + & 3\frac{\mathcal{G}M_{\saturn}}{||\vv{r_{\saturn}^c}||^5}\frac{A^c-C^c+A_b^o-C_b^o}{B^c}x_{\saturn}^cz_{\saturn}^c-\frac{\left(\vec{\omega^c}\times\vec{G^c}\right)\cdot\vv{f_2^c}}{B^c}, \label{eq:dotomega2c} \\
	\frac{d\omega_3^c}{dt} & = & \left(3\beta^on_1n_2+6\gamma^o\left(l_1l_2-m_1m_2\right)\right)\frac{\Delta_3}{C^c} \nonumber \\
	                       & + & 3\frac{\mathcal{G}M_{\saturn}}{||\vv{r_{\saturn}^c}||^5}\frac{B^c-A^c+B_b^o-A_b^o}{C^c}x_{\saturn}^cy_{\saturn}^c-\frac{\left(\vec{\omega^c}\times\vec{G^c}\right)\cdot\vv{f_3^c}}{C^c} \label{eq:dotomega3c}
\end{eqnarray}
pour le noyau, et

\begin{eqnarray}
	\frac{d\xi^s}{dt}      & = & \omega_1^s\left(\cos\theta^s\cos\frac{\epsilon^s}{2}\sin h^s+\frac{\sin\theta^s\cos h^s}{\cos\left(\epsilon^s/2\right)}\right) \nonumber \\
                               & + & \omega_2^s\left(\frac{\cos\theta^s\cos h^s}{\cos\left(\epsilon^s/2\right)}-\sin\theta^s\cos\frac{\epsilon^s}{2}\sin h^s\right), \label{eq:dotxis} \\
	\frac{d\eta^s}{dt}     & = & \omega_1^s\left(\cos\theta^s\cos\frac{\epsilon^s}{2}\cos h^s-\frac{\sin\theta^s\sin h^s}{\cos\left(\epsilon^s/2\right)}\right) \nonumber \\
                               & - & \omega_2^s\left(\sin\theta^s\cos\frac{\epsilon^s}{2}\cos h^s+\frac{\cos\theta^s\sin h^s}{\cos\left(\epsilon^s/2\right)}\right), \label{eq:dotetas} \\
	\frac{dp^s}{dt}        & = & \omega_3^s+\left(\omega_1^s\sin\theta^s+\omega_2^s\cos\theta^s\right)\tan\frac{\epsilon^s}{2}, \label{eq:dotps} \\
	\frac{d\omega_1^s}{dt} & = & -\frac{3\beta^on_2n_3+6\gamma^o\left(l_2l_3-m_2m_3\right)}{A^s}\left(\Delta_1 \vv{f_1^c}\cdot\vv{f_1^s} + \Delta_2 \vv{f_2^c}\cdot\vv{f_1^s} + \Delta_3 \vv{f_3^c}\cdot\vv{f_1^s} \right) \nonumber \\
	                       & + & 3\frac{\mathcal{G}M_{\saturn}}{||\vv{r_{\saturn}^s}||^5}\frac{C^s-B^s+C_t^o-B_t^o}{A^s}y_{\saturn}^sz_{\saturn}^s-\frac{\left(\vec{\omega^s}\times\vec{G^s}\right)\cdot\vv{f_1^s}}{A^s}, \label{eq:dotomega1s} \\
	\frac{d\omega_2^s}{dt} & = & \frac{3\beta^on_1n_3+6\gamma^o\left(l_1l_3-m_1m_3\right)}{B^s}\left(\Delta_1 \vv{f_1^c}\cdot\vv{f_2^s} + \Delta_2 \vv{f_2^c}\cdot\vv{f_2^s} + \Delta_3 \vv{f_3^c}\cdot\vv{f_2^s} \right) \nonumber \\
	                       & + & 3\frac{\mathcal{G}M_{\saturn}}{||\vv{r_{\saturn}^s}||^5}\frac{A^s-C^s+A_t^o-C_t^o}{B^s}x_{\saturn}^sz_{\saturn}^s-\frac{\left(\vec{\omega^s}\times\vec{G^s}\right)\cdot\vv{f_2^s}}{B^s}, \label{eq:dotomega2s} \\
	\frac{d\omega_3^s}{dt} & = & -\frac{3\beta^on_1n_2+6\gamma^o\left(l_1l_2-m_1m_2\right)}{C^s}\left(\Delta_1 \vv{f_1^c}\cdot\vv{f_3^s} + \Delta_2 \vv{f_2^c}\cdot\vv{f_3^s} + \Delta_3 \vv{f_3^c}\cdot\vv{f_3^s} \right) \nonumber \\
	                       & + & 3\frac{\mathcal{G}M_{\saturn}}{||\vv{r_{\saturn}^s}||^5}\frac{B^s-A^s+B_t^o-A_t^o}{C^s}x_{\saturn}^sy_{\saturn}^s-\frac{\left(\vec{\omega^s}\times\vec{G^s}\right)\cdot\vv{f_3^s}}{C^s} \label{eq:dotomega3s}
\end{eqnarray}
pour la cro\^ute, avec

\begin{eqnarray}
	\Delta_1 & = & C^c-B^c+C_b^o-B_b^o, \label{eq:Delta1} \\
	\Delta_2 & = & C^c-A^c+C_b^o-A_b^o, \label{eq:Delta2} \\
	\Delta_3 & = & B^c-A^c+B_b^o-A_b^o. \label{eq:Delta3}
\end{eqnarray}
  
\par On peut remarquer que le couple gravitationnel entre les couches rigides \'etant initialement \'ecrit dans le rep\`ere de figure du noyau, une projection suppl\'ementaire 
est n\'ecessaire pour l'\'ecrire dans le rep\`ere de figure de la cro\^ute.

  \section{Application \`a Titan}
  
  \par Le but de cette \'etude est de l'appliquer \`a Titan. Depuis mes \'etudes pr\'ec\'edentes \citep{nlv2008,n2008}, la mission \emph{Cassini} a donn\'e de nombreuses 
  informations sur Titan. Sa forme est connue ($(2575.15\pm0.02)\times(2574.78\pm0.06)\times(2574.47\pm0.06)$) km avec un rayon moyen de $2574.73\pm0.09$ km \citep{zshlkl2009}.
  Par contre, 2 solutions (Tab.\ref{tab:titangravity}) ont \'et\'e publi\'ees pour son champ de gravit\'e \citep{irjrstaa2010}. L'une, SOL1, vient des donn\'ees de 4 survols de
  Titan par \emph{Cassini} d\'edi\'es \`a la gravitation (T11, T22, T33 et T45), alors que SOL2 utilise plus de donn\'ees de \emph{Cassini}, ainsi que de \emph{Voyager} et 
  \emph{Pioneer}. Ces 2 solutions ne sont pas parfaitement compatibles. Notamment, si Titan est \`a l'\'equilibre hydrostatique, on devrait avoir $J_2/C_{22}\approx10/3$, ce qui 
  est le cas pour SOL2, mais pas pour SOL1. Par contre, la forme de Titan est trop aplatie par rapport \`a sa vitesse de rotation, Titan n'est donc pas \`a l'\'equilibre hydrostatique.

  \begin{table}[ht]
\centering
\caption[Les 2 solutions du champ de gravit\'e de Titan.]{Les 2 solutions du champ de gravit\'e de Titan \citep{irjrstaa2010}.\label{tab:titangravity}}
\begin{tabular}{lrr}
\hline
                  & SOL1                             & SOL2 \\
\hline
$\mathcal{G}M_6$  &  --                              & $8978.1394$ $\textrm{km}^3.\textrm{s}^{-2}$ \\
$J_2$             & $(3.1808\pm0.0404)\times10^{-5}$ & $(3.3462\pm0.0632)\times10^{-5}$ \\
$C_{21}$          & $(3.38\pm3.50)\times10^{-7}$     & $(4.8\pm11.5)\times10^{-8}$ \\
$S_{21}$          & $(-3.52\pm4.38)\times10^{-7}$    & $(6.20\pm4.96)\times10^{-7}$ \\
$C_{22}$          & $(9.983\pm0.039)\times10^{-6}$   & $(1.0022\pm0.0071)\times10^{-5}$ \\
$S_{22}$          & $(2.17\pm0.41)\times10^{-7}$     & $(2.56\pm0.72)\times10^{-7}$ \\
$J_3$             & $(-1.879\pm1.019)\times10^{-6}$  & $(-7.4\pm105.1)\times10^{-8}$ \\
$C_{31}$          & $(1.058\pm0.260)\times10^{-6}$   & $(1.805\pm0.297)\times10^{-6}$ \\
$S_{31}$          & $(5.09\pm2.02)\times10^{-7}$     & $(2.83\pm3.54)\times10^{-7}$ \\
$C_{32}$          & $(3.64\pm1.13)\times10^{-7}$     & $(1.36\pm1.58)\times10^{-7}$ \\
$S_{32}$          & $(3.47\pm0.80)\times10^{-7}$     & $(1.59\pm1.05)\times10^{-7}$ \\
$C_{33}$          & $(-1.99\pm0.09)\times10^{-7}$    & $(-1.85\pm0.12)\times10^{-7}$ \\
$S_{33}$          & $(-1.71\pm0.15)\times10^{-7}$    & $(-1.49\pm0.16)\times10^{-7}$ \\
$J_2/C_{22}$      & $3.186\pm0.042$                  & $3.339\pm0.067$ \\
\hline
\end{tabular}
\end{table}

  \par Plusieurs \'etudes sugg\`erent que la non-hydrostaticit\'e de Titan r\'eside essentiellement dans la cro\^ute. Cette cro\^ute flotte sur un oc\'ean et aurait des anomalies
  de masse partiellement compens\'ees par isostasie. Cette compensation se traduirait par des variations lat\'erales soit de l'\'epaisseur de la cro\^ute (isostasie d'Airy, \citet{nb2010}),
  soit de sa masse volumique (isostasie de Pratt, \citet{cs2012}). Une r\'ecente \'etude de la topographie de Titan au degr\'e 3 \citep{hnzi2013} plaide en faveur de variations lat\'erales.
  
  \par La rotation de Titan a \'et\'e observ\'ee par \emph{Cassini}. \citet{sklhloacgiphjw2008} ont annonc\'e une obliquit\'e importante, de l'ordre de $0.3^{\circ}$ ou $18$ 
  minutes d'arc, ainsi qu'une rotation l\'eg\`erement super-synchrone ($+0.36^{\circ}$ par an). 2 ans plus tard, un erratum \citep{sklhloacgiphjw2010} maintient qu'il y a un 
  \'ecart \`a la synchronicit\'e, mais 3 fois plus faible. Plus r\'ecemment, \citet{mi2012} ont sugg\'er\'e que Titan \'etait en fait synchrone, et confirment une obliquit\'e 
  de $(18.6\pm3)$ minutes d'arc \`a la date moyenne du 11 mars 2007, soit $\approx2007.2$. Une telle obliquit\'e, si elle correspond \`a un \'equilibre dynamique, ne peut pas 
  \^etre atteinte si Titan est enti\`erement rigide \citep{bn2008}. La premi\`ere explication satisfaisante pr\'esente dans la litt\'erature a \'et\'e publi\'ee par 
  \citet{bvyk2011} dans le cadre d'un mod\`ele avec oc\'ean global, moyenn\'e sur les courtes p\'eriodes, et ne repr\'esentant pas les autres degr\'es de libert\'e. Dans cette 
  \'etude, l'obliquit\'e est forc\'ee par une r\'esonance entre une fr\'equence d'oscillations libres et une fr\'equence de for\c{c}age. En g\'en\'eral, la communaut\'e 
  accueille froidement ce genre d'explications, car une r\'esonance est vue comme un ph\'enom\`ene exceptionnel, n\'ecessitant d'\^etre chanceux que les param\`etres du mod\`ele, 
  c'est-\`a-dire les param\`etres d'int\'erieur, permettent au syst\`eme d'\^etre en r\'esonance. J'esp\`ere convaincre le lecteur, dans les pages qui suivent, 
  que cette explication est en fait tout-\`a-fait acceptable.

  \subsection{Vers un mod\`ele d'int\'erieur r\'ealiste}
  
  \par Nous cherchons \`a \'elaborer des mod\`eles de Titan qui soient les plus r\'ealistes possibles, c'est-\`a-dire compatibles avec la forme observ\'ee \citep{zshlkl2009} et l'un 
  ou l'autre des champs de gravit\'e mesur\'es \citep{irjrstaa2010}. Nous construisons un ensemble de Titans en 3 \'etapes :
  
  \begin{enumerate}
  
   \item \'Elaboration de Titans hydrostatiques. Le choix de ce point de d\'epart vient de l'id\'ee que l'\'ecart \`a l'hydrostaticit\'e du vrai Titan est suffisamment petit
   pour \^etre consid\'er\'e comme une correction,
   
   \item Application d'une anomalie de masse en surface (\emph{top loading}) ou \`a l'interface cro\^ute-oc\'ean (\emph{bottom loading}), partiellement compens\'ee par
   des variations lat\'erales de l'\'epaisseur de la cro\^ute,
   
   \item Comparaison avec le champ de gravit\'e. Les mod\`eles incompatibles avec les mesures de \citet{irjrstaa2010} ne sont pas conserv\'es.
  
  \end{enumerate}

  \par Pour chacun de ces mod\`eles de Titan, on fixe au d\'epart les masses volumiques de la cro\^ute $\rho_s$ et de l'oc\'ean $\rho_o$, ainsi que les \'epaisseurs de la 
  cro\^ute $d_s$ et de l'oc\'ean $d_o$. De ces param\`etres on peut d\'eduire la taille et la masse volumique du noyau, la masse volumique de Titan $\rho_6 = 1881\,\textrm{kg}/\textrm{m}^3$ 
  \'etant connue \citep{irjrstaa2010}. Les intervalles de variation que nous consid\'erons comme admissibles pour nos param\`etres d'int\'erieur sont :
  
  \begin{itemize}
  
   \item \'epaisseur de la cro\^ute $d_s$ entre $50$ et $200$ km,
   
   \item \'epaisseur de l'oc\'ean $d_o$ entre $50$ et $400-d_s$ km,
   
   \item masse volumique de la cro\^ute $\rho_s$ entre $900$ et $950\,\textrm{kg}/\textrm{m}^3$,
   
   \item masse volumique de l'oc\'ean $\rho_o$ entre $950$ et $1200\,\textrm{kg}/\textrm{m}^3$.
  
  \end{itemize}

  \par Le choix des masses volumiques possibles de l'oc\'ean vient de \citep{f2012}. $\rho_o = 950\,\textrm{kg}/\textrm{m}^3$ est proche de la masse volumique d'une solution 
  p\'eritectique \footnote{D'apr\`es Wikip\'edia, un p\'eritectique est un m\'elange de 2 corps purs dans des proportions d\'efinies, et qui fond de mani\`ere particuli\`ere : il se 
  d\'ecompose en un liquide et en un solide, le nouveau solide \'etant d'une phase diff\'erente de celle du p\'eritectique.} d'eau et de m\'ethanol \citep{ysk1971}, alors que 
  $\rho_o = 1200\,\textrm{kg}/\textrm{m}^3$ correspond \`a la masse d'un m\'elange eutectique d'eau et de sulfate d'ammonium \citep{ns1988}. Les \'epaisseurs de cro\^ute admissibles 
  viennent de \citep{hnzi2013}, et la limite de $400$ km pour la profondeur de l'interface noyau-oc\'ean vient du fait qu'un oc\'ean liquide ne peut pas subsister au-del\`a de cette 
  profondeur, les valeurs de pression et de temp\'erature ne le permettant pas, comme le montrent les diagrammes de phase d'un oc\'ean d'eau \citep{smrss2010} et 
  d'ammoniaque \citep{shssl2003}. M\^eme pour un oc\'ean \`a 270 K, ce qui est trop chaud si de l'ammoniaque est pr\'esent, la pression \`a la base de l'oc\'ean doit \^etre 
  inf\'erieure \`a 0.6 GPa, ce qui correspond \`a une profondeur de 450 km. Ainsi 400 km semble \^etre une limite raisonnable.
  
  \par Nos Titans hydrostatiques (\'Etape 1) sont construits suivant la m\'ethode donn\'ee dans \citep{vrkdr2008}. Nous consid\'erons connus le rayon moyen $R=2574.73$ km et le rayon
  $b=2574.78$ km. L'\'equation de Radau donne (cf. e.g. \citet{j1952}) :
  
   \begin{eqnarray}
      h_2 & = & \frac{5}{1+\left(\frac{5}{2}-\frac{15}{4}\frac{C}{M_6R^2}\right)^2}, \label{eq:h2} \\
      q & = & \frac{n_6^2R^3}{\mathcal{G}M_6}, \label{eq:q} \\ 
    \tilde{c} & = & b-\frac{h_2}{2}qR, \label{eq:cradau} \\
    \tilde{a} & = & b+\frac{3}{2}qh_2R, \label{eq:aradau}
    \end{eqnarray}
  o\`u $h_2$ est le nombre de Love de d\'eplacement radial. $\tilde{a}$ et $\tilde{c}$ sont les rayons du Titan hydrostatique, ils doivent \^etre tr\`es proches de $a$ et $c$.
  $C = C^s+C^o+C^c$ est le moment d'inertie polaire total de Titan. Les 3 rayons de chacune des 2 interfaces solide-fluide sont d\'etermin\'es en propageant l'\'equation de
  Clairaut :
  
  \begin{eqnarray}
  \frac{d^2\alpha}{dr^2}+\frac{6}{r}\frac{\rho}{\bar{\rho}}-\frac{6}{r^2}\left(1-\frac{\rho}{\bar{\rho}}\right)\alpha & = & 0, \label{eq:clairaut} \\
  \frac{d\alpha}{dr}(R) & = & \frac{1}{R}\left[\frac{25}{4}q-2\alpha(R)\right], \label{eq:clairautci1} \\
  \alpha(R) & = & \frac{\tilde{a}+b-2\tilde{c}}{\tilde{a}+b}. \label{eq:clairautci0}
  \end{eqnarray}
  La condition initiale (\ref{eq:clairautci1}) n'est pas celle donn\'ee dans la th\'eorie de Clairaut o\`u un coefficient $5/2$ remplace $25/4$. L'origine de ce changement
  de coefficient est la prise en compte des mar\'ees dans le cas particulier d'un corps en rotation synchrone. De plus, on a $\beta=6\alpha/5=(\tilde{a}-\tilde{b})/\tilde{a}$
  \`a n'importe quelle distance radiale $r$ \citep{vrkdr2008}.
  
  \par Nous modifions ensuite (\'Etape 2) ces Titans hydrostatiques en introduisant une anomalie topographique $h_t$ \`a la surface :
  
  \begin{equation}
  \label{eq:anomaly}
  h_t(\psi,\phi) = h_{t1}Y_{20}(\psi)+h_{t2}Y_{22}(\psi,\phi)
\end{equation}
o\`u $Y_{20}$ et $Y_{22}$ sont les harmoniques sph\'eriques du second ordre d\'efinies par

\begin{equation}
  \label{eq:harmosferik}
  Y_{lm}(\psi,\phi) = \sqrt{\left(2-\delta_{0m}\right)(2l+1)\frac{(l-m)!}{(l+m)!}}P_l^m(\cos\psi)\cos m\phi
\end{equation}
pour $m\geq0$. $P_l^0=P_l$ sont les polyn\^omes de Legendre et $P_l^m$ les fonctions de Legendre associ\'ees pour $m\ne0$ (Eq.\ref{eq:legendreassociees}). $\psi$ et $\phi$ sont,
comme pr\'ec\'edemment, la colatitude et la longitude. L'\'equation (\ref{eq:anomaly}) devient ainsi :

\begin{equation}
  \label{eq:anomaly2}
  h_t(\psi,\phi) = \frac{\sqrt{5}}{2}h_{t1}\left(3\cos^2\psi-1\right)+3\sqrt{\frac{5}{12}}h_{t2}\sin^2\psi\cos2\phi.
\end{equation}

\par Soient $\Delta a = a-\tilde{a}$, $\Delta b = b-\tilde{b}$ et $\Delta c = c-\tilde{c}$ les corrections radiales correspondant respectivement \`a
$(\phi=k\pi,\psi=\frac{\pi}{2}+k'\pi)$, $(\phi=\frac{\pi}{2}+k\pi,\psi=\frac{\pi}{2}+k'\pi)$ et $(\psi=k'\pi)$ o\`u $k$ et $k'$ sont des entiers. On a 

\begin{eqnarray}
  \Delta c & = & \sqrt{5}h_{t1}, \label{eq:deltac} \\
  \Delta a & = & -\frac{\sqrt{5}}{2}h_{t1}+3\sqrt{\frac{5}{12}}h_{t2}, \label{eq:deltaa}
\end{eqnarray}
dont on d\'eduit $h_{t1}$ et $h_{t2}$. Dans ce cas le rayon $b$, initialement choisi \'egal \`a l'observation, est modifi\'e ainsi :

\begin{equation}
  \label{eq:deltab}
  \Delta b = -\frac{\sqrt{5}}{2}h_{t1}.
\end{equation}
Ceci n\'ecessite d'it\'erer le processus jusqu'\`a convergence. Notre crit\`ere d'arr\^et est $\left|\Delta b\right|<0.5$ m\`etre.

\par La compensation par isostasie implique une anomalie topographique $h_b$ \`a l'interface oc\'ean-cro\^ute, elle aussi d'ordre 2 :

\begin{equation}
  \label{eq:anomalyb}
  h_b(\psi,\phi) = h_{b1}Y_{20}(\psi)+h_{b2}Y_{22}(\psi,\phi).
\end{equation}

\par Une anomalie de masse \`a la surface de Titan r\'esulte en une anomalie topographique

\begin{equation}
  \label{eq:toploading}
  h_b = fh_t\left(\frac{R}{R-d_s}\right)^2\frac{\rho_s}{\rho_o-\rho_s},
\end{equation}
alors qu'une anomalie de masse \`a l'interface oc\'ean-cro\^ute donnera

\begin{equation}
  \label{eq:bottomloading}
  h_t = fh_b\left(\frac{R-d_s}{R}\right)^2\frac{\rho_o-\rho_s}{\rho_s}.
\end{equation}

\par Ici, $f\in[0:1]$ est le facteur de compensation isostatique. $f=0$ veut dire qu'il n'y a pas de compensation, la cro\^ute est parfaitement rigide, et une seule
de ses interfaces est d\'eform\'ee. Par contre, $f=1$ signifie que la compensation est totale, dans ce cas le champ de gravit\'e n'est pas modifi\'e.

\par
\begin{figure}[ht]
\centering
\begin{tabular}{cc}
  \includegraphics[width=0.47\textwidth]{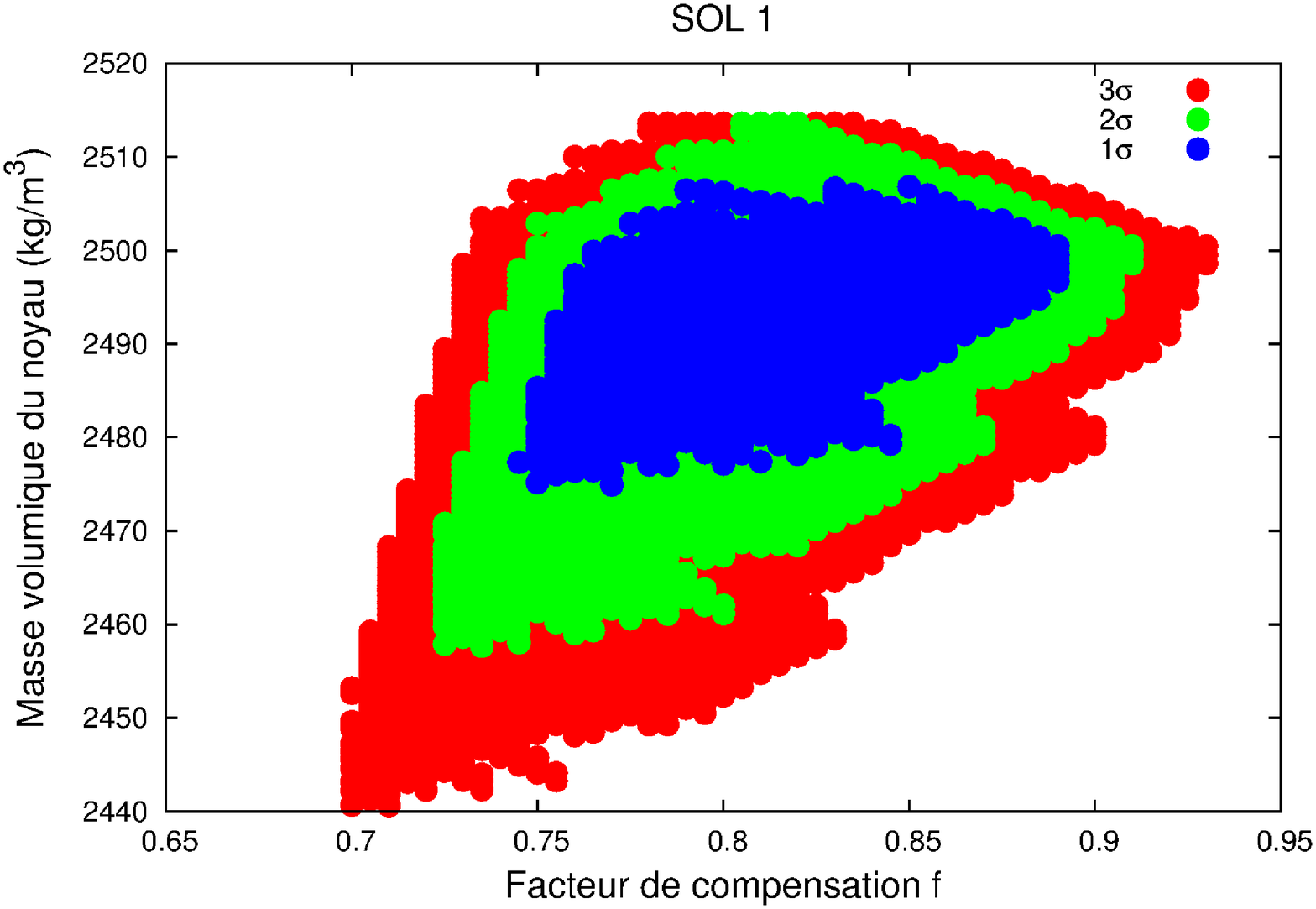} & \includegraphics[width=0.47\textwidth]{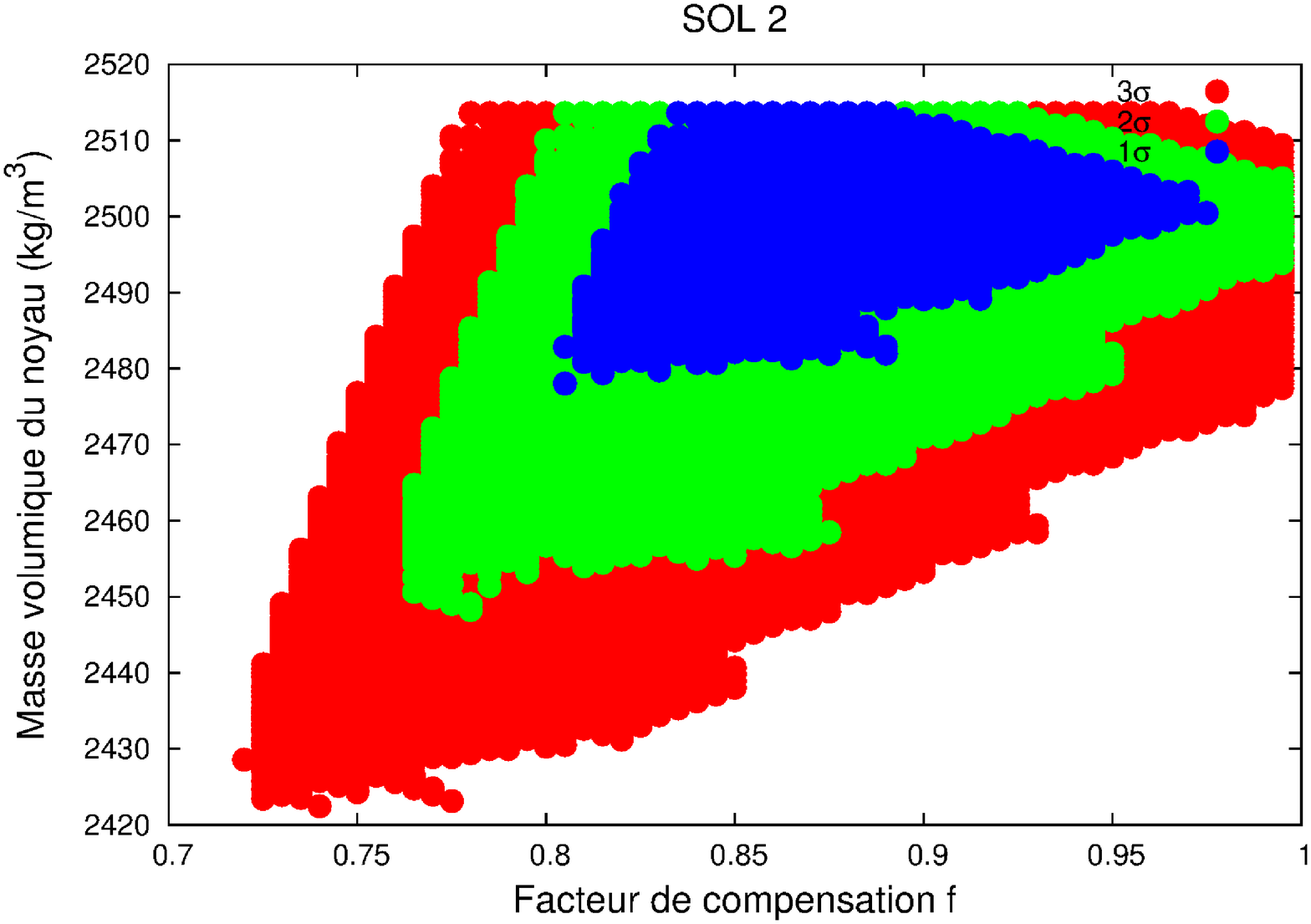}
\end{tabular}
\caption[Propri\'et\'es de nos Titans.]{Propri\'et\'es de nos Titans, en consid\'erant une anomalie de masse \`a l'interface cro\^ute-oc\'ean. Ici ne sont conserv\'es
que les mod\`eles compatibles avec les solutions du champ de gravit\'e publi\'ees SOL1 et SOL2, \`a $1\sigma$ en bleu, $2\sigma$ en vert et $3\sigma$ en rouge. On voit 
en particulier que SOL1 sugg\`ere entre $75$ et $88\%$ de compensation, alors que SOL2 sugg\`ere entre $80$ et $96\%$.\label{fig:iso2sol}}
\end{figure}

 Une fois ces mod\`eles \'elabor\'es, ne sont conserv\'es que ceux qui sont compatibles avec les champs de gravit\'e publi\'es par \citet{irjrstaa2010}. En pratique,
nous n'avons de bons r\'esultats qu'avec une anomalie de masse sous la cro\^ute (\emph{bottom loading}). Nous obtenons un facteur de compensation isostatique 
entre $75$ et $96\%$ (Fig.\ref{fig:iso2sol}), ce qui est compatible avec l'analyse de la topographie d'ordre 3 de \citet{hnzi2013}.

\par Une fois ces mod\`eles d'int\'erieur obtenus, nous simulons leur rotation, comme pr\'ec\'edemment. Le mouvement de Titan autour de Saturne est issu de TASS1.6 \citep{vd1995},
le plan inertiel $\left(\vv{e_1},\vv{e_2}\right)$ est le plan moyen de l'orbite de Titan comme sugg\'er\'e Sect.\ref{sec:propplanref}, et les conditions initiales sont choisies
proches de l'\'Etat de Cassini.

\par Nous avons obtenu 71046 mod\`eles d'int\'erieur. 20546 d'entre eux sont compatibles avec SOL1 \`a $3\sigma$ dont 2237 \`a $1\sigma$, et les 50500 autres sont compatibles avec 
SOL2 \`a $3\sigma$, dont 2665 \`a $1\sigma$. Les simulations num\'eriques ont \'et\'e faites sur 2300 ans.

  \subsection{Une obliquit\'e r\'esonnante}
  
  \par Nous avons obtenu des quantit\'es de rotation concernant les 3 degr\'es de libert\'e de chacune des couches rigides. Nos librations physiques en longitude de la cro\^ute 
  \`a la fr\'equence orbitale  varient entre 12 secondes d'arc soit 150 m\`etres pour une cro\^ute de 200 km d'\'epaisseur, et 50 secondes d'arc soit 600 m\`etres pour une 
  cro\^ute de 50 km. Cependant, ces chiffres doivent \^etre accueillis avec beaucoup de prudence car notre cro\^ute n'est pas d\'eformable. Les librations du noyau ont une 
  amplitude entre 2.1 et 2.6 secondes d'arc.
  
  \par Le mouvement polaire, habituellement n\'eglig\'e, se r\'ev\`ele culminer \`a $3.5$ km pour la cro\^ute et $4.5$ km pour le noyau. L'obliquit\'e moyenne du noyau,
  obtenue par moyenne arithm\'etique sur les 2300 ans couverts par chacune des simulations num\'eriques, va de 8 \`a 11.5 minutes d'arc.
  
  \par
\begin{figure}[ht]
\centering
\begin{tabular}{cc}
  \includegraphics[width=0.43\textwidth]{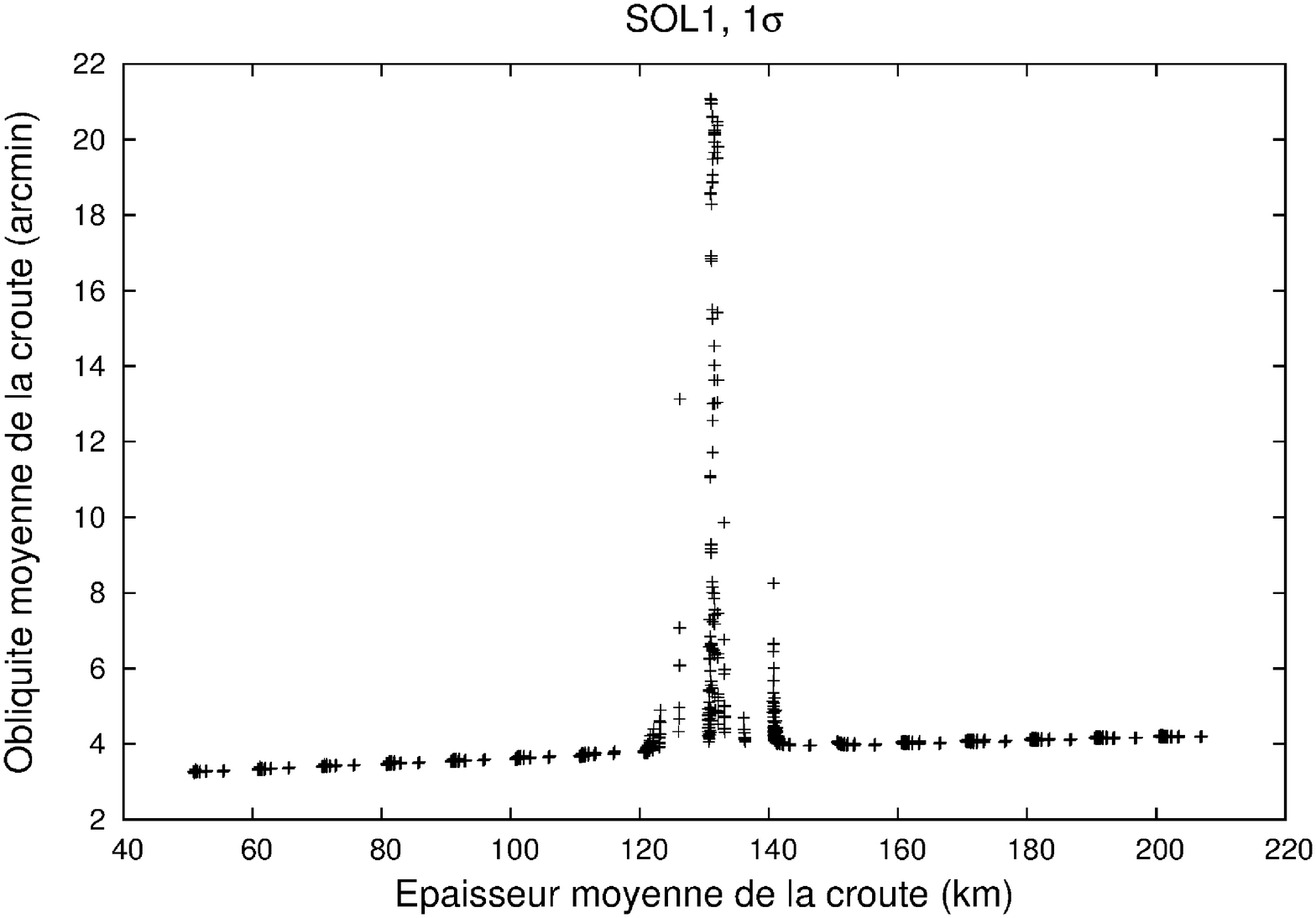} & \includegraphics[width=0.43\textwidth]{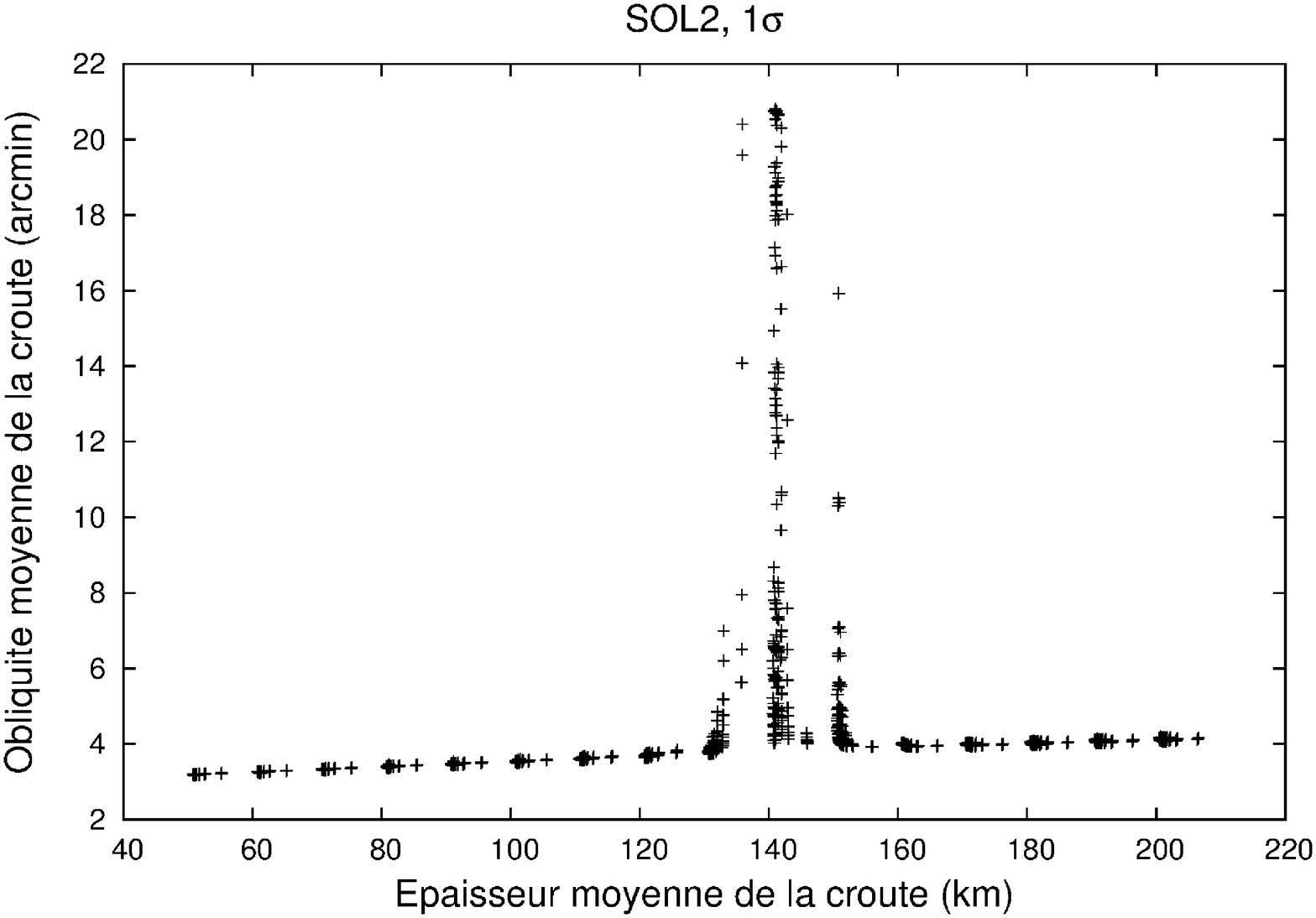} \\
  \includegraphics[width=0.43\textwidth]{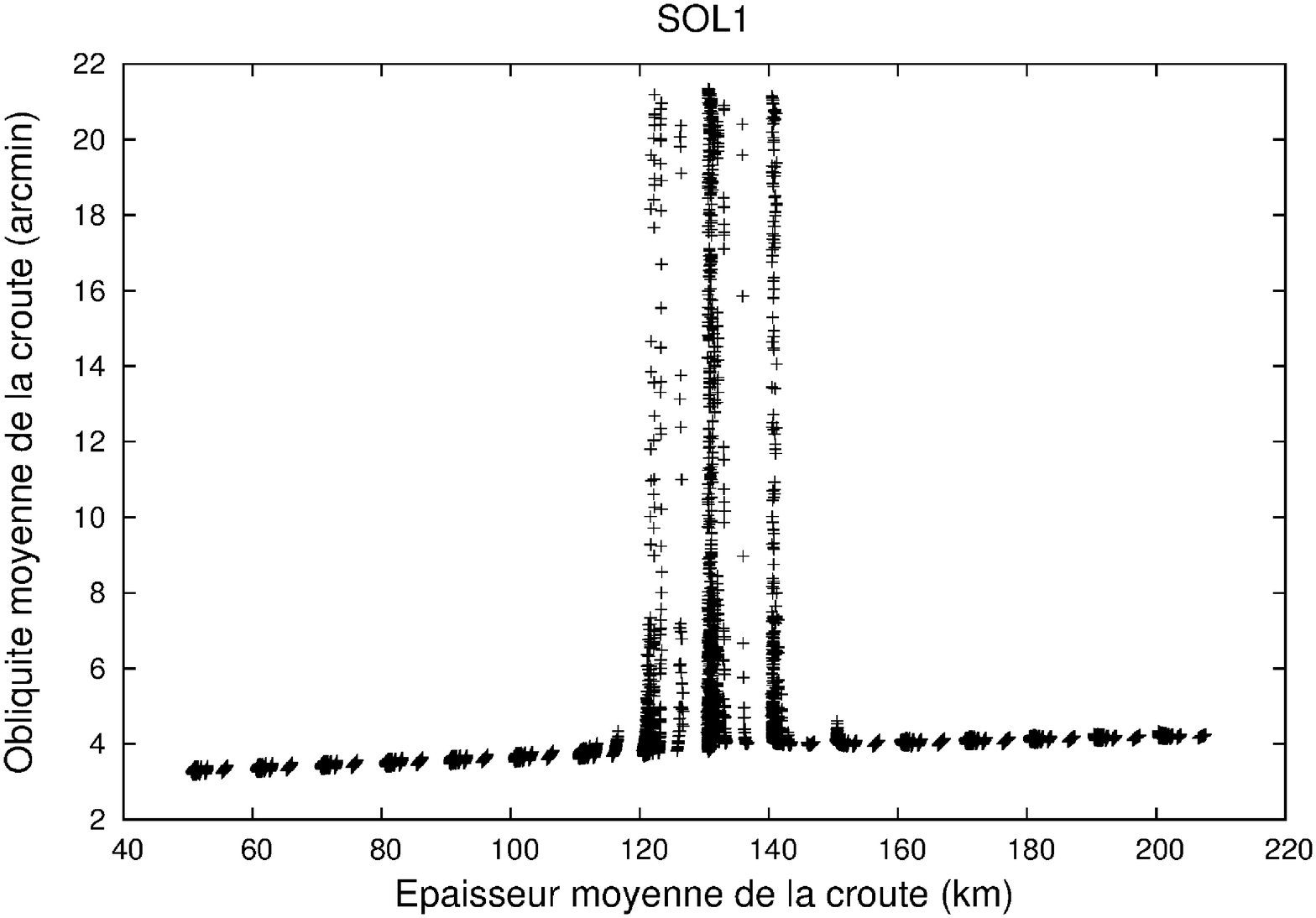} & \includegraphics[width=0.43\textwidth]{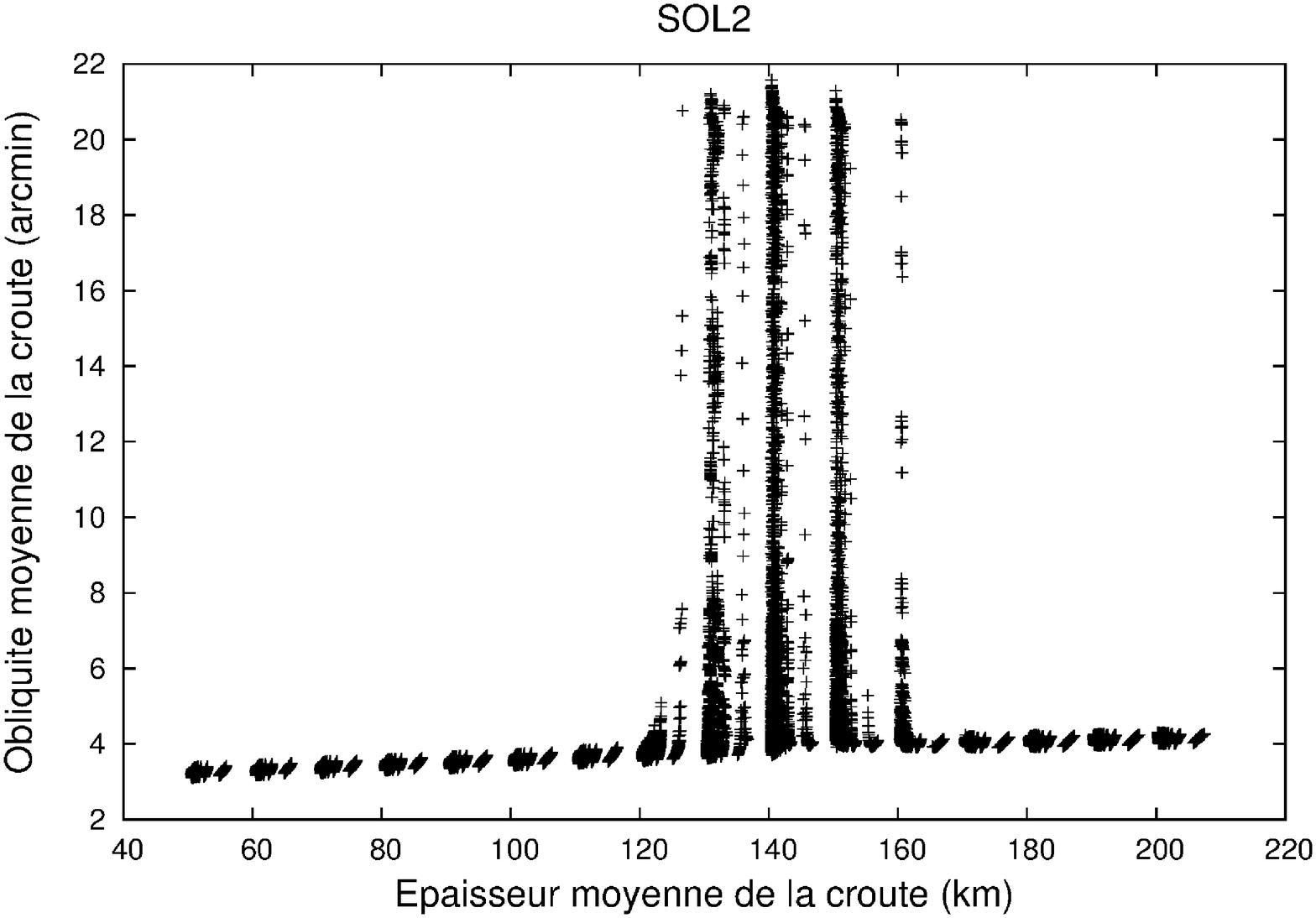}
\end{tabular}
\caption{Influence de l'\'epaisseur de la cro\^ute sur son obliquit\'e moyenne.\label{fig:oblishtsh}}
\end{figure}

\begin{figure}[ht]
\centering
\begin{tabular}{cc}
  \includegraphics[width=0.47\textwidth]{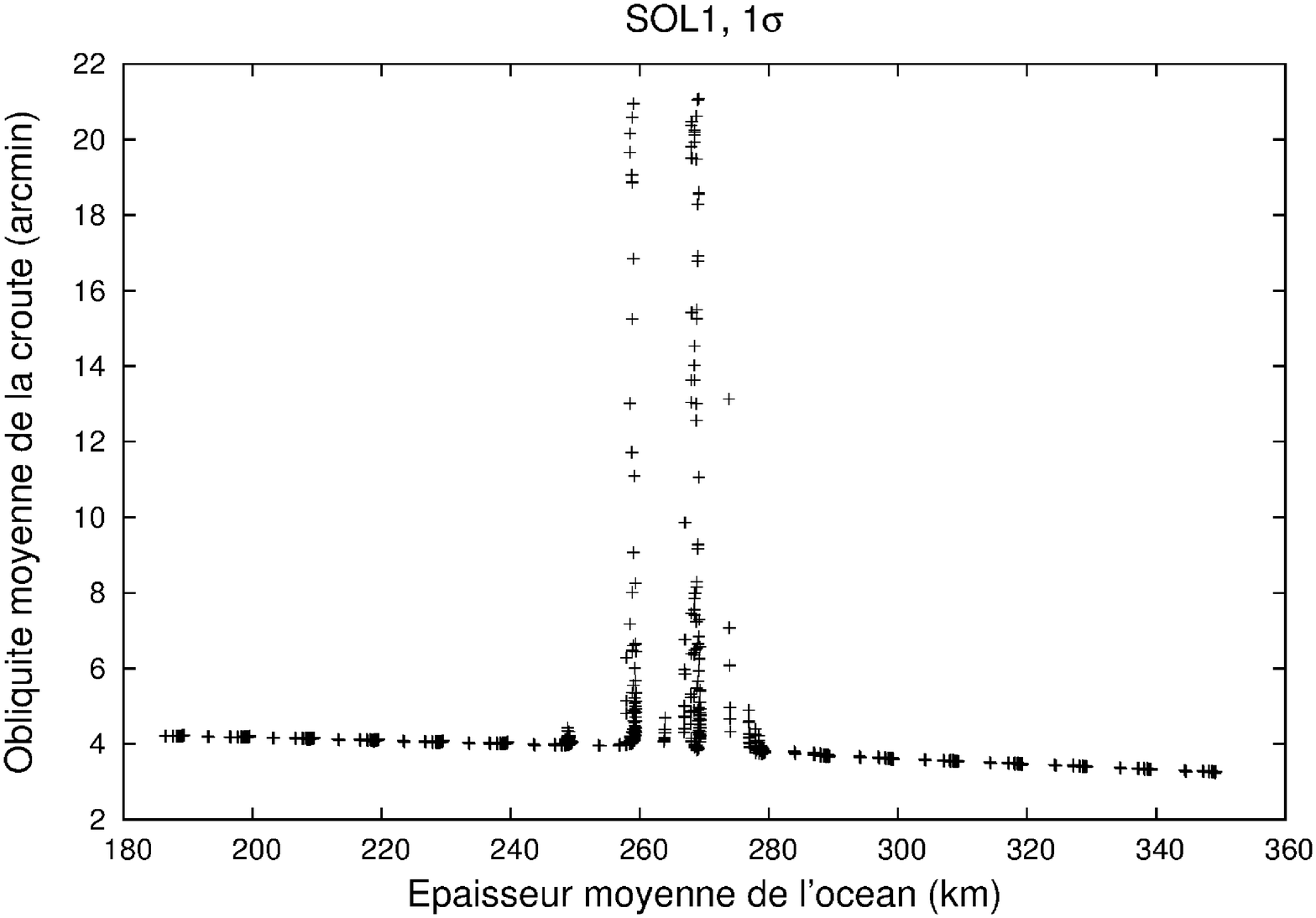} & \includegraphics[width=0.47\textwidth]{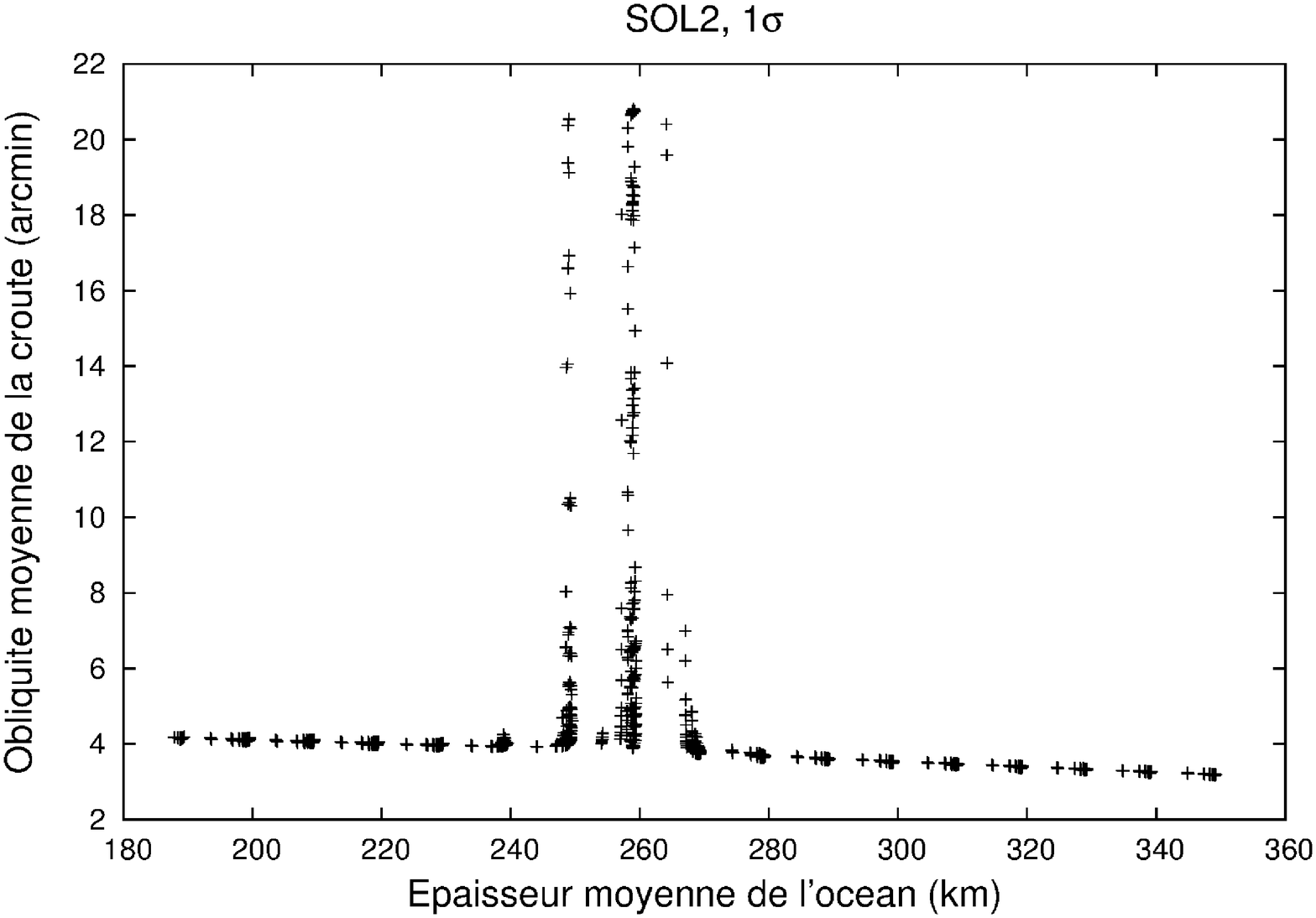} \\
  \includegraphics[width=0.47\textwidth]{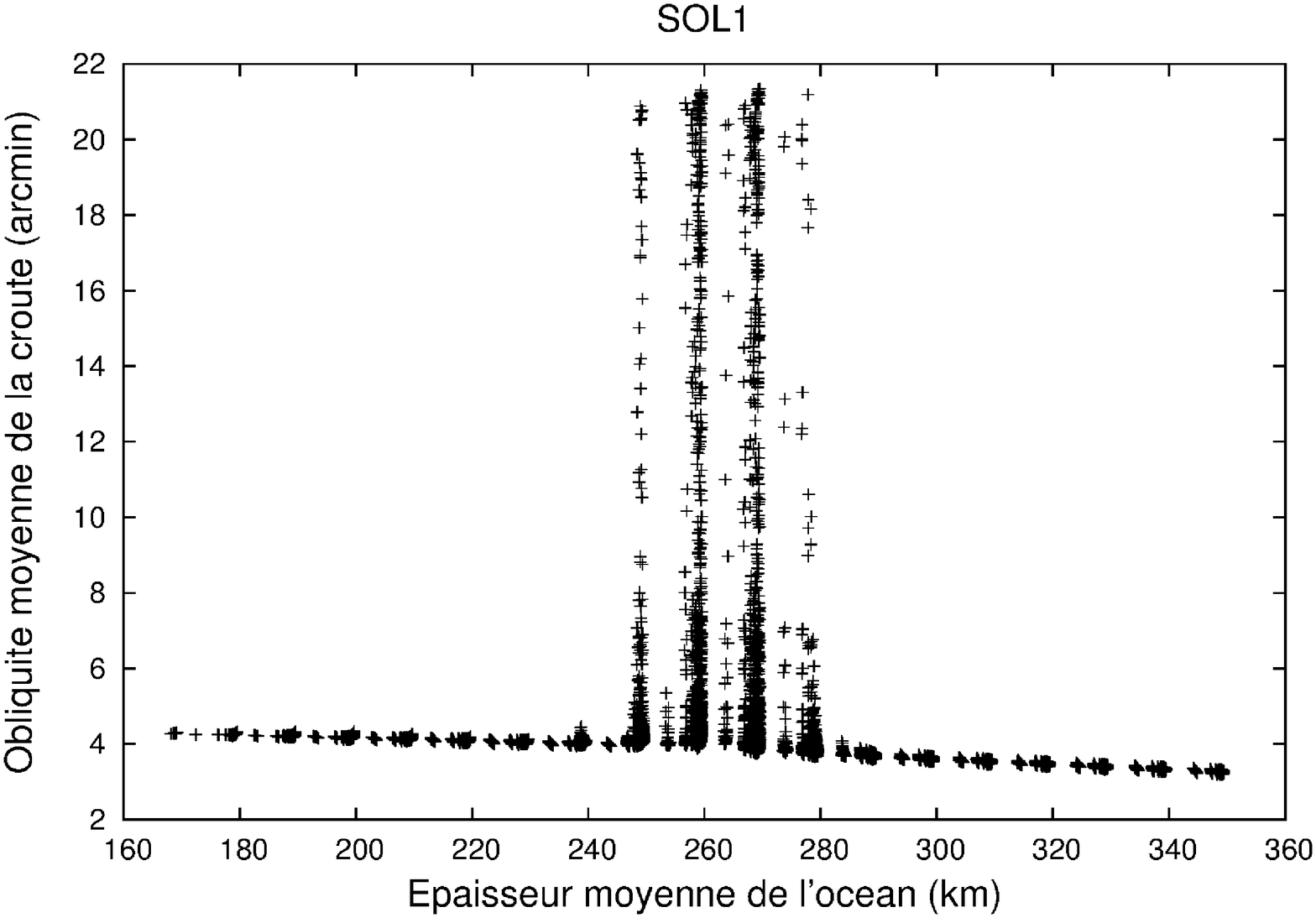} & \includegraphics[width=0.47\textwidth]{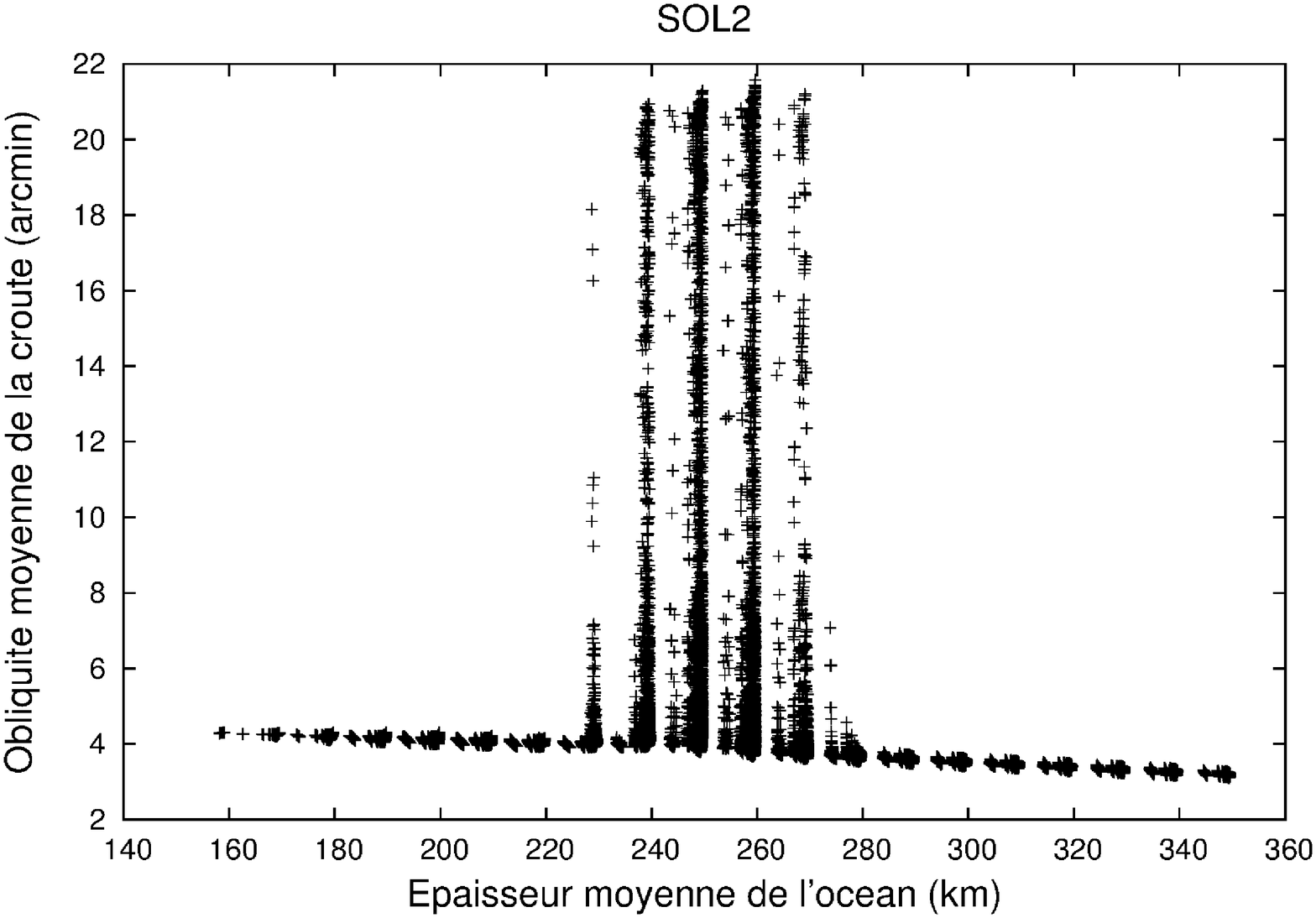}
\end{tabular}
\caption{Influence de l'\'epaisseur de l'oc\'ean sur l'obliquit\'e de la cro\^ute.\label{fig:oblishtoc}}
\end{figure}

 Le plus int\'eressant concerne l'obliquit\'e de la cro\^ute. Nous voyons (Fig.\ref{fig:oblishtsh} \& Fig.\ref{fig:oblishtoc}) en fait 2 r\'egimes. L'obliquit\'e 
  a g\'en\'eralement une valeur entre $3$ et $4$ minutes d'arc, sauf pour certaines valeurs des param\`etres d'int\'erieur o\`u elle peut d\'epasser $20$ minutes d'arc,
  ce qui peut expliquer les observations. Dans ce cas, on a une contrainte sur l'\'epaisseur de la cro\^ute, de l'ordre de 130 km pour SOL1 et 140 km pour SOL2, ainsi que sur 
  l'\'epaisseur de l'oc\'ean, de l'ordre de 260 km.
  
  \par Nous avons voulu v\'erifier l'explication du for\c{c}age r\'esonnant. Il s'est av\'er\'e impossible, dans la plupart des cas, d'extraire num\'eriquement la 
  fr\'equence propre quasi-r\'esonnante des simulations num\'eriques, d'une part car l'amplitude associ\'ee \'etait faible, ce qui est une cons\'equence de la 
  proximit\'e de l'\'equilibre dynamique, et aussi car elle \'etait tr\`es proche  d'une fr\'equence de for\c{c}age, ce qui plaide en faveur de l'explication 
  de la r\'esonance. Nous avons cherch\'e \`a d\'eterminer analytiquement les fr\'equences des petites oscillations susceptibles d'entrer en r\'esonance.
  
  \par Notre dynamique de rotation a 6 degr\'es de libert\'e, 3 par couche rigide. Autrement dit, il existe, pour chaque simulation, 6 fr\'equences de petites
  oscillations lin\'eairement ind\'ependantes. Cependant, en limitant notre \'etude aux obliquit\'es des 2 couches rigides, nous pouvons nous restreindre \`a 2
  fr\'equences. 
  
  \par La d\'etermination analytique de ces 2 fr\'equences, appel\'ees dans \citep{nn2014} $\sigma_3$ et $\sigma_4$, les indices 1 et 2 \'etant r\'eserv\'es au
  mouvement en longitude, est propos\'ee dans \citep{bvyk2011}. En adaptant leurs calculs \`a nos notations, nous obtenons
  
  \begin{eqnarray}
  \sigma_3 & = & -\frac{Z+\sqrt{\Delta}}{2C^cC^s}, \label{eq:sigma3} \\
  \sigma_4 & = & -\frac{Z-\sqrt{\Delta}}{2C^cC^s}, \label{eq:sigma4}
\end{eqnarray}
avec

\begin{eqnarray}
  Z        & = & K(C^c+C^s)-C^s\kappa^c-C^c\kappa^s,                                        \label{eq:Z} \\
  \Delta   & = & -4C^cC^s\left(\kappa^c\kappa^s-K\left(\kappa^c+\kappa^s\right)\right)+Z^2, \label{eq:Delta} \\
  K        & = & \frac{3\beta^o}{n_6}\left(C^c-A^c+C_b^o-A_b^o\right),                      \label{eq:K} \\
  \kappa^s & = & \frac{3n_6}{2}\left(C^s-A^s+C_t^o-A_t^o\right),                              \label{eq:kappas} \\
  \kappa^c & = & \frac{3n_6}{2}\left(C^c-A^c+C_b^o-A_b^o\right).                              \label{eq:kappac}
\end{eqnarray}
  Les p\'eriodes associ\'ees, c'est-\`a-dire $T_3=2\pi/\sigma_3$ et $T_4=2\pi/\sigma_4$, sont respectivement entre 200 et 260 ans, et entre 10 et 55 ans. Si 
  l'intervalle 200-260 ans ne correspond \`a aucun for\c{c}age \'evident, le for\c{c}age annuel, de p\'eriode $29.46$ ans, peut r\'esonner avec $\sigma_4$.
  
  \begin{table}[ht]
   \centering
   \caption[D\'ecomposition fr\'equentielle des variables d'inclinaison de Titan]{D\'ecomposition quasi-p\'eriodique des variables li\'ees \`a l'inclinaison orbitale de Titan
   $I_6\exp\imath\ascnode_6$, issue de TASS1.6 \citep{vd1995}. Le 1er terme est le terme constant qui donnera le plan inertiel de r\'ef\'erence $\left(\vv{e_1},\vv{e_2}\right)$, 
   le 5\`eme donnera la r\'esonance (Fig.\ref{fig:resonance}).\label{tab:inclitass}}
  \begin{tabular}{r|rrl}
  \hline
N & Amplitude &  P\'eriode & Cause \\
  & (arcmin)  &  (ans)&      \\
\hline
$1$ & $38.52$ &     $\infty$ &  Soleil \\
$2$ & $19.18$ &    $-703.51$ &  Aplatissement de Saturne\\
$3$ &  $0.90$ &   $-3263.07$ &  Pr\'ecession de Japet \\
$4$ &  $0.77$ &      $14.73$ &  For\c{c}age semi-annuel \\
$5$ &  $0.13$ &     $-29.46$ &  For\c{c}age annuel \\
$6$ &  $0.10$ &       $9.82$ &  For\c{c}age tri-annuel \\
\hline
\end{tabular}
\end{table}

  \par
\begin{figure}[ht]
\centering
\begin{tabular}{cc}
  \includegraphics[width=0.47\textwidth]{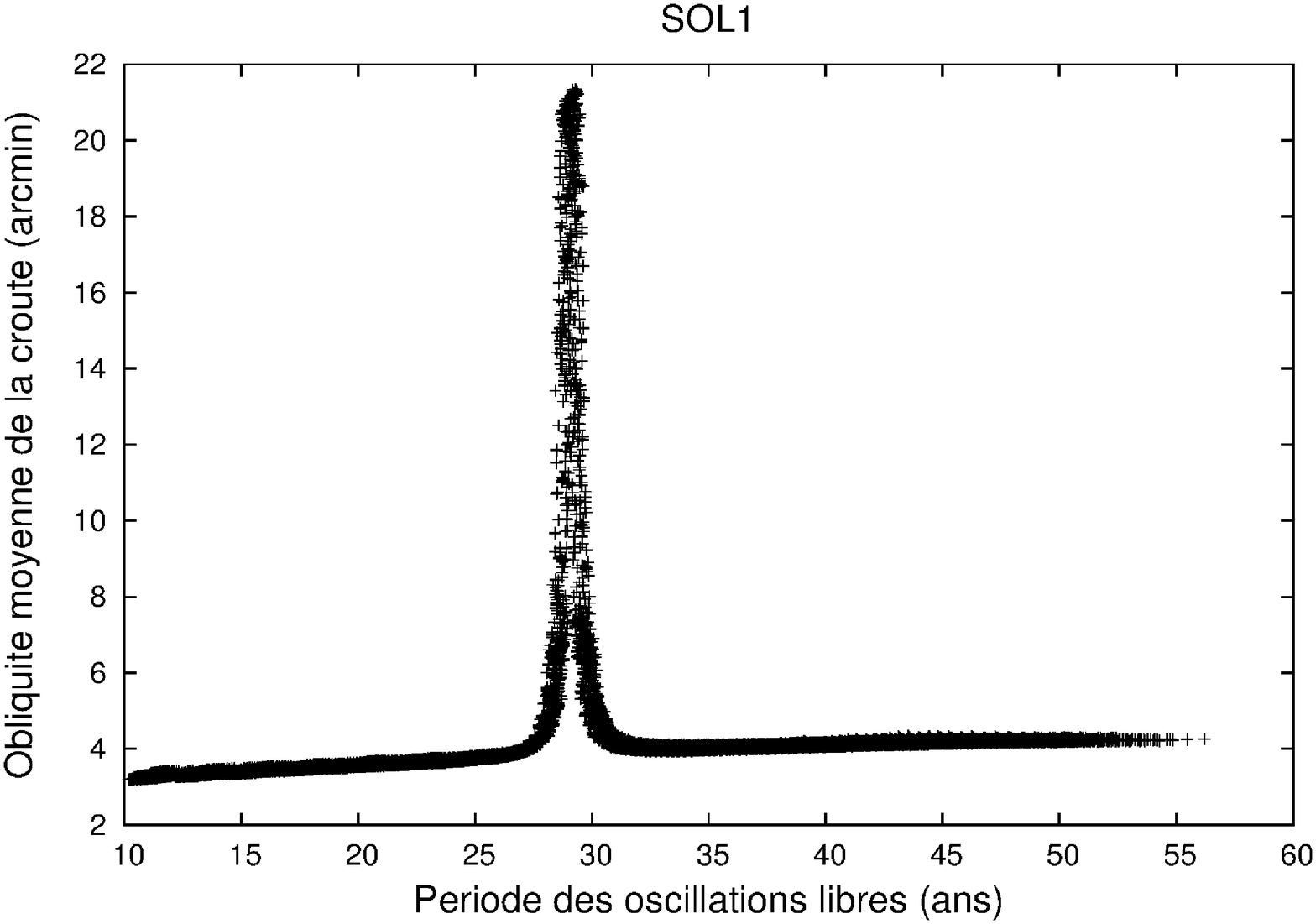} & \includegraphics[width=0.47\textwidth]{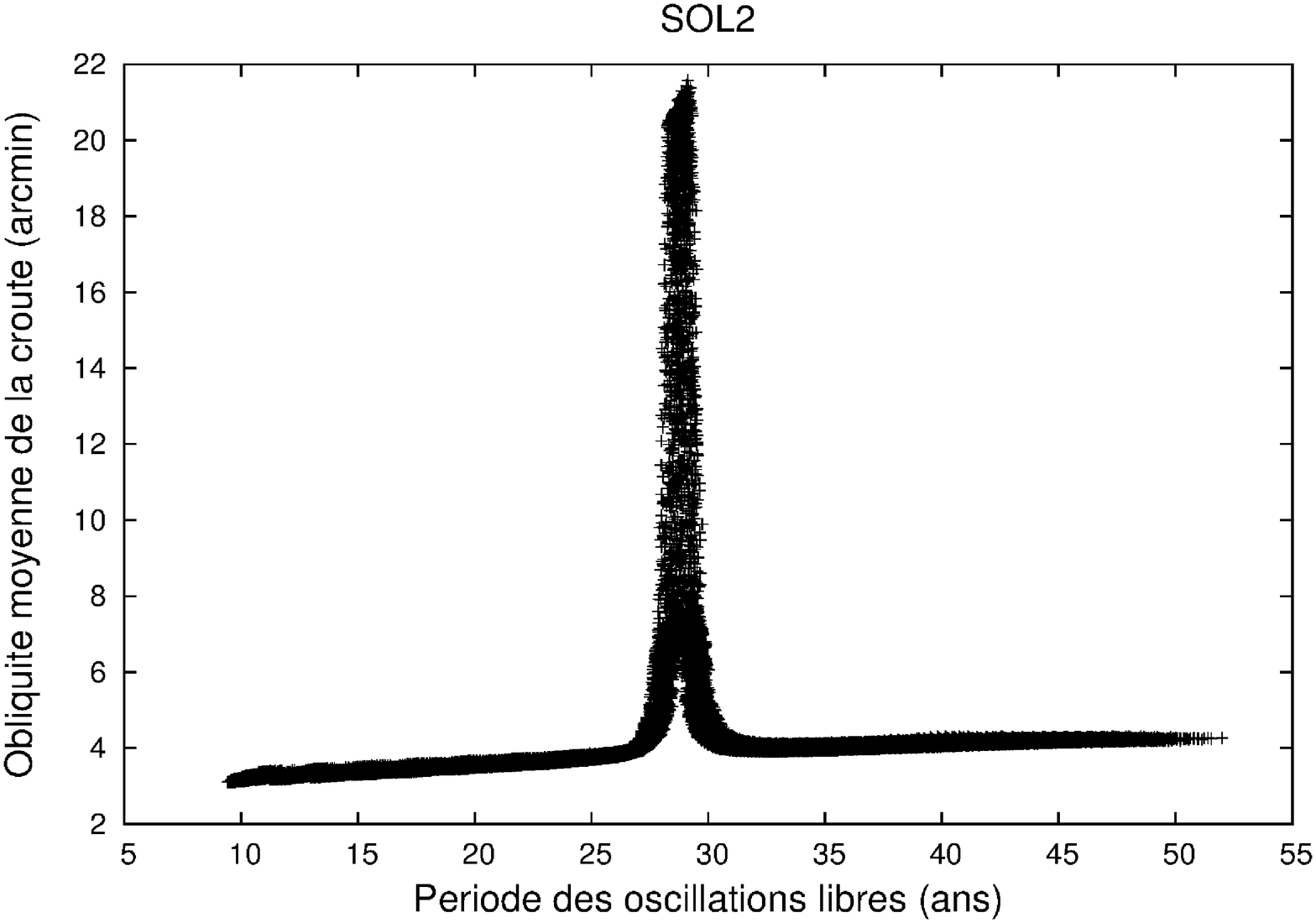}
\end{tabular}
\caption[For\c{c}age r\'esonnant de l'obliquit\'e de Titan.]{For\c{c}age r\'esonnant de l'obliquit\'e de Titan.\label{fig:resonance}}
\end{figure}

 La Fig.\ref{fig:resonance}, dont l'axe des abscisses a \'et\'e obtenu avec la formule (\ref{eq:sigma4}), et l'axe des ordonn\'ees contient les moyennes arithm\'etiques
  de l'obliquit\'e de la cro\^ute sur chaque simulation, confirme ce for\c{c}age r\'esonnant.
  
  \par Il appara\^it que pour SOL1, 2202 de nos Titans sur 20546 tombent dans la r\'esonance, soit $10.72\%$. Pour SOL2, il s'agit de 6540 Titans sur 50500, soit $12.95\%$.
  Si on se limite \`a $1\sigma$, les chiffres deviennent 234 sur 2237 pour SOL1, soit $10.46\%$, et 354 sur 2665 Titans pour SOL2, soit $13.28\%$. Ainsi, en construisant des 
  mod\`eles d'int\'erieur de Titan bas\'es uniquement sur des consid\'erations g\'eophysiques, des mesures du champ de gravit\'e et de la forme, on a entre $10$ et $13\%$
  de chances d'avoir une obliquit\'e r\'esonnante. C'est donc une explication tout-\`a-fait acceptable. \`A cette date, la litt\'erature n'en propose pas d'autre.

  \subsection{La variation temporelle de l'obliquit\'e}
  
  \par Le but est ici de simuler l'\'evolution future de l'obliquit\'e de Titan, afin de faire une pr\'ediction qui pourrait \^etre v\'erifiable par les observations. Nous
  cherchons \`a \'ecrire l'obliquit\'e de la cro\^ute sous la forme
  
  \begin{equation}
  \label{eq:Ks0}
  K^s(t) = A_0+\sum_i A_i \cos\phi_i(t),
  \end{equation}
  o\`u les $A_i$ sont des amplitudes r\'eelles et les $\phi_i(t)$ des fonctions lin\'eaires du temps. Pour obtenir ces coefficients nous utilisons notre syst\`eme d'\'equations
  pour simuler avec pr\'ecision l'obliquit\'e de la cro\^ute pour 13 de nos Titans, dont l'obliquit\'e a un comportement r\'esonnant. Simuler avec pr\'ecision veut notamment dire
  affiner les conditions initiales avec NAFFO (Chap.\ref{chap:naffo}) pour \^etre le plus pr\`es possible de l'\'equilibre, qui est difficile \`a localiser du fait du comportement 
  r\'esonnant\footnote{2 fr\'equences tr\`es proches sont pr\'esentes dans le syst\`eme, l'une, $\sigma_4$, qu'il faut enlever, et l'autre, le for\c{c}age annuel, qui restera.}.
  Nous avons d\^u simuler ce comportement sur 40000 ans.
  
  \par Apr\`es obtention des trajectoires \`a l'\'equilibre et analyse en fr\'equence des obliquit\'es, nous obtenons
  
  \begin{equation}
  \label{eq:Ksserie}
  K^s(t,\sigma_4) \approx A_0(\sigma_4)+A_1(\sigma_4)\cos\phi_1(t)+A_2(\sigma_4)\cos\phi_2(t),
\end{equation}
avec

\begin{eqnarray}
  \phi_1(t) & = & 0.20436788 \, t - 1.08611, \label{eq:phi1} \\
  \phi_2(t) & = & 0.63989736 \, t + 1.22080, \label{eq:phi2}
\end{eqnarray}
l'origine des temps \'etant J2000. Le temps $t$ est en ann\'ees, et les angles sont en radians. Les p\'eriodes de ces oscillations forc\'ees sont respectivement $30.74475$ et
$9.81906$ ans. Les termes $\phi_1$ et $\phi_2$ peuvent \^etre reconstruits \`a partir de la d\'ecomposition quasi-p\'eriodique de l'inclinaison de Titan donn\'ee par TASS1.6 
(Tab.\ref{tab:inclitass}), on a en effet $\phi_1\equiv(2)-(5)$ et $\phi_2\equiv(4)-(5)$. \`A partir des r\'esultats obtenus, qui sont donn\'es de fa\c{c}on exhaustive dans 
\citep{nn2014}, on obtient apr\`es ajustement par moindres carr\'es :

\begin{eqnarray}
  A_0 & = & \frac{\alpha}{\left|\sigma_4-\sigma\right|}, \label{eq:A0} \\
  A_1 & = & \beta \sgn(\sigma_4-\sigma), \label{eq:A1} \\
  A_2 & = & \gamma \sgn(\sigma-\sigma_4), \label{eq:A2}
\end{eqnarray}
avec

\begin{eqnarray}
  \alpha & = & (0.0176084 \pm 0.0002901)\, \textrm{arcmin}, \label{eq:alphatit} \\
  \beta  & = & (3.69780 \pm 0.01442)\, \textrm{arcmin}, \label{eq:betatit} \\
  \gamma & = & (0.573225 \pm 0.003937)\, \textrm{arcmin}, \label{eq:gammatit}
\end{eqnarray}
o\`u $\sigma = 2\pi/29.45716 = 0.2132991\,\textrm{rad}/\textrm{y}$ est la fr\'equence du for\c{c}age annuel (terme (5) dans la Tab.\ref{tab:inclitass}). Le changement de signe 
\`a la travers\'ee de la r\'esonance exacte est r\'ev\'elateur d'une discontinuit\'e topologique induite par la r\'esonance.

\par
\begin{figure}[ht]
\centering
\begin{tabular}{cc}
  \includegraphics[width=0.47\textwidth]{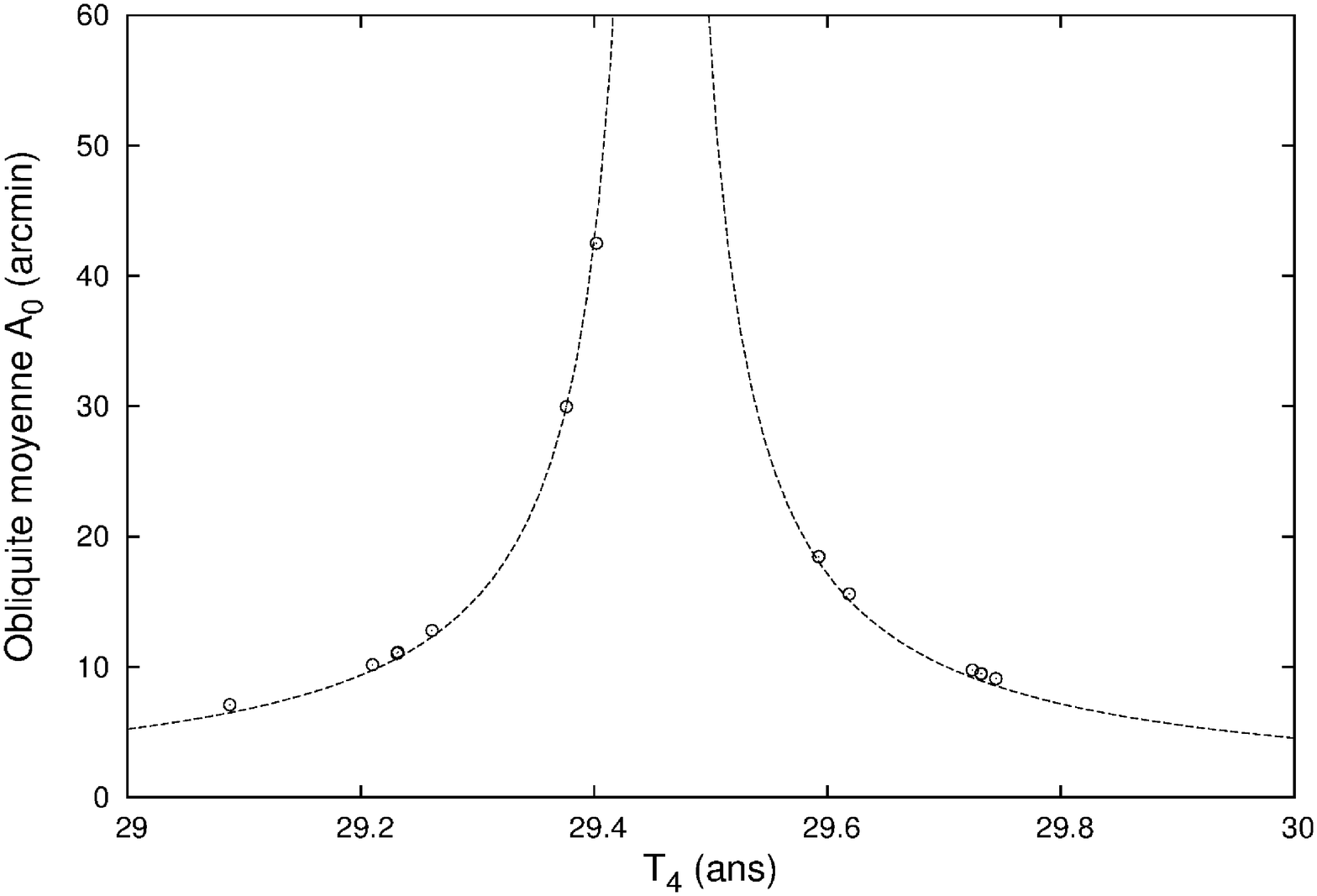} & \includegraphics[width=0.47\textwidth]{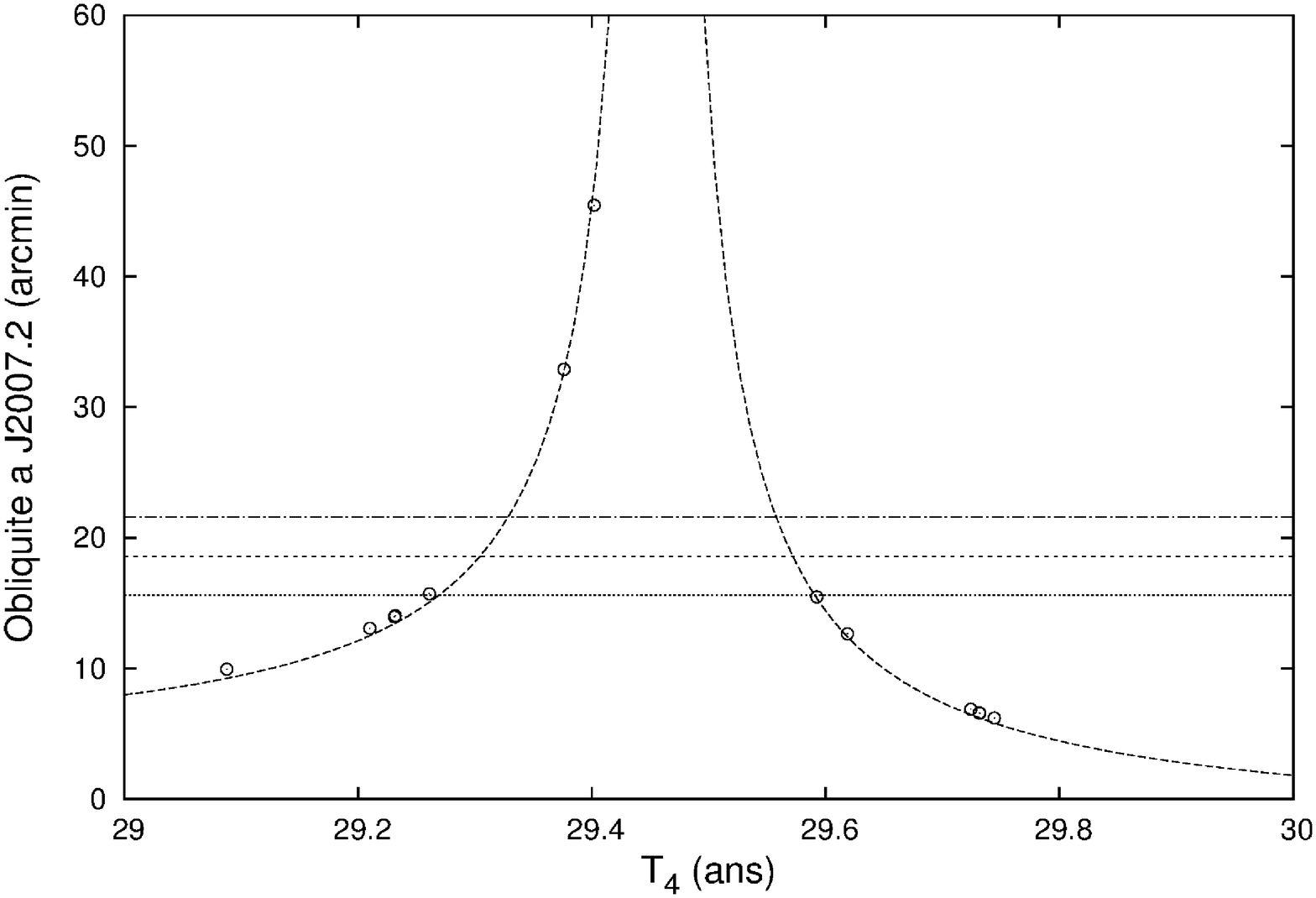}
\end{tabular}
\caption[Obliquit\'es moyenne et instantan\'ee]{Obliquit\'es moyenne (gauche) et instantan\'ee \`a J2007.2 (droite), donn\'ee par la formule (\ref{eq:Ksserie}). 
Les ronds indiquent les 13 trajectoires que nous avons simul\'ees, et les lignes horizontales repr\'esentent la mesure et ses barres d'erreur.\label{fig:obliquitemesuree}}
\end{figure}

 $A_0$ peut \^etre vu comme la valeur moyenne de l'obliquit\'e (Fig.\ref{fig:obliquitemesuree}, gauche). Pour obtenir la valeur de l'obliquit\'e \`a J2007.2, \`a comparer avec 
la mesure de \emph{Cassini}, nous utilisons la formule (\ref{eq:Ksserie}), ce qui donne la Fig.\ref{fig:obliquitemesuree}, droite.
  
\par
\begin{figure}[ht]
\centering
\begin{tabular}{cc}
  \includegraphics[width=0.47\textwidth]{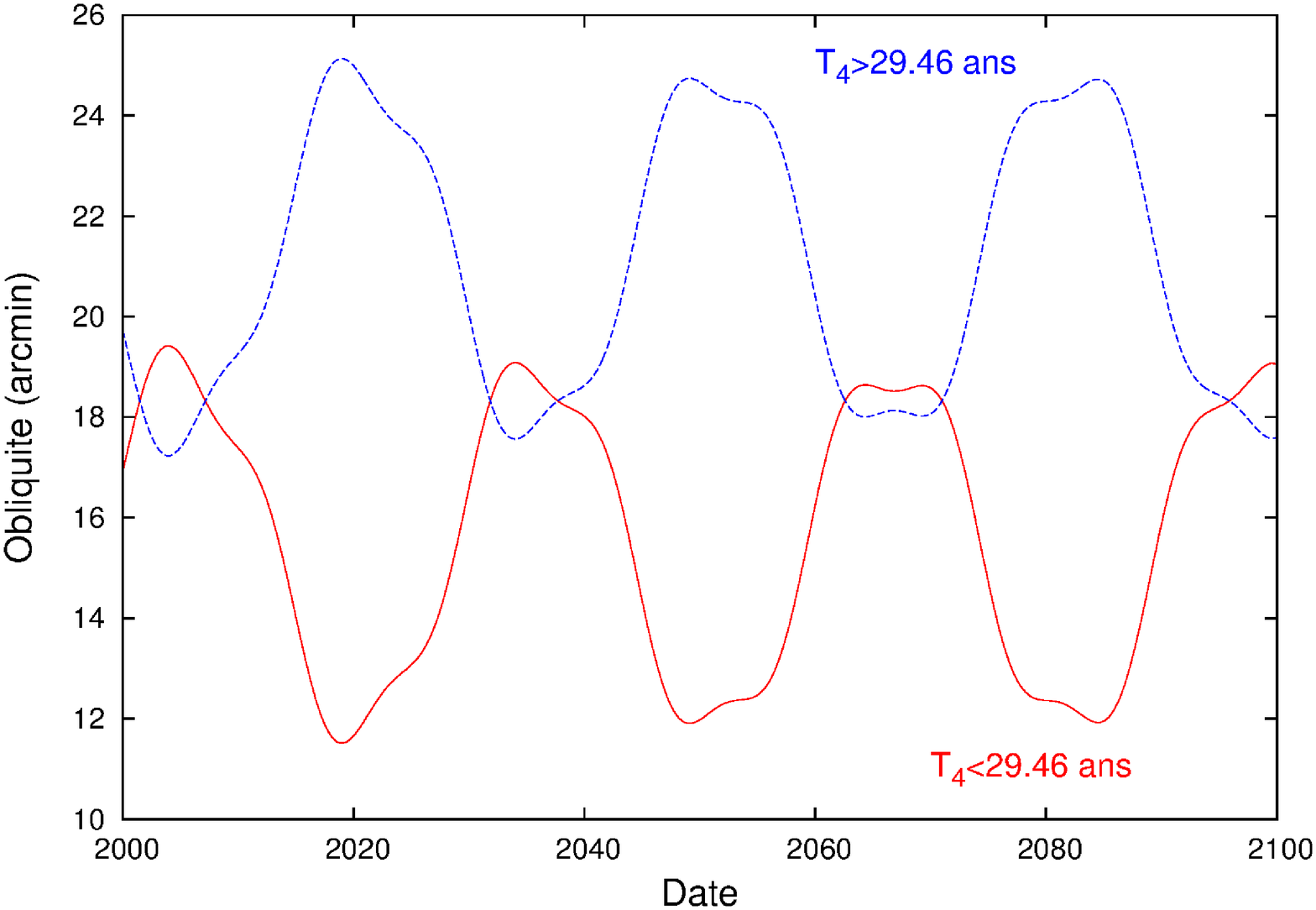} & \includegraphics[width=0.47\textwidth]{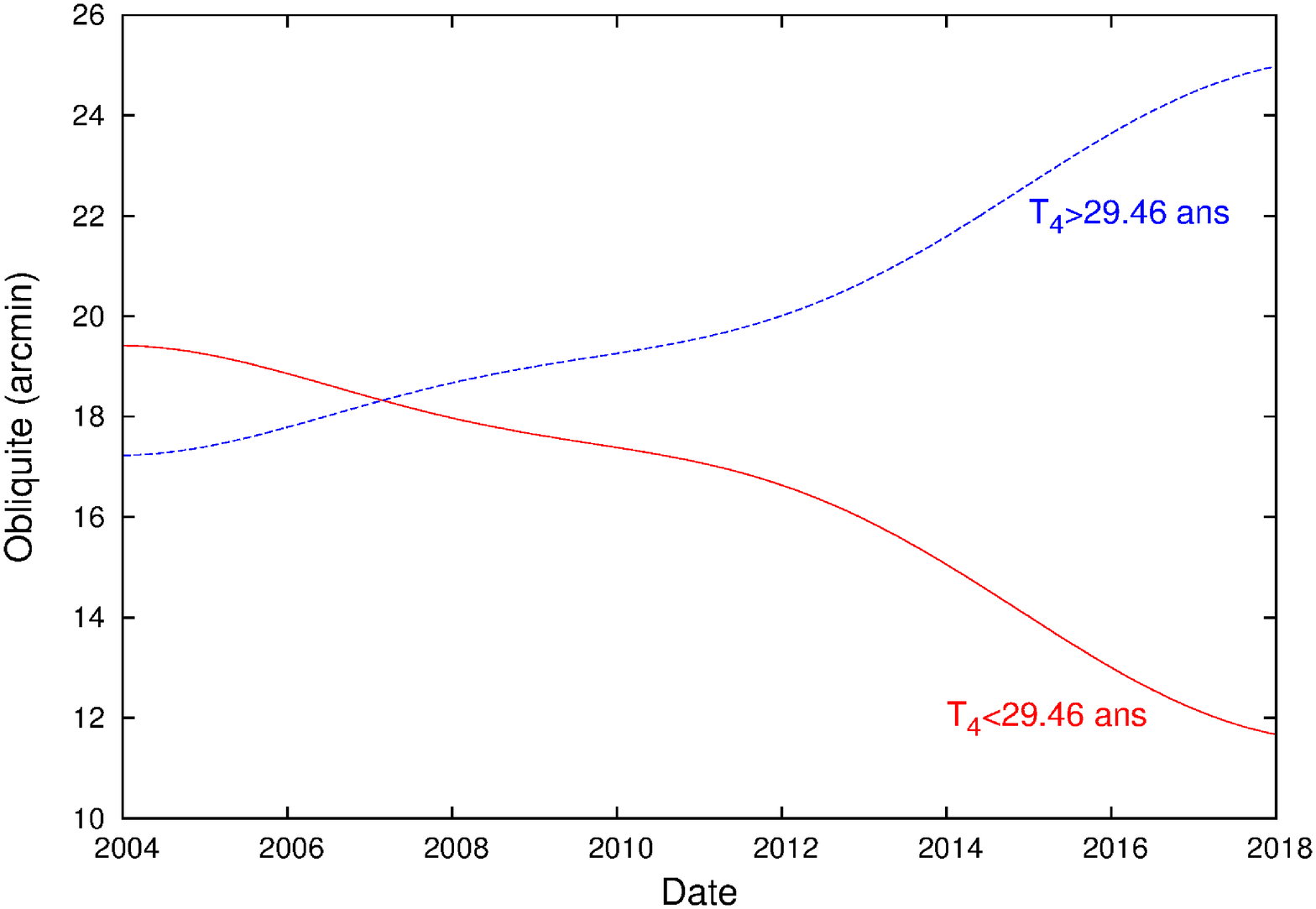}
\end{tabular}
\caption[Obliquit\'e de Titan \`a court terme]{Simulation de l'obliquit\'e de Titan sur 100 ans (\`a gauche) et sur la dur\'ee de la mission \emph{Cassini} (\`a droite).\label{fig:predictiontitan}}
\end{figure}

 Cette figure \ref{fig:obliquitemesuree} montre qu'il y a 2 intersections entre l'obliquit\'e th\'eorique et celle mesur\'ee \`a J2007.2, une \`a gauche et une \`a droite 
de la r\'esonance exacte. Les p\'eriodes des oscillations libres $T_4$ correspondantes sont respectivement $T_4 = 29.3\pm 0.03$ pour la solution dite de gauche et 
$T_4 = 29.572^{+0.019}_{-0.015}$ ans pour la solution dite de droite. Nous simulons l'\'evolution temporelle de l'obliquit\'e dans chacun des 2 cas (Fig.\ref{fig:predictiontitan})
et constatons 2 comportements diff\'erents : si $T_4<29.46$ ans, alors l'obliquit\'e de la cro\^ute de Titan est d\'ecroissante sur la dur\'ee de la mission \emph{Cassini}, et 
si $T_4>29.46$ ans alors elle est croissante.

 \par Nous nous attendons \`a une variation de l'obliquit\'e de Titan de 7 minutes d'arc. La mesure de \citet{mi2012} sugg\`ere une incertitude de 3 minutes 
 d'arc, a priori cette d\'etection est donc possible. Elle permettrait dans un premier temps de v\'erifier la validit\'e de la th\'eorie, et dans un deuxi\`eme 
 temps de d\'eterminer si Titan est \`a gauche ou \`a droite de la r\'esonance. On aurait ainsi de nouvelles informations sur l'int\'erieur.

  \section{Et si la cro\^ute est \'elastique\ldots\label{sec:elastic}}
  
  \par En 2010, un article de \citet{gm2010} a alert\'e la communaut\'e scientifique sur l'importance de la prise en compte de l'\'elasticit\'e de la cro\^ute des satellites naturels.
  En effet, elle induit un couple de rappel significatif. Ceci a motiv\'e sa prise en compte dans le calcul des librations en longitude \citep{vbt2013,jv2014,rrc2014}. Ces \'etudes ont
  montr\'e que les librations en longitude pouvaient \^etre r\'eduites d'un facteur $5$ si l'\'elasticit\'e \'etait consid\'er\'ee. \`A notre connaissance, il n'existe pas d'\'etude
  traitant la question de l'obliquit\'e d'une cro\^ute \'elastique. Dans cette Section, nous \'etendons 2 \'etudes existantes, \citep{gm2010} et \citep{rrc2014}, \`a ce probl\`eme.
  
  \subsection{Par le formalisme de \citet{gm2010}}
  
  \par L'id\'ee de cette \'etude \'etait d'estimer le d\'eplacement de l'axe le plus long de la cro\^ute \'elastique, en libration en longitude, sous l'effet de la 
  perturbation gravitationnelle du perturbateur, \'ecrite comme un couple de mar\'ee. Il s'agit donc d'un couple de rappel qui s'oppose \`a la libration en longitude. 
  Nous allons dans un premier temps retrouver le r\'esultat de \citet{gm2010} afin de valider la m\'ethode pour l'appliquer au mouvement latitudinal.
  
  \subsubsection{Couple \'elastique d\^u \`a un d\'eplacement radial en longitude}
  
  \par Comme pr\'ec\'edemment on utilise une param\'etrisation sph\'erique $(r,\psi,\phi)$ de la cro\^ute de Titan, o\`u $\phi$ est la longitude et $\psi$ la colatitude.
  En consid\'erant une libration en longitude $\phi_T$ comme une r\'eorientation de l'axe de mar\'ee, le tenseur des contraintes $(\sigma)$ nous est donn\'e par
  \citet{mn2008} :
  
  \begin{eqnarray}
  \sigma_{\psi\psi} & = &  3\left(\frac{1+\nu}{5+\nu}\right)\mu f\sin\phi_T\sin\left(2\phi-\phi_T\right)\left[\cos 2\psi-5\right], \label{eq:sigmatt} \\
  \sigma_{\phi\phi}     & = &  3\left(\frac{1+\nu}{5+\nu}\right)\mu f\sin\phi_T\sin\left(2\phi-\phi_T\right)\left[3\cos 2\psi+1\right], \label{eq:sigmaff} \\
  \sigma_{\psi\phi}   & = & 12\left(\frac{1+\nu}{5+\nu}\right)\mu f\sin\phi_T\cos\psi\cos\left(2\phi-\phi_T\right), \label{eq:sigmatf} \\
  \sigma_{\phi\psi}   & = & \sigma_{\psi\phi}, \label{eq:sigmaft}
\end{eqnarray}
o\`u $\nu$ est le coefficient de Poisson, $\mu$ est la rigidit\'e, et $f$ l'aplatissement de la cro\^ute.  En consid\'erant que l'angle de libration $\phi_T$ est petit, on a

\begin{eqnarray}
   \sin\phi_T\sin\left(2\phi-\phi_T\right) & =       & \frac{1}{2}\left(\cos 2\phi-\cos 2\phi \cos 2\phi_T +\sin 2\phi \sin 2\phi_T\right) \nonumber \\
                                           & \approx & \phi_T\sin 2\phi, \\
   \sin\phi_T\sin\left(2\phi-\phi_T\right) & =       & \frac{1}{2}\left(\sin 2\phi-\sin 2\phi \cos 2\phi_T +\cos 2\phi \sin 2\phi_T\right) \nonumber \\
                                           & \approx & \phi_T\cos 2\phi,
\end{eqnarray}
et de plus

\begin{eqnarray}
   \cos 2\psi-5    & =  & -2\left(2+\sin^2\psi\right), \\
   3\cos 2\psi+1   & =  & 2\left(2-3\sin^2\psi\right), \\
   f               & =  & (1+k_f)\frac{q}{2}, \\
   q               & =  & \frac{n_6^2R^3}{\mathcal{G}M_6},
\end{eqnarray}
on obtient

\begin{eqnarray}
  \sigma_{\psi\psi} & = & -3\left(\frac{1+\nu}{5+\nu}\right)\mu(1+k_f)q\phi_T\left(2+\sin^2\psi\right)\sin 2\phi , \label{eq:sigmattconst} \\
  \sigma_{\phi\phi} & = &  3\left(\frac{1+\nu}{5+\nu}\right)\mu(1+k_f)q\phi_T\left(2-3\sin^2\psi\right)\sin 2\phi, \label{eq:sigmaffconst} \\
  \sigma_{\psi\phi} & = &  6\left(\frac{1+\nu}{5+\nu}\right)\mu(1+k_f)q\phi_T\cos\psi\cos2\phi, \label{eq:sigmatfconst}
\end{eqnarray}
o\`u $k_f$ est le nombre de Love s\'eculaire.

\par Pour calculer la dissipation d'\'energie, nous avons besoin du tenseur des d\'eformations $(u)$. La quantit\'e $u_{\psi\psi}$ est compos\'ee 
d'une contrainte normale $\sigma_{\psi\psi}/E$ et d'une contrainte tangentielle $-\nu\sigma_{\phi\phi}/E$ o\`u $E=2\mu(1+\nu)$ est le module 
d'Young (voir par exemple \citet{l2011}). On obtient $u_{\phi\phi}$ de la m\^eme fa\c{c}on. Le cisaillement $u_{\psi\phi}=u_{\phi\psi}$ est 
proportionnel \`a la contrainte $\sigma_{\psi\phi}$, et on obtient :

\begin{eqnarray}
  u_{\psi\psi} & = & \frac{\sigma_{\psi\psi}-\nu\sigma_{\phi\phi}}{2\mu(1+\nu)}, \label{eq:utt} \\
  u_{\phi\phi} & = & \frac{\sigma_{\phi\phi}-\nu\sigma_{\psi\psi}}{2\mu(1+\nu)}, \label{eq:uff} \\
  u_{\psi\phi} & = & \frac{\sigma_{\psi\phi}}{2\mu}, \label{eq:utf} \\
  u_{\phi\psi} & = & u_{\psi\phi}. \label{eq:uft}
\end{eqnarray}

\par L'\'energie $\mathcal{E}$ associ\'ee \`a un \'el\'ement de la cro\^ute est ainsi

\begin{equation}
  \label{eq:Einf}
  \mathcal{E} = \frac{1}{2}\left(\sigma_{\psi\psi}u_{\psi\psi}+\sigma_{\phi\phi}u_{\phi\phi}+\sigma_{\psi\phi}u_{\psi\phi}+\sigma_{\phi\psi}u_{\phi\psi}\right),
\end{equation}
et l'\'energie \'elastique totale $E_{el}$ est obtenue par int\'egration sur la cro\^ute de Titan, de faible \'epaisseur $d_s$:

\begin{equation}
  \label{eq:Eel}
  E_{el} = \int_{r=R-d_s}^{r=R}\int_{\phi=0}^{\phi=2\pi}\int_{\psi=0}^{\psi=\pi}\mathcal{E} R^2 \sin\psi\,\textrm{d}r\,\textrm{d}\psi\,\textrm{d}\phi,
\end{equation}
ce qui donne la formule 9 de \citet{gm2010}:

\begin{equation}
  \label{eq:Eel2}
  E_{el} = \frac{48\pi}{5}\left(\frac{1+\nu}{5+\nu}\right)\left(1+k_f\right)^2q^2\mu d_sR_6^2\phi_T^2.
\end{equation}
Le couple de mar\'ee qui en r\'esulte $\vv{T_{el}}$ est donc :

\begin{eqnarray}
  \vv{T_{el}} & = & -\frac{\partial E_{el}}{\partial \phi}\bigg|_{\phi=\phi_T}\vv{e_{\phi}} \nonumber \\
              & = & -\frac{96\pi}{5}\left(\frac{1+\nu}{5+\nu}\right)\left(1+k_f\right)^2q^2\mu d_sR^2\phi_T\vv{e_{\phi}}. \label{eq:torqueel}
\end{eqnarray}

  \subsubsection{Couple de mar\'ee s'appliquant en longitude sur la cro\^ute}
  
  \par \`A partir de l'\'Eq.\ref{eq:sigmatt} on obtient :
  
  \begin{equation}
  \label{eq:sigmatt2}
  \sigma_{\psi\psi} = -6\left(\frac{1+\nu}{5+\nu}\right)\mu f\sin\phi_T\sin\left(2\phi-\phi_T\right)\left(2+\sin^2\psi\right).
  \end{equation}
  
  On a 
  
  \begin{equation}
  \label{eq:flattening}
  f = h_2\frac{R^3}{\mathcal{G}M_6}\left(\frac{\mathcal{G}M_{\saturn}}{r^3}\right) 
    = \frac{h_2}{2}\frac{R^3}{\mathcal{G}M_6}\left(\frac{\mathcal{G}M_{\saturn}}{a^3}\right)\left(\frac{a}{r}\right)^3
    = \frac{h_2}{2}q\left(\frac{a}{r}\right)^3,
\end{equation}
ce qui donne

\begin{eqnarray}
  \sigma_{\psi\psi} & = & -3\left(\frac{1+\nu}{5+\nu}\right)\mu h_2q\left(2+\sin^2\psi\right)\left(\frac{a}{r}\right)^3\sin\phi_T\sin\left(2\phi-\phi_T\right), \\
                    & = & -\frac{3}{2}\left(\frac{1+\nu}{5+\nu}\right)\mu h_2q\left(2+\sin^2\psi\right)\left(\frac{a}{r}\right)^3\left(\cos 2\phi-\cos \left(2(\phi-\phi_T)\right)\right). \label{eq:sigmattinter}
\end{eqnarray}
  
  \par La libration en longitude est tr\`es proche de la libration optique, on peut donc \'ecrire $\phi_T = \nu_6-\mathcal{M}_6$ o\`u $\nu_6$ est l'anomalie vraie et 
  $\mathcal{M}_6 = n_1t = (n_6-\dot{\varpi}_6)t$ l'anomalie moyenne. On obtient directement
  
  \begin{eqnarray}
  \left(\frac{a}{r}\right)^3\cos\left(2(\phi-\phi_T)\right) & = & \left(\frac{a}{r}\right)^3\cos\left(2(\phi+\mathcal{M}_6-\nu_6)\right) \nonumber \\
   & = & \cos\left(2(\phi+\mathcal{M}_6)\right)\left(\frac{a}{r}\right)^3\cos 2\nu_6 \label{eq:cosdevelop} \\
   & + & \sin\left(2(\phi+\mathcal{M}_6)\right)\left(\frac{a}{r}\right)^3\sin 2\nu_6. \nonumber
\end{eqnarray}

\par \`A partir de \citep{c1861,k1966}

\begin{eqnarray}
  \left(\frac{a}{r}\right)^3\cos 2\nu_6 & = & \sum_{i=-\infty}^{+\infty} G_{20i}(e)\cos\left((2+i)\mathcal{M}_6\right), \label{eq;kaulacos} \\
  \left(\frac{a}{r}\right)^3\sin 2\nu_6 & = & \sum_{i=-\infty}^{+\infty} G_{20i}(e)\sin\left((2+i)\mathcal{M}_6\right), \label{eq:kaulasin}
\end{eqnarray}
on obtient $\sigma_{\psi\psi}$ d\'evelopp\'e au degr\'e $1$ en excentricit\'e :

\begin{equation}
  \label{eq:sigmatttidlib}
  \sigma_{\psi\psi} = -\frac{3}{4}\left(\frac{1+\nu}{5+\nu}\right)\mu h_2eq\left(2+\sin^2\psi\right)\left(\cos(2\phi+n_1t)-7\cos(2\phi-n_1t)\right),
\end{equation}
et de la m\^eme fa\c{c}on :

\begin{equation}
  \label{eq:sigmafftidlib}
  \sigma_{\phi\phi} = \frac{3}{4}\left(\frac{1+\nu}{5+\nu}\right)\mu h_2eq\left(2-3\sin^2\psi\right)\left(\cos(2\phi+n_1t)-7\cos(2\phi-n_1t)\right).
\end{equation}
2 modes de mar\'ee sont impliqu\'es, comme le montrent les termes $\cos(2\phi-n_1t) = \cos(2(\phi-n_1t/2))$ et $\cos(2\phi+n_1t) = \cos(2(\phi+n_1t/2))$ qui
correspondent respectivement aux modes $n_1/2$ et $3n_1/2$. De la m\^eme fa\c{c}on, on a

\begin{equation}
  \label{eq:sigmatftidlib}
  \sigma_{\psi\phi} = \frac{3}{2}\left(\frac{1+\nu}{5+\nu}\right)\mu h_2 q e\cos\psi\left(7\sin(2\phi-n_1t)-\sin(2\phi+n_1t)\right)
\end{equation}
et $\sigma_{\psi\phi} = \sigma_{\phi\psi}$. Maintenant calculons l'\'energie associ\'ee \`a chacun de ces 2 modes de mar\'ee.

\par On a, pour le mode $n_1/2$ :

\begin{eqnarray}
  \sigma_{\psi\psi} & = & -\frac{3}{4}\left(\frac{1+\nu}{5+\nu}\right)\mu h_2 e q\left(2+\sin^2\psi\right)\cos(2\phi+n_1t), \label{eq:sigmattp1} \\
  \sigma_{\phi\phi} & = &  \frac{3}{4}\left(\frac{1+\nu}{5+\nu}\right)\mu h_2 e q\left(2-3\sin^2\psi\right)\cos(2\phi+n_1t), \label{eq:sigmaffp1} \\
  \sigma_{\psi\phi} & = & -\frac{3}{2}\left(\frac{1+\nu}{5+\nu}\right)\mu h_2 e q\cos\psi\sin(2\phi+n_1t), \label{eq:sigmatfp1}
\end{eqnarray}
et pour le mode $3n_1/2$:

\begin{eqnarray}
  \sigma_{\psi\psi} & = &  \frac{21}{4}\left(\frac{1+\nu}{5+\nu}\right)\mu h_2 e q\left(2+\sin^2\psi\right)\cos(2\phi-n_1t), \label{eq:sigmattp2} \\
  \sigma_{\phi\phi} & = & -\frac{21}{4}\left(\frac{1+\nu}{5+\nu}\right)\mu h_2 e q\left(2-3\sin^2\psi\right)\cos(2\phi-n_1t), \label{eq:sigmaffp2} \\
  \sigma_{\psi\phi} & = &  \frac{21}{2}\left(\frac{1+\nu}{5+\nu}\right)\mu h_2 e q\cos\psi\sin(2\phi-n_1t). \label{eq:sigmatfp2}
\end{eqnarray}
Ces formules se retrouvent dans \citep{gm2010} \`a l'exception de (\ref{eq:sigmatfp2}), probablement \`a cause d'une erreur typographique. Les \'energies associ\'ees 
sont

\begin{eqnarray}
  E_{n_1/2}  & = & \frac{3}{5}\pi\left(\frac{1+\nu}{5+\nu}\right)(h_2 e q)^2\mu d_sR^2, \label{eq:Ens2} \\
  E_{3n_1/2} & = & \frac{147}{5}\pi\left(\frac{1+\nu}{5+\nu}\right)(h_2 e q)^2\mu d_sR^2 = 49E_{n_1/2}, \label{eq:E3ns2}
\end{eqnarray}
l\`a encore je retrouve les formules de \citet{gm2010}. Le facteur $49$ entre ces 2 \'energies vient du fait qu'\`a chaque fr\'equence d'excitation correspond un aplatissement 
$f$ diff\'erent du corps consid\'er\'e, ici la cro\^ute de Titan, donc une r\'eponse diff\'erente. Sans cela, l'\'energie totale de mar\'ee serait nulle car la somme de
ses 2 composantes s'annulerait. Le couple de mar\'ee \`a la fr\'equence $\omega$ est $\vv{T_{\omega}} = -2n_1E_{\omega}/Q_{\omega}(\omega-n_1)\vv{e_{\phi}}$, c'est-\`a-dire

\begin{eqnarray}
  \vv{T_{n_1/2}}  & = &  \frac{12\pi}{5Q_{n_1/2}}\left(\frac{1+\nu}{5+\nu}\right)(h_2eq)^2\mu d_sR^2\vv{e_{\phi}}, \label{eq:torquens2} \\
  \vv{T_{3n_1/2}} & = & -\frac{588\pi}{5Q_{3n_1/2}}\left(\frac{1+\nu}{5+\nu}\right)(h_2eq)^2\mu d_sR^2\vv{e_{\phi}}, \label{eq:torque3ns2}
\end{eqnarray}
et le couple total est

\begin{equation}
  \label{eq:tidtorklib}
  \vec{T} = \frac{12\pi}{5}\left(\frac{1+\nu}{5+\nu}\right)(h_2eq)^2\mu d_sR^2\left(\frac{1}{Q_{n_1/2}}-\frac{49}{Q_{3n_1/2}}\right)\vv{e_{\phi}}.
\end{equation}
\citet{gm2010} ont suppos\'e que les 2 coefficients de dissipation $Q_{n_1/2}$ et $Q_{3n_1/2}$ \'etaient \'egaux.

\par Une fois le couple externe total $\vv{T_{ext}}$ obtenu, compos\'e du couple de la mar\'ee de Saturne et \'eventuellement du couple atmosph\'erique, on obtient
la r\'eponse de la cro\^ute $\lambda_{ext}$ avec $\vv{T_{ext}} = \vv{T_{el}}$ (\'Eq.\ref{eq:torqueel}), ce qui donne

\begin{equation}
  \label{eq:elresponse}
  \lambda_{ext}=\frac{5}{96\pi}\left(\frac{5+\nu}{1+\nu}\right)\frac{T_{ext}}{(1+k_f)^2q^2\mu d_sR^2},
\end{equation}
et l\`a encore on retrouve une formule de \citet{gm2010}, en l'occurence la (17).

  \subsubsection{Les mar\'ees en latitude}
  
  \par Je propose ici une extension du raisonnement de \citet{gm2010} au mouvement en latitude. Il s'agit donc d'une contribution originale.
  
  \par L'obliquit\'e provoque une r\'eorientation en latitude, d'angle $\psi_R$, donc orthogonale \`a l'axe de rotation. Dans ce cas le tenseur des contraintes $(\sigma)$
  s'\'ecrit \citep{mn2008} :
  
  \begin{eqnarray}
  \sigma_{\psi\psi} & =      & \frac{1}{2}\left(\frac{1+\nu}{5+\nu}\right)\mu f\sin\psi_R \nonumber \\
                    & \times & \left(4\cos\psi_R\sin(2\psi)\cos\phi-\sin\psi_R\left(\cos(2\psi)\left(3+\cos(2\phi)\right)+10\sin^2\phi\right)\right), \label{eq:sigmattr} \\
  \sigma_{\phi\phi} & =      & \frac{1}{2}\left(\frac{1+\nu}{5+\nu}\right)\mu f\sin\psi_R \nonumber \\
                    & \times & \left(2\sin\psi_R\sin^2\phi+12\cos\psi_R\sin(2\psi)\cos\phi-3\sin\psi_R\cos(2\psi)\left(3+\cos(2\phi)\right)\right), \label{eq:sigmaffr} \\
  \sigma_{\psi\phi} & =      & \frac{1}{2}\left(\frac{1+\nu}{5+\nu}\right)\mu f\sin\psi_R\sin\phi\left(\sin\psi_R\cos\psi\cos\phi-\cos\psi_R\sin\psi\right), \label{eq:sigmatfr}
  \end{eqnarray}
  et nous incluons les librations en latitude de la fa\c{c}on suivante :
  
  \begin{equation}
  \label{eq:nslib}
  \psi_R = K^s\sin n_2t,
  \end{equation}
  o\`u $K^s$ est l'obliquit\'e de la cro\^ute de Titan, constante dans cette Section. $n_2 = n_6-\dot{\ascnode}_6$ est la p\'eriode des librations en latitude lorsque 
  l'obliquit\'e est constante. On peut d\'ej\`a voir une diff\'erence fondamentale entre l'amplitude des librations en longitude, associ\'ee \`a une p\'eriode qui est 
  aussi celle du couple de mar\'ee, alors que l'obliquit\'e est une d\'eviation constante de l'axe de rotation par rapport \`a la normale \`a l'orbite, son mouvement 
  de pr\'ecession est \`a longue p\'eriode ($\approx700$ ans pour un Titan non r\'esonnant, $\approx29.5$ ans pour un Titan r\'esonnant), mais le couple de mar\'ee,
  susceptible de provoquer une d\'eviation de la cro\^ute si elle est \'elastique, a une p\'eriode tr\`es proche de la p\'eriode diurnale, l\'eg\`erement inf\'erieure 
  \`a $16$ jours. On peut donc s'attendre \`a ce que l'effet de l'\'elasticit\'e de la cro\^ute sur l'obliquit\'e soit une oscillation \`a courte p\'eriode qui s'ajouterait
  au mouvement rigide \`a longue p\'eriode. Une autre diff\'erence est que le param\`etre donnant l'amplitude des librations en latitude est l'obliquit\'e, qui n'induit pas 
  de variation de la distance Titan-Saturne, donc de l'aplatissement $f$ et du couple de mar\'ee, alors que l'excentricit\'e, principale cause des librations en longitude, 
  est associ\'ee \`a des variations de ce couple.
  
  \par Le tenseur de contraintes $(\sigma)$ s'\'ecrit ainsi : 
  
  \begin{eqnarray}
  \sigma_{\psi\psi} & = &   \left(\frac{1+\nu}{5+\nu}\right) \mu f K^s \cos\phi\left(\cos(2\psi-n_2t)-\cos(2\psi+n_2t)\right), \label{eq:sigmattr2} \\
  \sigma_{\phi\phi} & = &  3\left(\frac{1+\nu}{5+\nu}\right) \mu f K^s \cos\phi\left(\cos(2\psi-n_2t)-\cos(2\psi+n_2t)\right), \label{eq:sigmaffr2} \\
  \sigma_{\psi\phi} & = & -2\left(\frac{1+\nu}{5+\nu}\right) \mu f K^s \sin\phi\left(\cos(\psi-n_2t)-\cos(\psi+n_2t)\right). \label{eq:sigmatfr2}
\end{eqnarray}
  
  \par En gardant \`a l'esprit que l'angle $\psi$ est en moyenne constant puisqu'il correspond \`a une direction orthogonale \`a l'axe de rotation, nous voyons 
  4 modes appara\^itre : $-n_2$, $-n_2/2$, $n_2/2$ et $n_2$. Nous calculons l'\'energie comme pr\'ec\'edemment, ce qui donne :
  
  \begin{eqnarray}
  E_{-n_2}   & = & \frac{2\pi}{3}\left(\frac{1+\nu}{5+\nu}\right)^2\mu K^{s2}f^2\left(3-\cos 2n_2t\right), \label{eq:Emn} \\
  E_{-n_2/2} & = & \frac{\pi}{2}\left(\frac{1+\nu}{5+\nu}\right)^2\left(\frac{5-3\nu}{1+\nu}\right)\mu K^{s2}f^2\left(1-\frac{\cos 2n_2t}{15}\right), \label{eq:Emns2} \\
  E_{n_2/2}  & = & \frac{\pi}{2}\left(\frac{1+\nu}{5+\nu}\right)^2\left(\frac{5-3\nu}{1+\nu}\right)\mu K^{s2}f^2\left(1-\frac{\cos 2n_2t}{15}\right) = E_{-n_2/2}, \label{eq:Ens22} \\
  E_{n_2}    & = & \frac{2\pi}{3}\left(\frac{1+\nu}{5+\nu}\right)^2\mu K^{s2}f^2\left(3-\cos 2n_2t\right) = E_{-n_2}. \label{eq:En}
  \end{eqnarray}

  \par Une cons\'equence de la sym\'etrie de ces 4 \'energies est que le couple r\'esultant est nul. En effet, toutes les th\'eories de mar\'ee s'accordent \`a dire que la 
  dissipation de mar\'ee ne d\'epend pas du signe de la fr\'equence d'excitation, autrement dit, pour toute fr\'equence de mar\'ee $\omega$, on a $Q_{\omega} = Q_{-\omega}$.
  
  \par Cette m\'ethode ne permet donc pas de mettre en \'evidence un \'eventuel impact de l'\'elasticit\'e de la cro\^ute sur son obliquit\'e.

  \subsection{Par le formalisme de \citet{rrc2014}}

  \par Dans \citep{rrc2014}, les auteurs \'etudient l'impact de l'\'elasticit\'e sur les librations en longitude par le biais d'une composante variable dans le tenseur 
  d'inertie.
  
  \par Le potentiel de mar\'ee de Saturne agissant sur Titan s'\'ecrit \citep[Eq.1]{k1964}

\begin{eqnarray}
	U & = & \frac{\mathcal{G}M_{\saturn}}{a}\sum_{l=2}^{\infty}\left(\frac{R}{a}\right)^l\sum_{m=0}^l\frac{(l-m)!}{(l+m)!}\left(2-\delta_{0m}\right)P_l^m(\cos\psi)\sum_{p,q}F_{lmp}(K^s)G_{lpq}(e) \label{eq:kaula} \\
	& \times & \left(\cos m\phi\binom{\cos}{\sin}_{l-m\,\textrm{impair}}^{l-m\,\textrm{pair}}\left((l-2p)\omega_6+(l-2p+q)\mathcal{M}_6+m(\ascnode_6-\theta^*)\right)\right. \nonumber \\
	& & \left.+\sin m\phi\binom{\sin}{-\cos}_{l-m\,\textrm{impair}}^{l-m\,\textrm{pair}}\left((l-2p)\omega_6+(l-2p+q)\mathcal{M}_6+m(\ascnode_6-\theta^*)\right)\right), \nonumber
\end{eqnarray}
o\`u $\omega_6=\varpi_6-\ascnode_6$ est l'argument du p\'ericentre de Titan. $\theta^*$ est le temps sid\'eral du m\'eridien de r\'ef\'erence, il d\'ecoule de la rotation
synchrone $\theta^* = \lambda_6 = \mathcal{M}_6+\omega_6+\ascnode_6$. Les fonctions $G$ ont d\'ej\`a \'et\'e pr\'esent\'ees, et les fonctions $F$ sont les fonctions en
inclinaison de \citet{k1964}. Nous faisons les approximations suivantes :

\begin{itemize}

 \item l=2,
 
 \item nous nous limitons au degr\'e $1$ en excentricit\'e $e$ et en obliquit\'e $K^s$,
 
\end{itemize}
ce qui donne

\begin{eqnarray}
	U & = & \frac{\mathcal{G}M_{\saturn}}{a}\left(\frac{R}{a}\right)^2\sum_{m=0}^2\frac{(2-m)!}{(2+m)!}\left(2-\delta_{0m}\right)P_2^m(\cos\psi)\sum_{p,q}F_{2mp}(K^s)G_{2pq}(e) \label{eq:kaula2} \\
	& \times & \left(\cos m\phi\binom{\cos}{\sin}_{2-m\,\textrm{impair}}^{2-m\,\textrm{pair}}\left((2-2p-m)\omega_6+(2-2p+q-m)\mathcal{M}_6\right)\right. \nonumber \\
	& & \left.+\sin m\phi\binom{\sin}{-\cos}_{2-m\,\textrm{impair}}^{2-m\,\textrm{pair}}\left((2-2p-m)\omega_6+(2-2p+q-m)\mathcal{M}_6\right)\right), \nonumber
\end{eqnarray}
c'est-\`a-dire

\begin{eqnarray}
	U & = & n^2R^2\left(P_2^0(\cos\psi)F_{201}(K^s)\left(G_{210}(e)+\left(G_{21-1}(e)+G_{211}(e)\right)\cos\mathcal{M}_6\right)\right. \label{eq:kaula3} \\
	& & \left. +\frac{P_2^1(\cos\psi)}{3}\left(F_{210}(K^s)G_{200}(e)\sin(\mathcal{M}_6+\omega_6-\phi)+F_{211}(K^s)G_{210}(e)\sin(-\mathcal{M}_6-\omega_6-\phi)\right)\right. \nonumber \\
	& & \left. +\frac{P_2^2(\cos\psi)}{12}F_{220}(K^s)\left(G_{20-1}(e)\cos(2\phi+\mathcal{M}_6)+G_{200}(e)\cos 2\phi+G_{201}(e)\cos(2\phi-\mathcal{M}_6)\right)\right), \nonumber
\end{eqnarray}
avec

\begin{eqnarray}
	F_{201}(K^s) & = & \frac{3\sin^2 K^s}{4}-\frac{1}{2}, \label{eq:F201} \\
	F_{210}(K^s) & = & \frac{3}{4}\sin K^s \left(1+\cos K^s \right), \label{eq:F210} \\
	F_{211}(K^s) & = & -\frac{3}{2}\sin K^s \cos K^s, \label{eq:F211} \\
        F_{220}(K^s) & = & \frac{3}{4}\left(1+\cos K^s \right)^2, \label{eq:F220}
\end{eqnarray}
\begin{eqnarray}
	G_{20-1}(e) & = & -\frac{1}{2}e+\frac{1}{16}e^3+\mathcal{O}(e^4), \label{eq:G20m1} \\
	G_{200}(e)  & = & 1-\frac{5}{2}e^2+\frac{13}{16}e^4+\mathcal{O}(e^6), \label{eq:G200} \\
	G_{201}(e)  & = & \frac{7}{2}e-\frac{123}{16}e^3+\mathcal{O}(e^5), \label{eq:G201} \\
	G_{21-1}(e) & = & \frac{3}{2}e+\frac{27}{16}e^3+\mathcal{O}(e^5), \label{eq:G21m1} \\
	G_{210}(e)  & = & (1-e^2)^{-3/2}, \label{eq:G210} \\
	G_{211}(e)  & = & \frac{3}{2}e+\frac{27}{16}e^3+\mathcal{O}(e^5), \label{eq:G211}
\end{eqnarray}
ce qui donne

\begin{eqnarray}
	U & = & n^2R^2\left(-\frac{P_2^0(\cos\psi)}{2}\left(1+3e\cos\mathcal{M}_6\right)+P_2^1(\cos\psi)K^s\cos\phi\sin(\mathcal{M}_6+\omega_6)\right. \nonumber \\
	&   & \left.+P_2^2(\cos\psi)\left(\frac{\cos 2\phi}{4}+e\left(\frac{3}{4}\cos 2\phi\cos\mathcal{M}_6+\sin 2\phi\sin\mathcal{M}_6\right)\right)\right). \label{eq:kaula4}
\end{eqnarray}
  ce potentiel \'etant compatible au degr\'e 1 avec celui donn\'e par \citep[\'Eq.2]{jv2011} avec l'exception que cette derni\`ere r\'ef\'erence consid\`ere une rotation
  l\'eg\`erement super-synchrone, et avec celui donn\'e par \citep[\'Eq.9]{g2004}, qui ne consid\'erait pas l'obliquit\'e. On consid\`ere que la d\'eformation de la 
  cro\^ute est due au potentiel de mar\'ee et au potentiel centrifuge $U_r$ :
  
  \begin{equation}
	\label{eq:centrifugal}
	U_r = \frac{\mathcal{G}M_6}{R}\frac{q}{3}\left(1-P_2^0(\cos\psi)\right),
\end{equation}
avec

\begin{equation}
	\label{eq:qrrrc}
	q = \frac{n_6^2R^3}{\mathcal{G}M_6}.
\end{equation}
  
  \par \citet{rrc2014} ont consid\'er\'e une d\'ecomposition d'ordre 2 en harmoniques sph\'eriques du rayon de Titan, suivant \citet{hla1982}. Ils ont pos\'e
  
  \begin{equation}
	\label{eq:radiusandy}
	r(r_0,\psi,\phi) = r_0+u_r(r_0,\psi,\phi)
\end{equation}
avec

\begin{equation}
	\label{eq:urandy}
	u_r(r_0,\psi,\phi) = r_0 \left(\sum_{j=0}^2P_2^j(\cos\psi)(d_{2,j}\cos j\phi+e_{2,j}\sin j\phi)\right).
\end{equation}

\par On a

\begin{equation}
	\label{eq:lovenumber}
	u_r(r_0,\psi,\phi) = \frac{H(r_0)}{g(R)}V_2,
\end{equation}
  o\`u $g(R)$ repr\'esente la gravit\'e de surface de Titan ($1.35\,\textrm{m}/\textrm{s}^2$) et $V_2=U+U_r$ est le potentiel perturbateur. $H$ est une fonction radiale
  adimensionn\'ee correspondant au nombre de Love $h_2$ \`a la surface. On peut la consid\'erer constante sur la fine \'epaisseur de la cro\^ute. Nous obtenons directement :
  
  \begin{eqnarray}
	d_{2,0} & = & -n_6^2R^2\left(\frac{5}{6}+\frac{3}{2}e\cos\mathcal{M}_6\right)\frac{h_2}{r_0g(R)}, \label{eq:d20} \\
	d_{2,1} & = & n_6^2R^2\frac{h_2}{r_0g(R)}K^s\sin(\mathcal{M}_6+\omega_6), \label{eq:d21} \\
	d_{2,2} & = & \frac{n_6^2R^2}{4}\frac{h_2}{r_0g(R)}(1+3e\cos\mathcal{M}_6), \label{eq:d22} \\
	e_{2,1} & = & 0, \label{eq:e21} \\
	e_{2,2} & = & \frac{3n_6^2R^2}{4}\frac{h_2}{r_0g(R)}e\sin\mathcal{M}_6. \label{eq:e22}
\end{eqnarray}	

  \par Les composantes du tenseur d'inertie de la cro\^ute s'\'ecrivent
  
  \begin{eqnarray}
	I_{11} & = & \iiint \rho(r) r^4 (\sin^2\psi\sin^2\phi+\cos^2\psi)\sin\psi\,\textrm{d}r\,\textrm{d}\psi\,\textrm{d}\phi, \label{eq:I11rrc} \\
	I_{22} & = & \iiint \rho(r) r^4 (\sin^2\psi\cos^2\phi+\cos^2\psi)\sin\psi\,\textrm{d}r\,\textrm{d}\psi\,\textrm{d}\phi, \label{eq:I22rrc} \\
        I_{33} & = & \iiint \rho(r) r^4 \sin^3\psi\,\textrm{d}r\,\textrm{d}\psi\,\textrm{d}\phi, \label{eq:I33rrc} \\
        I_{12} & = & \frac{1}{2}\iiint \rho(r) r^4 \sin^3\psi\sin 2\phi\,\textrm{d}r\,\textrm{d}\psi\,\textrm{d}\phi, \label{eq:I12rrc} \\
        I_{13} & = & \iiint \rho(r) r^4 \sin^2\psi\cos\psi\cos \phi\,\textrm{d}r\,\textrm{d}\psi\,\textrm{d}\phi, \label{eq:I13rrc} \\
        I_{23} & = & \iiint \rho(r) r^4 \sin^2\psi\cos\psi\sin \phi\,\textrm{d}r\,\textrm{d}\psi\,\textrm{d}\phi. \label{eq:I23rrc}
\end{eqnarray}
  
  \par Pour inclure les d\'eformations radiales de la cro\^ute dans le tenseur d'inertie, faisons le changement de variables suivant :
  
  \begin{eqnarray}
	r & = & r_0\left(1+\sum_{j=0}^2P_2^j(\cos\psi)(d_{2,j}\cos j\phi+e_{2,j}\sin j\phi)\right), \label{eq:urandy2} \\
	\frac{\textrm{d}r}{\textrm{d}r_0} & = & \frac{r}{r_0}+r_0\sum_{j=0}^2P_2^j(\cos\psi)\left(\frac{\textrm{d}d_{2,j}}{\textrm{d}r_0}\cos j\phi+\frac{\textrm{d}e_{2,j}}{\textrm{d}r_0}\sin j\phi\right), \label{eq:durandy}
\end{eqnarray}	
et avec 

\begin{eqnarray}
	\rho(r) & \approx & \rho(r_0), \label{eq:rho} \\
	\frac{\textrm{d}}{\textrm{d}r_0}\left(r_0^5(1+u_r)^5\right) & = & 5r_0^4(1+u_r)^4\left(1+u_r+r_0\frac{\textrm{d}u_r}{\textrm{d}r_0}\right), \label{eq:deriv5} \\
\end{eqnarray}
nous avons, pour chacune des int\'egrales (\ref{eq:I11rrc}) \`a (\ref{eq:I23rrc}), la forme suivante

\begin{eqnarray}
	I_{\alpha\beta} & = & \int_{r_0=R-d_s}^{r_0=R}\int_{\psi=0}^{\psi=\pi}\int_{\phi=0}^{\phi=2\pi}\rho(r_0)\left[r_0^4+\frac{\textrm{d}(d_{20}r_0^5)}{\textrm{d}r_0}P_2^0(\cos\psi)+\frac{\textrm{d}(d_{21}r_0^5)}{\textrm{d}r_0}P_2^1(\cos\psi)\cos\phi\right. \nonumber \\
	& & \left.+\frac{\textrm{d}(d_{22}r_0^5)}{\textrm{d}r_0}P_2^2(\cos\psi)\cos 2\phi+\frac{\textrm{d}(e_{22}r_0^5)}{\textrm{d}r_0}P_2^2(\cos\psi)\sin 2\phi\right]\mathcal{F}_{\alpha\beta}(\psi,\phi)\,\textrm{d}r_0\,\textrm{d}\psi\,\textrm{d}\phi \label{eq:Iab}
\end{eqnarray}
avec

\begin{eqnarray}
	\mathcal{F}_{11}(\psi,\phi) & = & (\sin^2\psi\sin^2\phi+\cos^2\psi)\sin\psi, \label{eq:f11} \\
	\mathcal{F}_{22}(\psi,\phi) & = & (\sin^2\psi\cos^2\phi+\cos^2\psi)\sin\psi, \label{eq:f22} \\
	\mathcal{F}_{33}(\psi,\phi) & = & \sin^3\psi, \label{eq:f33} \\
        \mathcal{F}_{12}(\psi,\phi) & = & \frac{1}{2}\sin^3\psi\sin 2\phi, \label{eq:f12} \\
        \mathcal{F}_{13}(\psi,\phi) & = & \sin^2\psi\cos\psi\cos\phi, \label{eq:f13} \\
        \mathcal{F}_{23}(\psi,\phi) & = & \sin^2\psi\cos\psi\sin\phi, \label{eq:f23}
\end{eqnarray}
ce qui donne

\begin{eqnarray}
	I_{11} & = & \frac{8\pi}{3}\int_{R-d_s}^{R} \rho_s\left(r_0^4+\frac{1}{10}\frac{\textrm{d}(d_{2,0}r_0^5)}{\textrm{d}r_0}-\frac{3}{5}\frac{\textrm{d}(d_{2,2}r_0^5)}{\textrm{d}r_0}\right)\,\textrm{d}r_0, \label{eq:I11calc} \\
	I_{22} & = & \frac{8\pi}{3}\int_{R-d_s}^{R} \rho_s\left(r_0^4+\frac{1}{10}\frac{\textrm{d}(d_{2,0}r_0^5)}{\textrm{d}r_0}+\frac{3}{5}\frac{\textrm{d}(d_{2,2}r_0^5)}{\textrm{d}r_0}\right)\,\textrm{d}r_0, \label{eq:I22calc} \\
	I_{33} & = & \frac{8\pi}{3}\int_{R-d_s}^{R} \rho_s\left(r_0^4-\frac{1}{5}\frac{\textrm{d}(d_{2,0}r_0^5)}{\textrm{d}r_0}\right)\,\textrm{d}r_0, \label{eq:I33calc} \\
	I_{12} & = & \frac{8\pi}{5}\int_{R-d_s}^{R} \rho_s\frac{\textrm{d}(e_{2,2}r_0^5)}{\textrm{d}r_0}\,\textrm{d}r_0, \label{eq:I12calc} \\
	I_{13} & = & \frac{4\pi}{5}\int_{R-d_s}^{R} \rho_s\frac{\textrm{d}(d_{2,1}r_0^5)}{\textrm{d}r_0}\,\textrm{d}r_0, \label{eq:I13calc} \\
	I_{23} & = & 0, \label{eq:I23calc}
\end{eqnarray}	
  sans oublier bien s\^ur $I_{21}=I_{12}$, $I_{31}=I_{13}$ et $I_{32}=I_{23}$. Les formules (\ref{eq:I11calc}) \`a (\ref{eq:I33calc}) sont les m\^emes que \citep[\'Eq.13-15]{rrc2014},
  alors que $I_{12}$ y est donn\'e avec un coefficient $-8\pi/15$. En rempla\c{c}ant les coefficients $d_{i,j}$ et $e_{i,j}$ par leurs expressions (\ref{eq:d20}) \`a (\ref{eq:e22})
  et en lin\'earisant par rapport \`a l'\'epaisseur de la cro\^ute $d_s$, nous obtenons
  
  \begin{eqnarray}
	I_{11} & = & \frac{8\pi}{3}\rho_sd_sR^4\left(1-\frac{2}{5}n_6^2R\frac{h_2}{g(R)}\left(\frac{7}{3}+6e\cos\mathcal{M}_6\right)\right), \label{eq:I11calc2} \\
	I_{22} & = & \frac{8\pi}{3}\rho_sd_sR^4\left(1+\frac{2}{5}n_6^2R\frac{h_2}{g(R)}\left(\frac{2}{3}+3e\cos\mathcal{M}_6\right)\right), \label{eq:I22calc2} \\
	I_{33} & = & \frac{8\pi}{3}\rho_sd_sR^4\left(1+2n_6^2R\frac{h_2}{g(R)}\left(\frac{1}{3}+\frac{2}{5}e\cos\mathcal{M}_6\right)\right), \label{eq:I33calc2} \\
	I_{12} & = & \frac{24\pi}{5}\rho_sn_6^2d_sR^5\frac{h_2}{g(R)}e\sin\mathcal{M}_6, \label{eq:I12calc2} \\
	I_{13} & = & \frac{16\pi}{5}\rho_sn_6^2d_sR^5\frac{h_2}{g(R)}K^s\sin(\mathcal{M}_6+\omega_6), \label{eq:I13calc2} \\
	I_{23} & = & 0. \label{eq:I23calc2}
\end{eqnarray}	
  
  \par On voit donc que l'\'elasticit\'e de la cro\^ute la rend d\'eformable, et que les d\'eformations ont une p\'eriode proche de la p\'eriode diurne, ou p\'eriode orbitale de Titan.
  En fait, 2 fr\'equences interviennent au 1er degr\'e en excentricit\'e / obliquit\'e : celle de l'anomalie moyenne $\mathcal{M}_6$, not\'ee $n_1$, et celle de $\mathcal{M}_6+\omega_6$,
  not\'ee $n_2$. La premi\`ere, de p\'eriode $15.9464$ jours, intervient en longitude et est li\'ee \`a l'excentricit\'e de Titan, alors que la deuxi\`eme, de p\'eriode 
  $15.9445$ jours, intervient en latitude et est li\'ee \`a l'obliquit\'e.

  \subsection{Synth\`ese}
  
  \par Ces 2 m\'ethodes indiquent que si la perturbation en longitude due a l'\'elasticit\'e de la cro\^ute \`a la m\^eme p\'eriode que les librations non \'elastiques,
  et ainsi doivent affecter son amplitude, ce n'est pas le cas pour les librations en latitude, la perturbation ayant une p\'eriode bien plus courte que celles li\'ees au 
  mouvement de pr\'ecession du p\^ole de rotation. On s'attend donc \`a des librations \`a courte p\'eriode additionnelles, qui ne devraient pas affecter le mouvement de
  pr\'ecession rigide.
  
  \section{Conclusions et perspectives}
  
  \par Notre \'etude confirme que l'explication d'une obliquit\'e r\'esonnante pour Titan initialement donn\'ee par \citep{bvyk2011} est tout-\`a-fait acceptable. De plus,
  elle implique des contraintes sur l'int\'erieur, en particulier l'\'epaisseur de la cro\^ute, ou profondeur de l'oc\'ean, de l'ordre de 130 - 140 km.
  
  \par Au moins 2 observations peuvent \^etres interpr\'et\'ees comme une d\'etection de l'oc\'ean global. La premi\`ere est une r\'esonance de Schumann dans l'atmosph\`ere de 
  Titan \citep{sghlsbbbbcfffhjjmrstt2007,bsh2010,brhkswbgs2012}, qui sugg\`ere la pr\'esence d'une couche liquide conductrice, contenant de l'ammoniaque, \`a une profondeur
  entre 55 et 80 km. Il s'agit d'un ph\'enom\`ene non lin\'eaire qui n'est peut-\^etre, de nos jours, pas encore bien compris. La deuxi\`eme est la mesure d'un nombre de Love 
  \'elev\'e de Titan ($k_2\approx0.6$) par \citet{ijdslaarrt2012} \`a partir des mesures des variations diurnes du champ de gravit\'e de Titan. Cette mesure est interpr\'et\'ee 
  par \citet{mmhltglz2014} comme la preuve que la cro\^ute de Titan a moins de 100 km d'\'epaisseur. Par contre, une \'etude de la topographie d'ordre 3 par \citet{hnzi2013}
  sugg\`ere une cro\^ute plus \'epaisse, qui pourrait atteindre 200 km. Cette m\^eme \'etude montre qu'une cro\^ute aussi \'epaisse n'est pas incompatible avec la mesure de
  $k_2$ de \citet{ijdslaarrt2012}. Pour r\'esumer, il n'y a pas, pour l'instant, dans la litt\'erature, d'accord sur l'\'epaisseur de la cro\^ute de Titan. Notre \'etude 
  participe donc au d\'ebat.
  
  \par La communaut\'e attend la prise en compte future de plus en plus d'effets physiques, comme l'\'elasticit\'e de la cro\^ute ou les effets non lin\'eaires dans le fluide.
  Si l'\'elasticit\'e peut avoir une influence pr\'epond\'erante dans les librations en longitude \citep{gm2010,rrc2014}, nous pensons qu'elle ne ferait qu'ajouter une oscillation
  rapide \`a un mouvement lent. N\'eanmoins, une \'etude compl\`ete des effets de l'\'elasticit\'e permettrait de trancher cette question. Une piste pour traiter ce probl\`eme
  est de partir de la m\'ethode pr\'esent\'ee dans \citep{jv2014}, qui utilise des convolutions pour tenir compte d'une comportement de la cro\^ute qui d\'epend de la fr\'equence 
  d'excitation. Une prise en compte compl\`ete des effets non lin\'eaires dans le fluide n\'ecessiterait de r\'esoudre l'\'equation de Navier-Stokes. Actuellement, les approches
  les plus satisfaisantes utilisent des codes num\'eriques d'\'el\'ements finis.
  
  \par Pour ma part, je recommande \'egalement de se poser la question de l'existence, le nombre, la position et la stabilit\'e des \'Etats de Cassini d'un corps en
  rotation synchrone contenant un oc\'ean global. Ces \'Etats sont connus pour un corps rigide (Sect.\ref{sec:etatscassini}), mais pas pour un int\'erieur comme celui 
  de cette section. Par exemple, on ne sait pas si d'autres \'equilibres sont possibles. Une \'etude math\'ematique de ces \'equilibres repr\'esenterait donc une r\'eelle
  avanc\'ee. Il est envisageable de le faire \`a partir d'une formulation Hamiltonienne.
  
\part{Mercure en r\'esonance 3:2 \label{part:mercure}}

  \chapter[L'exploration de Mercure]{Introduction : l'exploration de Mercure\label{chap:explorMercure}}
 
  \par La plan\`ete Mercure est connue depuis l'antiquit\'e comme la plan\`ete des voyageurs. Sa proximit\'e au Soleil ne la rend en effet observable que le matin et le soir, 
  les voyageurs pouvaient donc la voir \`a leur d\'epart et \`a leur arriv\'ee. Ainsi, Mercure est essentiellement visible quand elle est basse sur l'horizon, quand le Soleil est 
  occult\'e, mais aussi quand l'atmosph\`ere, plus \'epaisse sur l'horizon, g\^ene le plus son observation. 
  
  \par Sa rotation en fait un cas unique dans le Syst\`eme Solaire car elle est en r\'esonance spin-orbite 3:2, sa p\'eriode de r\'evolution autour du Soleil \'etant de 88 jours
  et celle de rotation 58 jours. Le Chapitre \ref{chap:tides} l\`eve le voile sur l'origine de cette configuration. Cette r\'esonance a \'et\'e d\'ecouverte par des observations 
  radar de \citet{pd1965} faites \`a Arecibo (Porto-Rico). Cette vitesse de rotation a depuis \'et\'e maintes fois confirm\'ee \citep{mgr1965,s1967,c1967,sr1968,g1971,k1975}, et 
  des librations en longitude ont m\^eme \'et\'e d\'etect\'ees, toujours par observation radar depuis la Terre \citep{mpjsh2007,mpshgjygpc2012}.

  \par Ce sont les missions spatiales qui ont le plus permis de faire progresser notre connaissance de Mercure. \`A cette date, elle n'a \'et\'e visit\'ee que 2 fois, par les 
  missions am\'ericaines Mariner 10 (1974-1975) et MESSENGER, en orbite depuis le 18 mars 2011. Une mission europ\'eo-japonaise, BepiColombo, est pr\'evue pour \^etre lanc\'ee 
  en juillet 2016 et pour \^etre mise en orbite autour de Mercure en janvier 2024. La raison pour laquelle il y a si peu de missions est que Mercure est une cible difficile \`a
  atteindre avec une vitesse suffisamment faible pour acqu\'erir des donn\'ees, une mise en orbite \'etant encore plus difficile. La proximit\'e du Soleil acc\'el\`ere naturellement
  la sonde, il faut donc utiliser plusieurs assistances gravitationnelles afin de suffisamment la d\'ec\'el\'erer. De plus, les instruments doivent \^etre dimensionn\'es pour 
  r\'esister \`a de hautes temp\'eratures.
  
  \section{Mariner 10 (NASA)}
  
  \par La mission Mariner 10 est le premier cas d'utilisation de l'assistance gravitationnelle. Apr\`es son lancement le 3 novembre 1973 par une fus\'ee Atlas-Centaur depuis Cap
  Canaveral, elle a en effet survol\'e V\'enus le 5 f\'evrier 1974 \`a 5794 km d'altitude avant de faire un premier survol de Mercure le 29 mars 1974. Durant la conception de la 
  mission, il \'etait initialement pr\'evu de faire un seul survol de Mercure. Mais Giuseppe Colombo, alors en visite au JPL, a fait remarquer que la p\'eriode orbitale de la sonde 
  autour du Soleil serait alors le double de celle de Mercure, permettant d'autres survols. Il y a eu au total 3 survols \citep{bgsslkf2007}.
  
  \par Le premier survol de Mercure a eu lieu le 29 mars 1974 \`a 705 km d'altitude. Il a permis les premi\`eres images de crat\`eres, mais surtout la d\'ecouverte du champ 
  magn\'etique interne \citep{nblws1974,nblw1975,nblw1976}. Le deuxi\`eme survol a eu lieu le 21 septembre 1974, \`a une altitude bien plus \'elev\'ee. Il a permis la 
  cartographie de $45\%$ de la surface de la plan\`ete. Le troisi\`eme survol a eu lieu le 16 mars 1975 \`a 327 km d'altitude, et a confirm\'e les observations du champ 
  magn\'etique. Ces survols ont \'egalement permis une premi\`ere d\'etection des param\`etres du champ de gravit\'e $J_2$ et $C_{22}$ 
  (Tab.\ref{tab:gravitemercure} \& \citet{acelt1987}). Un bilan de Mariner 10 est propos\'e dans \citep{s1979}. L'essentiel de la connaissance de Mercure due \`a Mariner 10 est 
  resté d'actualit\'e pendant environ 35 ans.
  
  \section{MESSENGER (NASA)}
  
  \par La mission MESSENGER (MErcury Surface, Space ENvironment, GEochemistry, and Ranging) a \'et\'e s\'electionn\'ee par la NASA en 1999. Ses principaux objectifs
  sont de d\'eterminer la composition de Mercure, d'accumuler des informations sur son histoire g\'eologique, de caract\'eriser son champ magn\'etique et d'analyser
  son exosph\`ere \citep{smgabbccghkmmpprssstz2001,s2003,msglm2006}. En cons\'equence, on attendait notamment de MESSENGER une cartographie compl\`ete de Mercure,
  un inventaire de ses crat\`eres, ainsi qu'une bonne connaissance de son champ de gravit\'e. Elle a \'et\'e lanc\'ee le 3 ao\^ut 2004 par une fus\'ee Boeing Delta II
  depuis Cap Canaveral. Avant d'atteindre Mercure, elle a effectu\'e un survol de la Terre le 2 ao\^ut 2005 ainsi que 2 survols de V\'enus les 24 octobre 2006 et 5
  juin 2007. Elle a ensuite effectu\'e 3 survols de Mercure, les 14 janvier et 6 octobre 2008, et le 29 septembre 2009, avant d'\^etre finalement mise sur orbite le 
  18 mars 2011, \`a l'origine pour un an. MESSENGER est \`a cette date toujours en orbite autour de Mercure, et la fin de la mission est pr\'evue pour mars 2015.
  
  \par Une cartographie compl\`ete de Mercure a \'et\'e r\'ealis\'ee, qui a permis de d\'enombrer 46 crat\`eres d'impact, certains ou probables, de diam\`etre sup\'erieur 
  \`a 300 km \citep{fhbzsnskscppop2012}. Le plus gros d'entre eux, Caloris, a un diam\`etre de 1550 km. La densit\'e de ces crat\`eres sur la surface est inf\'erieure \`a
  celle de la Lune. De plus, on constate une claire assym\'etrie est-ouest. \citet{wcllr2012} expliquent cette assym\'etrie par une ancienne rotation synchrone de Mercure,
  alors que \citet{fhbzsnskscppop2012} semblent privil\'egier un renouvellement diff\'erentiel de la surface. Ceci semble confirm\'e par l'assym\'etrie des plaines 
  \citep{drsmbdmehwc2009}, susceptibles d'avoir enfoui les crat\`eres.
  
  \par Le champ magn\'etique de Mercure a \'et\'e confirm\'e, et mesur\'e comme \'etant 100 fois plus faible que celui de la Terre \citep{ajkpwssmrz2011,ajkwbpsszm2012}.
  De plus, il pr\'esente une anomalie nord-sud significative qui n'est pas pr\'esente sur Terre. \citet{cawdsr2014} expliquent cette anomalie par un champ magn\'etique
  g\'en\'er\'e par une dynamo fortement convective.
  
  \par Le champ de gravit\'e est maintenant connu avec une tr\`es bonne pr\'ecision jusqu'\`a l'ordre 2 \citep{szpshlmnpmjtprght2012}, et un signal sup\'erieur au bruit 
  a m\^eme \'et\'e d\'etect\'e jusqu'\`a l'harmonique 6 \citep{gim2013}. La Tab.\ref{tab:gravitemercure} r\'esume l'\'evolution de notre connaissance du champ de 
  gravit\'e de Mercure. On peut y voir une valeur surprenante du $J_2$ obtenue \`a partir des 2 premiers survols de MESSENGER. On peut aussi constater que Mercure n'est pas
  \`a l'\'equilibre hydrostatique. On a en effet, pour un corps en r\'esonance 3:2,  $C_{22}/J_2=7e/10+\mathcal{O}(e^2)\approx0.14$ \citep{mn2009}, et les derni\`eres mesures 
  donnent $C_{22}/J_2\approx0.16$.
  
  \begin{landscape}
  \begin{table}[ht]
   \centering
   \caption[Le champ de gravit\'e de Mercure]{Le champ de gravit\'e de Mercure. Les 3 derni\`eres colonnes sont issues de MESSENGER, et les 2 derni\`eres datent d'apr\`es la 
   mise en orbite. Le champ de gravit\'e publi\'e par \citet{szpshlmnpmjtprght2012} est connu sous le nom HgM002, il a \'et\'e obtenu \`a partir des d\'eviations de la sonde
   lors des 2 premiers survols ainsi que de son orbite autour de Mercure du 11 mars au 23 ao\^ut 2011. La solution de \citet{gim2013} r\'esulte d'une analyse ind\'ependante des 
   donn\'ees de navigation couvrant les 6 premiers mois de la mission.\label{tab:gravitemercure}}
  \begin{tabular}{l|rrrr}
                & Mariner 10 & 2 survols & MESSENGER & MESSENGER \\
                & \citep{acelt1987} & \citep{szpsnlpmtthhjpbmo2010} & \citep{szpshlmnpmjtprght2012} & \citep{gim2013} \\
  \hline
  $C_{20}=-J_2$ & $(-6.0\pm2.0)\times 10^{-5}$ & $(-1.92\pm0.67)\times 10^{-5}$ & $(-5.031\pm0.02)\times 10^{-5}$ & $(-5.048\pm0.02)\times 10^{-5}$\\
  $C_{21}$      & -- & -- & $(-5.99\pm6.5)\times 10^{-8}$ & $(-8.6\pm9.9)\times 10^{-8}$ \\
  $S_{21}$      & -- & -- & $(1.74\pm6.5)\times 10^{-8}$ & $(-1.37\pm1.24)\times 10^{-7}$ \\
  $C_{22}$      & $(1.0\pm0.5)\times 10^{-5}$ & $(8.1\pm0.8)\times 10^{-6}$ & $(8.088\pm0.065)\times 10^{-6}$ & $(8.081\pm0.067)\times 10^{-6}$ \\
  $S_{22}$      & -- & $(-0.3\pm1.2)\times 10^{-6}$ & $(3.22\pm6.5)\times 10^{-8}$ & $(0.006\pm6.6)\times 10^{-8}$ \\
  $C_{30}=-J_3$ & -- & -- & $(-1.188\pm0.08)\times 10^{-5}$ & $(-1.273\pm0.04)\times 10^{-5}$ \\
  $C_{40}=-J_4$ & -- & -- & $(-1.95\pm0.24)\times 10^{-5}$ & $(-1.698\pm0.10)\times 10^{-5}$ \\
  $C_{50}=-J_5$ & -- & -- & -- & $(0.19\pm2.40)\times 10^{-6}$ \\
  $C_{60}=-J_6$ & -- & -- & -- & $(7.04\pm4.79)\times 10^{-6}$ \\
  $C_{70}=-J_7$ & -- & -- & -- & $(-3.19\pm7.92)\times 10^{-6}$ \\
  \hline
  \end{tabular}
  \end{table}
  \end{landscape}

  \section{BepiColombo (ESA / JAXA)}
  
  \par La mission BepiColombo devrait \^etre lanc\'ee en ao\^ut 2016 par une Ariane 5 depuis Kourou, et \^etre mise en  orbite autour de Mercure en janvier 2024. La mission 
  nominale doit durer un an, mais on peut s'attendre \`a des extensions comme \c{c}a a \'et\'e le cas pour MESSENGER. L\`a encore, l'assistance gravitationnelle sera utilis\'ee,
  d'abord de la Lune pour quitter le champ gravitationnel de la Terre, puis de la Terre, de V\'enus, et enfin de Mercure \`a plusieurs reprises. Parmi les principaux objectifs
  de la mission, on trouve une analyse de la composition de l'exosph\`ere de Mercure, des mesures de son champ magn\'etique ainsi que l'\'etude de son interaction avec le vent solaire, une 
  cartographie plus pr\'ecise qu'avec MESSENGER, la recherche d'une activit\'e tectonique \'eventuelle, ainsi qu'un test de la relativit\'e g\'en\'erale \citep{gb2001}. 
  
  \par Pour cela, 2 sondes seront utilis\'ees. Le MPO (Mercury Planetary Orbiter), r\'ealis\'e par l'ESA, orbitera autour de Mercure en 2.3 heures sur une orbite polaire, 
  avec un p\'ericentre \`a 400 km et un apocentre \`a 1500 km. Il r\'ealisera notamment la cartographie avec l'instrument BELA (BepiColombo Laser Altimeter), embarquera plusieurs 
  spectrom\`etres \`a diff\'erentes longueurs d'onde, ainsi que l'exp\'erience de radio-science MORE (Mercury Orbiter Radio-science Experiment). Le MMO (Mercury Magnetospheric
  Orbiter), r\'ealis\'e par la JAXA, aura une orbite bien plus elliptique (p\'ericentre \`a 400 km et apocentre \`a 11800 km) qu'il parcourera en 9.2 heures. Il aura pour principal 
  objectif l'\'etude de l'atmosph\`ere et de la magn\'etosph\`ere de Mercure \citep{bvhflnfmz2010}.
  
  \par J'ai pour ma part \'et\'e associ\'e, \`a Namur, \`a l'exp\'erience de radioscience MORE \citep{ib2001}, en collaboration avec l'\'equipe d'Andrea Milani \`a Pise. 
  C\^ot\'e Namur, j'ai collabor\'e avec Anne Lema\^itre, Sandrine D'Hoedt, Julien Dufey, Julien Frouard, Christoph Lhotka et S\'ebastien Wailliez, dans le cadre
  du projet ROMEO (Rotation Of Mercury and Equations of an Orbiter), financ\'e par la politique scientifique belge BELSPO, qui elle-m\^eme le finan\c{c}ait avec des
  fonds de l'ESA. Notre travail a consist\'e 
  \`a mod\'eliser la rotation de Mercure en fonction de son int\'erieur (Chap.\ref{chap:mercobliq} \& \ref{chap:merclibra}). Nous avons fourni \`a l'\'equipe de Pise une 
  routine mod\'elisant cette rotation \`a tout instant de la mission. Cette routine est destin\'ee \`a faire partie d'un logiciel, fait par l'\'equipe de Pise, de simulation 
  de l'exp\'erience de radio-science, non seulement l'inversion de la rotation de Mercure, mais aussi les variations diurnes du champ de gravit\'e de Mercure \citep{mrvvb2001} 
  et le test de relativit\'e g\'en\'erale. 
  
  \par Le test de relativit\'e g\'en\'erale va consister \`a mesurer les param\`etres post-Newtoniens $\beta$ et $\gamma$. $\beta$ est une signature des non-lin\'earit\'es 
  dans le champ de gravit\'e, sa mesure est d\'eg\'en\'er\'ee avec celle de l'aplatissement solaire $J_2$, et $\gamma$ traduit la courbure de l'espace-temps. La th\'eorie de 
  la relativit\'e g\'en\'erale pr\'evoit que ces 2 param\`etres valent strictement 1. $\gamma=1$ a \'et\'e v\'erifi\'ee \`a la pr\'ecision $10^{-5}$ par la sonde Cassini 
  \citep{bit2003}, \`a partir de la deflexion de la lumi\`ere par le Soleil. $\beta=1$ a \'et\'e v\'erifi\'e \`a la pr\'ecision $10^{-4}$ \`a partir de mesures Lunar Laser 
  Ranging \citep{wtb2009}. Le test de BepiColombo devrait avoir une pr\'ecision de l'ordre de $10^{-5}-10^{-6}$ \citep{mvvbr2002,abw2007} pour $\beta$ et $\gamma$.
  
  \par L'exp\'erience de rotation va consister, comme pour Titan et Mimas, \`a rep\'erer des signes distinctifs sur la surface de Mercure et sur
  plusieurs images, afin d'ajuster un axe et une vitesse de rotation, pour d\'eterminer une amplitude de libration diurne ainsi qu'une obliquit\'e. Les contraintes 
  d'ensoleillement de Mercure ainsi que la r\'esonance spin-orbite rendent difficiles l'accumulation de telles donn\'ees, \citet{pvd2011} estiment que sur la dur\'ee 
  nominale de la mission d'un an, 25 de ces points devraient \^etre utilisables, pour donner une pr\'ecision de 1.4 seconde d'arc sur la libration diurne et 1 seconde d'arc 
  sur l'obliquit\'e. Bien s\^ur, une prolongation de la mission donnerait une meilleure pr\'ecision. \citet{cm2012} sugg\`erent de compl\'eter cette exp\'erience en utilisant 
  les variations diurnes du champ de gravit\'e, qui contiennent de l'information sur la rotation.

  \par La proximit\'e du Soleil rend les missions vers Mercure rares car ch\`eres. BepiColombo devait d'ailleurs initialement inclure un 
  atterrisseur\footnote{amercurisseur? ahermisseur? lander?} qui a \'et\'e supprim\'e pour raisons budg\'etaires. Il est malheureusement \`a craindre qu'\`a la suite de
  BepiColombo, Mercure ne soit plus explor\'ee pendant plusieurs d\'ecennies.

  \chapter[L'obliquit\'e de Mercure]{\'Etude \`a long terme : l'obliquit\'e\label{chap:mercobliq}}
  
  \section{Introduction}
  
  \par Depuis la d\'ecouverte de la r\'esonance spin-orbite de Mercure, cette plan\`ete est consid\'er\'ee comme \'etant \`a l'\'Etat de Cassini 1 \citep{c1966,p1969,b1972},
  ce qui a n\'ecessit\'e d'\'etendre la d\'efinition des \'Etats de Cassini aux corps non synchrones. Ceci a \'et\'e v\'erifi\'e observationnellement en 2007 \citep{mpjsh2007},
  une nouvelle fois \`a l'aide d'observations radar. L'avantage d'une rotation amortie, c'est-\`a-dire correspondant \`a un \'equilibre dynamique, est qu'elle peut \^etre
  calcul\'ee th\'eoriquement. Elle ne d\'epend pas, en effet, de conditions initiales, et donc ne n\'ecessite pas de mesures. \citet{p1969} a propos\'e la formule suivante
  pour l'obliquit\'e $\epsilon$ \`a l'\'Etat de Cassini 1 :
  
  \begin{equation}
    \label{eq:oblikpeale}
    \epsilon = -\frac{C\dot{\ascnode}\sin\iota}{C\dot{\ascnode}\cos\iota+2nM_{\mercury}R_{\mercury}^2\left(\frac{7}{2}e-\frac{123}{16}e^3\right)C_{22}-n M_{\mercury} R_{\mercury}^2 (1-e^2)^{-3/2} C_{20}},
  \end{equation}
  o\`u $C$ est le moment d'inertie polaire, $n \approx 26.0878\textrm{ rad}/\textrm{an}$ est le moyen mouvement moyen de Mercure autour du Soleil, $e \approx 0.2056$ son 
  excentricit\'e, $C_{20}=-J_2$ et $C_{22}$ les harmoniques d'ordre 2 de son champ de gravit\'e (cf. Tab.\ref{tab:gravitemercure}), $M_{\mercury}$ sa masse, et 
  $R_{\mercury} = 2439.7$ km \citep{aabcccfhhknossttw2011} son rayon moyen. $\iota$ et $\dot{\ascnode}$ sont respectivement l'inclinaison orbitale et la vitesse de pr\'ecession
  du n{\oe}ud ascendant par rapport au Plan de Laplace (Sect.\ref{sec:laplcall} \& \ref{sec:laplacemercure}), dont le choix doit minimiser les variations de l'inclinaison.
  Son choix d\'epend donc de la fa\c{c}on dont ces variations sont minimis\'ees, et de ce choix d\'ependront les valeurs de $\iota$ et $\dot{\ascnode}$. En pratique, les param\`etres
  orbitaux et de forme sont bien connus, et les coefficients du champ de gravit\'e $M_{\mercury}$, $C_{20}$ et $C_{22}$ peuvent \^etre calcul\'es \`a partir des d\'eviations de la 
  sonde lors des survols ou de la course orbitale. Ainsi, mesurer l'obliquit\'e $\epsilon$ revient \`a mesurer le moment d'inertie polaire $C$.
  
  \par Une d\'etermination num\'erique de l'obliquit\'e n\'ecessite, comme pour la rotation des corps synchrones, de d\'eterminer des conditions initiales les plus proches 
  possibles de l'\'Etat de Cassini. Pour cela, il faut \'eliminer les oscillations libres. Le probl\`eme est que les p\'eriodes impliqu\'ees sont tr\`es longues par rapport
  \`a la dur\'ee de validit\'e des \'eph\'em\'erides. En utilisant les valeurs du champ de gravit\'e de Mariner 10 \citep{acelt1987} et $C=0.34M_{\mercury}R_{\mercury}^2$
  comme sugg\'er\'e par \citet{mrvvb2001}, on trouve une p\'eriode des oscillations libres de l'ordre de 1066 ans \citep{dl2004,rb2004,br2007}. La p\'eriode de r\'egression
  du n{\oe}ud ascendant est, elle, de l'ordre de 250\,000 ans, sous l'effet des perturbations plan\'etaires, de l'aplatissement du Soleil et de la correction relativiste \citep{e1915}.
  Par contre, les \'eph\'em\'erides plan\'etaires ne sont valides que sur quelques milliers d'ann\'ees. En accord avec nos partenaires de Pise, nous utilisons la th\'eorie du JPL
  DE406 \citep{s1998}, valide sur 6000 ans. Un tel intervalle de validit\'e rend impossible l'extraction par analyse en fr\'equence d'un terme de 250\,000 ans, et donc sa s\'eparation
  d'un terme de 1066 ans. Une autre piste pour approcher l'\'Etat de Cassini serait d'introduire un amortissement num\'erique, artificiel, qui convergerait vers la solution 
  amortie. Un tel amortissement doit \^etre suffisamment lent (adiabatique) pour ne pas affecter la position d'\'equilibre, ce qui veut dire que le passage du r\'egime non amorti 
  au r\'egime amorti doit se faire sur un nombre relativement important de p\'eriodes de ces oscillations libres\ldots mais les \'eph\'em\'erides sont valides sur \`a peine 6 p\'eriodes.
  
  \par Il faut bien comprendre que le mouvement du n{\oe}ud n'est pas uniforme, ce qui complique significativement la mod\'elisation de l'obliquit\'e. La formule de Peale 
  (\'Eq.\ref{eq:oblikpeale}) suppose que $\dot{\ascnode}$ est constant, qu'il faut interpr\'eter par \emph{ne varie pas trop sur un intervalle donn\'e}. Un choix judicieux
  du plan de r\'ef\'erence, que nous appellerons Plan de Laplace, permet de renforcer ce caract\`ere constant. On comprend alors qu'il d\'epend de la th\'eorie orbitale
  ainsi que de l'intervalle d'\'etude. C'est pourquoi nous proposons une solution alternative \`a ce probl\`eme (Sec.\ref{sec:oblikmercure}).
  
  \section[Le Plan de Laplace]{Diff\'erentes fa\c{c}ons de d\'eterminer un Plan de Laplace\label{sec:laplacemercure}}
  
  \par Lorsque j'ai commenc\'e \`a travailler sur Mercure en octobre 2007, la d\'etermination du Plan de Laplace m'a \'et\'e pr\'esent\'ee comme un enjeu important dans la
  pr\'eparation de la mission BepiColombo. L'une des difficult\'es vient du fait qu'il y a plusieurs m\'ethodes, non \'equivalentes, de le d\'eterminer. Je pr\'esente ici 3 
  d'entre elles.
  
  \subsection[Moyennisation s\'eculaire]{Moyennisation s\'eculaire \citep{ym2006}}
  
  \par \citet{ym2006} proposent de se baser sur une th\'eorie orbitale s\'eculaire\footnote{c'est-\`a-dire moyenn\'ee sur les longitudes moyennes} de Mercure. Le Plan de Laplace peut 
  \^etre vu comme le plan sur lequel l'orbite de Mercure  pr\'ecesse. Pour l'obtenir, ils partent de l'expression de la perturbation s\'eculaire des autres plan\`etes sur l'orbite de Mercure :
  
  \begin{eqnarray}
    R_{sec} & = & \sum_{k=2}^8\frac{\mathcal{G}M_ka_1}{8a_k^2}\left(b_{3/2}^{(1)}\left(\frac{a_1}{a_k}\right)e_1^2-\left(p_1^2+q_1^2\right)b_{3/2}^{(1)}\left(\frac{a_1}{a_k}\right)-2e_1e_k\cos(\varpi_1-\varpi_k)b_{3/2}^{(2)}\left(\frac{a_1}{a_k}\right)\right. \nonumber \\
            & + & \left.2\left(p_kp_1+q_kq_0\right)b_{3/2}^{(1)}\left(\frac{a_1}{a_k}\right)\right) \label{eq:Rsec}
  \end{eqnarray}
o\`u les plan\`etes de Mercure \`a Neptune sont num\'erot\'ees de $1$ \`a $8$\footnote{Dans \citep{ym2006}, elles sont num\'erot\'ees de $0$ \`a $7$.}, $p_k = \sin i_k\sin\ascnode_k$,
$q_k = \sin i_k\cos\ascnode_k$, et les $b_{3/2}^{(i)}(x)$ sont les coefficients de Laplace (\'Eq.\ref{eq:coeflaplace}). Ils en d\'eduisent, pour chaque plan\`ete perturbatrice,
une fr\'equence de pr\'ecession $w_k$ d\'ej\`a d\'efinie par \citet{bhcd1979} :

\begin{equation}
  \label{eq:precburns}
  w_k = \frac{M_ka_1^2nb_{3/2}^{(1)}\left(\frac{a_1}{a_k}\right)}{4M_{\sun}a_k^2}
\end{equation}
o\`u $n$ est le moyen mouvement de Mercure et $M_{\sun}$ la masse du Soleil. La somme des $w_k$ donne une p\'eriode de r\'egression du n{\oe}ud ascendant d'environ 235\,000 ans, ici 
sans tenir compte de la relativit\'e g\'en\'erale. On obtient

\begin{eqnarray}
  \frac{dq_1}{dt} & = & \sum_{k=2}^8\left(p_k-p_1\right)w_k, \label{eq:dq1} \\
  \frac{dp_1}{dt} & = & \sum_{k=2}^8\left(q_k-q_1\right)w_k, \label{eq:dp1}
\end{eqnarray}
en consid\'erant que l'inclinaison de chaque plan\`ete est constante. Si le plan de r\'ef\'erence est le Plan de Laplace, alors on a $dq_1/dt = dp_1/dt = 0$, ce qui implique

\begin{eqnarray}
 p_1 & = & \frac{\sum_{k=2}^8p_kw_k}{\sum_{k=2}^8w_k}, \label{eq:pun} \\
 q_1 & = & \frac{\sum_{k=2}^8q_kw_k}{\sum_{k=2}^8w_k}. \label{eq:qun}
\end{eqnarray}
On a ainsi la normale au Plan de Laplace dont la longitude et la latitude ecliptiques sont respectivement $\lambda = -8.8^{\circ}$ et $\beta = 87.9^{\circ}$.

  \subsection[Moyennisation num\'erique]{Moyennisation num\'erique \citep{ym2006}}

 \par Les m\^emes auteurs ont compar\'e cette d\'etermination de l'orientation du P\^ole de Laplace avec celle obtenue num\'eriquement, \`a partir des \'eph\'em\'erides orbitales 
 DE408. Pour cela, ils ont consid\'er\'e un Plan de Laplace instantan\'e, bas\'e sur le mouvement instantan\'e de pr\'ecession de l'orbite, ce mouvement \'etant lui-m\^eme obtenu
 par moyennisation sur un certain intervalle de temps autour de J2000. Ils ont essay\'e sur des intervalles de plus ou moins 1000, 2000, 4000 et 8000 ans autour de J2000 et 
 confirment que l'approximation d'une pr\'ecession constante est de moins en moins pr\'ecise \`a mesure que l'intervalle cro\^it. Sur un intervalle de 2000 ans centr\'e sur J2000,
 ils obtiennent $\lambda = 66.6^{\circ}$ et $\beta = 86.725^{\circ}$.
  
  \subsection[Plan de Laplace instantan\'e]{Plan de Laplace instantan\'e \`a t=J2000 \citep{dndl2009}}
  
  \par \citet{dndl2009} proposent une d\'etermination analytique du Plan de Laplace instantan\'e \`a partir de d\'eveloppements paraboliques de l'inclinaison et du n{\oe}ud ascendant 
  par rapport \`a l'\'ecliptique :
  
  \begin{eqnarray}
   i(t)        & \approx & i_0+i_1t+i_2t^2, \label{eq:expandi} \\
   \ascnode(t) & \approx & \ascnode_0+\ascnode_1t+\ascnode_2t^2, \label{eq:expandasc}
  \end{eqnarray}
les coefficients \'etant obtenus num\'eriquement \`a partir d'une th\'eorie orbitale (DE406 dans notre cas). Apr\`es de nombreux calculs d\'etaill\'es dans l'article, les auteurs
obtiennent $\lambda = 48.1932^{\circ}$\footnote{Cette valeur est erron\'ee dans \citep{dndl2009} mais corrig\'ee dans \citep{nd2012}.} et $\beta = 87.2883^{\circ}$.

\par Dans les 3 cas, on peut remarquer que les valeurs de la latitude $\beta$ sont assez proches. Il semble par contre que la longitude $\lambda$ soit peu contrainte, ce qui avait
d\'ej\`a \'et\'e remarqu\'e par ces \'etudes.

  \section{Notre mod\'elisation de l'obliquit\'e\label{sec:oblikmercure}}
  
  \par L'\'etude que je pr\'esente maintenant, publi\'ee dans \citep{nd2012,nl2013}, a \'et\'e initialement motiv\'ee par notre obligation contractuelle
  \footnote{Projet ROMEO, pour Rotation Of Mercury and Equations of an Orbiter. Contrat Prodex C90253 de BELSPO, pour BELgian Scientific POlicy.}, \`a Namur, de fournir \`a
  l'\'equipe d'A.~Milani une routine donnant, pour chaque date concern\'ee par BepiColombo, la matrice de passage de l'\'ecliptique \`a J2000 au rep\`ere des axes principaux d'inertie 
  de Mercure. L'imp\'eratif de pr\'ecision exigeait l'utilisation d'\'eph\'em\'erides orbitales, dans notre cas DE406, ainsi que d'une r\'esolution num\'erique du probl\`eme. Les 
  oscillations en obliquit\'e ne pouvant \^etre amorties suffisamment lentement, nous devions conna\^itre avec une grande pr\'ecision les variations de l'orientation du p\^ole de 
  Mercure, un mod\`ele uniform\'emement pr\'ecessant \'etant insuffisant.
  
  \par Cette approximation n'\'etant plus utilis\'ee, le Plan de Laplace perd de son int\'er\^et\footnote{En fait, il peut \^etre utile pour v\'erifier la Troisi\`eme Loi de Cassini
  qui stipule que l'axe du moment cin\'etique, la normale \`a l'orbite et la normale au Plan de Laplace doivent \^etre coplanaires \citep{c1693,c1966}.}. C'est pourquoi nous utiliserons
  l'\'ecliptique \`a J2000 comme rep\`ere de r\'ef\'erence dans cette \'etude. Je pr\'esente d'abord la m\'ethode, num\'erique, \'elabor\'ee alors que le champ de gravit\'e de Mercure 
  connu \'etait celui donn\'e par Mariner 10, i.e. avec une incertitude de $33\%$ sur $J_2$ et $50\%$ sur $C_{22}$, puis je donne une solution plus pr\'ecise utilisant les valeurs 
  de \citep{szpshlmnpmjtprght2012} issues des premiers mois de la mission MESSENGER. Je terminerai par une estimation d'autres effets sur l'obliquit\'e.
  
  \subsection[La m\'ethode]{La m\'ethode \citep{nd2012}}
  
  \par Afin de d\'eterminer l'\'equilibre dynamique d'un syst\`eme perturb\'e, comme la rotation d'un corps en r\'esonance spin-orbite, j'obtiens d'excellents r\'esultats en
  utilisant un algorithme bas\'e sur l'analyse en fr\'equence (Chap.\ref{chap:naffo}). Ceci n\'ecessite de mod\'eliser le mouvement sur un intervalle de temps au minimum \'egal 
  \`a 2 fois la plus longue p\'eriode intervenant, sachant que le mouvement du p\^ole de Mercure est adiabatique \citep{bc2005,p2006,rld2007,dl2008}. Plus pr\'ecis\'ement, les \'echelles
  de temps associ\'ees se chiffrent en centaines de milliers d'ann\'ees. Malheureusement, les \'eph\'em\'erides ne sont disponibles que sur plusieurs milliers d'ann\'ees.
  
  \par L'id\'ee est d'extrapoler les \'eph\'em\'erides, pour obtenir une mod\'elisation du mouvement sur plusieurs millions d'ann\'ees. Nous obtiendrons ainsi des r\'esultats sur 
  plusieurs millions d'ann\'ees, en sachant qu'au maximum 30 ans nous int\'eressent, afin de couvrir les missions MESSENGER et BepiColombo. De ce fait, nous ne cherchons pas \`a 
  savoir si notre extrapolation est r\'ealiste, nous souhaitons simplement qu'elle soit au mieux manipulable et que le r\'esultat obtenu soit tangent \`a une solution r\'ealiste
  sur quelques ann\'ees.
  
  \par Nous savons que les termes \`a courte p\'eriode ont une influence n\'egligeable sur les variables d'obliquit\'e \citep{rld2007,dnrl2009}. Par cons\'equent nous pouvons 
  utiliser un syst\`eme d'\'equations moyenn\'ees sur le mouvement en longitude, apr\`es expression de l'argument de la r\'esonance 3:2, $2\sigma=2p-3\lambda+\varpi$, o\`u 
  l'angle de rotation $p=l+g+h$ a \'et\'e d\'efini pr\'ec\'edemment (\'Eq.\ref{eq:andoyermodif}), $\lambda$ est la longitude moyenne de Mercure, et $\varpi$ la longitude du 
  p\'ericentre. Pour cela nous partons du Hamiltonien moyenn\'e $<\mathcal{H}>$, d\'evelopp\'e au degr\'e 2 en excentricit\'e $e$ / inclinaison $I$ :
  
  \begin{equation}
\label{eq:aveHamil}
 <\mathcal{H}>=-\frac{\mathcal{G}M_{\sun}M_{\mercury}}{2a}+\frac{\Lambda_1^2}{2C}+<V_G>,      
\end{equation}
  avec
  
  \begin{equation}
  \label{eq:theVg}
<V_G>=-\frac{\mathcal{G}^4M_{\sun}^4 M_{\mercury}^7}{(\Lambda_0-\frac{3}{2}\Lambda_1)^6}R_{\mercury}^2\left(\frac{1}{2}C_{20}\,\gamma_1+3C_{22} \gamma_2\right),
\end{equation}
\begin{eqnarray}
\gamma_1 & = & \left(1+\frac{3e^2}{2}\right)\left(-\frac{1}{4}\left(-1+3\cos^2I\right)\left(-1+3\cos^2 K\right)\right. \nonumber \\
         & - & 3\cos I\cos K \cos (\varpi-\ascnode+r)\sin I\sin  K \label{eq:gamma1sandrine} \\
         & - & \left.\frac{3}{4}\left(1-\cos^2 I \right)\left(1-\cos^2 K \right) \cos (2\varpi-2\ascnode+2r)\right) \nonumber
\end{eqnarray}
et
 \begin{eqnarray}
  \gamma_2 & = & \frac{7}{32}e\left(1-\cos I\right)^2\left(1-\cos K\right)^2\cos (2\varpi+4r+2\sigma-4\ascnode) \nonumber \\
           & + & \frac{7}{8}e\left(1-\cos I\right)\left(1-\cos K\right)\sin I\sin K \cos(\varpi+3r+2\sigma-3\ascnode) \nonumber \\
           & + & \frac{21}{16}e\sin^2 I\sin^2 K\cos (2r+2\sigma-2\ascnode) \label{eq:gamma2sandrine} \\
           & + & \frac{7}{8}e\left(1+\cos I\right)\left(1+\cos K \right)\sin I\sin K\cos(\varpi-r-2\sigma+\ascnode) \nonumber \\
           & + & \frac{7}{32}e\left(1+\cos I\right)^2\left(1+\cos K\right)^2\cos(2\varpi-2\sigma). \nonumber
\end{eqnarray}

\par Les variables canoniques de notre Hamiltonien sont

\begin{equation}
\label{eq:variabsandrine}
\begin{array}{lll}
\sigma, & \hspace{2cm} & \Lambda_1=\frac{3}{2}nC, \\
r,      & \hspace{2cm} & \Lambda_3=\Lambda_1(1-\cos K),
\end{array}
\end{equation}
o\`u $r$ est l'angle d'Euler d\'efini pr\'ec\'edemment (\'Eq.\ref{eq:andoyermodif}), $K$ est l'obliquit\'e de Mercure par rapport \`a la normale \`a l'\'ecliptique, et
$\Lambda_0=M_{\mercury}\sqrt{\mathcal{G}M_{\sun}}$. $K$ est laiss\'e par commodit\'e d'\'ecriture, mais il doit \^etre remplac\'e par son expression en fonction de $\Lambda_1$
et $\Lambda_3$ pour calculer les \'equations de la dynamique. Les \'equations d\'eriv\'ees de ce Hamiltonien sont

\begin{equation}
\label{eq:equationssandrine}
\begin{array}{lll}
\dot{\sigma}=\frac{\partial<\mathcal{H}>}{\partial \Lambda_1},    & \hspace{2cm} & \dot{\Lambda}_1=-\frac{\partial <\mathcal{H}>}{\partial \sigma}, \\
\dot{\lambda}_3=\frac{\partial<\mathcal{H}>}{\partial \Lambda_3}, & \hspace{2cm} & \dot{\Lambda}_3=-\frac{\partial <\mathcal{H}>}{\partial \lambda_3}.
\end{array}
\end{equation}

\par Nous avons besoin des \'el\'ements orbitaux $e$, $I$, $\varpi$ et $\ascnode$. L'id\'ee maintenant est d'utiliser les \'el\'ements r\'eguliers de Poincar\'e :

\begin{equation}
\label{eq:orbitpoincare}
\begin{array}{lll}
k = e\cos\varpi,                              & \hspace{2cm} & h = e\sin\varpi, \\
q = \sin\left(\frac{I}{2}\right)\cos\ascnode, & \hspace{2cm} & p = \sin\left(\frac{I}{2}\right)\sin\ascnode.
\end{array}
\end{equation}

\par Dans le cadre d'une th\'eorie orbitale quasi-p\'eriodique, on aurait

\begin{eqnarray}
  k(t) & = & \sum \alpha_i\cos \left(\omega_i t+\phi_i\right), \label{eq:kqs} \\
  h(t) & = & \sum \alpha_i\sin \left(\omega_i t+\phi_i\right), \label{eq:hqs} \\
  q(t) & = & \sum  \beta_i\cos \left(\Omega_i t+\Phi_i\right), \label{eq:qqs} \\
  p(t) & = & \sum  \beta_i\sin \left(\Omega_i t+\Phi_i\right), \label{eq:pqs} 
\end{eqnarray}
o\`u les $\alpha_i$, $\beta_i$ sont des amplitudes, $\omega_i$, $\Omega_i$ des fr\'equences, et $\phi_i$, $\Phi_i$ des phases \`a l'origine du temps. Toutes ces quantit\'es
sont constantes. Ici on utilise le fait que ces 4 quantit\'es sont dynamiquement li\'ees 2 par 2, $h$ et $k$ d'un c\^ot\'e, et $q$ et $p$ de l'autre. Les p\'eriodes 
impliqu\'ees sont tr\`es longues par rapport \`a l'intervalle de validit\'e des \'eph\'em\'erides (cf. \citet{l1988} pour une d\'ecomposition quasi-p\'eriodique des mouvements
orbitaux des plan\`etes du Syst\`eme Solaire), les \'eph\'em\'erides permettent plut\^ot d'obtenir des \'evolutions de type parabolique, c'est-\`a-dire

\begin{eqnarray}
  k(t) & \approx a_k+b_kt+c_kt^2, \label{eq:kparabol} \\
  h(t) & \approx a_h+b_ht+c_ht^2, \label{eq:hparabol} \\
  q(t) & \approx a_q+b_qt+c_qt^2, \label{eq:qparabol} \\
  p(t) & \approx a_p+b_pt+c_pt^2. \label{eq:pparabol}  
\end{eqnarray}

\par On peut remarquer que pour une d\'ecomposition quasi-p\'eriodique \`a 2 termes, on a autant d'\'equations que d'inconnues. En effet, en \'ecrivant

\begin{eqnarray}
  k(t) & = & \alpha_1\cos \left(\omega_1 t+\phi_1\right)+\alpha_2\cos \left(\omega_2 t+\phi_2\right), \label{eq:k2qs} \\
  h(t) & = & \alpha_1\sin \left(\omega_1 t+\phi_1\right)+\alpha_2\sin \left(\omega_2 t+\phi_2\right), \label{eq:h2qs} \\
  q(t) & = & \beta_1\cos \left(\Omega_1 t+\Phi_1\right)+\beta_2\cos \left(\Omega_2 t+\Phi_2\right), \label{eq:q2qs} \\
  p(t) & = & \beta_1\sin \left(\Omega_1 t+\Phi_1\right)+\beta_2\sin \left(\Omega_2 t+\Phi_2\right), \label{eq:p2qs} 
\end{eqnarray}
et en d\'eveloppant ces \'equations autour de la m\^eme origine $t=0$, on obtient

\begin{eqnarray}
a_k & = & \alpha_1\cos\phi_1+\alpha_2\cos\phi_2, \label{eq:ak} \\
b_k & = & -\big(\alpha_1\omega_1\sin\phi_1+\alpha_2\omega_2\sin\phi_2\big), \label{eq:bk} \\
c_k & = & -\frac{1}{2}\left(\alpha_1\omega_1^2\cos\phi_1+\alpha_2\omega_2^2\cos\phi_2\right), \label{eq:ck} \\ 
a_h & = & \alpha_1\sin\phi_1+\alpha_2\sin\phi_2, \label{eq:ah} \\
b_h & = & \alpha_1\omega_1\cos\phi_1+\alpha_2\omega_2\cos\phi_2, \label{eq:bh} \\
c_h & = & -\frac{1}{2}\left(\alpha_1\omega_1^2\sin\phi_1+\alpha_2\omega_2^2\sin\phi_2\right), \label{eq:ch}
\end{eqnarray}

et 

\begin{eqnarray}
a_q & = & \beta_1\cos\Phi_1+\beta_2\cos\Phi_2, \label{eq:aq} \\
b_q & = & -\left(\beta_1\Omega_1\sin\Phi_1+\beta_2\Omega_2\sin\Phi_2\right), \label{eq:bq} \\
c_q & = & -\frac{1}{2}\left(\beta_1\Omega_1^2\cos\Phi_1+\beta_2\Omega_2^2\cos\Phi_2\right), \label{eq:cq} \\
a_p & = & \beta_1\sin\Phi_1+\beta_2\sin\Phi_2, \label{eq:ap} \\
b_p & = & \beta_1\Omega_1\cos\Phi_1+\beta_2\Omega_2\cos\Phi_2, \label{eq:bp} \\
c_p & = & -\frac{1}{2}\left(\beta_1\Omega_1^2\sin\Phi_1+\beta_2\Omega_2^2\sin\Phi_2\right). \label{eq:cp}
\end{eqnarray}

\par
\begin{figure}[ht]
\centering
\begin{tabular}{cc}
\includegraphics[width=.47\textwidth]{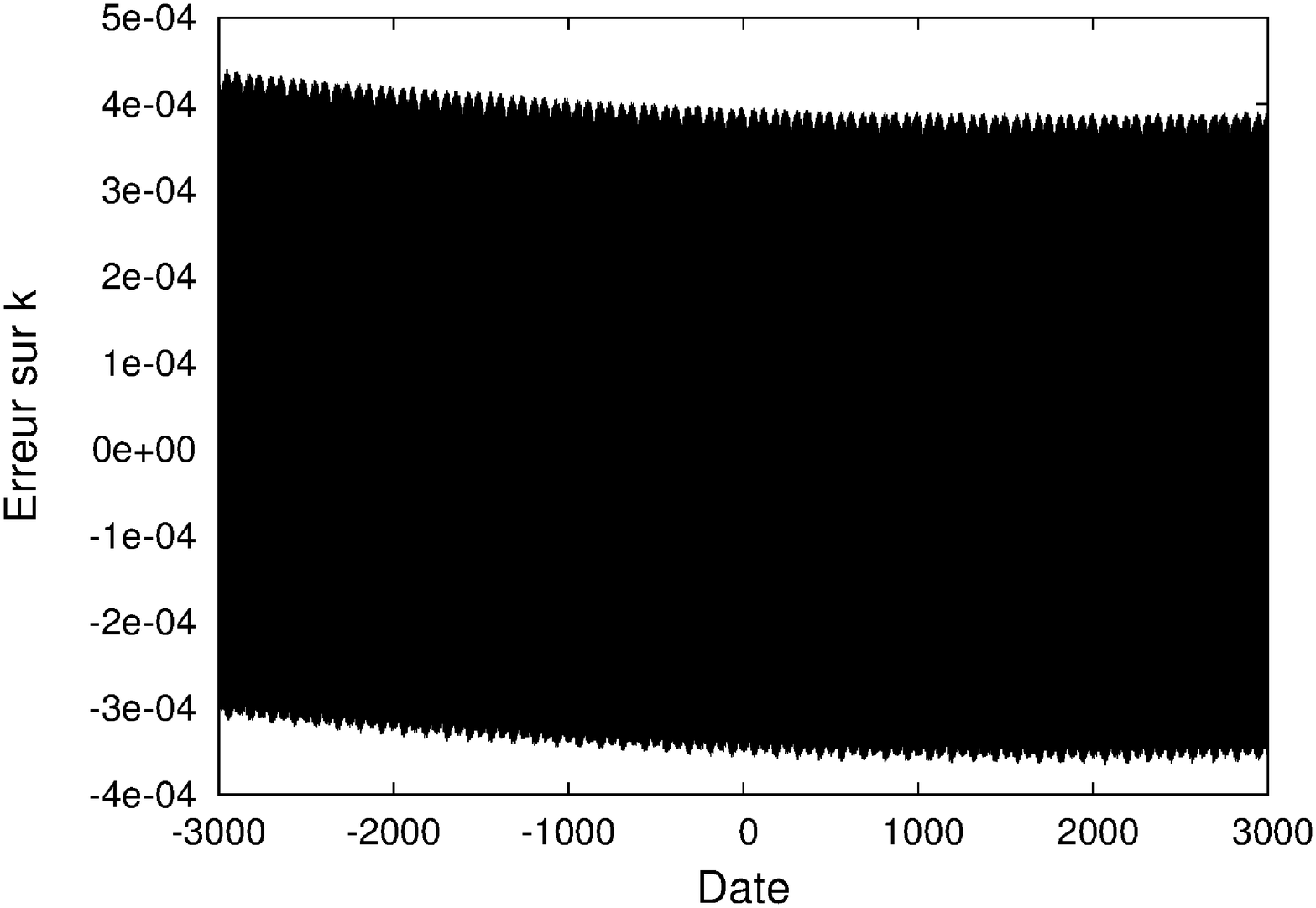} & \includegraphics[width=.47\textwidth]{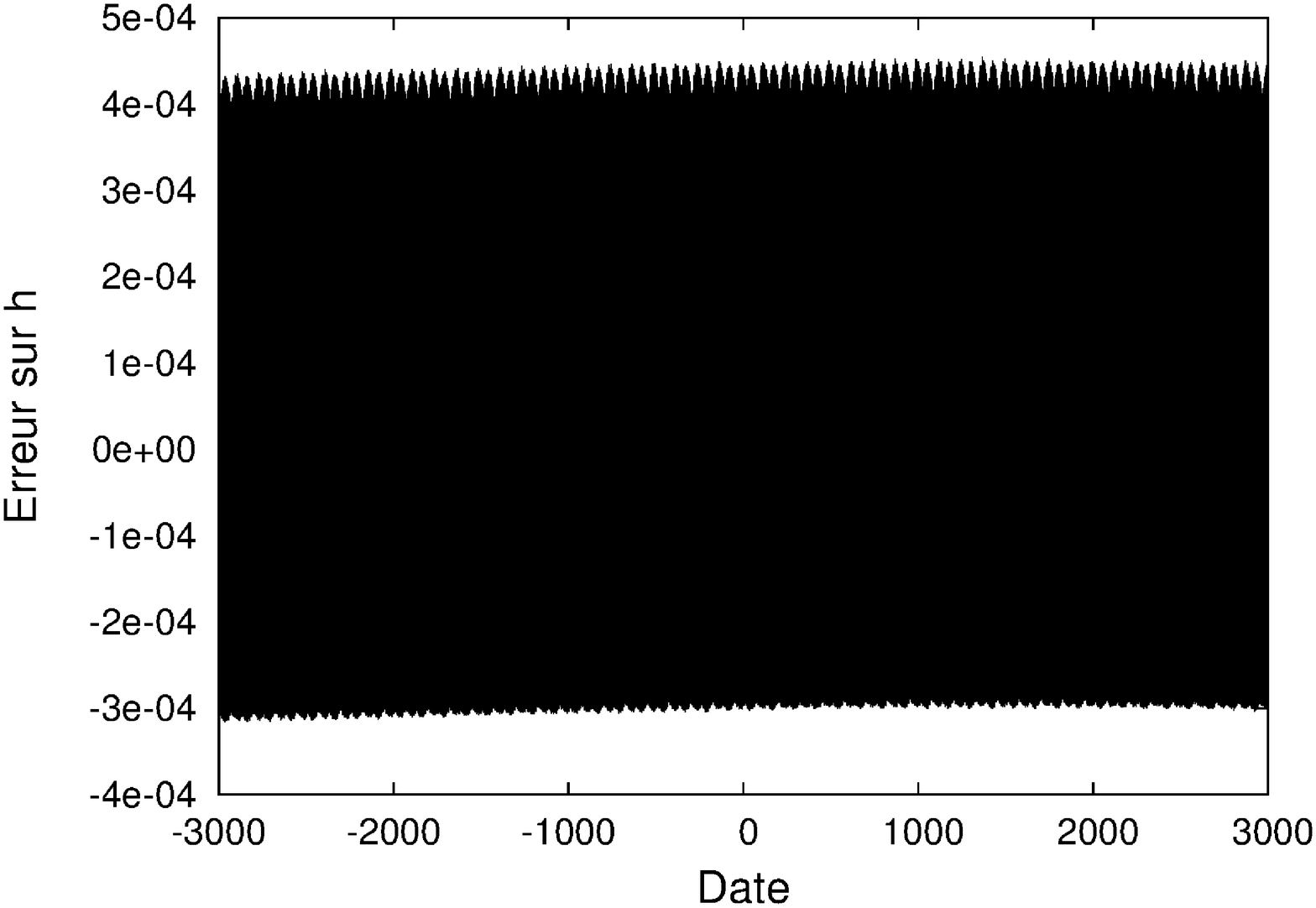} \\
\includegraphics[width=.47\textwidth]{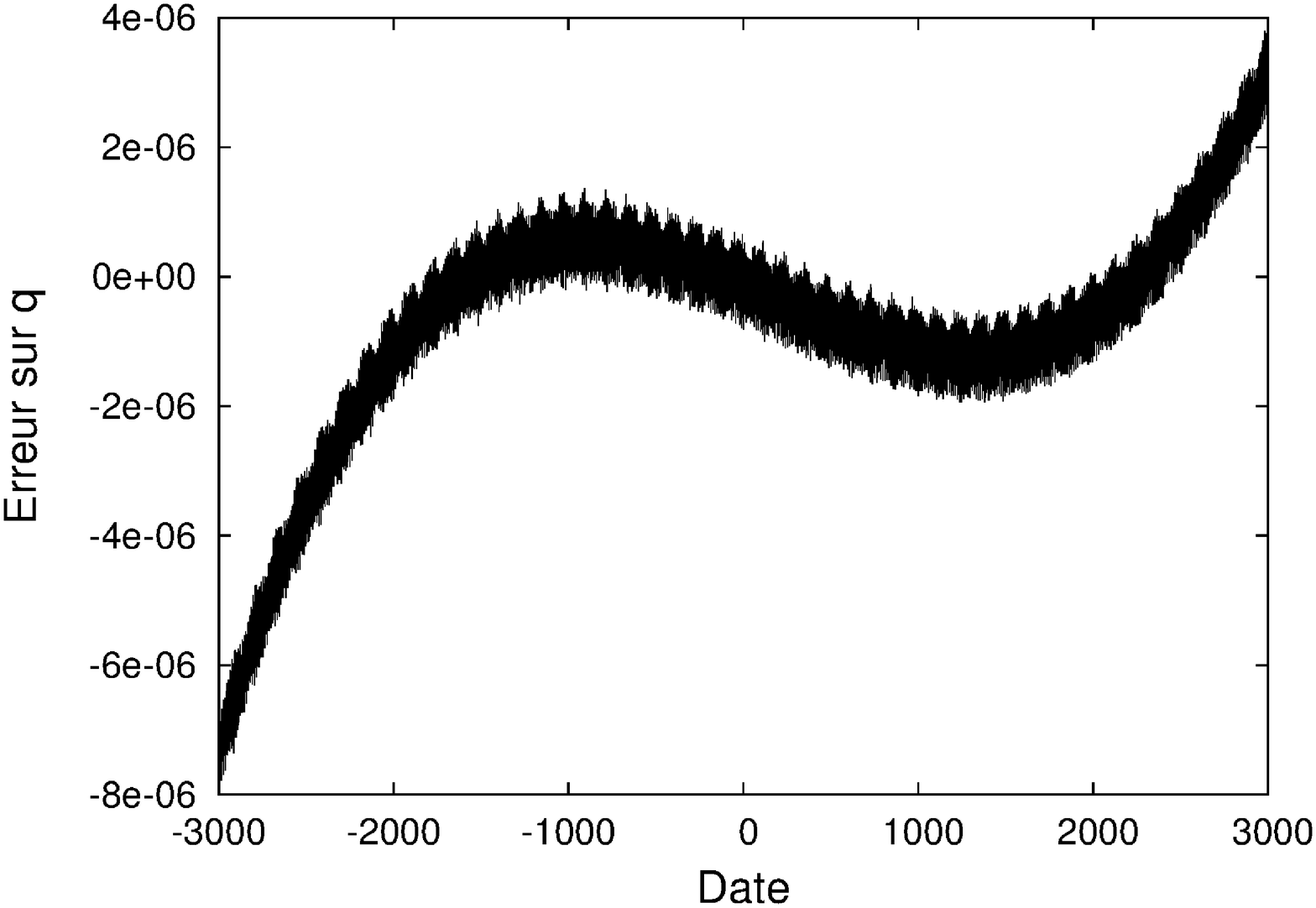} & \includegraphics[width=.47\textwidth]{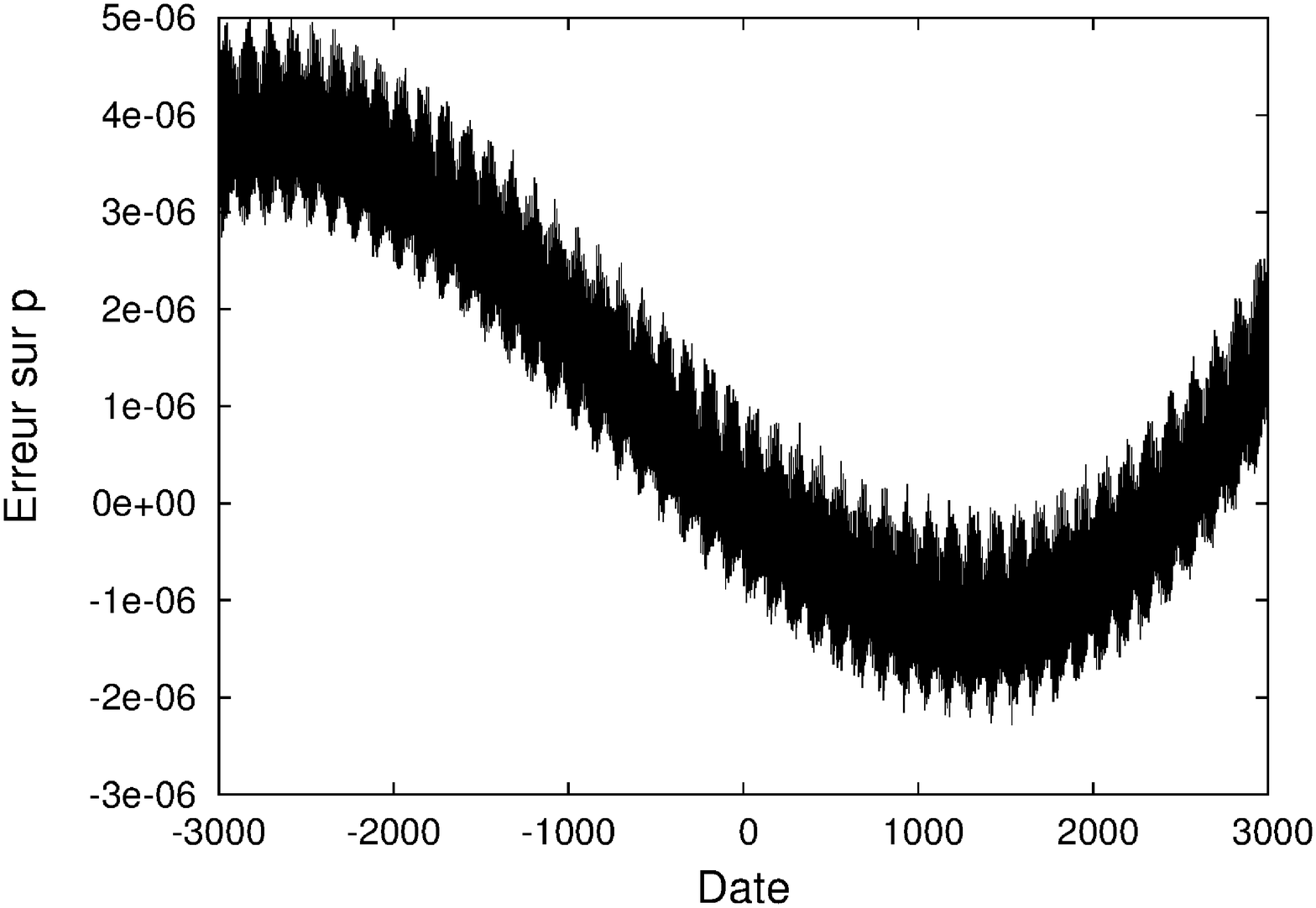}
\end{tabular}
\caption[Erreur sur les \'el\'ements orbitaux]{Erreur sur les \'el\'ements orbitaux. Ces courbes montrent la diff\'erence entre les \'el\'ements orbitaux donn\'es par
DE406 et les s\'eries trigonom\'etriques (\ref{eq:knum}) \`a (\ref{eq:pnum}).\label{fig:errororb}}
\end{figure}

 M\^eme si ces \'equations ne sont pas lin\'eaires, une r\'esolution num\'erique fonctionne tr\`es bien. J'ai pour ma part utilis\'e Maple\textregistered. 
En ajustant les paraboles (\'Eq.\ref{eq:kparabol} \`a \ref{eq:pparabol}) sur l'intervalle [J1000:J3000], on obtient

\begin{eqnarray}
k(t) & \approx & (-2.31417\times10^{-11}\pm1.785\times10^{-13})t^2 \nonumber \\
 & & +(-5.52628\times10^{-6}\pm7.656\times10^{-10})t \nonumber \\
 & & +0.0446629\pm7.53\times10^{-10}, \label{eq:kapp} \\
h(t) & \approx & (-7.76651\times10^{-11}\pm1.769\times10^{-13})t^2 \nonumber \\
 & & +(1.43999\times10^{-6}\pm7.586\times10^{-10})t \nonumber \\
 & & +0.200722\pm7.461\times10^{-7}, \label{eq:happ} \\
q(t) & \approx & (-1.21729\times10^{-11}\pm7.664\times10^{-16})t^2 \nonumber \\
 & & +(6.52656\times10^{-7}\pm3.287\times10^{-12})t \nonumber \\
 & & +0.0406163\pm3.233\times10^{-9}, \label{eq:qapp} \\
p(t) & \approx & (-1.04673\times10^{-11}\pm7.53\times10^{-16})t^2 \nonumber \\
 & & +(-1.27792\times10^{-6}\pm3.23\times10^{-12})t \nonumber \\
 & & +0.0456362\pm3.177\times10^{-9}, \label{eq:papp}
\end{eqnarray}
ce qui donne

\begin{eqnarray}
\alpha_1 & = & 0.1990903983, \nonumber \\
\omega_1 & = & 2.852011398\times10^{-5}\,\textrm{rad/an (p\'eriode : 220\,307.16 ans)}, \nonumber \\
\phi_1 & = & 1.30845314198\,\textrm{rad}\quad(74.969^{\circ}), \nonumber \\
\alpha_2 & = & 0.01094807206, \nonumber \\
\omega_2 & = & 4.767836272\times10^{-6}\,\textrm{rad/an (p\'eriode : 1\,317\,827.41 ans)}, \nonumber \\
\phi_2 & = & 2.26085090227\,\textrm{rad}\quad(129.537^{\circ}), \nonumber
\end{eqnarray}

et

\begin{eqnarray}
\beta_1 & = & 0.06094690052, \nonumber \\
\Omega_1 & = & -2.298222197\times10^{-5}\,\textrm{rad/an (p\'eriode : 273\,393.29 ans)}, \nonumber \\
\Phi_1 & = & 0.60658814513\,\textrm{rad}\quad(34.755^{\circ}), \nonumber \\
\beta_2 & = & 0.01442538649, \nonumber \\
\Omega_2 & = &  1.340719884\times10^{-5}\,\textrm{rad/an (p\'eriode : 468\,642.66 ans)}, \nonumber \\
\Phi_2 & = & 2.28580288184\,\textrm{rad}\quad(130.967^{\circ}). \nonumber
\end{eqnarray}
On peut \'ecrire ainsi :

\begin{eqnarray}
k(t) & \approx & 0.1990903983\cos(2.852011398\times10^{-5}t+1.30845314198) \nonumber \\
 & & +0.01094807206\cos(4.767836272\times10^{-6}t+2.26085090227), \label{eq:knum} \\
h(t) & \approx & 0.1990903983\sin(2.852011398\times10^{-5}t+1.30845314198) \nonumber \\
 & & +0.01094807206\sin(4.767836272\times10^{-6}t+2.26085090227), \label{eq:hnum} \\ 
q(t) & \approx & 0.06094690052\cos(-2.298222197\times10^{-5}t+0.60658814513) \nonumber \\
 & & +0.01442538649\cos(1.340719884\times10^{-5}t+2.28580288184), \label{eq:qnum} \\
p(t) & \approx & 0.06094690052\sin(-2.298222197\times10^{-5}t+0.60658814513) \nonumber \\
 & & +0.01442538649\sin(1.340719884\times10^{-5}t+2.28580288184), \label{eq:pnum}
 \end{eqnarray}
les phases \'etant donn\'ees \`a J2000. La Fig.\ref{fig:errororb} donne l'erreur induite par l'utilisation des s\'eries trigonom\'etriques (\ref{eq:knum}) \`a (\ref{eq:pnum}).
On peut voir pour les variables li\'ees \`a l'inclinaison, $q$ et $p$, un signal de degr\'e $3$. Il s'est en fait av\'er\'e que ces variables \'etaient au mieux mod\'elis\'ees
par des polyn\^omes de degr\'e 3, mais nous aurions eu alors 8 \'equations, alors qu'une s\'erie trigonom\'etrique \`a 3 \'el\'ements contient 9 inconnues. Il \'etait donc
plus pratique de se limiter au degr\'e 2, l'erreur \'etant bien plus faible que pour les variables d'excentricit\'e $k$ et $h$ sur l'intervalle de validit\'e des \'eph\'em\'erides.

  \par Une fois ces s\'eries trigonom\'etriques obtenues, elles peuvent \^etre extrapol\'ees sur plusieurs millions d'ann\'ees sans diverger. La mod\'elisation de l'obliquit\'e
  $K$ et de l'angle associ\'e $\sigma_3=r+\ascnode$ est obtenue avec :
  
  \begin{enumerate}
  
   \item Simulation num\'erique \`a long terme, en utilisant le syst\`eme d'\'equations moyenn\'ees (\ref{eq:equationssandrine}) et les s\'eries trigonom\'etriques (\ref{eq:knum})
   \`a (\ref{eq:pnum}) sur plusieurs millions d'ann\'ees, en couvrant les valeurs possibles des param\`etres du champ de gravit\'e,
   
   \item Identification des diff\'erents termes forc\'es et \'elimination des librations libres (cf. Chap.\ref{chap:naffo}),
   
   \item Ajustement, par moindres carr\'es, d'une loi reliant chaque amplitude aux param\`etres d'int\'erieur $C_{20}$, $C_{22}$ et $C/(M_{\mercury}R_{\mercury}^2)$,
   
   \item V\'erification, a posteriori, du bien fond\'e de cette m\'ethode et de la validit\'e du r\'esultat. Pour cela, les conditions intiales obtenues, utilisant les amplitudes
   ajust\'ees, sont utilis\'ees en propageant les \'equations de la rotation rigide de Mercure non moyenn\'ees. L'\'ecart \`a l'\'equilibre dynamique se traduira par des oscillations
   libres dont la p\'eriode sera de l'ordre de 1000 ans; leur amplitude permet d'estimer l'erreur induite par la m\'ethode.
   
  \end{enumerate}

  \par Dans \citep{nd2012}, nous avons appliqu\'e cette th\'eorie avec succ\`es mais en utilisant des valeurs du champ de gravit\'e de Mercure donn\'ees par 
  Mariner 10, qui sont maintenant obsol\`etes. C'est pourquoi je d\'eveloppe l'application pr\'esente dans \citep{nl2013}, qui utilise des donn\'ees MESSENGER.
  
  \par Il a \'et\'e difficile de publier cet article, l'un des reviewers \'etant si peu convaincu du bien fond\'e de l'\'etude qu'il insistait pour que toute r\'ef\'erence \`a 
  BepiColombo soit supprim\'ee. Comme nous \'etions financ\'es dans le cadre de la pr\'eparation de cette mission, nous ne pouvions pas accepter cette exigence. Ceci s'est
  finalement r\'esolu par la d\'esignation d'un autre reviewer.
  
  \subsection[Notre solution]{Notre solution \citep{nl2013}}
  
  \par Les d\'eterminations de param\`etres du champ de gravit\'e avec une bonne pr\'ecision (Tab.\ref{tab:gravitemercure}) par \citet{szpshlmnpmjtprght2012} et de l'obliquit\'e
  de Mercure par \citet{mpshgjygpc2012} nous ont incit\'es \`a reprendre cette \'etude pour l'appliquer aux valeurs actuelles. L'un des buts \'etait de d\'eterminer le moment
  d'inertie polaire de Mercure $C$ par une m\'ethode ind\'ependante de celle de Peale.
  
  \par Ce travail a \'et\'e fait en collaboration avec Christoph Lhotka qui a fait l'essentiel de la partie analytique, alors que j'ai fait l'essentiel de la partie num\'erique.
  C'est pourquoi j'insisterai sur le num\'erique. Le travail de Christoph a consist\'e \`a r\'e\'ecrire une formule dans l'id\'ee de celle de Peale, apr\`es moyennisation des 
  termes p\'eriodiques et sans consid\'erer a priori que la diff\'erence des n{\oe}uds $\sigma_3$ \'etait identiquement nulle. De plus, l'influence des harmoniques d'ordre
  sup\'erieur \'etait consid\'er\'ee. La formule obtenue est
  
  \begin{eqnarray}
    \epsilon & = & \left(1+\frac{2}{3}\frac{\dot{\ascnode}}{n}\cos I+\frac{2\dot{\omega}}{3n}\right) \label{eqF3} \\
             & \times & \frac{C/(M_{\mercury}R_{\mercury}^2) \dot{\ascnode} \sin I}{n\left(C_{22}f_1(e)-C_{20}f_2(e)+C_{40}\left(\frac{R_{\mercury}}{a}\right)^2f_3(e)-\frac{2}{3}\left(\frac{\dot{\ascnode}}{n}\right)^2\frac{C}{M_{\mercury}R_{\mercury}^2}\sin^2I\right)} \nonumber.
  \end{eqnarray}
  avec
  
  \begin{eqnarray}
    f_1(e) & = & 7e-\frac{123}{8}e^3, \label{eq:f1christoph} \\
    f_2(e) & = & 1+\frac{3}{2}e^2+\frac{15}{8}e^4, \label{eq:f2christoph} \\
    f_3(e) & = & \frac{5}{2}+\frac{25}{2}e^2+\frac{525}{16}e^4. \label{eq:f3christoph}
  \end{eqnarray}

  Cette formule (\ref{eqF3}) est \`a comparer avec celle de Peale (\'Eq.\ref{eq:oblikpeale}). On peut remarquer la pr\'esence de la pr\'ecession de l'argument du 
  p\'ericentre $\dot{\omega}$, ainsi que de l'harmonique $C_{40}$, dont l'influence sera toutefois tr\`es limit\'ee par la pr\'esence du pr\'efacteur $(R_{\mercury}/a)^2$.
  
  \par D'un point de vue num\'erique, nous sommes partis du Hamiltonien moyenn\'e $<\mathcal{H}>$ suivant :
  
  \begin{eqnarray}
    <\mathcal{H}>(\Sigma_1,\Sigma_3,\sigma_1,\sigma_3,t) & = & -\frac{2M_{\mercury}^3\mathcal{G}^2(M_{\mercury}+M_{\sun})^2}{(2\Sigma_4-3\Sigma_1)^2} \label{eq:Hamilchristoph} \\
    & + & \frac{\Sigma_1^2}{2C}-\Sigma_1\dot{\omega}+(\Sigma_3-\Sigma_1)\dot{\ascnode}-\mathcal{G}M_{\sun}M_{\mercury}<\mathcal{V}>, \nonumber
  \end{eqnarray}
o\`u

\begin{itemize}

 \item $\Sigma_1$ est la norme du moment cin\'etique de rotation de Mercure ($=3/2nCP$ si on veut se ref\'erer \`a des notations utilis\'ees pr\'ec\'edemment),
 
 \item $\Sigma_3=\Sigma_1(1-\cos K)$ o\`u $K$ est l'obliquit\'e de Mercure par rapport \`a l'\'ecliptique J2000,
 
 \item $\sigma_1$ est l'argument de la r\'esonance 3:2 tel que $2\sigma_1=2p-3\lambda+\varpi$ (voir plus haut),
 
 \item $\sigma_3=r+\ascnode$ est l'angle en libration du fait de la Troisi\`eme Loi de Cassini,
 
 \item $\Sigma_4=M_{\mercury}\sqrt{\mathcal{G}(M_{\mercury}+M_{\sun})a}+3/2\Sigma_1$. L'angle associ\'e, $\sigma_4=\lambda$, n'appara\^it pas dans notre Hamiltonien moyenn\'e,
 ce qui fait de $\Sigma_4$ une constante.

\end{itemize}
$<\mathcal{V}>$ est, \`a un pr\'efacteur pr\`es, le potentiel perturbateur moyenn\'e sur les courtes p\'eriodes. Il est de la forme

\begin{equation}
 \label{eq:Vchristoph}
 <\mathcal{V}> = C_{20}<\mathcal{V}_{20}>+C_{22}<\mathcal{V}_{22}>+C_{30}<\mathcal{V}_{30}>+C_{40}<\mathcal{V}_{40}>,
\end{equation}
l'expression des $<\mathcal{V}_{nm}>$ \'etant donn\'ee dans \citep{nl2013}. On peut remarquer dans l'expression du Hamiltonien $<\mathcal{H}>$, la vitesse de pr\'ecession de l'argument
du p\'ericentre $\dot{\omega}$, que nous consid\'erons dans notre mod\`ele. Nous allons \'egalement jusqu'au degr\'e 4 en excentricit\'e / inclinaison.

\par Nous obtenons ensuite les \'equations Hamiltoniennes 

\begin{equation}
\label{eq:equationschristoph}
\begin{array}{lll}
\dot{\sigma_1}=\frac{\partial<\mathcal{H}>}{\partial \Sigma_1}, & \hspace{2cm} & \dot{\Sigma}_1=-\frac{\partial <\mathcal{H}>}{\partial \sigma_1}, \\
\dot{\sigma}_3=\frac{\partial<\mathcal{H}>}{\partial \Sigma_3}, & \hspace{2cm} & \dot{\Sigma}_3=-\frac{\partial <\mathcal{H}>}{\partial \sigma_3},
\end{array}
\end{equation}
dans lesquelles nous injectons les expressions trigonom\'etriques \`a long terme de $k$, $h$, $q$, $p$ (\'Eq. \ref{eq:knum} \`a \ref{eq:pnum}). Nous avons num\'eriquement
localis\'e l'\'Etat de Cassini, d\'ependant du temps, pour 56 mod\`eles de Mercure coh\'erents avec le champ de gravit\'e mesur\'e par \citet{szpshlmnpmjtprght2012}, 
c'est-\`a-dire

\begin{itemize}

 \item $C_{20}\in [-5.1\times 10^{-5};-4.9\times 10^{-5}]$ (valeur nominale : $-5.031\times 10^{-5}$),
 
 \item $C_{22}\in [8.0\times 10^{-6};8.2\times 10^{-6}]$ (valeur nominale : $8.088\times 10^{-6}$),
 
 \item $C_{30}\in [-1.3\times 10^{-5};-1.1\times 10^{-5}]$ (valeur nominale : $-1.188\times 10^{-5}$),
 
 \item $C_{40}\in [-2.19\times 10^{-5};-1.71\times 10^{-5}]$ (valeur nominale : $-1.95\times 10^{-5}$),
 
 \item $C/(M_{\mercury}R_{\mercury}^2)\in [0.32;0.38]$ (valeur nominale : $0.35$).

\end{itemize}

\par Un exemple d'analyse en fr\'equence de la solution num\'erique est donn\'e dans la Tab.\ref{tab:Ksig1}. On peut voir des termes d\'ependant uniquement des fr\'equences donn\'ees
dans les expressions trigonom\'etriques (\ref{eq:knum}) \`a (\ref{eq:pnum}), et d'autres qui correspondent \`a des oscillations libres et ne doivent pas appara\^itre lorsque les
conditions initiales correspondent exactement \`a l'\'Etat de Cassini. 

\begin{table}[ht]
	\centering
	\caption[Analyse en fr\'equence d'une solution du syst\`eme]{Un exemple d'analyse en fr\'equence de la solution num\'erique de notre syst\`eme d'\'equations (\ref{eq:equationschristoph}) pour les 
	param\`etres d'int\'erieur nominaux $C_{20}=-5.031\times10^{-5}$, $C_{22}=8.088\times 10^{-6}$, $C_{30}=-1.188\times 10^{-5}$, $C_{40}-1.95\times 10^{-5}$
	et $C = 0.35M_{\mercury}R_{\mercury}^2$.\label{tab:Ksig1}}
	\begin{tabular}{rrrr}
		\hline
		$N$ & Amplitude & P\'eriode & Identification \\
		& (arcmin) & (ans) & \\
		\hline
		$K-i$ & & & \\
		\hline
		$1$ & $1.9755$ &  $\infty$ & $<K-i>$ \\
		$2$ & $0.2682$ & 172\,332.6 & $2\Omega_2-2\Omega_1$ \\
		$4$ & $0.0254$ & 264\,530.5 & $\omega_1-\omega_2$ \\
		$5$ & $0.0132$ &  60\,999.1 & $2\omega_1-2\Omega_1$ \\
		$6$ & $0.0121$ &  57\,555.3 & $3\Omega_2-3\Omega_1$ \\
		$7$ & $0.0026$ &   4\,167.1 & libre \\
		$8$ & $0.0025$ &  43\,166.2 & $4\Omega_2-4\Omega_1$ \\
		$9$ & $0.0025$ &   4\,302.6 & libre \\
		\hline\hline
		$\sigma_3$ & & & \\
		\hline
		  $1$ & $6.2509$ & 172\,665.2 & $\Omega_2-\Omega_1$ \\
		  $2$ & $1.4683$ &  86\,332.6 & $2\Omega_2-2\Omega_1$ \\
		  $3$ & $0.3462$ &  57\,555.1 & $3\Omega_2-3\Omega_1$ \\
		  $4$ & $0.1112$ &  60\,999.0 & $2\omega_1-2\Omega_1$ \\
		  $5$ & $0.0816$ &  43\,166.3 & $4\Omega_2-4\Omega_1$ \\
		  $6$ & $0.0400$ & 497\,291.8 & $\varpi_2-\varpi_1+\Omega_2-\Omega_1$ \\
		  $7$ & $0.0396$ & 104\,473.0 & $\varpi_1-\varpi_2+\Omega_2-\Omega_1$ \\
		  $8$ & $0.0261$ &  45\,075.0 & $2\varpi_1-3\Omega_1+\Omega_2$ \\
		  $9$ & $0.0214$ &   4\,167.1 & libre \\
		 $10$ & $0.0209$ &   4\,302.6 & libre \\
		 \hline 
 	\end{tabular}
\end{table}

\par Les termes forc\'es sont gard\'es, et nous avons finalement identifi\'e une solution \`a 34 termes :

\begin{eqnarray}
K & = & i+a_1-2a_2\cos\left(\Omega_2-\Omega_1\right)+2a_3\cos\left(2\Omega_2-2\Omega_1\right)-2a_4\cos\left(\varpi_1-\varpi_2\right) \nonumber \\ 
 & + & 2a_5\cos\left(2\varpi_1-2\Omega_1\right)-2a_6\cos\left(3\Omega_2-3\Omega_1\right)+2a_7\cos\left(4\Omega_2-4\Omega_1\right)\nonumber \\
 & + & 2a_8\cos\left(\varpi_2-\varpi_1+\Omega_2-\Omega_1\right)+2a_9\cos\left(\varpi_1-\varpi_2+\Omega_2-\Omega_1\right)+2a_{10}\cos\left(\varpi_1+\varpi_2-2\Omega_1\right) \nonumber \\
 & - & 2a_{11}\cos\left(2\varpi_1-3\Omega_1+\Omega_2\right)-2a_{12}\cos\left(5\Omega_2-5\Omega_1\right)+2a_{13}\cos\left(2\varpi_1-2\varpi_2\right) \nonumber \\
 & - & 2a_{14}\cos\left(\varpi_1-\varpi_2-2\Omega_1+2\Omega_2\right)-2a_{15}\cos\left(\varpi_2-\varpi_1-2\Omega_1+2\Omega_2\right) \nonumber \\
 & - & 2a_{16}\cos\left(2\varpi_1-2\Omega_2\right), \label{eq:KmI} \\
\sigma_3 & = & 2a_{17}\sin\left(\Omega_2-\Omega_1\right)-2a_{18}\sin\left(2\Omega_2-2\Omega_1\right)+2a_{19}\sin\left(3\Omega_2-3\Omega_1\right) \nonumber \\
 & - & 2a_{20}\sin\left(2\varpi_1-2\Omega_1\right)-2a_{21}\sin\left(4\Omega_2-4\Omega_1\right)-2a_{22}\sin\left(\varpi_2-\varpi_1+\Omega_2-\Omega_1\right) \nonumber \\
 & - & 2a_{23}\sin\left(\varpi_1-\varpi_2+\Omega_2-\Omega_1\right)+2a_{24}\sin\left(2\varpi_1-3\Omega_1+\Omega_2\right)+2a_{25}\sin\left(5\Omega_2-5\Omega_1\right) \nonumber \\
 & + & 2a_{26}\sin\left(2\varpi_1-\Omega_1-\Omega_2\right)-2a_{27}\sin\left(\varpi_1+\varpi_2-2\Omega_1\right)+2a_{28}\sin\left(-\varpi_1+\varpi_2-2\Omega_1+2\Omega_2\right) \nonumber \\
 & + & 2a_{29}\sin\left(\varpi_1-\varpi_2-2\Omega_1+2\Omega_2\right)-2a_{30}\sin\left(2\varpi_1-4\Omega_1+2\Omega_2\right)-2a_{31}\sin\left(6\Omega_2-6\Omega_1\right) \nonumber \\
 & + & 2a_{32}\sin\left(\varpi_1+\varpi_2-3\Omega_1+\Omega_2\right)-2a_{33}\sin\left(\varpi_1-\varpi_2-3\Omega_1+3\Omega_2\right) \nonumber \\
 & - & 2a_{34}\sin\left(\varpi_2-\varpi_1-3\Omega_1+3\Omega_2\right), \label{eq:sig3}
\end{eqnarray}
avec

\begin{equation}
\label{eq:ai1}
a_i  =  \frac{C/\left(M_{\mercury}R_{\mercury}^2\right)}{\alpha_i C/\left(M_{\mercury}R_{\mercury}^2\right)+\beta_i C_{20}+\gamma_i C_{22}+\delta_i}
\end{equation}
pour $i=1,2,5,6,17,18,19,20,21,22,23,24,26,27$,

\begin{equation}
\label{eq:ai2}
a_i  =  \frac{C/\left(M_{\mercury}R_{\mercury}^2\right)}{\alpha_i+\beta_i C_{20}+\gamma_i C_{22}}
\end{equation}
pour $i=3,4$

\begin{equation}
\label{eq:ai3}
a_i  =  \frac{C/\left(M_{\mercury}R_{\mercury}^2\right)}{\alpha_i+\beta_i C_{20}}
\end{equation}
pour $i=7,8,9,10,11,12,13,14,15,16,34$, et

\begin{equation}
\label{eq:ai4}
a_i  =  \frac{C/\left(M_{\mercury}R_{\mercury}^2\right)}{\alpha_i C/\left(M_{\mercury}R_{\mercury}^2\right)+\beta_i C_{20}+\gamma_i}
\end{equation}
pour $i=25,28,29,30,31,32,33$. Ces formules sont ajust\'ees sur les amplitudes obtenues num\'eriquement pour les 56 mod\`eles de Mercure, et leur forme est inspir\'ee de l'\'equation
de Peale (\'Eq.\ref{eq:oblikpeale}). On peut remarquer qu'on n'a d\'etect\'e aucune influence des harmoniques $J_3=-C_{30}$ et $J_4=-C_{40}$ sur les variables li\'ees \`a l'obliquit\'e.

\begin{table}[!ht]
\centering
\caption{Les coefficients intervant dans les \'Eq.\ref{eq:ai1} \`a \ref{eq:ai4}.\label{tab:coeff}}
\begin{tabular}{r|rrrr}
i & $\alpha_i$ & $\beta_i$ & $\gamma_i$ & $\delta_i$ \\
\hline
 1 & $-9.6916394157$ &     $-1.022791\times10^7$ &      $1.224118\times10^7$ &  $-0.041284174514$ \\
 2 &  $57.319667133$ &     $-1.495053\times10^8$ &     $1.8067315\times10^8$ & $-15.290923597$ \\
 3 &   $-288.588048$ &      $-6.75794\times10^8$ &      $8.648745\times10^8$ & --  \\
 4 &   $109.5839605$ &    $-2.3327045\times10^9$ &    $-2.8315945\times10^9$ & -- \\
 5 &  $9618.2490875$ &     $-5.967955\times10^9$ & $-1.4620515\times10^{10}$ & $-4270.5575456$ \\
 6 &  $5690.3083952$ &      $-3.29021\times10^9$ &       $4.59872\times10^9$ & $-5791.2841683$ \\
 7 &     $160882.75$ &  $-1.563842\times10^{10}$ &  --                       & -- \\
 8 &    $-330146.25$ & $-3.3956545\times10^{10}$ &  --                       & -- \\
 9 &    $-307545.35$ &  $-3.441592\times10^{10}$ &  --                       & -- \\
10 &     $-879441.5$ &   $-4.59746\times10^{10}$ &  --                       & -- \\
11 &    $-1025209.5$ &   $-5.12722\times10^{10}$ &  --                       & -- \\
12 &        $760424$ &  $-7.318535\times10^{10}$ &  --                       & -- \\
13 &      $-3583265$ & $-1.7097535\times10^{11}$ &  --                       & -- \\
14 &      $-1352596$ &  $-1.522164\times10^{11}$ &  --                       & -- \\
15 &    $-1475953.5$ &  $-1.554497\times10^{11}$ &  --                       & -- \\
16 &      $-4788455$ & $-2.4530905\times10^{11}$ &  --                       & -- \\
17 &  $0.0459703076$ &               $-111936.3$ &               $133840.35$ & $0.00068014043987$ \\
18 &  $0.5097512707$ &                 $-474698$ &                  $568708$ & $-0.0063487867442$ \\
19 &  $3.6758306831$ &                $-2005542$ &                 $2431534$ & $-0.32882074509$ \\
20 &  $20.470309592$ &               $-12322835$ &               $-31382330$ & $1.6246447377$ \\
21 &  $22.351343806$ &                $-8470000$ &                $10602865$ & $-4.3502813922$ \\
22 &  $1.8968370715$ &               $-25710160$ &               $-28807205$ & $-14.991580755$ \\
23 &  $28.007891612$ &               $-25704595$ &               $-30518425$ & $0.75276141087$ \\
24 &  $151.69560968$ &               $-52410050$ &              $-127924300$ & $-54.254151937$ \\
25 &  $11.320467298$ &               $-35729050$ &            $384.87733645$ & -- \\
26 &  $60.949766585$ &               $-91723450$ &              $-262171000$ & $275.67522095$ \\
27 &  $93.149257619$ &               $-96422550$ &              $-265414800$ & $257.09346578$ \\
28 &  $256.01109782$ &              $-108893750$ &           $-1098.8078842$ & -- \\
29 &  $93.896968203$ &              $-109386550$ &           $-1017.8654389$ & -- \\
30 &  $880.39601125$ &              $-223772150$ &           $-4722.8636039$ & -- \\
31 & $-390.02956574$ &              $-145710250$ &            $2105.3653480$ & -- \\
32 &  $1929.3930214$ &              $-404092500$ &           $-8327.8625575$ & -- \\
33 &  $2613.8902647$ &              $-456302000$ &           $-4855.3715926$ & -- \\
34 &    $-2905.1715$ &              $-438245500$ &  --                       & -- \\
\hline
\end{tabular}
\end{table}

\newpage

\subsection{Test de la solution}

\par Le but de cette \'etude est d'obtenir des conditions initiales proches de l'\'Etat de Cassini pour un Mercure r\'ealiste. Notamment, pour une dynamique de rotation non 
moyenn\'ee et de vraies \'eph\'em\'erides. Tout \'ecart \`a l'\'Etat de Cassini se traduira par des oscillations libres, dont nous souhaitons que l'amplitude soit la plus petite possible.

\par
\begin{figure}[ht]
\centering
\includegraphics[width=0.6\textwidth]{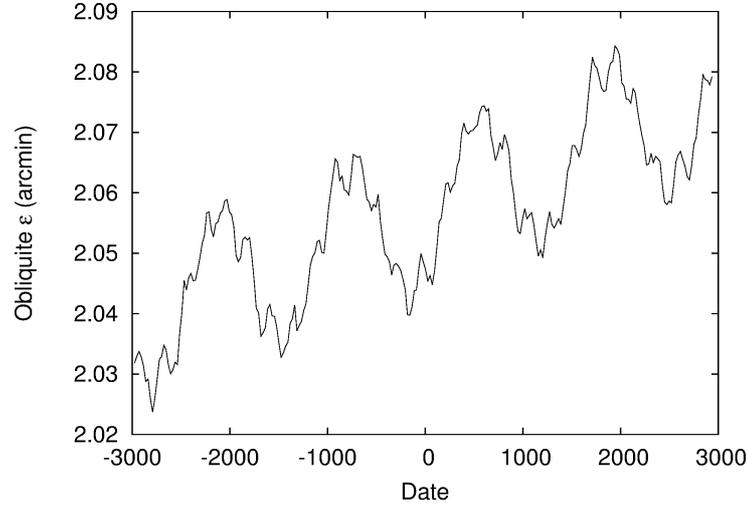}
\caption[Obliquit\'e de Mercure simul\'ee \`a l'aide des \'equations non moyenn\'ees.]{Obliquit\'e de Mercure simul\'ee en propageant les \'equations non
moyenn\'ees, mais en partant des conditions initiales donn\'ees par le syst\`eme moyenn\'e, avec $C_{20}=-5.031\times10^{-5}$, $C_{22}=8.088\times10^{-6}$, 
$C_{30}=C_{40}=0$ et $C=0.35M_{\mercury}R_{\mercury}^2$. Les oscillations de p\'eriode environ 1000 ans sont des librations libres, qui ne seraient pas pr\'esentes
si les conditions initiales \'etaient id\'eales.\label{fig:obliq}}
\end{figure}

 Nous avons fait ces tests \`a partir des \'equations issues d'un Hamiltonien non moyenn\'e, en tenant compte des harmoniques $C_{20}$, $C_{22}$, $C_{30}$ et $C_{40}$.
Un exemple est donn\'e par la Fig.\ref{fig:obliq}. On peut y voir le terme libre de p\'eriode de l'ordre de 1000 ans.

\par
\begin{figure}[ht]
   \centering
   \begin{tabular}{cc}
   \includegraphics[width=0.47\textwidth]{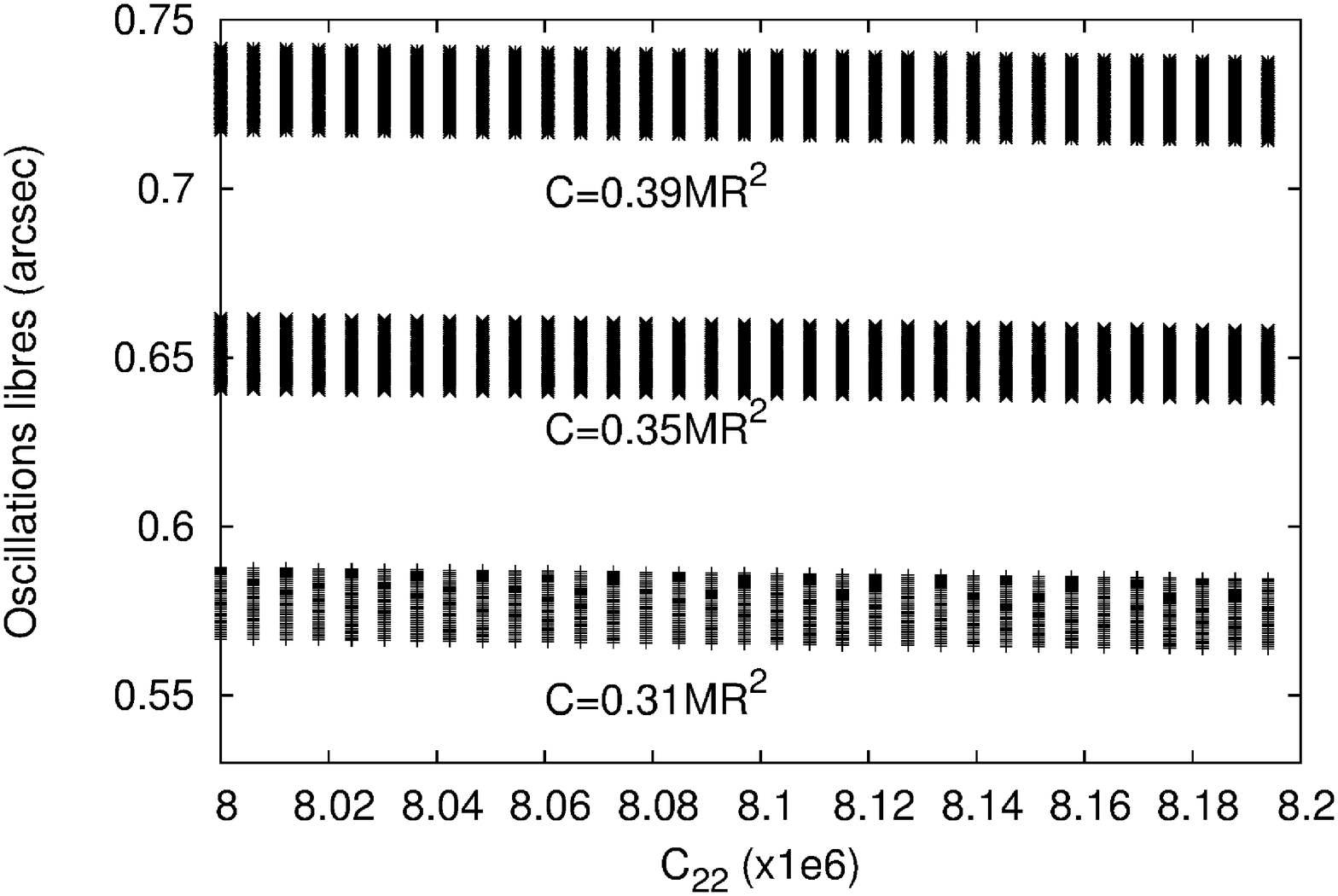} & \includegraphics[width=0.47\textwidth]{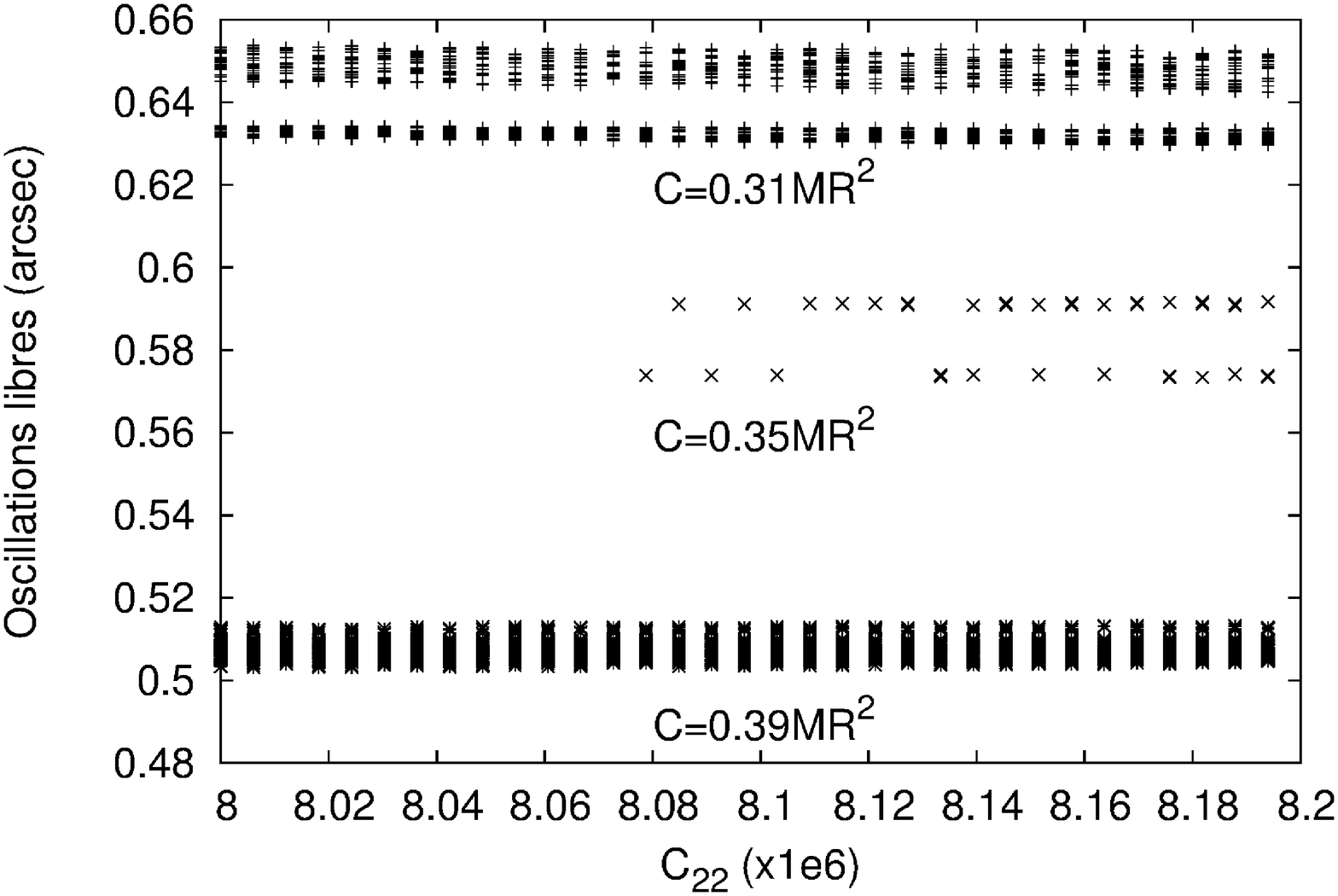} \\
   avec DE406 & avec INPOP10a
   \end{tabular}
   \caption[Amplitude des oscillations libres]{Amplitude des oscillations libres dues \`a nos conditions initiales.\label{fig:freeeps}}
  \end{figure}

 Nous avons en fait test\'e la validit\'e de nos conditions initiales (\ref{eq:KmI} \& \ref{eq:sig3}) avec les 56 jeux de param\`etres d'int\'erieur, des \'equations non moyenn\'ees,
et les \'eph\'em\'erides orbitales DE406 et INPOP10a \citep{flkmdgct2011}. L'amplitude des oscillations libres a \'et\'e obtenue par analyse en fr\'equences (Fig.\ref{fig:freeeps}).
  
  \par Nous pouvons voir qu'en aucun cas, l'amplitude des oscillations libres n'est sup\'erieure \`a 750 millisecondes d'arc. Pour r\'ef\'erence, la pr\'ecision de la mesure
  de l'obliquit\'e est d'environ 5 secondes d'arc \citep{mpshgjygpc2012}. On peut remarquer que, dans le cas d'INPOP10a, certains points sont absents. La raison est qu'INPOP10a
  n'est disponible que sur 2000 ans, p\'eriode sur laquelle il peut \^etre difficile de d\'etecter un terme p\'eriodique dont la p\'eriode est proche de 1000 ans.
  
  \subsection{Influence des harmoniques d'ordre sup\'erieur}
  
  \par
\begin{figure}[ht]
   \centering
   \includegraphics[width=0.6\textwidth]{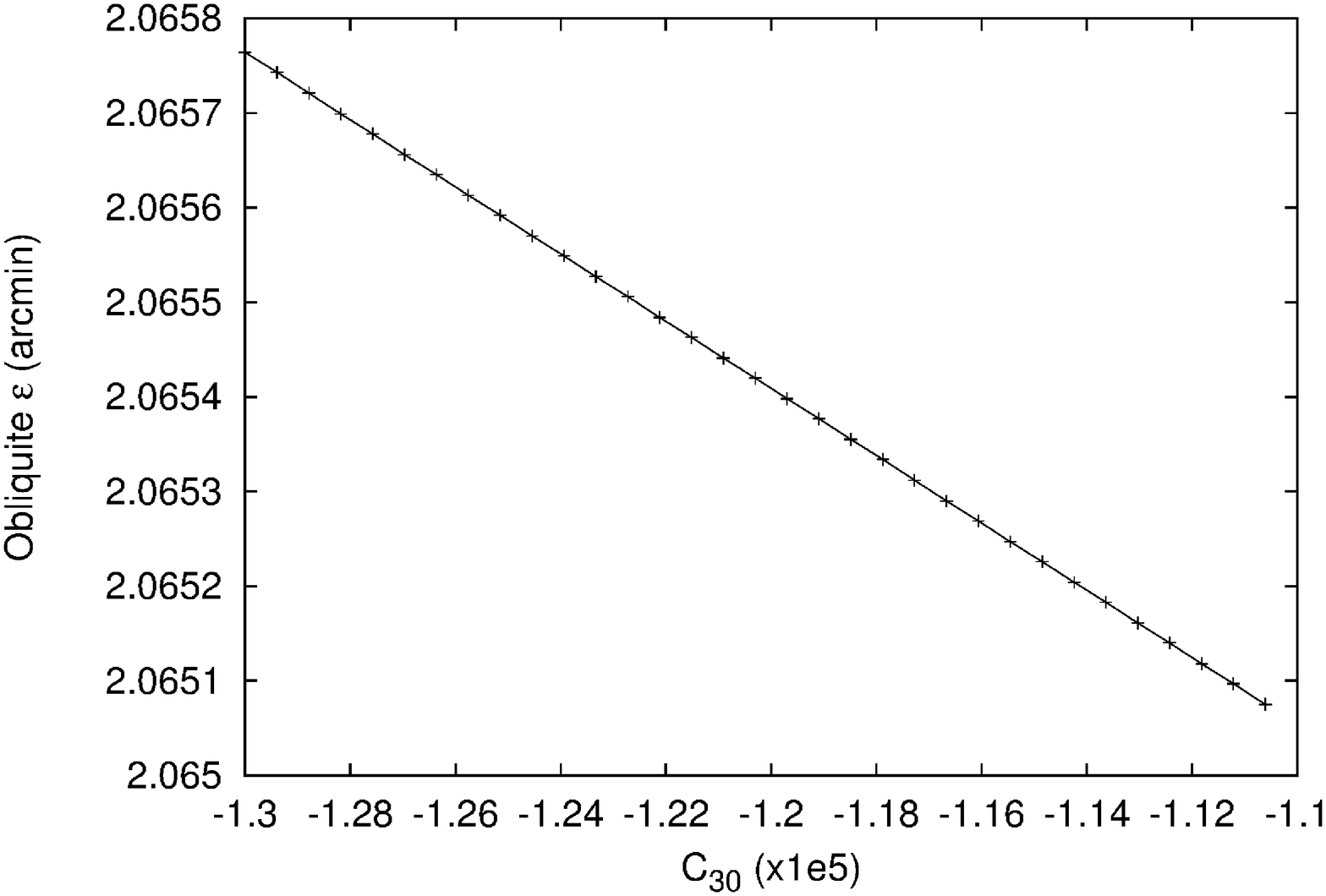}
   \caption[Influence de $C_{30}$ sur l'obliquit\'e moyenne]{Influence de $C_{30}$ sur l'obliquit\'e moyenne, obtenue apr\`es int\'egration num\'erique des \'equations non moyenn\'ees.
   Les autres param\`etres d'int\'erieur sont les nominaux.\label{fig:J3}}
  \end{figure}

 Nous avons test\'e l'influence des harmoniques d'ordre sup\'erieur \`a partir du syst\`eme non moyenn\'e. Le syst\`eme moyenn\'e sugg\`ere que les harmoniques $J_3$ et $J_4$
  n'ont pas d'influence. Dans le cas du syst\`eme non moyenn\'e, nous avons fait varier les param\`etres $J_3$ et $J_4$ et avons extrait des simulations num\'eriques l'obliquit\'e
  moyenne. Si $J_4$ ne semble pas avoir d'influence, $J_3$ en a (Fig.\ref{fig:J3}).
  
  \par On peut voir que la d\'ependance en $J_3$ semble lin\'eaire dans le domaine compatible avec les observations. Un ajustement par moindres carr\'es donne
  
  \begin{equation}
	\label{eq:fitJ3}
	\epsilon=(-355.197C_{30}+2.06115)\,\textrm{arcmin},
  \end{equation}
  ce qui veut dire que n\'egliger $J_3$ peut induire une erreur de $\approx253$ millisecondes d'arc pour la valeur nominale $J_3=1.188\times10^{-5}$. Ceci est en contradiction
  avec la formule (\ref{eqF3}) et l'\'etude num\'erique \`a long terme. Nous avons cherch\'e \`a am\'eliorer les calculs pr\'ec\'edents en poussant l'ordre des d\'eveloppements en 
  excentricit\'e / inclinaison et n'avons pas r\'eussi \`a trouver une influence de $J_3$. Une piste d'am\'elioration est d'introduire une moyennisation d'ordre 2 en les 
  param\`etres du champ de gravit\'e. Ceci est une t\^ache tr\`es lourde, nous ne l'avons pas faite dans le cadre de cette \'etude.
  
  \subsection{Influence des mar\'ees}
  
  \par L'attraction diff\'erentielle du Soleil sur Mercure, c'est-\`a-dire les mar\'ees, contribue \`a la dissipation de l'\'energie de rotation, ce qui a pour cons\'equences
  l'\'etablissement de la r\'esonance spin-orbite (Chap.\ref{chap:tides}) et de l'\'Etat de Cassini, mais aussi la forme triaxiale de Mercure. Une autre cons\'equence des mar\'ees 
  est un effet p\'eriodique, non dissipatif, qui alt\`ere le champ de gravit\'e et la forme \`a la p\'eriode orbitale et ses harmoniques. En effet, le couple de mar\'ee d\'epend 
  de la distance Soleil-Mercure, qui varie d'un facteur $\approx1.5$ sur une orbite du fait de l'excentricit\'e importante $\approx 0.206$. Il d\'epend \'egalement de la vitesse de 
  rotation de Mercure car le couple subi par un \'el\'ement de masse d\'epend aussi de son orientation par rapport au Soleil.
  
  \par On a, d'apr\`es \citet{g2004} :

  \begin{eqnarray}
	C_{20}(t) & = & C_{20}^{static}+\frac{k_2}{2}q_te\cos \mathcal{M}, \label{eq:C20rap} \\
	C_{22}(t) & = & C_{22}^{static}-\frac{k_2}{24}q_t\left(2\cos \mathcal{M}-e\cos 2\mathcal{M}\right), \label{eq:C22rap}
  \end{eqnarray}
  ce r\'esultat \'etant obtenu \`a partir du d\'eveloppement de Kaula du potentiel de mar\'ee (\ref{eq:kaula}) dans lequel l'angle d'heure locale $\theta^*$ doit tenir compte 
  de la rotation en r\'esonance 3:2. De plus, \citet{vrkdr2008} nous donnent
  
  \begin{equation}
   \label{eq:Cvh}
   C(t) = C^{static}-\frac{k_2}{3}q_teM_{\mercury}R_{\mercury}^2\cos \mathcal{M},
  \end{equation}
  o\`u $k_2$ est le nombre de Love classique d'ordre 2 li\'e au champ de gravit\'e, et 
  
  \begin{equation}
   \label{eq:qtmercure}
   q_t = -3\frac{M_{\sun}}{M_{\mercury}}\left(\frac{R_{\mercury}}{a}\right)^3.
  \end{equation}

  \par
\begin{figure}[ht]
   \centering
   \begin{tabular}{cc}
   \includegraphics[width=0.47\textwidth]{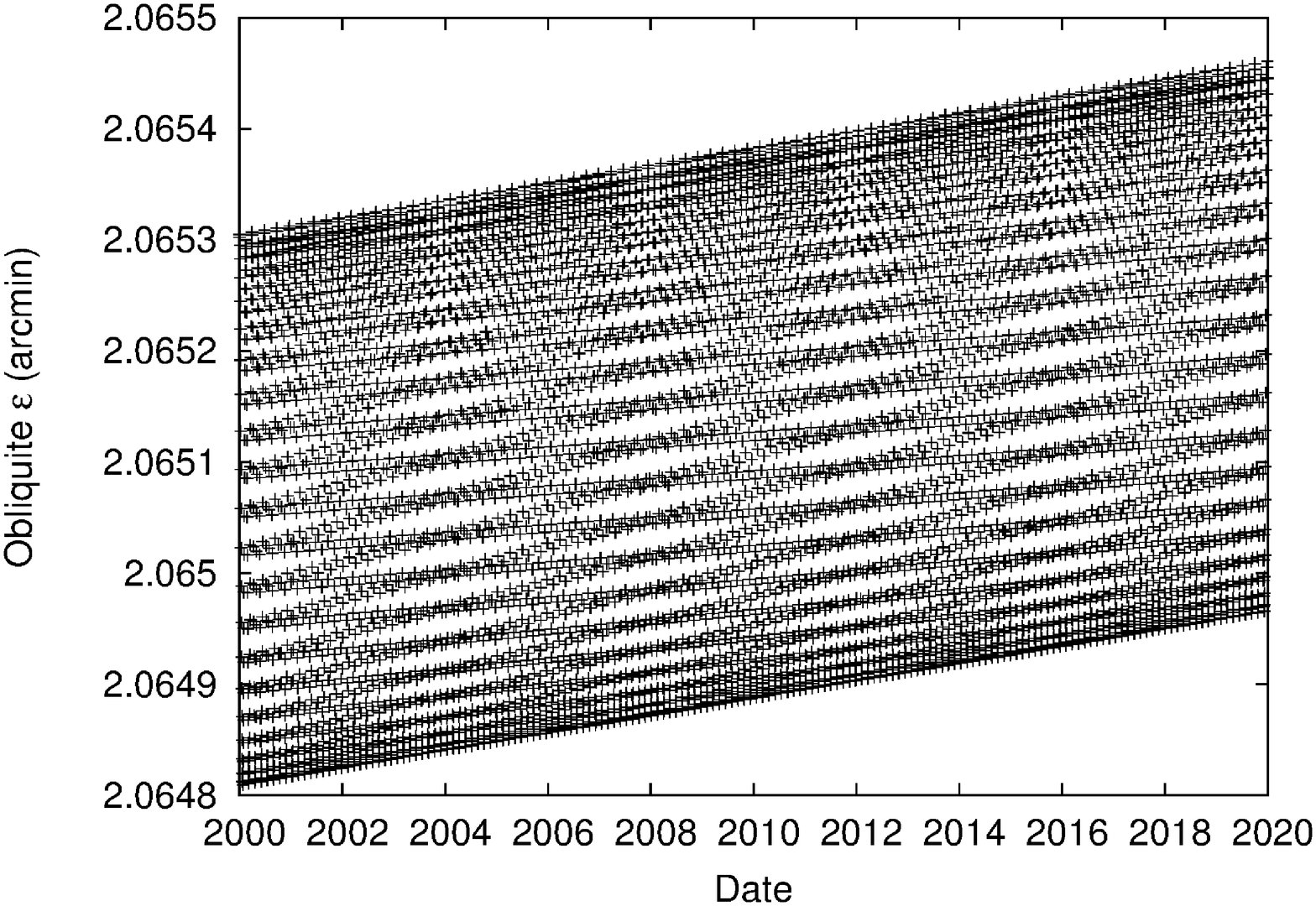} & \includegraphics[width=0.47\textwidth]{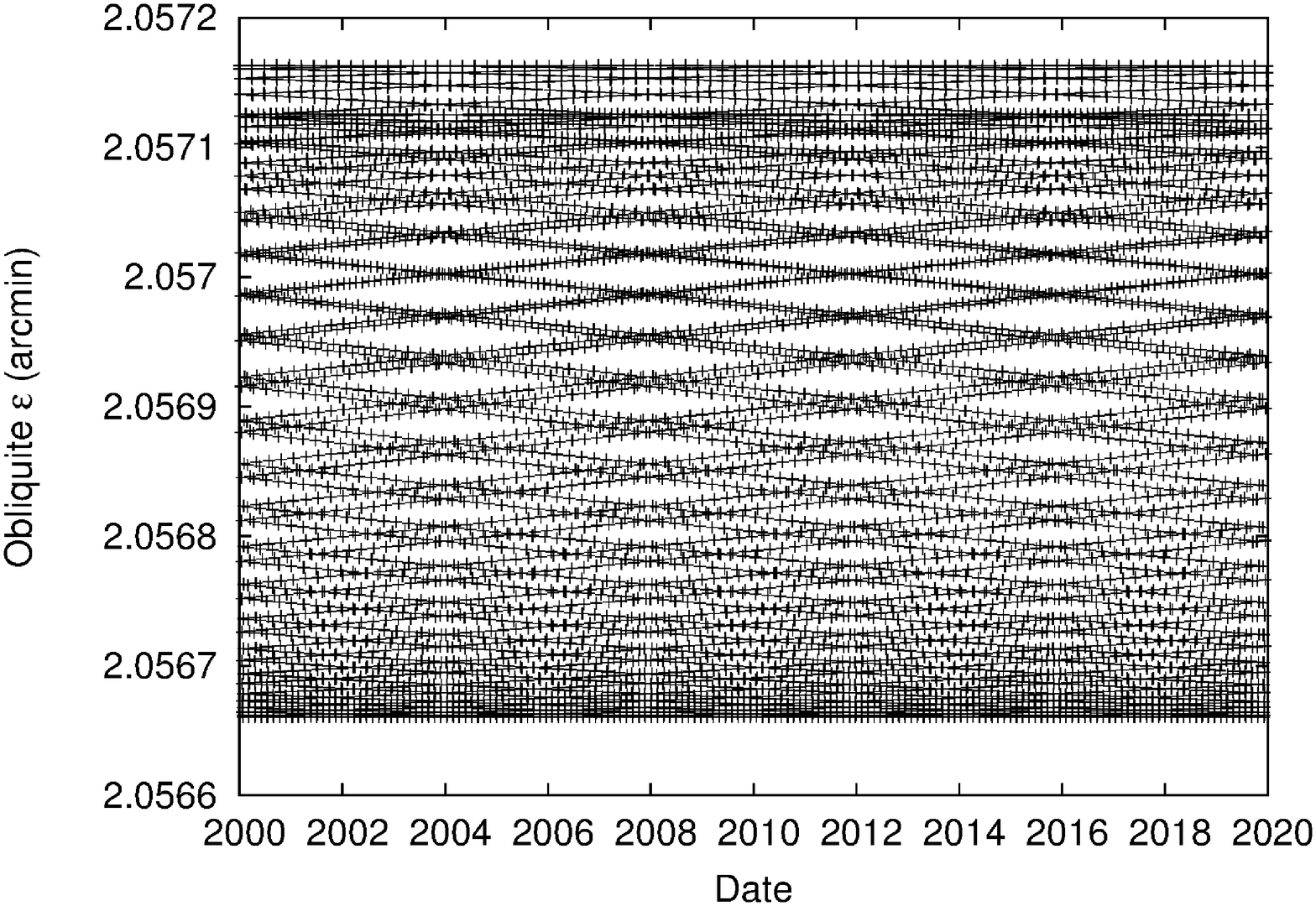}
   \end{tabular}
   \caption[Influence des mar\'ees sur l'obliquit\'e]{Variations de l'obliquit\'e de Mercure dues aux mar\'ees, avec la formule num\'erique \`a gauche et la formule 
   analytique \`a droite. La p\'eriode des variations est la p\'eriode orbitale de Mercure, soit 88 jours.\label{fig:tides}}
  \end{figure}

 Nous disposons de peu d'informations concernant $k_2$. Il n'a jamais \'et\'e mesur\'e, nous devons donc nous contenter de mod\`eles th\'eoriques. \citet{sswc2001} 
  l'estiment entre $0.3$ et $0.45$ si Mercure n'a pas de graine rigide, et entre $0.1$ et $0.4$ si elle en a une. \citet{rvv2009} l'estiment entre $0.2$ et $0.8$ en faisant
  diff\'erentes hypoth\`eses sur la composition de la plan\`ete. La plus r\'ecente estimation, qui utilise des donn\'ees de MESSENGER, est entre $0.45$ et $0.52$ \citep{pmhms2014}.
  Nous avons consid\'er\'e $k_2=0.5$ et ajout\'e les composantes p\'eriodiques des \'el\'ements du champ de gravit\'e dans nos formules, analytique (\ref{eqF3}) et num\'erique 
  (\ref{eq:KmI} \& \ref{eq:sig3}). Le r\'esultat est donn\'e Fig.\ref{fig:tides}.
  
  \par L'\'epaisseur des courbes, de l'ordre de 30 millisecondes d'arc, vient des oscillations \`a courte p\'eriode, dues aux mar\'ees. On peut remarquer, dans le mod\`ele
  num\'erique, une pente qui n'est pas pr\'esente dans les mod\`eles analytiques. Ceci est la signature des termes \`a tr\`es longue p\'eriode introduits dans les 
  expressions trigonom\'etriques extrapol\'ees des \'el\'ements orbitaux (\'Eq.\ref{eq:knum} \`a \ref{eq:pnum}), et qui ne sont pas pr\'esents dans la formule analytique.
  
  \par La Tab.\ref{tab:infoblik} compare diff\'erentes influences sur l'obliquit\'e. La premi\`ere ligne, les librations libres, ne vient pas de la physique de Mercure
  mais d'une erreur sur la mod\'elisation de l'obliquit\'e. Cette erreur est \`a comparer avec l'incertitude de 5 secondes d'arc sur la mesure de l'obliquit\'e \citep{mpshgjygpc2012}.
  Compar\'e aux autres effets, $J_3$ semble important, donc son influence doit \^etre clarifi\'ee. Le mouvement polaire et les courtes p\'eriodes viennent de 2 \'etudes dont je
  parle au Chap.\ref{chap:merclibra}.

  \begin{table}[ht]
   \centering
   \caption{Les diff\'erents effets agissant sur l'obliquit\'e de Mercure.\label{tab:infoblik}}
   \begin{tabular}{ll}
   \hline
    Effet                             & Influence sur l'obliquit\'e \\
    \hline
    Librations libres                 & $<750$ mas \\
    $C_{30}$                          & $\approx250$ mas \\
    Mouvement polaire                 & $\approx80$ mas \citep{ndl2010} \\
    Mar\'ees                          & $\approx30$ mas \\
    Librations \`a courtes p\'eriodes & $<20$ mas \citep{dnrl2009} \\
    D\'erive s\'eculaire              & $\approx10$ mas sur 20 ans \\
    $C_{40}$                            & non d\'etect\'ee \\
    \hline
   \end{tabular}
  \end{table}

  \subsection{Le moment d'inertie polaire $C$}
  
  \par Le but de la mod\'elisation de l'obliquit\'e est d'inverser la mesure de l'obliquit\'e $\epsilon=(2.04\pm0.08)$ minutes d'arc. Nous disposons de 2 formules, analytique et 
  num\'erique, auxquelles nous pouvons ajouter l'effet du coefficient $J_3$. Dans la formule analytique, nous utilisons, comme \citet{ym2006}, $I=8.6^{\circ}$ et une p\'eriode du
  n{\oe}ud ascendant n\'egative de 328\,000 ans. De plus, nous consid\'erons une p\'eriode positive de pr\'ecession de l'argument du p\'ericentre de 128\,000 ans. Et ceci nous donne
  
  \begin{itemize}
  
   \item \citet{mpshgjygpc2012} : $C/(M_{\mercury}R_{\mercury}^2) = 0.346 \pm 0.014$,
   
   \item Formule analytique (\'Eq.\ref{eqF3}) : $C/(M_{\mercury}R_{\mercury}^2) = 0.34712 \pm 0.01361$,
   
   \item Formule num\'erique (\'Eq.\ref{eq:KmI} \& \ref{eq:sig3}) : $C/(M_{\mercury}R_{\mercury}^2) = 0.34576 \pm 0.01349$,
   
   \item Formule analytique (\'Eq.\ref{eqF3}) + $J_3$ (\'Eq.\ref{eq:fitJ3}) : $C/(M_{\mercury}R_{\mercury}^2) = 0.34640 \pm 0.01361$,
   
   \item Formule num\'erique (\'Eq.\ref{eq:KmI} \& \ref{eq:sig3}) + $J_3$ (\'Eq.\ref{eq:fitJ3}) : $C/(M_{\mercury}R_{\mercury}^2) = 0.34506 \pm 0.01348$.
  
  \end{itemize}

  \par Cela n'a physiquement pas de sens de donner autant de chiffres significatifs. La raison est que les diff\'erences entre les r\'esultats sont faibles, il faut donc donner 
  plus de chiffres que la mesure physique le sugg\`ere pour pouvoir comparer les r\'esultats. Nous constatons que tous les r\'esultats convergent. Avec la formule de Peale 
  (\'Eq.\ref{eq:oblikpeale}), \citet{mpshgjygpc2012} trouvent pour valeur la plus probable de $C/(M_{\mercury}R_{\mercury}^2)$ $0.346$. Cela pourrait \^etre en fait $0.345$
  ou $0.347$.
  
  \section{Conclusion}
  
  \par Le probl\`eme de l'obliquit\'e de Mercure est fondamental dans le cadre de la mission MESSENGER et de la pr\'eparation de BepiColombo. C'est en effet la seule mesure 
  qui donnera son moment d'inertie polaire, porteur d'information sur la structure interne. Nous verrons dans le chapitre qui arrive que cette information est utile pour 
  conna\^itre la taille de la partie fluide du noyau.
  
  \par Avant ces \'etudes \citep{nd2012,nl2013}, seule la formule de Peale existait dans la litt\'erature pour inverser l'obliquit\'e. En renon\c{c}ant \`a certaines approximations
  qui y sont faites, notamment en ne consid\'erant pas la pr\'ecession du n{\oe}ud ascendant comme uniforme, en recourant \`a l'outil num\'erique, et en renon\c{c}ant \`a l'utilisation
  du Plan de Laplace, nous avons \'elabor\'e une m\'ethode compl\`etement ind\'ependante de celle de Peale. Nos r\'esultats convergent, ce qui permet de dire que l'erreur due \`a la 
  m\'ethode est tr\`es faible.
  
  \par Mercure est consid\'er\'ee comme rigide pour l'estimation de son obliquit\'e. En effet, sur des \'echelles de temps aussi longues, la viscosit\'e est suppos\'ee solidariser
  le manteau du reste de l'int\'erieur. Une perspective pourrait \^etre d'utiliser un mod\`ele d'int\'erieur plus complet, en consid\'erant un noyau interne (une graine), et une 
  viscosit\'e chiffr\'ee dans le noyau externe, fluide. Ce sujet a \'et\'e abord\'e r\'ecemment dans \citep{pmhs2014}, o\`u les auteurs ont pris en compte une graine dont 
  l'orientation est diff\'erente du noyau, et une viscosit\'e intervenant dans le couplage de pression aux interfaces rigide-fluide. Par exemple avec une graine dont l'axe d'orientation
  est d\'esax\'e de $3.55$ minutes d'arc par rapport au manteau, les auteurs trouvent que l'axe d'orientation du manteau est d\'eplac\'e de 55 millisecondes d'arc. Cet effet est donc 5
  fois plus faible que celui du $J_3$.
  
  \chapter[Les librations en longitude]{\'Etude \`a court terme : les librations en longitude\label{chap:merclibra}}
  
  \section{Introduction}
  
  \par Les premi\`eres mesures du champ magn\'etique de Mercure \citep{nblws1974,nblw1975,nblw1976} par MESSENGER ont sugg\'er\'e la pr\'esence d'une couche globale 
  fluide. La pr\'esence de cette couche se devait d'\^etre confirm\'ee par des observations compl\`etement ind\'ependantes, et qui si possible pourraient \^etre 
  invers\'ees de fa\c{c}on assez directe. Ceci permettrait \'egalement d'en savoir plus sur cette couche fluide, notamment d'estimer sa taille.
  
  \par \citet{p1972} a propos\'e une exp\'erience se basant sur les mesures de l'obliquit\'e (cf. Chap.\ref{chap:mercobliq}) et des librations en longitude \`a la p\'eriode 
  orbitale (88 jours)\footnote{les librations diurnes}. Ceci a motiv\'e un certain nombre d'\'etudes th\'eoriques \'evaluant ces librations en fonction des param\`etres d'int\'erieur. Les amplitudes des 
  librations dues aux perturbations plan\'etaires, dont les p\'eriodes sont de plusieurs ann\'ees, sont \'egalement \'etudi\'ees, car elles sont susceptibles d'\^etre d\'etectables.
  
  \section{L'exp\'erience de Peale (1976)}
  
  \par L'id\'ee principale de l'exp\'erience de Peale est que les librations en longitude et l'obliquit\'e sont 2 r\'eponses de la rotation de Mercure \`a des excitations 
  sur des \'echelles de temps diff\'erentes, la signature de l'interface noyau-manteau n'\'etant pr\'esente que dans les librations en longitude.
  
  \subsection{Th\'eorie}
  
  \par Nous avons d\'ej\`a vu que pour la mesure de l'obliquit\'e de Mercure, nous pouvions, avec une bonne approximation, utiliser la formule suivante, due \`a \citet{p1969} :
  
    \begin{equation}
    \label{eq:oblikpealebis}
    \epsilon = -\frac{C\dot{\ascnode}\sin\iota}{C\dot{\ascnode}\cos\iota+2nM_{\mercury}R_{\mercury}^2\left(\frac{7}{2}e-\frac{123}{16}e^3\right)C_{22}-n M_{\mercury} R_{\mercury}^2 (1-e^2)^{-3/2} C_{20}}.
  \end{equation}
  
  \par Les librations en longitude peuvent \^etre calcul\'ees \`a partir du Hamiltonien en longitude (\ref{eq:hamilplan2}) d\'evelopp\'e \`a l'ordre 2 en excentricit\'e, mais dans lequel
  l'argument r\'esonnant est $2\sigma=2p-3\lambda$. On en d\'eduit l'\'equation suivante :
  
  \begin{equation}
    \label{eq:ddotmercure}
    \ddot{\sigma}+\frac{21}{2}n^2e\frac{B-A}{C}\sigma+\frac{3}{2}n^2\frac{B-A}{C}(1-11e^2)\sin\lambda-\frac{3}{4}n^2e\frac{B-A}{C}\sin 2\lambda=0,
  \end{equation}
ce qui donne
  
  \begin{eqnarray}
    \sigma(t) & = & \mathcal{A}\cos(\omega_0t+\alpha)-\frac{3}{2}\frac{n^2}{n^2-\omega_0^2}\frac{B-A}{C}(1-11e^2)\sin\lambda+\frac{3}{4}\frac{n^2}{4n^2-\omega_0^2}\frac{B-A}{C}e\sin 2\lambda \label{eq:sigmamercure} \\
              & \approx & \mathcal{A}\cos(\omega_0t+\alpha)-\frac{3}{2}\frac{B-A}{C}(1-11e^2)\sin\lambda+\frac{3}{16}\frac{B-A}{C}e\sin 2\lambda, \label{eq:sigmamercure2}
  \end{eqnarray}
  avec 
  
  \begin{equation}
   \label{eq:omega0mercure}
   \omega_0^2 = \frac{21}{2}n^2e\frac{B-A}{C},
  \end{equation}
$\mathcal{A}$ et $\alpha$ \'etant des constantes d'int\'egration. L'approximation g\'en\'eralement faite $\omega_0^2 \ll n^2$ est bas\'ee sur la consid\'eration que le coefficient de 
triaxialit\'e $(B-A)/C$ est tr\`es petit, non seulement par rapport \`a 1, mais aussi par rapport \`a l'excentricit\'e $e$ qui est l'autre petit param\`etre. Cette approximation n'est 
en g\'en\'eral pas faite dans le cas de la rotation synchrone, par exemple \'Epim\'eth\'ee est si triaxial que son mouvement en longitude se rapproche d'un comportement r\'esonnant 
(Fig. \ref{fig:errtwojanepim}). Dans le cas de Mercure, on a $e\approx0.2056$ et $(B-A)/C_m = (2.18\pm0.04)\times10^{-4}$ o\`u $C_m$ est le moment d'inertie polaire du manteau 
(son utilisation sera justifi\'ee dans les lignes suivantes), donc notre approximation est valide. Mercure est le seul cas connu de r\'esonance spin-orbite 3:2, toutefois cette 
configuration r\'esonnante est probablement assez courante dans les syst\`emes plan\'etaires \citep{m2012}. N\'eanmoins, la triaxialit\'e d'une plan\`ete vient en partie du couple de 
mar\'ee s'appliquant. Dans le cas de la r\'esonance 3:2, il est bien plus faible que dans celui de la r\'esonance synchrone. On a notamment, \`a l'\'equilibre hydrostatique, 
$C_{22}/J_2\approx3/10$ pour la rotation synchrone, et $C_{22}/J_2\approx 7e/10$ pour la r\'esonance 3:2 \citep{mn2009}. Il est donc raisonnable de penser que le param\`etre de 
triaxialit\'e $(B-A)/C$ est g\'en\'eralement petit pour la r\'esonance 3:2. \citet{m2009} donne une expression plus g\'en\'erale en excentricit\'e pour l'amplitude des librations diurnes 
et de ses harmoniques :

\begin{equation}
  \label{eq:librmargot}
  \gamma = \frac{3}{2}\frac{B-A}{C}\sum_k f_k(e)\sin(k\mathcal{M})
\end{equation}
o\`u $\mathcal{M}$ est l'anomalie moyenne, et

\begin{eqnarray}
  f_1(e) & = & 1-11e^2+\frac{959}{48}e^4-\frac{3641}{288}e^6+\mathcal{O}(e^8), \label{eq:f1margot} \\
  f_2(e) & = & -\frac{e}{8}-\frac{421}{96}e^3+\frac{32515}{3072}e^5+\mathcal{O}(e^7), \label{eq:f2margot} \\
  f_3(e) & = & -\frac{533}{144}e^4+\frac{4609}{480}e^6+\mathcal{O}(e^8), \label{eq:f3margot} \\
  f_4(e) & = & \frac{e^3}{768}-\frac{57073}{15360}e^5+\mathcal{O}(e^7), \label{eq:f4margot} \\
  f_5(e) & = & \frac{e^4}{600}-\frac{18337}{4500}e^6+\mathcal{O}(e^8). \label{eq:f5margot}
\end{eqnarray}

\par Une ambigu\"it\'e appara\^it entre les formules (\ref{eq:sigmamercure2}) et (\ref{eq:librmargot}) \`a propos de l'angle \`a utiliser, la longitude moyenne $\lambda$ d'un c\^ot\'e, 
l'anomalie moyenne $\mathcal{M}$ de l'autre. Il est en fait plus exact de consid\'erer l'anomalie moyenne. La formule (\ref{eq:sigmamercure2}) consid\`ere que la vitesse de pr\'ecession
du p\'ericentre est n\'egligeable, ce qui est vrai dans le cas de Mercure\footnote{la p\'eriode de l'anomalie moyenne est de 88 jours, celle du p\'ericentre se chiffre en centaines
de milliers d'ann\'ees}. Cependant, cette formule est dangereuse \`a g\'en\'eraliser car elle incite \`a oublier les variations du p\'ericentre, qui ont un impact si on consid\`ere les
perturbations plan\'etaires. Ceci a induit une erreur dans \citep{pym2007}.

\par Dans le cadre de l'exp\'erience, on s'int\'eresse \`a l'amplitude la plus facilement d\'etectable, soit celle des librations diurnes, $\phi\approx3/2\times (B-A)/C$. L'id\'ee forte de cette 
exp\'erience est que seul le manteau suit les librations diurnes, il faudrait donc utiliser ses moments d'inertie $A_m$, $B_m$ et $C_m$ en lieu et place des moments d'inertie de 
Mercure $A$, $B$ et $C$.

\par Une autre approximation sugg\'er\'ee est de consid\'erer que le noyau, de moments d'inertie $A_c$, $B_c$ et $C_c$, est sph\'erique. Ses moments d'inertie sont donc \'egaux.
Dans ce cas on a $B-A = (B_m+B_c)-(A_m+A_c) = B_m-A_m$. On peut donc \'ecrire

\begin{eqnarray}
  \phi & \approx & \frac{3}{2}\frac{B-A}{C_m}\left(1-11e^2+\frac{959}{48}e^4\right) \label{eq:librationsmercure} \\
       & \approx & 6C_{22}\frac{M_{\mercury}R_{\mercury}^2}{C}\frac{C}{C_m}\left(1-11e^2+\frac{959}{48}e^4\right). \label{eq:librationsmercure2}
\end{eqnarray}

\par Le fonctionnement de l'exp\'erience de Peale est donc le suivant : les \'eph\'em\'erides de Mercure donnent son mouvement orbital, donc son excentricit\'e, et 
les sondes les coefficients du champ de gravit\'e $J_2$ et $C_{22}$. Une fois l'obliquit\'e mesur\'ee, le moment d'inertie polaire $C$ sera connu 
(\'Eq.\ref{eq:oblikpealebis}). Ensuite la mesure de l'amplitude de la libration diurne $\phi$ donnera directement le manteau d'inertie du manteau $C_m$, donc 
celui du manteau $C_c=C-C_m$, ce qui permettra d'en estimer la taille.

  \subsection{Observations radar}

  \par Le moyen le plus efficace pour observer la rotation de Mercure depuis la Terre est le radar. Ce sont des observations radar qui ont permis de mesurer la vitesse de 
  rotation de Mercure \citep{pd1965}. Les signaux radar permettent de mesurer la vitesse de rotation, et l'instant o\`u ils sont d\'etect\'es, correspondant \`a une orientation
  spatiale de l'antenne radar, contraint l'orientation de l'axe de rotation de Mercure.
  
  \par
\begin{figure}[ht]
   \centering
   \begin{tabular}{m{.47\textwidth} m{.47\textwidth}}
   \includegraphics[width=0.47\textwidth]{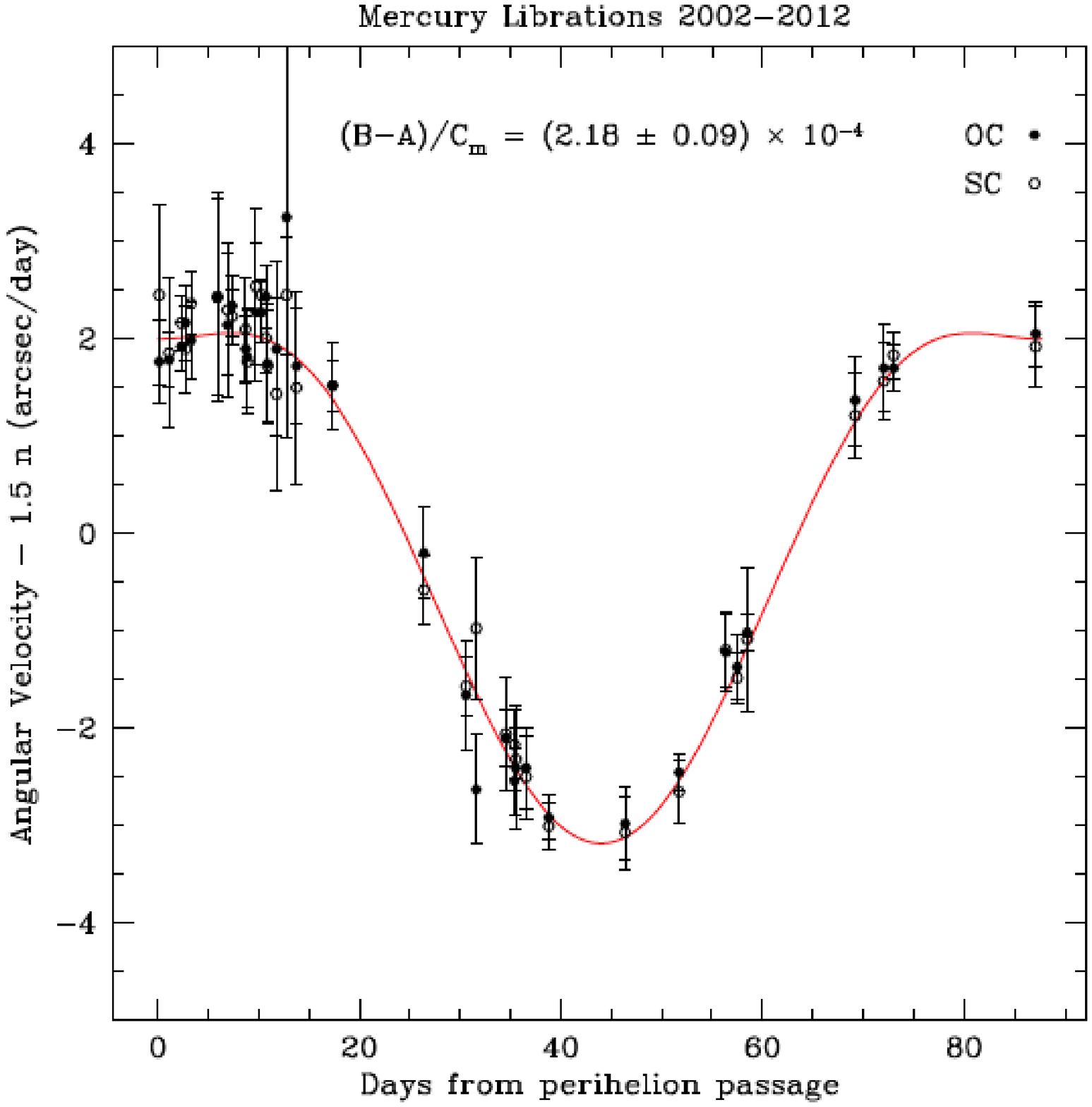} & \includegraphics[width=0.47\textwidth]{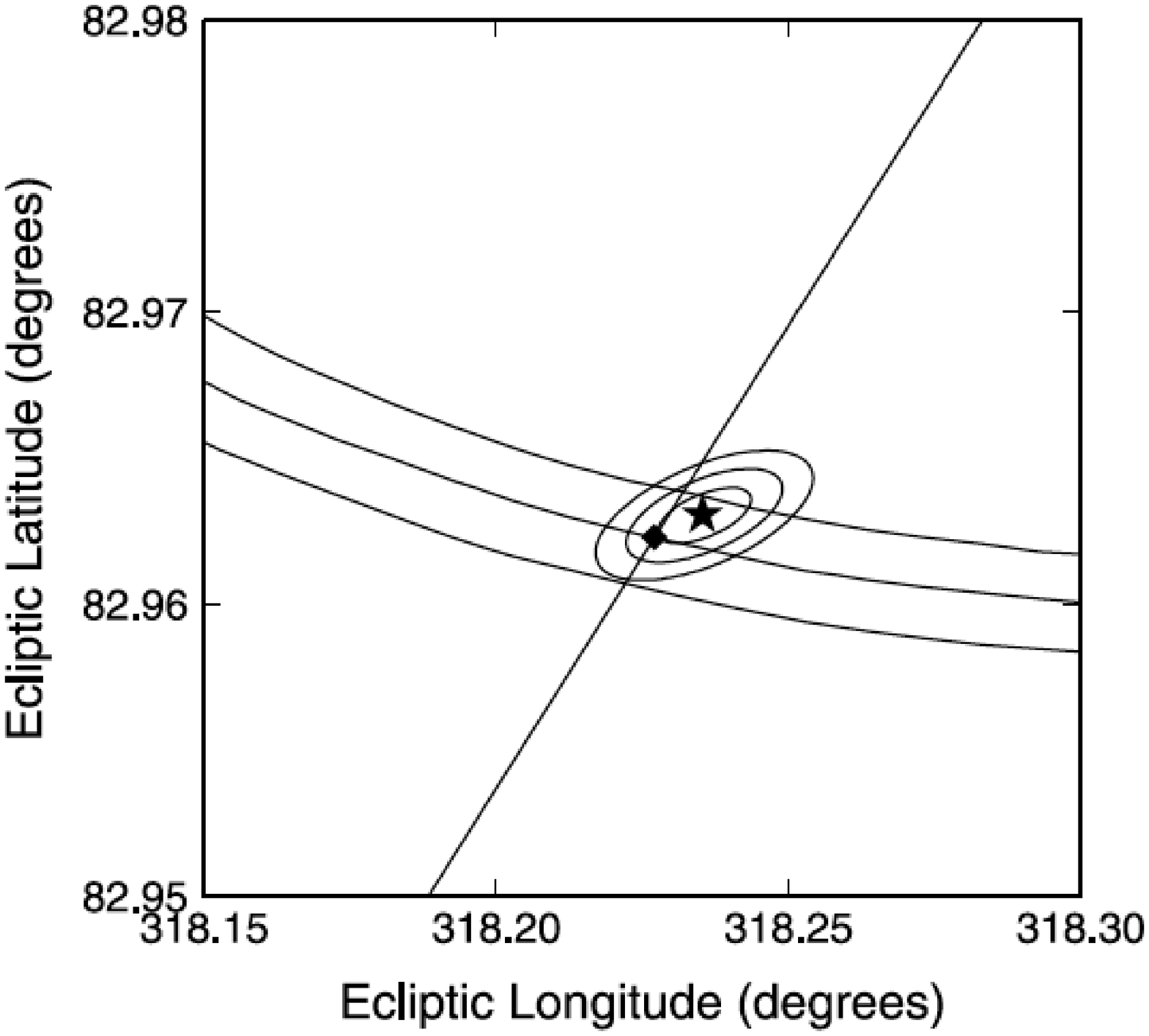}
   \end{tabular}
   \caption[Libration et obliquit\'e de Mercure mesur\'ees]{Libration (\`a gauche) et obliquit\'e (\`a droite) de Mercure mesur\'ees. Les mesures de librations ont \'et\'e
   regroup\'ees sur une p\'eriode orbitale de Mercure. Pour l'obliquit\'e, l'\'etoile indique la mesure, et les ellipses autour sont les incertitudes \`a $\sigma$, $2\sigma$
   et $3\sigma$. Le carreau est la mesure de \citep{mpjsh2007}, et la ligne oblique est la position de l'\'Etat de Cassini 1 th\'eorique. Ces figures sont reproduites de 
   \citep{mpshgjygpc2012}.\label{fig:mesuremargot}}
  \end{figure}

 La libration et l'obliquit\'e de Mercure ont \'et\'e pour la premi\`ere fois mesur\'ees par \citet{mpjsh2007} par des observations radar terrestres, essentiellement
  faites \`a Goldstone, Californie, d'autres ayant \'et\'e r\'ealis\'ees au Green Bank Telescope en Virginie Occidentale. Ces mesures ont \'et\'e mises \`a jour en 2012 
  (Fig.\ref{fig:mesuremargot} \& Tab.\ref{tab:mesuremargot}).

  \par La Fig.\ref{fig:mesuremargot} illustre les mesures radar. \`A gauche, on voit qu'en repliant toutes les mesures sur une m\^eme p\'eriode orbitale, on a effectivement
  un signal. \`A droite on voit que la mesure (l'\'etoile) est proche de l'\'Etat de Cassini 1. Les derni\`eres mesures (Tab.\ref{tab:mesuremargot}) sugg\`erent 
  $C/(M_{\mercury}R_{\mercury}^2) = 0.346 \pm 0.014$ et $C_m/C = 0.431 \pm 0.025$.
  
  \begin{table}[ht]
   \centering
   \caption{Les mesures radar de libration et d'obliquit\'e de Mercure.\label{tab:mesuremargot}}
   \begin{tabular}{l|cc}
   \hline
    & \citet{mpjsh2007} & \citet{mpshgjygpc2012} \\
   \hline
   Libration   &   $35.8\pm2$ arcsec &  $38.5\pm1.6$ arcsec \\
   Obliquit\'e & $2.11\pm0.1$ arcmin & $2.04\pm0.08$ arcmin \\
   \hline
   \end{tabular}
  \end{table}

  \section{Mod\'elisation th\'eorique}
  
  \par Je pr\'esente ici diff\'erentes \'etudes sur les librations en longitude de Mercure. Il s'agit aussi bien de librations diurnes et de ses harmoniques, que de perturbations
  plan\'etaires. Dans ce dernier cas, les librations ont une p\'eriode de plusieurs ann\'ees.
  
  \subsection{Si le noyau est sph\'erique}
  
  \par Les premi\`eres \'etudes ont consid\'er\'e que le noyau de Mercure \'etait sph\'erique. Dans ce cas, le couple de pression appliqu\'e par le fluide \`a l'interface 
  noyau-manteau a une r\'esultante nulle. On peut donc consid\'erer Mercure comme une coquille vide. L'historique de l'\'etude th\'eorique des librations est le suivant :
  
  \begin{enumerate}
  
   \item \citet{pym2007} ont fait une premi\`ere \'etude en n\'egligeant les variations du p\'ericentre,
   
   \item \citet{dlr2008} ont fait une \'etude Hamiltonienne en longitude, dont le r\'esultat \'etait diff\'erent de \citep{pym2007} (Tab.\ref{tab:analongi}). Ces r\'esultats ont \'et\'e 
   obtenus de 2 mani\`eres : par m\'ethode de perturbation \citep{d1969}, et par int\'egration num\'erique. La m\'ethode de perturbation a \'et\'e rendue possible par la mise 
   \`a disposition par Jean-Louis Simon d'une \'eph\'em\'eride de Mercure sous forme de s\'eries de Poisson. L'int\'egration num\'erique a notamment l'avantage de n\'ecessiter 
   moins d'approximations que la m\'ethode de perturbation, qui peut \^etre vue comme semi-analytique ou semi-num\'erique dans le sens o\`u elle consiste \`a manipuler des 
   s\'eries trigonom\'etriques dont les coefficients sont num\'eriques. Dans le cas pr\'esent, l'\'etude num\'erique a fait appara\^itre la perturbation de Saturne, qui avait 
   \'et\'e initialement omise,
   
   \item \citet{pmy2009} ont revu leurs calculs, reconnu leur erreur, et confirm\'e les r\'esultats de \citep{dlr2008}. Ils ont en outre mis l'accent sur la possibilit\'e
   d'une r\'esonance secondaire entre la perturbation jovienne, de p\'eriode 11.86 ans, et la libration libre en longitude. La cons\'equence serait une amplitude importante 
   de la composante \`a 11.86 ans de la libration en longitude,
   
   \item \citet{dnrl2009} ont confirm\'e ces r\'esultats avec un mod\`ele \`a 2 dimensions, c'est-\`a-dire sans n\'egliger les couplages entre longitude et obliquit\'e.
   Il s'av\`ere qu'en fait ces couplages sont n\'egligeables, ainsi que les librations plan\'etaires en latitude. L\`a encore, les r\'esultats ont \'et\'e confirm\'es 
   num\'eriquement. Pour l'anecdote, nous avons rep\'er\'e, pour un noyau sph\'erique, la proximit\'e d'une r\'esonance en obliquit\'e avec la Grande In\'egalit\'e 
   Jupiter - Saturne\footnote{il s'agit de la proximit\'e d'une r\'esonance de moyen mouvement 5:2, qui a une influence tr\`es significative dans la dynamique orbitale 
   des plan\`etes du Syst\`eme Solaire \citep{l1785}}, de p\'eriode $\approx900$ ans, et avons \'etudi\'e cette r\'esonance dans \citep{dndl2010}. Mais l'int\'er\^et de 
   cette r\'esonance n'est probablement qu'acad\'emique; en effet la prise en compte d'un noyau est discutable pour les longues p\'eriodes,
   
   \item \citet{ymp2010} ont donn\'e une expression analytique des amplitudes des perturbations plan\'etaires.
   
  \end{enumerate}

  \begin{table}[ht]
   \centering
   \caption[Librations en longitude en Mercure]{D\'ecomposition quasi-p\'eriodique des librations en longitude de Mercure, d'apr\`es \citep{dnrl2009}.
   $l_o$ est le terme diurne, $l_v$ le terme de V\'enus, $l_e$ celui de la Terre, $l_j$ celui de Jupiter et $l_s$ celui de Saturne. Ce sont des librations physiques,
   obtenues apr\`es la soustraction d'une pente \`a l'angle de spin de Mercure.\label{tab:analongi}}
   \begin{tabular}{c|rrrrrrrr}
   N & $l_o$ & $l_v$ & $l_e$ & $l_j$ & $l_s$ & P\'eriode & Amplitude & Rapport \\
   \hline
    1 & - & -  & -  & 1  & - & $11.862$ a & $43.712$ as & $1.2193$ \\
    2 & 1 & -  & -  & -  & - & $87.970$ j & $35.849$ as & $1.0000$ \\
    3 & 2 & -  & -  & -  & - & $43.985$ j &  $3.754$ as & $0.1047$ \\
    4 & 2 & -5 & -  & -  & - &  $5.664$ a &  $3.597$ as & $0.1003$ \\
    5 & - & -  & -  & -  & 2 & $14.729$ a &  $1.568$ as & $0.0437$ \\
    6 & - & -  & -  & 2  & - &  $5.931$ a &  $1.379$ as & $0.0385$ \\
    7 & 1 & -  & -4 & -  & - &  $6.575$ a &  $0.578$ as & $0.0161$ \\
    8 & 3 & -  & -  & -  & - & $29.323$ j &  $0.386$ as & $0.0108$ \\
    9 & 1 & -  & -  & -2 & - & $91.692$ j &  $0.201$ as & $0.0056$ \\
   10 & 1 & -  & -  & 2  & - & $84.537$ j &  $0.191$ as & $0.0053$ \\
   11 & - & -  & -  & 2  & -5& $883.28$ a &  $0.103$ as & $0.0029$ \\
   12 & 2 & -  & -  & -1 & - & $44.436$ j &  $0.069$ as & $0.0019$ \\
   13 & 2 & -  & -  & 1  & - & $43.541$ j &  $0.067$ as & $0.0019$ \\
   14 & 1 & -  & -  & -1 & - & $89.793$ j &  $0.044$ as & $0.0012$ \\
   15 & 1 & -  & -  & 1  & - & $86.217$ j &  $0.043$ as & $0.0012$ \\
   16 & 2 & -  & -  & -2 & - & $44.897$ j &  $0.041$ as & $0.0011$ \\
   17 & 2 & -  & -  & 2  & - & $43.110$ j &  $0.040$ as & $0.0011$ \\
   \hline
\end{tabular}
\end{table}

\par La Tab.\ref{tab:analongi} donne les amplitudes des diff\'erentes composantes p\'eriodiques de la libration physique en longitude, pour le champ de gravit\'e de 
  Mariner 10 ($J_2 = 6\times 10^{-5}$, $C_{22} = 2\times 10^{-5}$), $C = 0.34M_{\mercury}R_{\mercury}^2$ et $C_m = 0.549C$, ce qui permettait d'obtenir la libration diurne
  mesur\'ee par \citep{mpjsh2007}. Il s'agit d'un r\'esultat th\'eorique, seul le terme 2 (le terme diurne) ayant \'et\'e mesur\'e. Le terme 1 (le for\c{c}age jovien) pourrait 
  avoir une amplitude plus importante, mais il \'evolue nettement moins vite, ce qui le rend beaucoup plus difficile \`a d\'etecter. De plus, il est forc\'e par la proximit\'e 
  de la r\'esonance avec la fr\'equence des librations libres $\omega_0$, ce qui rend son amplitude tr\`es sensible \`a la taille du noyau, alors que les autres termes 
  plan\'etaires peuvent \^etre consid\'er\'es comme constants. En effet, \citet{ymp2010} ont montr\'e, en ajoutant un for\c{c}age p\'eriodique \`a l'\'equation 
  (\ref{eq:ddotmercure}) d'amplitude $\lambda_i$ et de pulsation $\omega_i$, que l'amplitude de la r\'eponse en libration $\phi_i$ \'etait :
  
  \begin{equation}
   \label{eq:marie}
   \phi_i = \frac{\lambda_i\omega_i^2}{\omega_0^2-\omega_i^2}.
  \end{equation}
  Pour les perturbations plan\'etaires \`a l'exception de celle \`a $11.86$ ans, l'\'ecart entre la fr\'equence de la perturbation, $\omega_i$, qui est fixe,
  et la fr\'equence propre $\omega_0$ est suffisant important pour \^etre insensible \`a une \'eventuelle mise \`a jour du param\`etre de triaxialit\'e $(B-A)/C_m$.

  \subsection{Si le noyau est ellipso\"idal}
  
  \par Nous avons appliqu\'e le mod\`ele de Poincar\'e-Hough (Chap.\ref{chap:poincarehough}) \`a Mercure \citep{ndl2010}. Chronologiquement cette application est ant\'erieure 
  \`a celle d'Io \citep{n2013}. J'ai appliqu\'e ce mod\`ele num\'eriquement pendant que Julien Dufey le r\'esolvait par une m\'ethode de perturbation, ce qui s'est r\'ev\'el\'e 
  particuli\`erement ardu du fait de la proximit\'e d'une fr\'equence propre avec celle de rotation de Mercure.
  
  \par Physiquement, on consid\`ere que Mercure est constitu\'ee d'un manteau rigide et d'une cavit\'e triaxiale remplie d'un fluide non visqueux, qui a un comportement
  non turbulent. Le couplage noyau-manteau se traduit par un couplage de pression \`a l'interface fluide-rigide. Math\'ematiquement, l'\'etude num\'erique a consist\'e \`a
  int\'egrer les \'equations d\'eriv\'ees du Hamiltonien (\ref{eq:HG3}), et l'\'etude par perturbation est partie du Hamiltonien (\ref{eq:HG4}), dont la partie libre 
  s'\'ecrit, apr\`es quelques transformations canoniques, sous la forme
  
  \begin{equation}
   \label{eq:quadramercureph}
   \mathcal{N} = \omega_uU+\omega_vV+\omega_wW+\omega_zZ,
  \end{equation}
  il s'agit donc d'un mod\`ele \`a 4 degr\'es de libert\'e dont les fr\'equences propres sont $\omega_u$, $\omega_v$, $\omega_w$ et $\omega_z$. En pratique, $\omega_u$ est
  proche de la libration libre en longitude, $\omega_v$ de celle en obliquit\'e, $\omega_w$ est associ\'ee au mouvement polaire du moment cin\'etique total de Mercure autour 
  de l'axe de figure, et $\omega_z$ est li\'ee \`a l'orientation du moment cin\'etique du fluide contenu dans la cavit\'e. C'est cette fr\'equence qui est proche de la 
  fr\'equence de rotation. Nous consid\'erons que le champ de gravit\'e, le moment d'inertie polaire et du noyau sont connus, nous faisons donc juste varier les param\`etres de 
  forme du noyau $\epsilon_3=(2C_c-A_c-B_c)/(2C_c)$ (aplatissement du noyau, \'Eq.\ref{eq:epsiloncm3}) et $\epsilon_4 = (B_c-A_c)/(2C_c)$ 
  (triaxialit\'e du noyau, \'Eq.\ref{eq:epsiloncm4}). Par exemple, pour $\epsilon_3=\epsilon_1$ et $\epsilon_4=\epsilon_2$, nous obtenons :
  
  \begin{eqnarray}
    T_u & = & 12.0568 \textrm{ ans}, \nonumber \\
    T_v & = & 1626.51 \textrm{ ans}, \nonumber \\
    T_w & = & 337.853 \textrm{ ans}, \nonumber \\
    T_z & = & 58.6189 \textrm{ jours}, \nonumber
  \end{eqnarray}
  aussi bien analytiquement que num\'eriquement. Dans la suite, nous faisons varier les param\`etres de forme du noyau $\epsilon_3$ et $\epsilon_4$.
  
  \par Les r\'esultats en longitude se trouvent confirm\'es, la consid\'eration de la triaxialit\'e du noyau n'affecte donc pas ce mouvement de fa\c{c}on significative,
  comme si le couplage de pression s'auto-compensait en longitude. Ce r\'esultat avait d\'ej\`a \'et\'e obtenu par \citep{rvdb2007}. Je vais donc plut\^ot 
  d\'etailler les effets sur les autres degr\'es de libert\'e.
  
  \par En fait, les r\'esultats que nous avons obtenus pour la r\'esonance 3:2 sont relativement proches de ceux obtenus pour la r\'esonance synchrone 
  (Chap.\ref{chap:poincarehough}). Le coefficient de triaxialit\'e du noyau, $\epsilon_4$, ne semble pas avoir d'influence significative. Par contre, le 
  coefficient d'aplatissement du noyau, $\epsilon_3$, semble critique pour le degr\'e de libert\'e li\'e \`a l'obliquit\'e de Mercure, et pour celui li\'e au
  champ de vitesses dans le fluide. Ceci est illustr\'e par la Tab.\ref{tab:periodnum}, donnant les valeurs des p\'eriodes des librations libres associ\'ees,
  $T_v$ et $T_z$.

  \begin{table}[ht]
\caption[Variation des p\'eriodes propres $T_v$ and $T_z$]{Variation des p\'eriodes propres d'obliquit\'e $T_v$ et d'orientation du fluide $T_z$ en fonction de l'aplatissement
	du noyau $\epsilon_3$, avec $\epsilon_4=0$. La derni\`ere colonne repr\'esente l'\'ecart \`a la fr\'equence de rotation de Mercure $\omega=3n/2$. Ces p\'eriodes ont \'et\'e d\'etermin\'ees 
num\'eriquement, par analyse en fr\'equences.\label{tab:periodnum}}
\centering
\begin{tabular}{l|rrr}
\hline
$\epsilon_3/\epsilon_1$ & $T_v$ (a) & $T_z$ (j) & $T_{z-\omega}$ (a) \\
\hline
$0.33$ & $3335.16$ & $58.628$ & $511.17$ \\
$0.7$  & $1966.31$ & $58.623$ & $409.08$ \\
$0.8$  & $1823.63$ & $58.622$ & $385.35$ \\
$0.9$  & $1718.34$ & $58.620$ & $363.50$ \\
$1.0$  & $1636.35$ & $58.619$ & $343.46$ \\
$1.1$  & $1570.86$ & $58.617$ & $325.10$ \\
$1.2$  & $1519.36$ & $58.616$ & $308.30$ \\
$1.5$  & $1408.10$ & $58.611$ & $266.01$ \\
$2$    & $1313.11$ & $58.602$ & $214.85$ \\
$2.5$  & $1250.26$ & $58.594$ & $179.64$ \\
$3$    & $1216.09$ & $58.585$ & $154.01$ \\
$3.5$  & $1198.68$ & $58.576$ & $134.72$ \\
$5$    & $1149.35$ & $58.550$ & $97.69$ \\
$10$   & $1107.62$ & $58.462$ & $50.83$ \\
\hline
\end{tabular}
\end{table}

\par
\begin{figure}[ht]
\centering
\includegraphics[width=.6\textwidth]{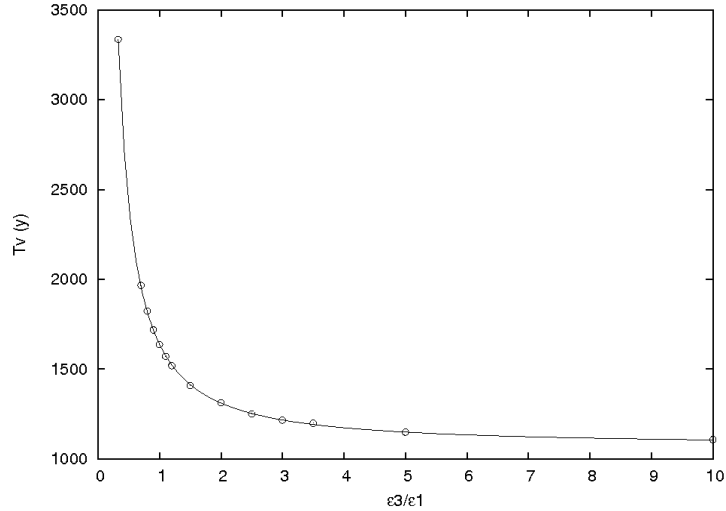}
\caption[La p\'eriode propre $T_v$]{La p\'eriode propre $T_v$, li\'ee \`a l'obliquit\'e, en fonction de l'aplatissement du noyau $\epsilon_3$.  Un ajustement par 
moindres carr\'es donne  $T_v(\epsilon_3/\epsilon_1)=A\times (\epsilon_3/\epsilon_1)^B+C$ avec $A=564.488$ ans, $B=-1.25224\approx -5/4$ et $C=1074.3$ ans, cette 
derni\`ere valeur \'etant tr\`es proche de la p\'eriode rigide. Les ronds sont des valeurs extraites des simulations num\'eriques par analyse en fr\'equence, 
alors que la ligne solide est l'ajustement.\label{fig:graphtv}}
\end{figure}

 La p\'eriode associ\'ee \`a l'obliquit\'e, $T_v$, peut devenir tr\`es grande lorsque le noyau est tr\`es peu aplati. Par contre, lorsque le noyau est tr\`es
aplati, la p\'eriode propre associ\'ee tend vers une p\'eriode tr\`es proche de celle correspondant \`a un Mercure rigide, soit $1066$ ans. Ce ph\'enom\`ene 
avait d\'ej\`a \'et\'e mentionn\'e par \citep{p1910} sous le nom de \emph{rigidit\'e gyrostatique}. La Fig.\ref{fig:graphtv} montre le comportement de cette p\'eriode en fonction 
de l'aplatissement du noyau $\epsilon_3$.

\par Un ajustement num\'erique, par moindres carr\'es, donne $T_v\approx A\times (\epsilon_3/\epsilon_1)^B+C$ avec $A=564.488\pm 4.146$, 
$B=-1.25224\pm 6.003\times 10^{-3}$ et $C=1074.3\pm 3.233$ ans. Ces chiffres rendent tentante l'id\'ee que $T_v$ suit une loi en $\epsilon_3^{-5/4}$ et que sa
valeur asymptotique tend vers la valeur rigide, mais l'absence de confirmation analytique\footnote{Avis aux amateurs.} incite \`a rester prudent quant \`a 
l'interpr\'etation de cette \'evolution.

\par
\begin{figure}[ht]
\centering
\begin{tabular}{cc}
\includegraphics[width=.47\textwidth]{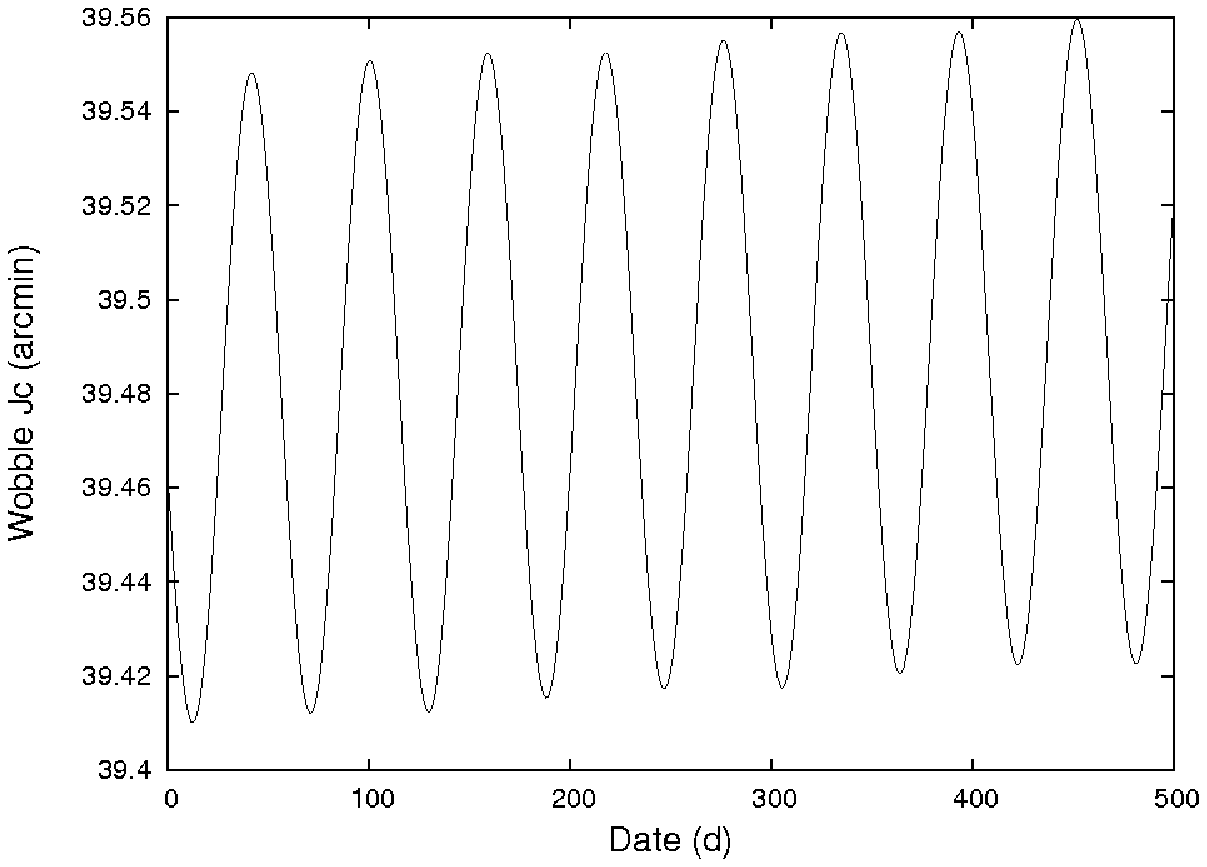} & \includegraphics[width=.47\textwidth]{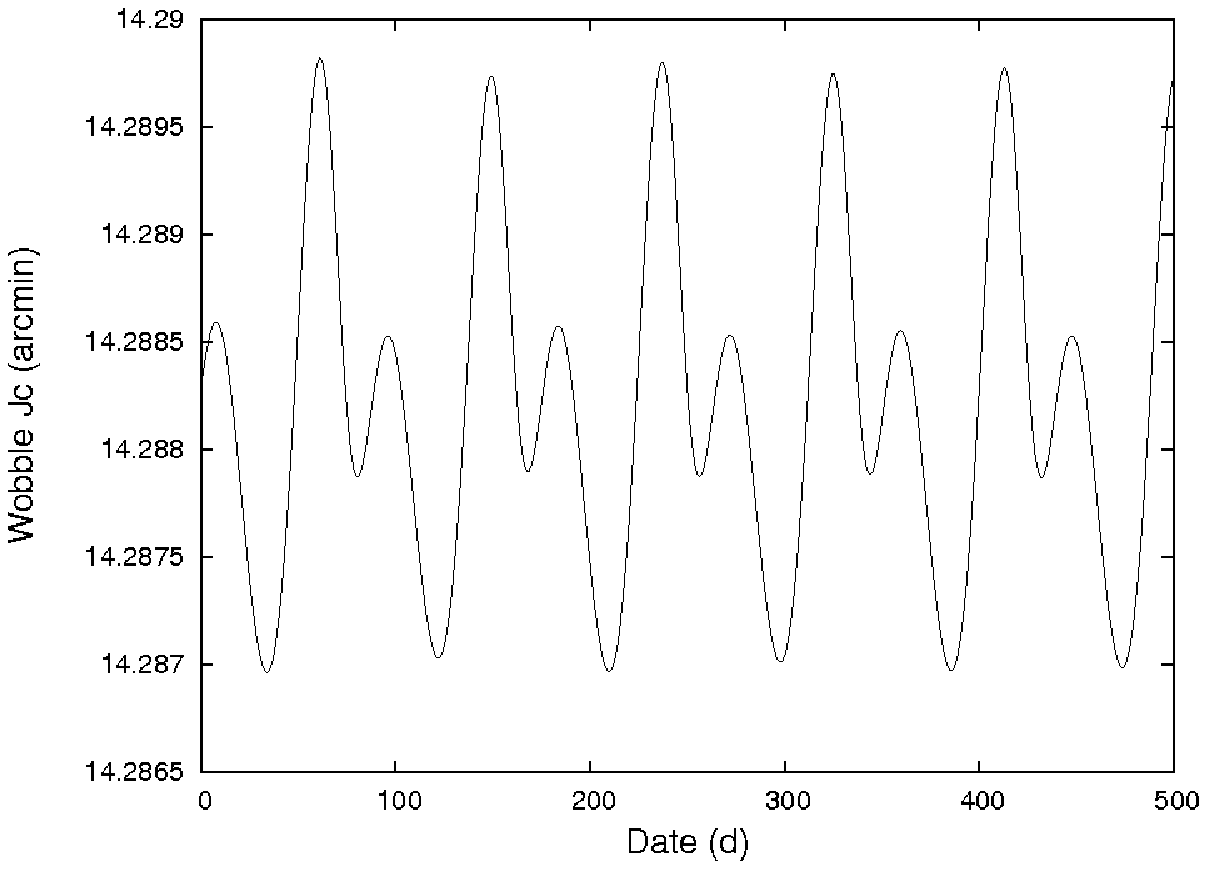} \\
$\epsilon_3=\epsilon_4=0$ & $\epsilon_3=\epsilon_1/10,\epsilon_4=0$ \\
\includegraphics[width=.47\textwidth]{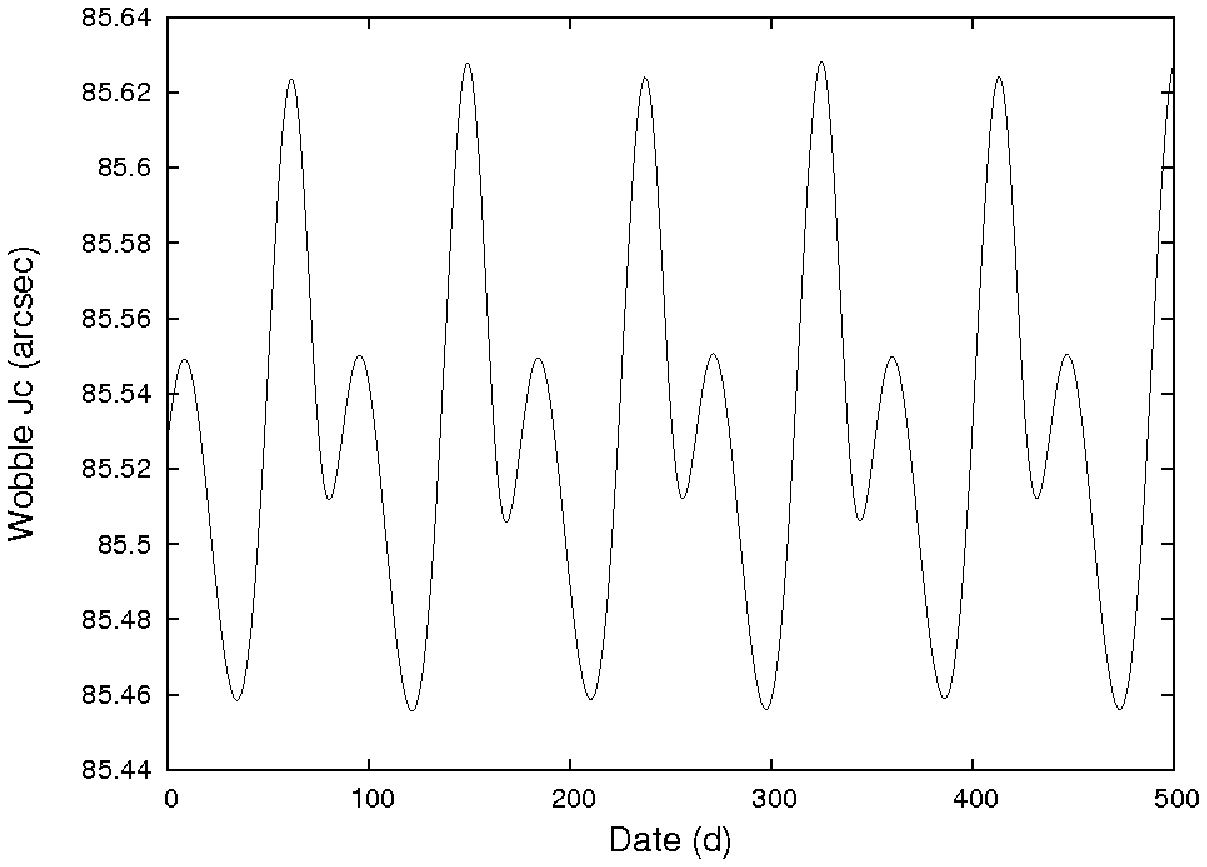} & \includegraphics[width=.47\textwidth]{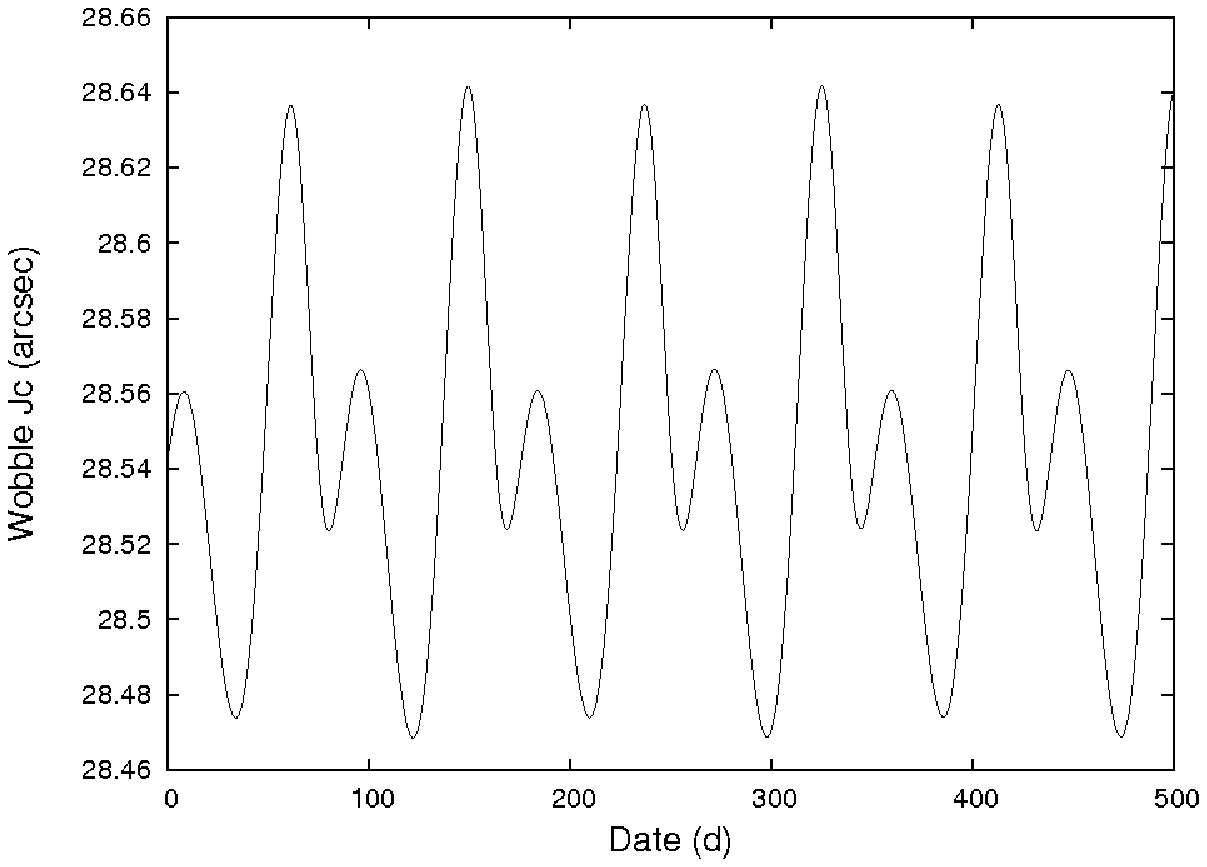} \\
$\epsilon_3/\epsilon_1=\epsilon_4/\epsilon_2=1$ & $\epsilon_3=3\epsilon_1,\epsilon_4=0$ \\
\end{tabular}
\caption[Orientation du moment cin\'etique du fluide dans le noyau]{Orientation du moment cin\'etique du fluide dans le noyau, pour diff\'erents param\`etres d'int\'erieur. 
Nous voyons que l'amplitude est d'autant plus grande que l'aplatissement $\epsilon_3$ est petit.\label{fig:polarcore}}
\end{figure}

 La p\'eriode de rotation de Mercure est 58.646 jours. On peut remarquer que la p\'eriode libre $T_z$ en est tr\`es proche (Tab.\ref{tab:periodnum}), ce qui 
laisse penser \`a la possibilit\'e d'un ph\'enom\`ene r\'esonnant d\'esaxant le champ de vitesses dans le fluide par rapport \`a l'axe de figure. On peut
voir que c'est bien le cas (Fig.\ref{fig:polarcore}).

\par
\begin{figure}[ht]
\centering
\begin{tabular}{cc}
\includegraphics[width=.47\textwidth]{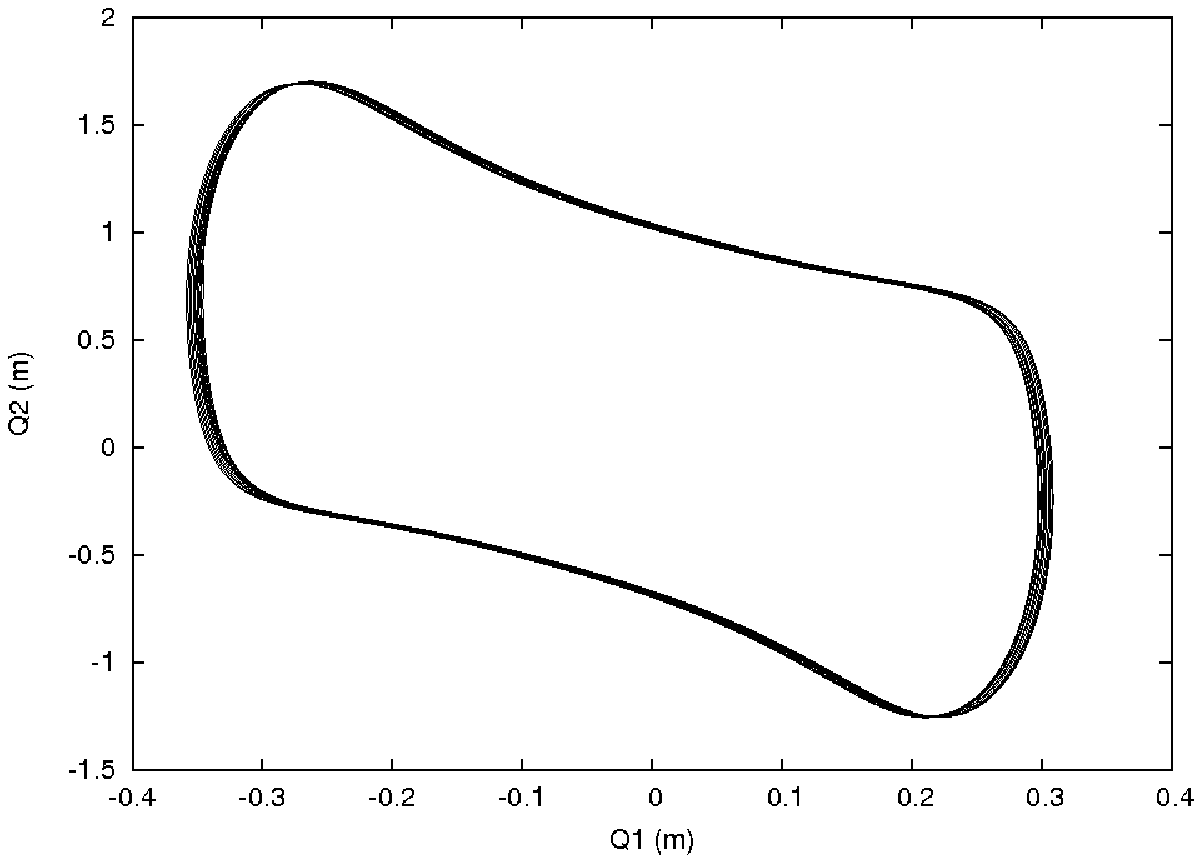} & \includegraphics[width=.47\textwidth]{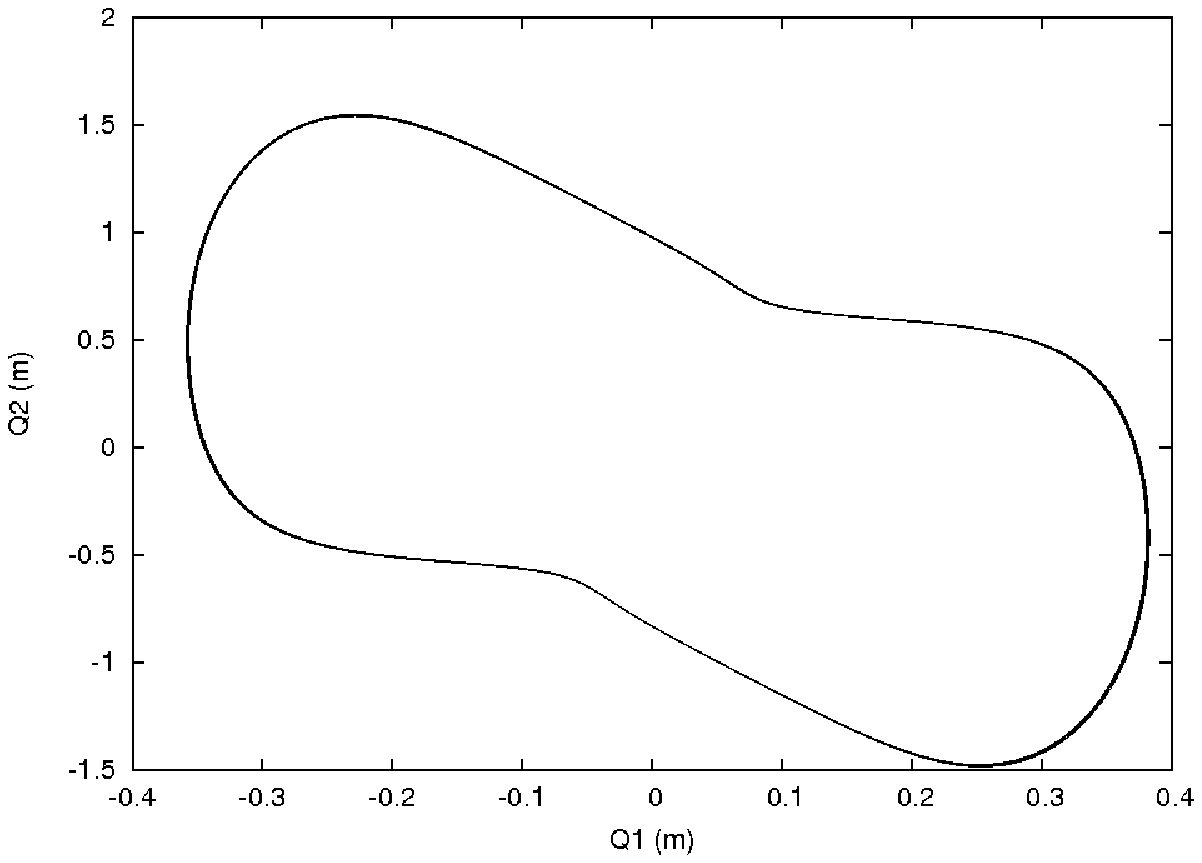} \\
$\epsilon_3/\epsilon_1=\epsilon_4/\epsilon_2=1$ & $\epsilon_3=3\epsilon_1,\epsilon_4=0$ \\
\end{tabular}
\caption[Mouvement polaire du manteau]{Mouvement polaire du manteau, trac\'e sur 5 ans \`a partir de J2000. $Q_1$ et $Q_2$ sont les amplitudes des mouvements
est-ouest et nord-sud, au niveau du p\^ole nord g\'eom\'etrique de Mercure. La principale p\'eriode de ce mouvement est $175.9$ jours, soit 2 p\'eriodes orbitales
ou 3 p\'eriodes de rotation.\label{fig:polarmantle}}
\end{figure}

 Cette figure montre une amplitude du d\'esaxage, $J_c$, d'autant plus grande que le noyau est faiblement aplati ($\epsilon_3$ petit). On n'observe par
contre rien de tr\`es int\'eressant sur le mouvement polaire de Mercure, dont l'amplitude est de l'ordre du m\`etre (Fig.\ref{fig:polarmantle}).

\par Il semblerait donc qu'un noyau de Mercure proche d'\^etre sph\'erique aurait des effets r\'esonnants dans le noyau, mais qui resteraient dans le noyau dans
le sens o\`u une \'etude de la rotation de sa surface n'en r\'ev\'elerait pas la signature. Ceci dit, le champ de vitesses dans le fluide doit affecter le champ
magn\'etique de la plan\`ete. Pour relier ces 2 objets d'\'etudes, il faudrait introduire dans le mod\`ele de rotation les mouvements convectifs dans le fluide. 

  \subsection{Avec un noyau interne rigide}
  
  \par Ce mod\`ele de Mercure \`a 2 couches n\'eglige l'influence d'une \'eventuelle graine, soit un noyau rigide interne. Il sera probablement possible
  avant longtemps d'inverser la rotation observ\'ee avec un mod\`ele \`a 2 couches, n\'eanmoins la pr\'esence ou non d'une graine n'implique pas la m\^eme
  interpr\'etation physique de l'int\'erieur. Notamment le couplage gravitationnel entre le manteau et la graine ajoute de l'inertie au manteau, donc \`a la 
  surface observ\'ee de Mercure (Chap.\ref{chap:oceanglobal}). Une mesure de libration donnera donc un manteau plus \'epais si la graine est n\'eglig\'ee.
  
  \par L'id\'ee d'introduire une graine est pr\'esente dans la litt\'erature d\`es \citep{ppssz2002}, mais l'influence de cette graine avait \'et\'e 
  alors vue comme n\'egligeable. Les premi\`eres mesures de \citep{mpjsh2007} ont sugg\'er\'e que le noyau fluide \'etait plus large que pr\'evu, ce qui
  a ouvert la possibilit\'e d'une graine de taille significative, de rayon pouvant d\'epasser les 1000 km. L'id\'ee a \'et\'e remise au go\^ut du jour par
  une \'equipe canadienne \citep{vd2011}, ce qui a entra\^in\'e une succession de publications. Une \'equipe de l'Observatoire Royal de Belgique \citep{vrby2012} 
  y a introduit le couplage de pression en plus du couplage gravitationnel, puis 2 articles de collaboration entre ces 2 \'equipes \citep{drvy2013,yrvd2013}
  ont estim\'e les librations en longitude \`a l'aide de mod\`eles d'int\'erieur les plus r\'ealistes possibles.
  
  \par Dans un tel cas, la mod\'elisation des librations en longitude est plus complexe, car 2 degr\'es de libert\'e apparaissent (les librations, coupl\'ees,
   du noyau et du manteau), et donc 2 p\'eriodes de libration, l'une plus longue et l'autre plus courte que celle du mod\`ele sans graine, soit $\approx12$
  ans. Ceci multiplie les possibilit\'es de r\'esonance avec les diff\'erentes perturbations plan\'etaires. N\'eanmoins, les perturbations diurnes restent les
  plus d\'etectables du fait de leur vitesse, et leur amplitude variera au mieux de $5\%$.

  \section{Conclusion}
  
  \par Les librations diurnes ont \'et\'e mesur\'ees avec une pr\'ecision de l'ordre de $5\%$ \citep{mpshgjygpc2012}, ce qui en permet une premi\`ere 
  exploitation. Le d\'efi pour les ann\'ees \`a venir sera d'am\'eliorer cette mesure, ainsi que de mesurer les termes de perturbation plan\'etaire et
  \'eventuellement les librations propres, si elles ne sont pas tout-\`a-fait amorties. Dans ce cas, il faudra essayer de comprendre pourquoi elles ne le
  sont pas, \citet{p2005} estimant le temps d'amortissement \`a quelques centaines de milliers d'ann\'ees.
  
  \par L'observation radar a toujours fourni les premi\`eres mesures de la rotation de Mercure, aussi bien la vitesse de rotation moyenne, que l'obliquit\'e
  et les librations diurnes. Je ne sais pas si l'accumulation de donn\'ees sur plusieurs ann\'ees permettrait de d\'etecter les librations dues aux perturbations
  plan\'etaires. La mesure de rotation par BepiColombo semble difficile mais possible \citep{pvd2011,cm2012}.

  \chapter[La capture de Mercure en r\'esonance]{\'Etude \`a tr\`es long terme : la capture de Mercure en r\'esonance 3:2\label{chap:tides}}
  
  \section{Introduction}
  
  \par Je pr\'esente ici une \'etude r\'ecemment publi\'ee \citep{nfme2014}. Il s'agit d'une collaboration entre l'Universit\'e de Namur, o\`u Julien Frouard
  et moi-m\^eme travaillions sur la rotation de Mercure, et l'US Naval Observatory (USNO), o\`u Michael Efroimsky et Valeri Makarov revisitent la th\'eorie des 
  mar\'ees.
  
  \par La rotation 3:2 de Mercure en fait un cas unique dans le Syst\`eme Solaire, et depuis les ann\'ees 60, plusieurs \'etudes ont \'et\'e publi\'ees pour 
  tenter d'expliquer cet \'etat de rotation, en partant d'une rotation initiale inconnue, et en faisant agir des dissipations pour finalement obtenir un 
  \'etat d'\'equilibre, qui doit correspondre \`a la r\'esonance spin-orbite 3:2. Ces dissipations incluent syst\'ematiquement les forces de mar\'ee, qui 
  peuvent \^etre assist\'ees d'un frottement \`a l'interface fluide-rigide si une couche rigide globale existe au moment de la capture en r\'esonance. Il faut
  bien comprendre que cette \'etude ne concerne pas la plan\`ete Mercure telle qu'on l'observe actuellement, o\`u la pr\'esence de la couche fluide me 
  para\^it unanimement accept\'ee, mais telle qu'elle \'etait au moment de sa capture en r\'esonance. Le processus de capture d\'epend donc de la fa\c{c}on
  dont Mercure s'est form\'ee, et aussi de la fa\c{c}on dont ses param\`etres orbitaux ont \'evolu\'e depuis sa formation.
  
  \par Ce qui a motiv\'e cette \'etude est l'emploi syst\'ematique, par toutes les \'etudes pr\'ec\'edentes, d'un mod\`ele de mar\'ee dit Constant Time Lag (CTL),
  qui consid\`ere que le temps de retard avec lequel la plan\`ete r\'epond \`a l'excitation de mar\'ee est ind\'ependant de sa fr\'equence d'excitation. Ce mod\`ele physiquement
  irr\'ealiste peut \^etre appropri\'e pour tenter de mesurer ou quantifier l'effet des mar\'ees \`a l'\'echelle de temps des observations, c'est-\`a-dire
  de l'ordre du si\`ecle pour les satellites naturels des plan\`etes g\'eantes, du fait de sa simplicit\'e. Par contre il alt\`ere fortement l'\'evolution 
  temporelle de l'amortissement de la rotation au cours des \^ages, et de plus il pr\'esente une configuration d'\'equilibre stable pseudosynchrone, qui semble
  r\'esulter d'un artefact math\'ematique plus que de la physique du probl\`eme, et qui n'a jamais \'et\'e observ\'ee, m\^eme pour la Lune\footnote{Certains 
  consid\`erent que la rotation de V\'enus est pseudosynchrone, mais V\'enus a la particularit\'e d'avoir une atmosph\`ere \'epaisse. Les fractures \`a la
  surface d'Europe semblent \^etre la preuve d'une rotation tr\`es l\'eg\`erement super-synchrone \citep{gghhmptosgbdcbv1998}, mais l\`a encore, la physique
  n'est pas la m\^eme. Europe a vraisemblablement un oc\'ean global sous une surface visco-\'elastique.}.
  
  \par Je pr\'esente d'abord les diff\'erents scenarii d'\'evolution pr\'esents dans la litt\'erature (Sec.\ref{sec:diffscen}) avant de pr\'esenter notre 
  mod\`ele de mar\'ees (Sec.\ref{sec:ourtides}). Puis j'expose notre m\'ethode de r\'esolution du probl\`eme (Sec.\ref{sec:ourmethode}), bas\'ee sur de nombreuses simulations 
  num\'eriques, avant de revisiter les principaux scenarii d'\'evolution. 
  
  \section{Diff\'erents scenarii d'\'evolution \label{sec:diffscen}}
  
  \par
\begin{figure}[ht]
  \centering
  \includegraphics[width=.5\textwidth]{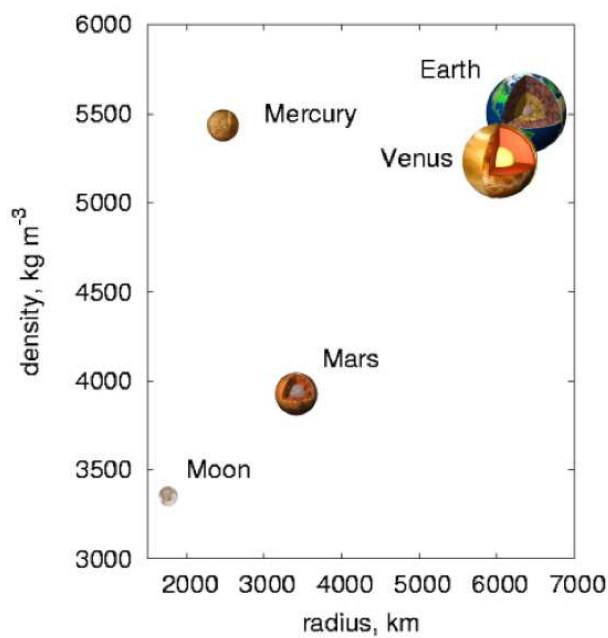}
  \caption[Masses volumiques des plan\`etes telluriques]{Masses volumiques des plan\`etes telluriques, illustration reprise de \citep{vshvds2007}. On peut voir
  que Mercure semble anormalement dense.\label{fig:mercuredense}}
  \end{figure}

 Comme je l'ai \'ecrit pr\'ec\'edemment, il n'y a pas unanimit\'e dans la communaut\'e scientifique sur la fa\c{c}on dont la structure interne de Mercure
  a \'evolu\'e depuis sa formation. Mercure est un myst\`ere car anormalement dense par rapport \`a sa taille (Fig.\ref{fig:mercuredense}).
  
  \par On s'attend en g\'en\'eral \`a ce qu'une plan\`ete soit relativement homog\`ene juste apr\`es sa formation, puis que les \'el\'ements lourds migrent vers 
  le centre, afin de cr\'eer un noyau. La masse volumique de Mercure sugg\`ere que les \'el\'ements lourds sont en proportion anormalement importante, comme
  si Mercure \'etait en fait le noyau d'un proto-Mercure plus large, dont le manteau d'\'el\'ements l\'egers aurait disparu.
  
  \subsection{Diff\'erents mod\`eles de formation de Mercure}
  
  \par On peut trouver dans la litt\'erature au moins 4 scenarii d'\'evolution de la structure interne de Mercure.
  
  \begin{enumerate}
  
  \item \emph{Volatilisation des \'el\'ements l\'egers dans la n\'ebuleuse proto-plan\'etaire} Il s'agit de la premi\`ere explication, due \`a \citet{w1978}. Elle
  suppose qu'une combinaison des forces gravitationnelles et non-gravitationnelles enl\`eve de fa\c{c}on plus efficace les silicates que le fer.
  
  \item \emph{Volatilisation des \'el\'ements l\'egers par le vent solaire} Dans ce cas, le processus se passe apr\`es la formation de Mercure \citep{c1985,fc1987}.
  
  \item \emph{\'Enorme impact qui aurait \'eject\'e le manteau \citep{bsc1988}} Il semble que ce soit la th\'eorie actuellement la plus populaire dans la 
  communaut\'e scientifique. La violence de cet impact aurait aussi consid\'erablement chauff\'e Mercure, ce qui aurait acc\'el\'er\'e la formation de l'interface 
  noyau-manteau, et donc de la couche fluide globale. N\'eanmoins, ce scenario semble \^etre mis \`a mal par des donn\'ees MESSENGER.
  En effet, \citet{pehmbgeghlmnsrsss2011} pr\'etendent que le chauffage tr\`es important induit par cet impact est incompatible avec l'abondance de potassium
  par rapport au thorium pr\'esent dans la surface de l'h\'emisph\`ere nord de Mercure. 
  
  \item \emph{Photophor\`ese} \citet{wtr2013} ont montr\'e que sous l'action de la chaleur d'une \'etoile dans un disque proto-plan\'etaire, les \'el\'ements l\'egers
  avaient tendance \`a \^etre repouss\'es vers l'ext\'erieur. En cons\'equence, les plan\`etes internes doivent \^etre relativement denses.
  
  \end{enumerate}
  
  \par Trois de ces scenarii sugg\`erent que Mercure \'etait initialement froid, alors que l'impact sugg\`ere un proto-Mercure chaud\footnote{Je parle ici de 
  temp\'erature, pas d'excentricit\'e.}. Si Mercure \'etait initialement froid, alors sa principale source de chauffage \'etait le chauffage radiog\'enique.
  En n\'egligeant la diffusion, l'\'el\'evation de la temp\'erature $\Delta T$ pendant un intervalle de temps $\Delta t$ nous est donn\'ee par l'\'equation
  de la chaleur :
  
  \begin{equation}
  \label{eq:chaleur}
  \Delta T = \frac{H}{C_p} \Delta t,
  \end{equation}
  o\`u $H$ est le chauffage radiog\'enique et $C_p$ la capacit\'e de chaleur sp\'ecifique. Avec les valeurs actuelles $H=3.5\times 10^{-12}\,W/kg$ et 
  $C_p=1200\,J/(K\,kg)$ \citep{ts2002}, on aurait une \'el\'evation de temp\'erature de $1.84\,K$ sur 20 millions d'ann\'ees\footnote{nous verrons plus tard la
  justification de cette \'echelle de temps}. Dans le Syst\`eme Solaire primordial, $H$ aurait pu \^etre 4 fois plus important, mais le chauffage r\'esultant 
  aurait de toute fa\c{c}on \'et\'e insuffisant pour cr\'eer l'interface noyau-manteau avant plusieurs centaines de millions d'ann\'ees. Dans ce cas, le 
  proto-Mercure que nous consid\'erons doit \^etre solide.
  
  \par Si par contre la th\'eorie de l'impact est la bonne, alors le proto-Mercure \'etait chaud. Dans ce cas l'interface noyau-manteau a pu se former beaucoup
  plus rapidement, et la friction noyau-manteau doit \^etre prise en compte pour simuler l'\'evolution de la rotation.
  
  \par Les donn\'ees dont nous disposons sur les syst\`emes exoplan\'etaires ne permettent pas de savoir si la masse volumique de Mercure est une r\`egle 
  ou une exception. Si elle s'av\'erait \^etre une exception, alors la th\'eorie de l'impact serait renforc\'ee.

  \subsection{Si Mercure \'etait solide}
  
  \par La premi\`ere \'etude de la capture en r\'esonance 3:2 est due \`a \citet{gp1966} qui ont consid\'er\'e que Mercure tournait sur elle-m\^eme initialement
  de fa\c{c}on prograde, et assez vite, puis qu'elle s'\'etait ralentie sous l'effet des mar\'ees solaires. Ils ont consid\'er\'e que l'excentricit\'e de la 
  plan\`ete avait toujours eu sa valeur actuelle, soit $\approx0.206$, et ont obtenu, de fa\c{c}on analytique, une probabilit\'e de capture dans la r\'esonance
  3:2 d'environ $7\%$. Autrement dit, la situation observ\'ee avait $7\%$ de chances de se produire. Une telle conclusion a ceci de frustrant qu'elle sugg\`ere
  que le mod\`ele est insuffisamment pr\'ecis, mais ne le prouve pas.
  
  \par Quelques ann\'ees plus tard, \citet{c1969} a propos\'e que l'excentricit\'e de Mercure pouvait avoir \'et\'e initialement diff\'erente, mais ne voyait
  pas pourquoi cette diff\'erence aurait pu \^etre radicale. Il a consid\'er\'e qu'elle ne pouvait pas d\'epasser $0.25$. Ceci n'a que peu chang\'e les 
  r\'esultats.
  
  \par Cette question a \'et\'e revisit\'ee plus de 3 d\'ecennies plus tard, lorsque \citet{cl2004} ont simul\'e l'\'evolution orbitale de Mercure \`a rebours sur 4 
  milliards d'ann\'ees avec 1000 conditions initiales, correspondant \`a notre \'epoque, proches mais diff\'erentes, puis ont utilis\'e ces 1000 trajectoires
  pour extraire 1000 \'evolutions temporelles de l'excentricit\'e de Mercure, qui pouvait avoir atteint 0.45. Ces 1000 \'evolutions de l'excentricit\'e ont
  ensuite \'et\'e utilis\'ees pour simuler l'amortissement de la rotation, et les auteurs obtiennent 554 captures dans la r\'esonance 3:2 sur 1000 trajectoires.
  Ils observent en fait 3 types de capture :
  
  \begin{itemize}
  
    \item{\emph{Type I:}} Mercure est captur\'ee dans la r\'esonance 3:2 d\`es sa premi\`ere travers\'ee (31 trajectoires),
    
    \item{\emph{Type II:}} Mercure n'est pas initialement captur\'ee, mais du fait de la forme du couple de mar\'ee, sa rotation ne ralentit plus; elle finira
    captur\'ee (168 trajectoires),
    
    \item{\emph{Type III:}} Mercure traverse la r\'esonance sans \^etre captur\'ee, et sa rotation continue \`a se ralentir. Bien plus tard, soit quelques 
    dizaines ou centaines de millions d'ann\'ees, sa rotation s'acc\'el\`ere du fait des variations s\'eculaires de son excentricit\'e, jusqu'\`a de 
    nouveau croiser la r\'esonance 3:2. Et elle finira par \^etre captur\'ee (355 trajectoires).
  
  \end{itemize}
  
  \par Toutes ces \'etudes consid\`erent que le mouvement de rotation ne se fait qu'en longitude, c'est-\`a-dire que l'obliquit\'e et le mouvement polaire ont
  d\'ej\`a \'et\'e amortis. L'\'equation \`a int\'egrer \'etait donc
  
  \begin{equation}
  \label{eq:ddotlaskar1}
  \ddot{\theta} = \frac{\mathcal{T}_z^{(TRI)}+\mathcal{T}_z^{(TIDE)}}{C}
  \end{equation}
  o\`u $\mathcal{T}_z^{(TRI)}$ est la composante en $z$ du couple triaxial, et $\mathcal{T}_z^{(TIDE)}$ est la composante du couple de mar\'ees. Contrairement
  \`a la plupart des chapitres pr\'ec\'edents, nous ne travaillons pas en formulation Hamiltonienne, ceci ayant moins de sens lorsqu'une dissipation est
  pr\'esente. On peut montrer que le couple triaxial s'\'ecrit
  
 \begin{equation} 
 \label{eq:tri4}
 \mathcal{T}_z^{^{(TRI)}} = -\frac{3}{2}(B-A)n^2\sum_{q=-\infty}^{+\infty}G_{20q}(e)\sin\left(2\left[\theta-\left(1+\frac{q}{2}\right)\cal{M}\right]\right).
 \end{equation}
  
 \par Les \'etudes sus-cit\'ees ont toutes consid\'er\'e un couple de mar\'ee s'exprimant ainsi, ou de fa\c{c}on \'equivalente \`a :
 
 \begin{equation}
 \label{eq:correialaskar}
 \mathcal{T}_z^{^{(TIDE)}} = - {3n^2M_{\sun}k_{2}}{\Delta t}\frac{R^5}{a^3}\left[\dot{\theta}{\cal{A}}(e)-n\mathcal{N}(e)\right],
 \end{equation}
 avec
 
 \begin{eqnarray}
 {\cal{A}}(e) & = & \frac{1+3e^2+\frac{3}{8}e^4}{\left(1-e^2\right)^{9/2}}, \label{eq:acorreialaskar} \\
 \mathcal{N}(e) & = & \frac{1+\frac{15}{2}e^2+\frac{45}{8}e^4+\frac{5}{16}e^6}{\left(1-e^2\right)^{6}}, \label{eq:ncorreialaskar}
 \end{eqnarray}
 $k_2=0.4$ et $n\Delta t=1/50$. On peut remarquer que dans cette derni\`ere quantit\'e, le d\'ephasage temporel $\Delta t$, soit le temps de r\'eponse du mat\'eriau
 \`a l'excitation de mar\'ee, ne d\'epend pas de la fr\'equence d'excitation, d'o\`u le nom de Constant Time Lag (CTL). Une cons\'equence du CTL est une vitesse
 de rotation d'\'equilibre super-synchrone (ou pseudosynchrone) :
 
 \begin{equation}
 \label{eq:pseudosynchrone}
 \dot{\theta} = n+6ne^2+\frac{3}{8}ne^4+\mathcal{O}(e^6).
 \end{equation}

 \par Avec l'excentricit\'e actuelle $e=0.206$, on a une vitesse de rotation d'\'equilibre \`a $1.26\,n$. Avec $e=0.285$, cet \'equilibre se confond avec la 
 r\'esonance 3:2. C'est cet \'equilibre qui a permis \`a \citet{cl2004} d'observer des acc\'el\'erations de la rotation de Mercure. Leur explication repose
 donc sur la croyance en la rotation pseudosynchrone.
 
  \subsection{Avec un noyau fluide}
  
  \par \citet{gp1967} ont \'et\'e les premiers \`a introduire un terme de friction noyau-manteau, \`a l'\'epoque pour \'etudier la rotation de V\'enus. En 
  exprimant l'angle $\gamma$ de libration autour de la r\'esonance q':2:
  
  \begin{equation}
  \gamma\equiv\theta-\left(1+\frac{q'}{2}\right){\cal{M}},
  \label{eq:libr}
  \end{equation}
 les \'equations du mouvement sont, dans le cadre d'un r\'egime laminaire dans le fluide :

 \begin{eqnarray}
 \ddot\gamma_m & = & -\omega_0^2\sin2\gamma_m+\frac{ \langle\,{\cal{T}}_z^{\rm{_{\,(TIDE)}}}\rangle }{C_m}-\frac{k}{C_m}\left(\dot\gamma_m-\dot\gamma_c\right), \label{eq:first} \nonumber \\
 \ddot\gamma_c & = & \frac{k}{C_c}\left(\dot\gamma_m-\dot\gamma_c\right), \label{eq:second}
 \end{eqnarray}
 o\`u les indices $m$ et $c$ se rapportent respectivement au manteau et au noyau, $\omega_0$ est la pulsation des oscillations libres en longitude, et $k$ est 
 une constante de friction qui couple les 2 mouvements.
 
 \par \citet{c1969} a appliqu\'e ce mod\`ele pour Mercure et a trouv\'e que la probabilit\'e de capture locale \'etait plus importante, mais pas la probabilit\'e
 de capture globale. En d'autres termes : si le syst\`eme traversait la r\'esonance 3:2, alors il avait des chances raisonnables d'y \^etre captur\'e. Mais en fait
 la plupart des trajectoires \'etaient captur\'ees en r\'esonance 2:1 et donc n'atteignaient jamais la r\'esonance 3:2. Ces r\'esultats ont \'et\'e confirm\'es
 et enrichis par \citet{pb1977} qui avaient conclu que, pour que Mercure \'evite d'\^etre captur\'ee dans la r\'esonance 2:1, alors, en supposant que la viscosit\'e
 cin\'ematique du fluide constituant le noyau externe \'etait de l'ordre de celle de l'eau, soit $\approx0.01\,cm^2/s$, le facteur de qualit\'e $Q\equiv (n\Delta t)^{-1}$
 devait \^etre plus petit que 100.
 
 \par En consid\'erant les variations s\'eculaires de l'excentricit\'e, \citet{cl2009} ont estim\'e que si un noyau fluide devait \^etre consid\'er\'e, alors il 
 \'etait vraisemblable que Mercure avait \'et\'e captur\'ee en r\'esonance 2:1 avant qu'une forte baisse de son excentricit\'e rompe cette r\'esonance, permettant
 ainsi la rencontre avec la r\'esonance 3:2, puis sa capture. Dans \citep{cl2012}, ils envisagent \'egalement que la r\'esonance 2:1 ait \'et\'e rompue par un impact.
 
  \subsection{Si Mercure \'etait initialement r\'etrograde}
  
  \par Toutes les \'etudes que j'ai cit\'ees dans ce chapitre jusqu'\`a pr\'esent ont consid\'er\'e un proto-Mercure ayant une rotation prograde. L'id\'ee que
  les rotations primordiales sont progrades \'etait largement r\'epandue jusqu'il y a peu, peut-\^etre car cela semble intuitif si on imagine le mouvement 
  d'ensemble de la n\'ebuleuse proto-plan\'etaire. Ceci n\'eglige le fait qu'au stade primordial de l'\'evolution du Syst\`eme Solaire, les collisions \'etaient
  fr\'equentes. La prise en compte de ces collisions rend al\'eatoire la distribution des obliquit\'es initiales des corps du Syst\`eme Solaire \citep{dt1993,ki2007}.
  
  \par Si Mercure a \'et\'e initialement r\'etrograde, alors elle a \'et\'e bloqu\'ee en rotation synchrone. \citet{wcllr2012}, par un comptage des crat\`eres
  utilisant des donn\'ees de Mariner 10 et des 2 premiers survols de MESSENGER, observent une assym\'etrie est-ouest, qu'ils interpr\'etent comme la preuve
  que Mercure a \'et\'e initialement synchrone. Ils supposent qu'ensuite un impact, laissant un crat\`ere de plus de 300 km, a \'et\'e assez important pour que
  Mercure quitte l'\'etat synchrone et atteigne l'\'etat pseudosynchrone, qui aurait fini par l'amener \`a la r\'esonance 3:2.
  
  \par Ce scenario est bas\'e sur 4 fondements : que Mercure a pu \^etre initialement r\'etrograde, que le proto-Mercure a \'et\'e fortement impact\'e, que le 
  comptage des crat\`eres est bien une preuve d'une ancienne rotation synchrone, et que la rotation pseudosynchrone est stable. Nous avons vu que Mercure a pu 
  \^etre initialement r\'etrograde, et que la rotation pseudosynchrone est induite par le CTL.
  
  \par Plusieurs \'etudes confirment que Mercure a \'et\'e fortement impact\'ee dans son premier milliard d'ann\'ees, notamment lors du Late Heavy Bombardment
  (LHB) \citep{scmsh2008,sbcffhmps2011}. Par contre, l'interpr\'etation du comptage des crat\`eres ne fait pas l'unanimit\'e. \citet{fhbzsnskscppop2012}, en 
  utilisant notamment des donn\'ees acquises apr\`es la mise en orbite de MESSENGER, confirment cette distribution assym\'etrique et reconnaissent qu'elle 
  peut venir d'une ancienne rotation synchrone, mais disent aussi qu'elle peut \^etre due \`a un renouvellement dissym\'etrique de la surface, ce qui 
  expliquerait \'egalement la distribution h\'emisph\'erique des plaines observ\'ee par \citet{drsmbdmehwc2009}.
  
  \section{Pourquoi utiliser un nouveau mod\`ele de mar\'ee \label{sec:ourtides}}
  
  \par Je pr\'esente ici bri\`evement notre mod\`ele de mar\'ee, \'elabor\'e par \citet{e2012a,e2012b}. Ce mod\`ele a l'avantage d'\^etre bas\'e sur des 
  consid\'erations physiques, notamment il d\'ecompose le couple de mar\'ee \`a la mani\`ere de \citep{k1964}, pour pouvoir prendre en compte la r\'eponse du 
  mat\'eriau aux diff\'erentes fr\'equences d'excitation possibles.
  
  \subsection{Th\'eorie}
  
  \par Il faut bien comprendre que le comportement d'un mat\'eriau diff\`ere selon la fr\'equence d'excitation. \`A basse fr\'equence, le mat\'eriau aura un 
  comportement relativement \'elastique, et plus visqueux \`a plus haute fr\'equence. Il est donc n\'ecessaire de d\'evelopper le potentiel de mar\'ee selon 
  les diff\'erents modes d'excitation.
  
  \par Ce travail a \'et\'e fait notamment par \citep{e2012a,e2012b} et fait appara\^itre des sommations sur 4 indices, ainsi qu'une partie oscillante dans le
  couple de mar\'ee. Apr\`es de nombreuses simplifications, consistant notamment \`a n\'egliger l'obliquit\'e et l'inclinaison, \`a se limiter aux modes les
  plus significatifs, et \`a faire dispara\^itre le couple oscillant par moyennisation, il reste:
  
  \begin{equation}
   \mathcal{T}_{tide} = \frac{3}{2}\frac{\mathcal{G}M_{\sun}^2}{a}\left(\frac{R}{a}\right)^5\sum_{q=-\infty}^{\infty}(G_{20q}(e))^2k_2(\omega_{220q})\sin\epsilon_2(\omega_{220q})+\mathcal{O}(i^2\epsilon)+\mathcal{O}((R/a)^7\epsilon). 
   \label{eq:torquemichael}
  \end{equation}

  \par Le terme $k_2(\omega_{220q})\sin\epsilon_2(\omega_{220q})$ contient la d\'ependance en la fr\'equence d'excitation, qui en fait ne d\'epend pas de son signe. 
  On d\'efinit la fr\'equence de mar\'ee $\chi$ par
  
  \begin{equation}
  \label{eq:chimichael}
  \chi_{220q} = |\omega_{220q}|\approx(2+q)n-2\dot{\theta}.
  \end{equation}
  
  \par Le mod\`ele dit de Maxwell \citep{m1867} est tr\`es populaire pour repr\'esenter les 2 types de comportement du mat\'eriau. Il utilise un temps de Maxwell
  $\tau_M=\eta/\mu$ o\`u $\eta$ est la viscosit\'e du mat\'eriau et $\mu$ sa rigidit\'e, qui signifie, grossi\`erement, que si la p\'eriode de l'excitation est 
  sup\'erieure au temps de Maxwell alors le mat\'eriau r\'epondra de fa\c{c}on \'elastique, sinon il aura un comportement visqueux.
  
  \par Des tests exp\'erimentaux, notamment sur la glace, ont montr\'e que le mod\`ele de Maxwell \'etait de moins en moins pr\'ecis \`a mesure que la 
  fr\'equence d'excitation augmentait. Pour cela, le mod\`ele d'Andrade donne de meilleurs r\'esultats \citep{a1910}. C'est la raison pour laquelle nous 
  utilisons la d\'ependance en fr\'equence suivante, reprise de \citep{e2012a,e2012b} :

\begin{eqnarray}
k_2(\omega_{220q})\sin\epsilon_2(\omega_{220q}) & = & \frac{3}{2}\frac{-A_2\mathcal{I}[\bar{J}(\chi)]}{\left(\mathcal{R}[\bar{J}(\chi)]+A_2\right)^2+\left(\mathcal{I}[\bar{J}(\chi)]\right)^2}Sgn(\omega_{220q}), \label{eq:klsemichael} \\
A_2 & = & \frac{57\mu}{8\pi G\rho^2R^2}, \label{eq:al} \\
\mathcal{R}[\bar{J}(\chi)] & = & 1+\left(\chi\tau_A\right)^{-\alpha}\cos\left(\frac{\alpha\pi}{2}\right)\Gamma(\alpha+1), \label{eq:rmichael} \\
\mathcal{I}[\bar{J}(\chi)] & = & -\left(\chi\tau_M\right)-\left(\chi\tau_A\right)^{-\alpha}\sin\left(\frac{\alpha\pi}{2}\right)\Gamma(\alpha+1), \label{eq:imichael}
\end{eqnarray}
o\`u $\bar{J}(\chi)$ est la compliance complexe \`a la fr\'equence de mar\'ee $\chi$, $\tau_A$ le temps d'Andrade, et $\alpha$ le param\`etre d'Andrade. Nous 
prenons $\alpha = 0.2$ et $\tau_M=\tau_A$.

\par
\begin{figure}[ht]
\centering
\includegraphics[width=.6\textwidth]{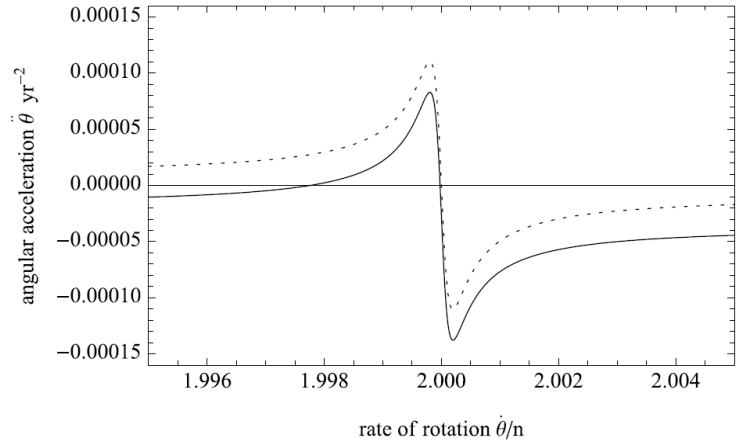}
\caption[Acc\'el\'eration de mar\'ee au voisinage d'une r\'esonance]{Acc\'el\'eration de mar\'ee au voisinage d'une r\'esonance. Figure reprise de \citep{makarov2013}.\label{fig:kink}}
\end{figure}

 Dans la pratique, la prise en compte de la fr\'equence d'excitation a une importance cruciale au moment de la travers\'ee d'une r\'esonance spin-orbite. 
En effet, pour chacun des mod\`eles physiques existant dans la litt\'erature, la prise en compte de la fr\'equence d'excitation induit un saut comme celui
montr\'e \`a la Fig.\ref{fig:kink}, alors que le CTL aura une forme affine \`a faible pente. Le comportement qualitatif du syst\`eme est donc fondamentalement
diff\'erent \`a la travers\'ee et au voisinage d'une r\'esonance spin-orbite.

  \subsection{Une rotation pseudosynchrone impossible}
  
  \par L'une des principales caract\'eristiques du CTL est la pr\'esence d'un \'equilibre pseudosynchrone, \'etat final le plus probable d'apr\`es les 
  simulations num\'eriques, lorsque l'excentricit\'e est constante. Lorsque l'excentricit\'e varie et que la plan\`ete est dans cet \'etat d'\'equilibre 
  pseudosynchrone, alors sa rotation peut s'acc\'el\'erer selon la loi (\ref{eq:pseudosynchrone}). Lorsque la rotation initiale est sup\'erieure \`a la 
  rotation pseudosynchrone, alors la rotation synchrone est en g\'en\'eral inatteignable, sauf ph\'enom\`ene catastrophique. Par \emph{\'equilibre}, il faut 
  comprendre que c'est un \'equilibre au sens pratique du terme, c'est-\`a-dire un \'etat final raisonnablement probable. Math\'ematiquement, on parle 
  d'\emph{\'equilibre stable}, c'est-\`a-dire que non seulement il correspond \`a un \'equilibre des couples agissant sur Mercure (le couple gravitationnel
  et le couple de mar\'ees), mais en plus, tout petit \'ecart \`a cet \'equilibre aura pour cons\'equence sa restauration, autrement dit une perturbation
  ne peut pas le rompre d\`es lors qu'elle est suffisamment petite.
  
  \par
\begin{figure}[ht]
\centering
\includegraphics[width=.6\textwidth]{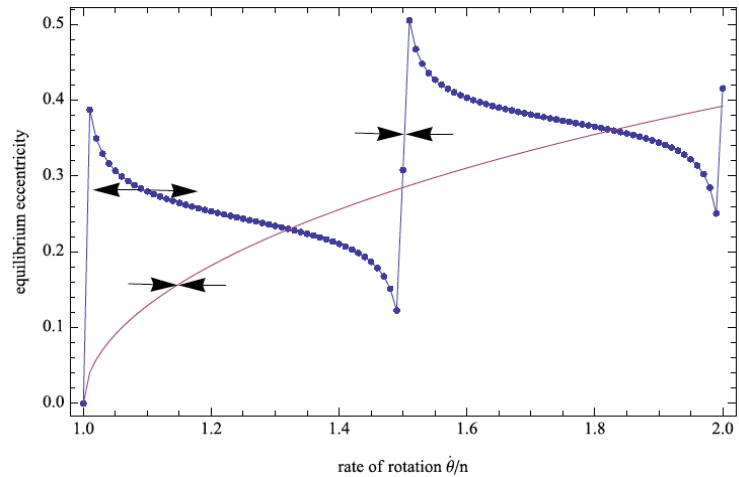}
\caption[Instabilit\'e de la rotation pseudosynchrone]{Instabilit\'e de la rotation pseudosynchrone. La rotation d'\'equilibre induite par le CTL est en mauve, 
celle due \`a notre mod\`ele est en bleu. Chacune de ces 2 courbes partage le plan en 2, le corps consid\'er\'e acc\'el\`ere au-dessus de la courbe, et 
d\'ec\'el\`ere en-dessous. Figure reprise de \citep{me2013}.\label{fig:nopseudo}}
\end{figure}

 La Fig.\ref{fig:nopseudo} montre l'\'equilibre d\^u au couple de mar\'ee dans les 2 cas. Les param\`etres utilis\'es sont ici ceux de la Lune, mais
  l'explication reste qualitativement valide. Sur la courbe, la rotation du corps est \`a l'\'equilibre. Chacune de ces courbes s\'epare le plan en 2, le corps
  est acc\'el\'er\'e au-dessus de la courbe, et d\'ec\'el\'er\'e en-dessous. Ceci a pour cons\'equence que lorsque la courbe a une pente positive, alors
  l'\'equilibre est stable, et il est instable si la pente est n\'egative. Dans le cas du CTL, tous les \'equilibres sont stables, c'est pour \c{c}a qu'on
  peut avoir une rotation d'\'equilibre pseudosynchrone (ou super-synchrone). Par contre, dans le cas de notre mod\`ele de mar\'ee, la pente n'est positive 
  qu'au voisinage des r\'esonances spin-orbite. En cons\'equence, ce sont les seuls \'equilibres stables, \'equilibres qui se trouvent renforc\'es par notre 
  mod\`ele de mar\'ee du fait qu'ils ne sont pas pr\'esents que dans le couple gravitationnel. Ceci sugg\`ere \'egalement, comme l'indiquera la suite, que
  les probabilit\'es de capture dans les r\'esonances spin-orbite sont plus grandes qu'avec le CTL.
  
  \subsection{R\'esultats pr\'eliminaires}
  
  \par
\begin{figure}[ht]
  \centering
  \includegraphics[width=.6\textwidth]{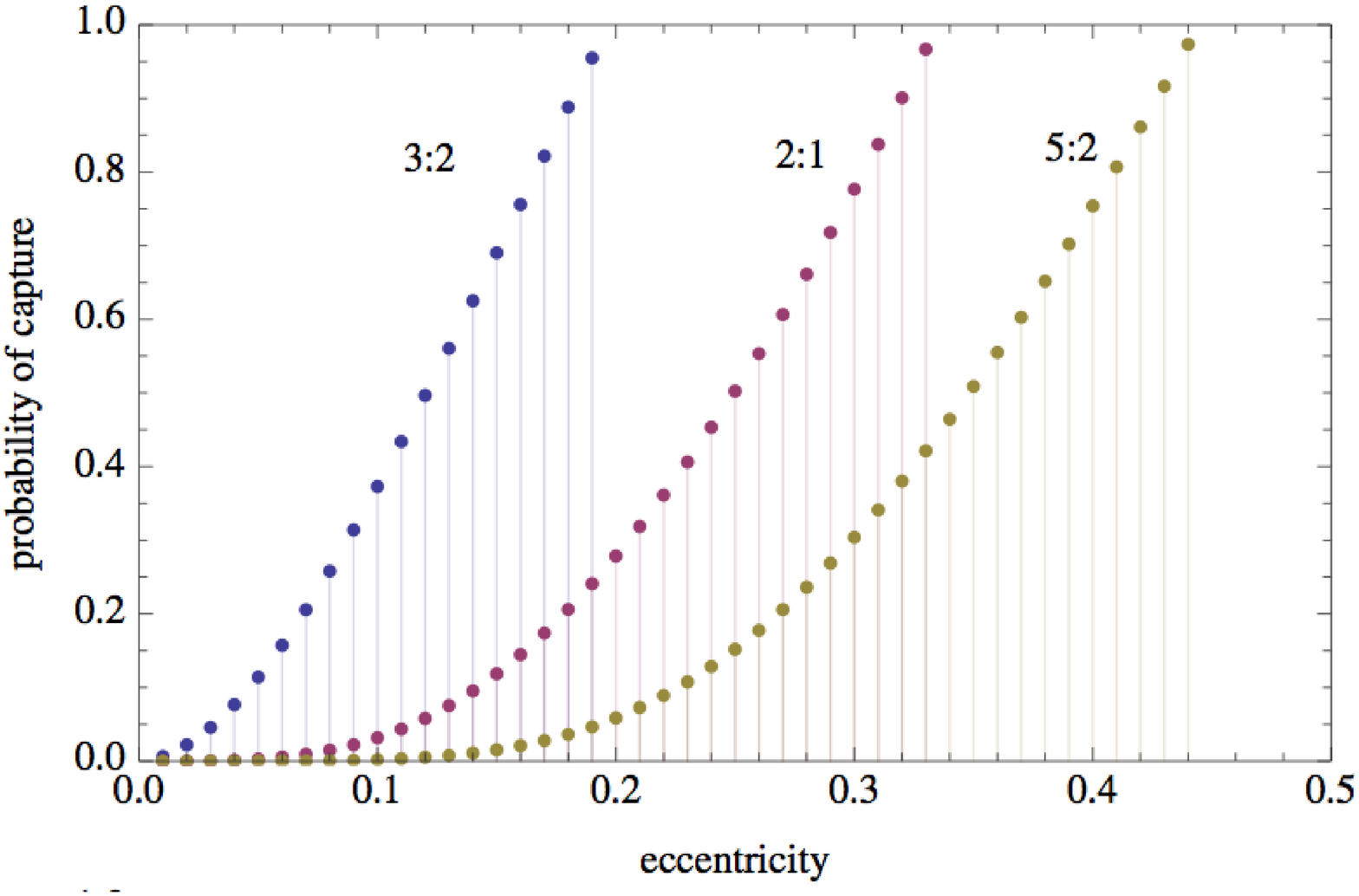}
  \caption[Probabilit\'es de capture]{Probabilit\'es de capture obtenues par \citep{m2012}. On constate qu'avec l'excentricit\'e actuelle $e=0.206$, alors 
  Mercure est certaine d'\^etre captur\'ee dans la r\'esonance 3:2 si elle la traverse.\label{fig:makarov2012}}
  \end{figure}

 \citet{m2012} a fait des simulations num\'eriques des probabilit\'es locales de capture dans les principales r\'esonances spin-orbite, c'est-\`a-dire
  en lan\c{c}ant chaque simulation juste avant la rencontre avec la r\'esonance, pour diff\'erentes excentricit\'es, consid\'er\'ees comme constantes. Les 
  simulations diff\'eraient par la condition initiale de l'angle de la rotation. Les r\'esultats sont donn\'es Fig.\ref{fig:makarov2012}, les param\`etres 
  d'int\'erieur utilis\'es \'etant ceux de Mercure, avec le temps de Maxwell $\tau_M=500$ ans.
  
  \par On constate que les probabilit\'es de capture d\'ependent de l'excentricit\'e, ce qui \'etait attendu car elle intervient dans le couple de mar\'ee par le 
  biais de la quantit\'e $G_{20q}(e)$, ainsi que dans le couple gravitationnel. Il est int\'eressant de constater que certaines r\'esonances deviennent certaines
  au-del\`a d'une certaine excentricit\'e. Notamment la r\'esonance 3:2 est certaine pour $e>0.18$, et la 2:1 est certaine pour $e>0.3$. Ce r\'esultat tranche
  radicalement avec le CTL, qui donne une probabilit\'e de capture dans la r\'esonance 3:2 de $7\%$. Ce r\'esultat justifie que les scenarii d'\'evolution
  spin-orbite de Mercure soient revisit\'es en utilisant un mod\`ele de mar\'ee plus r\'ealiste.

  \section{La m\'ethode \label{sec:ourmethode}}
  
  \par Tenir compte d'un Mercure primordial, ou proto-Mercure, n\'ecessite notamment de consid\'erer son orbite primordiale. Nous nous basons pour cela sur les 
  travaux de Jacques Laskar et ses collaborateurs sur l'\'evolution \`a long terme du Syst\`eme Solaire interne, qui est chaotique avec un temps de Lyapunov de 
  l'ordre de 10 millions d'ann\'ees \citep{l1989}. En cons\'equence, les quantit\'es orbitales primordiales des plan\`etes doivent \^etre exprim\'ees de 
  fa\c{c}on statistique. Si leurs demi-grands axes peuvent \^etre consid\'er\'es comme constants, l'excentricit\'e de Mercure peut avoir \'et\'e tr\`es basse, 
  comme elle peut avoir atteint $0.45$ \citep{l2008}. Une \'etude d\'edi\'ee \`a Mercure doit tenir compte de ces aspects.
  
  \par Il faut \'egalement consid\'erer que les conditions initiales de la rotation de Mercure sont m\'econnues. Si on consid\`ere que l'obliquit\'e et le 
  mouvement polaire se sont vite amortis, l'angle de rotation et la vitesse initiales sont incertains.
  
  \par Tout ceci incite \`a consid\'erer l'\'evolution de la rotation de Mercure d'un point de vue statistique. En gardant \`a l'esprit que des simulations
  num\'eriques sur le tr\`es long terme seront n\'ecessaires, le temps se mesurant en millions d'ann\'ees, il devient \'evident que des simulations 
  num\'eriques utilisant des outils de calcul performants seront n\'ecessaires. Nous avons utilis\'e, comme pr\'ec\'edemment, l'int\'egrateur 
  d'Adams-Bashforth-Moulton d'ordre 10.
  
  \subsection{\'Evolution de l'excentricit\'e}
  
  \par
\begin{figure}[ht]
  \centering
  \includegraphics[width=.6\textwidth]{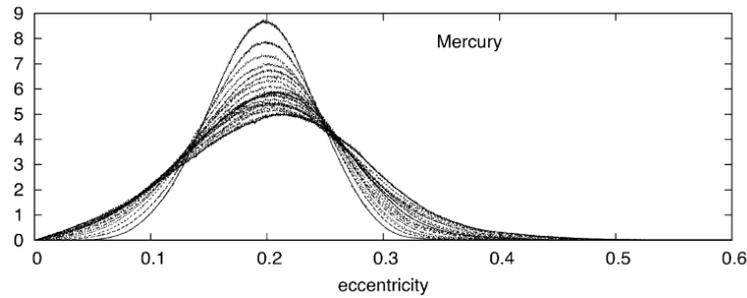}
  \caption[\'Evolution statistique de l'excentricit\'e de Mercure]{\'Evolution statistique de l'excentricit\'e de Mercure, d'apr\`es \citep{l2008}. Les courbes 
  sont espac\'ees de 250 millions d'ann\'ees, et elles s'\'evasent \`a mesure qu'on s'approche de l'origine du Syst\`eme Solaire.\label{fig:ecc_mercury}}
  \end{figure}

 \citet{l2008} a montr\'e, par des simulations num\'eriques sur 5 milliards d'ann\'ees, que l'\'evolution pass\'ee de l'excentricit\'e de Mercure 
  devait \^etre consid\'er\'ee d'un point de vue statistique, pour tenir compte de la diffusion chaotique (Fig.\ref{fig:ecc_mercury}).
  
  \par Il a sugg\'er\'e de d\'ecrire cette statistique par une loi de Rice 
  
  \begin{equation}
  f_{T,s}(e) = \frac{e}{\sigma^2} \exp \left( - \frac{e^2+s^2}{2\sigma^2}\right) I_0\left(\frac{e s }{\sigma^2}\right),
  \label{eq:rice}
  \end{equation}
  avec
  
  \begin{equation}
  \sigma^2(T) = b_0 + b_1 T
  \label{eq:sigmarice}
  \end{equation}
  pour mod\'eliser un ph\'enom\`ene de diffusion, o\`u $b_0 = 2.07\times 10^{-3}$, $b_1 = 1.043\times 10^{-3}$, $s = 0.1875$, et le temps $T$ est en milliards 
  d'ann\'ees. $I_0$ est la fonction de Bessel de premi\`ere esp\`ece.
  
  \par
\begin{figure}[ht]
  \centering
  \begin{tabular}{cc}
  \includegraphics[angle=270,width=.46\textwidth]{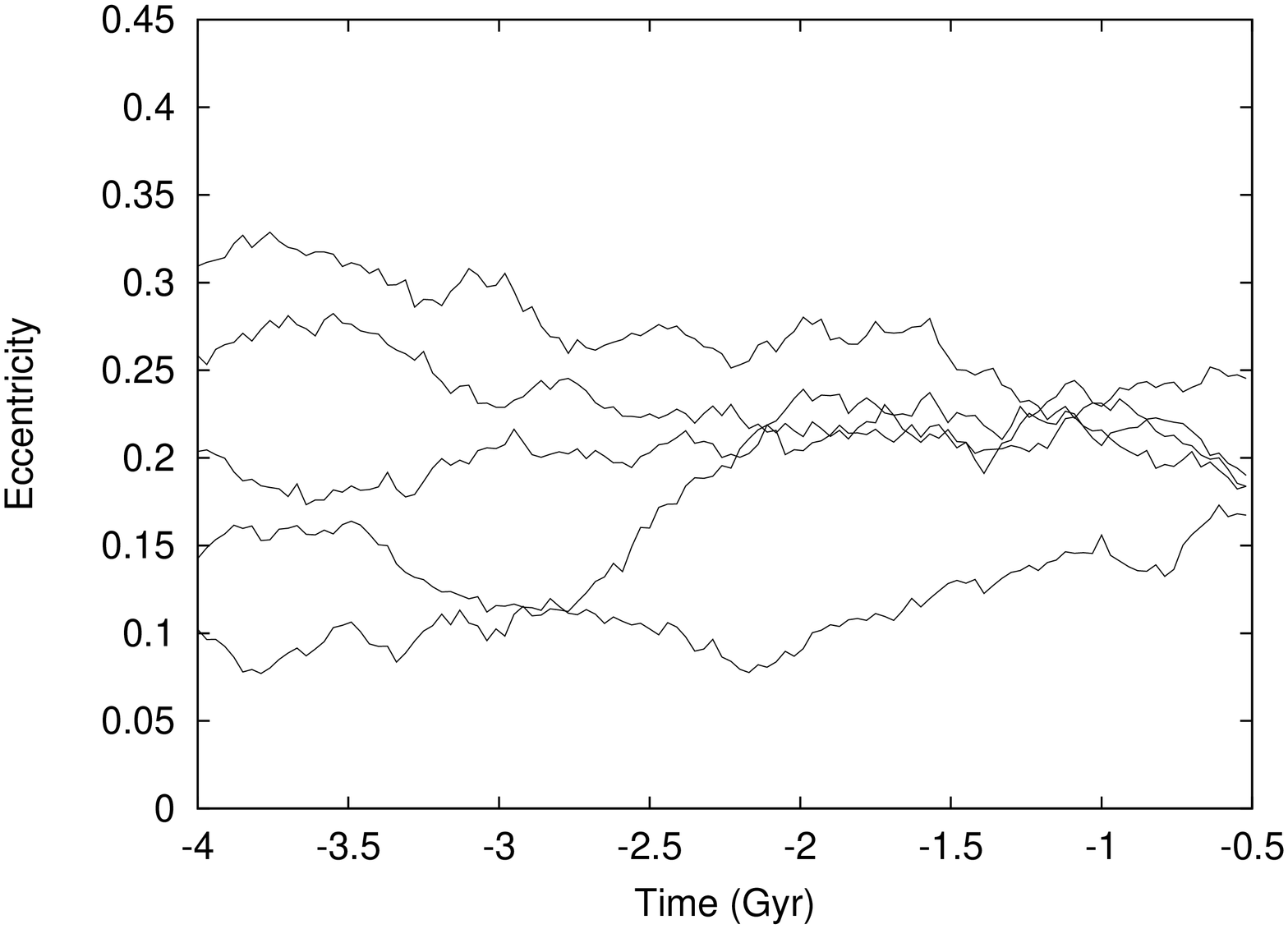} & \includegraphics[angle=270,width=.46\textwidth]{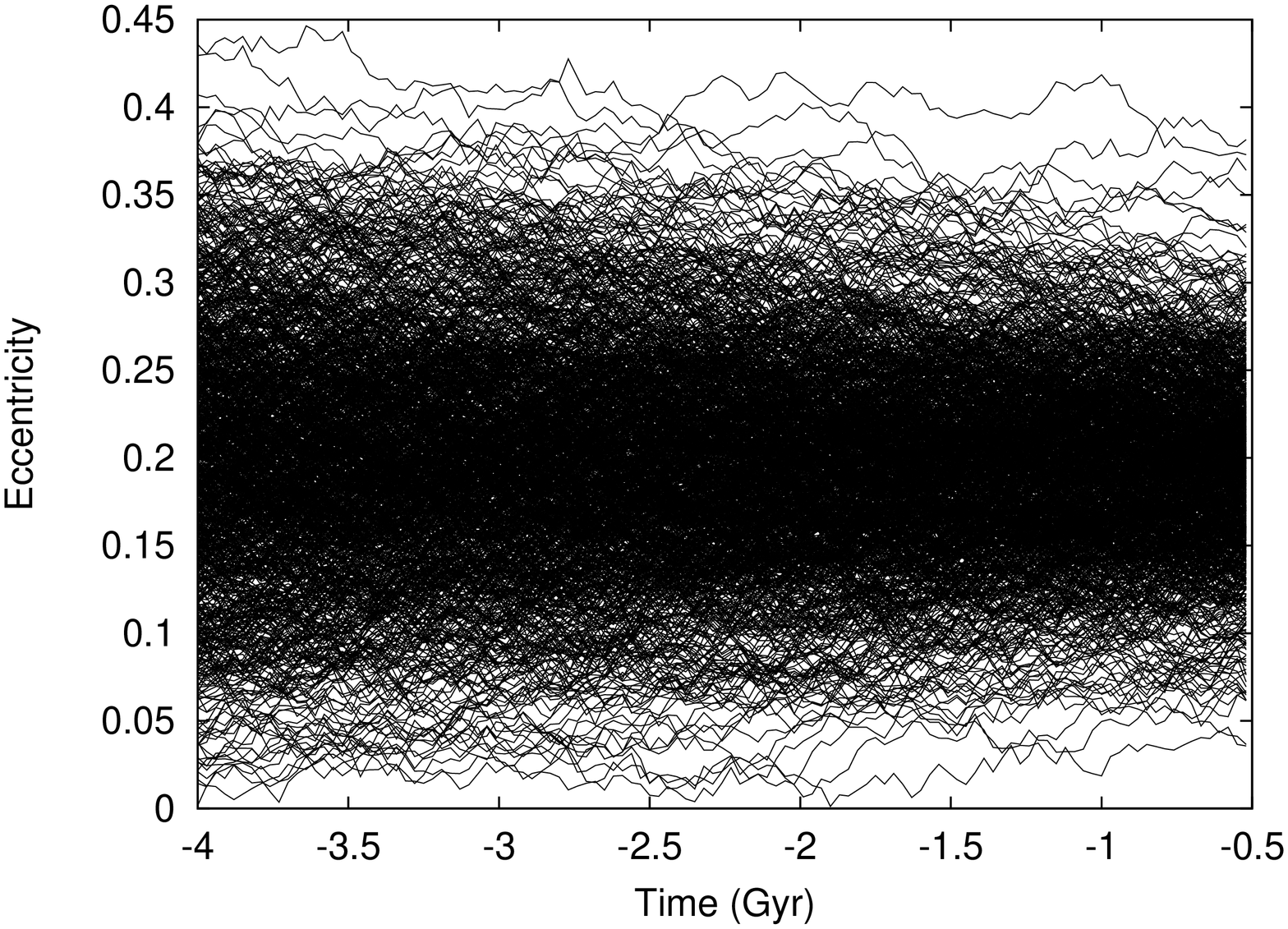}
  \end{tabular}
  \caption[\'Evolution temporelle de l'excentricit\'e de Mercure]{\'Evolution temporelle de 5 (\`a gauche) et 1000 (\`a droite) excentricit\'es synth\'etiques
  de Mercure.\label{fig:tempecc}}
  \end{figure}

 Julien Frouard a fait un travail formidable d'\'elaboration d'\'evolutions temporelles synth\'etiques de l'excentricit\'e de Mercure, afin de reproduire
  ces statistiques et d'obtenir ces trajectoires sans refaire les simulations num\'eriques de \citep{l2008}. Pour cela il a utilis\'e un processus de Wiener
  du type 
  
  \begin{equation}
  e(T+\delta_t) = e(T) + \sigma(\delta_t)\delta_e,
  \label{eq:wiener}
  \end{equation}
  o\`u $e(0)$ est l'excentricit\'e initiale et $\sigma$ la d\'eviation standard. $\delta_t$ est un incr\'ement en temps, et $\delta_e$ un incr\'ement 
  ind\'ependant, ob\'eissant \`a une loi gaussienne. La d\'eviation standard utilis\'ee dans l'\'equation (\ref{eq:wiener}) est obtenue en ajustant une loi 
  gaussienne \`a la loi de Rice (\ref{eq:rice}), ce qui donne $\sigma^2(t) = 0.0009\,t$. Apr\`es quelques autres \'etapes d\'etaill\'ees dans \citep{nfme2014},
  visant notamment \`a supprimer les excentricit\'es n\'egatives donn\'ees par la loi gaussienne, et \`a retourner le sens du temps car on raisonne \`a rebours,
  on obtient 1000 trajectoires (Fig.\ref{fig:tempecc}), c'est-\`a-dire 1000 \'evolutions temporelles synth\'etiques de l'excentricit\'e, dont les statistiques 
  v\'erifient la loi (\ref{eq:rice}).
  
  \par Nous avons ainsi obtenu des \'evolutions temporelles de l'excentricit\'e de Mercure que nous avons pu injecter dans l'\'equation r\'egissant la rotation
  (\ref{eq:ddotlaskar1}).

  \subsection{Validation de la m\'ethode}
  
  \par
\begin{figure}[ht]
  \centering
  \includegraphics[width=.5\textwidth]{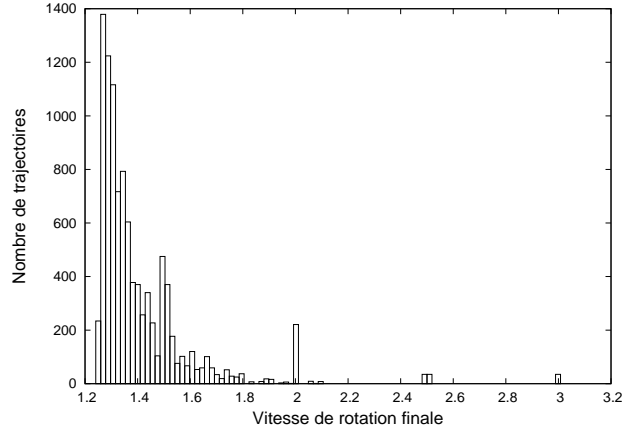}
  \caption[R\'epartition des vitesses de rotation apr\`es 3 millions d'ann\'ees, avec le CTL]{R\'epartition des vitesses de rotation de 10\,000 proto-Mercures
  apr\`es 3 millions d'ann\'ees, avec le CTL. On peut constater une accumulation au niveau de la rotation pseudosynchrone.\label{fig:histolaskar}}
  \end{figure}

 Une bonne fa\c{c}on de valider notre m\'ethode \'etait de tenter de reproduire les r\'esultats statistiques de \citep{cl2004}. Afin de limiter le temps
  de calcul, j'ai limit\'e chacune des simulations \`a 3 millions d'ann\'ees, ce qui permet de dire si Mercure a \'et\'e captur\'ee \`a la premi\`ere 
  travers\'ee de la r\'esonance 3:2. On peut donc comparer avec les 31 captures de Type I sur 1000 trajectoires obtenues par \citet{cl2004}. \`A partir des 1000 
  \'evolutions synth\'etiques de l'excentricit\'e, nous cr\'eons 10 ensembles de 1000 simulations num\'eriques avec une p\'eriode de rotation initiale de 20 
  jours, et des angles de rotation initiaux diff\'erents, le mod\`ele de mar\'ee \'etant le CTL (\'Eq.\ref{eq:correialaskar}). La r\'epartition des vitesses
  de rotation finales est donn\'ee Fig.\ref{fig:histolaskar}.
  
  \par On constate des pics de densit\'e aux r\'esonances spin-orbite 3:2, 2:1, 5:2 et 3:1, correspondant \`a des captures. Le pic le plus important correspond
  \`a la rotation pseudosynchrone, nous voyons bien que la r\'esonance synchrone n'est pas atteinte.
  
  \par Afin d'estimer le nombre de trajectoires effectivement captur\'ees dans chaque r\'esonance, en particulier la 3:2, nous ajustons une fonction de 
  r\'epartition par moindres carr\'es pour supprimer les trajectoires en rotation pseudosynchrone. En supprimant cette fonction 
  $f(\dot{\theta})=a/\dot{\theta}^b$ avec $a=31\,476.4$ et $b=12.9944$ des statistiques globales, nous pouvons estimer le nombre de trajectoires captur\'ees
  dans chacun des \'echantillons (Tab.\ref{tab:r10sets}).
  
   \begin{table}[ht!]
 \centering
 \caption[Probabilit\'es de capture avec le CTL]{Estimation du nombre de captures pour chacun des 10 \'echantillons de 1000 simulations simulations num\'eriques, 
 avec le CTL. Du fait d'erreurs d'arrondi, la ligne \emph{Total} ne correspond pas exactement \`a la somme des 10 lignes pr\'ec\'ecentes.\label{tab:r10sets}}
 \begin{tabular}{l|cccc}
      & 3:2 & 2:1 & 5:2 & 3:1 \\
\hline
\'Echantillon 0 &  37 &  20 &   8 & 5 \\
\'Echantillon 1 &  47 &  15 &   5 & 1 \\
\'Echantillon 2 &  51 &  22 &  10 & 4 \\
\'Echantillon 3 &  34 &  26 &   8 & 8 \\
\'Echantillon 4 &  56 &  21 &   5 & 1 \\
\'Echantillon 5 &  35 &  28 &   6 & 4 \\
\'Echantillon 6 &  40 &  23 &   6 & 1 \\
\'Echantillon 7 &  29 &  17 &  11 & 3 \\
\'Echantillon 8 &  39 &  26 &   5 & 7 \\
\'Echantillon 9 &  49 &  23 &   6 & 1 \\
\hline
Total & 415 & 217 &  70 & 35 \\
\hline
\end{tabular}
\end{table}
  
\par On a 415 captures de Type I (pour reprendre la terminologie de \citep{cl2004}) en r\'esonance 3:2 pour 10\,000 trajectoires, alors que \citep{cl2004} en 
avaient 31 pour 1000 trajectoires. Il est int\'eressant de constater la variabilit\'e du nombre de trajectoires captur\'ees selon les \'echantillons, qui 
utilisent les m\^emes \'evolutions temporelles de l'excentricit\'e. On a en effet entre 29 et 56 captures, donc nos r\'esultats sont coh\'erents avec ceux de 
l'\'etude sus-cit\'ee. Ceci dit, ils illustrent la difficult\'e de l'estimation de probabilit\'es de capture \`a partir de r\'esultats num\'eriques.

\par Nous pouvons consid\'erer la m\'ethode comme valid\'ee et pouvons maintenant utiliser notre couple de mar\'ee pour revisiter l'histoire de la rotation de
Mercure.

  \section[Revisite du scenario 1]{Revisite du scenario 1 : Mercure \'etait solide au moment de la capture}
  
  \par Ici nous revisitons le premier scenario, qui consid\`ere que Mercure \'etait initialement prograde, et que l'interface noyau-manteau n'\'etait pas 
  encore form\'ee. En fixant le param\`etre d'Andrade $\alpha=0.2$ et les moments d'inertie de Mercure, et en imposant $\tau_A=\tau_M$, alors le mod\`ele ne 
  d\'epend plus que d'un seul param\`etre : le temps de Maxwell $\tau_M$. Nous allons le faire varier.
  
  \subsection{Premiers essais}
  
  \par
\begin{figure}[ht]
\centering
\includegraphics[width=.6\textwidth]{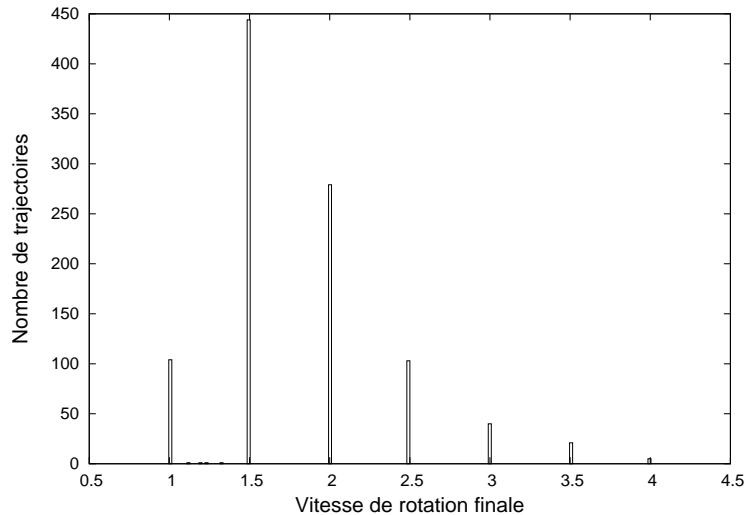}
\caption[Statistiques de capture avec $\tau_M=500$ ans]{Statistiques de capture avec $\tau_M=500$ ans. Ces r\'esultats ont \'et\'e obtenus sur 20 millions 
d'ann\'ees. On peut voir que la r\'esonance 3:2 est la plus probable.\label{fig:tau500}}
\end{figure}

 Nos premiers essais ont \'et\'e faits avec $\tau_M=500$ ans, et il s'av\`ere que 20 millions d'ann\'ees suffisent pour obtenir l'\'etat final de Mercure.
  Cette valeur du temps de Maxwell est parfois utilis\'ee pour la Terre. Nous obtenons les statistiques suivantes :
  
  \begin{itemize}
	
\item R\'esonance 4:1 : ~~~~5 captures ~($0.5\%$)$\,$,

\item R\'esonance 7:2 : ~~~21 captures ~($2.1\%$)$\,$,

\item R\'esonance 3:1 : ~~~40 captures ~($4\%$)$\,$,

\item R\'esonance 5:2 : ~~103 captures ~($10.3\%$)$\,$,

\item R\'esonance 2:1 : ~~279 captures ~($27.9\%$)$\,$,

\item R\'esonance 3:2 : ~~444 captures ~($44.4\%$)$\,$,

\item R\'esonance 1:1 : ~~104 captures ~($10.4\%$)$\,$,

\end{itemize}
rassembl\'ees dans la Fig.\ref{fig:tau500}.

\par La diff\'erence avec le CTL est spectaculaire. Comme pr\'evu, on n'a plus l'\'etat de rotation pseudosynchrone. De plus, la r\'esonance spin-orbite
3:2 est l'\'etat final le plus probable. Ainsi on apporte une solution au probl\`eme pos\'e apr\`es la probabilit\'e de capture de $7\%$ par \citep{gp1966},
sans avoir recours aux variations s\'eculaires de l'excentricit\'e. Ceci signifie aussi que dans ce cas, Mercure a \'et\'e captur\'ee en r\'esonance 3:2
en moins de 20 millions d'ann\'ees, un temps bien trop court pour former l'interface noyau-manteau si le proto-Mercure \'etait froid, c'est-\`a-dire s'il ne
r\'esulte pas d'un impact suffisamment violent pour expulser le manteau.

  \subsection{Avec un proto-Mercure plus chaud}
  
  \par
\begin{figure}[ht]
\centering
\begin{tabular}{cc}
\includegraphics[width=.46\textwidth]{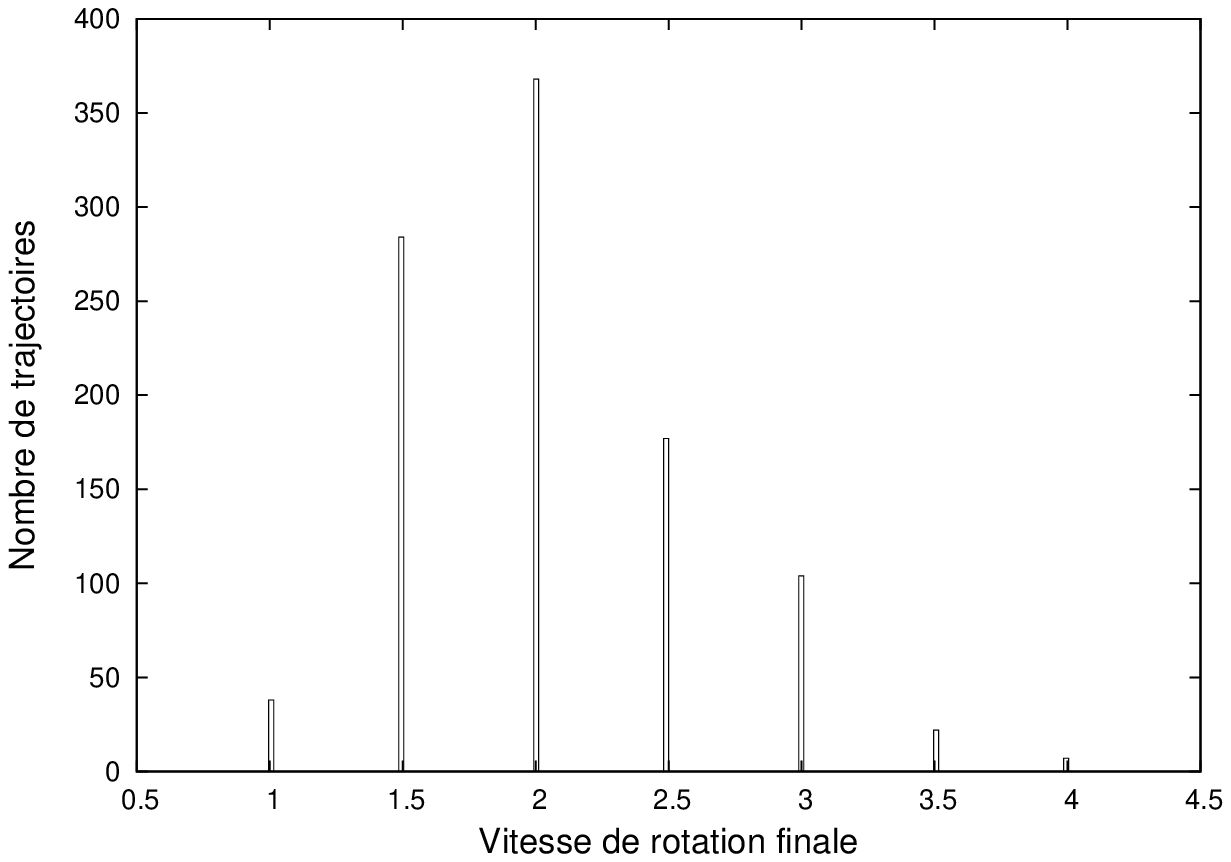} & \includegraphics[width=.46\textwidth]{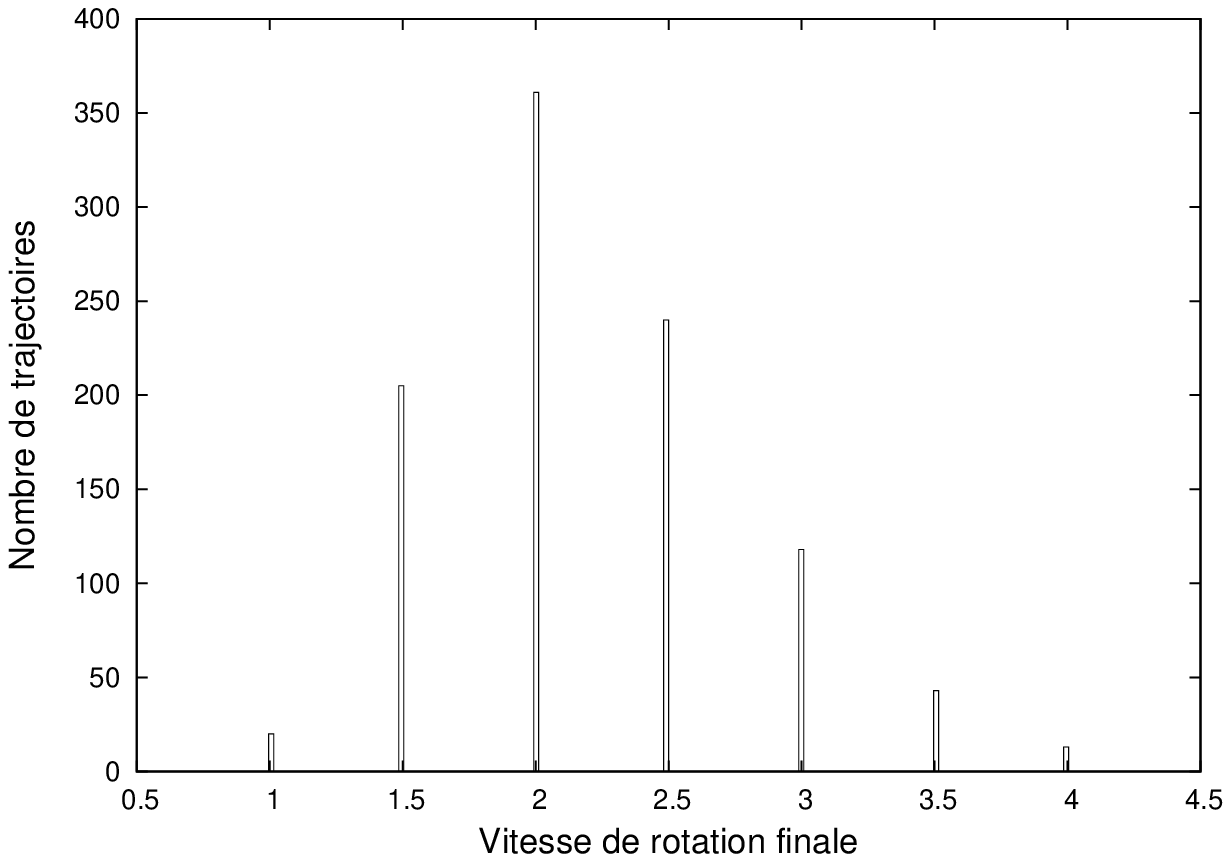} \\
$\tau_M=15$ ans & $\tau_M=5$ ans
\end{tabular}
\caption[Statistiques de captures pour un Mercure plus chaud]{Statistiques de captures pour un proto-Mercure plus chaud. On constate que la diminution du temps 
de Maxwell $\tau_M$ favorise la r\'esonance 2:1 par rapport \`a la r\'esonance 3:2.\label{fig:tau515}}
\end{figure}

 Le Mercure primordial \'etant tr\`es mal connu, nous avons test\'e le mod\`ele avec des temps de Maxwell plus courts, correspondants \`a des mod\`eles plus
  chauds. Plus pr\'ecis\'ement, nous avons test\'e $\tau_M=5$ ans et $\tau_M=15$ ans. Les r\'esultats sont rassembl\'es Tab.\ref{tab:newmaxwell} et 
  Fig.\ref{fig:tau515}.
  
    \begin{table}[ht]
	\centering
	\caption{Statististiques de capture pour un temps de Maxwell plus court.\label{tab:newmaxwell}}
	\begin{tabular}{l|rr}
\hline
		& $\tau_M = 5$ ans & $\tau_M = 15$ ans \\
		\hline
		R\'esonance 4:1 &  13 captures ~($1.3\%$)    &   7 captures ~($0.7\%$) \\
		R\'esonance 7:2 &  43 captures ~($4.3\%$)    &  22 captures ~($2.2\%$) \\
		R\'esonance 3:1 & 118 captures ~($11.8\%$)   & 104 captures ~($10.4\%$) \\
		R\'esonance 5:2 & 240 captures ~($24\%$)     & 177 captures ~($17.7\%$) \\
		R\'esonance 2:1 & 361 captures ~($36.1\%$)   & 368 captures ~($36.8\%$) \\
		R\'esonance 3:2 & 205 captures ~($20.5\%$)   & 284 captures ~($28.4\%$) \\
		R\'esonance 1:1 &  20 captures ~($2\%$)      &  38 captures ~($3.8\%$)\\		
\hline
	\end{tabular}
\end{table}

\par Nous constatons qu'un temps de Maxwell plus court favorise les r\'esonances d'ordre plus \'elev\'e. En cons\'equence, le proto-Mercure a moins de chances
de finir captur\'e dans la r\'esonance 3:2 du fait qu'il a moins de chances de l'atteindre. Ceci semble \^etre un argument en faveur d'un Mercure froid, mais 
n'oublions pas que les r\'esultats sont statistiques. Ce n'est pas parce qu'un scenario est peu probable qu'il ne s'est pas produit.

  \section[Revisite du scenario 2]{Revisite du scenario 2 : le noyau de Mercure existait d\'ej\`a}
  
  \par
\begin{figure}[ht]
  \centering
  \includegraphics[width=.6\textwidth]{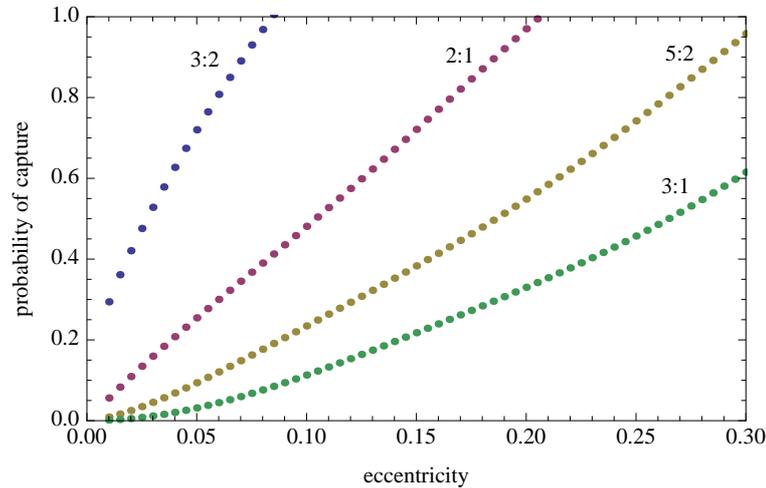}
  \caption[Probabilit\'es de capture avec la friction noyau-manteau]{Probabilit\'es de capture dans les r\'esonances spin-orbite pour un grand noyau 
  v\'erifiant $C_m/C=0.5$, et un coefficient de friction $k$ tel que $k/C_m=10^{-5}$.\label{fig:probavaleri}}
  \end{figure}

 Si Mercure a subi un impact qui aurait expuls\'e le manteau d'\'el\'ements l\'egers, alors il aurait \'et\'e suffisamment chaud pour cr\'eer une interface
  fluide-solide rapidement. Il faut donc tenir compte de la friction noyau-manteau dans l'\'evolution de la rotation. Valeri Makarov a fait une \'etude 
  analytique du probl\`eme, en g\'en\'eralisant le formalisme de \citep{gp1967} \`a tout type de couple de mar\'ee, pour un fluide laminaire. Il en a d\'eduit
  une formule de probabilit\'e de capture dont l'application num\'erique est donn\'ee Fig.\ref{fig:probavaleri}.
  
  \par On constate, comme \citep{pb1977} et \citep{cl2009}, que la consid\'eration de la friction noyau-manteau augmente consid\'erablement les probabilit\'es
  de capture. \`A l'excentricit\'e actuelle, la capture dans la r\'esonance 2:1 est certaine, rendant la r\'esonance 3:2 impossible. Il faut donc envisager
  que l'excentricit\'e de Mercure ait \'et\'e plus petite.

  \section[Revisite du scenario 3]{Revisite du scenario 3 : une r\'esonance synchrone destabilis\'ee}
  
  \par Ce scenario, issu de \citep{wcllr2012}, consid\`ere que la r\'epartition des crat\`eres \`a la surface de Mercure est la preuve d'une ancienne r\'esonance
  synchrone, qui aurait \'et\'e destabilis\'ee par un impact. Il faut donc examiner 2 probl\`emes :
  
  \begin{enumerate}
  
  \item Comment Mercure peut avoir \'et\'e en rotation synchrone,
  
  \item Comment un impact peut rompre cette rotation synchrone et provoquer la capture en r\'esonance 3:2.
  
  \end{enumerate}
  
  \par Avec le CTL, la rotation synchrone n'est possible que si Mercure est initialement r\'etrograde, l'\'etat d'\'equilibre pseudosynchrone emp\^echant la 
  r\'esonance synchrone si Mercure est initialement prograde. Ce n'est pas le cas avec notre mod\`ele, comme nous l'avons montr\'e dans les sections 
  pr\'ec\'edentes, o\`u la rotation synchrone est possible avec un Mercure prograde, et plus facile sans noyau et \`a faible excentricit\'e.
  
  \par
\begin{figure}[ht]
 	 \centering
 	 \includegraphics[width=.6\textwidth]{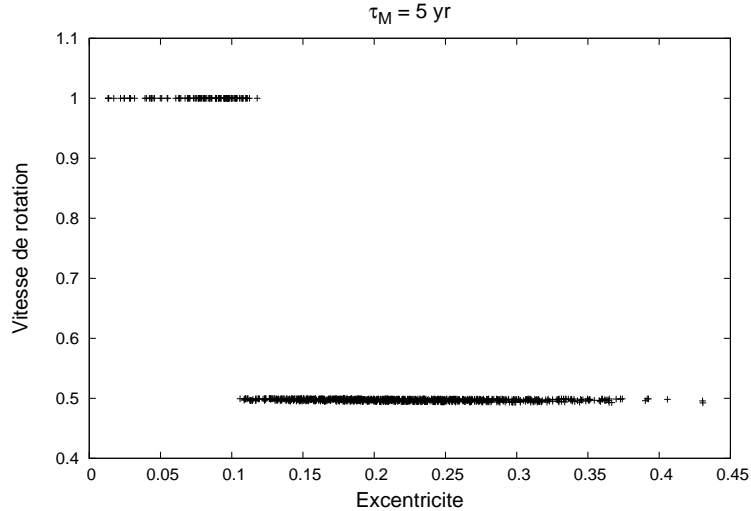}
 	 \caption[Rotation finale pour un proto-Mercure chaud initialement r\'etrograde]{Vitesses de rotation finales d'un proto-Mercure chaud 
 	 ($\tau_M=\tau_A=5$ ans). On observe que la r\'esonance 1:2 est certaine pour une excentricit\'e sup\'erieure \`a $0.12$. Lorsque le temps de
 	 Maxwell est plus long, on n'observe que la rotation synchrone.\label{fig:res12back}}
 \end{figure}

 Si le proto-Mercure \'etait initialement r\'etrograde, alors il a travers\'e des r\'esonances -5:2,-2:1,-3:2, et -1:1 avant de devenir prograde. J'ai 
  r\'ealis\'e des simulations num\'eriques en utilisant notre mod\`ele de mar\'ee et un proto-Mercure solide initialement r\'etrograde, et toutes mes
  trajectoires ont fini en rotation synchrone pour des temps de Maxwell de 15 et 500 ans. Pour $\tau_M=5$ ans, j'obtiens des captures en r\'esonance 1:2 
  pour une excentricit\'e sup\'erieure \`a $0.12$ (Fig.\ref{fig:res12back}).
  
 \par La question de l'impact est plus compliqu\'ee qu'avec le CTL. Avec le CTL, il suffisait que l'impact rompe la rotation synchrone, et l'\'etat 
 pseudosynchrone prenait le relais, pouvant \'eventuellement acc\'el\'erer la rotation de Mercure. Dans notre cas, un impact qui simplement romprait la rotation 
 synchrone verrait Mercure \^etre de nouveau captur\'ee en rotation synchrone, \`a moins que son excentricit\'e soit suffisamment importante pour 
 \^etre dans la zone d'acc\'el\'eration (Fig.\ref{fig:nopseudo}). Il faut donc, \`a excentricit\'e raisonnable, que l'impact soit suffisamment violent pour 
 atteindre la r\'esonance 3:2. Il faut \'egalement que la direction de
 l'impact soit favorable pour modifier la vitesse de la rotation pr\'ef\'erentiellement en longitude.
 
 \par Nous estimons qu'un tel impacteur aurait laiss\'e un bassin d'au moins 600 km de diam\`etre; il y a probablement environ une douzaine de crat\`eres de
 cette taille \`a la surface de Mercure. Ce scenario reste donc possible, m\^eme s'il est moins envisageable qu'avec le CTL.

  \section{Conclusion}
  
  \par Ce chapitre a propos\'e de revisiter le processus de capture de Mercure dans la r\'esonance spin-orbite 3:2. Nous avons vu l'importance du mod\`ele de
  mar\'ee, le populaire Constant Time Lag sous-estimant largement les probabilit\'es de capture et permettant des acc\'el\'erations fictives de la rotation.
  
  \par Nous avons examin\'e les 3 scenarii pr\'esents dans la litt\'erature : Mercure initialement solide et prograde, Mercure initialement diff\'erenci\'ee et
  prograde, et Mercure synchrone dans le pass\'e puis impact\'ee. Le cas d'un proto-Mercure initialement diff\'erenci\'e para\^it peu probable, car la r\'esonance
  3:2 n'est alors possible qu'\`a faible excentricit\'e. \citet{bwn2013} soutiennent d'ailleurs que Mercure \'etait initialement quasi-circulaire. La th\'eorie 
  de l'impact qui aurait destabilis\'e la rotation synchrone n'est pas impossible non plus, mais elle est moins probable qu'auparavant car elle n\'ecessite un plus gros
  impacteur et / ou une excentricit\'e plus grande.
  
  \par Nous pensons que le scenario le plus probable est celui d'un Mercure initialement froid, solide et prograde; dans ce cas la r\'esonance spin-orbite
  3:2 est l'\'etat final le plus probable. On a aussi une capture qui se produit en moins de 20 millions d'ann\'ees, ce qui est tr\`es rapide.
  
  \par Une extension de cette \'etude serait de partir d'un proto-Mercure tournant sur 3 axes. L'effet de l'obliquit\'e a \'et\'e abord\'e dans \citep{pb1977a} et 
  \citep{cl2010}, mais beaucoup reste \`a faire. 
  
\part{Un outil math\'ematique pour la rotation}

  \chapter[L'algorithme NAFFO]{Utilisation de l'analyse en fr\'equence pour simuler un \'equilibre dynamique\label{chap:naffo}}
  
  \section{Introduction}
  
  \par La mod\'elisation de la rotation r\'esonnante des corps du Syst\`eme Solaire revient \`a d\'ecrire un \'etat d'\'equilibre dynamique, qui est 
  math\'ematiquement proche d'une superposition d'oscillations. C'est en tout cas le cas pour la plupart des satellites naturels, o\`u leurs mouvements
  orbitaux peuvent \^etre d\'ecrits de fa\c{c}on quasip\'eriodique au moins sur 
  quelques si\`ecles \citep{ldv2006,vd1995}. Dans le cas de Mercure, cette description est valable en longitude, et nous avons utilis\'e une extrapolation 
  sinuso\"idale pour pouvoir aussi l'utiliser en obliquit\'e (Chap.\ref{chap:mercobliq}).
  
  \par Math\'ematiquement on peut \'ecrire les choses de la fa\c{c}on suivante : on part d'un syst\`eme r\'egi par l'\'equation diff\'erentielle ordinaire :
  
  \begin{equation}
  \label{eq:edo}
  \dot{\vec{X}}(t) = f(\vec{X})+g(\vec{X},t),
  \end{equation}
  avec $f:\mathbb{B}\subset \mathbb{C}^n \rightarrow \mathbb{C}^n$, $g:\mathbb{B}\subset \mathbb{C}^n \times \mathbb{R} \rightarrow \mathbb{C}^n$ est une 
  perturbation externe, et le vecteur d'\'etats $\vec{X}\in \mathbb{B}\subset \mathbb{C}^n$, $\mathbb{B}$ \'etant un ouvert de $\mathbb{C}^n$ contenant 
  l'\'equilibre qui nous int\'eresse. Nous supposons que chaque composante $x_i \in \mathbb{C}$ de $\vec{X}$ est quasip\'eriodique, c'est-\`a-dire peut
  s'exprimer sous la forme d'une somme convergente de mon\^omes trigonom\'etriques p\'eriodiques, autrement dit :
  
  \begin{equation}
  \label{eq:lesxi}
  x_i(t) = \sum_l a_l\exp(\imath f_lt),
  \end{equation}
  o\`u les $a_l$ sont des amplitudes complexes et les $f_l$ des fr\'equences r\'eelles. On peut s\'eparer ces fr\'equences $f_l$ en 2 groupes : des fr\'equences
  libres, ou propres, qui sont les seules existant dans le syst\`eme non perturb\'e, c'est-\`a-dire lorsque $g\equiv 0$, et les fr\'equences de for\c{c}age dues
  \`a la perturbation $g$.
  
  \par La l\'egitimit\'e de la d\'ecomposition (\ref{eq:lesxi}) est donn\'ee par 2 th\'eor\`emes dans le cadre Hamiltonien pour des syst\`emes quasi-int\'egrables. 
  D'une part, le th\'eor\`eme KAM\footnote{Kolmogorov - Arnold - Moser} \citep{a1963,m1962} garantit l'existence de tores quasip\'eriodiques invariants si la 
  partie int\'egrable de la fonction Hamiltonienne est non d\'eg\'en\'er\'ee, si la condition Diophantienne est v\'erifi\'ee\footnote{c'est-\`a-dire si on est 
  suffisamment loin de toute r\'esonance}, et si la perturbation est suffisamment petite. D'autre part, le th\'eor\`eme de Nekhoroshev \citep{n1977,n1979}
  donne un r\'esultat de stabilit\'e. Il prouve que les orbites d'un syst\`eme l\'eg\`erement perturb\'e restent proches de celle du syst\`eme non perturb\'e
  sur un temps exponentiellement long.
  
  \par Par exemple, dans la mod\'elisation de la dynamique de rotation d'un corps en r\'esonance spin-orbite, les fr\'equence de for\c{c}age viennent des 
  \'eph\'em\'erides orbitales, et les fr\'equences libres ou propres sont celles des petites oscillations autour de l'\'equilibre, not\'ees $\omega_{u,v,w,z}$
  dans les chapitres pr\'ec\'edents. De plus, des termes crois\'es apparaissent, combinaisons enti\`eres d'oscillations libres et forc\'ees. Nous consid\'erons
  ces termes comme libres car nous devons nous en d\'ebarrasser, ils ne sont pas pr\'esents dans la solution \`a \'energie minimale, correspondant \`a l'\'equilibre. 
  La solution que nous cherchons ne doit contenir que des termes forc\'es, il faut donc se rapprocher le plus possible des conditions initiales donnant une amplitude nulle aux oscillations libres.
  
  \par Une fa\c{c}on intuitive d'atteindre les conditions initiales est de proc\'eder it\'erativement \`a partir de l'analyse en fr\'equence. C'est-\`a-dire 
  analyser une solution obtenue num\'eriquement, identifier les termes libres, et les retirer des conditions initiales. Cette fa\c{c}on de faire n'est pas 
  nouvelle, nous l'avons trouv\'ee dans la litt\'erature d\`es 2000 dans le cadre d'une preuve du th\'eor\`eme KAM assist\'ee par ordinateur \citep{lg2000}.
  
  \par Ce que je pr\'esente ici est une tentative d'aller plus loin, en s'attaquant \`a la preuve de la convergence de l'algorithme. Ce travail, fait en 
  collaboration avec Nicolas Delsate et Timoteo Carletti \citep{ndc20xx}, n'a pour l'instant pas \'et\'e publi\'e autrement que sur arXiv, et est l'un de mes 
  travaux les plus cit\'es.
  
  \section{Notre algorithme}
  
  \par Je pr\'esente ici le principe de notre algorithme, visant \`a d\'eterminer la solution forc\'ee. Nous supposons ici qu'elle existe, nous ne cherchons
  pas \`a prouver son existence\footnote{C'est probablement ce qu'il manque \`a l'\'etude pour \^etre publi\'ee.}.
  
  \subsection{Principe}
  
  \par Consid\'erons un syst\`eme d\'ecrit par l'\'equation diff\'erentielle ordinaire 
  
  \begin{equation}
  \label{eq:edom}
  \dot{\vec{X}}(t) = f(\vec{X}),
  \end{equation}
  dans le voisinage d'une solution quasip\'eriodique de fr\'equences $(\omega_i)_{1\leq i\leq n}$, les fr\'equences libres du syst\`eme. Notons $\vec{\omega}$
  le vecteur de $\mathbb{R}^n$ dont les $\omega_i \in \mathbb{R}$ sont les composantes. Le vecteur d'\'etat $\vec{X}$ appartient \`a $\mathbb{C}^n$. Modifions 
  maintenant le syst\`eme en ajoutant un for\c{c}age quasip\'eriodique $g$ :
  
  \begin{equation}
  \label{eq:edo2}
  \dot{\vec{X}}(t) = f(\vec{X})+g(\vec{X},t),
  \end{equation}
  o\`u les fr\'equences de $g$ sont les composantes $(\nu_j)_{1\leq j\leq p}$ du vecteur $\vec{\nu} \in \mathbb{R}^p$. Les fonctions $f$ et $g$ sont d\'efinies
  d'un ouvert de $\mathbb{C}^n$ et de $\mathbb{C}^n\times\mathbb{R}$ respectivement, \`a valeurs dans $\mathbb{C}^n$, et $n$ et $p$ sont des entiers 
  strictement positifs.
  
  \par Consid\'erons qu'il n'y a pas de r\'esonance entre les $(\omega_i)$ et les $(\nu_j)$, autrement dit que ces fr\'equences sont lin\'eairement 
  ind\'ependantes par combinaisons enti\`eres. Appelons $\vec{\phi}(t;\vec{X}):\mathbb{R}\times\mathbb{C}^n\rightarrow\mathbb{C}^n$ la solution avec une 
  condition initiale $\vec{X}$. La d\'ecomposition quasip\'eriodique de $\phi$ s'\'ecrit
  
  \begin{equation}
  \label{eq:sol}
  \vec{\phi}\left(t;\vec{X}\right)=\sum_{\vec{l}\in\mathbb{Z}^n,\vec{m}\in\mathbb{Z}^p}\vec{\phi}_{\vec{l}\vec{m}}\left(\vec{X}\right)e^{\imath (\vec{\omega}\cdot\vec{l} +\vec{\nu}\cdot\vec{m})t}\, ,
  \end{equation}
  on peut la r\'e\'ecrire en s\'eparant les termes libres des termes forc\'es :
  
  \begin{eqnarray}
  \vec{\phi}\left(t;\vec{X}\right) & = & \sum_{\vec{m}\in\mathbb{Z}^p}\vec{\phi}_{\vec{0},\vec{m}}\left(\vec{X}\right)e^{\imath\vec{\nu}\cdot\vec{m}t}+\sum_{\vec{l}\ne \vec{0}, \vec{m}\in\mathbb{Z}^p}\vec{\phi}_{\vec{l},\vec{m}}\left(\vec{X}\right)e^{\imath(\vec{\omega}\cdot\vec{l}+\vec{\nu}\cdot\vec{m}) t} \nonumber \\
                                   & =: & \vec{S}(t;\vec{X})+\vec{L}(t;\vec{X})\, \label{eq:sol2} ,
  \end{eqnarray}
  les fonctions $S$ et $L$ \'etant d\'efinies d'un ouvert de $\mathbb{R}\times\mathbb{C}^n$ \`a valeurs dans $\mathbb{C}^n$. 
  
  \par L'utilisation de l'algorithme se base sur l'hypoth\`ese de l'existence d'une orbite $(2\pi/\nu_j)_{1\leq j\leq p}$-quasip\'eriodique, donc appartenant
  \`a un tore d'ordre $p$, compos\'ee uniquement de fr\'equences de for\c{c}age. Cette hypoth\`ese peut de fa\c{c}on \'equivalente s'\'ecrire comme l'existence
  d'une condition initiale $\vec{X}_{\infty}$ telle que la solution associ\'ee est $(2\pi/\nu_j)_{1\leq j\leq p}$-quasip\'eriodique, donc
  
  \begin{equation}
  \label{eq:phiqp}
  \vec{\phi}\left(t;\vec{X}_{\infty}\right)=\vec{S}(t;\vec{X}_{\infty}),
  \end{equation}
  ou encore $\vec{L}(t;\vec{X}_{\infty})\equiv \vec{0}$.
  
  \par Le principe de notre algorithme est le suivant :
  
  \begin{enumerate}
  
    \item Prenons la condition initiale $\vec{X}_0$ relativement proche de $\vec{X}_{\infty}$,
    
    \item Int\'egrons num\'eriquement l'\'equation diff\'erentielle ordinaire (\ref{eq:edo2}) pour d\'eterminer $\vec{\phi}\left(t;\vec{X}\right)$ et ensuite
    d\'eterminer la d\'ecomposition quasi-p\'eriodique $\vec{\phi}\left(t;\vec{X}_0\right)=\vec{S}(t;\vec{X}_0)+\vec{L}(t;\vec{X}_0)$,
    
    \item Posons $\vec{X}_1=\vec{S}(0;\vec{X}_0)$ et r\'eit\'erons le proc\'ed\'e en utilisant $\vec{X}_1$ en lieu et place de $\vec{X}_0$,
    
    \item De cette fa\c{c}on on construit une suite $(\vec{X}_n)$ d\'efinie par la relation de r\'ecurrence suivante
    
    \begin{equation}
    \label{eq:iteration}
    \vec{X}_{n+1}=\vec{S}(0;\vec{X}_n),   
   \end{equation}
de telle fa\c{c}on que

    \begin{equation}
    \label{eq:limit1}
    \lim_{n \rightarrow \infty} \vec{X}_n = \vec{X}_{\infty},
    \end{equation}
    ou encore 
    
    \begin{equation}
    \label{eq:limit2}
    \lim_{n \rightarrow \infty} \vec{L}(t;\vec{X}_n) = \vec{0}.
    \end{equation}

  \end{enumerate}
  
  \par Cet algorithme a d\'ej\`a \'et\'e utilis\'e dans les circonstances suivantes :
  
  \begin{itemize}
  
    \item \citep{n2009} : Rotation synchrone de Callisto ($n=3$, $p=13$),
    
    \item \citep{dnrl2009} : Librations de Mercure en r\'esonance 3:2 ($n=2$, $p=5$),
  
    \item \citep{clcmu2010} : Dynamique orbitale d'un syst\`eme exoplan\'etaire ($n=1$, $p=3$),
    
    \item \citep{hl2010,hsl2014} : Dynamique orbitale des ast\'ero\"ides troyens,
    
    \item \citep{ndl2010} : Rotation de Mercure avec interactions noyau-manteau ($n=4$, $p=5$),
    
    \item \citep{rrc2011} : Rotation en longitude de Janus et \'Epim\'eth\'ee ($n=1$, $p=6$),
    
    \item \citep{d2011} : R\'esonances gravitationnelles\footnote{Ce sont des r\'esonances entre la rotation du primaire et l'orbite du secondaire.} d'un 
    orbiteur autour de Vesta ($n=1$, $p=3$),
    
    \item \citep{nkr2011} : Rotation rigide de Mimas ($n=3$, $p=5$),
    
    \item \citep{n2012} : Mod\`ele de Poincar\'e-Hough pour la rotation d'un satellite synchrone fictif ($n=4$, $p=3$),
    
    \item \citep{n2013} : Mod\`ele de Poincar\'e-Hough pour Io ($n=4$, $p=13$),
    
    \item \citep{nd2012,nl2013} : Extrapolation sinuso\"idale de l'obliquit\'e de Mercure ($n=1$, $p=4$),
    
    \item \citep{nn2014} : Rotation de Titan \`a 3 couches ($n=6$, $p=9$).
    
  \end{itemize}
  
  \par Cet algorithme est d'autant plus facile \`a utiliser que les fr\'equences propres varient peu au voisinage de l'\'equilibre, ce qui est particuli\`erement
  vrai pour les r\'esonances spin-orbite. Cela l'est moins pour les r\'esonances de moyen mouvement.

  \subsection{L'algorithme NAFF}
  
  \par Notre algorithme n\'ecessite l'utilisation d'un algorithme d'analyse en fr\'equence le plus pr\'ecis possible, qui donnera la d\'ecomposition
  quasip\'eriodique (\ref{eq:sol}). Nous avons utilis\'e l'algorithme Numerical Analysis of the Fundamental Frequencies (NAFF) \citep{l1993,l2005}, dont
  voici le principe.
  
  \par Le but est d'\'ecrire un signal complexe quasip\'eriodique $x(t)$ sous la forme
  
  \begin{equation}
  \label{eq:naff}
  x(t) \approx \sum_{l=1}^Na_l^{\bullet}\exp\left(\imath \nu_l^{\bullet}t\right),
  \end{equation}
  sur l'intervalle de temps $[-T;T]$ o\`u $\nu_l^{\bullet}$ et $a_l^{\bullet}$ sont les fr\'equences r\'eelles et amplitudes complexes d\'etermin\'ees 
  num\'eriquement. Les fr\'equences et amplitudes sont d\'etermin\'ees par un processus it\'eratif. Pour d\'eterminer la premi\`ere fr\'equence 
  $\nu_1^{\bullet}$, qui en th\'eorie doit correspondre \`a l'amplitude de module le plus important, on cherche le maximum de la quantit\'e
  
  \begin{equation}
  \label{eq:philas}
  \phi(\nu)=\langle x(t),\exp(\imath \nu t)\rangle,
  \end{equation}
  o\`u le produit scalaire $\langle f(t),g(t)\rangle$ est d\'efini par
  
  \begin{equation}
  \label{eq:prodscal}
  \langle f(t),g(t)\rangle =\frac{1}{2T}\int_{-T}^T f(t)\,\overline{g(t)}\,\chi(t)\, dt, 
  \end{equation}
  $\overline{g(t)}$ \'etant le complexe conjugu\'e de $g(t)$ et $\chi(t)$ est une fonction de poids, c'est-\`a-dire une fonction positive qui v\'erifie
  
  \begin{equation}
  \label{eq:poids}
  \frac{1}{2T}\int_{-T}^T \chi(t)\,dt=1.
  \end{equation}

  \par Laskar conseille d'utiliser
  
  \begin{equation}
  \label{eq:chip}
  \chi(t)=\frac{2^p(p!)^2}{(2p)!}\Big(1+\cos (\pi t)\Big)^p,
  \end{equation}
  o\`u $p$ est un entier positif. En pratique, l'algorithme est plus efficace avec $p=1$ ou $p=2$. Nous avons utilis\'e $p=2$.
  
  \par Une fois que le premier terme p\'eriodique $a_1\exp(\imath\nu_1t)$ est trouv\'e, il est retir\'e de la fonction $x(t)$, et l'algorithme est r\'eit\'er\'e
  pour trouver le terme suivant. L'algorithme s'arr\^ete lorsqu'une fr\'equence d\'etect\'ee est tr\`es proche d'une autre d\'ej\`a d\'etect\'ee, ou lorsque
  le nombre de termes d\'etect\'es atteint un maximum pr\'ed\'efini par l'utilisateur.
  
  \par Plusieurs m\'ethodes existent dans la litt\'erature pour am\'eliorer la pr\'ecision de l'algorithme. Notamment, lorsque 2 fr\'equences sont s\'epar\'ees
  par moins de 2 fois la fr\'equence fondamentale, correspondant \`a la longueur de l'intervalle sur lequel le signal est analys\'e, alors leurs oscillations
  peuvent bruiter la d\'etermination de chacune. Laskar recommande d'orthogonaliser au fur et \`a mesure la base sur laquelle le signal est exprim\'e, alors 
  que \citet{c1998} recommande de r\'eit\'erer num\'eriquement la d\'etermination de chaque terme ind\'ependamment des autres. \citet{s2014} a montr\'e que
  ces 2 m\'ethodes \'etaient math\'ematiquement \'equivalentes. \citet{sn1997} ont propos\'e une correction analytique des termes d\'etermin\'es.

  \section{Approches de la preuve}
  
  \par Je pr\'esente ici des id\'ees de preuve de convergence de l'algorithme, dans un contexte assez g\'en\'eral, puis dans un cadre Hamiltonien, dont je 
  montre qu'il n\'ecessite moins d'hypoth\`eses. 
  
  \par La convergence de l'algorithme peut \^etre prouv\'ee en montrant la convergence de la suite $(\vec{X}_k)_{k \in \mathbb{N}}$ (\'Eq.\ref{eq:iteration}).
  Si $\vec{X}_k$ converge vers $\vec{P} \in \mathbb{C}^n$, alors en r\'e\'ecrivant l'\'Eq.(\ref{eq:iteration}) de la fa\c{c}on suivante
  
  \begin{equation}
  \label{eq:iteration2}
  \vec{X}_{k+1}=\vec{S}(0;\vec{X}_k)=\vec{\phi}(0;\vec{X}_k)-\vec{L}(0;\vec{X}_k)=\vec{X}_k-\vec{L}(0;\vec{X}_k),   
\end{equation}
la continuit\'e de la fonction $\vec{L}$, garantie par l'\'Eq.\ref{eq:sol2} ainsi que l'hypoth\`ese que la suite $(\vec{X})_k$ converge donnent

\begin{equation}
  \vec{0}=  \lim_{n\rightarrow \infty}\left(\vec{X}_{n+1}-\vec{X}_n\right)= -\lim_{n\rightarrow\infty}\vec{L}(0;\vec{X}_n)=-\vec{L}(0;\vec{P}),
\end{equation}
alors, par unicit\'e de la solution $(2\pi/\nu_j)_{1\leq j\leq p}$-quasip\'eriodique, on a $\vec{P}=\vec{X}_{\infty}$.

  \subsection{Dans un cadre g\'en\'eral}
  
  \par Nous avons besoin de l'hypoth\`ese suppl\'ementaire suivante : les matrices jacobiennes $\Sigma$ et $\Lambda$ des fonctions $\vec{S}$ et $\vec{L}$
  v\'erifient 
  
  \begin{equation}
  \label{eq:hypo}
   \lim_{\vec{X}\rightarrow \vec{X}_{\infty}}\left[\Lambda(\vec{X})\right]^{-1}\Sigma(\vec{X})=\mathbb{O}_n,
  \end{equation}
  o\`u $\mathbb{O}_n$ est la matrice nulle de dimension $n\times n$. Nous allons voir qu'il s'agit d'une condition suffisante pour la convergence de notre 
  algorithme.
  
  \par \`A partir de la d\'efinition (\ref{eq:iteration}) et de la d\'efinition premi\`ere de $\vec{X}_{k+1}$ comme condition initiale d'une orbite, nous 
  avons
  
  \begin{equation}
  \label{eq:sls}
  \vec{S}(0;\vec{X}_k)= \vec{X}_{k+1}=\vec{\phi}(0;\vec{X}_{k+1})=\vec{S}(0;\vec{X}_{k+1})+\vec{L}(0;\vec{X}_{k+1}).
  \end{equation}
  Cette relation d\'efinit implicitement l'application qui lie les 2 it\'erations successives $\vec{X}_k$ et $\vec{X}_{k+1}$. Appelons la $\vec{F}$, on a alors
  $\vec{X}_{k+1}=\vec{F}(\vec{X}_k)$. Nous supposons depuis le d\'ebut l'existence d'une orbite $(2\pi/\nu_j)_{1\leq j\leq p}$-quasip\'eriodique issue de la
  condition initiale $\vec{X}_{\infty}$, ceci revient \`a supposer que $\vec{X}_{\infty}$ est un point fixe de $\vec{F}$, autrement dit
  
  \begin{equation}
  \label{eq:pointfixe}
  \vec{F}\left(\vec{X}_{\infty}\right) = \vec{X}_{\infty}.
  \end{equation}
  Ainsi, prouver que la suite $(\vec{X}_k)$ converge vers $\vec{X}_{\infty}$ revient \`a prouver que le point fixe de $\vec{F}$ $\vec{X}_{\infty}$ est un 
  attracteur, donc que chaque valeur propre de la matrice jacobienne de $\vec{F}$ $\Phi$ \'evalu\'ee en $\vec{X}_{\infty}$ a un module inf\'erieur \`a $1$.
  
  \par On peut r\'e\'ecrire la relation (\ref{eq:sls}) sous la forme
  
  \begin{equation}
  \label{eq:sls2}
  \vec{S}(0;\vec{F}(\vec{X}_{k}))+\vec{L}(0;\vec{F}(\vec{X}_{k}))=\vec{S}(0;\vec{X}_k),
\end{equation}
puis en diff\'erentier chaque composante $S_l$ ($1\leq l \leq n$) par rapport aux $x_i$ ($1\leq i \leq n$), composantes de $\vec{X}_k$:

\begin{equation}
  \label{eq:sls3}
  \frac{\partial S_l}{\partial x_i}=\sum_{h=1}^n\left(\frac{\partial S_l}{\partial x_h}\frac{\partial F(\vec{X_k})_h}{\partial x_i}+\frac{\partial L_l}{\partial x_h}\frac{\partial F(\vec{X_k})_h}{\partial x_i}\right).
\end{equation}
On peut \'ecrire cette derni\`ere relation \`a l'aide de matrices jacobiennes, c'est-\`a-dire

\begin{equation}
  \label{eq:sls4}
  \Sigma(\vec{X}_k)=\Sigma\left(\vec{F}(\vec{X}_k)\right)\Phi(\vec{X}_k)+\Lambda\left(\vec{F}(\vec{X}_k)\right)\Phi(\vec{X}_k),
\end{equation}
ce qui donne

\begin{eqnarray}
  \Phi(\vec{X}_k) & = & \left(\Sigma\left(\vec{F}(\vec{X}_k)\right)+\Lambda\left(\vec{F}(\vec{X}_k)\right)\right)^{-1}\Sigma(\vec{X}_k) \nonumber \\
                  & = & \left(\Sigma\left(\vec{F}(\vec{X}_k)\right)+\Lambda\left(\vec{F}(\vec{X}_k)\right)\right)^{-1}\Lambda\left(\vec{F}(\vec{X}_k\right)\left[\Lambda\left(\vec{F}(\vec{X}_k\right)\right]^{-1}\Sigma(\vec{X}_k).
\end{eqnarray}
Il en d\'ecoule directement, avec l'aide de l'hypoth\`ese (\ref{eq:hypo}) et en gardant \`a l'esprit que $\vec{X}_{\infty}$ est un point fixe de $\vec{F}$, que 

\begin{equation}
  \lim_{\vec{X}_k\rightarrow \vec{X}_{\infty}}\Phi(\vec{X}_k)=\vec{0},
\end{equation}
donc les valeurs propres de la matrice $\Phi(\vec{X}_{\infty})$ ont un module inf\'erieur \`a $1$.
  
  \subsection{Dans un contexte Hamiltonien}
  
  \par Consid\'erons un syst\`eme Hamiltonien \`a $(n+p)$ degr\'es de libert\'e $\mathcal{H}(y_i,x_i,\Lambda_j,\lambda_j)$ avec $1\leq i\leq n$ et 
  $1\leq j\leq p$, o\`u les $y_i$ et $\Lambda_j$ sont les variables d'action et les $x_i$ et $\lambda_j$ les angles. Le for\c{c}age est contenu dans les 
  $\Lambda_j$ et $\lambda_j$. Supposons de plus que le syst\`eme peut \^etre localement d\'ecrit par un oscillateur harmonique perturb\'e\footnote{Ce qui est
  classiquement le cas pour un probl\`eme de r\'esonance spin-orbite, cf. \'Eq.\ref{eq:hamilkcallisto}.} :
  
  \begin{equation}
  \label{eq:birkhoff}
  \mathcal{H}(\vec{U},\vec{u},\vec{\Lambda},\vec{\lambda})=\vec{\omega}\cdot \vec{U} + \epsilon \mathcal{H}_1(\vec{U},\vec{u},\vec{\Lambda},\vec{\lambda}),
\end{equation}
avec $\vec{U}=\left(U_i\right)_{1\leq i\leq n}$, $\vec{u}=\left(u_i\right)_{1\leq i\leq n}$, $\vec{\Lambda}=\left(\Lambda_j\right)_{1\leq j\leq p}$, et 
$\vec{\lambda}=\left(\lambda_j\right)_{1\leq j\leq p}$. Les composantes du vecteur $\vec{\omega}=\left(\omega_i\right)_{1\leq i\leq n}$ sont les fr\'equences
(ou pulsations) des oscillations propres, et $\epsilon \mathcal{H}_1$ est une petite perturbation qui contient le for\c{c}age. On est ici dans un cas d\'eg\'en\'er\'e,
l'existence de tores invariants peut n\'eanmoins \^etre \'etudi\'ee gr\^ace \`a la th\'eorie de Birkhoff \citep{m1968}. Le Hamiltonien $\mathcal{H}$ a \'et\'e
obtenu plusieurs transformations canoniques, notamment une visant \`a supprimer les termes crois\'es de la forme $y_jy_k$ avec $j\ne k$ \citep{hl2005}, et la
transformation polaire canonique classique :

\begin{eqnarray}
  x_i  & = & \sqrt{2U_iZ_i}\sin u_i, \nonumber \\
  y_i  & = &  \sqrt{2U_i/Z_i}\cos u_i, \label{eq:polar}
\end{eqnarray}
o\`u $Z_i$ est une constante, choisie de fa\c{c}on \`a ce que les termes de degr\'e 1 en $\sqrt{U_i}$ disparaissent. La transformation (\ref{eq:polar}) peut 
\^etre vue comme l'expression des variables $x_i$ et $y_i$ en l'absence du for\c{c}age perturbatif $\epsilon\mathcal{H}_1$, c'est-\`a-dire

\begin{eqnarray}
\begin{cases}
  x_i(U_i,u_i,\Lambda_j=0,\lambda_j) & =  \sqrt{2U_iZ_i}\sin u_i, \nonumber \\
  y_i(U_i,u_i,\Lambda_j=0,\lambda_j) & =  \sqrt{2U_i/Z_i}\cos u_i\, . \label{eq:polar2}
\end{cases}
\end{eqnarray}
  
\par On peut voir dans l'expression de la transformation polaire (\ref{eq:polar}) la caract\'eristique de d'Alembert, qui dit notamment qu'une expression
contenant l'angle $ku_i$ o\`u $k$ est un entier doit avoir un pr\'efacteur dont la d\'ependance en $U_i$ est de la forme $U_i^{|k|/2}(1+\sum_j\alpha_jU_i^j)$.
\citet{h1974} a montr\'e que cette relation \'etait conserv\'ee \`a tout ordre si on r\'esolvait le probl\`eme par la m\'ethode des Transform\'ees de Lie 
\citep{d1969} si le Hamiltonien $\mathcal{H}$ est analytique\footnote{d\'eveloppable en s\'erie enti\`ere au voisinage de tout point de son 
ensemble de d\'efinition}. Consid\'erons ici que nous utilisons la m\'ethode des Transform\'ees de Lie pour r\'esoudre le probl\`eme.

\par Ainsi, toute fonction $x_q$ ou $y_q$ pour $1\leq q\leq n$, composante du vecteur d'\'etat $\vec{X}$ ou $\vec{Y}$, suit la caract\'eristique de d'Alembert
pour $(U_i,u_i)$, et donc peuvent s'\'ecrire sous la forme

\begin{equation}
  \label{eq:xfour}
  \begin{split}
  x_q(\vec{U},\vec{u},\vec{\Lambda},\vec{\lambda})=\sum_{\underset{\vec{k}\in\mathbb{Z}^n}{\vec{h}\in\mathbb{Z}^p}}\alpha_{\vec{k},\vec{h}}\prod_{i=1}^n U_i^{\frac{|k_i|}{2}}\Big(1+\sum_{\vec{m}\in\mathbb{N}^n}\gamma_{\vec{m},\vec{k},\vec{h}} \prod_{s=1}^{n} \sqrt{U_s}^{m_s}\Big) \\
\times \exp\Big(\imath\big(\sum_{i=1}^n k_iu_i+\sum_{j=1}^p h_j\lambda_j\big)\Big)
  \end{split}
\end{equation}
et

\begin{equation}
  \label{eq:yfour}
  \begin{split}
  y_q(\vec{U},\vec{u},\vec{\Lambda},\vec{\lambda})=\sum_{\underset{\vec{k}\in\mathbb{Z}^n}{\vec{h}\in\mathbb{Z}^p}}\beta_{\vec{k},\vec{h}}\prod_{i=1}^n U_i^{\frac{|k_i|}{2}}\Big(1+\sum_{\vec{m}\in\mathbb{N}^n}\delta_{\vec{m},\vec{k},\vec{h}} \prod_{s=1}^{n} \sqrt{U_s}^{m_s}\Big) \\
\times \exp\Big(\imath\big(\sum_{i=1}^n k_iu_i+\sum_{j=1}^p h_j\lambda_j\big)\Big),
  \end{split}
\end{equation}
o\`u $\alpha_{\vec{k},\vec{h}}$, $\beta_{\vec{k},\vec{h}}$, $\gamma_{\vec{m},\vec{k},\vec{h}}$ et $\delta_{\vec{m},\vec{k},\vec{h}}$ sont des constantes 
complexes, $m_i$, $k_i$ et $h_j$ sont des entiers, et $m=||\vec{m}||=\sum_s m_s$ est pair. On suppose que les membres de droite convergent absolument pour 
$U_i$ inf\'erieur \`a un certain $U_{i,0}$ strictement positif. On a ainsi

\begin{equation}
  \begin{split}
  x_q(t)=\sum_{\underset{\vec{k}\in\mathbb{Z}^n}{\vec{h}\in\mathbb{Z}^p}}\alpha_{\vec{k},\vec{h}}\prod_{i=1}^n U_i(t)^{\frac{|k_i|}{2}}\Big(1+\sum_{\vec{m}\in\mathbb{N}^n}\gamma_{\vec{m},\vec{k},\vec{h}} \prod_{s=1}^{n} \sqrt{U_s(t)}^{m_s}\Big) \\
\times \exp\Big(\imath\big(\sum_{i=1}^n k_iu_i(t)+\sum_{j=1}^p h_j\lambda_j(t)\big)\Big),
  \end{split}
\end{equation}

\begin{equation}
  \begin{split}
  y_q(t)=\sum_{\underset{\vec{k}\in\mathbb{Z}^n}{\vec{h}\in\mathbb{Z}^p}}\beta_{\vec{k},\vec{h}}\prod_{i=1}^n U_i(t)^{\frac{|k_i|}{2}}\Big(1+\sum_{\vec{m}\in\mathbb{N}^n}\delta_{\vec{m},\vec{k},\vec{h}} \prod_{s=1}^{n} \sqrt{U_s(t)}^{m_s}\Big) \\
\times \exp\Big(\imath\big(\sum_{i=1}^n k_iu_i(t)+\sum_{j=1}^p h_j\lambda_j(t)\big)\Big).
  \end{split}
\end{equation}
  
\par En supposant que l'ordre des Transform\'ees de Lie $M$ auquel on s'arr\^ete pour obtenir $x_q$ et $y_q$ est suffisamment grand pour que l'on puisse consid\'erer
les quantit\'es $U_i(t)$ comme des constantes comme ce serait le cas si $M$ \'etait $\infty$, on peut d\'ecomposer $x_q(t)$ de la fa\c{c}on suivante :

\begin{equation}
  \begin{split}
x_q(t)=\underbrace{\sum_{\vec{h}\in\mathbb{Z}^p}\alpha_{\vec{0},\vec{h}}\Big(1+\sum_{\vec{m}\in\mathbb{N}^n}\gamma_{\vec{m},\vec{0},\vec{h}} \prod_{s=1}^n \sqrt{U_s'}^{m_s}\Big)
\exp\Big(\imath\big(\sum_{j=1}^ph_j\lambda_j(t)\big)\Big)}_{S_q}+\\
\underbrace{\sum_{\underset{\vec{k}\in\mathbb{Z}^n\backslash\{\vec{0}\}}{\vec{h}\in\mathbb{Z}^p}}\alpha_{\vec{k},\vec{h}}\prod_{i=1}^n U_i'^{\frac{|k_i|}{2}}
\Big(1+\sum_{\vec{m}\in\mathbb{N}^n}\gamma_{\vec{m},\vec{k},\vec{h}} \prod_{s=1}^n \sqrt{U_s'}^{m_s}\Big)\exp\Big(\imath\big(\sum_{i=1}^n k_iu_i'+\sum_{j=1}^p h_j\lambda_j(t)\big)\Big)}_{L_q}\, ,
   \end{split}
\end{equation}
en une partie libre $L_q$ et une partie forc\'ee $S_q$. $y_q(t)$ peut \^etre d\'ecompos\'ee de la m\^eme mani\`ere. On voit directement

\begin{eqnarray}
S_q & \sim & A+\sum_{i,j}B_{i,j}\sqrt{U'_i}\sqrt{U'_j}+\ldots, \\
L_q & \sim & \sum_iC_i\sqrt{U'_i}+\ldots.
\end{eqnarray}

\par En appelant $\Sigma$ et $\Lambda$ les matrices jacobiennes des fonctions $\vec{S}$ et $\vec{L}$ limit\'ees aux variables d'actions $U'_i$ (ce qui est 
diff\'erent du cas non-Hamiltonien, puisqu'ici notre syst\`eme est d\'eg\'en\'er\'e), on a 

\begin{eqnarray}
  \Sigma & = & \left(\frac{\partial S_q}{\partial U_i'}\right)_{1\leq q,i\leq n}, \\
  \Lambda & = & \left(\frac{\partial L_q}{\partial U_i'}\right)_{1\leq q,i\leq n},
\end{eqnarray}
ce qui donne directement $\frac{\partial S_q}{\partial U_i'}=\mathcal{O}(1)$ et $\frac{\partial L_q}{\partial U_i'}=\mathcal{O}(1/\sqrt{U_i'})$, donc,
lorsque $U'_i$ tend vers $0$, $\frac{\partial S_q}{\partial U_i'}$ converge vers une valeur complexe finie tandis que $\frac{\partial L_q}{\partial U_i'}$
diverge vers l'infinie. Donc l'inverse de $\Lambda$, soit $\Lambda^{-1}$, converge vers la matrice nulle $\mathbb{O}_n$, ainsi que le produit 
$\Lambda^{-1}\Sigma$. Nous v\'erifions donc la condition suffisante (\ref{eq:hypo}) pour que notre algorithme converge.

  \subsection{Vitesse de convergence}
  
  \par Dans le cas o\`u on pourrait \'ecrire
  
  \begin{eqnarray}
  S(0;x) & \sim & x_{\infty}+a(x-x_{\infty})^{\alpha}+\ldots, \\
  L(0;x) & \sim & b(x-x_{\infty})^{\beta}+\ldots,
  \end{eqnarray}
  o\`u $\ldots$ signifierait \emph{termes d'ordre sup\'erieur}, pour $\alpha>\beta\geq0$, on a aussi la vitesse de convergence de notre algorithme. Si on pose 
  $\theta_n=x_n-x_{\infty}$, alors l'\'Eq.(\ref{eq:sls}) s'\'ecrit
  
  \begin{equation}
  a\theta_{n+1}^{\alpha}+b\theta_{n+1}^{\beta}\sim a\theta_{n}^{\alpha},
  \end{equation}
  ce qui peut \^etre r\'esolu de fa\c{c}on approximative pour $\theta_{n+1}$, par exemple en utilisant la formule d'inversion de Lagrange \citep{c2003}, ce qui
  donne
  
  \begin{equation}
   \theta_{n+1}\sim\left(\frac{a}{b}\right)^{1/\beta}\theta_n^{\alpha/\beta}+\dots,
  \end{equation}
  et sugg\`ere ainsi que notre algorithme converge \`a la vitesse $\alpha/\beta$. Il d\'ecoule de la r\`egle de d'Alembert que $\alpha=1$ et $\beta=1/2$,
  donc notre algorithme aurait une convergence quadratique dans le cas Hamiltonien.
  
  \par J'\'ecris bien \emph{aurait}, car il s'agit d'une vitesse de convergence th\'eorique. Elle suppose notamment que la pr\'ecision machine est infinie;
  en pratique elle est de l'ordre de $10^{-15}-10^{-16}$ en double pr\'ecision, qui est un ordre rapidement atteint \`a la vitesse quadratique, la pr\'ecision 
  machine se r\'ev\`ele donc plus limitante que la pr\'ecision math\'ematique.
  
  \section{2 exemples num\'eriques}
  
  \par Dans le cadre de cette \'etude, Nicolas Delsate a fait des essais num\'eriques pouss\'es pour illustrer notre algorithme et tenter de confirmer la 
  convergence quadratique. Des exemples sont pr\'esents tout au long de ce manuscrit, ici je me limiterai \`a 2 : la r\'esonance spin-orbite dans un cas 
  extr\^eme, car quasi-r\'esonnant, et un exemple de proie-pr\'edateur avec for\c{c}age p\'eriodique, dans un cadre non Hamiltonien.
  
  \subsection{La r\'esonance spin-orbite 1:1 en longitude}
  
  \par La th\'eorie associ\'ee \`a ce cas a d\'ej\`a \'et\'e largement d\'evelopp\'ee. Nous partons du Hamiltonien
  
  \begin{equation}
  \label{eq:alessandra}
  \begin{split}
    \mathcal{H}(y,x,\Lambda,\lambda)=\frac{y^2}{2}-\epsilon\Big(\alpha_1(e)\cos(2x-\lambda)+\alpha_2(e)\cos(2x-2\lambda) \\
    +\alpha_3(e)\cos(2x-3\lambda)+\alpha_4(e)\cos(2x-4\lambda)+\alpha_5(e)\cos(2x-5\lambda)\Big)+\Lambda
  \end{split}
\end{equation}
avec $\lambda=t$ et
\begin{eqnarray*}
  \alpha_1(e) & = & -\frac{e}{4}+\frac{e^3}{32} \quad , \quad
  \alpha_2(e) = \frac{1}{2}-\frac{5}{4}e^2+\frac{13}{32}e^4  \quad , \quad
  \alpha_3(e) =  \frac{7}{4}e-\frac{123}{32}e^3 \\
  \alpha_4(e) &=& \frac{17}{4}e^2-\frac{115}{12}e^4 \quad \text{et} \quad
  \alpha_5(e)  =  \frac{845}{96}e^3-\frac{32\,525}{1\,536}e^5.
\end{eqnarray*}
Nous fixons $\epsilon=0.3$ et $e=5.49\times 10^{-2}$ afin de se rapprocher d'une r\'esonance entre la fr\'equence orbitale (normalis\'ee \`a 1) et la fr\'equence
propre $\omega$. Un tel cas d\'efavorise la convergence de notre algorithme, ceci nous permet de maximiser le nombre d'it\'erations avant d'\^etre bloqu\'es
par la pr\'ecision machine, et donc d'\'etudier la fa\c{c}on dont il converge. Les \'equations issues de ce Hamiltonien sont

\begin{eqnarray}
  \frac{dy}{dt} & = & -\frac{\partial\mathcal{H}}{\partial x} \nonumber \\
                & = & -2\epsilon\Big(\alpha_1(e)\sin(2x-\lambda)+\alpha_2(e)\sin(2x-2\lambda)+\alpha_3(e)\sin(2x-3\lambda) \nonumber \\ 
                & + & \alpha_4(e)\sin(2x-4\lambda)+\alpha_5(e)\sin(2x-5\lambda)\Big), \label{eq:dy} \\
  \frac{dx}{dt} & = & \frac{\partial\mathcal{H}}{\partial y} = y, \label{eq:dx} \\
  \frac{d\Lambda}{dt} & = & -\frac{\partial\mathcal{H}}{\partial \lambda}, \\
  \frac{d\lambda}{dt} & = & \frac{\partial\mathcal{H}}{\partial \Lambda} = 1,
\end{eqnarray}
et bien s\^ur on n'int\`egre num\'eriquement que les 2 donnant $y$ et $x$. On est dans un cas \`a 2 degr\'es de libert\'e, l'un libre et l'autre forc\'e, la 
fr\'equence de for\c{c}age \'etant ici normalis\'ee \`a $1$. On a donc $n=p=1$. En d\'efinissant l'argument r\'esonnant $\sigma = x-\lambda$, une Transform\'ee
de Lie \`a l'ordre 2 donne les expressions approch\'ees suivantes :

  \begin{eqnarray}
\label{eq:sigmaqres}
  \sigma & \approx & (-0.159\,123+1.015\,533\,U)\sin\lambda \nonumber \\
  & & +(-4.534\,205\dexp{-3}+1.511\,041\dexp{-2}U)\sin 2\lambda \nonumber \\
  & & +(-2.633\,611\dexp{-4}+1.140\,088\dexp{-3}U)\sin 3\lambda \nonumber \\
  & & +(-1.055\,803\dexp{-5}+5.937\,715\dexp{-5}U)\sin 4\lambda \nonumber \\
  & & +(-1.119\,934\dexp{-7}+1.662\,371\dexp{-6}U)\sin 5\lambda \nonumber \\
  & & +(-2.529\,386\dexp{-9}+2.976\,143\dexp{-8}U)\sin 6\lambda \nonumber \\
  & & +1.627\,894\sqrt{U}\sin u +2.058\,034\dexp{-3}U\sin 2u \label{eq:sigmaqr} \\
  & & +2.976\,133\dexp{-2}\sqrt{U}\sin(\lambda+u) +0.142\,827\sqrt{U}\sin(\lambda-u) \nonumber \\
  & & +4.307\,322\dexp{-3}\sqrt{U}\sin(2\lambda+u)-2.413\,618\dexp{-2}\sqrt{U}\sin(2\lambda-u) \nonumber \\
  & & +2.894\,153\dexp{-4}\sqrt{U}\sin(3\lambda+u)-7.262\,346\dexp{-4}\sqrt{U}\sin(3\lambda-u) \nonumber \\
  & & +1.396\,981\dexp{-5}\sqrt{U}\sin(4\lambda+u)-2.614\,574\dexp{-5}\sqrt{U}\sin(4\lambda-u) \nonumber \\
  & & +2.043\,814\dexp{-7}\sqrt{U}\sin(5\lambda+u)-1.071\,171\dexp{-6}\sqrt{U}\sin(5\lambda-u) \nonumber \\
  & & -3.552\,194\dexp{-2}U\sin(\lambda+2u) +0.685\,948\,U\sin(\lambda-2u) \nonumber \\
  & & -2.153\,015\dexp{-3}U\sin(2\lambda+2u)-1.915\,010\dexp{-2}U\sin(2\lambda-2u) \nonumber \\
  & & -1.827\,299\dexp{-4}U\sin(3\lambda+2u)+5.436\,345\dexp{-4}U\sin(3\lambda-2u) \nonumber \\
  & & -1.021\,550\dexp{-5}U\sin(4\lambda+2u)+3.627\,801\dexp{-5}U\sin(4\lambda-2u) \nonumber \\
  & & +\mathcal{O}\left(U^{3/2}\right), \nonumber
\end{eqnarray}
et
\begin{eqnarray}
\label{eq:yqres}
  y & \approx & 1+(-0.159\,123+1.015\,533\,U)\cos\lambda \nonumber \\
  & & +(-9.068\,410\dexp{-3}+3.022\,082\dexp{-2}U)\cos 2\lambda \nonumber \\
  & & +(-7.900\,833\dexp{-4}+3.420\,266\dexp{-3}U)\cos 3\lambda \nonumber \\
  & & +(-4.223\,213\dexp{-5}+2.375\,086\dexp{-4}U)\cos 4\lambda \nonumber \\
  & & +(-5.599\,670\dexp{-7}+8.311\,856\dexp{-6}U)\cos 5\lambda \nonumber \\
  & & +(-1.517\,631\dexp{-8}+1.392\,351\dexp{-7}U)\cos 6\lambda \nonumber \\
  & & +1.230\,493\sqrt{U}\cos u +3.176\,270\dexp{-3}U\cos 2u  \label{eq:yqr} \\
  & & +5.272\,743\dexp{-2}\sqrt{U}\cos(\lambda+u) +3.261\,102\dexp{-2}\sqrt{U}\cos(\lambda-u) \nonumber \\
  & & +1.193\,850\dexp{-2}\sqrt{U}\cos(2\lambda+u)-2.964\,706\dexp{-2}\sqrt{U}\cos(2\lambda-u) \nonumber \\
  & & +1.091\,580\dexp{-3}\sqrt{U}\cos(3\lambda+u)-1.618\,286\dexp{-3}\sqrt{U}\cos(3\lambda-u) \nonumber \\
  & & +6.665\,941\dexp{-5}\sqrt{U}\cos(4\lambda+u)-8.440\,693\dexp{-5}\sqrt{U}\cos(4\lambda-u) \nonumber \\
  & & +1.179\,623\dexp{-6}\sqrt{U}\cos(5\lambda+u)-4.529\,258\dexp{-6}\sqrt{U}\cos(5\lambda-u) \nonumber \\
  & & -9.034\,478\dexp{-2}U\cos(\lambda+2u) -0.372\,710\,U\cos(\lambda-2u) \nonumber \\
  & & -7.628\,889\dexp{-3}U\cos(2\lambda+2u)-8.744\,867\dexp{-2}U\cos(2\lambda-2u) \nonumber \\
  & & -8.302\,064\dexp{-4}U\cos(3\lambda+2u)+7.918\,844\dexp{-4}U\cos(3\lambda-2u) \nonumber \\
  & & -5.662\,812\dexp{-5}U\cos(4\lambda+2u)+8.912\,232\dexp{-5}U\cos(4\lambda-2u) \nonumber \\
  & & +\mathcal{O}\left(U^{3/2}\right), \nonumber
\end{eqnarray}
avec 
\begin{equation}\label{eq:omegaqr}
  \omega=0.739\,732\,052\,1,
\end{equation}
$U$ \'etant l'amplitude de l'oscillation libre, que l'on souhaite voir nulle. Dans ce cas les conditions initiales id\'eales sont
\begin{equation}\label{eq:condqr} 
  \sigma(0)  =  0 \quad \text{et} \quad y(0)  =  0.830\,975\,124\,73\,. 
\end{equation}
  
\par La Tab.\ref{tab:quasireson} donne le comportement de notre algorithme. Les \'equations ont \'et\'e int\'egr\'ees \`a l'aide d'un int\'egrateur de
Bulirsh-Stoer, sur un intervalle \'egal \`a 238 p\'eriodes de for\c{c}age. Les calculs ont \'et\'e faits en Fortran 90, compil\'es avec un compilateur
Ifort 11.0. On peut remarquer plusieurs choses :

\begin{itemize}

  \item L'analyse en fr\'equence d\'etecte de moins en moins de termes \`a mesure que les it\'erations augmentent. C'est normal car la partie libre 
  du signal est de plus en plus faible.
  
  \item La position du terme de fr\'equence $\omega$ descend dans l'ordre des termes d\'etect\'es.
  
  \item La p\'eriode du terme libre, \'evoluant comme l'inverse de la fr\'equence, diminue \`a mesure que l'algorithme converge. C'est un r\'esultat classique
  pour les r\'esonances en m\'ecanique c\'eleste, o\`u la p\'eriode du terme r\'esonnant va tendre vers l'infini \`a mesure qu'on s'approche de la s\'eparatrice.
  
  \item On ne peut pas r\'eellement dire si la quantit\'e $U_{n+1}/U_n^2$ reste constante, donc si la convergence de l'algorithme est vraiment quadratique. 
  Le r\^ole de la pr\'ecision machine n'est sans doute pas \`a n\'egliger.

\end{itemize}

  \begin{landscape}
    \begin{table}[ht]
    \begin{center}
      \caption[Convergence de notre algorithme dans le cas quasi-r\'esonnant]{Convergence de notre algorithme dans le cas quasi-r\'esonnant. Les colonnes
      sont respectivement : nombre d'it\'erations n, variables, conditions initiales (C.I.), nombre de fr\'equences d\'etect\'ees, rang du terme libre
      dans la d\'etermination, d'amplitude \emph{Ampl.} et de fr\'equence d\'etermin\'ee num\'eriquement $\omega^{\bullet}$. La derni\`ere colonne
      permet de tester la convergence quadratique.\label{tab:quasireson}} 
      \begin{tabular}{ >{$}l<{$} c >{$}l<{$} | >{$}c<{$}>{$}c<{$} | >{$}c<{$} >{$}c<{$} | >{$}c<{$} >{$}c<{$}}
        \text{n} &  & \text{C.I.} & \# \text{freq.} & \text{Rang} & \text{Ampl.} & \omega^{\bullet} & U_{n+1}/U_n^2 (y) & U_{n+1}/U_n^2 (\sigma)\\
        \hline
        \hline
        \multirow{2}{*}{0} & y &  1.000\,000\,000 & 30 & 1 & 7.368\,530\dexp{-2} & 0.741\,860& \multirow{4}{*}{$0.481\,3$} & \multirow{4}{*}{$0.456\,3$}\\
        & $\sigma$ &   0.000\,000\,000 & 30 & 1 & 9.932\,502\dexp{-2} & 0.741\,860\\
        \cline{1-7}
        \multirow{2}{*}{1} & y &  0.851\,607\,082 & 21 & 2 & 6.122\,476\dexp{-3} & 0.747\,751\\
        & $\sigma$ & -3.170\,672\,140\dexp{-9} & 16 & 2 & 8.187\,849\dexp{-3}  & 0.747\,751& \multirow{2}{*}{$0.506\,6$} & \multirow{2}{*}{$0.495\,7$}\\
        \cline{1-7}
        \multirow{2}{*}{2} & y &  0.838\,740\,063 & 26 & 4 & 4.336\,535\dexp{-5} & 0.747\,791 \\
        & $\sigma$ & -9.855\,863\,550\dexp{-10} & 22 & 4 & 5.799\,125\dexp{-5} & 0.747\,791 & \multirow{2}{*}{$0.527\,4$} & \multirow{2}{*}{$0.516\,1$}\\
        \cline{1-7}
        \multirow{2}{*}{3} & y &  0.838\,648\,575 & 23 & 9 & 2.219\,788\dexp{-9} & 0.747\,791 \\
        & $\sigma$ & -1.102\,439\,070\dexp{-9} & 20 & 8 & 2.968\,459\dexp{-9} & 0.747\,791 & \multirow{2}{*}{--} & \multirow{2}{*}{--}\\
        \cline{1-7}
        \multirow{2}{*}{4} & y &  0.838\,648\,571 & 13 & 13 & 1.158\,263\dexp{-14} & 0.747\,791\\
        & $\sigma$ & -1.638\,333\,711\dexp{-14} & 12 & 12 & 1.548\,912\dexp{-14} & 0.747\,791
      \end{tabular}
    \end{center}
  \end{table}
  \end{landscape}
  
  \par
\begin{figure}[ht]
  \centering
  \includegraphics[width=.95\textwidth]{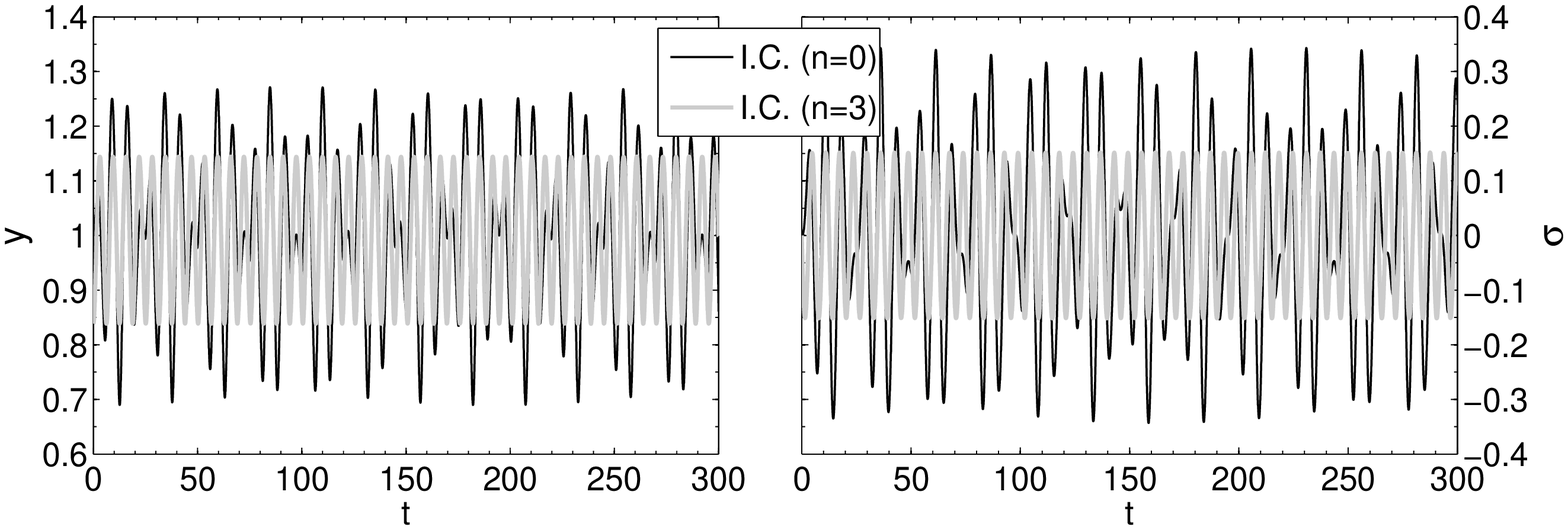}
  \caption[Convergence de notre algorithme pour la rotation synchrone quasi-r\'esonnante]{Illustration de la convergence de notre algorithme pour la rotation synchrone quasi-r\'esonnante. En noir on voit le r\'esultat de la premi\`ere
  int\'egration num\'erique, avec pour conditions initiales ($y=1$, $\sigma=0$), alors qu'en gris on voit le r\'esultat apr\`es 3 it\'erations de notre algorithme.
  Les oscillations \`a la fr\'equence $\omega$ ont disparu.\label{fig:QuasiResn}} 
\end{figure}

 La Fig.\ref{fig:QuasiResn} illustre graphiquement le r\'esultat.
  
  \subsection{Un syst\`eme proie-pr\'edateur avec for\c{c}age p\'eriodique}
  
  \par Pour les derniers calculs de cette th\`ese d'HDR, oublions l'astronomie, revenons sur Terre. Nous avons tenu \`a montrer que notre algorithme peut 
  s'utiliser dans un cadre tr\`es large, donc nous avons choisi un exemple de biologie math\'ematique, plus pr\'ecis\'ement un exemple de proie-pr\'edateur avec
  for\c{c}age p\'eriodique repris de \citep{bbgv1981}, qui utilise les c\'el\`ebres \'equations de Lotka-Volterra. Nous sortons aussi du cadre Hamiltonien.
  
  \par Les \'equations sont 
  
  \begin{eqnarray}
  \frac{dx_1}{dt} & = & \alpha x_1\big(1+\gamma\cos(2\pi t)-x_2-\eta x_1\big), \\
  \frac{dx_2}{dt} & = & \beta x_2(-1+x_1),
  \end{eqnarray}
  $\alpha$, $\beta$, $\gamma$ et $\eta$ \'etant des constantes strictement positives. L'id\'ee est la suivante : les pr\'edateurs $x_2$ (des loups) se nourrissent de
  proies $x_1$ (des lapins), qui eux-m\^eme ont besoin pour vivre d'une ressource (la carotte), dont la quantit\'e disponible varie p\'eriodiquement 
  (au cycle des saisons). Avec $\gamma=0$ (pas de for\c{c}age, syst\`eme autonome) et $\eta=0$, on peut prouver facilement l'existence de solutions p\'eriodiques
  de p\'eriode $2\pi/\sqrt{\alpha\beta}$ oscillant autour de l'\'equilibre $(x_1,x_2)=(1,1)$. Si $\eta>0$ l'\'equilibre devient $(x_1,x_2)=(1,1-\eta)$ et les 
  oscillations sont amorties. Si $\gamma\ne0$, le syst\`eme pr\'esente des oscillations $2\pi$-p\'eriodiques.
  
  \par Nous avons int\'egr\'e num\'eriquement les \'equations du syst\`eme avec dissipation ($\eta=0.025$), et sans ($\eta=0$). Pour ce dernier cas, nous avons 
  appliqu\'e notre algorithme (Tab.\ref{tab:preypredator}). Nous voyons que cet algorithme converge assez efficacement vers la solution forc\'ee.
  
  \begin{figure}[ht]
  \centering
  \includegraphics[width=\textwidth]{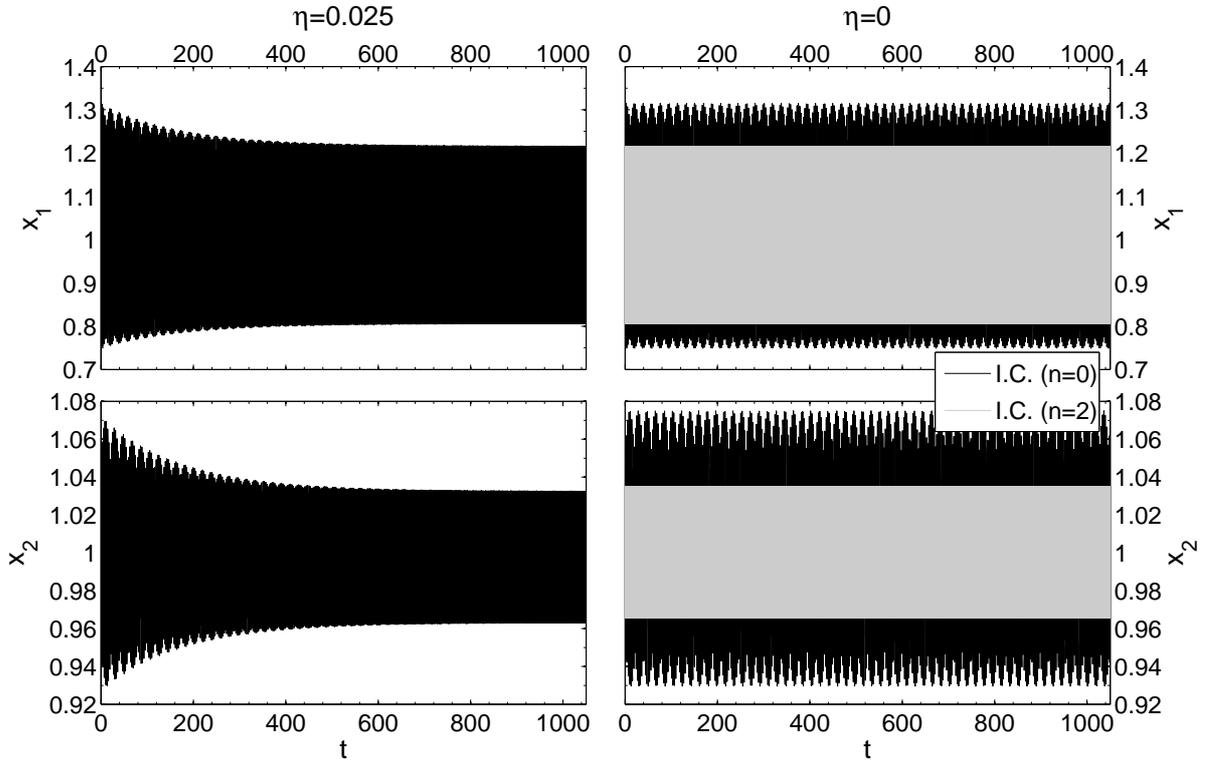}
  \caption[Int\'egrations num\'eriques du syst\`eme proie-pr\'edateur]{Int\'egrations num\'eriques du syst\`eme proie-pr\'edateur pour $\alpha=4.539$, 
  $\beta=1.068$ et $\gamma=0.25$, avec pour conditions initiales $x_1=1$ et $x_2=1-\eta$ (en noir), la p\'eriode propre \'etant $T=2.853\,74$. On voit un 
  amortissement des oscillations pour $\eta=0.025$. \`A droite, les courbes grises r\'esultent d'une int\'egration num\'erique des \'equations du syst\`eme
  avec des conditions initiales obtenues apr\`es 2 it\'erations de notre algorithme.\label{fig:x1x2}}
\end{figure}

\begin{table}[ht]
\centering
    \caption[Utilisation de notre algorithme sur le syst\`eme proie-pr\'edateur]{Utilisation de notre algorithme sur le syst\`eme proie-pr\'edateur, avec 
    $\alpha=4.539$, $\beta=1.068$, $\gamma=0.25$ et $\eta=0$.\label{tab:preypredator}} 
    \begin{tabular}{ >{$}l<{$} >{$}c<{$} >{$}l<{$} | >{$}c<{$} >{$}c<{$} | >{$}c<{$} >{$}c<{$} }
      \text{n} & & \text{I.C.} & \# \text{freq.} & \text{Rang} & \text{Ampl.} & \omega^{\bullet}\\
      \hline
      \hline
      \multirow{2}{*}{0} & x_1 & 1.000\,000\,000\,000\,000 & 50 & 2 & 3.831\,163\dexp{-2} & 2.206\,634\\
      & x_2 & 1.000\,000\,000\,000\,000 & 50 & 1 & 1.854\,280\dexp{-2} & 2.206\,634 \\
      \hline
      \multirow{2}{*}{1} & x_1 & 0.989\,166\,714\,745\,100 & 22 & 4 & 3.573\,335\dexp{-5} & 2.207\,483 \\
      & x_2 & 0.965\,514\,795\,157\,481 & 21 & 3 & 1.729\,063\dexp{-5} & 2.207\,483\\
      \hline
      \multirow{2}{*}{2} & x_1 & 0.989\,186\,585\,234\,344\,297\,1 & 26 & 6 & 4.508\,632\dexp{-9} & 2.207\,483 \\
      & x_2 & 0.965\,545\,142\,090\,197\,513\,7 & 26 & 6 & 2.181\,634\dexp{-9} & 2.207\,483\\
      \hline
      \multirow{2}{*}{3} & x_1 & 0.989\,186\,576\,347\,806\,470\,6 & 14 & 11 & 6.524\,090\dexp{-17} & 2.207\,483 \\
      & x_2 & 0.965\,545\,142\,191\,326\,750\,5 & 14 & 11 & 3.156\,872\dexp{-17} & 2.207\,483\\
      \hline
      \multirow{2}{*}{4} & x_1 & 0.989\,186\,576\,347\,806\,470\,2 & 14 & 11 & 6.503\,088\dexp{-17} & 2.207\,483 \\
      & x_2 & 0.965\,545\,142\,191\,326\,750\,4 & 14 & 11 & 3.146\,710\dexp{-17} & 2.207\,483\\
    \end{tabular}
\end{table}

  \section{Conclusion}
  
  \par Nous avons explor\'e la convergence d'un algorithme donnant un \'equilibre quasi-p\'eriodique, cas souvent rencontr\'e dans les probl\`emes spin-orbite.
  D'autres algorithmes existent, par exemple, dans le cadre de la dynamique galactique, \citep{rs2006,ro2008,ras2009,ra2011} proposent d'inclure une dissipation
  dans le syst\`eme, de le laisser \'evoluer lentement, d'obtenir ainsi de nouvelles conditions initiales qui seront r\'eutilis\'ees\ldots et ainsi d'approcher 
  it\'erativement l'\'equilibre. On peut difficilement, de cette fa\c{c}on, \'eviter de d\'eplacer l'\'equilibre, comme l'illustre le param\`etre $\eta$ dans
  notre probl\`eme proie-pr\'edateur. Dans un autre contexte, celui de l'obliquit\'e de Mercure, \citet{br2007} proposent d'ajuster un centre de libration \`a
  leur signal, qui servira de nouvelle condition initiale.
  
  \par En interne, nous appelons cet algorithme NAFFO, pour Numerical Algorithm For Forced Oscillations. Ce nom a l'avantage de commencer comme Namur, et il 
  rend hommage \`a l'algorithme NAFF, sans lequel notre algorithme n'existerait pas. Comme cet algorithme ne vient pas originellement de nous, nous n'avons pas 
  de l\'egitimit\'e pour le faire conna\^itre sous un nom que nous aurions nous-m\^eme choisi.
  
  \par Dans le Chapitre \ref{chap:mercobliq}, je pr\'esente un algorithme d'extrapolation sinuso\"idale des \'eph\'em\'erides orbitales. J'envisage de faire un
  jour une \'etude math\'ematique de cet algorithme, pour essayer de le publier en tant que tel.
  
\part{Conclusions}

\chapter{Conclusion}

\par Comment conclure un tel document? J'esp\`ere qu'il aura int\'eress\'e le lecteur. Il existe de nombreuses mani\`eres de r\'ediger une th\`ese d'Habilitation
\`a Diriger des Recherches, j'ai personnellement fait le choix de r\'ediger un document sur les r\'esonances spin-orbite, car mon int\'er\^et pour le sujet
a co\"incid\'e avec mon arriv\'ee \`a Namur. Notamment je ne m'y \'etais absolument pas int\'eress\'e au moment o\`u j'ai pass\'e ma th\`ese de doctorat, en 2005.
J'ai voulu y rassembler mes principaux r\'esultats sur le sujet, quitte \`a parfois tomber dans l'\'ecueil du catalogue. Mon id\'ee premi\`ere \'etait d'\'ecrire
un \emph{textbook} sur le sujet, t\^ache qui s'est av\'er\'ee trop ambitieuse\ldots pour l'instant. En tout cas, r\'ediger ce document m'a permis de prendre du 
recul sur mes travaux, qui devrait m'\^etre profitable.

\par Ma th\`ese de doctorat a port\'e sur la dynamique orbitale des satellites naturels, sujet sur lequel je travaille en fait encore, notamment dans le cadre 
du groupe de travail Encelade. Je travaille \'egalement sur les observations de ph\'enom\`enes mutuels, qui seront, je l'esp\`ere, nombreuses cette ann\'ee \`a Lille.

\par Ces ann\'ees de recherche se sont \'egalement accompagn\'ees d'enseignement et d'encadrement. J'\'etais notamment en charge, pendant 5 ans \`a l'Universit\'e 
de Namur, du cours de m\'ecanique c\'eleste de Master 1. J'ai \'egalement co-encadr\'e le stage de Master 1 de Nathalie B. \`a Lille, ainsi que les m\'emoires de
Master de Marianne L. \& Aur\'elie H \`a Namur. Mais ma plus grande exp\'erience d'encadrement, et ma plus grande fiert\'e pour l'instant, est la co-direction de la th\`ese
de doctorat d'\'Emilie Verheylewegen, soutenue \`a Namur le 4 avril 2014, sur la dynamique des satellites d'Uranus. J'ai longtemps cru que le rapport entre 
enseignant et \'etudiant \'etait \`a sens unique, je sais maintenant que c'est un \'echange, et toutes ces personnes m'ont beaucoup appris.


\end{document}